\documentclass[twoside,twocolumn,9pt]{article}
\usepackage{extsizes}
\usepackage[super,sort&compress,comma]{natbib} 
\usepackage[left=1.5cm, right=1.5cm, top=1.785cm, bottom=2.0cm]{geometry}
\usepackage{balance}
\usepackage{braket}
\usepackage{mathptmx}
\usepackage{sectsty}
\usepackage{amssymb}
\usepackage{graphicx} 
\usepackage{lastpage}
\usepackage[format=plain,justification=justified,singlelinecheck=false,font={stretch=1.125,small,sf},labelfont=bf,labelsep=space]{caption}
\usepackage{float}
\usepackage{fancyhdr}
\usepackage{fnpos}
\usepackage{caption}
\usepackage{subcaption}
\usepackage{graphicx}
\usepackage{amsmath}
\usepackage{geometry}
\usepackage{authblk}
\usepackage{hyperref}
\usepackage[T1]{fontenc}
\usepackage[utf8]{inputenc}
\usepackage{float}
\usepackage{csquotes}
\usepackage[version=4]{mhchem}
\usepackage{array}
\usepackage{droidsans}
\usepackage{charter}
\usepackage[T1]{fontenc}
\usepackage[usenames,dvipsnames]{xcolor}
\usepackage{setspace}
\usepackage[compact]{titlesec}
\usepackage{hyperref}
\hypersetup{colorlinks,
	citecolor=blue
}

\usepackage{epstopdf}

\definecolor{cream}{RGB}{222,217,201}

\begin{document}

\pagestyle{fancy}
\thispagestyle{plain}
\fancypagestyle{plain}{
\renewcommand{\headrulewidth}{0pt}
}

\makeFNbottom
\makeatletter
\renewcommand\LARGE{\@setfontsize\LARGE{15pt}{17}}
\renewcommand\Large{\@setfontsize\Large{12pt}{14}}
\renewcommand\large{\@setfontsize\large{10pt}{12}}
\renewcommand\footnotesize{\@setfontsize\footnotesize{7pt}{10}}
\makeatother

\renewcommand{\thefootnote}{\fnsymbol{footnote}}
\renewcommand\footnoterule{\vspace*{1pt}%
\color{cream}\hrule width 3.5in height 0.4pt \color{black}\vspace*{5pt}} 
\setcounter{secnumdepth}{5}

\makeatletter 
\renewcommand\@biblabel[1]{#1}            
\renewcommand\@makefntext[1]%
{\noindent\makebox[0pt][r]{\@thefnmark\,}#1}
\makeatother 
\renewcommand{\figurename}{\small{Fig.}~}
\sectionfont{\sffamily\Large}
\subsectionfont{\normalsize}
\subsubsectionfont{\bf}
\setstretch{1.125} 
\setlength{\skip\footins}{0.8cm}
\setlength{\footnotesep}{0.25cm}
\setlength{\jot}{10pt}
\titlespacing*{\section}{0pt}{4pt}{4pt}
\titlespacing*{\subsection}{0pt}{15pt}{1pt}

\fancyfoot{}
\fancyfoot[LO,RE]{\vspace{-7.1pt}\includegraphics[height=9pt]{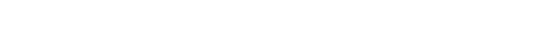}}
\fancyfoot[CO]{\vspace{-7.1pt}\hspace{13.2cm}\includegraphics{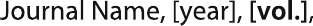}}
\fancyfoot[CE]{\vspace{-7.2pt}\hspace{-14.2cm}\includegraphics{head_foot/RF}}
\fancyfoot[RO]{\footnotesize{\sffamily{1--\pageref{LastPage} ~\textbar  \hspace{2pt}\thepage}}}
\fancyfoot[LE]{\footnotesize{\sffamily{\thepage~\textbar\hspace{3.45cm} 1--\pageref{LastPage}}}}
\fancyhead{}
\renewcommand{\headrulewidth}{0pt} 
\renewcommand{\footrulewidth}{0pt}
\setlength{\arrayrulewidth}{1pt}
\setlength{\columnsep}{6.5mm}
\setlength\bibsep{1pt}

\makeatletter 
\newlength{\figrulesep} 
\setlength{\figrulesep}{0.5\textfloatsep} 

\newcommand{\topfigrule}{\vspace*{-1pt}%
\noindent{\color{cream}\rule[-\figrulesep]{\columnwidth}{1.5pt}} }

\newcommand{\botfigrule}{\vspace*{-2pt}%
\noindent{\color{cream}\rule[\figrulesep]{\columnwidth}{1.5pt}} }

\newcommand{\dblfigrule}{\vspace*{-1pt}%
\noindent{\color{cream}\rule[-\figrulesep]{\textwidth}{1.5pt}} }

\makeatother

\twocolumn[
  \begin{@twocolumnfalse}
{\includegraphics[height=30pt]{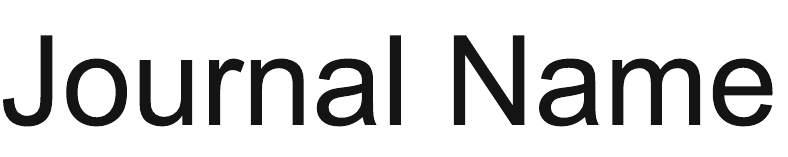}\hfill\raisebox{0pt}[0pt][0pt]{\includegraphics[height=55pt]{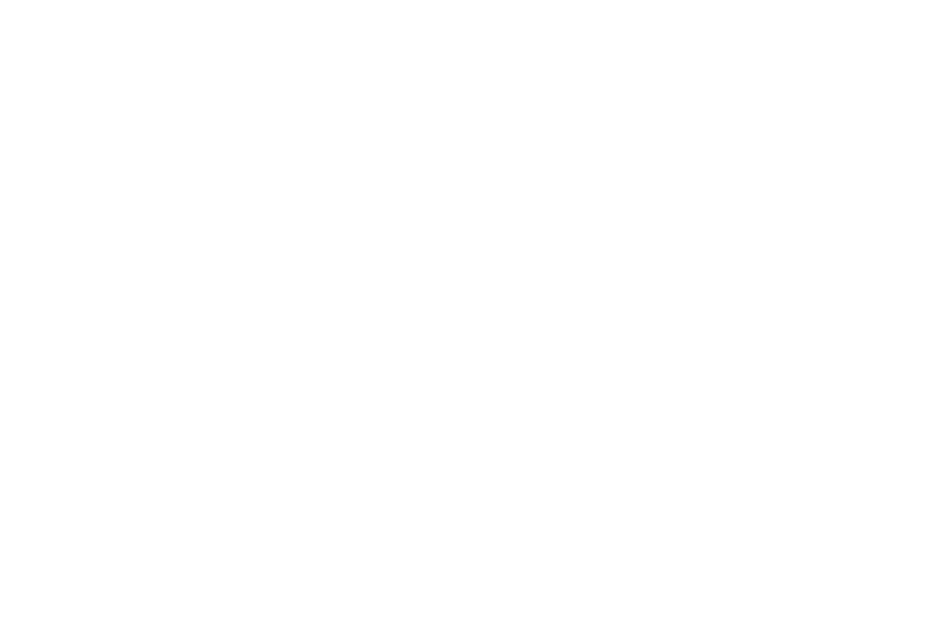}}\\[1ex]
\includegraphics[width=18.5cm]{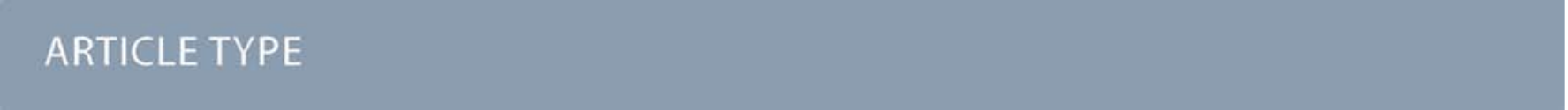}}\par
\vspace{1em}
\sffamily
\begin{tabular}{m{4.5cm} p{13.5cm} }

\includegraphics{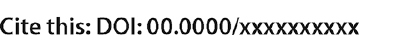} & \noindent\LARGE{\textbf{Quantum  Machine Learning for Chemistry and Physics}} \\
\vspace{0.3cm} & \vspace{0.3cm} \\

 & \noindent\large{Manas Sajjan,\textit{$^{a, d}$}
 Junxu Li\textit{$^{b, d\ddag}$}, Raja Selvarajan\textit{$^{b, d\ddag}$}, Shree Hari Sureshbabu\textit{$^{c, d\P}$}, Sumit Suresh Kale\textit{$^{a, d\P}$}, Rishabh Gupta\textit{$^{a, d\P}$}, Vinit Singh\textit{$^{a, d\P}$} and Sabre Kais$^{\ast}$\textit{$^{a,b,c,d}$}} \\

\includegraphics{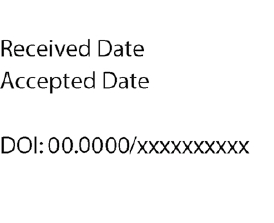} & \noindent\normalsize{Machine learning (ML) has emerged into formidable force for identifying hidden but pertinent patterns within a given data set with the objective of subsequent generation of automated predictive behavior. In the recent years, it is safe to conclude that ML and its close cousin deep learning (DL) have ushered unprecedented developments in all areas of physical sciences especially chemistry. Not only the classical variants of ML , even those trainable on near-term quantum hardwares have been developed with promising outcomes. Such algorithms have revolutionzed material design and performance of photo-voltaics, electronic structure calculations of ground and excited states of correlated matter, computation of force-fields and potential energy surfaces informing chemical reaction dynamics, reactivity inspired rational strategies of drug designing and even classification of phases of matter with accurate identification of emergent criticality. In this review we shall explicate a subset of such topics and delineate the contributions made by both classical and quantum computing enhanced machine learning algorithms over the past few years. We shall not only present a brief overview of the well-known techniques but also highlight their learning strategies using statistical physical insight. The objective of the review is to not only to foster exposition to the aforesaid techniques but also to empower and promote cross-pollination among future-research in all areas of chemistry which can benefit from ML and in turn can potentially accelerate the growth of such algorithms.} \\

\end{tabular}

 \end{@twocolumnfalse} \vspace{0.6cm}

  ]

\renewcommand*\rmdefault{bch}\normalfont\upshape
\rmfamily
\section*{}
\vspace{-1cm}


\footnotetext{\textit{$^{a}$~Department of Chemistry, Purdue University,
West Lafayette, IN-47907, USA}}
\footnotetext{\textit{$^{b}$~Department of Physics and Astronomy, Purdue University, West Lafayette, IN-47907, USA}}
\footnotetext{\textit{$^{c}$~Elmore Family School of Electrical and Computer Engineering, Purdue University, West Lafayette, IN-47907, USA}}

\footnotetext{\textit{$^{d}$~Purdue Quantum Science and
Engineering Institute, Purdue University, West Lafayette,
Indiana 47907, United States}}

\footnotetext{\ddag~These authors contributed equally to this work} 

\footnotetext{\P~These authors contributed equally to this work}

\footnotetext{$\ast$~Corresponding author. E-mail - kais@purdue.edu}


\tableofcontents

\section{Introduction}\label{Intro}

The 21st century data revolution sparked by machine learning (ML) have yielded unprecedented applications in several domains of technology like natural language processing \cite{li2017deep,torfi2021natural,nagarhalli2021impact}, translation \cite{wu2016google,lopez2008statistical}, autonomous vehicles \cite{tuncali2018simulation,janai2020computer,daily2017self}, robotics \cite{siau2018building,pierson2017deep}, image-recognition \cite{he2015delving, he2016deep,wu2015image}, recommender system \cite{covington2016deep}, web-searching \cite{pazzani1997learning} and fraudulent email filtering \cite{dada2019machine, guzella2009review} and in medical science like bio-informatics \cite{lengauer2007bioinformatics, larranaga2006machine}, medical imaging \cite{erickson2017machine}, brain-computer interfacing \cite{sakhavi2018learning} and in social sciences \cite{grimmer2021machine} and finance \cite{heaton2017deep} and even in problems like refugee integration \cite{bansak2018improving}. The primary reason for such prodigious advances is the uncanny ability of ML based protocols to detect and recognize unforeseen patterns in a data and integrate the acquired knowledge into decision-making, a process fancifully coined as `learning'. The fruitful use of this ability has been further accelerated by not only large-scale availability of shared databases and exponential growth of computing resources but also ingenuous algorithmic advances over the past few decades that precipitated in efficient dimensionality reduction \cite{van2009dimensionality} and data-manipulation. Needless to say that such a positively disruptive methodology have also fruitfully impacted several domains of physical sciences\cite{carleo2019machine}. Applications ranging from astronomy \cite{ball2010data, ivezic2014statistics}, particle-physics\cite{radovic2018machine}, atomic and molecular physics \cite{wigley2016fast}, optical manipulation of matter\cite{zhou2019emerging}, forecasting of weather patterns and climate dynamics \cite{chattopadhyay2020analog,ren2021deep} and even identification of evolutionary information from fossil records in paleontology \cite{monson2018using, spradley2019mammalian, romero2020improving} have been recorded with unforeseen success ratio. Chemical applications like understanding electronic properties of matter\cite{schleder2019dft,behler2016perspective}, material discovery with optimal properties \cite{von2020retrospective,liu2020machine}, retrosynthetic design and control of chemical reactions \cite{strieth2020machine, coley2018machine,fu2020optimizing,C7ME00107J}, understanding reaction pathways \cite{hu2018inclusion, amabilino2019training} on a potential energy surface, cheminformatics\cite{kumar2021exploiting} have been analyzed using the newly acquired lens of ML and continues to register a meteoric rise. Simulations performed in a recent review \cite{keith2021combining} bears testimony to this fact by highlighting that keywords based on ML have made steady appearances ($\ge 10^2$) across all divisions of chemistry over the last 20 years in technical journals of a particular publishing house. The number of such occurrences have specifically increased steadily for applications in physical chemistry/chemical physics. While ML based algorithms was enjoying this attention, much along the same time, the world was also witnessing the rapid emergence of another computing revolution which is fundamentally different from the familiar classical bit-based architecture. The new paradigm, called quantum computing\cite{nielson}, leverages the power of quantum parallelism and non-classical correlations like quantum entanglement to offer a platform that has shown algorithmic speed-up over the classical version in many instances \cite{bennett1993teleporting,harrow2004superdense,PhysRevLett.83.5162, shor1999polynomial, grover1996fast}. The natural question which has been posed in the community is whether quantum computing can also expand the horizon for predicting and identifying relevant features in a given data-set\cite{biamonte2017quantum}, lead to newer more efficient algorithms for machine learning or even record algorithmic speed-up for some of the established toolkits that are now routinely employed by ML practitioners in physics and chemistry\cite{wiebe2014quantum}. In this review, we shall try to explore this exciting intersection.

\subsection{Scope of the Review}

The scope and philosophy of this review would thus be the following
\begin{enumerate}
    \item Ref \cite{keith2021combining} highlights that a survey has indicated that ML algorithms are increasingly becoming opaque to human comprehension. We feel that a part of the reason for this is the under-emphasis on the various  methodologies that inform the basic building blocks of ML in recent reviews. While such topics are usually covered elaborately in data-science textbooks\cite{10.5555/1734076, Goodfellow-et-al-2016, ML_kubat_book} yet the latter resources lack domain-specific examples/applications which a new researcher in the field may find beneficial. Thus a holistic yet detailed account which focuses on both the basic tools used by ML practitioners as well as how such tools are enabling various physico-chemical applications, consolidated in one place for researchers to use synergistically is lacking. This review will try to address that gap. 
    
    \item We shall not only discuss the common tools as used by traditional ML practitioners in theoretical and computational physics and chemistry but also delineate the analogues of these algorithms trainable on a quantum computer. This will be attained in two steps. Firstly, we shall discuss the underlying theoretical framework of quantum versions of each of the vanilla ML algorithms in detail along with their classical counterparts. Secondly, the contributions made by both the classical and the quantum versions would be discussed separately while exploring each of the respective applications in subsequent parts of the review. To this end, it is important to clearly define certain terms which will set the tone for the review. All applications to be discussed in this review will entail deploying ML based algorithms on dataset involving features or representations of molecules/atoms and/or nanomaterials. Due to the specific nature of the data, we shall broadly call all such examples as instances of quantum machine learning (as is commonly done in this domain \cite{von2018quantum, Huang2018}) even if the analysis is performed on a classical computer. However to distinguish examples wherein quantum computers have been used as a part of the training process for the ML algorithm we shall specifically call such applications as `quantum computing enhanced'. To the best of our knowledge, explicating such quantum computing enhanced variants in the physico-chemical domain have not been attempted in any of the recent reviews which distinctively sets this one apart from the rest in terms of coverage and focus.
    
    \item We shall also discuss five different domains of physico-chemical applications which includes - tomographic preparation of quantum states in matter, classification of states and phases of matter, electronic structure of matter, force field parameterization for molecular dynamics and drug discovery pipeline. For each such application we shall discuss ML algorithms (both the classical and quantum computing enhanced variety) that has been successfully used in recent literature focusing on as many different architectures as possible. The objective of treating such a diverse portfolio of applications is to ensure the reader is aware of the many distinct domains in physical chemistry that have benefited immensely from ML over the past decade. Since the quantum-computing variants are still a nascent variety, bulk of the applications to be discussed will involve classical ML algorithms on quantum data even though the focus will certainly be how the capabilities in each domain can be augmented with the former in the arsenal. To the best of our knowledge, such a diverse and comprehensive portfolio of applications consolidated in one place have not been presented in any single review most of which have been topical and focused on a single domain only.
    It must also be emphasized that the aforesaid list is by no means exhaustive. Indeed we shall enlist several other domains later which have not been discussed in this review. Topical reviews on such applications will be duly referenced which the interested reader may consult.
    
    \item Lastly, another reason for the obscurity in the interpretation of machine learning algorithms especially those involving neural networks is the lack of clarity in the underlying learning process. Indeed, physicists and chemists are motivated to design computational tools which explicitly uses physical laws and scientifically guided domain intuition to understand a given natural phenomenon. However in most of machine learning algorithms, the models are initially agnostic to such physical principles. Instead they identify pertinent features and/or strategies directly from the data without the need for human intervention. While this process is intriguing, certain researchers may be reluctant to reap the full benefits of ML due to this fundamental difference in the operational paradigm. In this review we strive to address this issue by discussing several statistical physical tools which have been used in recent years to demystify the learning process. This is either completely absent or is less emphasized in recent reviews which we believe also fuels the increasing opacity as highlighted in Ref \cite{keith2021combining}

\end{enumerate}
\subsection{Organization of the Review}

The organization of the review is as follows. In Section \ref{Primer_QComp} we offer a glimpse of some basic notions in quantum computing to be used for understanding the subsequent portions of the review. In Section \ref{ML_toolkits} we discuss in detail each of the commonly used architectures in ML and DL (both the classical and the quantum computing enhanced variant). The basic theoretical framework discussed in this section for each of the methods will be frequently referred to the subsequently.
In Section \ref{Case_QML} we enlist and discuss in detail some of the recent reports wherein the power of quantum computers for machine learning tasks has been explicitly demonstrated or theoretically proven to be superior to classical models. In Section
\ref{Applications} we discuss applications of ML in five different domains of physics and chemistry. In Section \ref{Learnability} we discuss the several different models for explaining the learning mechanisms of deep learning algorithms using statistical physics. In Section \ref{conclusion} we conclude with an foray into emerging domains not discussed in this review.

\section{Short Primer on Quantum Computing}
\label{Primer_QComp}
In this section we shall discuss some of the basic terminologies and conceptual foundations of quantum computing that will be useful for the rest of the review. This is not only done for completeness but with the motivation that since quantum computing as a paradigm is relatively new, it may be unfamiliar to traditional ML practitioners and/or new entrants in the field. To appreciate the quantum analogue of commonly used machine learning algorithms, a basic understanding of some of the operational concepts and terms used in this domain would be beneficial. This section would attempt to familiarize the reader with that knowledge. 
We shall visit the common operational paradigms of computing using quantum devices that are widely used. Just as in classical computers where one has binary bits encoded as $\{0,1\}$ used for all logical operations, on a quantum computer the primary unit of information is commonly encoded within a qubit. To define a qubit, one would need two two-dimensional vectors commonly denoted as $|0\rangle$ and $|1\rangle$ and are referred to as computational basis states. Physically this two states can be the two hyperfine energy levels of an ion like in trapped-ion based quantum computing platforms\cite{haffner2008quantum, bruzewicz2019trapped} or can be energy levels corresponding to different number of Cooper pairs in a superconducting island created between a Josephson junction and a capacitor plate \cite{kjaergaard2020superconducting} or can be the highly excited electronic energy levels of Rydberg atom based cold atomic-ensembles \cite{saffman2010quantum, saffman2016quantum} or polar molecules in pendular states \cite{Wei2016QuantumCU,Karra2016ProspectsFQ,Impl_pend_gates, entangle_pen_states, polar_sym_top} to name a few. Mathematically the two states can be represented as $|0\rangle = \begin{pmatrix} 1 \:\: 0 \end{pmatrix}^T$ and $|1\rangle = \begin{pmatrix} 0 \:\: 1 \end{pmatrix}^T$ and collectively forms a basis for the two-dimensional state space ($\mathbb{H}$) of the system.

\subsection{Single qubit state} \label{single_qubs}
The state of the qubit in the two-dimensional basis of $(|0\rangle, |1\rangle)$ is defined by unit trace positive semi-definite operator (denoted as $\rho \in \mathcal{L(\mathbb{H})}$) as follows
\begin{eqnarray}
    \rho &= 
    (\frac{1}{2}+n_z) |0\rangle \langle 0| + (n_x + i n_y) |0\rangle \langle 1| \nonumber \\
    &+ (n_x - i n_y) |1\rangle \langle 0|  + (\frac{1}{2}-n_z) |1\rangle \langle 1|\label{rho_one_qubit}
\end{eqnarray}
wherein $\begin{pmatrix} n_x, n_y, n_z \end{pmatrix}^T \in \mathbb{R}^3$ and $i = \sqrt{-1}$ and the operators of the form $|i\rangle \langle j| \:\: \forall \:\: (i,j)\:\: \in (0,1)$ corresponds to familiar outer-product of two vectors \cite{nielson} . Positive semi-definiteness of the matrix in Eq.\ref{rho_one_qubit} guarantees that $n_x^2 + n_y^2 + n_z^2 \le 1$ which allows the vector $\begin{pmatrix} n_x, n_y, n_z \end{pmatrix}^T$ to reside within a Bloch sphere \cite{nielson}. 
For pure states which are defined by the additional idempotency constraint of $\rho^2 = \rho$, the inequality is saturated. One can then parameterize the $\begin{pmatrix} n_x, n_y, n_z \end{pmatrix}^T = \begin{pmatrix} \cos{\theta}\sin{\phi}, \sin{\theta}\cos{\phi}, \cos{\theta} \end{pmatrix}^T$ and
establish a bijective correspondence with a vector (say $|\psi\rangle \in \mathbb{H}$) defined as 
\begin{eqnarray}
    |\psi\rangle &=
    \cos(\frac{\theta}{2}) |0\rangle +
    e^{i\phi}\sin({\frac{\theta}{2}}) |1\rangle \label{psi_one_qubit}
\end{eqnarray}

\begin{figure}[ht!]
    \centering
    \includegraphics[width=0.50\textwidth]{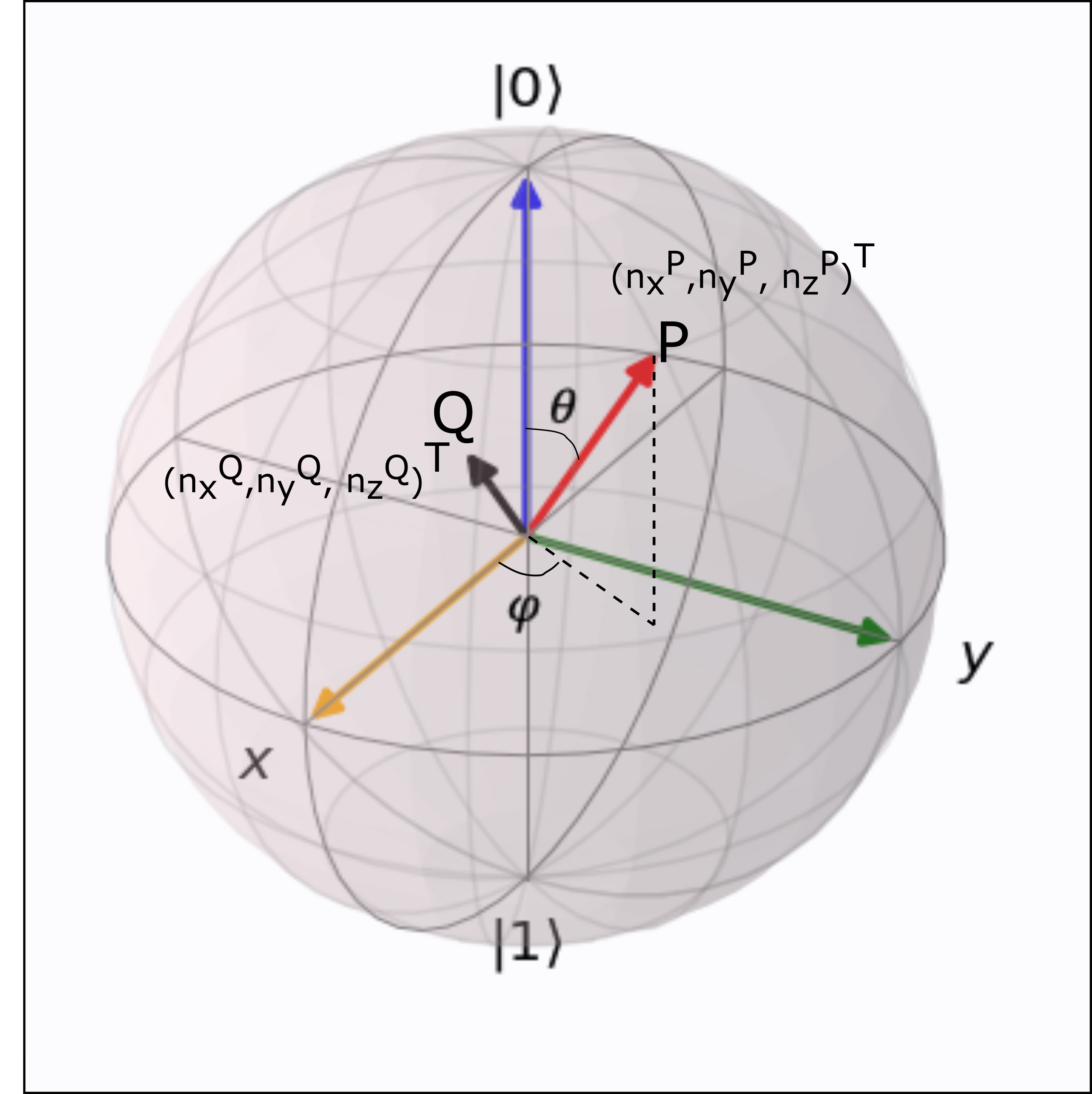}
    \caption{ The Bloch sphere (blue) and the parametric angles $\theta$ and $\phi$ as used in Eq. \ref{psi_one_qubit}. The point P marked in red lies on the surface of the Bloch sphere and has $(n_x, n_y, n_z)^T = \begin{pmatrix} \cos{\theta}\sin{\phi}, \sin{\theta}\cos{\phi}, \cos{\theta} \end{pmatrix}^T$. Such states can be represented as in Eq. \ref{psi_one_qubit}. On the other hand, states like point Q (marked in black) lies inside the Bloch sphere $n_x^2+ n_y^2 + n_z^2 \le 1$ and cannot be represented as in Eq. \ref{psi_one_qubit}. Only way to denote such states would be using Eq. \ref{rho_one_qubit}
    }
    \label{fig_bloch_sph}
\end{figure}
The parametric angles $\{\theta \in [0,\pi], \phi \in [0, 2\pi] \}$ are geometrically defined in the Bloch sphere in Fig.\ref{fig_bloch_sph}  
Such states of the system defined by Eq.\ref{psi_one_qubit} will be extensively used in this review and will be exclusively referred to for single-qubit states unless otherwise specified. For certain magnitudes of the parameter $\theta$ wherein both $(\cos{\frac{\theta}{2}}, \sin{\frac{\theta}{2}})$ acquires non-zero values, the state of the system in Eq.\ref{psi_one_qubit} is said to be in a  superposition of the two basis states. Realization of such superpositions presents one of the fundamental differences between qubit paradigm of computing and the bit paradigm of computing as used in classical processors. The parametric angle $\phi$ controls the relative phase difference between the computational basis in such superposition states. However a superposition even though is responsible for quantum parallelism would not survive a projective measurement protocol \cite{watrous2018theory, nielson}. Such measurements would collapse the state in Eq. \ref{psi_one_qubit} in either computational basis state $|0\rangle$ with probability $\cos^2{\frac{\theta}{2}}$ or in the computational basis state $|1\rangle$ with probability $\sin^2{\frac{\theta}{2}}$.

\subsection{Multi-qubit state} \label{multi_qubs}

For multiple qubits (say N), the corresponding state space is $\mathbb{H_A}\otimes\mathbb{H_B}\otimes\mathbb{H_C}....\mathbb{H_N}$ \cite{nielson}. One can thus define a computational basis using Kronecker product such as $|i_A\rangle \otimes |i_B\rangle ......|i_N\rangle$ where the labels $(A,B,C...N)$ are physically used to demarcate the state-space of each qubit. There are now $2^N$ basis states generated from two choices $(|0\rangle, |1\rangle)$ for each of $i_j\:\:, j \in \{A,B,...N\}$. Let us denote this set collectively as $\{|\xi_i\rangle\}_{i=0}^{2^N-1}$. For notational convenience such multi-qubit basis states will often be abbreviated in this review such as $|i_A, i_B....,i_N\rangle \equiv |i_A\rangle|i_B\rangle....|i_N\rangle \equiv |i_A\rangle \otimes |i_B\rangle ......|i_N\rangle$. A general state $\rho_{A,B,C...N} \in \mathcal{L}(\mathbb{H_A}\otimes\mathbb{H_B}\otimes\mathbb{H_C}....\mathbb{H_N})$ of the multi-qubit system would again correspond to a positive semi-definite operator with unit trace defined as
\begin{eqnarray}
    \rho_{A,B,C...N} = \sum_{i=0}^{2^n-1}\sum_{j=0}^{2^n-1} \rho_{ij} |\xi_i\rangle \langle\xi_j|  \label{many_qub_state_mixed}
\end{eqnarray}

where the elements $\rho_{ij} \in \mathbb{C}^2 \:\: \forall\:\: (i,j)$. 
One can also define a reduced state for each of sub-system qubits (say for the K-th qubit) through partial tracing of the state $\rho_{A,B,C...N}$ over computational basis states of the remaining qubits as follows
\begin{eqnarray}
    \rho_{K} = Tr_{A,B,...J, L....N} \:\:(\rho_{A,B,C...N}) \label{red_K_dens_op}
\end{eqnarray}
where $\rho_{K} \in \mathcal{L}(\mathbb{H}_{K})$. Such operations are completely positive trace-preserving (CPTP) maps and hence generates valid states\cite{watrous2018theory,nielson} of the sub-system (often called reduced density operator of $K$-th qubit). Just like in the case for single qubits, if the general state in Eq.\ref{many_qub_state_mixed} is pure ($\rho_{A,B,C...N}^2 = \rho_{A,B,C...N}$) one can associate a vector (say $|\psi\rangle_{A,B,C...N} \in \mathbb{H_A}\otimes\mathbb{H_B}\otimes\mathbb{H_C}....\mathbb{H_N}$) which in the multi-qubit computational basis is denoted as 
\begin{eqnarray}
    |\psi\rangle_{A,B,C...N} &=& \sum_{i=0}^{2^N-1} C_i |\xi_i\rangle \nonumber \\
    &=& \sum_{i_A=0}^{1}\sum_{i_B=0}^{1}...\sum_{i_N=0}^{1} C_{i_A i_B i_C...i_N} |i_A i_B ...i_N\rangle
    \label{state_many_qub}
\end{eqnarray}
The coefficients $C_{i_A i_B i_C...i_N} \in \mathbb{C}^2 \:\: \forall\: i_j\:\:, j \in \{A,B,...N\}$. For a normalized state as is usually considered in this review, $\sum_{i_A=0}^{1}\sum_{i_B=0}^{1}...\sum_{i_N=0}^{1}|C_{i_A i_B i_C...i_N}|^2 =1$.

\begin{table}[ht!]
\small
  \caption{Commonly used single and multi-qubit gates in quantum circuits and the corresponding matrix representations in the computational basis. }
  \label{gate_tab}
  \begin{tabular*}{0.48\textwidth}{@{\extracolsep{\fill}}lll}
    \hline
    Gate type & Number of qubit(s) & Matrix representation \\
    \hline\\
    $R_x(\theta)$ & 1 &  $\begin{pmatrix} \cos{\frac{\theta}{2}} & -i\sin{\frac{\theta}{2}} \\ -i\sin{\frac{\theta}{2}} & \cos{\frac{\theta}{2}}\end{pmatrix}$\\\\
     $R_y(\theta)$ & 1 &  $\begin{pmatrix} \cos{\frac{\theta}{2}} & \sin{\frac{\theta}{2}} \\ \sin{\frac{\theta}{2}} & \cos{\frac{\theta}{2}}\end{pmatrix}$\\\\
    $R_z(\theta)$ & 1 &  $\begin{pmatrix} e^{-i\frac{\theta}{2}} & 0 \\ 0 & e^{i\frac{\theta}{2}}\end{pmatrix}$\\\\
    X & 1 &  $\begin{pmatrix} 0 & 1 \\ 1 & 0\end{pmatrix}$\\\\
    Y & 1 &  $\begin{pmatrix} 0 & -i \\ i & 0 \end{pmatrix}$\\\\
    Z & 1 &  $\begin{pmatrix} 1 & 0 \\ 0 & -1\end{pmatrix}$\\\\
    H & 1 &  $\frac{1}{\sqrt{2}}\begin{pmatrix} 1 & 1 \\ 1 & -1\end{pmatrix}$\\\\
    $P(\alpha)$ & 1 & $\begin{pmatrix} 1 & 0 \\ 0 & e^{i\alpha}\end{pmatrix}$ \\\\
    $T=P(\frac{\pi}{4})$ & 1 & $\begin{pmatrix} 1 & 0 \\ 0 & e^{i\frac{\pi}{4}}\end{pmatrix}$\\\\
    $S=P(\frac{\pi}{2})$ & 1 & $\begin{pmatrix} 1 & 0 \\ 0 & i\end{pmatrix}$\\\\
    CNOT & 2 & $\begin{pmatrix} 1 & 0 & 0 & 0\\ 0 & 1 & 0 & 0\\ 0 & 0 & 0 & 1 \\ 0 & 0 & 1 & 0 
    \end{pmatrix}$\\\\
    CPHASE ($\alpha$) & 2 & $\begin{pmatrix} 1 & 0 & 0 & 0\\ 0 & 1 & 0 & 0\\ 0 & 0 & 1 & 0 \\ 0 & 0 & 0 & e^{i\alpha} \end{pmatrix}$\\\\
    SWAP & 2 & $\begin{pmatrix} 1 & 0 & 0 & 0\\ 0 & 0 & 1 & 0\\ 0 & 1 & 0 & 0 \\ 0 & 0 & 0 & 1 
    \end{pmatrix}$\\\\
    CZ = CPHASE($\pi$) & 2 & $\begin{pmatrix} 1 & 0 & 0 & 0\\ 0 & 1 & 0 & 0\\ 0 & 0 & 1 & 0 \\ 0 & 0 & 0 & -1 \end{pmatrix}$\\\\
    Toffoli & 3 & $\begin{pmatrix} 1 & 0 & 0 & 0 & 0 & 0 &0 & 0\\ 0 & 1 & 0 & 0 & 0 & 0 & 0 & 0\\ 0 & 0 & 1 & 0 & 0 & 0 & 0 & 0 \\ 0 & 0 & 0 & 1 & 0 & 0 & 0 & 0 \\
    0 & 0 & 0 & 0 & 1 & 0 & 0 & 0 \\ 0 & 0 & 0 & 0 & 0 & 1& 0 & 0\\ 0 & 0 & 0 & 0& 0 & 0 & 0 & 1 \\ 0 & 0 & 0 & 0 & 0 & 0 & 1 & 0
    \end{pmatrix}$\\\\
    \hline
  \end{tabular*}
\end{table}

Other than the possibility of superposition over all basis states similar to the case of single-qubit as discussed in previous section, it is also possible now to encounter a new phenomenon which has to do with non-classical correlation. The pure state in Eq.\ref{state_many_qub} will be termed separable if $\exists$ scalars $\zeta_{i_A}, \gamma_{i_B} .....\omega_{i_N}$ for each sub-system such that $C_{i_A i_B i_C...i_N} = \zeta_{i_A}\gamma_{i_B} .....\omega_{i_N} \:\: \forall\:\: (i_A, i_B...i_N)^T \:\in \{0,1\}^N$ i.e. if \textit{every} coefficient is multiplicatively factorizable into scalars characterizing the 2D basis states of each sub-system qubit \cite{Preskill_notes,watrous2018theory,nielson}. For such pure separable state it is possible to express Eq.\ref{state_many_qub} as $|\psi\rangle_{A,B,C...N} = |\phi_1\rangle_A \otimes |\phi_2\rangle_B ....\otimes |\phi_N\rangle_N$ wherein $|\phi_1\rangle_A \in \mathbb{H_A}$, $|\phi_2\rangle_B \in \mathbb{H_B}$,.... $|\phi_2\rangle_N \in \mathbb{H_N}$. If a state in Eq.\ref{state_many_qub} is not separable then it is said to be entangled which is a non-classical correlation.

\begin{figure*}[ht!]
    \centering
    \includegraphics[width=0.98\textwidth]{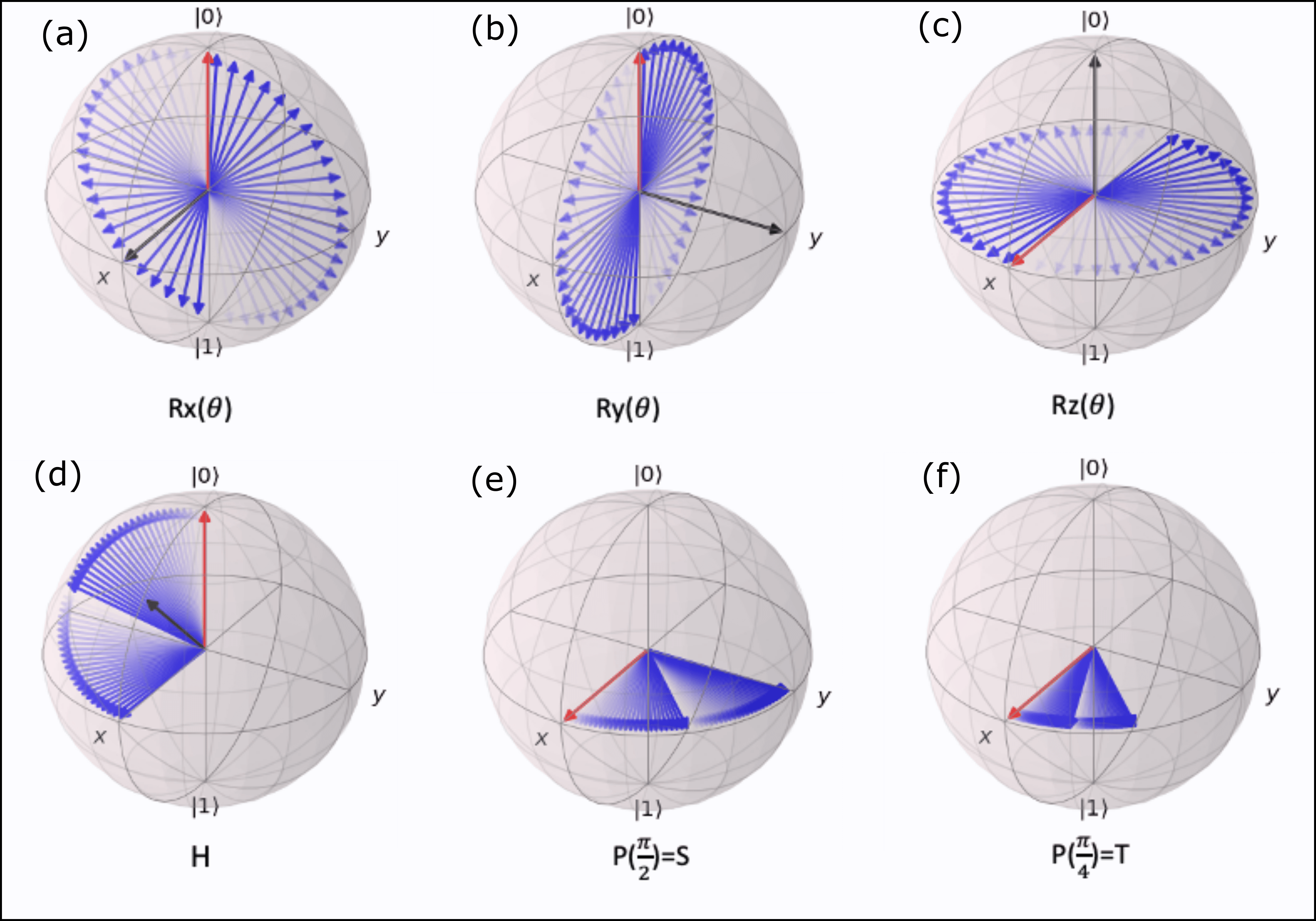}
    \caption{{\color{black}(a) The operation $R_x(\theta)$ which involves rotation about the x-axis (marked in black) as shown in the Bloch sphere. The initial state is $|0\rangle$ (marked in red). (b) The operation $R_y(\theta)$ which involves rotation about the y-axis (marked in black) as shown in the Bloch sphere. The initial state is $|0\rangle$ (marked in red) (c) Same as in (a), (b) but with $R_z(\theta)$ wherein the axis of rotation is z (marked in black). The initial state chosen here is (marked in red) $\frac{|0\rangle + |1\rangle}{\sqrt{2}}$ (d) Hadamard transformation of the initial state $|0\rangle$ (marked in red) as visualized in the Bloch sphere.
    The operation can be viewed as rotation around the axis $[\frac{1}{\sqrt{2}}, 0, \frac{1}{\sqrt{2}}]^T$ shown in black through an angle of $\pi$. Note that unlike the rotation gates in (a)-(c), Hadamard transformation do not have a variable user-defined angle of rotation and hence the final state starting from the said initial state is always fixed i.e. $\frac{|0\rangle + |1\rangle}{\sqrt{2}}$ (e) The transformation of the initial state $\frac{|0\rangle + |1\rangle}{\sqrt{2}}$ under phase-gate $S=P(\alpha = \frac{\pi}{2})$ (see Table \ref{gate_tab}). The operation produces a final state $\frac{|0\rangle + i|1\rangle}{\sqrt{2}}$ (f) The transformation of the initial state $\frac{|0\rangle + |1\rangle}{\sqrt{2}}$ under T-gate where $T=P(\alpha = \frac{\pi}{4})$ (see Table \ref{gate_tab}). The operation produces a final state $\frac{|0\rangle + e^{\frac{i\pi}{4}}|1\rangle}{\sqrt{2}}$. The matrix representations of the operators in (a)-(f) is given in Table.\ref{gate_tab}
    }}
    \label{fig_bloch_sph_single}
\end{figure*}

\begin{figure*}[ht!]
    \centering
    \includegraphics[width=0.75\textwidth]{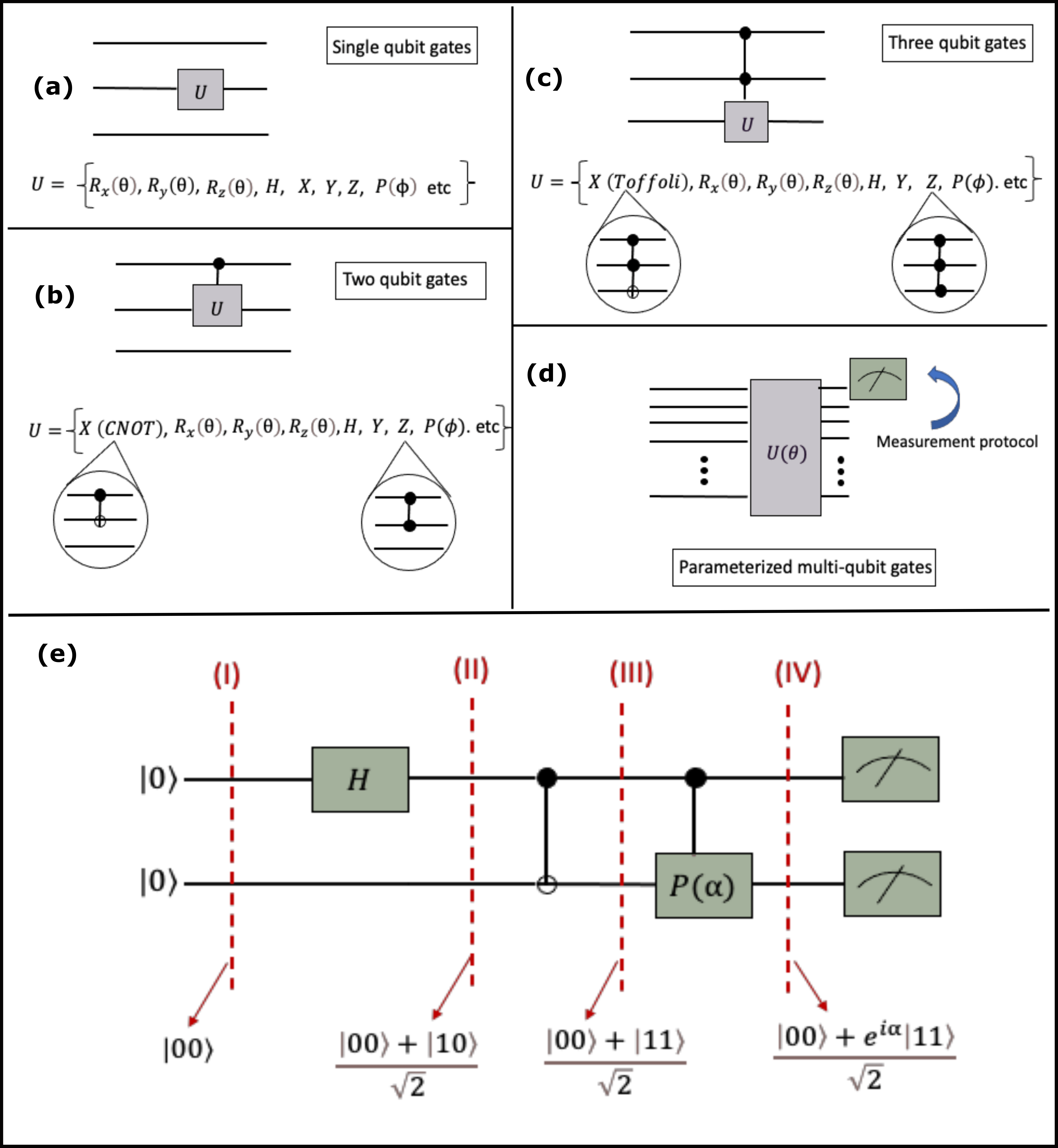}
    \caption{ Commonly used circuit representation of (a) 1-qubit gates (b) 2-qubit gates. Special gates in this category like CNOT and CZ gate have slightly different representation than the rest as has been highlighted within the oval windows. {\color{black} One must note that the solid dot indicates the control qubit and the hollow dot with a plus i.e $\oplus$ indicates the target qubit. Its the target qubit whose state is actually altered conditioned on the state of the control qubit being $|1\rangle$ in this case. The operation need not always be controlled on state $|1\rangle$ for the control qubit. Indeed two-qubit gates with the non-trivial operations on the target initiated by the control is $|0\rangle$ is also routinely used. (See text for more details)}. (c) 3-qubit gates. Special gates in this category like Toffoli gate and CCZ gate have slightly different representation than the rest as has been highlighted within the oval window. {\color{black} Similar interpretation as in (b) for the solid and hollow dots ($\oplus$) must be followed in terms of the control and target qubits.} (d) A generic n-qubit parameterized unitary. This is very often used to describe quantum circuits as we shall see later in the review. The explicit construction of gates in $U(\theta)$ is often omitted but is implied to be made up of elementary gates from (a)-(b) and occasionally even (c). The measurement protocol for any qubit will be denoted by boxes of the kind shown in green symbolically representing a monitoring device/meter. (e) A simple representative quantum circuit for the preparation of Bell state $\frac{|00\rangle + e^{i\alpha}|11\rangle}{\sqrt{2}}$. To facilitate interpretation of the circuit, the state of both the two-qubits is illustrated at junctions (I), (II), (III), (IV) after the operation of each elementary gate. To evaluate the states one can also use the matrix representation of the respective gates given in Table \ref{gate_tab} and apply it on the initial state $|00\rangle$ with the unitaries on the left acting first.}
    \label{fig_qcirc}
\end{figure*}

The presence of entanglement is another feature wherein computation using qubits can be different from that of the classical bit counterparts and is often leveraged in many different algorithms as a useful resource \cite{bennett1993teleporting,harrow2004superdense}. Similar to that of the case of a single qubit, the probabilistic interpretation of a projective measurement on the state in Eq.\ref{state_many_qub} is retained with the probability of collapsing onto a computational basis state $|i_A i_B ...i_N\rangle$ is $|C_{i_A i_B i_C...i_N}|^2$. Unless, otherwise stated by multi-qubit states in this review we shall almost always exclusively mean pure states of the kind given in Eq.\ref{state_many_qub}. {\color{black} Such states as we shall see can not only provide an efficient representation of the many-body states of any interacting quantum system in quantum simulations of stationary/time-independent processes but also for real and imaginary time evolution \cite{PRXQuantum.2.010342, PhysRevA.105.012412} in quantum dynamics either through Lie-Trotter Suzuki expansion \cite{PhysRevA.104.052603} or through variational frameworks \cite{PhysRevLett.125.010501}.}

\subsection{Quantum gates and Quantum circuit based paradigm} \label{q_gates}

Now that we know how to define quantum states of single and many qubits, it is important to learn how such states are transformed or manipulated. In the gate-model of quantum computing paradigm, transformations between states is achieved using unitary matrices which are represented as `Quantum Gates'. Since all quantum gates are unitary, the inverse of such gates necessarily exists and hence transformations using quantum gates alone is always reversible. The way to incorporate irreversibility in the paradigm is through making projective measurements as that disturbs the state vector irrevocably making it loose its present memory (interactions with the environment induces irreversibility too in the form of qubit decoherence \cite{nielson}. We shall return to this point later). Commonly used quantum gates and their matrix representation in the computational basis is given in Table \ref{gate_tab}. These gates acts on either one, two or three qubits as has been indicated in the table. {\color{black} For visualization of operations of single-qubit gates, in Fig.\ref{fig_bloch_sph_single} we plot the corresponding operations for most commonly-used single qubit gates in the Bloch sphere. We see that for $R_n(\theta)$ the axis of rotation $n$ can be either $\{x,y,z\}$ and that decides the accessible state-space for a given initial state. For Hadamard transformation, the operation can be viewed as rotation about the axis $(n_x, n_y, n_z)^T = (\frac{1}{\sqrt{2}}, 0, \frac{1}{\sqrt{2}})$ through an angle of $\pi$ and hence creates the state $\frac{|0\rangle + |1\rangle}{2}$ starting from $|0\rangle$. The S-gate ($P(\frac{\pi}{2})$) and T-gate ($P(\frac{\pi}{4})$) controls the relative phases of $|0\rangle$ and $|1\rangle$ as has been shown in Fig. \ref{fig_bloch_sph_single}. Table \ref{gate_tab} also discusses several commonly used multi-qubit gates. These operations are commonly used to entangle two or more qubits in a quantum circuit. For example  one of the most celebrated two-qubit gate CNOT (see Table \ref{gate_tab}) can be interpreted as the following
\begin{eqnarray}
    CNOT = |0\rangle \langle 0 |_c \otimes I_t + |1\rangle \langle 1|_c \otimes X_t
\end{eqnarray}
wherein the subscript $c$ indicates the control qubit whose state is not changed and the subscript $t$ indicates the target qubit whose state is altered conditioned on the state of the controlling qubit. In this case if the state of the control is $|1\rangle_c$ the target qubit is flipped but it is left unchanged if the state of the control is $|0\rangle_c$. Similarly using CPHASE($\alpha$) (see Table \ref{gate_tab}) one imparts a relative phase of $\alpha$ between the basis states of the target qubit if the state of the control qubit is $|1\rangle_c$.
It must be emphasized that gates wherein a non-trivial operation on a target qubit is initiated if the control qubit is in state $|0\rangle$ are also routinely used in quantum algorithms. Such a controlled two qubit gate ($CU_0$) for an arbitrary single qubit operation $U_t$ on the target qubit is written as 
\begin{eqnarray}
 CU_0 = |0\rangle \langle 0 |_c \otimes U_t + |1\rangle \langle 1|_c \otimes I_t
\end{eqnarray}

This interpretation extends to multi-qubit gates beyond two as well, except the size of the control register now is more than one and many more possibilities of multiple controls are realizable (for example for a three qubit control unitary, the two controlling qubits can be in any of the four states $|00\rangle_c, |01\rangle_c, |10\rangle_c, |11\rangle_c$ to initiate a non-trivial operation on the target). In the well-known Toffoli gate (see Table \ref{gate_tab}) the state of the target is flipped by a X operation conditioned on the joint state of two-qubits instead of one in unlike in CNOT gate. This means the operation is non-trivial only if this joint state is $11\rangle_c$. Intuitively, one-qubit gates are required to initiate superposition between the 
two-basis states of individual qubits as depicted within the Bloch sphere in Fig.\ref{fig_bloch_sph_single} but multi-qubit gates are required to initiate correlation between the joint-states of several qubits. Both these operations are therefore necessary to create non-trivial many-body quantum states.}

A certain subset of gates forms a universal set \cite{nielson}in the sense that any arbitrary $n$-qubit unitary operation can be approximately modelled as a finite sequence of gates from this set within a preset user-defined precision. The choice of this set is not unique and is largely determined by which gates are operationally convenient for implementation on a given platform used for constructing the quantum hardware. One popular choice is the (\{$R_x(\theta),R_y(\theta), R_z(\theta), P(\alpha), CNOT$\}) gate-set. Equivalent yet a minimalistic choice can be (\{$T=P(\frac{\pi}{4}), H, ,CNOT, S= P(\frac{\pi}{2})$\}) \cite{Preskill_notes, nielson}. One must emphasize that the use of universal gate-sets only guarantees reachability i.e. the ability to approximately implement any desired unitary using a finite-sequence of gates from the set without placing any restriction on the number of gates inhabiting the sequence\cite{Preskill_notes}. Indeed it may so happen that implementation of certain $n$-qubit unitaries would require gate-sequences from the universal set with length scaling as $O(c^n)$ i.e exponential. On the other hand for certain other operations length of gate-sequences scaling as $O(n^k)$ (polynomial) is seen. Only the latter kind of unitaries can be hoped to be efficiently simulated on a quantum computer.

A quantum circuit is essentially an assembly of quantum gates which transforms an initial state of a multi-qubit system to a final desired state. The set of quantum gates operationally represents a user-defined unitary transformation. Such operations are frequently followed by measurement either in the computational basis or in the basis of the operator whose statistics in the prepared state is desired \cite{nielson}.
The circuit representation of the commonly used gates is given in Fig.\ref{fig_qcirc} (a)-(d). A representative example of a quantum circuit built using some of the gates in Table. \ref{gate_tab} is given in Fig. \ref{fig_qcirc} (e). The circuit shows the preparation of a typical Bell state of the kind $\frac{|00\rangle + e^{i\alpha} |11\rangle}{\sqrt{2}}$ in a 2-qubit systems  with $\alpha$ being the relative phase difference between the two basis states ($|00\rangle, |11\rangle$). One practically useful way to interpret such a circuit is to probe the state of the system at various junctions. We have divided the circuit into four junctions. At the first junction labelled as (I), the joint state of the two qubits is the initial computational basis $|00\rangle$. At the junction (II), the effect of the Hadamard (H) on first qubit yields a separable state wherein the first qubit is in an equal superposition of the single-qubit basis states and the second qubit is still in $|0\rangle$. The CNOT gate with the first qubit as the control and the second qubit as the target yields the state $\frac{|00\rangle +  |11\rangle}{\sqrt{2}}$ at junction (III). At the junction (IV), the controlled-phase gate (CPHASE($\alpha$)) selectively creates a phase difference of $\alpha$ between the states  $|00\rangle$ and $|11\rangle$ which results in the target state. Measurements on the target state in the computational basis would yield equal probability ($\frac{1}{2}$) of observing either the state $|00\rangle$ or $|11\rangle$  and zero probability of observing $|01\rangle$ or $|10\rangle$. Circuit representations of the quantum-enhanced machine learning algorithms shall appear throughout the review. Interpretations of each of them can be done analogously.

{\color{black} Development of quantum computing is underway since the 1980s \cite{deutsch1985quantum, benioff1980computer} but it gained unprecedented attention with the exponential speed-up reported in prime factorization by Peter Shor in the last decade of 20th century \cite{shor1999polynomial}. It was quickly realized however that uncontrolled interactions of the qubit register with the environmental bath leads to loss of coherence of the initialized state. Moreover for experimental implementation of a digital quantum-computing platform, the gate-operations (unitary gates defined above) may be imperfect too \cite{hu2002gate}. The collective effect of both of these would be to introduce noise or errors thereby hampering the performance of the algorithm. Quantum error-correction (QEC) schemes were proposed \cite{RevModPhys.87.307} which can act to mitigate the effect of these noises. However scalable implementation of such protocols are challenging \cite{KITAEV20032,knill1997theory} and is currently under development. In the current era, operational quantum devices are already a reality consisting of around 10-100 qubits but without any error-correction. This era of quantum computers is therefore termed as noisy-intermediate-scale quantum devices (NISQ) \cite{preskill2018quantum}. Due to the inherently erroneous gate operations, the algorithms developed for NISQ devices are designed to use shallow-circuit depth and usually variational and delegate a part of the computation to a classical processor\cite{bharti2022noisy}. Such algorithms are meant to reap the maximum benefits from noisy hardwares and look for potential advantages. Such algorithms will be a key player in this review for understanding some of the near-term ML applications. These algorithms have proven to be advantageous for applications in chemistry/chemical physics\cite{google2020hartree, cao2019quantum, head2020quantum,yuan2019theory, cerezo2021variational}, in condensed-matter physics and material science \cite{bauer2020quantum}, atomic physics\cite{dumitrescu2018cloud}, high-energy physics\cite{wu2021application,guan2021quantum}, bio-chemistry\cite{cheng2020application}, finance\cite{orus2019quantum}.
In contrast, there are algorithms like quantum phase estimation\cite{Lloyd2003, aspuru2005simulated} which have provable exponential advantage but requires high-circuit depth and hence is amenable to be implemented in fault-tolerant devices.     
}

\subsection{Quantum Annealing based paradigm}

This paradigm is particularly useful for solving optimization problems wherein the optimal solution can be encoded within the ground state of a given Hamiltonian of a system (say $H_2$). The key working principle of the hardware operating under the annealing model is to prepare the ground state of a system which is efficiently prepared (say for a Hamiltonian $H_1$) from which the ground state of the target Hamiltonian $H_2$ is subsequently retrieved. 
\begin{eqnarray}
    H(s) = A(s) H_1 + B(s) H_2 \label{Ham_qadia}
\end{eqnarray}

To be more specific, at time $t=0$, the  the Hamiltonian of the system be $H_1$ i.e. $(A(s)=1, B(s)=0)$ in Eq. \ref{Ham_qadia} whose ground state can be easily constructed. Thereafter the switching parameter $s$ is varied until $(A(s)=0, B(s)=1)$. If the variations is sufficiently `slow' then the quantum adiabatic theorem \cite{albash2018adiabatic} guarantees that the evolution trajectory would be traversing the instantaneous ground states of Hamiltonian $H(s)$  with high probability. Under such circumstances this implies that one would yield the ground state of the target Hamiltonian $H_2$ at the end of the protocol with high fidelity (see Eq.\ref{Ham_qadia}). A popular quantum annealer D-wave uses ground states of Ising type Hamiltonians \cite{Dwave_sys} for encoding the solution to the problem being investigated. Optimization schemes like quadratic unconstrained binary optimization (QUBO), combinatoric problems etc which can be mapped to such Hamiltonians can thus be efficiently solved using this paradigm \cite{2020_pers_ann, djidjev2018efficient, li2018quantum, neukart2017traffic}. Except in a very small number of examples, this paradigm of quantum computing will not be explored much in this review. Interested reader may consult topical reviews like \cite{nath2021review, 2020_pers_ann}.

\section{Short Primer on the commonly used Toolkits in Machine Learning}\label{ML_toolkits}

{\color{black}
\subsection{Overview}

Broadly problems tackled in machine learning can be categorized into 3 classes: Supervised, unsupervised and reinforcement learning. We start of by discussing each of the categories independently and introduce commonly used terminologies within the machine learning community. 

\subsubsection{Supervised learning}

We are given a dataset of the form $\{(x_i, y_i)| i \in [N] \}$ , where $x_i$'s are inputs sampled from some fixed distribution, $y_i$ is the corresponding label and $N$ is the size of the dataset. Typically $x_i$ is an element in $\mathbb{R}^d$ and $y_i$ belongs to $\mathbb{R}$. The task is to identify the correct label for $y^{*}$ for a randomly chosen sample $x^{*}$ from that distribution. The dataset $\{(x_i, y_i)| i \in [N] \}$ is referred to the training dataset. A loss function $L(h(x_i,w),y_i)$ is defined based on the problem at hand that quantifies the error in the learning. Here $h(x,w)$ refers to the hypothesis function that the learning procedure outputs and $w$ refers to the parameters or weights over which the optimization is performed. An empirical risk minimization is carried over $\sum_i L(h(x_i,w),y_i)$ to output $h(x,w^{*})$, where $w{*}$ are the parameters output at the end of learning. A test data set is finally used to output the performance of $h(x,w^{*})$ and used as metric of comparison across several learning methods. In general, the labelled dataset that one seeks to learn is partitioned manually into training and test data sets. The process of learning thus comprises of 2 parts: trainability (empirical risk minimization over training data) and generalization (how well it performs on unseen data). Typically the optimization of the parameters involves computing gradients of the loss function with respect to these parameters. 

Apart from the parameters that are trained definitively through optimization schemes, other parameters referred to as hyperparameters {\color{black} become critically important for neural-network based supervised learning schemes (to be explored soon). Such parameters/variables are fixed manually by hand apriori}. These may include, the learning technique employed, the number of parameters, the optimization procedure \cite{gradient_methods} (standard gradient descent, stochastic gradient descent, Adam optimizer) the parameter initialization scheme, the learning rate, the stopping condition for training (threshold for convergence or the number of parameter update iterations), batch sizes, choice of loss function etc \cite{DBLP:journals/corr/Breuel15a}. 

{\color{black} Examples of problems in supervised learning procedure include classification, where the labels $y_i$ are discrete and regression, where the labels $y_i$ are continuous and extrapolation to unknown cases is sought. Some of the techniques used exclusively for problems in this domain include, support vector machine, kernel ridge regression wherein data is mapped to a higher-dimensional space for manipulation, Gaussian process regression, decision trees, Naive Bayes classifier. Other techniques not exclusive to this learning model, include Neural networks whose applications have spanned every field of industry and research. We will discuss more about each of the above learning models in the subsequent section.}

\subsubsection{Unsupervised learning}
Unlike Supervised learning, here we are provided with data points that do not have any continuous or discrete labels associated with it. The task is to learn something intrinsic to the distribution of the data points. Some commonly tackled tasks under this learning scheme include, clustering with respect to some metric on the space, learning the given probability distribution by training latent variables (eg: Boltzmann Machine, Restricted Boltzmann Machine(RBM), Generative Adversarial Networks), dimensionality reduction that allows reduction in size of the feature space with little information loss (eg: autoencoders, principal component analysis, RBM's). Unsupervised learning mainly tries to solve the problem of learning an arbitrary probability distribution by minimizing some loss function that quantifies the divergence between the given distribution to be learnt and model distribution being trained(eg: cross entropy, KL divergence, Renyi entropy) \cite{entropy-types}. We would like to point that the techniques mentioned above are different variations to making use of a neural networks, whose functionality depends on the exact form of cost function being employed in the training.

One is not restricted to using methods from either supervised or unsupervised learning exclusively for solving a problem. In practice we notice, a mix of the methods are employed to solve a given problem. For instance, one might require dimensionality reduction or noise filtering or distribution learning using unsupervised methods prior to introducing labels and solving a classification problem with supervised methods. These methods are commonly referred to as semi-supervised learning \cite{learning_semisupervised, supervising_unsupervised} or hybrid learning methods.

\subsubsection{Reinforcement learning}
Unlike the above two learning models, here we take a totally different stand on the setting in which the learning happens. An artificial agent is made to interact with an environment through actions so as to maximize the reward function that has been identified. This type of learning is employed when the agent can learn about its surroundings only through interaction which is limited by finite set of actions that the agent is provided with. Due to the unbounded sequence of actions that the agent can explore, one needs to employ good heuristics in regards to designing reward functions that  help accept or reject the outcome of a certain action in exploring this space. Thus optimal control theory \cite{DBLP:journals/corr/abs-1912-03513} plays an important role in this learning method. Some of the most popular applications involve self driving cars (Tesla Autopilot), training bots in a game (Alpha zero for chess, Alpha Go zero for Go) and smart home robots (vacuum cleaning bots and companion bots) For an extensive introduction to Reinforcement learning refer \cite{2018_reinforcement}. 
}

\subsection{Classical and Quantum variants of commonly used algorithms}

In this section we shall elaborate some of the commonly encountered machine and deep learning techniques that has been used extensively for physico-chemical studies. We shall discuss both the classical implementation of the algorithms and also the appropriate quantum versions. 

{\color{black}
\subsubsection{Kernel based learning theory} \label{Kernel_learning}
 
The concept of kernels is very important in machine learning , both quantum and classical \cite{10.5555/559923, genton2001classes, azim2018kernel}. Let us imagine a dataset D= $\{(\bf{x}_i, y_i)|\: \bf{x}_i \in \chi$, $\bf{y}_i$ $\in \Omega  \:\:\forall \:\:i \:\:\in [m]\}$ as described in the supervised learning section. In the set D, $\bf{x}_i$ are the feature vectors sampled from the set $\chi$ whereas the labels $\bf{y}_i$ are sampled from another set $\Omega$. In the cases frequently encountered, one usually finds $\chi \subseteq \mathbb{R}^d$
and $\Omega \subseteq \mathbb{R}$. $m$ is the sample size of the training data-set $D$ or the number of observations. It is often convenient to define a map $\phi$ such that $\phi : \chi \mapsto \mathcal{F}$ such that the new feature-space $\mathcal{F}$ is usually a higher-dimensional space equipped with an inner product. For example if $\chi \subseteq \mathbb{R}^d$ and $\mathcal{F} \subseteq \mathbb{R}^p$ then 
p $\ge$ d. The Kernel $K : \chi \times \chi \mapsto \mathbb{R}$ of the map $\phi(\bf{x})$ is then defined as the following
\begin{eqnarray}
    K(\bf{x},\bf{x}') = (\phi(\bf{x}), \phi(\bf{x}'))_\mathcal{F} \label{kernel_def}
\end{eqnarray}
where $(\cdot, \cdot)_F$ symbolizes an inner product on  $\mathcal{F}$. For example, if $\mathcal{F} \subseteq \mathbb{R}^p$ then the inner product can be familiar $(\phi(\bf{x}), \phi(\bf{x}'))_\mathcal{F} = \phi(\bf{x})^T \phi(\bf{x}')$.

The importance of kernels lies in the fact that since the space $\mathcal{F}$ is high-dimensional, direct computation of feature map $\phi(x)$ in that space might be intractable and/or expensive. However most algorithms using the kernel trick are designed such that the only quantity required would be the inner product $K(\bf{x},\bf{x}')$ (See Eq.\ref{kernel_def}) without explicit construction or manipulation of $\phi(\bf{x})$ or $\phi(\bf{x}')$. Thus several popular kernel functions have been designed in literature \cite {10.5555/559923,genton2001classes} which can be computed directly from the entries $\bf{x}$ in the dataset D. Some of them are displayed below:
\begin{equation*}
    \begin{split}
        Linear
        \qquad
        &K({\bf x}, {\bf x}')={\bf x}\cdot{\bf x}'
        \\
        Polynomial
        \qquad
        &K({\bf x}, {\bf x}', \gamma, d)=\left(
        r + \gamma\cdot{\bf x}\cdot{\bf x}'
        \right)^{d}
        \\
        Gaussian
        \qquad
        &K({\bf x}, {\bf x}', \sigma)=
        \exp{\left(
        -\frac{||{\bf x}-{\bf x}'||^2}{2\sigma^2}
        \right)}
        \\
        Sigmoid 
        \qquad
        &K({\bf x}, {\bf x}', r, \gamma)=
        \tanh\left(
        r + \gamma\cdot{\bf x}\cdot{\bf x}'
        \right)
    \end{split}
\end{equation*}

The success of the kernel trick has been extended to several important supervised machine learning algorithms like kernel-ridge regression, dimensionality reduction techniques like kernel-based principal component analysis, classification routines like k-nearest neighbor (see Section \ref{k-NN_section}) and support-vector machines (SVM) (see Section \ref{Supp_vec_machine}) etc. For classification tasks like in SVM the effect is more conventionally described as inability of a hyperplane for linearly discriminating the data entries which can be ameliorated through the kernel trick of transporting the feature vectors $\bf{x}$ to higher dimension $\phi(\bf{x})$ wherein such a separability is easily attainable. Both regression and classification algorithms will be discussed in detail in appropriate sections. In this section, we shall first discuss the kernel theory developed recently for quantum-computing enhanced machine learning techniques.

\paragraph{Quantum enhanced variants \\}

The theory of quantum kernels has been formalized in Ref \cite{schuld2018supervised, PhysRevLett.122.040504}. For a given classical data set D= $\{(\bf{x}_i, y_i)|\: \bf{x}_i \in \chi$, $\bf{y}_i$ $\in \Omega  \:\:\forall \:\:i \:\:\in [m]\}$ as defined above wherein $\bf{x}_i \in \chi$, a data domain, Ref \cite{schuld2018supervised} defines a data-encoding feature map as a quantum state $\rho(\bf{x}_i) = |\phi(\bf{x}_i\rangle \langle \phi (\bf{x}_i)|$ which is created from a data-encoding unitary ($U(\bf{x}_i) \in \mathbb{C}^{2^n \times 2^n})$ as $|\phi(\bf{x}_i\rangle = U(\bf{x}_i)|0\rangle^n$. This unitary $U(\bf{x}_i)$ thus embeds each feature vector of the dataset within a quantum state $\rho(\bf{x}_i)$. The state $\rho(\bf{x}_i)$ is part of a Hilbert space $\mathbb{L}(\mathbb{C}^n)$ which is thereby equipped with an inner product defined as $\langle\rho, \tau\rangle = Tr(\rho\tau) \:\:\forall \:\:\rho, \:\tau \:\:\in \:\: \mathbb{L}(\mathbb{C}^n)$. The quantum variant of the kernel matrix entries from the dataset $D$ is thus computed from this inner product as 
\begin{eqnarray}
   K(\bf{x_i}, \bf{x_j}) = Tr(\rho(\bf{x})_i \rho(\bf{x}_j)) \label{QKernel}
\end{eqnarray}

The authors prove that such a definition of a quantum kernel indeed satisfies Mercer's condition \cite{ghojogh2021reproducing} of positive-semi definiteness. The authors then define a reproducing kernel Hilbert space (RKHS) which is a span of basis functions $f : \chi \mapsto \mathbb{R}$
where the function $f(\bf{x})=K(\bf{x}_i, \bf{x})$ i.e each such basis function in the spanning set
comprises of quantum kernel matrix elements $K(\bf{x}_i, \bf{x})$ as defined in Eq.\ref{QKernel} with one input argument of the matrix element being made from a particular datum (say $\bf{x}_i \in \chi$) of the dataset $D$. Any arbitrary function (say $g(\bf{x})$) that lives in the RKHS is thus a linear combination of such basis functions and is expressed as
\begin{eqnarray}
g(\bf{x}) = \sum_i \alpha_i K(\bf{x}_i, \bf{x})
\end{eqnarray}
where $\alpha_i$ are the linear combination coefficients. The author proves that any hypothesis function (say $h(\bf{x}) = Tr(M\rho(\bf{x}))$ where $M$ is the measurement operator) which the supervised learning task `learns' on the quantum computer by minimizing a loss function are essentially members of RKHS. In Ref \cite{PhysRevLett.122.040504}, the authors proposes two different approaches for utilizing quantum Kernel entries as defined in Eq.\ref{QKernel}. The first approach which the authors call the implicit approach requires the quantum processor to just estimate entries of the Kernel matrix. The classical processor then performs the usual machine learning algorithm using this quantum-enhanced kernel. The second approach which the authors call as explicit involves performing the entire machine learning algorithm on the quantum computer using parameterized unitaries. We shall analyze examples of these approaches in Section \ref{Case_QML} and \ref{State_class_sec}.


\subsubsection {Ridge Regression (RR)-Linear and Kernel based} \label{KKR_sec}

This is a form of supervised machine learning which allows us to determine and construct an explicit functional dependence of the variates/labels and the feature vectors $x_i$ based on certain tunable parameters \cite{brunton2019data, marquardt1975ridge,mcdonald2009ridge, hoerl1975ridge}. The dependence can later be extrapolated, interpolated to learn values associated with unknown feature vectors not a part of the training set. Let us start with the feature vectors $x_i \in \chi \subseteq \mathbb{R}^d$ in dataset D defined in the above section. Using these vectors, one can define a design matrix often designated as $X$ as follows:
\begin{eqnarray}
    X=\begin{pmatrix}
     x_1^T\\
     x_2^T\\
     . \\
     . \\
     x_m^T
    \end{pmatrix} \label{x_matrix_PCA}
\end{eqnarray} 
Using the design matrix above and the training data label $Y=[y_1, y_2, y_3,....y_m]^T \:\:\in \:\: \mathbb{R}^m$, the objective is to fit a linear model of the kind $X\vec{\alpha}$ where $\vec{\alpha}\:\:\in \:\: \mathbb{R}^d$ to the data and obtain the optimal fitting parameters. This can be done through the minimization of the following mean-squared error loss (MSE)
\begin{eqnarray}
   MSE &=& (Y-X\vec{\alpha})^T  (Y-X\vec{\alpha})\label{MSE_loss_RR} + \lambda \frac{||\vec{\alpha}||^2}{2} \label{MSE_RR}
\end{eqnarray}
In the expression above, the second term is the regularization to prevent over-fitting and also in case if the column space of the design matrix $X$ is not linearly -independent, the presence of this term can facilitate inversion of $X^TX$. The solution to Eq.\ref{MSE_RR} (say $\vec{\alpha}^*$) is the following:
\begin{eqnarray}
    \vec{\alpha}^* = (X^T X + \lambda \mathbb{I})^{-1}X^T Y
\end{eqnarray}
One must emphasize, the formulation is quite general and can be extended to cases wherein a constant term within the $\vec{\alpha}$ is necessary. That can be tackled by augmenting the design matrix as $X \rightarrow [\vec{1}|X]^T$. Also extension to polynomial regression is straightforward as one can create a design matrix treating higher powers of $x_i$ as independent variables in each row of the design matrix as $X_i \rightarrow [x_i^T\:\: (x_i^2)^T\:\: (x_i^3)^T \:\:.....(x_i^k)^T]$ where $x_i^k$ denotes raising $x_i$ element-wise to $kth$ power\cite{ostertagova2012modelling, poly_reg_book}.

For the kernel variant of ridge-regression, if the prediction from the model is designated as $\tilde{y}(x, \alpha)$, then the formalism represents the function $\tilde{y}(\vec{x}, \vec{\alpha})$ as \cite{vovk2013kernel, vu2015understanding}
\begin{eqnarray}
\tilde{y}(\vec{x}, \vec{\alpha}) = \sum_{j=1}^m \alpha_j K(x, x_j, \vec{b}) \label{Ker_func_KRR}
\end{eqnarray}
where $\vec{\alpha}$ are trainable parameters and $\vec{b}$ are the hyper-parameters associated with the kernel $K(x,x_j,\vec{b})$. These hyperparameters are fixed at the beginning of the optimization and can be tuned for separate runs to modify accuracy. Using the labels $y_i$ of the training data set $D$ defined before and Eq. \ref{Ker_func_KRR} one can now formulate a  mean-squared error loss (MSE) (similar to Eq.\ref{MSE_loss_RR}) to learn the parameters $\vec{\alpha}$ as below
\begin{eqnarray}
    MSE &=& (Y-\tilde{K}\tilde{\alpha})^T  (Y-\tilde{K}\tilde{\alpha}) + \lambda \frac{||\vec{\alpha}||^2}{2} \label{MSE_loss_KRR}
\end{eqnarray} 
The matrix $\tilde{K}$ is called the Gram matrix of the kernel \cite{vovk2013kernel} with entries as follows:
\begin{eqnarray}
    \tilde{K}_{ij} = K(x_i, x_j, \vec{b}) \label{Gram_matrix}
\end{eqnarray}
The minimizer of Eq. \ref{MSE_loss_KRR} can be easily shown to be 
\begin{eqnarray}
    \vec{\alpha^*} = (\tilde{K} + \lambda \mathbb{I})^{-1} Y
    \label{KRR_minimizer}
\end{eqnarray}
Another alternative formulation which leads to the same minimizer $\vec{\alpha^*}$ is the dual formulation of the problem which involves minimizing the following Langrangian \cite{saunders1998ridge}
\begin{eqnarray}
    L(\vec{\alpha}) = \frac{\alpha^T\alpha}{2} + \frac{1}{2\lambda}{\alpha^T \tilde{K}\alpha} - \alpha^TY
    \label{Lang_KRR}
\end{eqnarray}
One can prove that the minimzer of Eq.\ref{Lang_KRR} is actually Eq.\ref{KRR_minimizer}.
This is a form of supervised machine learning which allows us to determine and construct an explicit functional dependence of the variates/labels and the feature vectors $x_i$ based on certain tunable parameters \cite{brunton2019data, marquardt1975ridge,hoerl1975ridge}. The dependence can later be extrapolated, interpolated to learn values associated with unknown feature vectors not a part of the time dynamics \cite{ullah2021speeding} or excited state dynamics \cite{Westermayr_2020}. We shall return to a subset of these topics in Section \ref{MBS_sec}.\\

\paragraph{Quantum enhanced variants \\}

Several quantum algorithms have been proposed in the last decade for solving linear systems which can be directly extended to the solution of the vanilla linear least-square fitting. The earliest was by Weibe $etal$ \cite{PhysRevLett.109.050505} and is based on the Harrow, Hassidim, Lloyd (HHL) algorithm \cite{PhysRevLett.103.150502, lee2019hybrid}. The technique starts with a non-hermitian $X$ matrix (say $m\times d$ as in the design matrix in our example above) which is required to be sparse. The algorithm assumes oracular access to a quantum state encoding its row space and also a state encoding the $\vec{y}$. The key point of the routine is to expand the non-hermitian design matrix into a $(m+d) \times (m+d)$ dimensional matrix with which the quantum-phase estimation algorithm \cite{PhysRevLett.83.5162} is performed. The ultimate product of this algorithm is a quantum state that encodes the fitted values. Even though extraction of the exact fitting parameters from the state might be exponentially hard yet prediction of new $y$ value for a given test-input can be made effortlessly through overlap with the register containing the fitted values. Variants of this algorithm for detecting statistic leverage score and matrix coherence have also been reported \cite{LIU201738}. Wang reported a quantum algorithm which can actually yield the fitted values as vector just as in classic least squares \cite{PhysRevA.96.012335}. Both the method have query complexity which is $O(log(m))$. A subset of these algorithms have also been experimentally implemented on a variety of platforms like NMR\cite{PhysRevA.89.022313}, superconducting qubits \cite{PhysRevLett.118.210504}, photonic\cite{doi:10.1063/1.5115814} etc. Schuld $et al$ \cite{PhysRevA.94.022342} have also designed an algorithm which does not require the design matrix $X$ to be sparse. The only requirement is that $X^\dagger X$ should be well-represented by a low rank approximation i.e. should be dominated by few eigenvalues only. The key point in the technique is to perform quantum-phase estimation with a density matrix $\rho$ encoding $X^\dagger X$. The algorithm also returns the fitted values encoded within a quantum state with which efficient overlap of a new input can be initiated. An algorithm by Yigit $etal$ \cite{subacsi2019quantum} which solves for a linear system of equation through adiabatic hamiltonian evolution has also been demonstrated recently . A variational algorithm amenable to the NISQ era for linear equation solver has also been reported \cite{bravo2019variational}. The algorithm takes as input a gate sequence $U$ that prepares and encodes the state $\vec{y}$, a design matrix $X$ that is decomposable into implementable unitaries. The method implements a trainable unitary $V(\vec{\vec{\gamma}}) |0\rangle$ where $\vec{\vec{\gamma}}$ are variational parameters. The aim of the unitary is to prepare a candidate state $|\alpha(\vec{\gamma})\rangle$ which encodes a prospective solution to the least-square problem. The prospective solution is tuned using a cost-function which measures the overlap of the state $X|\alpha(\vec{\gamma})\rangle$  with the orthogonal subspace of the vector $\vec{y}$ as follows:
\begin{eqnarray}
C(\vec{\gamma}) &=& Tr(X|\alpha(\vec{\gamma})\rangle \langle \alpha(\vec{\gamma})| X^\dagger (\mathbb{I} - |\vec{y}\rangle \langle \vec{y}|)) 
\end{eqnarray}
The cost function above is minimized with respect to $\vec{\gamma}$ on a classical computer and the parameter vector is fed into the trainable unitary $V$ for the next iteration until the desired convergence is met. The authors show that the above cost-function being a global one suffers from barren plateaus and is rendered untrainable for the size of design matrix $X$ being close to $2^{50} \times 2^{50}$. To evade the issue, they define local merit functions which remain faithful throughout. The ansatz used for encoding $V(\vec{\gamma})$ is the hardware-efficient ansatz and the algorithm showed logarithmic dependance on the error tolerance but near linear dependance on the condition number of the design matrix. The dependance on qubit requirements was found to be poly-logarithmic. The algorithm was implemented on an actual hardware for a design matrix of size $2^{10}\times 2^{10}$.
Recently, Yu $et al$ reported an algorithm for Ridge-Regression (linear variant) \cite{yu2019improved} which like the one reported by Weibe requires oracular access to elements of the design matrix and the $\vec{y}$. The design matrix is expanded to make it hermitian and quantum-phase estimation is performed as before with respect to $e^{-iXt}$ as the unitary to encode the eigenvalues onto a extra register. The difference comes at this stage when an ancillary qubit is added and rotated to invert the eigenvalues. The rotation angles are dependant on the Ridge parameter $\lambda$. Like previous algorithm this also yields the final optimal parameters as a quantum state. The authors also propose another quantum algorithm (which can be used along with this) for the choice of the Ridge parameter which is similar in principle to K-fold cross validation technique \cite{BERRAR2019542}. To the best of our knowledge, no quantum algorithm has been proposed that directly attempts to implement the kernelized variant of Ridge-regression but any of the aforesaid ones can be trivially extended with the replacement of the design matrix with the Gram matrix of the appropriate kernel.   

\subsubsection{Principal Component Analysis-Linear and Kernel based} \label{PCA_sec}

Dimensionality reduction without sacrificing the variance of the data-set is very important for most machine learning tasks that has large number of features and comparatively fewer training samples. One starts with a dataset (D as discussed before where D= $\{(x_i, y_i)| x_i \in \mathbb{R}^d, y_i \in \mathbb{R} \:\:\forall i \:\:\in [m]\}$ ). Here we define a design matrix (say $X \:\:\in \:\: \mathbb{R}^{m\times d}$) as in Eq.\ref{x_matrix_PCA} 
Formally the goal of PCA is to replace the matrix $X$ with another matrix $Z$ such that $Z \:\in \: \mathbb{R}^{m\times R}$ where $R \le d$ \cite{jolliffe2016principal, jolliffe2002principal, brunton2019data}. To do that for the usual linear variant of the PCA one defines a mean-centered data matrix (say $B = X-\hat{X}$) where $\hat{X}$ is the stacked row-wise mean of the data matrix (mean of each feature over the samples). One then constructs the covariance matrix \cite{jolliffe2002principal} ($Cov(B) = B^T B$, $Cov(B) \in \mathbb{R}^{d\times d}$) and diagonalizes it to get the $d$ eigenvectors  $\{\nu_i\}_{i=1}^d$ ($\nu_i \in \mathbb{R}^d$). From the set $\{\nu_i\}_{i=1}^d$ one picks up the $R$ eigenvectors with the largest eigenvalues to form a new matrix (say $V \:\:\in \:\:\mathbb{R}^{m \times R}$) as 
\begin{eqnarray}
    V =\begin{pmatrix}
     \nu_1 \nu_2 \nu_3 ....\nu_R
    \end{pmatrix}
\end{eqnarray}
The principal component matrix $Z$ defined before is the projection of the data matrix onto the space of matrix $V$ as 
\begin{eqnarray}
Z = X V
\end{eqnarray}

The kernel-based variant of PCA \cite{scholkopf1997kernel} becomes important when the data needs to expressed in a high-dimensional subspace induced by the map $\phi$ as defined before. i.e. $\forall$ $x_i \in \mathbb{R}^d$ , $\phi(x_i) \in \mathbb{R}^p$ where $p \ge d$. One can then construct the covariance matrix in this new feature space of the data as follows:
\begin{eqnarray}
    Cov(\phi(X)) = \frac{1}{m}\sum_{i}^m \phi(x_i) \phi(x_i)^T
\end{eqnarray}
In principle, one can simply do a PCA in the feature space $\phi(X)$ to get the eigenvectors $\{\nu_k\}_{k=1}^{p}$ as follows
\begin{eqnarray}
    Cov(\phi(X))\nu_k &=& \lambda_k \nu_k 
    \label{Cov_phi_eig}
\end{eqnarray}
wherein $\nu_k \in \mathbb{R}^p \:\: \forall \:\: k$.
However, since this space is high-dimensional, computation can be expensive. It is thus desirable to use the power of the kernel and design an algorithm wherein the explicit construction and/or manipulation of $\phi(X)$ is evaded. To do so, one can expand the eigenvectors $\nu_k$ as follows 
\begin{eqnarray}
    \nu_k &=& \frac{1}{m\lambda_k}\sum_{i=1}^m (\phi(x_i)^T \nu_k) \phi(x_i) \\
    &=&\frac{1}{m\lambda_k}\sum_{i=1}^m \hat{\alpha}_{ki} \phi(x_i)
\end{eqnarray}
It is easy to show that the coefficient vector $\hat{\alpha_k} \:\: \forall k\:\: \in \{1,2,...p\}$ satisfies the eigenvalue equation for the Gram matrix (see Eq.\ref{Gram_matrix}) of the kernel as
\begin{eqnarray}
    \tilde{K} \hat{\alpha_k} = (\lambda_k m ) \hat{\alpha_k} \label{eigvec_Gram}
\end{eqnarray}
Thus one can simply diagonalize the Gram matrix of the kernel $\tilde{K} \in \mathbb{R}^{m\times m}$ to get the coefficients of the eigenvectors of $Cov(\phi(X))$ (see Eq.\ref{Cov_phi_eig}) without explicitly constructing $\phi(X)$ or even the covariance .Since valid kernels need to be positive-semi-definite as a condition imposed by Mercer's theorem \cite{minh2006mercer, campbell2002kernel}, one can choose a subset (say R) from ${\hat\{\alpha_k\}_{k=1}^p}$ in decreasing order of their eigenvalues $\lambda_k$ and construct a matrix $V \in  \mathbb{R}^{m \times R}$ as
\begin{eqnarray}
V = \begin{pmatrix}
\hat{\alpha}_1 \hat{\alpha}_2....\hat{\alpha}_R
\end{pmatrix} \label{Z_feature_mat}
\end{eqnarray}
thereby affording dimensionality reduction. Any projection onto the $\nu_k$ can be computed using the vector $\hat{\alpha}_k$ and the Kernel Gram matrix only as follows \cite{wang2012kernel}
\begin{eqnarray}
    y_k(x) = \sum_j \tilde{K}(x, x_j, \vec{b}) \hat{\alpha}_{kj}
\end{eqnarray}
We shall return to applications of PCA in Section \ref{Drug_discovery}

\paragraph{Quantum enhanced variants \\}

The very first instance of performing principal component analysis on a quantum computer was put forward by Lloyd $etal$ \cite{lloyd2014quantum}. The algorithm starts with a matrix say $P$ which is positive-semi-definite. Even the usual linear variant of the PCA using the covariance matrices of mean-centered data was discussed as an application for the method, yet the algorithm can be extended to any positive-semi-definite matrix including the Gram matrices of Kernels. Any such matrix $P$ can be written as $\sum_{ij}|a_j||a_j\rangle \langle e_i|$ where $|e_i\rangle$ is the computational basis and $|a_j\rangle$ are column vectors of P normalized to 1 \cite{lloyd2014quantum} and $|a_j|$ is the corresponding norm. The algorithm assumes that a oracular object exists 
that encode the column space onto a quantum state as 
$\sum_{ij}|a_j|e_i\rangle |a_j\rangle$ where $|e_i\rangle$ is an orthonormal basis usually the standard computational basis. The corresponding reduced density matrix for the first register in this state is exactly the positive semi-definite matrix $P$.
\begin{figure}[hpt]
    \centering
    \includegraphics[width=0.5\textwidth]{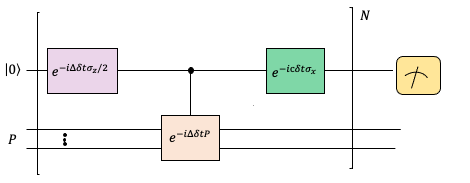}
    \caption{ {\color{black}The schematic of the quantum circuit used for Trotterized evolution as illustrated in Ref \cite{li2021resonant}. The eigenvalues and eigenvectors of the positive-semi definite matrix $P$ is desired. At the end of the evolution , the ancilla qubit is measured and the algorithm is considered successful if it collapses to state $|1\rangle$. (See text for more details)}
    }
    \label{fig:PCA_res}
\end{figure}
The crux of the method is to prepare the unitary $e^{-iPt}$. This is done on a qubit register that encodes the reduced density matrix $P$ as described above and also another density matrix (say $\sigma$). With repeated applications of $e^{-i S \delta}$ on the joint state $P\otimes \sigma$ with n copies of $P$ yields $e^{-i P n\delta} \sigma e^{-i P n\delta}$ where $S$ is the efficiently implementable SWAP operator. Once efficient preparation of the unitary $e^{-i P n\delta}$ has been conducted one can thereafter use standard phase estimation algorithms \cite{PhysRevLett.83.5162} to measure the first say $R$ eigenvalues and eigenvectors of the desired matrix $K$ and cast the data matrix $X$ in the form of Eq. \ref{Z_feature_mat}. The algorithm produces the $R$ eigenvectors with a time complexity of $O(Rlog(m))$ where $m$ is the column dimension of the Gram matrix defined in \ref{Gram_matrix}.
Since the above formulation relies on quantum-phase estimation to extract eigenvalues and eigenvectors of the Gram matrix, application of the procedure to near-term quantum devices is cost-prohibitive due to high-qubit and gate requirements. Recently Li $et al$ \cite{li2021resonant} reported a new algorithm for extraction of principal components of any positive-semi-definite matrix using a single ancillary qubit. The algorithm encodes a joint initial state of an ancillary qubit and the n-qubit positive semi-definite matrix ($P$ in the notation and it could be a Gram matrix of the kernel as well as defined in Eq.\ref{Gram_matrix}) as $|0\rangle\langle 0|\otimes P$. This state is evolved under the effect of the following Hamiltonian for a time $\delta t$ (see circuit for implementing Trotterized evolution in Fig. \ref{fig:PCA_res}
\begin{eqnarray}
    H= \frac{\Delta}{2}\sigma_z\otimes \mathbb{I}_n + c\sigma_x\otimes \mathbb{I}_n + |1\rangle\langle 1|\otimes P
\end{eqnarray}
where $c$ is the strength of the drive on the probe ancillary qubit and $\Delta$ is its natural frequency. The probability of finding the probe ancillary qubit in state $|1\rangle$ after $\delta t$ is given by
\begin{eqnarray}
    P_i(\Delta, \delta_t) = \lambda_i D_i^2 \sin^2(\frac{c\delta t}{D_i})
\end{eqnarray}
where the index $i$ is for any of the eigenvalues $\omega_i$ of the matrix $P$. The quantity $D_i$ is defined as
\begin{eqnarray}
    D_i = \sqrt{\frac{(2c)^2}{(2c)^2 + (\Delta-\omega_i)^2}} \label{prob_res_PCA}
\end{eqnarray}
So from Eq.\ref{prob_res_PCA} one can directly see that a resonance condition is reached when $\omega_i \approx \Delta$ i.e. by sweeping the probe qubit frequency $\Delta$ one can enhance the probability of finding the probe qubit in state
$|1\rangle$ which gives an estimate of the eigenvalue $\omega_i$. Near such a resonance if the qubit is measured then it would collapse to $|1\rangle$ with high probability and the corresponding state in the n-qubit register would be $|\nu_i\rangle$. Thus the entire spectrum of matrix $P$ can be ascertained from which PCA procedure as described above can be performed. The group conducted successful experimental implementation of the algorithm on a nitrogen vacancy center for a system of 2-qubits \cite{li2021resonant} with dynamical decoupling sequences to prevent qubit de-phasing. 
Recently a direct kernel-based quantum-enhanced algorithm for PCA has been demonstrated too \cite{li2020quantum}. The algorithm starts with an initial state which encodes the elements of the Gram matrix of the kernel (See Eq.\ref{Gram_matrix}). The register encoding the row vector of the Kernel is then conceived to be expanded in the basis of its eigenvectors (preparation of which is the target, see Eq.\ref{eigvec_Gram}). Use of quantum-phase estimation followed by controlled rotations then prepare an auxillary qubit in a superposition state. Measurement of this ancillary qubit (with a probability proportional to $\sum_k\frac{1}{\lambda_k}$ where $\lambda_k$ are the eigenvalues of the Gram matrix) encodes columns of the target matrix $V$ (see Eq.\ref{Z_feature_mat}) onto the register scaled with $\sqrt{\lambda_k}$ due to phase-kickback from the measurement. Note the final state thus prepared is not entirely the columns of Eq.\ref{Z_feature_mat} but are scaled by the square root of the corresponding eigenvalues $\lambda_k$.
}

\subsubsection{k-nearest neighbors algorithm (k-NN)}
\label{k-NN_section}
\begin{figure}[hpt]
    \centering
    \includegraphics[width=0.45\textwidth]{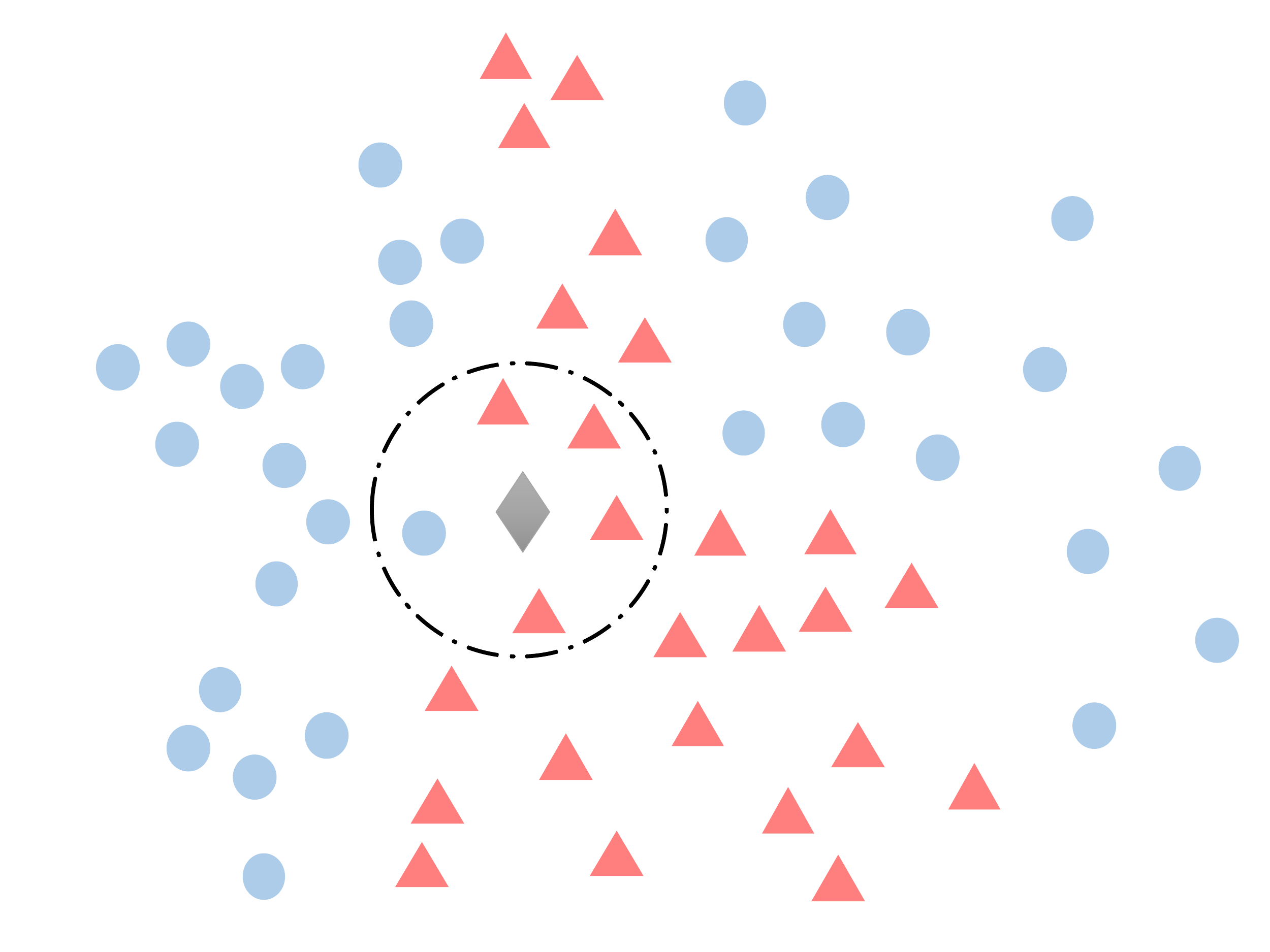}
    \caption{
    {\bf A simple example of the kNN algorithm approach.}
    \\
    the blue dots and red triangles represent the training instances with two labels, and the grey diamond is a new test instance.
    In this example k is set as 5, and the 5 nearest instances are included in the black circle.
    As there are 4 red triangles and only 1 blue dot, the class of the test instance is classified as red triangle.
    }
    \label{fig_kNN}
\end{figure}

The kNN approach is based on the principle that the instances within a dataset will generally exist in close proximity to other instances that have similar properties.
If the instances are tagged with a classification label, then the value of the label of an unclassified instance can be determined by observing the class of its nearest neighbours.
The kNN locates the k nearest instances to the query instance and determines its class by identifying the single most frequent class label\cite{cover1967nearest}. 
In ML, instances are generally be considered as points within an n-dimensional instance space, where each of the n-dimensions corresponds to one of the n-features.
To classify a new test instance with kNN method, the first step is to find the k most nearest instances of the training set according to {\color{black} some} distance metric. Then the resulting class is the most frequent class label of the k nearest instances. Fig.(\ref{fig_kNN}) is a simple example of kNN algorithm, where the blue dots and red triangles represent the training instances with two labels, and the grey diamond is a new test instance. In this example k is set as 5, and the 5 nearest instances are included in the black circle.
For simplicity, here the relative distance $D(x,x')$ between two instances $x$ and $x'$ is calculated by Euclidean metric,
\begin{equation}
    D(x,x')=\left(
    \sum_{i=1}^{n}{\left|x_i-x'_i\right|^2}
    \right)^{\frac{1}{2}}
\end{equation}
As there are 4 red triangles and only 1 blue dot, the class of the test instance is classified as red triangle.

Generally, the relative distance is determined by using a distance metric instead of the absolute position of the instances. Apart from {\color{black} the} Euclidean metric, some significant metrics are presented as following:
\begin{equation*}
    \begin{split}
        Minkowsky
        \qquad
        &D(x,x')=\left(
        \sum_{i=1}^{n}{\left|x_i-x'_i\right|^r}
        \right)^{\frac{1}{r}}
        \\
        Manhattan
        \qquad
        &D(x,x')=
        \sum_{i=1}^{n}{\left|x_i-x'_i\right|}
        \\
        Chebychev
        \qquad
        &D(x,x')=
        \max_{i=1}^{n}{\left|x_i-x'_i\right|}
        \\
        Camberra
        \qquad
        &D(x,x')=
        \sum_{i=1}^{n}\frac{\left|x_i-x'_i\right|}{\left|x_i+x'_i\right|}
    \end{split}
\end{equation*}
{\color{black} An ideal distance metric should be chosen to minimize the distance between two similarly classified instances, meanwhile maximizing the distance between instances of different classes. Sometimes even these typical metrics do not lead to satisfying results, then one might consider to learn a distance metric for kNN classification\cite{ruan2017quantum}. In other words, the metric is optimized with the goal that k-nearest neighbors always belong to the same class while examples from different classes are separated}.
When there are plenty of training instances, centroid method\cite{gou2012local} could be applied initially, where the instances in different labels are clustered into several groups, and the kNN approach works on the centroid of each group instead of the original instances.
Additionally, for more accurate classification, various weighting schemes\cite{geler2016comparison} could be included that alter the distance measurements and voting influence of each instance. We shall return to examples of k-NN in Section \ref{State_class_sec}.

\paragraph{Quantum enhanced variants \\}

In 2013, Lloyd and coworkers proposed a quantum clustering algorithm for supervised or unsupervised QML\cite{lloyd2013quantum}, {\color{black} relying on the fact that estimating distances and inner products between post-processed vectors in $N$-dimensional vector spaces takes time $O(log N)$ on a quantum computer whereas on a classical computer it would take $O(N)$ time for sampling and estimating such distances and inner products  thereby apparently providing an exponentially advantage\cite{aaronson2010bqp}.More discussion about this speedup on a quantum computer can be found in Section \ref{State_class_sec}}.
The significant speedup of {\color{black} estimating} distances provokes enormous enthusiasm studying QML, particularly the quantum instance-based learning algorithms.
Wiebe and coworkers developed the quantum nearest neighbor algorithm based on he Euclidean distance, and studied the performance on several real-world binary classification tasks\cite{wiebe2014quantum}.
Moreover, assorted quantum kNN methods\cite{ruan2017quantum, wisniewska2018recognizing, wang2019improved} are proposed with heterogeneous distance metrics, assisted solving a variety of pattern recognition problems.

\begin{figure}[h!]
    \centering
    \includegraphics[width=0.5\textwidth]{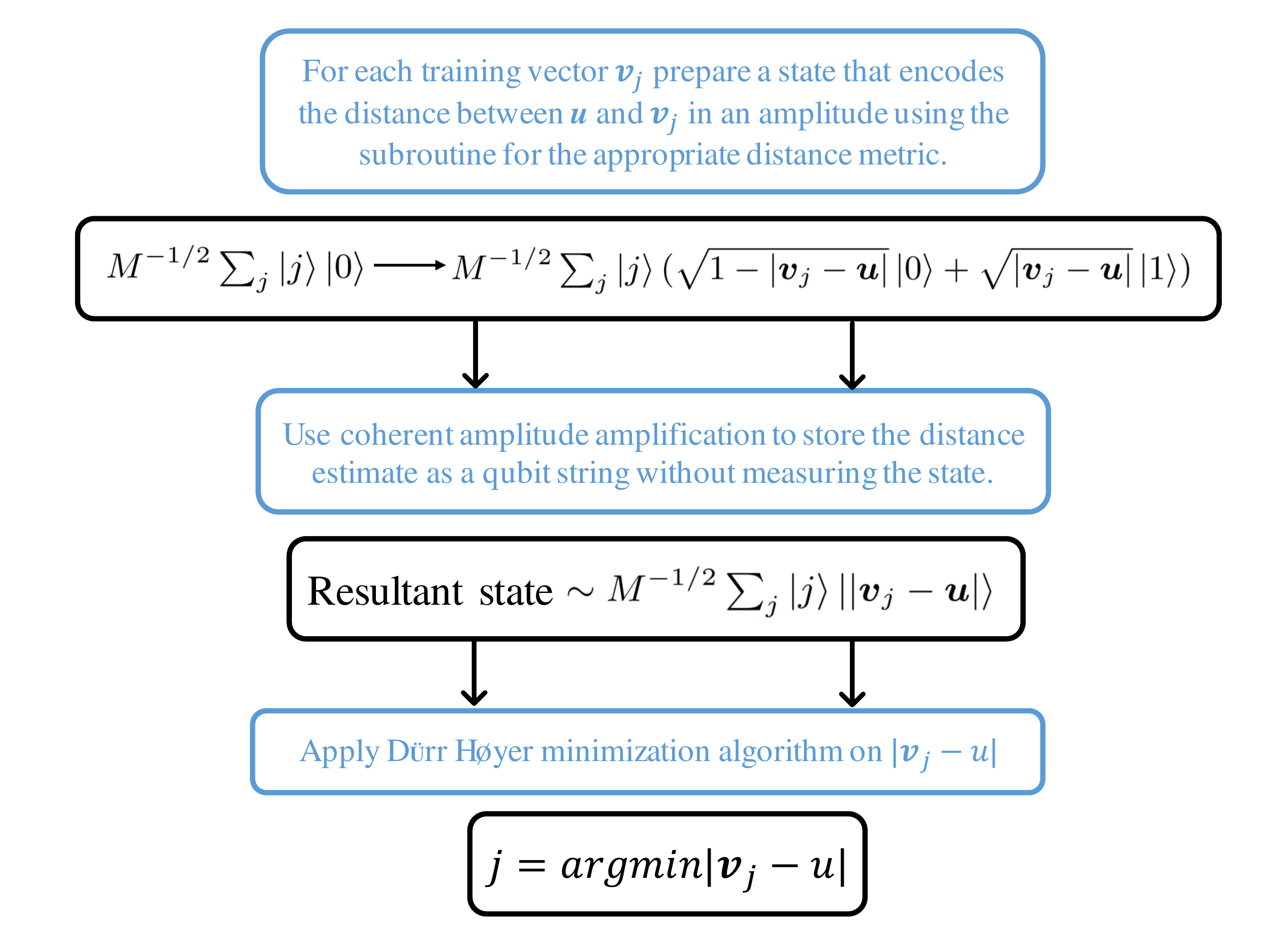}
    \caption{
    {\bf Scheme of structure of the quantum nearest–neighbor algorithm.\cite{wiebe2014quantum}}
    \\
    Firstly for each training vector ${\bf v}_j$, prepare a state that encodes the distance between the test instance ${\bf u}$ and ${\bf v}_j$ in {\color{black} the amplitudes} using the subroutine for the appropriate distance metric.
    Then, use coherent amplitude amplification to store the distance estimate as a qubit string without measuring the state.
    Finally, find the ${\bf v}_j$ {\color{black} that} minimize the distance under certain distance metrics, and ${\bf v}_j$ is the nearest instance.
    Label of the test instance ${\bf u}$ is thus predicted as the same label as ${\bf v}_j$.
    }
    \label{fig_qNN}
\end{figure}

The structure of the quantum nearest–neighbor algorithm is shown in Fig.(\ref{fig_qNN})\cite{wiebe2014quantum}.
The quantum nearest neighbor algorithm can be implemented in briefly three steps\cite{wiebe2014quantum}.
Firstly for each training vector ${\bf v}_j$, prepare a state that encodes the distance between the test instance ${\bf u}$ and ${\bf v}_j$ in an amplitude using the subroutine for the appropriate distance metric.
Then, use coherent amplitude amplification to store the distance estimate as a qubit string without measuring the state.
Finally, find the ${\bf v}_j$ {\color{black} that} minimizes the distance under certain distance metrics, and ${\bf v}_j$ is the nearest instance.
Label of the test instance ${\bf u}$ is thus predicted as the same label as ${\bf v}_j$.

\subsubsection{Decision Trees}\label{Dec_trees}
Decision trees are a way to represent rules underlying data with hierarchical, sequential structures that recursively partition the data\cite{murthy1998automatic}.
In other words, decision trees are trees classifying instances by sorting
them based on their features. 
Each node in a decision tree represents a feature in an instance to be classified, and each branch represents a value that the node can
assume. Instances are classified starting from the root node and sorted based on their specific feature values.

\begin{figure}[hpt]
    \centering
    \includegraphics[width=0.5\textwidth]{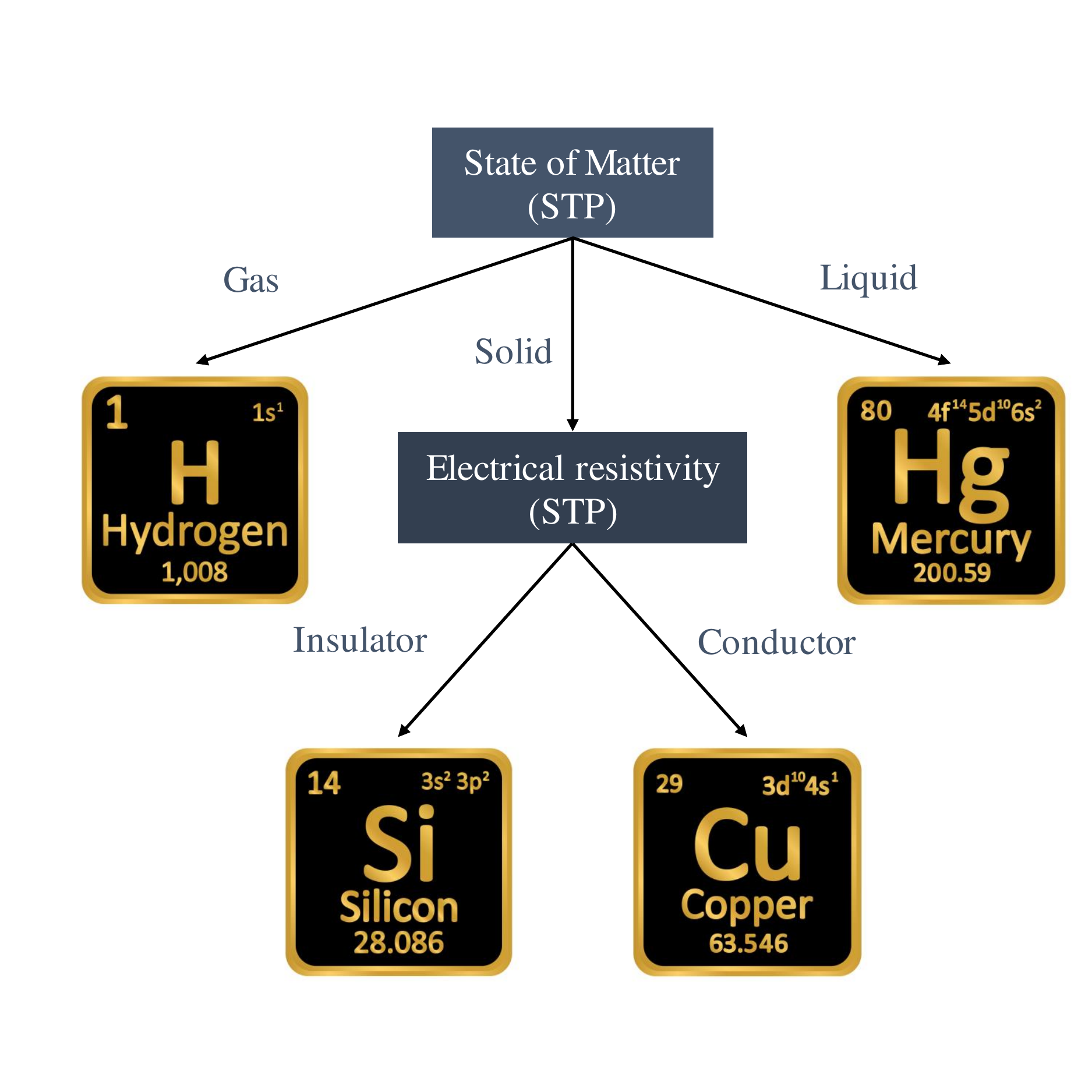}
    \caption{
    {\bf Scheme of the classification process with decision tree.}
    \\
    Instances are classified starting from the first node, or the root node, where we study the state of matter at standard temperature and pressure (STP).
    If the instance is gas, then it will be assigned as Hydrogen.
    If it is liquid, then it will be assigned as Mercury.
    For solid state, we further go to the next node, where we study its electrical resistivity (STP).
    Instance as conductor will be classified as copper, while insulator is classified as silicon.
    For simplicity, we only consider these four chemical substances.
    }
    \label{fig_CDecisionTree}
\end{figure}

A simple example is shown in Fig.(\ref{fig_CDecisionTree}), where four chemical substances are classified with a decision tree model.
Instances are classified starting from the first node, or the root node, where we study the state of matter at standard temperature and pressure (STP).
If the instance is gas, then it will be assigned as Hydrogen. If it is liquid, then it will be assigned as Mercury. For solid state, we further go to the next node, where we study its electrical resistivity (STP).
Instance as conductor will be classified as copper, while insulator is classified as silicon.

Constructing optimal binary decision trees is an NP complete problem, making it possible to find efficient heuristics for constructing near-optimal decision trees\cite{kotsiantis2007supervised}.
The feature that best divides the training data should be assigned as the root node of the tree. 
There are numerous methods for finding the feature that best divides the training data such as information gain\cite{hunt1966experiments} and gini index \cite{breiman2017classification}.
Comparison of individual methods may still be important when deciding which metric should be used in a particular dataset. We shall return to examples of decision trees in Section \ref{State_class_sec}, Section \ref{Drug_discovery}.

\paragraph{Quantum enhanced variants \\}

In 1998, Farhi and coworkers proposed a design of quantum decision tree, which can be experimentally implemented on a quantum computer that consists of enough spin-$\frac{1}{2}$ particles.
They further studied a single time-independent Hamiltonian that evolves a quantum state through the nodes of a decision tree\cite{farhi1998quantum}.
It is proved that if the classical strategy succeeds in reaching level $n$ in time polynomial in $n$, then so does the quantum algorithm.
Moreover, they found examples where the interference allows a class of trees to be penetrated exponentially faster by quantum evolution than by a classical random walk.
Even though, these examples could also be solved in polynomial time by different classical algorithms.

A quantum training dataset with n quantum data pairs can be described as
\begin{equation}
    D = \{(|x_1\rangle, |y_1\rangle), (|x_2\rangle,|y_2\rangle),
    \cdots,(|x_n\rangle, |y_n\rangle)\} 
\end{equation}
where the quantum state $|x_i\rangle$ denotes the $i$th quantum object of the training dataset, and state $|y_i\rangle$ denotes the corresponding label. Due to the existence of superposition, the classical node splitting criteria can hardly work in quantum world. Instead, criterion such as quantum entropy impurity\cite{lu2014quantum} are required to find the optimal features when designing the quantum decision trees.
Recently, a quantum version of the classification decision tree constructing algorithm is proposed\cite{khadiev2019quantum}, which is designed based on the classical version C5.0\cite{wu2008top}.

\subsubsection{Bayesian networks (BN)}
\label{Bayesian_networks}

Bayesian networks (BN) are the most well known representative of statistical learning algorithms, which are graphical models of causal relationships in a given domain.
As definition, BN consists of the following\cite{jensen2007bayesian}:

{\bf 1.}A set of variables and a set of directed edges between variables.

{\bf 2.}Each variable has a finite set of mutually exclusive states. 

{\bf 3.}The variables together with the directed edges form a directed acyclic
graph (DAG).

{\bf 4.}To each variable $A$ with parents $B_1, B_2, \cdots, B_n$, there is attached the
conditional probability table (CPT) $P(A|B_1, B_2, \cdots, B_n)$.

\begin{figure*}[ht!]
    \centering
    \includegraphics[width=0.75\textwidth]{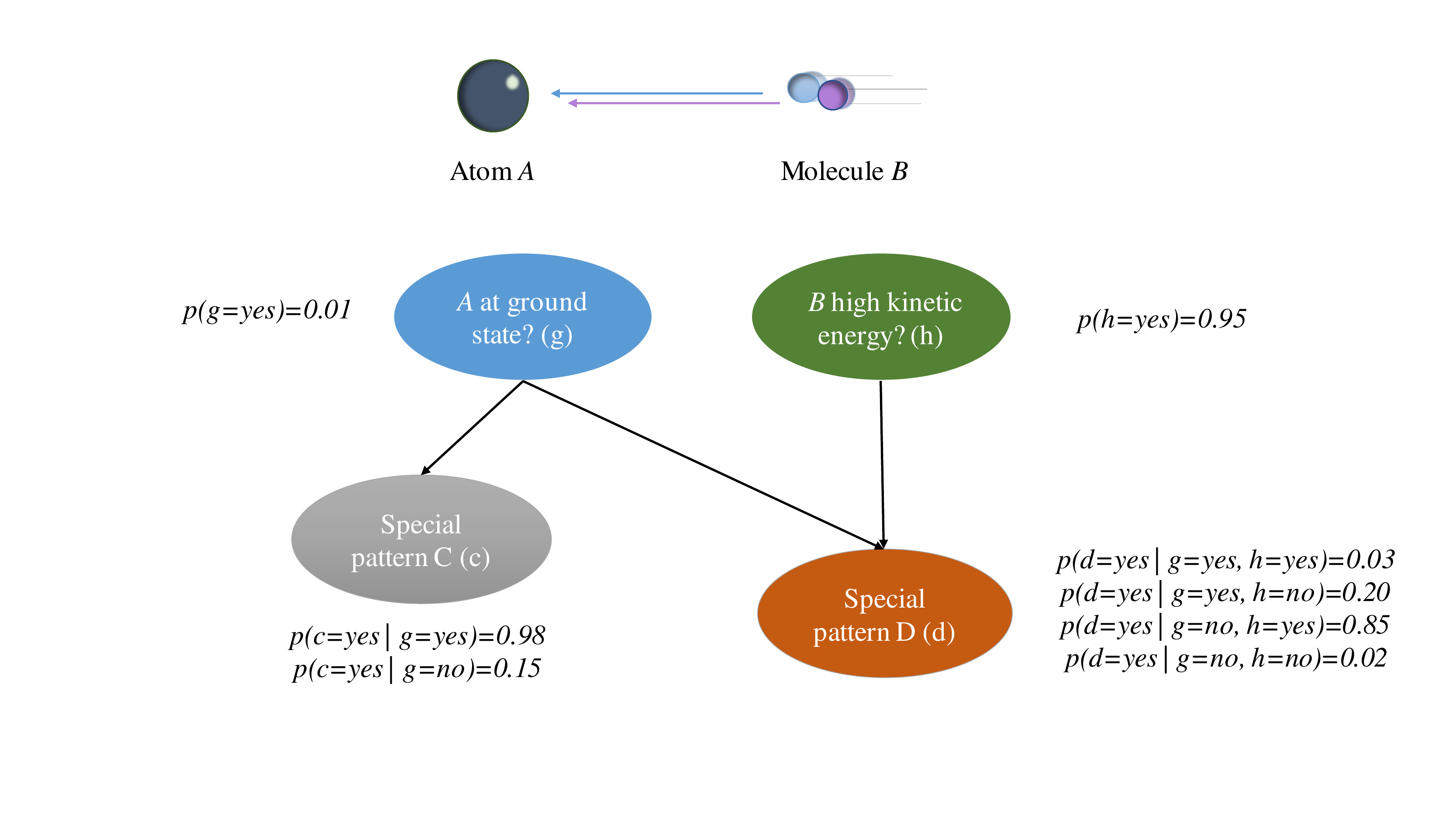}
    \caption{
    {\bf Scheme of the BN assisted study of scattering experiment between atom $A$ and molecule beams $B$.}
    \\
    Atoms $A$ and molecule beams $B$ are initially prepared at certain initial states before the collision.
    The special patterns observed in the scattering experiment results are denoted as pattern $C$ and $D$.
    In the network, Arcs are drawn from cause to effect.
    In chemical reactions we know that the initial states are causes, while the collected results are effects.
    The local probability distribution(s) associated with a node are shown adjacent to the node.
    For simplicity, here we assume that all the features (nodes) are binary, such as the feature $g$ will be set as 'true' or 'yes' as long as the kinetic energy of molecule beams $B$ is equal or greater than some certain threshold.
    }
    \label{fig_BN}
\end{figure*}

The learning process of BN methods generally contains two subtasks, the construction of the DAG network, and the determination of parameters.
The approach to design the structure is based on two observations\cite{heckerman2008tutorial}.
Firstly, people can often readily assert causal relationships among variables.
Secondly, causal relationships typically correspond to assertions of conditional dependence.
In particular, to construct a Bayesian network for a given set of variables, we simply draw arcs from cause variables to their immediate effects.
Sometimes structure of the network is given, then parameters in the CPT usually be learnt by estimating a locally exponential number of parameters from the data provided\cite{jensen2007bayesian}.

Fig.(\ref{fig_BN}) is an example of the BN assisted study of scattering experiment between atom $A$ and molecule beams $B$.
Arcs should be drawn from cause to effect in the network.
In chemical reactions we know that the initial states are causes, while the collected results are effects.
The local probability distribution(s) associated with a node are shown adjacent to the node.
For simplicity, here we assume that all the features (nodes) are binary, such as the feature $g$ will be set as 'true' or 'yes' as long as the kinetic energy of molecule beams $B$ is equal or greater than some certain threshold. We shall return to applications of BN in Section \ref{State_class_sec}.

\paragraph{Quantum enhanced variants \\}

In 1995, Tucci proposed the first design of quantum BN, which could be constructed by replacing real probabilities in classical BN with quantum complex amplitudes\cite{tucci1995quantum}.
Leifer and Poulin proposed another model in 2008, constructing quantum BN based on probability distributions, quantum marginal probabilities and quantum conditional probabilities\cite{leifer2008quantum}.
However, neither of these models could provide any advantage comparing with the classical models, due that they cannot take into account interference effects between random variables\cite{moreira2016quantum}.
A quantum-like BN based on quantum probability amplitudes was proposed by Moreira and Wichert in 2016\cite{moreira2016quantum}, where a similarity heuristic method was required to determine the parameters.

On the other hand, in 2014, Low and coworkers discussed the principles of quantum circuit design to represent a Bayesian network with discrete nodes that have two states.
Notably, it is reported that the graph structure of BN is able to efficiently construct a quantum state representing the intended classical distribution,
and a square-root speedup time can be obtained per sample by implementing a quantum version of rejection sampling\cite{low2014quantum}. 
Recently, Borujeni and coworkers further expanded the quantum representation of generic discrete BN with nodes that may have two or more states\cite{borujeni2021quantum}.
\begin{figure*}[ht!]
    \centering
    \includegraphics[width=0.65\textwidth]{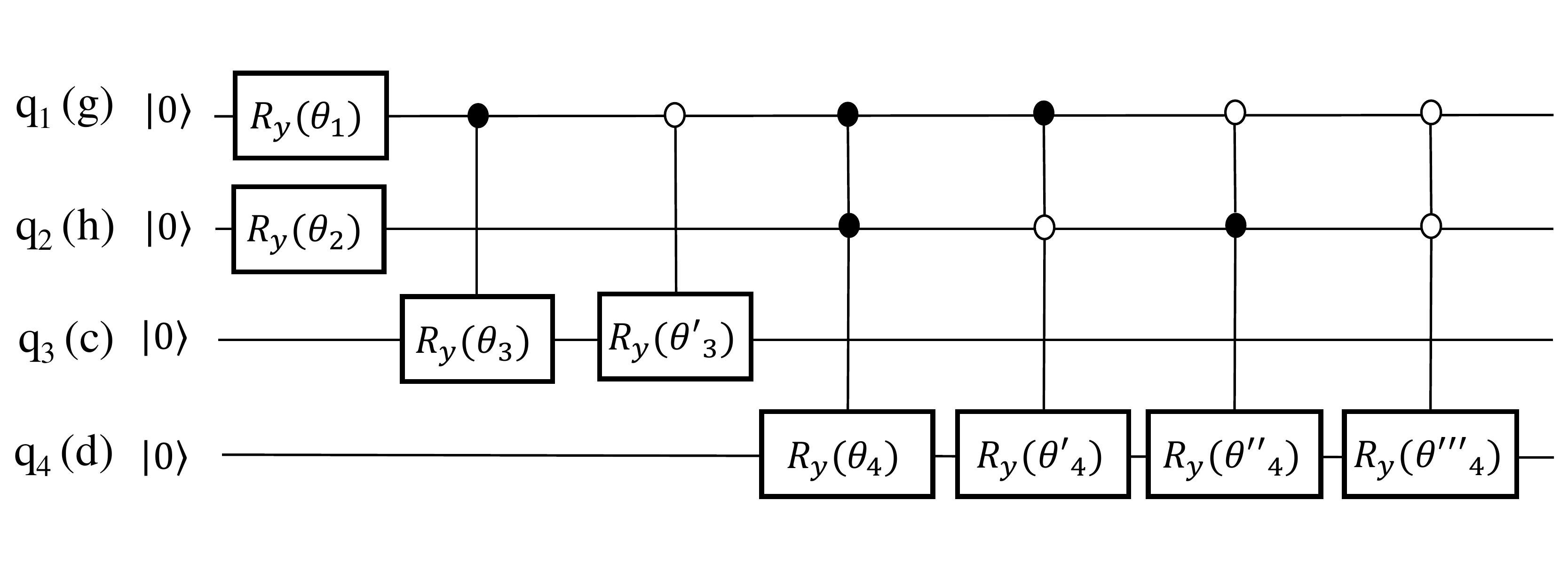}
    \caption{
    {\bf Quantum circuit for the BN shown in Fig.(\ref{fig_BN}). For quantum versions of more complex BN structures refer to \cite{borujeni2021quantum}}
    \\
    There are in total four qubits $q_{1,2,3,4}$ corresponding to the four nodes $g,h,c,d$, all of which are initially set at $|0\rangle$ state.
    $R_y$ gates are applied directly on root nodes to prepare the quantum states corresponding to the probability amplitudes.
    For example, node $g$ is a root node (In other words, there is no arcs pointing to node $g$), so that a single $R_y(\theta_1)$ gate is applied on $q_1$.
    Control-$R_y$ gates corresponds to the arcs in the BN.
    For example, there is only a single arc pointing to node $c$, which comes from node $g$.
    Thus, in the circuit there are two control-$R_y$ gates where $q_1$ is the control qubit and $q_3$ is the gate qubit.
    All the parameters can be derived from the DAG.
    }
    \label{fig_qBN}
\end{figure*}

There are mainly three steps to construct the quantum circuit representing a BN.
Firstly, map a BN node to one or more qubits depending on the number of states.
The next step is to map the marginal or conditional probabilities of nodes to probability amplitudes associated with the qubits to be in $|0\rangle$ and $|1\rangle$ states.
The final step is to realize the required probability amplitudes using single-qubit and controlled rotation gates.

{\color{black}
Fig.(\ref{fig_qBN}) is the quantum circuit representing the BN shown in Fig.(\ref{fig_BN}). The quantum circuit shown in
Fig.(\ref{fig_qBN}) is constructed based on the three steps as the above discussion. The first step is to assign qubits for each node shown in Fig.(\ref{fig_BN}). Here for simplicity, we only assign one qubit for the corresponding node. There are in total four qubits $q_1,q_2,q_3,q_4$ corresponding to the four nodes g, h, c, d, all of which are initially set at $|0\rangle$ state. Next, we need to map the conditional probabilities of nodes to probability amplitudes.
In the BN shown in Fig.(\ref{fig_BN}), there are only two possible results for each node, ’yes’ or ’no’. Here we use quantum state $|0\rangle$ to represent ’no’, and state $|1\rangle$ to represent ’yes’. Then we need to realize the
required probability amplitudes using single-qubit and controlled rotation gates. Single-qubit rotation gates
are applied to construct the independent probability for nodes g and h, as p(g = yes) and p(h = yes)
has nothing to do with the states of other nodes. 
Node $g$ is a root node (In other words, there is no arcs pointing to node $g$ in Fig. (\ref{fig_BN}), so that a single $R_y(\theta_1)$ gate is applied on $q_1$. The value of $\theta_1$ can be derived from the constraint,
\begin{equation}
    p(g=yes)=|\langle1|_{q_1}|\Psi\rangle|^2
\end{equation}
where we denote the final state as $|\Psi\rangle$, and use state $|1\rangle$ to represent 'yes' as said before. Similarly the value of $\theta_2$ can be calculated, as $h$ is also a root node.
The controlled rotation gates are used to construct the
conditional probabilities. For example, to construct $p(c = yes |g = yes)$, we need to build a controlled rotation gates between $q_1$ (control qubit, representing node g) and $q_3$ (target qubit, representing node c).
As the condition is g = yes,the controlled rotation gate works when the control qubit is at state $|1\rangle$, thus there is a solid dot in the corresponding operation in Fig.(\ref{fig_qBN}). On the other hand, when the condition is $g = no$, then controlled rotation gate will work when the control qubit is at state $|0\rangle$, leading to a hollow dot in the quantum circuit. As there are only two arcs in Fig.\ref{fig_BN}, there are two controlled-Ry gates involving $c$ and $g$ in one of which $g=yes$ and in the other $g=no$. The value of $\theta_3$ can be obtained from,
\begin{equation}
    p(c=yes|g=yes)=
    \left|(\langle1|_{q_1}\otimes\langle1|_{q_3})
    |\Psi\rangle
    \right|^2
\end{equation}
Similarly we can obtain $\theta'_3$. To construct the condition probabilities with more than one conditions, we
need to include control rotation gates with more than one control qubits. For example, there are two arcs pointing to node $d$ in Fig.\ref{fig_BN}, one of which comes from node $g$ and the other comes from $h$. Thus, in the circuit there are four control-control-$R_y$ gates where $q_1 (g)$ and $q_2 (h)$ are the control qubits and $q_4 (d)$ is the target qubit corresponding to 4 different choices of configurations between $g$, $h$ i.e. when both are 'yes', both 'no', and one of them is 'yes' and the other is 'no' and vice versa.
The value of $\theta_4$ can be obtained from,
\begin{equation}
    p(d=yes|g=yes,h=yes)=
    \left|(\langle1|_{q_1}\otimes\langle1|_{q_2}\otimes\langle1|_{q_4})
    |\Psi\rangle
    \right|^2
\end{equation}
So that all parameters in the quantum gates are determined by the probability distribution from the DAG in Fig.\ref{fig_BN}.
On the other hand, one could obtain the conditional probability from a given quantum BN by measuring all qubits and estimating the corresponding frequency.
For instance, the probability $p(d=yes|g=yes,h=yes)$ could be estimated by the frequency that $q_1, q_2, q_4$ are all at state $|1\rangle$.
}
For simplicity, in the example we demonstrate a quantum representation of BN with nodes that have only two states ('yes' or 'no').
The quantum representation of more intricate BN structure are discussed thoroughly in Borujeni and coworkers' recent work\cite{borujeni2021quantum}.

\subsubsection{Support vector machine (SVM)}
\label{Supp_vec_machine}

Support vector machines (SVM) revolve around the margin that separates two data classes. Implementation of SVM contains two main steps:
Firstly map the input data into a high-dimensional feature space through some nonlinear methods, and then construct a optimal separating hyperplane.
Support vector machine (SVM) can deal with both regression and classification tasks.
Mathematically, if the dataset ${\bf x}$ is linearly separable and is capable of being assigned into groups denoted by two labels $A$ and $B$, there exist a weight vector ${\bf w}$ and a bias constant $b$, ensuring that
\begin{equation}
    \begin{split}
        &{\bf w}^T{\bf x}_i + b \geq 1,\qquad \forall {\bf x}_i\in A
        \\
        &{\bf w}^T{\bf x}_i + b \leq -1,\qquad \forall {\bf x}_i\in B
    \end{split}
    \label{eq_svm_class}
\end{equation}
Thereby, the classification rule for test instance ${\bf x}_t$ is given by
\begin{equation}
    y_{{\bf w},b}({\bf x}_t) = 
    sgn\left(
    {\bf w}^T{\bf x}_t + b
    \right)
    \label{eq_svm_predict}
\end{equation}
Finding the optimal hyperplane is equivalent to a a convex quadratic programming problem that minimizes the functional\cite{vapnik2013nature}
\begin{equation}
    \Phi({\bf w}) = \frac{1}{2}||{\bf w}||^2
\end{equation}

\begin{figure}[hpt]
    \centering
    \includegraphics[width=0.45\textwidth]{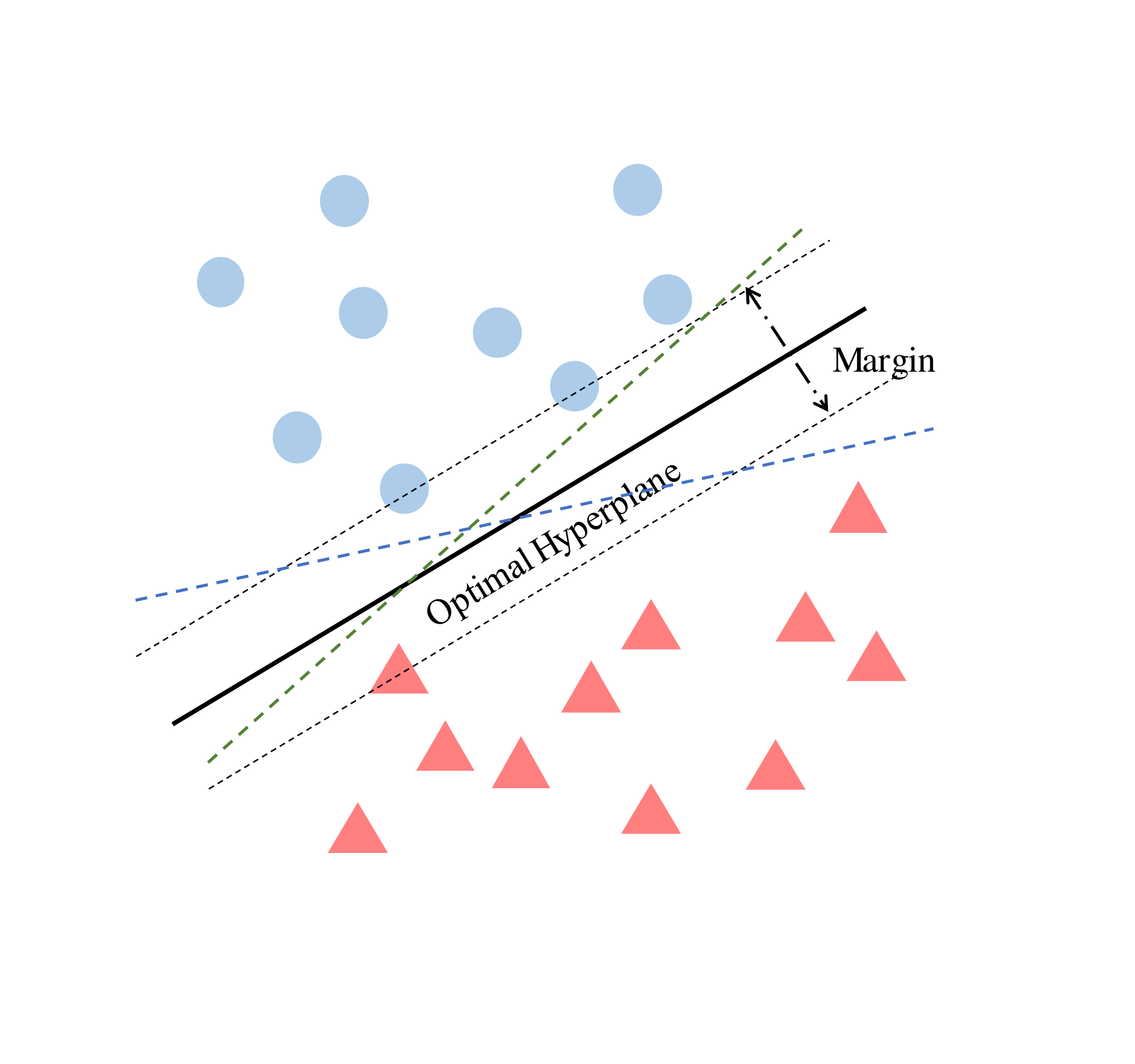}
    \caption{
    {\bf Scheme of the SVM classification with hyperplane.}
    \\
    Blue dots and red triangles represent the training instances with two labels.
    The black line represents the optimal hyperplane, which maximize the margin between the blue and red instances.
    The red and blue dash line are hyperlines that can separate the two groups apart, though the corresponding margin in less than the optimal one.
    }
    \label{fig_SVM}
\end{figure}

Fig.(\ref{fig_SVM}) is a scheme of the SVM classification with hyperplane.
Blue dots and red triangles represent the training instances with two labels.
The black line represents the optimal hyperplane, which maximize the margin between the blue and red instances.
The red and blue dash line are hyperlines that can separate the two groups apart, though the corresponding margin is less than the optimal one.

Sometimes due to the misclassified instances SVMs are not able to find any separating hyperplane that can perfectly separate two groups apart.
Then the soft margin and penalty functions could be applied where some misclassifications of the training instances are accepted\cite{vapnik2013nature}.

\begin{figure*}[ht!]
    \centering
    \includegraphics[width=0.75\textwidth]{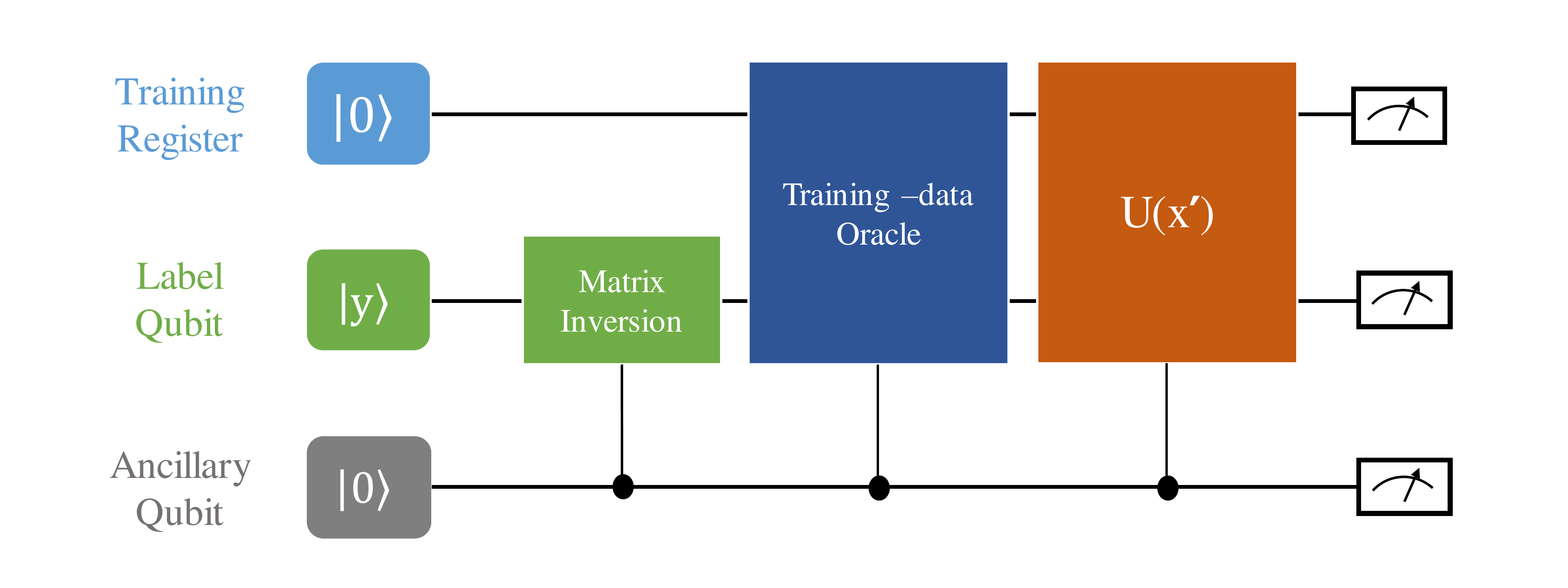}
    \caption{
    {\bf The schematic diagram of quantum SVM illustrated in Ref \cite{li2015experimental} }
    \\
    The qubits can be assigned into three groups: training registers (blue) that represent the training instances, label qubit (green) that takes the label, and ancillary qubit(grey).
    The matrix inversion is employed to acquire the hyperplane parameters.
    Then, the training-data oracle is applied to prepare the training-data state.
    Classification of new test instance ${\bf x'}$ is introduced by operation $U({\bf x'})$.
    }
    \label{fig_qSVM}
\end{figure*}

Moreover, real-world problems often involve non-separable data, where there is no separating hyperplane initially even without misclassifications. Then the training data should be firstly mapped onto a higher dimensional space, where the separating hyperplane would be constructed.
This higher-dimensional space is generally denoted as the transformed feature space, while the training instances occupy the input space.
Instead of repeating the mapping process $\Phi({\bf x})$ explicitly, the more popular approach is to calculate the Kernel functions defined in Eq. \ref{kernel_def}
which allow inner products to be calculated directly in feature space\cite{vapnik2013nature}.
After successfully constructing the hyperplane, new instances are mapped into the feature space by Kernel functions for classification.

SVM methods are binary classification, thus to solve multi-class problems we must reduce the problem into a set of multiple binary classification problems.
A core advantage of SVM is that training the optimization problem of the SVM necessarily reaches a global minimum, instead of being trapped in a local minimum. We shall return to applications in Section \ref{State_class_sec}, Section \ref{Drug_discovery} and in Section\ref{Case_QML}.

\paragraph{Quantum enhanced variants \\}

Enthused by the success of SVM assisted big data classification, Rebentrost and coworkers proposed the implementation of quantum SVM\cite{rebentrost2014quantum}.

Rewrite the weight vector ${\bf w}$ in Eq.(\ref{eq_svm_class}) and Eq.(\ref{eq_svm_predict}) as
\begin{equation}
    {\bf w} = \sum_{j=1}^M \alpha_j {\bf x}_j
\end{equation}
where $\alpha_j$ is the weight of the $i$th training instance ${\bf x}_j$, and there are $M$ training instances in total.
In the SVM with least-squares approximation, the optimal parameters $\alpha_j,b$ can be obtained by solving a linear equation\cite{rebentrost2014quantum}
\begin{equation}
    F(b, \alpha_1, \alpha_2,\cdots,\alpha_M)^T
    =
    (0, y_1, y_2, \cdots, y_M)^T
\end{equation}
where $F$ is a $(M+1)\times(M+1)$ matrix with the essential part as the
kernel. Fig.(\ref{fig_qSVM}) is a diagram of the quantum SVM\cite{li2015experimental}.
{\color{black}We can rewrite the classification rule as
\begin{eqnarray}
 y(x^\prime) = sgn[
\langle\psi|\hat{O}|\psi\rangle]
\end{eqnarray}
where $|\psi\rangle$ is the final quantum state. The big picture is that if the expectation value in the above equation is greater than zero, then the test instance $x^\prime$ will be assigned as label positive ($y(x^\prime) = 1$). Otherwise, it will be predicted with negative label ($y(x^\prime) = -1$). 

The circuit has three primary components in a nutshell as seen in Fig.\ref{fig_qSVM}: Matrix inversion operation
(green) is designed to acquire the hyperplane parameters; training-data oracle (blue) is included to prepare
the training-data state; and $U(x^\prime)$ is to map the test instance $x^\prime$ into quantum states. In classical SVM, the hyperplane is obtained by minimizing the functional as shown in Eq.\ref{eq_svm_predict}, while in qSVM, the hyperplane is obtained via solving linear equations, which leads to an exponential speedup.}

Let us now get into the details of the quantum version of the algorithm. The qubits can be assigned into three groups: training registers (blue) that represent the training instances, label qubit (green) that takes the label, and ancillary qubit(grey).
The matrix inversion is employed to acquire the hyperplane parameters. Then, the training-data oracle is applied to prepare the training-data state.
Classification of new test instance ${\bf x'}$ is introduced by operation $U({\bf x'})$.

The the training-data oracles are designed to return the quantum counterpart of the training data ${\bf x}_i$,
\begin{equation}
    |{\bf x}_i\rangle = 
    \frac{1}{|{\bf x}_i|}\sum_{j=1}^N
    \left(
    {\bf x}_i
    \right)_j|j\rangle
\end{equation}
where $\left({\bf x}_i\right)_j$ is the $j$th component of the training instance ${\bf x}_i$.
The training-data oracles will convert the initial state $1/\sqrt{M}\sum_{i=1}^M|i\rangle$ into the state $|\chi\rangle$, where
\begin{equation}
    |\chi\rangle=\frac{1}{\sqrt{N_\chi}}
    \sum_{i=1}^M|{\bf x}_i||i\rangle|{\bf x}_i\rangle
\end{equation}
with $N_\chi=\sum_{i=1}^N|{\bf x}_i|^2$ is normalization factor.

Optimization is implemented by the quantum algorithm solving linear equations, which provide exponential speedup comparing to the classical version\cite{harrow2009quantum}.
Registers are initialized into state $|0,{\bf y}\rangle = (1/\sqrt{N_{0,y}})(|0\rangle + \sum_{i=1}^My_i|i\rangle)$.
After applying the matrix inversion operation, the quantum state is transformed to
\begin{equation}
    |b,{\bf\alpha}\rangle
    =
    \frac{1}{\sqrt{N_{b,\alpha}}}
    \left(
    b|0\rangle + \sum_{i=1}^M\alpha_i|i\rangle
    \right)
\end{equation}
With the optimal parameters $\alpha_j,b$, the classification rule corresponding to Eq.(\ref{eq_svm_predict}) can be written as
\begin{equation}
    y({\bf x}') = 
    sgn\left[
    \sum_{i=1}^M
    \alpha_i ({\bf x}_i\cdot{\bf x}')
    + b
    \right]
    \label{eq_qsvm_predict}
\end{equation}
where for simplicity, the linear Kernel is considered.
Classification result will be derived by measuring the expectation value of
the coherent term $\hat{O}=|00\rangle\langle\otimes(|1\rangle\langle0|)_A$,
where the subscript $A$ denotes the ancillary qubit.

{
\color{black}
In spite of constructing quantum circuits to acquire the hyperplane, researchers further developed quantum kernel methods which harnesses the computational power of quantum devices.
In 2019, researchers from Xanadu proposed to compute a classically intractable kernel by estimating inner products of quantum states\cite{schuld2019quantum}, while the kernel can then be fed into any classical kernel method such as the SVM.
The crucial component of quantum kernel methods is quantum feature maps, which map a classical data point $x$ as an n-qubit quantum state $|\phi(x)\rangle$ nonlinearly, where the feature state $|\phi(x)\rangle = U(x)|0\rangle$ is obtained by a parameterized circuit family $\{U(x)\}$\cite{liu2021rigorous}.
In the learning process, quantum feature maps take the position of pattern recognition.
More details about the quantum kernel methods could be found in Sec.(\ref{Case_QML}).
}

\subsubsection{Gaussian Process Regression} \label{GPR_section}
\begin{figure*}[ht!]
    \centering
    \includegraphics[width=0.85\textwidth]{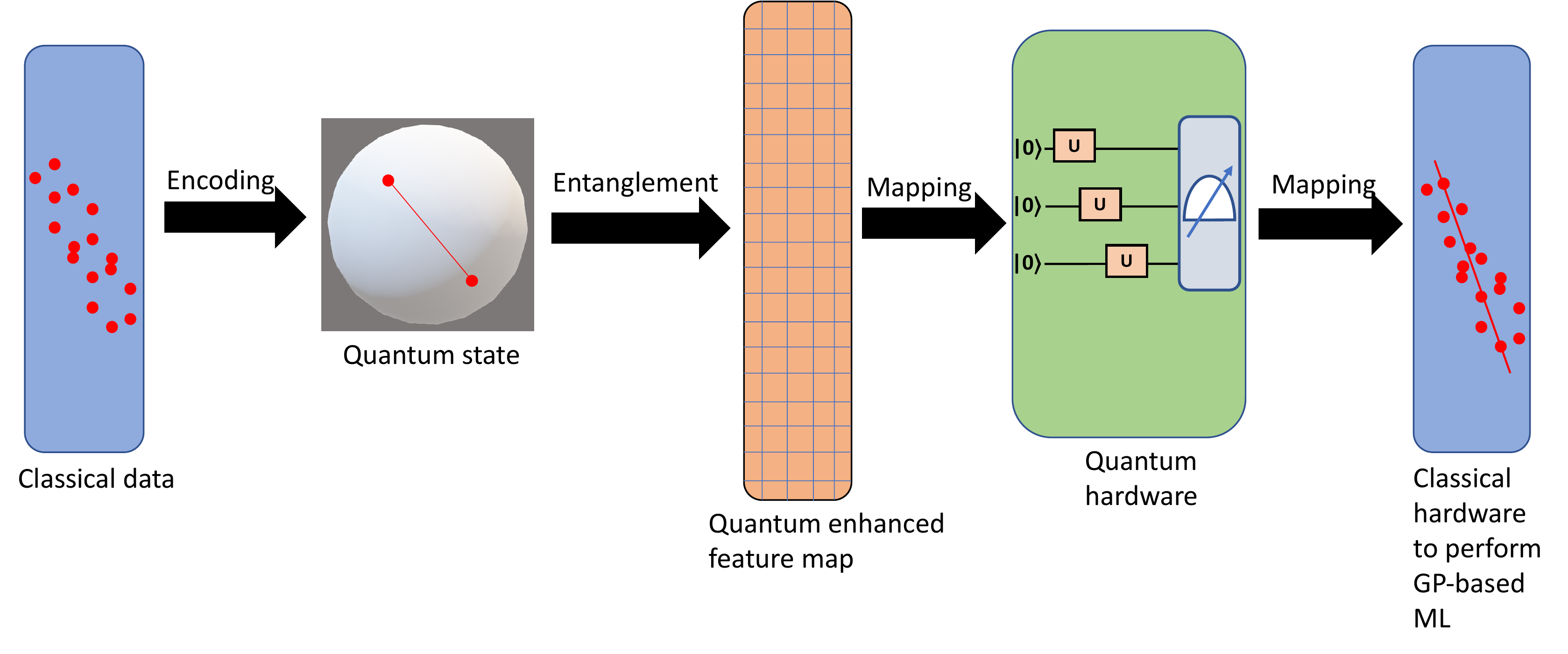}
    \caption{\textbf{Schematic representing quantum-enhanced kernel for Gaussian Process regression as described in Ref.\cite{otten2020quantum}.} }

\end{figure*}
Gaussian Process Regression (GPR) \cite{williams2006gaussian}, a non-parametric and supervised learning method that has become quite popular in the ML setting for Chemistry applications \cite{deringer2021gaussian}. It is based on the Bayesian approach, where a probability distribution over all possible values for the parameters is inferred by the ML model. Considering the input vector x and output y, a function of x, $f(x)$ with its functional form unknown, maps the d-dimensional vector to a scalar value: $y : \mathbb{R}^d \rightarrow \mathbb{R}$. The training set $D$ is made up of $n$ observations, $D = {\{(x_i, y_i) | i=1,....,n}\}$  Performing regression to predict the form of y can be obtained in two ways:

\begin{itemize}
    \item \textbf{Weight-space picture:} Having parameterized the function $f$, a prior is placed on the parameters of the model. Using the Bayes' Rule, the probabilities are modified based on the observed data and the distribution is updated (called the posterior distribution). Then, the predictive posterior distribution on the points $x_n$ is calculated by weighting all the possible predictions by their respective calculated posterior distributions. In order to improve the expressiveness of the model, the inputs are projected into a high dimensional space using a set of $M$ basis functions to approximate $y(x)$ by $\Tilde{y}(x)$:
    \begin{equation}
        \Tilde{y}(x) = \sum_{m=1}^M c_m k(x, x_m)
    \end{equation}
    where, $k$ is the kernel of choice placed on the representative set of input points, and $c_m$ are the associated weights. By choosing the Gaussian kernel, the model is fit to the data by finding the coefficients $c = (c_1, ...., c_M)$, that minimizes the loss:
    \begin{equation}
        L = \frac{\sum_{n=1}^N [y_n - \Tilde{y}(x_n)]^2}{\sigma_N^2} + \sum_{m, {m'}}^M c_m k(x_m, x_{m'})c_{m'}
    \end{equation}
    where the second term is the Tikhonov regularization.

    \item \textbf{Function-space picture:} The prior in this case is specified in the function space. For every $x \in \mathbb{R}^d$, the distribution of $f(x)$ along with the structure of covariance $k(x, x') = cov(f(x), f(x'))$ are characterized. A Gaussian process (GP) is used to describe a distribution over functions. A GP is completely specified by its mean function $m(x)$ and covariance function $(k(x, x')$):
    \begin{equation}
        m(x) = \mathbb{E}[f(x)], 
    \end{equation}
    \begin{equation}
        k(x, x') = \mathbb{E}[(f(x) - m(x))(f(x') - m(x'))]
    \end{equation}
    The GP can be written as:
    \begin{equation}
        f(x) \sim GP(m(x), k(x, x'))
    \end{equation}
    $\Tilde{y}(x)$ in this case is written as:
    \begin{equation}
        \Tilde{y}(x) = \sum_{h}^H w_h \phi_h(x)
    \end{equation}
    where, $\phi$ represent the basis functions that are fixed, which are independent of data and indicate the probability distribution of functions, and $w$ are the weights drawn from independent, identically distributed (i.i.d) Gaussian probability distributions. 
    Considering the squared exponential as the covariance function:
    \begin{equation}
        cov(f(x), f(x')) = exp(-\frac{1}{2} |x - x'|^2)
    \end{equation}
    which corresponds to a Bayesian linear regression model with infinite number of basis functions. Samples are drawn from the distribution of functions evaluated at a specified number of points and the corresponding covariance matrix is written elementwise. Then, a random Gaussian vector is generated with the covariance matrix and values are generated as a function of inputs. 
\end{itemize}
We shall return to applications of SVM in Section \ref{State_class_sec}, Section \ref{Drug_discovery}

{\color{black}
\paragraph{Quantum enhanced variants \\}

Matthew Otten et al. \cite{otten2020quantum} proposed a procedure to build quantum enhanced kernels while still capturing the relevant features of the classical kernels. As can be seen from the weight-space picture above, the quality of regression results is directly influenced by the choice of the kernel. Quantum computing enhanced kernels have the potential of being powerful in terms of performing higher dimensional regression tasks since quantum computers can represent functions that classical computers might not calculate efficiently. As coherent states approximate the squared exponential kernel, the classical feature maps corresponding to the squared exponential kernel can be first approximated using coherent states, which leads to a corresponding quantum kernel. A generic coherent state with parameter $\alpha$ can be written as: 
\begin{equation}
\ket{\alpha} = e^{|\alpha|^2/2} \sum_{n=0}^{\infty} \frac{\alpha^n}{\sqrt{n!}} \ket{n}
\end{equation}

The input data is encoded as $\alpha_i = x_i/(\sqrt{2}c_i)$, leading to the coherent state kernel: 
\begin{equation}
k(x,x') = s\prod_i |\braket{\frac{x_i}{\sqrt{2}c_i}|\frac{x'_i}{\sqrt{2}c_i}}|^2
\end{equation}

Since the coherent state can be written in terms of the displacement operator applied to the vacuum state, and truncating the Hilbert space at some maximum number of levels $N$, gives rise to the s finite-dimensional displacement operator $\tilde{D}_N(\alpha) = e^{\alpha(\tilde{b}^\dagger_N - \tilde{b}_N)}$, where $\tilde{b}^\dagger_N$ is the bosonic creation operator in the finite-dimensional Hibert space. 

The finite-dimensional coherent state based kernels are first prepared on a qubit system by decomposing the N level displacement operator into $log_2(N)$ Pauli operators and then using Trotterization upto $m$ steps on the qubit Hamiltonian. This defines the quantum feature map that approximates the feature map of the classical exponential squared kernel. Classically insipred quantum feature maps can then be applied to solve the requisite regression task.

In order to show a quantum advantage, entanglement enhanced kernel can be prepared by using multi-mode squeezing operator to entangle the different data dimensions for a multi-dimensional regression problem. Thereby, smaller quantum devices with only a few operations can perform higher-dimensional regression tasks.
Following this, the GP-based ML task is performed on the classical hardware.}

\subsection{Artificial Neural networks}\label{ANN_section}
In this section we briefly review the various architectures of neural network or deep learning algorithms that are commonly used
for applications in physics and chemistry. As before, we not only focus on the training of each such architecture on a classical computer but also on the quantum algorithms proposed wherever applicable. Applications of NN are discussed in all sections from Section\ref{Case_QML}, Section \ref{state_prep}, Section \ref{State_class_sec}, Section \ref{MBS_sec}, Section \ref{FF_sec} and in Section \ref{Drug_discovery}.

\subsubsection{Perceptron and Feed forward-neural networks}\label{DNN_section}


A perceptron is a single artificial neuron which models a non-linear function of the kind $f : x \mapsto y$ where $x \in \mathbb{R}^d$ and $y \in \mathbb{R}$ \cite{bishop2006pattern, Freuend_percept, al2017development}. The $d$ - dimensional vector $x$ is an input and the single number $y$ is the output. The perceptron layer in between first makes an affine transformation on the input $x$ using tunable parameters $w \in \mathbb{R}^d$ (often called \textit{weights}) and $b \in \mathbb{R}$ (often called \textit{bias}). This transformation is as follows:
\begin{eqnarray}
    z = w^T x + b
\end{eqnarray}
From the above transformation, it is clear that the weight vector $w$ strengthens or weakens the importance of each element in the input through multiplicative scaling. The bias $b$ physically sets a threshold when the neuron would `fire' as would be clarified soon. Non-linearity is thereafter introduced by passing this affine transformed variable $z$ as an input argument through an activation function (say $\sigma$). The output so obtained is the final output of the perceptron $y$ as follows:
\begin{eqnarray}
    y &=& f(x) = \sigma (z) \nonumber \\
      &=& \sigma (w^T x + b)
\end{eqnarray}
In Rosenblatt's model of perceptron \cite{rosenblatt1958perceptron} the activation function used was a step function i.e. $\sigma(z)=1$ if $z \ge 0$ but $\sigma(z)=0$ otherwise. It was essentially a linear classifier. However, more sophisticated and continuous activation functions commonly used now are as follows:
\begin{figure*}[ht!]
    \centering
    \includegraphics[width=1.0\textwidth]{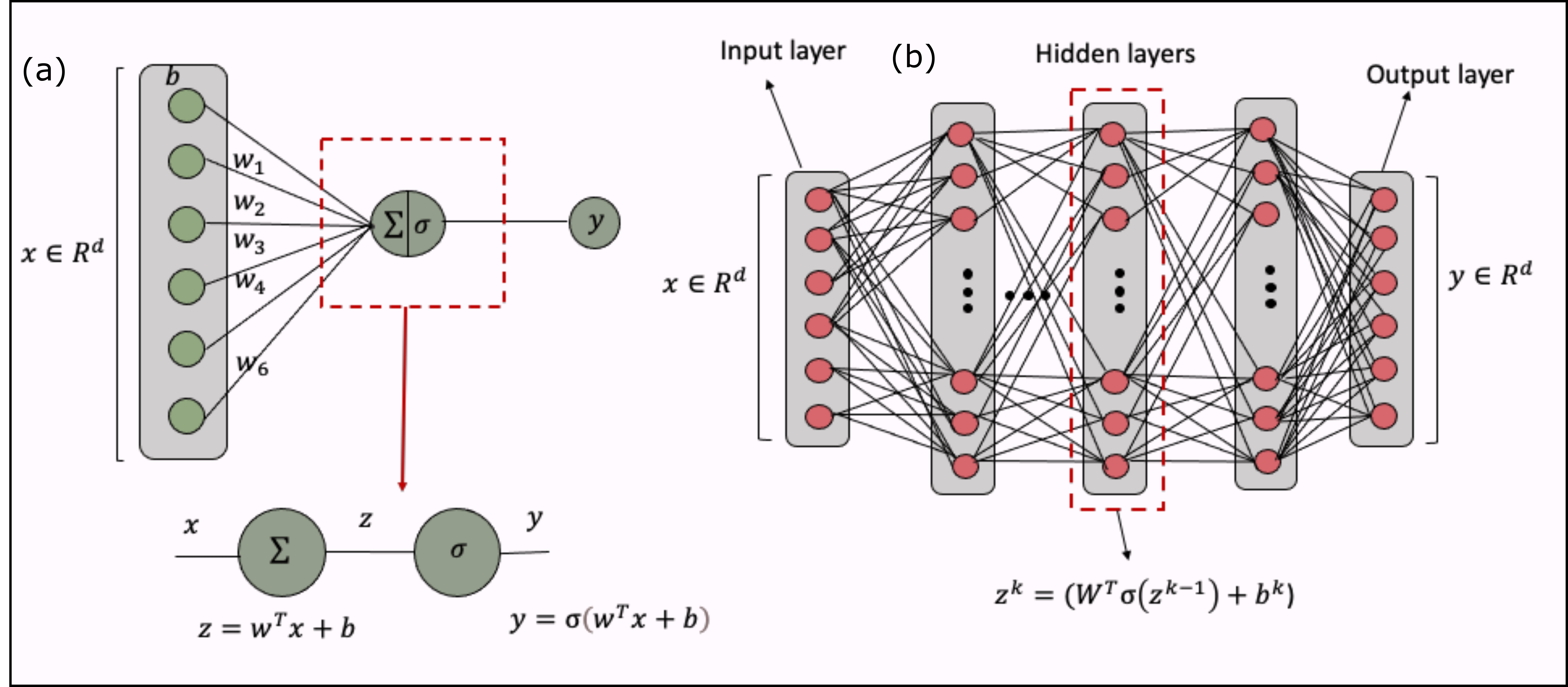}
    \caption{(a) A schematic of a generalized perceptron. The input is a vector $x \in \mathbb{R}^d$ and the output $y \in \mathbb{R}$. The parameters of the model are the $w \in \mathbb{R}^d$ (often called \textit{weights}) and $b \in \mathbb{R}$ (often called \textit{bias}). The layer in between performs an affine transformation to yield variable $z$ and passes $z$ as the argument of the non-linear activation function $\sigma$. Note that for Rosenblatt's perceptron\cite{rosenblatt1958perceptron} $\sigma(z) = 1$ if $z \ge 0$ and 0 otherwise but any generalized activation function would be fine (see text for more details). (b) A feed-forward neural network obtained by stacking many neurons in several layers. The layers have all to all connectivity pattern that may not necessarily be the case. The input is  $x \in \mathbb{R}^d$ and the output , unlike in the case of a perceptron is  $y \in \mathbb{R}^m$ (for the figure m=d is shown but that may not be necessarily true). Each layer much like in the case of a perceptron performs an affine transform and then a non-linear activation. The case for the $k$-th layer is shown wherein the affine transformed variable is $z^k$ which is subsequently passed into an activation function (see text for details)}
    \label{ANN_fig}
\end{figure*}

{\noindent\bf Logistic activation function:}

The original idea of a perceptron is to model a biological neuron in the central nervous system. The activation function serves the purpose of mimicking the biological neuron’s activation rate. Logistic functions are typical activation functions having a similar representation to a biological neuron’s activation rate\cite{haykin2004comprehensive}. Sigmoid function is a traditional type of logistic functions. Sigmoid function is an increasing function with `S' shape, assuming a continuous range of values from 0 to 1, as described with Eq.(\ref{eq_sigmoid}).
\begin{equation}
    \sigma_{sigmoid}(z) = \frac{1}{1+\exp(-\alpha z)}
    \label{eq_sigmoid}
\end{equation}
where $\alpha$ is the slope parameter.
Notice that the sigmoid function centers at 0.5, which might slow down the learning process.
Besides, the gradient of sigmoid function for the data fallen in the region of either 0 or 1 are almost zero, which causes the network performance degrades\cite{lau2018review}.
Therefore, hyperbolic tangent (tanh) function is introduced as another type of logistic activation function, which is the rescaled and biased version of the sigmoid function.
The tanh function is defined as follows,
\begin{equation}
    \sigma_{tanh}(z) = \frac{\exp(2z)-1}{\exp(2z)+1}
    \label{eq_tanh}
\end{equation}
Furthermore, there is adaptive hyperbolic tangent activation function with two trainable parameters $\beta$ and $\alpha$ to adjust the slope and amplitude of the $\tanh$ activation function throughout the training process.
The adaptive $\tanh$ activation function is defined as
\begin{equation}
    \sigma_{adaptive tanh}(z) = \alpha\frac{\exp(2\beta z)-1}{\exp(2\beta z)+1}
\end{equation}
Both the sigmoid function and tanh function are saturated activation function, as they squeeze the input (Sigmoid function squashes real numbers to range between $[0, 1]$, while tanh function squashes real numbers to range between $[-1, 1]$).\\

{\noindent\bf Rectified linear unit (ReLU):}
Rectified linear unit (ReLU) is defined as
\begin{equation}
    \sigma_{ReLU}(z) = \max(0,z)
\end{equation}
Due to its simplicity, ReLU is a popular activation function in ANN. ReLU is more efficient than other functions because as all the neurons are not activated at the same time, rather a certain number of neurons are activated at a time\cite{sharma2017activation}.
If we would like to activate the neuron in the negative region, the Leaky ReLU (LReLU) might be an appropriate choice, where we could set the negative region with a small constant value\cite{maas2013rectifier}.
The LReLU is defined as,
\begin{equation}
    \sigma_{LReLU}(z) =\left\{
    \begin{split}
        &z, &\qquad z\geq0
        \\
        &bz, &\qquad z \le 0
    \end{split}
    \right.
\end{equation}
Both the ReLU and LReLU are non-saturating activation functions.\\

{\noindent\bf Exponential linear unit:}
Exponential linear unit(ELU) is defined as
\begin{equation}
    \sigma_{ELU}(z) =\left\{
    \begin{split}
        &a(e^z-1), &\qquad z\geq0
        \\
        &z, &\qquad z<0
    \end{split}
    \right.
\end{equation}
ELU has a similar shape with LReLU, while it performs better than ReLU in batch normalization.\\

{\noindent\bf Multistate activation function (MSAF):}
Instead of combining numerous perceptrons with simple logistic activation functions, it is a simple way to achieve a N-state neuron by using a N-level activation function for real-valued neuronal states.
Thus multistate activation functions (MSAF) are applied in ANN, which are generally multilevel step functions.
As an example of MSAF, the N-level complex-signum activation function is defined as\cite{jankowski1996complex}
\begin{equation}
    \sigma_{csign}(z) =  CSIGN_N\left[\exp(\frac{i\theta_N}{2})\cdot z\right]
\end{equation}
where $\theta_N=2\pi/N$, and
\begin{equation}
    CSIGN_N(z) = \exp(in\theta_N), \qquad \arg(z)\in[(n-1)\theta_N, n\theta_N),\quad n=1,2,\cdots, N
\end{equation}
The complex-signum activation function is often applied in the associative memory models based on Hopfield-type neural networks\cite{tanaka2009complex}. Picking up the appropriate activation function is always essential in the classical ML.
More discussion of the performance analysis of various activation functions can be found in Refs\cite{karlik2011performance, agostinelli2014learning}.

A perceptron is trained by seeing if the output value $y$ matches with the true or expected value. If such a matching did not happen based on some pre-defined metric then the parameters of the neuron $(w,b)$ are optimized so that the output of the network matches up to the desired value. In Rosenblatt's perceptron \cite{rosenblatt1958perceptron}, this optimization was done by simply adding the input vector $x$ to the weights $w$ if the perceptron underestimated the output value compared to the true label and subtracting the $x$ if the perceptron over-estimated the output value compared to the true label. The bias $b$ was updated by $\pm 1$ in the two cases respectively as well. Once the output of the perceptron agrees with the label the neuron is said to have `learnt' to perform the task.

Each such perceptron described is essentially equivalent to a biological neuron. A feed-forward neural network is obtained by stacking many such neurons, layer by layer such that the neuron in one layer are connected to those in the other layer. Operationally, the network models a non-linear function of the kind $f : x \mapsto y$ where $x \in \mathbb{R}^d$ and $y \in \mathbb{R}^m$. The $d$ - dimensional vector $x$ is an input and the $m$ dimensional vector $y$ is the output. If the network has $L$ layers of stacked neurons, this would mean that the first (input) and the last (output) layer will have $d$ and $m$ neurons respectively. The layers in between are called hidden layers. Let us concentrate on the $k$-th and ($k-1$)-th layers ($(k, k-1) \in \{1,2,...L\}$) only. The affine transformation defined at the $k$-th layer will be parameterized by a \textit{weight} matrix $W \in \mathbb{R}^{p \times q}$ where $q$ is the number of neurons in the $k$-th layer and $p$ is the number of neurons in the ($k-1$)-th layer and also by a \textit{bias} vector $b^k \in \mathbb{R}^{q}$ \cite{bishop2006pattern, Goodfellow-et-al-2016}. The transformation acts on the activation response of the ($k-1$)-th layer i.e. $\sigma(z^{k-1}) \in \mathbb{R}^{p}$ as follows :
\begin{eqnarray}
    z^{k} = W^T \sigma(z^{k-1}) + b^k
\end{eqnarray}
The transformation thus yields a new vector $z^{k} \in \mathbb{R}^{q}$ which is passed through an activation process using any of the activation functions defined before and fed into the next layer. This process is repeated until one reaches the last layer. At the last $L$-th layer the activation response is $z^{L}= y$. This is compared with the true values /labels of the data (say $y^{*}$). Many such metric for the comparison can be defined, one simple example being $L^2$ norm of the difference vector $||y-y^{*}||_2$ or even cross-entropy \cite{de2005tutorial} if both $y$ and $y^{*}$ are probability distributions etc. Such metrics are often called merit-functions or cost functions. Once a cost-function is defined, the error in the metric is decided and that is used to evaluate the gradient of the cost-function with respect to the \textit{bias} parameters of the $L$-th layer and the interconnecting \textit{weights} between the $L$-th and  $(L-1)$-th layer. The process is then repeated for all the layers up until one reaches the first layer. At the end , one then has access to gradient of the cost function with respect to the tunable parameters of all the layers. This method of acquiring the gradient is called back-propagation \cite{rumelhart1986learning, lecun2015deep}. Once all such gradients are obtained, one can update the parameters of the entire network using simple gradient descent \cite{ruder2016overview} or sophisticated optimizers like ADAGRAD \cite{duchi2011adaptive}, RMSprop \cite{ruder2016overview}, ADAM\cite{kingma2014adam, ruder2016overview}, NADAM\cite{dozat2016incorporating} etc. When the error metric has decreased below a certain preset threshold, the network is said to have been `trained' to perform the task. At this point, predictions of the network are usually cross-validated using data outside that of the labelled training examples. It must be noted that often the term multi-layer perceptron is used interchangeably for feed-forward neural networks even though historically as mentioned above the training algorithm of perceptrons are slightly different. For fairly large neural-networks with many neurons stacked within each layer, the risk of overfitting the data exists. This can be handled using appropriate regularization techniques \cite{wan2013regularization, girosi1995regularization} or dropout \cite{srivastava2014dropout}.

\begin{figure*}[ht!]
    \centering
    \includegraphics[width=0.85\textwidth]{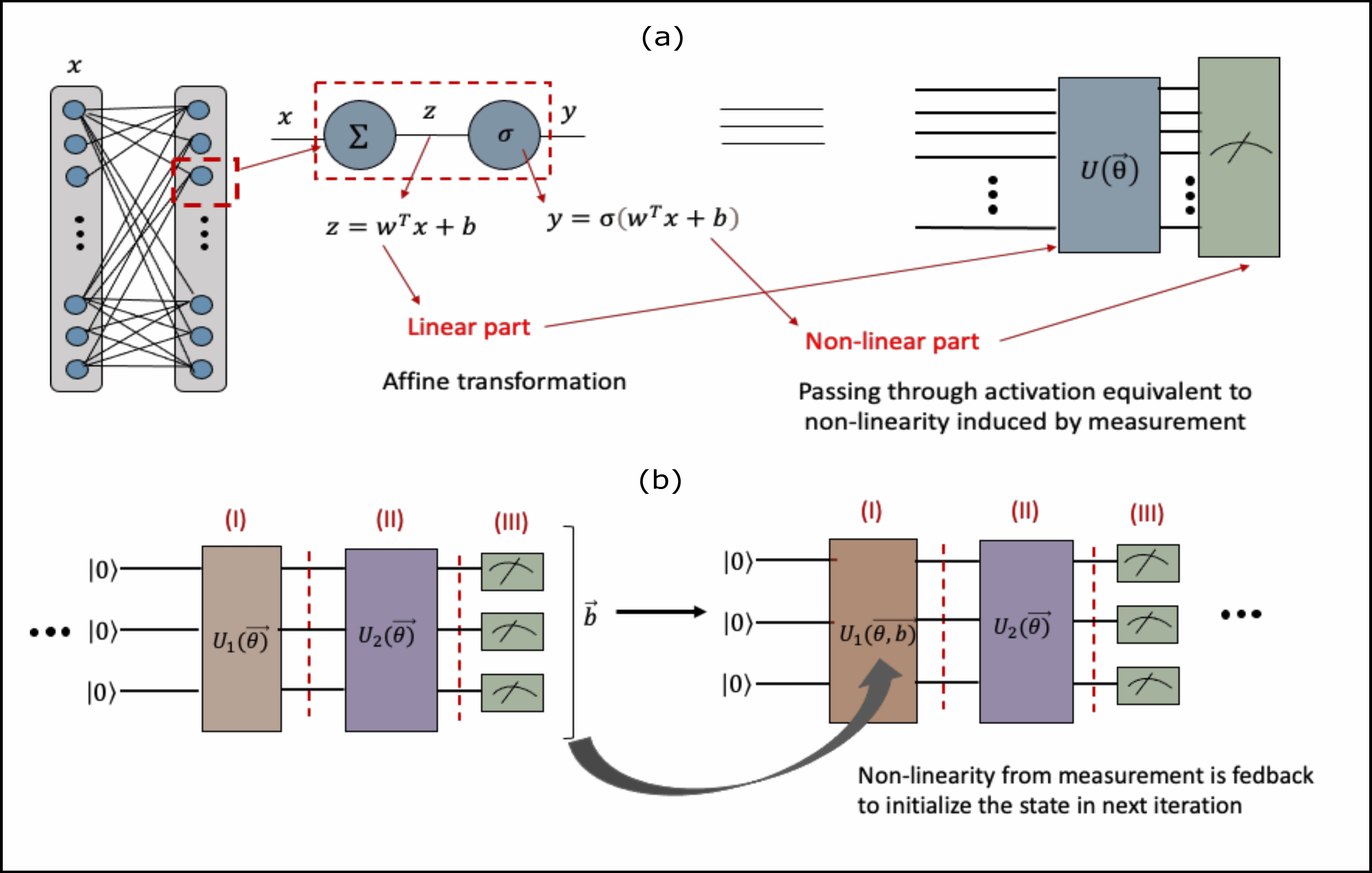}
    \caption{(a)shows a scheme of the structure of quantum-classical hybrid neural network for realization of ANN. The linear part is accomplished with parameterized unitaries which defines the tunable \textit{weights} and \textit{biases} of the network whereas the non-linear activation is obtained from measurements in the quantum-classical hybrid neural network. (b) shows an example of the construction of the hybrid quantum-classical neural network for 3 qubits. Stage (I) refers to state-encoding with unitary $U_1$
    , stage 2 is the actual variational circuit with parameterized unitary $U_2$ and stage 3 is measurement to reproduce the effect of non-linear activation. In the next iteration the measurement results $\vec{b}$ is used in the unitary $U_1$ for state-encoding. This way the full feed-forward neural network proceeds. Training is done by variation of the parameters of $U_2$. (See Ref\cite{xia2020hybrid} for details)}
    \label{fig_activation_1}
\end{figure*}

\paragraph{Quantum enhanced variant \\}

\begin{figure*}[ht!]
    \centering
    \includegraphics[width=0.65\textwidth]{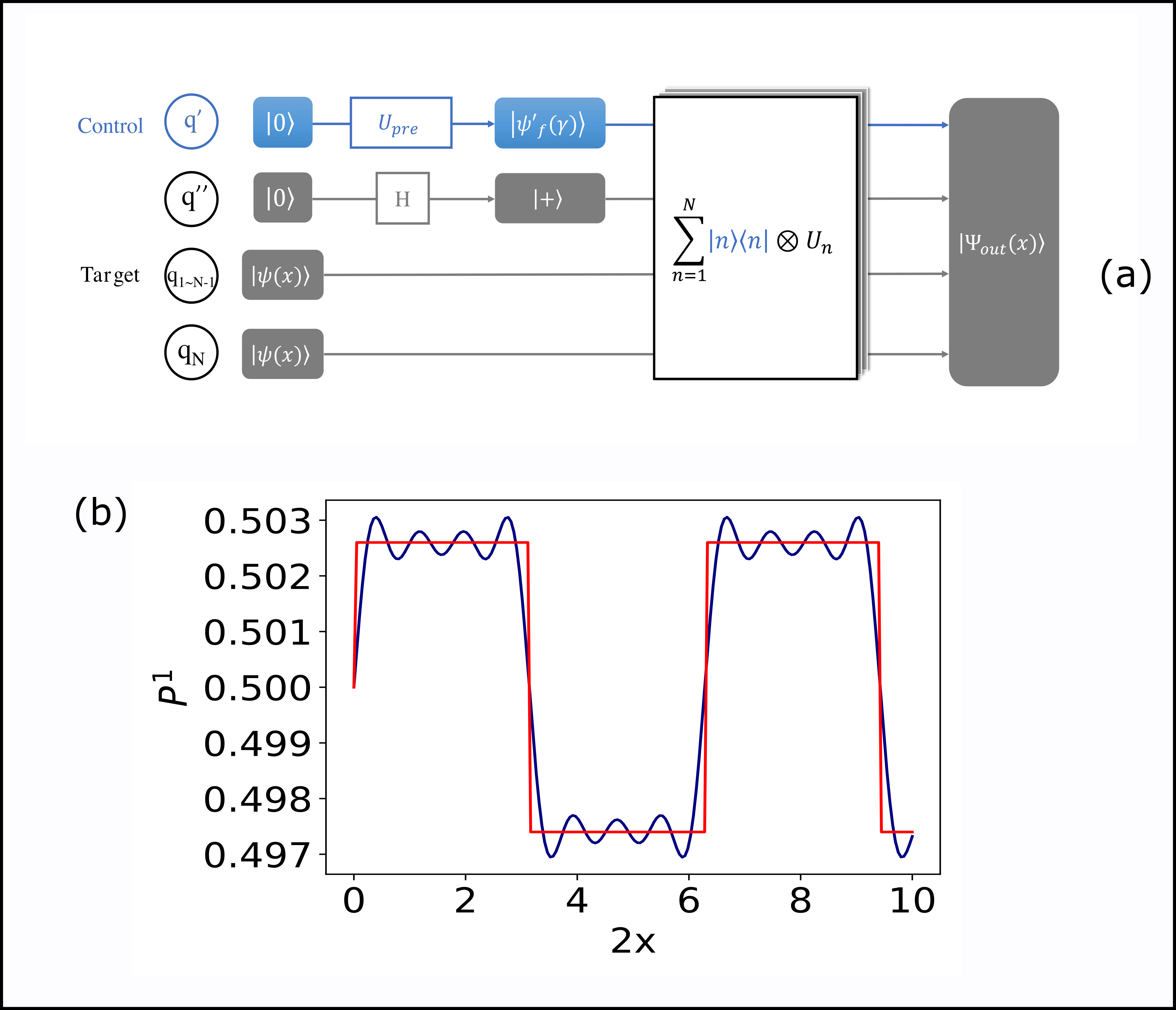}
    \caption{Fig.(\ref{fig_activation_2}) (a) shows the main structure of quantum circuits estimating arbitrary periodic functions and.There are two main modules.
    The first one contains $U_{pre}$ acting on the auxiliary qubits $q'$, and Hadamard gates acting on $q''$.
    The succeeding module is formed by $N$ controlled unitary operations denoted as $U_n$. $q'$ (blue color) are control qubits. $q'$ are converted to state $|\psi_f'(\gamma)\rangle$ under the operation $U_{pre}$, where $\gamma$ is determined by $F_N$. In Fig.(\ref{fig_activation_2}) (b), the blue curve represents the final output of the quantum circuit estimating square wave functions. Meanwhile, the red curve is the original shape of square wave functions. Fig.(\ref{fig_activation_2}) (a) and Fig.(\ref{fig_activation_2}) (b) are reproduced from Ref\cite{10.1088/1367-2630/ac2cb4} under Creative Common CC BY license.}
    \label{fig_activation_2}
\end{figure*}

Difficulties arise inevitably when attempting to include nonlinear activation functions into quantum circuits. The nonlinear activation functions do not immediately correspond to the mathematical framework of quantum theory, which describes system evolution with linear operations and probabilistic observation.
Conventionally, it is thus extremely difficult to generate these nonlinearities with a simple quantum circuit. Researchers could build up quantum-classical hybrid neural networks, where the linear part corresponds to the quantum unitary operations in quantum layers, while the nonlinear part corresponds to the classical layers. In other words, the classical layer in the quantum-classical hybrid neural network is to serve as the activation function connecting different quantum layers. Fig.(\ref{fig_activation_1}) (a) is a scheme of the quantum-classical hybrid neural networks,
where the linear part in the classical neural network is replaced by the quantum circuits.
Fig.(\ref{fig_activation_2}) (b) shows an example constructions of the hybrid quantum-classical neural network for 3 qubits. Fig.(\ref{fig_activation_1}) (see Ref\cite{xia2020hybrid}).
The quantum-classical hybrid neural networks generally work as follows.
Firstly, the classical data in converted to the quantum state via certain mapping process. Then, the quantum unitary operations will implement the linear calculation.
Next, the qubits are all measured and the estimation value is sent out to the classical layer. The classical layer will implement the nonlinear calculation (serve as the activation function),
and the output will be sent to the next quantum layer to repeat the steps above.
Based on the hybrid quantum-classical neural networks, researchers could construct quantum deep neural networks to calculate ground state energies of molecules\cite{xia2020hybrid}, to study the barren plateaus in training process\cite{mcclean2018barren}, and to recognition figures with transfer learning\cite{mari2020transfer}.
More details of the hybrid quantum-classical neural networks for various tasks could be found in these applications.

Though the unitary operations always correspond to linear calculation, the measurements could lead to nonlinearity. The repeat-until-success (RUS) circuit is a typical method implementing activation functions based on special measurements\cite{wiebe2013floating, paetznick2013repeat}. In the RUS circuit, an ancillary qubit is connected with the input and the output qubit.
After the certain operations, the ancillary qubit is measured. If result $|0\rangle$ is obtained, then the desired output is generated. Otherwise, we need to correct the operation and apply it on the qubits, then measure the ancillary qubit once again. The steps above should be repeated until we get result $|0\rangle$. Thus the circuit is named as repeat-until-success (RUS) circuit. In 2017, Cao and coworkers developed both the quantum feedforward neural network and quantum Hopfield network based on the RUS circuit\cite{cao2017quantum}.
Sometimes in the hybrid quantum-classical neural networks, researchers also use special intermediate measurements to implement certain nonlinear functions\cite{xia2020hybrid}.

There are some other approaches to implement the activation functions in quantum neural networks.
Activation functions can be implemented via the mapping process. In 2018, Daskin developed a simple quantum neural network with a periodic activation function\cite{daskin2018simple},
where the simple $\cos$ function is used as activation function. There are also methods to implement activation function with assistance of the phase estimation algorithm\cite{schuld2015simulating}.
Recently, our group also developed a quantum circuit to implement periodic nonlinear activation functions with multi copies of input\cite{10.1088/1367-2630/ac2cb4}.
Fig.(\ref{fig_activation_2}) (a) is a scheme of the circuit structure, and Fig.(\ref{fig_activation_2}) (b) shows the approximation of periodic square wave functions. The quantum circuit contains N-qubits to store the information on the  different N-Fourier components and $M+2$ auxiliary qubits with $M = \lceil{\log_2{N}}\rceil$ for control operations. The  desired  output  will be measured in the last qubit $q_N$ with a time complexity of the computation of $O(N^2\lceil \log_2N\rceil^2)$, which leads to polynomial speedup under certain circumstances.
In conclusion, it is an essential but still open question to find the optimal approach implementing nonlinear activation functions in the quantum neural network.


\subsubsection{Convolutional Neural Network (CNN)}\label{CNN_section}
\begin{figure*}[ht!]
    \centering
    \includegraphics[width=0.72\textwidth]{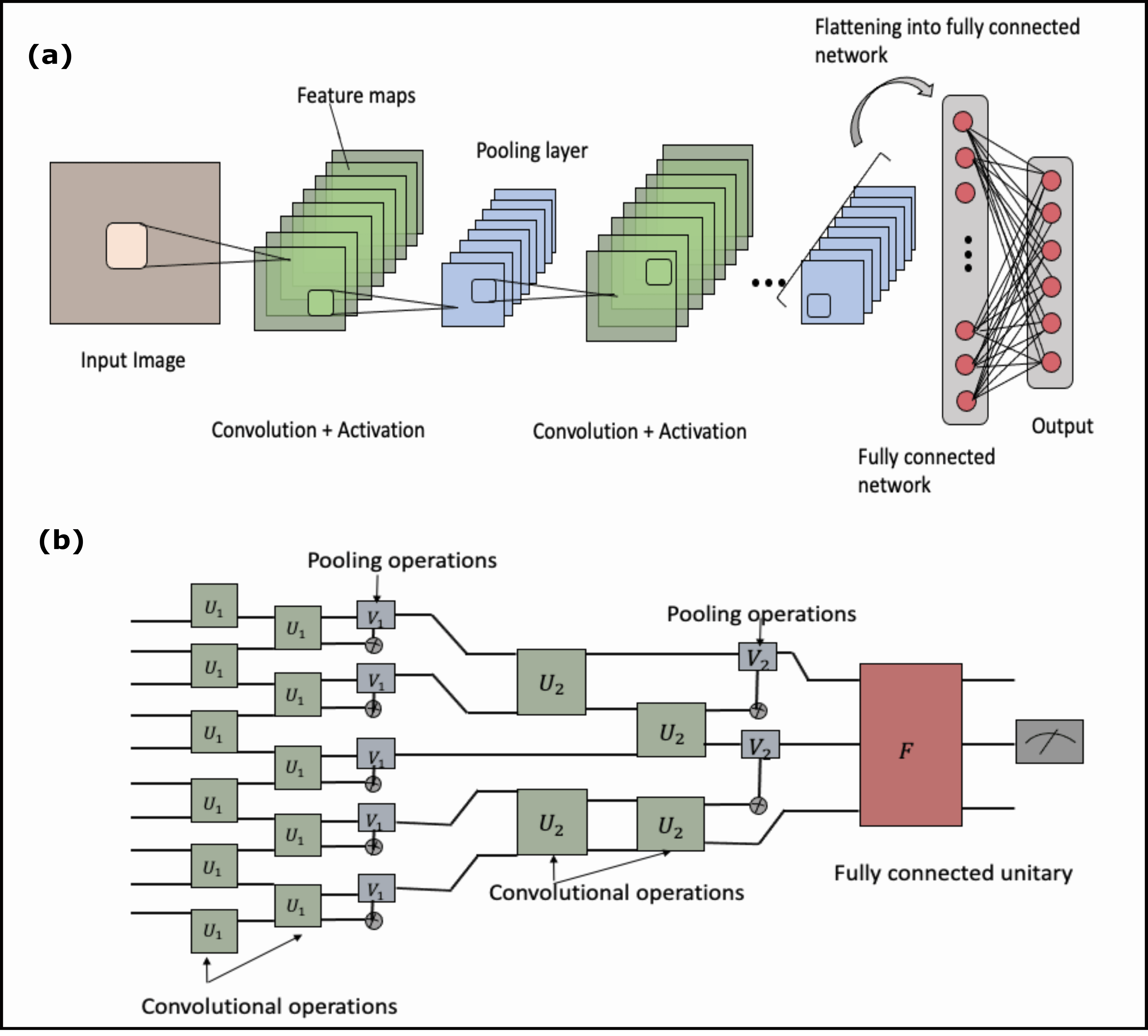}
    \caption{(a) The schematic of a typical convolutional neural network (CNN) is illustrated. The process starts with an input image from which feature maps are extracted through element-wise product with a kernel and then activation through any of the activation functions discussed in the previous section. Such feature maps are depicted in green. The pooling operation (blue layer) thereafter reduces the size of the feature maps by preserving values of a prescribed choice by the user within a certain window of each feature map. The two layers are repeated many times and then fed into a fully-connected neural network as discussed in the previous section. The output is read and back-propagation is used to train the parameters of the entire network. 
    (b) The schematic of the quantum circuit as illustrated in Ref \cite{cong2019quantum} for the realization of a CNN. The circuit receives an arbitrary input state say $\rho_{0}$. The unitaries designated as $U_i$ is responsible for convolutional operation whereas the unitaries designated as $V_i$ are responsible for controlled operations in the pooling layer. The unitaries $V_i$ is conditioned on the measurement results of neighboring unitaries. Such measurements reduces the qubit pool and is similar to dimensional reduction in conventional pooling layers. The operations are repeated several times until a fully-connected unitary (denoted as $F$) acts on. Certain qubits are measured subsequently to process the output.}
    \label{CNN_fig}
\end{figure*}

This is a specific kind of neural network architecture that is widely used for image classification and computer-vision problems \cite{rawat2017deep, voulodimos2018deep}. To understand the basic algorithm let us consider a grayscale image composed of ($n_1 \times n_2$) pixels. The image can be numerically represented as a matrix of intensity I such that $\rm{I} : \{1,2,,,n_1\} \times \{1,2,,,n_2\} \mapsto \mathbb{R}^{n_1 \times n_2}$. For colored images the only difference will be the intensity distribution at $(i,j)$ th pixel (position of the intensity matrix) will be a vector of $[R, G, B]^T$ entries. If the total pixel count ($n_1 n_2$) is too big, then converting the matrix into a 1D vector and using a feed-forward neural network as discussed before may be cumbersome and would require a large number of tunable parameters with the possibility of over-fitting. Besides, a 1D encoding looses the spatial correlation in the intensity pattern among the neighboring pixels. CNN is designed to use as input the entire 2D matrix and hinges on understanding and identifying the spatial information (often called feature maps) and then condensing the information into feature vectors of reduced sizes which is then fed into a fully-connected feed-forward network for usual operations as described before \cite{albawi2017understanding, lecun1999object, dhillon2020convolutional, aloysius2017review}. In other words, CNN is basically a simple neural network defined before in the final layer equipped with a robust feature-extractor before the final layer to remove redundancies and decrease parameter count. 

The key components which facilitates the CNN architecture are thus grouped into two parts : (a) Feature extractor (b) Fully-connected neural network. 
The component (a) is further made up of the repeated use of the following categories of layers. 
\begin{enumerate}
    \item Convolutional layer \cite{Goodfellow-et-al-2016} - This is where the magic of CNN happens. For each feature the user wants to identify and extract from the image, this layer uses a spatial filter (kernel) denoted as K which is essentially a matrix that can slide over the output of the previous layer (or the intensity matrix of the input image if one is looking right after the first layer) and define a new feature map. In other words the kernel acts on a chunk of the input matrix every time and the process is essentially a convolution. This feature map is obtained by a Frobenius inner product between the kernel and the chunk of the input it is acting on such that the resulting map has large entries only over the pixels (or $(i,j)$ positions) wherein the kernel entries 'are similar' with the entries of the chunk i.e. the feature is present. This is done for every kernel (one corresponding to every feature that needs extraction) and for every feature map from the previous layer. Operationally let the input to the $l$ th layer from the $(l-1)$ th layer  comprise feature maps denoted as $y_p^{l-1}$ each where $p=1,2,....\alpha^{l-1}$ features. Each such map is of sizes $\beta^{l-1} \times \gamma^{l-1}$. Then each of the output from the $l$ th layer denoted as $y_p^{l}$ ($p=1,2,....\alpha^{l-1}$) is a feature map of size $\beta^{l} \times \gamma^{l}$ obtained by convolving against kernels as follows:
    \begin{eqnarray}
        (y_p^{l})_{i,j} &= b^{l}_{i,j} + (\sum_q^{\alpha^{l-1}} K_{p,q}^l \circledast  y_q^{l-1})_{i,j} \\
        (\sum_q^{\alpha^{l-1}} K_{p,q}^l \circledast  y_q^{l-1})_{i,j} &= \sum_q^{\alpha^{l-1}}\sum_a\sum_b (K_{p,q}^l)_{a,b} (y_q^{l-1})_{i+a,j+b}
    \end{eqnarray}
    where $b^{l}_{i,j}$ are the elements of the bias matrix of the $l$ th layer. The tunable parameters within this layer are the bias matrix elements and the parameters within the kernel $K$. This convoluted feature map may be obtained by passing the kernel over the entire input without missing any row or column (without using any stride \cite{kong2017take}) or otherwise. The corresponding map so obtained may also be padded with zeros for dimensional consistency. All the feature maps so generated serve as input to the $(l+1)$ th layer. The early convolutional layers in the network usually extracts simple features with complexity increasing along the way. Feature maps can also be stacked along the third dimension to obtain compound features.
    \item Activation Layer - This layer is responsible for introducing non-linearity into the model using the input of the previous layer through the following expression 
    \begin{equation}
        (y_q^{l})_{i,j} = \sigma(y_q^{l-1})_{i,j}
    \end{equation}
    wherein $\sigma$ is an activation function like ReLU, sigmoid, tanh etc and $(y_q^{l})_{i,j}$ are defined as in the previous point. Sometimes rectification layers are also used which computes the absolute values of the input.
    \item Pooling layer- This is where dimensional reduction or downsampling of the feature maps happen. This layer takes in the feature maps from the previous layer and uses windows of pre-defined sizes within each chunk of the feature map and preserve only one value within each such window to generate a new feature map with reduced dimension. Number of feature maps remain unchanged. The one value so selected can be the maximum value of all features within the window (max-pooling \cite{alzubaidi2021review, yu2014mixed, li2019teeth}) or may be the average value (average pooling \cite{yu2014mixed, li2019teeth}). 
\end{enumerate}

The architecture has repeated applications of these layers to ensure parameter sharing and efficient feature extraction. The output at the last layer of the feature-extractor is vectorized into a 1D format and fed into a completely connected deep-neural network at the end. This network then processes the input and generates the final output. For example if  the final desired output is a multi-label classification task, the connected neural network will have as many neurons as the number of labels with each neuron being a placeholder for a 1 or 0 denoting classification into the corresponding label or not. Fig. \ref{CNN_fig}(a) illustrates pictorially all of these components.

\paragraph{Quantum enhanced variant}

In 2019, Cong $et al$ \cite{cong2019quantum} developed a quantum circuit based on CNN architecture which is used to classify an N-qubit quantum state with M-labels. In other words given a training data set of M states $\{(|\psi_i\rangle, y_i\}$ where $y_i$ are the binary classification labels associated with the states, the circuit can decide which of the training vectors the input unknown state resembles the most. This is useful in understanding whether a given state belongs to a particular phase with phase labels and shall be discussed later. The circuit architecture involves the following steps:
\begin{enumerate}
    \item The initial inputs are mapped to a quantum state
    \item In the convolutional layer, the quantum state is transformed using a set of quasi-local unitaries labelled as $U_i$ where $i\in \{1,2..\}$ 
    \item In the pooling layer , some of the qubits are measured and conditioned on this measurement outcome, the unitaries for the remaining qubits are decided. This reduces the width of the circuit as certain qubits whose state has been readout are no longer a part of subsequent operations. Such controlled entangling unitaries are labelled as $V_i$ where $i\in \{1,2..\}$
    \item The convolutional and the pooling layers are applied many times until the width of the circuit is reduced sufficiently.
    \item Finally a fully-connected layer of single and two-qubit gates (labelled as say $F$) are applied on the remaining qubits analogous to the fully-connected layer in classical CNN
    \item The final prediction from the algorithm is read by measuring certain qubits at the very end.
\end{enumerate}
The circuit is described in Fig.\ref{CNN_fig}(b) schematically.  

\subsubsection{Recurrent Neural networks}\label{RNN_section}

For data that involves time-ordered sequence as what appears frequently in natural-language processing \cite{yin2017comparative}, stock-price prediction \cite{selvin2017stock}, translation \cite{cho2014learning} or any simple time-series prediction \cite{qin2017dual} it is important for the neural network architecture to preserve information about the previous entries in the sequence i.e. the notion of building memory in the architecture becomes essential. Recurrent Neural networks (RNN) are specifically built to handle such tasks. The task such networks perform are usually supervised in which one has access to a sequence $\{x_i\}_{i=1}^{T}$ where each entry in the sequence $x_i \in \mathbb{R}^d$ and a corresponding label $\{y^{*}_i\}_{i=1}^{T}$ where $y^{*}_i \in \mathbb{R}^m \:\:\forall\:\: i$. The primary goal of the network is to produce a new sequence $\{y_i\}_{i=1}^{T}$ as output such that each $y_i$ is close enough to $y^{*}_i$ $\forall i$ as computed from a chosen metric and a preset threshold. 
In the vanilla RNN architecture \cite{yu2019review}, the primary functional unit which is used repeatedly for each input entry in the sequence consists of three layers of stacked neurons. The first layer is an input layer having $d$ neurons (as the input entries $x_i$ are $d$-dimensional). The next layer is the hidden layer having say $p$ neurons and is parameterized by \textit{weight} matrices $(W_{z}, W_{x})$ and bias vector $b_{z} \in \mathbb{R}^{p}$. The \textit{weight} matrix $W_{x} \in \mathbb{R}^{d\times p}$ is responsible for the affine transformation on the input $x_i$ as discussed for the case of feed-forward neural networks. However the primary difference from an usual feed-forward network is the presence of second set of \textit{weight} matrix $W_{z} \in \mathbb{R}^{p\times p}$ which performs an affine transform on the hidden layer activation response corresponding to the entry in the previous step i.e. for the last but one input entry $x_{i-1}$. If the activation response from the hidden layer for $x_i$ is denoted as $\sigma(z_i)$ and the activation response for the previous entry $x_{i-1}$ is denoted as $\sigma(z_{i-1})$ then the two are related as 
\begin{eqnarray}
    \sigma(z_{i}) = \sigma(W_{x}^T x_{i} + W_{z}^T \sigma(z_{i-1}) + b_z)
\end{eqnarray}
\begin{figure}[ht!]
    \centering
    \includegraphics[width=0.5\textwidth]{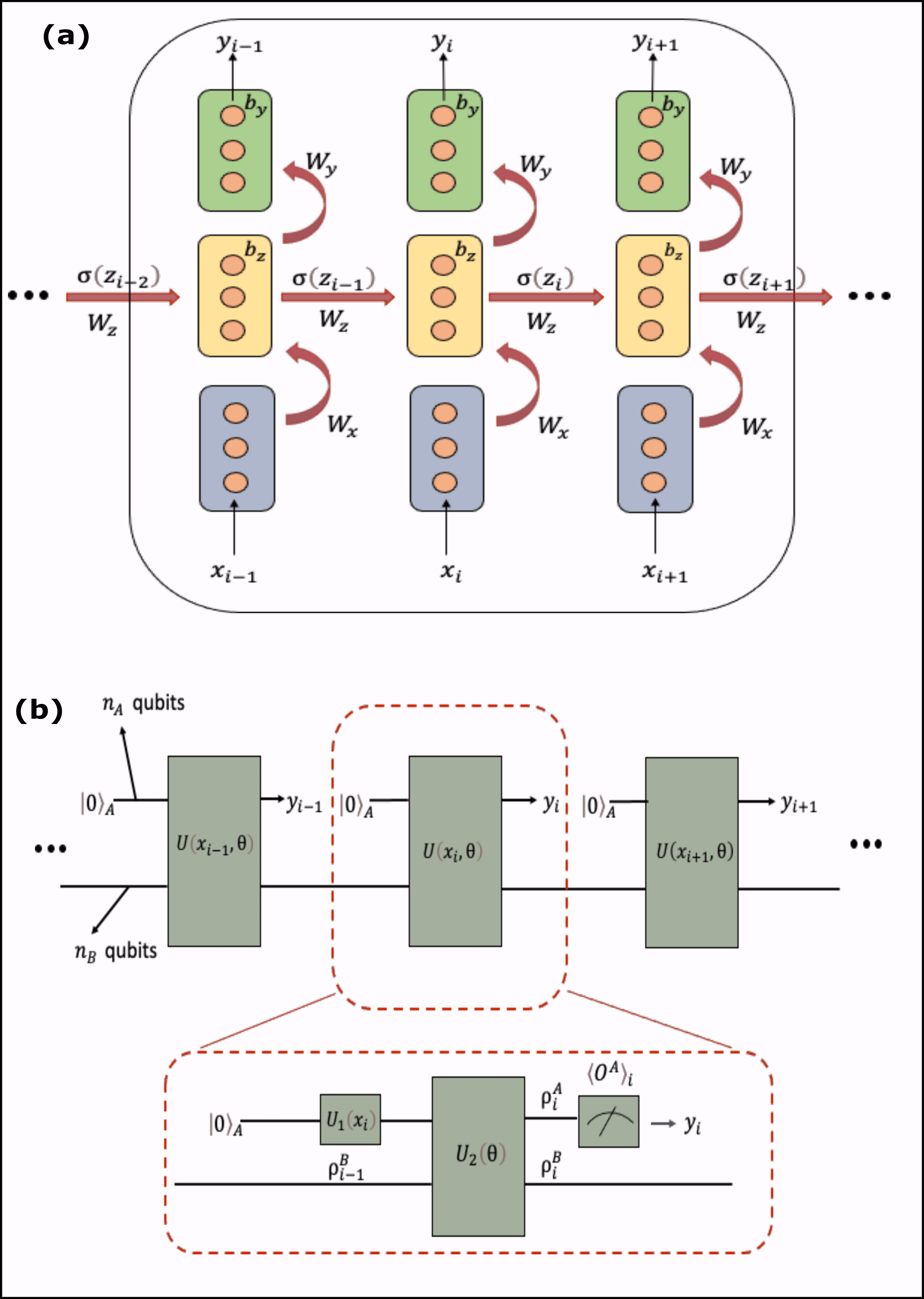}
    \caption{(a) A schematic of Recurrent Neural Network(RNN) architecture. The blue layer encodes the input sequence $\{x_1,x_2...x_{i-1},x_i,x_{i+1}..\}$. The yellow layer is the hidden layer which processes the input and generates an activated response $\sigma(x_i)$ for the input $x_i$. The key difference of RNN for other neural network architecture is that $\sigma(x_i)$ is fed onto the network when the next entry in the time-ordered sequence $x_{i+1}$ is the input. This forms the core of the memory retention process in RNN. $\sigma(x_i)$ is also used by the final layer (green) to generate the output $y_i$. The parameters of the network are the biases $(b_y, b_z)$ and the interconnecting weights $(W_x, W_z, W_y)$ between a pair of layers which are highlighted in the figure. (b) The schematic of the quantum circuit for processing time-ordered sequence using RNN architecture as illustrated in Ref \cite{takaki2021learning}. Two qubit registers are invoked with $n_A$ and $n_B$ qubits and the input entry $x_i$ is encoded within the first register using $U_1(x_i)$. The first and second register is entangled using a parameterized unitary $U_2(\theta)$ and thereafter an observable $O^A$ on the first register is measured to yield the output $y_i$. The second register is left untouched and is responsible for carrying the memory for the next input $x_{i+1}$. The circuit is adapted from Ref \cite{}. The parameters of the entangling unitary is optimized to match the output sequence $\{y_1,y_2...y_{i-1},y_i,y_{i+1}..\}$ to the desired.}
    \label{RNN_fig}
\end{figure}

Using this activation response(usually tanh) the last output layer now performs another affine transformation followed by the usual introduction of non-linearity through input to activation process as follows :
\begin{eqnarray}
    y_{i} = \sigma(W_{y}^T \sigma(z_{i}) + b_y)
\end{eqnarray}
where the \textit{weight} matrix $W_{y} \in \mathbb{R}^{p\times m}$ defines the inter-connections between the hidden layer and output layer and the \textit{bias} vector $b_y \in \mathbb{R}^{m}$. The total number of tunable parameters for this unit are thus $(W_{z}, W_{x}, W_{y}, b_{z}, b_{y})$. For all subsequent input entries (i.e. $x_{i+1}, x_{i+2}$ etc the functional unit is repeatedly queried using the activation response (usually tanh) of the previous step as explained above. The total number of parameters $(W_{z}, W_{x}, W_{y}, b_{z}, b_{y})$ are kept the same for all such steps which leads to reduction in the number of variables through efficient sharing. Each such iteration generates a $y_{i}$ as explained. After the first pass through the entire data-set (a subset of the data-set can also be used depending on user preference)
an ordered sequence $\{y\}_i$ is generated and a cost-function is defined to compare this sequence with the labelled sequence $\{y^{*}\}_i$. The error in this cost-function is minimized by updating the parameter set $(W_{z}, W_{x}, W_{y}, b_{z}, b_{y})$ using the gradient of the cost-function or any other optimizer as has been described for the case of feed-forward neural network. The architecture is pictorially depicted in Fig. \ref{RNN_fig} (a). 

{\color{black} During back-propagation for long RNNs it is possible to encounter a situation wherein the gradient vector accrues a zero value (or grow unbounded in magnitude)  with respect to parameters of nodes appearing at the beginning of the network. Such a situation is known as vanishing (exploding) gradient and if happens can render the model untrainable. Apart from changing the activation function from logistic ones like tanh to ReLU, one can adopt architectures of RNN like
long-short term memory (LSTM) or gated recurrent unit (GRU) \cite{sherstinsky2020fundamentals, yu2019review,yang2020lstm, dey2017gate,chung2014empirical} in such a case. LSTM networks introduced in Ref \cite{hochreiter1997long} also have successive repeating units/cells wherein the input to each cell are one entry of the ordered sequence $x_i$ (defined before) as well as the activation response (say $h_{i-1})$ which was denoted as $\sigma(z_{i-1})$ for vanilla RNN. The change in notation will be clarified soon. The two quantities are conceptually similar though.) However the key difference with the vanilla version of RNN lies in the presence of memory channel/carousel. The response of this memory channel from the previous cell (often denoted as $c_{i-1}$) is also fed as input into to the current cell. Inside the current cell there are three different networks/gates which work to erase, update the memory of the carousel entry and to generate a new output $y_i$ as well $h_{i})$. The latter is fed back into the next cell as before in the case of vanilla RNN. The primary components inside each cell are the following :

\begin{figure*}[ht!]
    \centering
    \includegraphics[width=0.90\textwidth]{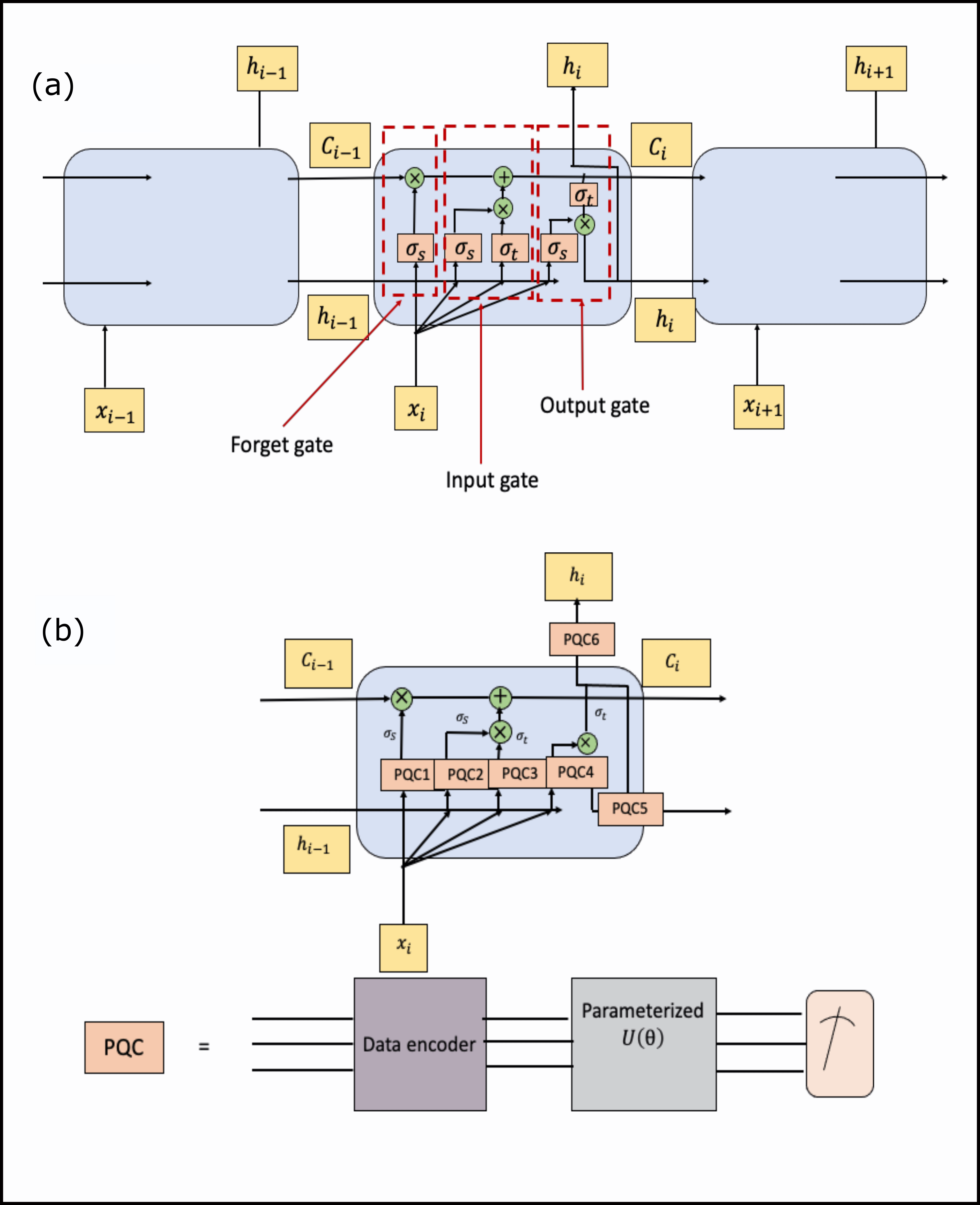}
    \caption{{\color{black}(a) A schematic representation of a typical LSTM network as implemented on a classical processor. The three different processing gates - forget gate, input gate and output gate are illustrated. (See text for details) along with the memory channel encoding $C_i$. $\otimes$ indicates elementwise multiplication whereas $\oplus$ indicates elementwise addition. $\sigma_t$ indicates $tanh$ activation and $\sigma_s$ indicates sigmoid activation.
    (b) The hybrid quantum-classical LSTM network implemented using parameterized quantum unitaries (PQC) in Ref \cite{chen2020quantum}. Each PQC has a data-encoding circuit and a variational circuit parameterized by angles (say $\vec{theta}$). For say PQC1 to PQC4 , the data encoder loads the concatenated vector $(x_i, h_{i-1})^T$. For PQC5 and PQC6, the output from PQC4 is processed and the memory channel is loaded. (See text for more details )
    }}
    \label{fig_LSTM}
\end{figure*}

(a) The forget gate- This takes in input $(x_i$, $h_{i-1})^T$, and performs an affine transformation with weight matrices and bias $(W_{xf}, W_{hf}, b_f)^T$ wherein the subscript $f$ stands for forget gate. This is passed onto a sigmoid activation which outputs values between 0 and 1 only. The purpose of this gate is to read from the present input entries $(x_i$, $h_{i-1})^T$ what features in the memory channel needs to be erased (hence the name forget gate). If the output of the forget gate is denoted as $f_i$ the transformation is abbreviated as 
\begin{eqnarray}
   f_i = \sigma_{s}(W_{xf}^T x_i + W_{hf}^T h_{i-1} + b_f) \label{for_gate}
\end{eqnarray}
where $\sigma_{s}$ is sigmoid activation.

b) The next important gate is the input gate whose purpose is to decide what new information needs to be updated into the memory channel and at what places. Two operations are performed herein. The first involves creating an affine transformation of the input 
$(x_i$, $h_{i-1})^T$ followed by sigmoid activation. The weights and biases in the process are $(W_{xI}, W_{hI}, b_I)^T$ wherein $I$ is for input gate. This accomplishes the task of where to update the new information through the 0's and 1's of the sigmoid activation. The next operation is to create a candidate memory $\tilde{C}_i$ for the memory channel using the input entries and parameters $(W_{xIc}, W_{hIc}, b_{Ic})^T$ using a tanh activation to acquire values between $\pm$ 1. 
The operations are as follows:
\begin{eqnarray}
   I_i &=& \sigma_{s}(W_{xI}^T x_i + W_{hI}^T h_{i-1} + b_I) \\
   \tilde{C}_i &=& \sigma_{tanh}(W_{xIc}^T x_i + W_{hIc}^T h_{i-1} + b_{Ic}) \label{in_gate}
\end{eqnarray}
The state in the memory channel is then updated using Eq.\ref{for_gate} and Eq. \ref{in_gate} as
\begin{eqnarray}
C_{i} = f_i*C_{i-1} + I_i*\tilde{C}_i
\end{eqnarray} \label{C_i_new}
where the first term erases the memory from the previous state $C_{i-1}$ using location in $f_i$ and the second term re-builds it with new information in $\tilde{C}_i$ at the location specified by the $I_i$ vector.

(c) The third component is the output gate which uses Eq.\ref{C_i_new} to create an output $h_{i}$ which will be fed into the next cell with $x_{i+1}$. The transformation is
\begin{eqnarray}
    h_i = (\sigma_{s}(W_{ox}^T x_i + W_{oh}^T h_{i-1} + b_o))*\sigma_{tanh}(C_i)
\end{eqnarray}
This operation can be interpreted as returning the tanh of the state of the memory channel $\sigma_{tanh}(C_i)$ as output at locations filtered by the vector $(\sigma_{s}(W_{ox}^T x_i + W_{oh}^T h_{i-1})$ which explains why the symbol was changed from $\sigma(z_{i-1})$ as it is not just an activation output but a scaled one. The weights and the biases $(W_{ox}, W_{oh}, b_{o})^T$ are parameters of this output gate.
Fig. \ref{fig_LSTM}(a) displays a schematic version of a typical LSTM network.}

\paragraph{Quantum enhanced variants}

Recently a quantum algorithm have been designed \cite{takaki2021learning}
to implement the vanilla RNN architecture using a hybrid-variational framework. The algorithm uses a quantum circuit of two set of qubits (say $n_A$ and $n_B$). The input sequence of data is stored within the quantum states of one of the two registers through appropriate unitary operations. Both registers are then processed  through unitaries with parameterized angles. The state of one of the register is measured subsequently to obtain the output whereas the other is untouched and passes onto the subsequent state to carry memory of previous steps. The key ingredients of the protocol are:
\begin{enumerate}
    \item Both the registers with $n_A$ and $n_B$ qubits are initialized to null kets.
    \item The first input entry $x_0$ is encoded onto the state of the register with $n_A$ qubits. Thereafter controlled unitaries are used to entangle the register with $n_A$ qubits and $n_B$ qubits. Such unitaries are parameterized by variational angles.
    \item The expectation value of an operator $O^A$ is measured using the reduced density matrix of the first set of qubits (say $\rho_A^0$). This measurement yields $y_0$. The second set of qubits (in register with $n_B$ qubits) remain untouched. The first register is re-initialized to null kets
    \item For subsequent input entries (say $x_1, x_2....x_t$ etc) the second step above is repeated with the input $x_i$ encoded within the first register. This is followed by the third step. The state of second register which is left pristine at each step retains the memory and information about previous input entries. This information is shared with the qubits in the first register through the parameterized entangling unitaries for each input
    \item The sequence of $\{y\}_i$ values so generated is fed into a cost function and the parameters of the entangling unitaries are updated for the next cycle from the knowledge of the errors 
\end{enumerate}
The circuit is schematically illustrated in Fig. \ref{RNN_fig} (b)
{\color{black}  A quantum version of the LSTM network have also been implemented recently using hybrid-variational circuits \cite{chen2020quantum}. The schema of the algorithm consists of 4 major components
\begin{enumerate}
    \item A data loader circuit- This component serves to map the concatenated form of the input vectors of the sequence $x_i$ and $h_{i-1}$ (defined before in the classical variant) to quantum states. The circuit consists of $R_z$ and $R_y$ gates after a conversion to an equal superposition state using Hadamard transforms
    
    \item The next step involves parameterized unitaries with CNOT gates and single-qubit rotations. This block of parameterized unitary is used inside the input gate, forget gate as well as the output gates (see Fig. \ref{fig_LSTM} (b)) for optimizing the response from each gate.
    
    \item Measurement protocol on each such block of parameterized unitary in (2) using the input encoded within the state-preparation circuit in (1) yields the necessary affine transformation which is subsequently passed through the appropriate activation function for each gate as defined before in the classical variant. (see Fig. \ref{fig_LSTM} (b))
    
    \item The parameters of the unitary in step 2) are updated through gradient estimates of a cost-function involving the error between the actual and the output from the network using a classical computer. The network was applied on many different problems including dynamics of damped harmonic oscillators with good results.
\end{enumerate}
The circuit is schematically illustrated in Fig. \ref{fig_LSTM} (b)
}

\subsubsection{Autoencoders}\label{Auto_section}
A typical autoencoder is a type of neural network which is used to generate useful representations of the input data, to be used for unsupervised learning. The data can be thought of being generated from some distribution that represents a class spanning a subspace of the vector space in which they are represented. This is usually the case in most practically used image datasets and thus allows for dimensionality reduction. Autoencoders have helped in providing sparse representations\cite{makhzani2014ksparse}, denoising \cite{10.5555/1756006.1953039}, generating compact representations, information retrieval\cite{SALAKHUTDINOV2009969}, anomaly detection \cite{RIBEIRO201813}, image processing as a precursor to classification tasks. An autoencoder can be thought of as a feed forward neural network composed of an encoder and decoder with a bottleneck layer providing a minimal representation separating them. The output of the encoder constructs a compact representation of the input data and is fed to the decoder which reconstructs it back. 

\begin{figure}[ht!]
    \centering
    \includegraphics[width=0.5\textwidth]{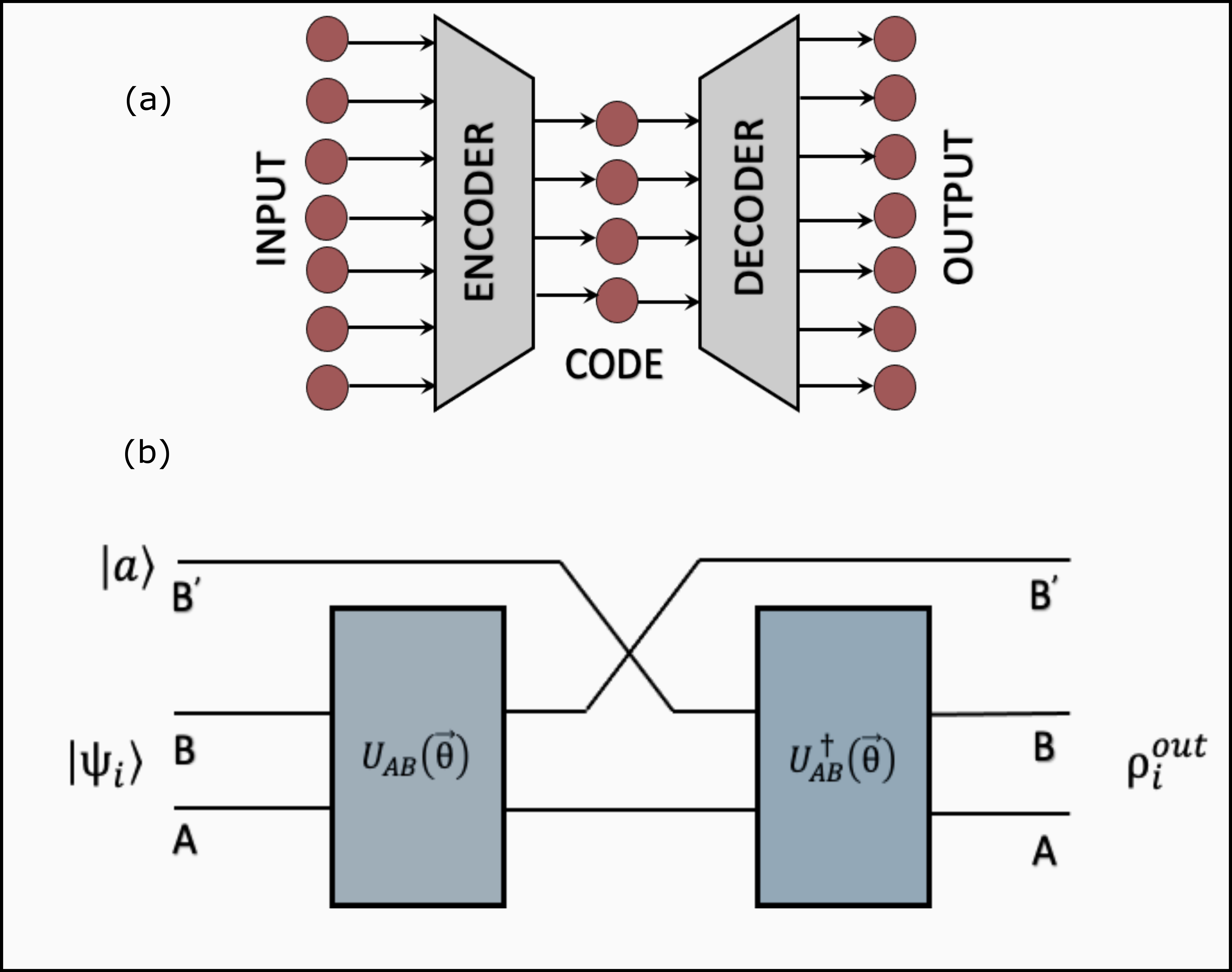}
    \caption{(a) Schematic representation of classical autoencoder. The encoder takes classical input to create a compact space representation. The decoder acts on the code to output the a representation  from the input space
    (b) Schematic representation of circuit used in quantum autoencoder as illustrated in Ref \cite{2017_romero}. The encoder $U_{AB}$ acts on circuit to create a code. The trash qubits are swapped out with a reference state and the decoder circuit works to reconstruct the input}
    \label{autoencoder_fig}
\end{figure}

 Like any other feed forward neural network it is trained through back-propagation to minimize the reconstruction error defined over it. For the simplest one layer encoder decoder circuit, with weights, biases and element wise activation function $W,b,\sigma$ and $W',b',{\sigma}'$, we can construct the $L_2$ norm loss function as follows,
\begin{equation}
    Loss(W,b,W',b') = \sum_{i=1}^{N} \left\lVert x_i- \sigma'(W'(\sigma(Wx_i+b))+b')\right\rVert^2
\end{equation}
where $N$ is the size of the training data set. Using a standard gradient descent one can train the parameters of the circuit to minimize the loss output. A regularization term might be added to ensure that network isn't overfitting to the training dataset. Here we described an undercomplete autoencoder that made use of no regularization term. Overfitting here is avoided by ensuring a small latent code size. Depending on the error function, inputs, and size of the latent space we can construct autoencoders that have different functionality. A sparse encoder for instance has the same latent space size as the input and minimizes the number of activations in the latent space, implemented by a $L1$-regularization term on the latent space. A denoising encoder, takes inputs that are overlayed with minimal perturbations to reconstruct the original image. A contractive autoencoder tries to ensure that samples that are close in the input space have a similar encoding representation.

\paragraph{Quantum enhanced variants}

To generalize a classical encoder to the quantum setting we start with building a unitary circuit that allows information to be compressed into a smaller set of qubits with a garbage state in the remaining qubits that can be replaced with a reference state. We start with an ensemble of $N$ pure states $\{\ket{\psi_i}_{AB}\}$, where $A$ is an $n$ qubit system, $B$ is a $k$ qubit system. Let $U$ be the encoding unitary that takes as input a pure state from the ensemble. The system $B$ in the output is then swapped with a reference state and we try reconstructing the input state with a decoder given by the unitary $U^{\dagger}$. The objective function to maximize is given by
\begin{equation}
    C(\vec{\theta}) = \sum_{i} F(\ket{\psi},\rho_i^{out}(\vec{\theta}))
\end{equation}
where
\begin{equation}
    \rho_{i}^{out} = U^{\dagger}_{AB}(\vec{\theta}) S_{BB'} Tr_B \bigg[ U_{AB}(\vec{\theta})\rho^{in}_{i} U^{\dagger}_{AB}(\vec{\theta}) \bigg] S_{BB'} U^{\dagger}_{AB}(\vec{\theta})
\end{equation}
Here $\rho^{in}_{i}= \ket{\psi_i}\bra{\psi_i}_{AB} \otimes \ket{a}\bra{a}_{B'}$, $F$ denotes the fidelity between the states, $S_{BB'}$ is a swap gate that swaps the corresponding qubits and $\vec{\theta}$ represents the parameters of the unitary circuit that needs to be trained. It can be show that the perfect fidelity is obtained when the output state of the encoder circuit produces a product circuit, i.e,  $U \ket{\psi}_{AB} = \ket{\tilde{\psi}_i}_A \otimes \ket{a}_B$ \cite{2017_romero}. Thus we could alternatively define the maximizing objective function as the fidelity over the trash system B with respect to the reference state as,
\begin{equation}
    \tilde{C}(\vec{\theta}) = \sum_i F\bigg(Tr_A \Big[ U_{AB}(\vec{\theta}) \ket{\psi_i}\bra{\psi_i}_{AB} U^{\dagger}_{AB}(\vec{\theta}) \Big],\ket{a}_B \bigg)   
\end{equation}
This problem can thus be framed within the context of developing unitary circuits that works to disentangle qubits. It has been shown that using circuits of exponential depth it is always possible to disentangle qubits \cite{841961}. Other alternative implementation include using approximate quantum adders trained with genetic algorithms \cite{2018_qaegen} and generalization of feed forward neural networks as quantum circuits to implement autoencoders \cite{2017_ffnn}

\subsubsection{Variational Encoders}\label{Vauto_section}
Unlike autoencoders that try and provide useful latent space representation, variational autoencoders (VAEs) are used to learn the distribution that models the latent space. The decoder thus generated can be used to sample the input space working similar to Generative Adversarial Networks (to be discussed shortly). They have found their use in unsupervised \cite{dilokthanakul2017deep} and semi-supervised learning \cite{Xu_Sun_Deng_Tan_2017}. Let $p_{\theta}(x|y)$ be the conditional likelihood of the decoder and $q_{\phi}(y|x)$ be the the approximated posterior distribution the the encoder computes, where $x$ is the input vector and $y$ is the latent vector. We train the network on the parameters $\theta$,$\phi$ to reduce the reconstruction error on the input and have $q_{\phi}(y|x)$ as close possible to $p_{\theta}(y|x)$. Thus we would like to minimize the following evidence lower bound loss function (ELBO),

\begin{eqnarray}
    Loss(\theta,\phi) &= \sum_x D_{KL}(q_{\phi}(y|x)||p_{\theta}(y|x) -log(p_{\theta}(x) ) \nonumber \\
    &= \sum_x  D_{KL}(q_{\phi}(y|x) ||p_{\theta}(z)) \nonumber \\   
    & - E_{z \sim q_{\phi}(y|x)}(log(p_{\theta}(x|y)))
\end{eqnarray}

where $E$ is the expectation value with respect to the specified distribution and $D_{KL}$ is the KL divergence between the distributions and x is input from the training set. The later equality of the above expression is obtained by expressing $p_{\theta}(y|x)$ using Bayes theorem and regrouping terms.  The KL divergence regularizes the expression allowing for continuity (neighbouring points in latent space are mapped to neighbouring points in input space) and completeness (points in latent space map to meaningful points in input space for any chosen distribution). At this point 2 assumptions are made to allow for training. Firstly, $p_{\theta}(x|y)$ is a Gaussian distribution and $q_{\theta}(y|x)$ is a multivariate Gaussian that can be re-expressed as $\mu + \sigma \odot \epsilon$ to allow for gradient back-propagation (reparametrization trick) , where $\epsilon \sim N(0,I)$ and $\odot$ is an element wise product. 

\begin{figure*}[ht!]
    \centering

    \includegraphics[width=0.6\textwidth]{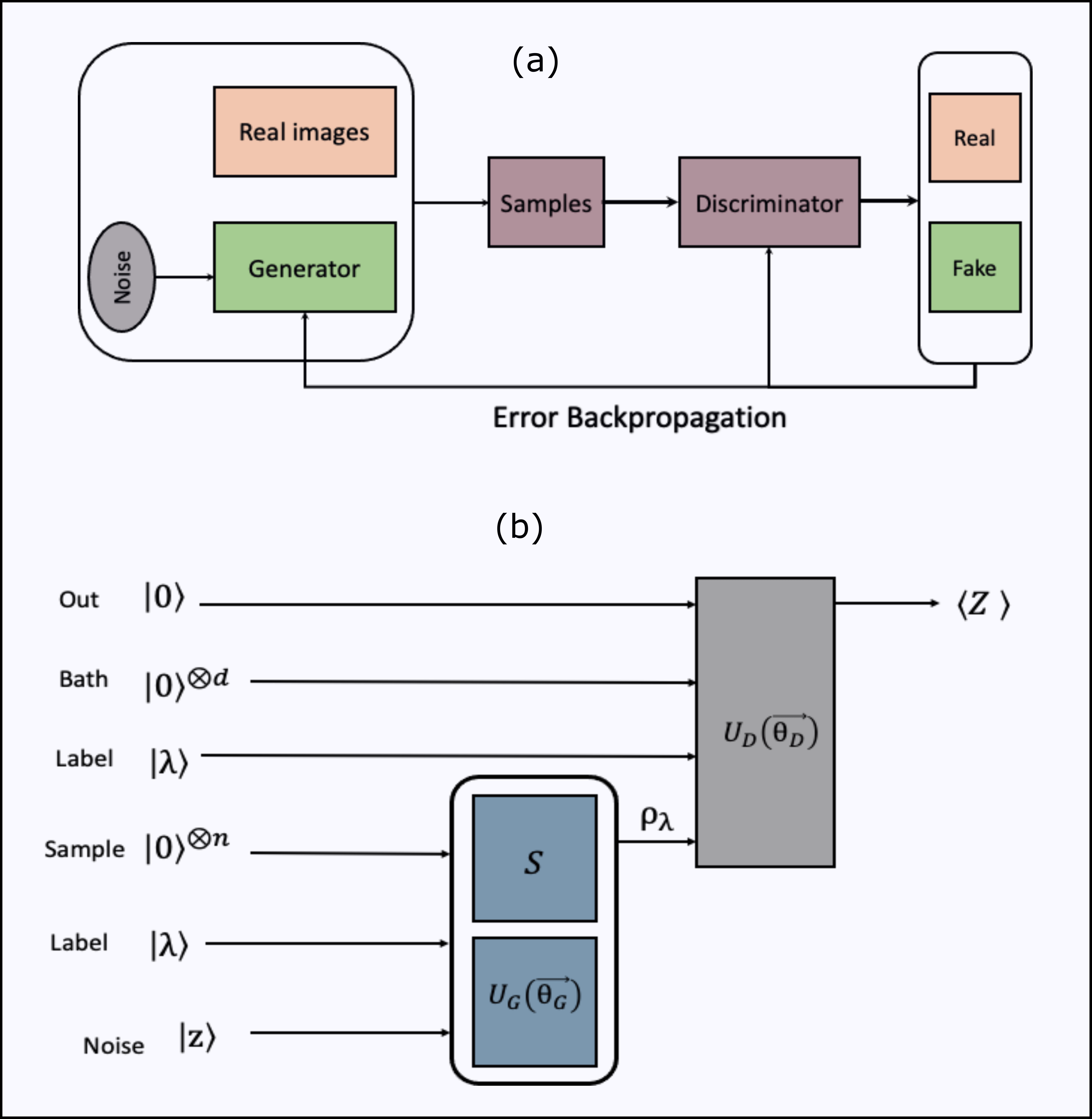}
    \caption{Schematic representation of classical GAN. The generator takes as input some noise to produce  a sample. Samples from real source and generator are fed to the discriminator. These work as labelled data for the discriminator to distinguish. The errors are used to backpropagate and train the generator and discriminator 
    (b) Schematic representation of the quantum circuit used in quantum GAN as illustrated in Ref \cite{2018_qgan}. Samples are output from the real source $S$ or the generator $U_G(\vec{\theta_D})$ that takes as input some noise and label. This is then fed to the discriminator $U_D(\vec{\theta_D})$ along with the label qubits and scratch space(bath) to work on. The measured output qubit is used to backpropagate the errors through classical updates to give new circuit parameters }
    \label{GAN_fig}
\end{figure*}

\paragraph{Quantum enhanced variants}

Khoshaman et al \cite{2018_qvae} developed a quantum variational autoencoder that makes use of a Quantum Boltzmann Machine (QBM) \cite{2018_qbm}  to evaluate the gradient updates used in training. A QBM is an energy model defined as,
\begin{eqnarray}
    & p_{\theta}(z) = Tr [\Lambda_z e^{-H_{\theta}}]\\
    & Z_{\theta} = Tr[e^{-H_{\theta}}] \\
    & H_{\theta} = \sum_l \sigma_l^x \Gamma_l  + \sum_l \sigma_l^zh_l  + \sum_{l<m} W_{lm} \sigma_l^z \sigma_m^z
\end{eqnarray}

where $\theta = {\Gamma,h,W}$, $\Lambda = \ket{z}\bra{z}$, $\sigma_l^{z,x}$ are pauli operators and $p_theta(z)$ governs the distribution of the states $\ket{z}$. The ELBO is defined with a cross entropy term as follows,

\begin{eqnarray}
    H(p_{\theta},q_{\phi}) &= -E_{z\sim q_{\phi}}\Big[ log(Tr \big[\Lambda_z e^{-H_\theta}\big]) \Big] + log(Z_{\theta}) \\ 
    & \geq E_{z \sim q_{\phi}}\Big[ log(Tr \big[e^{-H_\theta + ln\Lambda_z}\big]) \Big] + log(Z_{\theta})
\end{eqnarray}

where in the second line we have used Golden Thompson-Inequality( $Tr[e^Ae^B] \geq Tr[e^{A+B}] $ ) to express the intractable first term with a lower bound. Similar to the classical case, a reparametrization trick is employed to effectively evaluate gradients and the trace is taken to be concentrated at the state $\ket{z}$. See \cite{2018_qvae} for reparametrization trick in the continuous and discrete case setting.

\subsubsection{Generative Adversarial Network (GAN)}\label{GAN_section}

Generative adversarial network (GAN) was introduced by Ian Goodfellow et al in 2014 \cite{goodfellow2014generative} and is considered to be one of the major milestones of machine learning in the last 10 years. Its applications extends to art \cite{yu2018generative}, science \cite{2017_gen}, video games \cite{wang2018esrgan}, deepfakes and transfer learning \cite{li2021domain}. A GAN consists of generator and discriminator that are trained simultaneously to learn a given distribution by competing against each other. The goal of the generator is to generate fake samples that cannot be distinguished from the true samples of input data. The goal of the discriminator is to correctly distinguish the fake samples from true samples, thus solving a well understood classification problem. This game has a Nash equilibrium point that is attained when the generator is able to generate samples are distinguished with a probability of $1/2$, making it no better than a random guess. Let the generator $G$ take a random input from a distribution $p_z$ (usually taken to be a gaussian distribution) to generate samples from the distribution $p_f$.  Let $D$ be the discriminator that takes inputs equally likely sampled from  $p_t$ (true distribution) and $p_f$ to output the probability of data coming from $p_t$. The objective function for the discriminator is thus given by,
\begin{equation}
   \min_{D} \frac{1}{2} E_{x \sim p_t}[1-D(x)] + \frac{1}{2} E_{z \sim p_z}[D(G(z))]
\end{equation}
where the first term is the error in determining the true samples to be fake and the second term is the error in determining the generated samples to be true. The generator on the other hand tries to maximize this loss function against the trained discriminator, i.e, the objective function of the generator is given by
\begin{equation}
    \max_{G}  \min_{D} \frac{1}{2} E_{x \sim p_t}[1-D(x)] + \frac{1}{2} E_{z \sim p_z}[D(G(z))]
\end{equation}

The parameters of the generator and discriminator are trained alternatively till the discriminator no longer is able to differentiate $p_f$ from $p_t$. Thus the distributions are equal in the eyes of the discriminator and we have managed to create a generative model for the given training samples. The discriminator is discarded after the training. $p_z$ can be thought of the distribution that represents the domain set of the problem and thus the generator works to extract features of the input vector. The trained generator can be further re-purposed for transfer learning on similar input data. A conditional extension referred to as cGAN (conditional GAN) allows for generating inputs from a specific class by imposing additional restrictions on the random input vector provided. This restriction can be thought of selecting from classes within the domain set. To train a cGAN the discriminator also needs to be provided with this additional label input to constrain classification within the chosen class.

\paragraph{Quantum enhanced variants \\}

In the quantum generative adversarial network (QGAN) \cite{2018_seth}, we have a source $U_S$ that outputs true samples and a generator $U_G$ that outputs fake samples. Both $U_S$ and $U_G$ takes an input state $\ket{0}^{\otimes}n$, label $\ket{\lambda}$ and random noise $\ket{z}$ to output a density matrix in the system qubits. The noise vector supports to provide a distribution for the generator on the output qubit state corresponding to a given input label. With equal probability we choose between $U_G$ and $U_S$ to create a density matrix that is fed into the discriminator. Alongside the sample output from $U_S$ or $U_G$ provided, the discriminator takes as input the label $\ket{\lambda}$ used to generate the state, a bath $\ket{0}^{\otimes}d$ that works as scratchpad and an output qubit to measure the probability of the source of the sample. Figure \ref{GAN_fig} provides a sketch for the working of the QGAN. The objective function can thus be given by,

\begin{equation}
    \min_{\vec{\theta_G}}\max_{\vec{\theta_D}} \frac{1}{2} + \frac{1}{4N} \sum_{\lambda=1}^{N} tr[U_D(\vec{\theta}_D) \rho^{S}_{\lambda} U_D^{\dagger}(\vec{\theta}_D) Z] - tr[U_D(\vec{\theta}_D) \rho^{G}_{\lambda} U_D^{\dagger}(\vec{\theta}_D) Z] 
\end{equation}

where $\rho^{S}_{\lambda} = U_S \rho^{0}_{\lambda}U^{\dagger}_{S}$ is the state output by the source and $\rho^{G}_{\lambda}= U_{G}(\vec{\theta}_G)\rho^{0}_{\lambda}(z)U^{\dagger}_{G}(\vec{\theta}_G)$ is the state output by the generator and $Z$ represents the measurement made at the output qubit of the discriminator. The first and second term in the trace comes from the discriminators success in correctly predicting states from source and generator respectively. The cost function has been derived using measurement probabilities to keep the expression linear, unlike the maximum likelihood optimization used for the classical case. Given the optimization function, gradients can be computed using parameter shift trick or re-expressing it as a sum of simple unitary operations \cite{2019_maria}. For a complete discussion on the derivation of cost function, analyzing limiting cases and computing gradients corresponding to the parameters, refer \cite{2018_qgan}.\\

{\color{black}
\subsection{Tensor Networks}
\label{Tensor_Network}

\begin{figure*}
    \centering
    \includegraphics[width=14cm]{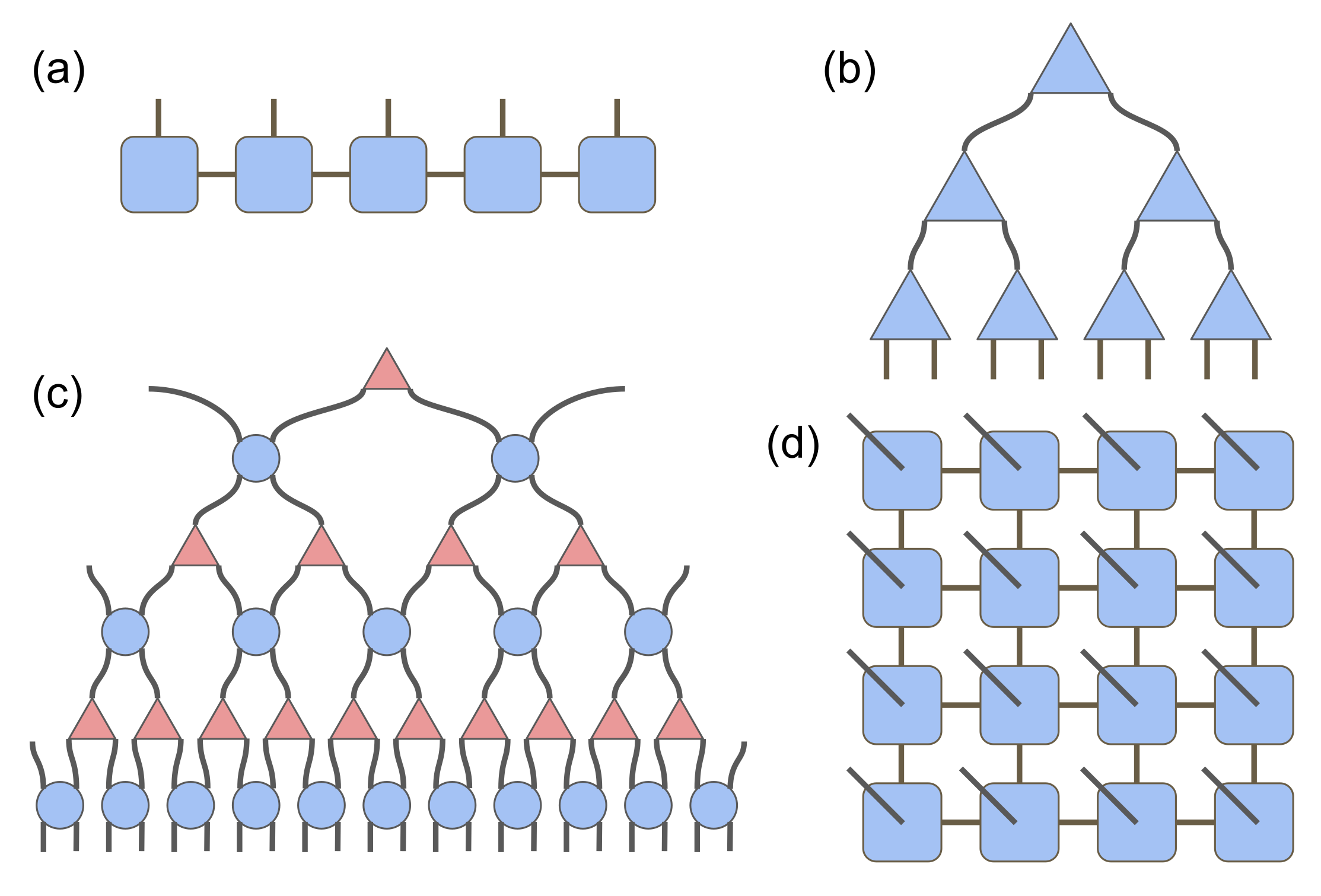}
    \caption{Schematic Representation of different types of Tensor Networks: (a) Matrix Product State, (b) Tree-Tensor Networks, (c) Multi-scale Entanglement Renormalization Ansatz, (d) Projected Entangled Pair States on a square lattice. Each solid object represents a tensor while the black lines denote the indices. The triangles in (b) and (c) are Isometric tensor while the circles in (c) are Unitary disentanglers.}
    \label{fig:TN_Diagrams}
\end{figure*}

Tensor network states constitute an important set of variational quantum states for numerical studies of strongly correlated systems in physics and chemistry as they attempt to construct global quantum states from tensors associated with local degrees of freedom. \cite{Biamonte2017-ip, Bridgeman2017-kq}. Expressing a quantum many-body systems defined on $n$ qubits requires $2^{n}$ complex coefficients. Storing and manipulating these numbers of coefficients on a classical computer poses a big challenge while simulating strongly correlated systems. Luckily physically relevant quantum states often possess limited amount of entanglement wherein only a subset of these coefficients are necessary to describe these states efficiently. Tensor networks provide a natural language to model complex quantum systems (states and operators) on which amount of entanglement (or correlations in the case of mixed-state dynamics) is conveniently restricted. 
The representation is such that the complexity of the structure grows linearly with qubit but exponentially with the amount of entanglement in the system. It thereby allows manipulation of quantum states residing in large Hilbert spaces with polynomial amount of resources. The classical simulation of low entangled systems (whose entanglement grows at most polynomially with system size $n$) becomes tractable using Tensor Network algorithms like Density Matrix Renormalization Group (DMRG) [Discussed in detail in Section \ref{DMRG_sec}] and Time-Evolving Block Decimation (TEBD)\cite{vidal2004efficient, Vidal2003-on}.

Tensor Networks are graphic representation of tensors in Einstein notation such that a rank-n tensor is represented by a box with n indices projecting out of it. The connections between tensor signify the set of indices of tensors which are contracted. Hence the final rank of a tensor network is determined by number of free edges. A quantum state $|\psi\rangle$ in n-dimensional Hilbert space is a basically a rank-n tensor and can be written as, 
\begin{eqnarray}
|\psi\rangle=\sum_{l_1 l_2 \cdots l_n} m_{1_1 l_2 \cdots l_n} |l_1\rangle |l_2\rangle \cdots |l_n\rangle
\end{eqnarray}
where  $|l_i\rangle$ represent that local basis states and the coefficients $m_{1_1 l_2 \cdots l_n}$ are the amplitude of wave function in a given basis state $|l_1\rangle |l_2\rangle \cdots |l_n\rangle$. 

The idea to using Tensor Networks to represent quantum systems is motivated from the very famous \textbf{Area Law}\cite{Eisert2010-xu} which states that the ground state of the Hamiltonian resides in the low entangled space such that entanglement entropy between any two partitions of the system grows as the area of the surface separating them. The entangled entropy is usually quantified in terms of the \textbf{von Neumann entropy} of a quantum many-body systems which is defined as $S(\rho) = -tr[\rho \log{\rho}]$, where $\rho$ is the density matrix of the state. It serves as a befitting measure of the degree of quantum correlation that exists between any of the two-partitions of the system under consideration. Area Law has been proven only for gapped Hamiltonians in one-dimension by Hastings \cite{Hastings2007-pl} and has been studied intensively for higher dimensions [See section IV of review \cite{Eisert2010-xu} for detailed discussion on Area Law in higher dimension]. Area Law guarantees an efficient description of ground states by a Matrix Product State and justifies the Density-Matrix Renormalization Group algorithm. Each of these two methods will be discussed in detail in Section \ref{MPS_sec} and Section \ref{DMRG_sec} respectively. 

Tensor Networks can be broadly classified into two main groups: those based on matrix product states (MPS), tree tensor network state (TTN) and their higher dimensions analogues (ex. PEPS); and those based on the multiscale entanglement renormalization ansatz (MERA). We shall discuss applications of TN in Section \ref{State_class_sec} and in Section\ref{Case_QML} and Section \ref{MBS_sec}

\subsubsection{Matrix Product State (MPS)}
\label{MPS_sec}

A pure quantum state $|\psi\rangle$ of a $n$-qubit system can be described by the sum of tensor products of orthogonal states in two subsystems. The \textbf{Schmidt decomposition} \cite{nielson} of $\psi$ with respect to the partition reads,

\begin{eqnarray}
|\Psi_{AB}\rangle =\sum_{i=1}^{m} \lambda_i |u_{i}\rangle_A \otimes |v_{i}\rangle_B
\end{eqnarray}
where $|u_i\rangle_A$ and $|v_i\rangle_B$ are the state of the subsystems A and B and $\lambda_i$'s are the Schmidt coefficients of the quantum state with respect to the partition. The Schmidt rank $\chi_A$ is defined by the number of non-zero Schmidt coefficients. It is a natural measure of the entanglement between the qubits in A and B, popularly known as bond dimensions in the Tensor Networks community. The von-Neumann entropy between the two partitions is given by,
\begin{eqnarray}
S=-\sum_i |\lambda_i|^2 \ln |\lambda_i|^2
\end{eqnarray}

Sometimes entanglement entropy is measured in ebits where one ebit is the amount of entanglement possessed by a maximally entangled two-qubit Bell state. Now if the states $|u\rangle_A$ and $|v\rangle_B$ are further partitioned in a similar fashion such that in the end the states we are only left with small states defined on single qubits. Let these states be denoted by $\Lambda^{[i]}$ and the diagonal matrix containing Schmidt coefficient be denoted by $\lambda^{[i]}$. Then the quantum state reads as,
\begin{eqnarray}
|\psi\rangle = \sum_\alpha |\Lambda^{\alpha_1}_{s_1}\rangle \lambda^{[1]} |\Lambda^{\alpha_1, \alpha_2}_{s_2}\rangle \lambda^{[2]} \cdots |\Lambda^{\alpha_{n-2}, \alpha_{n-1}}_{s_{n-1}}\rangle \lambda^{[n-1]} |\Lambda^{\alpha_{n-1}, \alpha_{n}}_{s_n}\rangle
\label{Eq:MPS_state}
\end{eqnarray}

where $|\Lambda_{s_i}\rangle$ are complex square matrices of order $\chi$ (the bond dimension). $s_i$ represent the state indices in the computational basis (physical indices). These format of representing quantum states is known as Matrix Product State.\cite{Orus2014-bg} In the tensor network notation it can be described as [Fig. \ref{fig:TN_Diagrams} (a)]. Operators can similarly be represented in Matrix Product form known as Matrix Product Operator (MPO).

Matrix Product States in theory can represent a Maximally Entangled state but the bond dimension at middle cut would grows as $O(2^n)$ \cite{Orus2014-bg}. For a MPS with fixed bond dimension $\chi$, the quantum state residing in $n$-qubit Hilbert space can now be represented using just $O(n\chi^2)$ parameters. Area Law limits the bond dimension of the ground state of local gapped Hamiltonians making them the best candidates for MPS representation. Evaluating inner products of two quantum states in MPS form takes $O(n \chi^2)$ 

$\lambda^{[i]}$ matrices in the Eq. \ref{Eq:MPS_state} are usually absorbed into the nearby local tensor $(\Lambda)$. The Matrix Product state is invariant of the contraction of the $\lambda^{[i]}$ either to left or right. This gives MPS a gauge degree of freedom. Usually the gauge is fixed by choosing either one direction for multiplying $\lambda^{[i]}$ giving rise to the left and right canonical forms of MPS. The process is known as canonicalization \cite{Vidal2007-by, Orus2014-bg}. There is another kind of canonical form known as the Mixed canonical form \cite{Schollwock2011-al} which is obtained by combining each $(\lambda^{[i]})$ to the left (right) of a given special site to its left (right) neighbouring $(\Lambda)$. 

\subsubsection{Tree Tensor Networks (TTN)}
\label{TTN}

Tree Tensor Networks provide another approach to model quantum states by arranging the local tensors in a tree-like pattern (See Fig \ref{fig:TN_Diagrams}(b)). A TTN can be formed from a $n$-qubit quantum state using the Tree-Tucker Decomposition \cite{Oseledets2009-xd, Oseledets2011-te}. Like other tensor networks, TTNs are used as ansatz to simulate ground state of local Hamiltonian. \cite{Shi2005-gm, Nagaj2008-kr, Friedman1997-zp} Tensors in TTN form the nodes of the tree which are connected to each other through bond indices. The physical indices appear on the leaf nodes. On contracting the bond indices, the TTN has $n$ free indices which represent the physical degree of freedom of the state. {\color{black} TTNs are a generalization of MPS and can in principle be non-binary as well \cite{doi:10.1063/1.4798639,https://doi.org/10.48550/arxiv.2011.00860}. A MPS can be thought of as a flattened TTN such that each parent node has one successor (bond indices of MPS) and another leaf node (physical indices of MPS).}

The structure of TTN is inspired from spatial Renormalization Group. \cite{Fisher1998-xg} At every layer of TTN, a coarse-graining is carried out between neighbouring sub-trees. Unlike MPS, the local tensors with access to physical indices in TTN are not connected directly to each other, the correlation between qubits is represented through the layers. The local correlation information is stored in the lower layers while the upper layers store long-range correlation information. 

Each node in a TTN is three-dimensional tensors (except the root/uppermost node) with at most one upper index $\alpha$ and two lower index $\beta_1$ and $\beta_2$. The tensors can be written as $w^\alpha_{\beta_1, \beta_2}$. The space required to store a TTN grows as $O(ND^3)$ (See Theorem 4.1  \cite{Oseledets2009-xd}), where $N$ is the number of physical indices and $D$ is the bond dimension of local tensors. Each tensor in TTN is an isometry satisfying the following condition: 
\begin{eqnarray}
\sum_{\beta_1, \beta_2} (w)^\alpha_{\beta_1, \beta_2} (w^\dagger)_{\alpha'}^{\beta_1, \beta_2} \label{Eq:TTN_tensor}
\end{eqnarray}

Choosing isometric tensor as in Eq.\ref{Eq:TTN_tensor} is advantageous in numerous ways. It simplifies the optimization of TTN and calculation of expectation values of local observables and it is also known to provide numerical stability to TTN algorithms.\cite{Tagliacozzo2009-xp}. TTN can very well be generalized to higher dimensions by appropriately placing isometries across local physical indices and hierarchically merging sub-trees through more isometries. Tagliacozzo et al. \cite{Tagliacozzo2009-xp} in their works demonstrate simulation of the Transverse-field Ising model on square lattice using a two-dimensional TTN. Their approach takes advantage of the Area Law which reduces their simulation cost to $exp(N)$ instead of $exp(N^2)$. 

{\color{black} 
Tree Tensor Networks form the basis of the multi-layer multi-configuration time-dependent Hartree (ML-MCTDH) methods which are used to perform quantum molecular dynamics simulations. In the report \cite{larsson2019computing} authors compute the vibrational eigenstates of acetonitrile using TTNs. ML-MCTDH methods are a generalization of MCTDH methods which can be optimized using MPS as shown in the report \cite{kurashige2018matrix}.Authors make use of the DMRG algorithm to efficiently evaluate the mean-field operators represented in MPS format. The runtime of MCTDH methods scales exponentially with system size, hence the multi-layered MCTDH is used which makes use of Tucker decomposition to reduce the dimensionality of the problem and enables it to simulate larger systems. }

\subsubsection{Projected Entangled Pair States (PEPS)}
\label{PEPS}
PEPS is a generalization of MPS in higher dimensions or for arbitrary graphs.\cite{Orus2014-bg} It get its name from the way it is constructed. Let a vertex of a graph contain $k$ edges, each edge can be represented by a virtual spin of dimension $D$ (bond dimension). The edges are described by a maximally entangled state $|I\rangle = \sum_{i=1}^D |ii\rangle$. Now the vertex can be defined by a $k$-rank tensor containing entangled states. Ultimately this tensor is projected onto the physical spin through a linear map, $(\mathbb{C}^D \otimes \mathbb{C}^D \otimes \cdots \otimes \mathbb{C}^D) \longrightarrow \mathbb{C}^d$, 
where d is the local dimension of the physical state. 

In one dimension (k=2), an entangled pair of states are projected onto the physical index. While in a square lattice, each local tensor has at most four neighbours. [See Fig \ref{fig:TN_Diagrams}. (d)] Hence the local tensor can be written as $\Lambda_s^{\alpha, \beta, \gamma, \delta}$, where the $s$ is the physical index and $\alpha$, $\beta$, $\gamma$, and $\delta$ are bond indices. Hence storing a PEPS requires $O(N^2 d D^4)$ space, where $N$ is the number of qubits along a side of the square, $d$ is the dimension of the physical local state and $D$ is the bond dimension. Performing computations on PEPS is difficult,\cite{Kliesch2014-rd} for instance evaluating the inner products of PEPS scales exponentially with the $D$. This is because any partition which divides PEPS into two equal parts always cuts $O(N)$ bonds, hence while evaluating the inner product one has to form a rank-$O(N)$ tensor as an intermediate. 

PEPS can theoretically represent any state due it's generic structure given that its bond dimension can be arbitrarily large. Due to this universality, PEPS serve as variational ansatz in numerical simulation of a wide variety of quantum systems. It can easily prepare physically important states like GHZ and Cluster State \cite{Raussendorf2001-ib} using $D=2$.  With D=3, PEPS can prepare a Resonance Valence Bond states. \cite{Verstraete2006-rp} Kitaev's Toric code which finds its applications in quantum error correction and demonstrates non-trivial topological properties can be prepared using PEPS with D=2. \cite{Kitaev2003-pu} It is widely known that PEPS can efficiently approximate ground states of gapped local Hamiltonian which satisfy the area law. In the report \cite{Schwarz2017-wy} author show that they can compute the expectation values of local observables in quasi-polynomial time. Jordan $et al.$ propose algorithms to compute the ground states and time evolution of two-dimensional Hamiltonians defined on infinite-size lattice using PEPS. \cite{Jordan2008-qw} It is known that it is difficult to simulate systems with long-range correlations on PEPS, but Gu et al. extensively studied these systems to demonstrate the power and versatility of PEPS. \cite{Gu2008-rz} They studied both systems which exhibit symmetry breaking phase transition (transverse field Ising model) and those that show topological phase transition ($Z_2$ gauge model, Double-semion model).

While PEPS have been designed to study quantum systems on classical computers, there have been approaches to simulate them on a quantum computer for faster computations. Schwarz et al. in their report \cite{Schwarz2012-oz} presented an algorithm to prepare a PEPS on a quantum computer which scale only polynomially with the spectral gap and the minimum condition number of the PEPS projectors. In the consecutive year they came up with another algorithm to prepare topologically projected entangled pair states on a quantum computer with similar runtime. \cite{Schwarz2013-rl} Specifically they simulated the resonance valence bond state which is hypothesized to contain topological spin liquid phase.

There also exist infinite version of MPS (PEPS) known as iMPS (iPEPS). \cite{Jordan2008-qw} They allow working directly in the thermodynamic limit without encountering the finite size or boundary effects. There have been accurate studies of continuous quantum phase transitions using iPEPS. \cite{Corboz2018-ut}

\subsubsection{Multi-scale Entanglement Renormalisation Ansatz (MERA)}
\label{MERA_sec}
MERA \cite{Milsted_undated-fs} is a powerful class of Tensor Networks which can be used to study gapless ground states and properties of systems near criticality. Despite its huge success in representing a wide variety of states, MPS is scalable only for gapped systems with exponentially decaying correlations and the Area Law is strictly satisfied. Owing to its hierarchical structure, MERA allow long-range correlations and shows a polynomially decaying correlations [Shown in Eq. 5 of Ref \cite{Vidal2008-vt}]. The entanglement entropy of a $N$-qubit 1D gapless system grow as $O(\log(N))$ and hence they can be naturally represented by a MERA. 

The architecture of MERA is inspired from the Renormalization Group. \cite{Vidal2007-by, Evenbly2009-am}
Its structure is very similar to TTN with an additional type of tensors known as disentanglers ($U$) [shown in Fig. \ref{fig:TN_Diagrams}. (c) using blue circles]. These are unitary tensors satisfying,
\begin{eqnarray}
\sum_{\beta_1, \beta_2} (U)_{\alpha_1, \alpha_2}^{\beta_1,\beta_2} (U^\dagger)_{\beta_1,\beta_2}^{\alpha'_1, \alpha'_2} = \delta(\alpha_1,\alpha'_1)\delta(\alpha_2,\alpha'_2)
\end{eqnarray}
whereas the isometries ($W$) [depicted using red triangles in Fig. \ref{fig:TN_Diagrams}. (c)] satisfies Eq. \ref{Eq:TTN_tensor}. To recover a TTN from MERA one can simply replace the disentangler with an Identity tensor. 

Entanglement in MERA builds up due its layered structure. To dissect a sub-system of $n$-qubits from the system requires at least $O(\log(n))$ bonds to be broken. Hence the maximum entanglement entropy generated by the MERA goes as $O(\log(n)\log(D))$. That is why MERA allows logarithmic divergence from the Area Law. \cite{Vidal2007-by, Vidal2008-vt} 

Storing a MERA on a classical computer requires space polynomial in number of qubits and the bond dimension. Performing computations using MERA is simplified due to it structure. It can perform efficient computation of local expectation values and correlators by only contracting over the shadow (causal cone) of the local operator, i.e, the tensor which are directly connected to the operator and those tensors on higher levels which are further connected to these tensors. The isometries and disentanglers which lie outside this shadow contract themselves with their conjugates to give unity. 


\subsubsection{Density Matrix Renormalization Group (DMRG)}
\label{DMRG_sec}
DMRG is one of the most successful algorithm for simulation of condensed matter systems. It was introduced by White \cite{White1993-cj} in the pre-Tensor Network era. The algorithm has changed a lot over the years and has been simplified by adapting to the language of tensor networks. In the following discussion, we will be describing the modern DMRG algorithm using Matrix Product State formulation. \cite{Schollwock2011-al}

Finding the ground state of a Hamiltonian is challenging problem and yet is one of core problem in physics, chemistry, and material sciences. Even for one-dimensional k-local Hamiltonian it is known to be QMA-complete \cite{Kempe2006-li} i.e. it's difficult to solve it in polynomial time even with access to fully functional quantum computer. 

Since the ground state of a gapped local Hamiltonian is known to reside in low entanglement regime by the Area Law. DMRG algorithm makes use of this property to find the solution of the local Hamiltonian problem. The algorithm makes use of an ansatz which succinctly represents the state with bounded entanglement (Matrix Product State). The ground state of the Hamiltonian is attained by minimizing the energy of the system,
\begin{eqnarray}
E=\frac{\langle \psi| \hat{H} | \psi \rangle}{\langle \psi| \psi \rangle}
\end{eqnarray}

Before starting the algorithm the Hamiltonian has to be converted into a Matrix Product Operator so that it is compatible with MPS. Since its a k-local Hamiltonians which can be written as, $\hat{H}=\sum_i h_i$, where $h_i$ are local hermitian operators acting on at most k-qubits. Each $h_i$ can be converted into a MPO defined on k physical indices using recursive Singular Valued Decomposition as explained in Section \ref{MPS_sec}. Once local operators are converted to MPO, they can be added using MPO addition operation, which is basically a direct sum operation over the bond indices. (See section 5.2 in Ref. \cite{Schollwock2011-al} for details)

The initial MPS can be created using random tensors of desired dimension. At each step of DMRG a contiguous set of sites are chosen which is to be optimized and is designated as the system. While everything outside the system is called the environment which is kept fixed. By performing local optimization over the system states iteratively, the ground state of the given Hamiltonian is attained. Usually it requires several sweeps over the complete lattice to reach convergence which depend on the complexity of the Hamiltonian and also the choice of the initial state. 

To perform local optimization over system, the environment qubits are contracted to form a reduced Hamiltonian ($H_S$) whose energy is then minimized. 
\begin{eqnarray}
\hat{H}_S = \frac{\langle \psi_E| \hat{H} | \psi_E \rangle}{\langle \psi_E| \psi_E \rangle}
\end{eqnarray}
Energy minimization of $\hat{H}_S$ can be analogously performed by solving the following eigenvalue problem,
$\hat{H}_S |\psi_S\rangle = E |\psi_S\rangle$

The system state $|\psi_S\rangle$ so obtained updates the current system state. The system can be defined by any number of qubits. For single qubit systems the bond dimension remains fixed while working with two or more site system can allow the bond dimensions to be changed dynamically. Basically, the local optimization procedure for multi-site system returns the state defined on multiple qubits. This state has to decomposed into a MPS using recursive Singular Value Decomposition before replacing them at the current state. Since SVD gives us the complete set of singular values we can choose to trim the bond dimensions which are below the threshold of the accuracy required. Usually larger system size means more accurate results and the trial state converges to the ground state in lesser number number of sweeps. But it also increase the overall computation cost. The computational cost heavily depends on the local optimization procedure which can be improved by using iterative algorithms like Lanczos which only computes smallest set of eigenvalues and eigenvector of a given operator. Since we are only interested in the ground state of reduced Hamiltonian, Lanczos algorithm can massively cut down the computation cost.

\subsubsection{Quantum enhanced Tensor Networks}
There are numerous connections between tensor networks and quantum circuits. These relations lead to interest into two broad research directions. First one is the classical simulation of quantum circuits using Tensor Network. There are works demonstrating implementation of quantum algorithms like Grover's algorithm and Shor's algorithm in Matrix Product State (MPS) \cite{Kawaguchi2004-kj, Dang2019-yp} and Tree Tensor Network (TTN) framework \cite{Dumitrescu2017-yk}. Recently a report showed classical simulation of the random quantum circuit using Tensor Network.\cite{Huang2020-nz} The same circuit which was implemented on the Sycamore quantum processor to demonstrate "Quantum Supremacy" by Google.\cite{Arute2019-mu} There has been massive improvement over the years in the runtimes for evaluating Tensor Network classically. In a recent report by Huang et al. \cite{Huang2021-uw}, they demonstrated a new method called Index Slicing which can accelerate the simulation of random quantum circuits through tensor network contraction process by up to five orders of magnitude using parallelization. Markov et al. \cite{Markov2008-kk} theorized the time complexity of simulating quantum circuits using Tensor Network. A quantum circuit with a treewidth $d$ (a measure of how far a graph is from being a tree) and $T$ gates can be deterministically simulated in $O(poly(T) exp(d))$ time. 
Another research direction which has been gaining traction due to advents of noisy intermediate scale quantum (NISQ) computers is the optimization of Tensor Networks using quantum computers. \cite{Liu2019-mi, Ran2020-hr}.
There are Quantum Machine Learning models which uses ansatz inspired from Tensor Network. \cite{cong2019quantum, Huggins_2019}
\textcolor{red}
The analogy between TN and quantum circuits can be exploited to develop an efficient state preparation mechanism on a quantum computer. There have been efforts in creating quantum states in MPS\cite{Kardashin2018-ds}, TTN\cite{Huggins_2019}, PEPS\cite{Schwarz2012-oz}, and MERA\cite{Kim2017-cc} format using quantum circuits. 

Since the dimensions of the associated tensor grow exponentially with the depth of the quantum circuit associated with it, it is possible to prepare certain tensor networks with large bond dimension on a quantum computer that cannot be efficiently simulated on a classical computer. These states are of utmost importance because there is a definite quantum advantage associated with them. Authors in the report \cite{Kim2017-cc} demonstrated preparation of such a state called Deep-MERA which can be represented by a local quantum circuit of depth $D$ consisting of two-qubit gates. The expectation values of local observables of a DMERA can be computed in time $O(\frac{D \log L}{\eta^2})$ on a quantum computer while a classical computer would take $O(e^{O(D)} \log L \log (1/\eta)$ time, where $\eta$ is the desired precision and L is the number of qubits.

Schwarz et al. \cite{Schwarz2012-oz} demonstrate a procedure to efficiently prepare PEPS on a quantum computer that scales polynomially with the inverse of the spectral gap of Hamiltonian. 
\textcolor{black}{There also have been efforts to use the advantages of Tensor Networks and Quantum Circuit simultaneously by fusing them. In the report \cite{Yuan2021-kv}, authors introduced a hybrid tree tensor network architecture to perform quantum simulation of spin lattice Hamiltonian with short-range interactions. They simulated two-dimensional spin systems as large as $9 \times 8$ qubits which require operations acting on at most 9 qubits. Their method can be generalized to arbitrary trees to represent $N=O(g^{D-1})$ qubit system, where $D$ and $g$ are the maximal depth and degree of the tree. It would require  $O(N k^2)$ circuits for computation and the cost for measuring local expectation values would be $O(N g k^4)$, where $k$ is the bond dimension of the TTN. They provide an efficient representation of a quantum state whose elements can be evaluated on a near-term quantum devices. When compared against standard DMRG on MPS and imaginary-TEBD on PEPS, they produce results more accurate by up to two orders.}

}

{\color{black}

\section{Case for Quantum computing enhanced Machine Learning}\label{Case_QML}

\subsection{Universal function approximation through supervised learning on a quantum computer}

In this section, we shall specifically highlight some of the recent claims that propose a theoretical guarantee for supervised machine learning tasks on a quantum computer, with these claims being the successful mimicking of arbitrary unknown functional dependence with high accuracy. It is thus needless to say that the benefits of these claims if realized can enhance the learning capabilities of supervised models for both quantum and classical data even beyond the precincts of physical sciences.

Several significant proposals have been reported recently that attempts to approximate a function (say $f(\vec{x})$) using a quantum circuit where $x \in \mathbb{R}^d$ are classical data entries. Intuitively this can be framed as a supervised learning task where one has access to a dataset $D={(\vec{x_i}, \vec{y_i})_{i=1}^t}$ which is assumed to follow the functional inter-relationship $f(\vec{x})$. The crux of the problem is therefore to learn a hypothesis $h(\vec{x})$ which closely mimics the actual function $f(\vec{x})$ within a certain error tolerance. To perform such tasks on a quantum computer and learn the hypothesis $h(\vec{x})$ one needs to encode classical data onto a quantum state first. Mitarai $et al$ \cite{mitarai2018quantum} proposed a data-uploading scheme on a quantum circuit for such a scenario. The scheme maps $\vec{x} \in \mathbb{R}^d$ with $-1 \le x_i \le 1\:\: \forall \:\: i$ wherein one requires access to $n_k$ th power for each datum $x_i$ with $k \in \{1,2,3..d\}$ into an $N=\sum_k n_k$ qubit state as $\rho (\vec{x}) \propto \otimes_{k=1}^d \:\otimes_{j=1}^{n_k} (I + x_k X_j + \sqrt{1-x_{k}^2}Z_j) $. The tensor product structure of the many-qubit state creates non-linear cross terms of the kind $x_mx_n$. Following this data-encoding, the state can be acted upon by any parameterized unitary (say $U(\vec{\theta})$). This will be followed by a measurement protocol using a pre-defined operator (say $M$) to learn the hypothesis function $h(\vec{x, \theta})= Tr (M U(\vec{\theta})\rho(\vec{x})U(\vec{\theta})^\dagger)$. The hypothesis function is optimized with respect to the parameters $\vec{\theta}$ using an appropriate loss function $L(h(\vec{x, \theta}), \vec{y}, \vec{x})$ until the desired tolerance is reached i.e. at $h(\vec{x}) = h(\vec{x}, \vec{\theta}^*) \approx f(\
\vec{x})$ where $\vec{\theta}^* = arg\:\: min_{\vec{\theta}} \:\: L(h(\vec{x, \theta}), \vec{y}, \vec{x})$ The authors claim that the encoding above is capable of approximating a larger class of functions than what classical supervised learning tasks can achieve. To substantiate this claim the authors argue that if classical processors could mimic every kind of function, which can be realized from such quantum data-encoding, then that would mean the classical device in principle learn the input-output relationship of complex computational models like quantum cellular automata \cite{arrighi2019overview} which is known to not being achievable using polynomial resources $(poly(N))$ on a classical device. Numerical experiments for fitting the time evolution of transverse Ising model and a binary classification task of a non-linearly separable data were performed with the above encoding with great success. 

Perez-Salinas \cite{perez2020data} demonstrated how to construct single-qubit classfiers using efficient data-reuploading which is essentially sequential loading of classical data entries. Many powerful insights into the function learning ability of a quantum circuit through data-encoders has been recently elaborated in Ref \cite{schuld2021effect}. The work explicates if the data-encoding unitary is expressed as $S(\vec{x})=e^{iH_1x_1}\otimes e^{iH_2x_2}....\otimes e^{iH_dx_d}$ and $r$ repetitions of such unitaries in the circuit are made along with parameterized unitaries (say $U(\vec{\theta})$ as above) for training the model then the frequency components of the hypothesis function $h(\vec{x})$ when resolved in the Fourier basis are entirely controlled by the encoding Hamiltonian family $\{H_m\}_{m=1}^d$. However, the Fourier coefficients are influenced by the remaining part of the circuit i.e. the trainable unitaries $U(\vec{\theta})$ as well as the measurement operator $M$. The authors further show that repeating the encoding in parallel or in a sequence would lead to a similar frequency spectrum. Under the assumption that the trainable part of the circuit is general enough to realize any arbitrary unitary, then it is possible to choose encoding Hamiltonians $\{H_m\}_{m=1}^d$ that can generate any arbitrary frequency range asymptotically. Using this fact the authors prove that it is possible for the hypothesis function $h(\vec{x})$ learnt by such a quantum circuit to mimic any square integrable function within an arbitrarily preset tolerance. This thereby lends to universal expressibility to such hypothesis functions. The importance of this result is many-fold as it allows one to not only realize that expressive power of the family of functions learnt from supervised learning task on a quantum circuit is extremely high but also allows one to design unitaries, set number of necessary repetitions etc to augment the learning process. Universality in  discriminative learning wherein a hybrid quantum model to learn the parameters of an unknown unitary was used, have also been illustrated recently \cite{chen2021universal}. Other than these expressive capacity of parameterized quantum circuits have been thoroughly investigated recently \cite{https://doi.org/10.1002/qute.201900070}. Since most of NISQ era quantum ML models are indeed variational, much of the insight from these studies are directly transferable.

\subsection{Power of Kernel estimation and data-classification from Quantum computers.\\}

In this section we shall highlight some of the key results that has been demonstrated in recent years regarding the superiority of constructing kernel matrix elements from the quantum computer as opposed to a classical processor. Such kernel estimates are necessary for a variety of supervised learning algorithms like Kernel-Ridge Regression (see Section \ref{KKR_sec}) or for classification tasks like in  Support-Vector Machine or SVM (see Section \ref{Supp_vec_machine}) to name a few. Kernel-Ridge Regression on a classical processor has been extensively used in chemistry for estimating density functionals \cite{snyder2012finding}, simulating non-adiabatic dynamics across potential energy surfaces \cite{hu2018inclusion}, dissipative quantum dynamics \cite{ullah2021speeding} and even procuring molecular and atomic properties like atomization energies \cite{PhysRevLett.108.058301,doi:10.1063/1.5126701}. We shall return to a subset of these applications and explore them in detail in Section \ref{MBS_sec}. Even for classification, kernelized variants of SVM on a classical processor have been useful for demarcating phases of matter, or for delineating malignant tumors from non-malignant ones \cite{huang2017svm} which would be of use to biochemists and oncologists. We shall return to a subset of these applications in Section \ref{State_class_sec}. Thus the learning capabilities of all the aforementioned algorithms can be augmented if quantum computing-enhanced kernel estimates are used. Kernelized SVM has also been used extensively for drug-designing process \cite{}, in drug-induced toxicity classification \cite{sun2012structure} etc. We shall discuss some of these in Section \ref{Drug_discovery}. In fact a study has already demonstrated quantum advantage recently \cite{batra2021quantum} wherein kernel SVM on an actual quantum device ($ibmq\_rochester$) was used with classical processing to delineate active vs inactive drug candidates for several diseases. The authors note faster training time on a quantum processor than on the classical processor for larger dataset sizes. We shall discuss this specific example in detail in Section \ref{drug_QC}.

It must be emphasized that for classification tasks, apart from quantum Kernel methods, the quantum instance-based learning algorithms could also outperform classical learners. Estimating the distance between the test data and the training ones is always crucial in the instance-based learning algorithms. For instance, in the nearest neighbor algorithm, one of the most typical instance-based learning frameworks, the label of the test data is determined by the nearest training data. In 2013, Lloyd and coworkers proposed a quantum clustering algorithm for unsupervised QML\cite{lloyd2013quantum}, showing that estimating distances and inner products between post-processed vectors in N-dimensional vector spaces then takes time $O(log N)$ on a quantum computer.
On the contrast, sampling and estimating distances and inner products between post-processed vectors on a classical computer is exponentially hard\cite{aaronson2010bqp}.The significant speedup yields considerable power of the quantum instance-based learning algorithms as well. In fact a specific example of this class of algorithms which inherits the aforesaid advantage has also been recently designed by one of the authors \cite{li2021quantum} and applied for phase classification of material $VO_2$ which will be of importance to material scientists. More details on such examples can be found in Sec.(\ref{State_class_sec}) and will not be elaborated herein. Here we shall specifically discuss the advantages of estimating the kernel on a quantum processors that has been noted recently for certain tasks which thereby promises exciting opportunities for kernelized quantum supervised learning with applications in physics and chemistry.

\begin{enumerate}

    \item Quantum-enhanced feature maps and kernels were defined in Section \ref{Kernel_learning}. As mentioned therein, Ref \cite{PhysRevLett.122.040504} provides two strategies for efficiently performing kernel-based machine learning algorithms using a quantum computer. The first is an implicit approach wherein the kernel matrix is estimated through an inner product once a quantum circuit for state preparation with encoding classical data is in place. With access to the entries of the kernel-matrix from the quantum computer, the actual ML algorithm is then performed classically. The other approach is the explicit approach, where the full ML task is performed on the quantum computer itself. Ref \cite{PhysRevLett.122.040504} adopts the first approach by encoding each entry $x_i$ of a given feature vector $x \in \mathbf{R}^d$ in the phase information of a multi-mode squeezed state and shows the corresponding kernel obtained through inner product of such states is expressive enough for classification tasks. To exemplify the second approach, it also used the two-mode squeezed state as a data-encoder and then applied a variational circuit (say $W(\theta)$) followed by photon-number measurement and assigned the probability of a binary classification task to obtain two specific Fock states in the two-modes. Using the distribution obtained from the QC, the authors could linearly separate a dataset with 100 \% accuracy. However, the authors note that the primary data-encoding strategy adopted in the paper is through the preparation of squeezed states in continuous variable quantum computing which can be efficiently simulated classically \cite{Bartlett2003, RevModPhys.77.513}. They further mention that inclusion of non-Gaussian elements like cubic-phase gates \cite{Lloyd2003}, non-linearity in photon-number measurements\cite{PhysRevA.65.042304}, classically intractable continuous-variable instantaneous quantum computing or CV-IQP circuits \cite{PhysRevLett.118.070503} may lead to non-trivial kernel estimation task wherein the power of quantum computers can be better used. Similar results as these are also reported in Ref \cite{li2021quantum} wherein classical data was not-only encoded within the phase information of a multi-mode squeezed state but also in the amplitude. Proper comparisons of such squeezed state encoded kernels with Gaussian kernels were also investigated using standard datasets from scikit learn \cite{pedregosa2011scikit}.

    \item The first work to exemplify an advantage is Ref \cite{havlivcek2019supervised}. The algorithm in Ref \cite{havlivcek2019supervised} performs a standard support-vector machine classification task (discussed in Section \ref{Supp_vec_machine}) with $x_{t} \in \mathcal{T} \subseteq \mathbf{R}^d$ (training feature vectors for input), and $x_{tt} \in \mathcal{S} \subseteq \mathbf{R}^d$ (testing feature vectors). The labels $ y : T \cup S \mapsto \{+1, -1\}$ where the set $y = \{y_t, y_{tt}\}$. The algorithm only had access to the training labels ($y_t$) and its job was to evaluate an approximation to the testing labels i.e obtain $\Tilde{y}_{tt} \:\:\forall \:\: x_{tt} \in S$ which matches with $y_{tt}$ with high probability. Unlike in the previous reference, the data-encoding feature map used did not produce product states. The specific data-encoding unitary used is the following:
    \begin{eqnarray}
       U(x) = \Tilde{U}_{\phi(x)} H^{\otimes n}\Tilde{U}_{\phi(x)}H^{\otimes n}  \nonumber \\
       \Tilde{U}_{\phi(x)} = e^{i\sum_{S \subseteq n} \phi_S(x) \prod_{i} Z_i} \label{data_encod_SVM}
    \end{eqnarray}
    where $n$ is the number of qubits, $S$ denotes the nature of the unitary i.e if the unitary is $S-$ local. For simulations the work used $S=2$. Ref \cite{havlivcek2019supervised} argued that the above mentioned data-encoding is hard to simulate classically. The feature vector size $d=2$ i.e $x_t=[x_1, x_2]^T$ and the feature maps are defining the unitaries in Eq.\ref{data_encod_SVM} are
    \begin{eqnarray}
    \phi_{S=1}(x) = x_1 \nonumber \\
    \phi_{S=2}(x) = (\pi - x_1) (\pi - x_2)
    \end{eqnarray}
     The first classification protocol which the authors in Ref \cite{havlivcek2019supervised} implemented is the explicit approach wherein after the above-mentioned data-encoding a variational circuit (say $W(\theta)$) was also implemented followed by a measurement protocol. If the probability for certain specific bit-strings were higher than a tunable threshold, the algorithm was said to yield $\tilde{y}_t =1$ (or -1 otherwise). Numerical experiments were conducted on a 5-qubit superconducting circuit and the depth of the variational circuit was varied from 0-4. The training set had 20 data points for each label and so did the testing set. The success ratio as seen in Fig 3 of Ref \cite{havlivcek2019supervised} was close to 100\% for 4 layers of the variational circuit. In the second part of the numerical experiment, the authors followed the implicit scheme in which only the estimates of the kernel matrix were obtained from the quantum computer. The rest of the classification task was performed classically once that was done. The constructed kernel matrix from the actual hardware agreed fairly well with the ideal one (see Fig. 4 in Ref \cite{havlivcek2019supervised}) and classification task using it was of 100 \% accuracy. After this demonstration of a data-encoding scheme which is hard to simulate classically, several other numerical experiments have been initiated to validate kernelized SVM on a quantum computer in different platforms \cite{kusumoto2021experimental,bartkiewicz2020experimental}.
     
     \item A recent report \cite{guo2022where} using the same feature-space encoding scheme as in Ref \cite{havlivcek2019supervised} above establishes that quantum enhanced kernels perform better for complex data classification tasks like geometric data patterns distributed according to Mersenne Twister distribution \cite{10.1145/272991.272995}. Classical methods cannot achieve similar accuracy. However, if the data distribution is simple such that large differences exist between data that belongs to the separating classes then classical kernels would perform as well. Also, the study claims that simpler data encoding circuits for computing entries of quantum kernels may be less effective for certain data-classification tasks. Another study \cite{Wang2021towards} has actually systematically studied the effect of noise and finite measurement samples and concluded that high noise content may expectedly be detrimental to the estimation of kernel entries on the quantum computer. However, the report \cite{Wang2021towards} also proposes to mitigate the effect of the noise by classical pre-processing of the estimated noisy kernel like discarding the negative eigenvalues.
     
     \item A clear and most decisive exhibition of the power of kernelized variant of support-vector machine on a quantum computer for a classification task was highlighted in Ref \cite{liu2021rigorous}. The motivation for the work was to demonstrate a specific example wherein estimation of the kernel Gram matrix on a classical processor would not only be not efficient, but the classification task itself would be provably disadvantageous. Also, the quantum advantage would be retained even in presence of finite sampling errors. The classification task chosen was based on the famous discrete-logarithm problem. The problem entails finding $log_g (x) \:\: \forall x \in \mathbf{Z}_p^*$ where $\mathbf{Z}_p^* = \{1,2,....p-1\}$ with $p$ being a large prime number and $g$ being the generator of the multiplicative cyclic group $\mathbf{Z}_p^*$. By generator one means an element $g \in \mathbf{Z}_p^*$ such that for every element $x \in \mathbf{Z}_p^*$, one can write $x=g^m$. In such a case $m$ is said to be the discrete-logarithm to base $g$ of $x$ in $\mathbf{Z}_p^*$ and is denoted by $m=log_g(x)$. It is believed that no classical algorithm can compute the discrete-logarithm in time which is polynomial in $n=log_2(p)$ even though quantum algorithms like Shor's algorithm is known to do it \cite{shor1999polynomial}. The classifier function makes the following decision
     \begin{eqnarray}
     f_s(x) &=& +1\:\:\:\: \rm{if}\:\:\:\: log_g(x) \in [s, s+(p-3)/2] \nonumber \\
    &=&-1\:\:\:(\rm{otherwise})
     \end{eqnarray}
     This classifier thus divides the set $\mathbf{Z}_p^*$ into two equal halves by mapping each of the halves to $\{+1,-1\}$. The authors prove that a classical learning algorithm for this task cannot achieve an accuracy more than $0.5 + 1/poly(n)$ indicating that the best classical algorithm can only do random guessing for unseen test data. For the kernelized quantum SVM, however the authors propose the following quantum feature map/state 
     \begin{eqnarray}
        |\phi(x)\rangle = \frac{1}{\sqrt{2^k}} \sum_{i=0}^{2^k -1} |xg^i\rangle
     \end{eqnarray}
     where $x \in \mathbf{Z}^*_p$, $k= n -tlog(n)$ for $t$ being some constant \cite{liu2021rigorous}. The state-preparation circuit which prepares the above state is shown to be efficient using Shor's algorithm \cite{shor1999polynomial}. Using the above state-preparation strategy, the authors can estimate the kernel matrix for each entry in the training set as $K(x, x^\prime) = |\langle \phi(x)| \phi(x^\prime)\rangle|^2$. Using this kernel, the authors rely on the usual SVM algorithm on a classical processor to construct a separating hyperplane and optimize the parameters for it. Once the parameters for the hyperplane are determined, classification of new test data also requires kernel matrix elements when new kernel estimates from the QC are invoked again. The authors call this procedure support vector machine with quantum kernel estimation (SVM-QKE) indicating that the quantum computer is only involved in constructing the entries of the kernel. The authors prove that SVM-QKE yields a classifier that can segregate the data in testing and training set with an accuracy of 0.99 in polynomial time and with a probability of at least $\frac{2}{3}$ over even random training samples. They further show that even when the QKE entries have a small additive perturbation due to finite sampling, the separating hyperplane so obtained is close to the exact one with high probability and so is the accuracy of the corresponding classifier. Since the kernel estimates from a classical processor cannot provably do better than random guessing, this classification task clearly explicates the superiority of quantum feature maps. Since then, several interesting demonstrations of kernel estimates on the quantum computer have emerged like using a linear combination of multiple quantum enhanced kernels with a variational circuit to enhance accuracy and expressivity over and beyond a single kernel for complex datasets \cite{vedaie2020quantum}, a fidelity based quantum kernel estimate between the members of the training dataset and the testing samples \cite{blank2020quantum} or even distinguishing the classical data entries directly after mapping to quantum states in the quantum feature space using metrics with shallow circuit depth in a process which the authors call quantum metric learning \cite{lloyd2020quantum} . Like in the above cases, numerical experiments have also been reported on real devices like a 17-qubit classification task \cite{peters2021machine} performed on Google's Sycamore to segregate data in a 67-dimensional space with appropriate noise-mitigation strategies.
     
\end{enumerate}     
\subsection{Power of Quantum-Neural Networks}     

In this section we shall highlight some of the recent reports wherein superiority of quantum computing enhanced neural network models have been demonstrated or theoretically proven in terms of its generalizability and expressive power, training capacity, resource and parameter requirements to mention a few. Neural networks in a classical processor have become the standard go-to method for many applications in chemistry and physics like in efficient state-preparation protocols using generative adversarial networks (see Section \ref{GAN_section}) \cite{yang2020tomographic, liu2020tomogan}. Networks like CNN (see Section \ref{CNN_section}) have been used for the classification of phases of matter \cite{carrasquilla2017machine, gao2018experimental}, in quantum state-tomography, \cite{lohani2020machine} and in structure and ligand based drug-designing protocols \cite{ragoza2017protein}. 
Deep neural networks (see Section \ref{ANN_section}) have been also used for predicting molecular properties even with non-bonding interactions \cite{yao2018tensormol}, in drug-induced toxicity detection \cite{hughes2015modeling}, many-body structure of correlated quantum matter like molecules and materials, \cite{caetano2011using, schutt2019unifying} and even in molecular dynamics \cite{galvelis2017neural,zeng2020complex}. Generative models like Restricted Boltzmann Machine based neural-network representation of many-body quantum states \cite{carleo2017solving} has been used for classification and understanding ground and excited state properties of quantum systems. We shall return to a subset of these applications in Section \ref{state_prep}, \ref{State_class_sec} \ref{MBS_sec}, \ref{FF_sec} and \ref{Drug_discovery}. It is thus apparent that all the aforesaid algorithms stand to benefit from any quantum advantage seen in the development of neural network based models on a quantum processor. In fact, in certain cases, direct advantages have already been reported. For example, the authors have reported a quantum circuit-based implementation of a Restricted Boltzmann Machine based ansatz for any of the electronic states of molecules and materials\cite{sajjan2021quantum} which requires polynomial resources for its construction. Similarly for quantum version of CNN which has been used for classification of phases in the Ising model \cite{cong2019quantum}, the authors claim a more parameter reduction. We shall return to these applications and their description in Section \ref{State_class_sec} and \ref{MBS_sec}. Herein we enlist some of the recent examples wherein quantum superiority has been seen or theoretically conjectured thereby promising many novel applications in chemistry and physics which can be realized in future. More theoretical insight into the learning mechanisms and generalizability of quantum computing enhanced neural networks are discussed in detail in Section \ref{learnability_QNN}.

\begin{enumerate}     
    \item Quantum-neural networks (QNN) have been discussed in Section \ref{ANN_section}. Each such network has three generic components - a data encoding circuit (often called feature map) which accepts classical data as input and usually encodes them into the amplitudes of a quantum state (Other encoding schemes are also possible. See ref \cite{schuld2018supervised}) followed by a layer of parameterized unitaries. Finally, measurement protocol is exercised whose outcome is post-processed on a classical computer to minimize a loss function and alter the parameters of the last layer of unitaries variationally until the desired convergence is reached. A recent report has suggested that such networks can be more expressive and faster trainable than corresponding classical networks if the data encoding circuit possesses non-trivial entangling gates which can identify hidden correlation among data entries \cite{abbas2021power}. The work used an input data-set $(x_i, y_i) \forall x \:\: \in \:\: \chi \subseteq \mathbf{R}^{s_i}, \:\: y \in \mathcal{Y} \subseteq \mathbf{R}^{s_o}$ and a parameter vector $\theta \subseteq [-1, 1]^d$. The input distribution $p(x)$ is the prior distribution and $p(y|x;\theta)$ is the output distribution from the QNN given the input and specific parameter set. Using this they constructed the empirical Fisher information matrix ($\in \mathbf{R}^{d\times d})$ \cite{kunstner2019limitations} as follows:
    \begin{eqnarray}
     F_k(\theta) = \frac{1}{k}\sum_{j=1}^k \frac{\partial log(p(x_j, y_j, \theta)}{\partial \theta} \frac{\partial log(p(x_j, y_j, \theta)^T }{d\theta} 
    \end{eqnarray}
    where $k$ denotes the sample size. The authors found that the eigenvalues of the Fisher information matrix for 100 samples with $(d=40, s_{i}=4, s_{o}=2) $ in the case of the QNN were fairly uniformly distributed contrary to that in the classical neural network wherein eigenvalues were largely concentrated near zero indicating the relative flatness of the optimization surface and difficulty in trainablity of the model with gradient-based schemes \cite{karakida2019universal}. They used an `easy-quantum' model as well with data-encoding scheme without any entangling gates and found the Fisher information spectrum to be within the two limiting cases of a classical NN and a full quantum NN. The results are retained for $(d=60, s_{i}=6, s_{o}=2), (d=80, s_{i}=8, s_{o}=2), (d=100, s_{i}=10, s_{o}=2) $ .The authors in Ref \cite{abbas2021power} thereafter promised a metric for effective dimension defined below as 
    \begin{eqnarray}
    d_{y,n} = \frac{2log(\frac{1}{V}\int \sqrt{det(\mathbf{I}_d + \frac{\gamma n F(\theta)}{2\pi log n}}) d\theta}{log(\frac{\gamma n}{2\pi log n})}
    \end{eqnarray}
where $n$ is the number of data samples, $F(\theta)$ is the normalized Fisher information matrix and $V = \int d\theta$ is the volume in parameter space and $\gamma \in (0,1]$. The physical motivation of defining an effective dimension is to quantify the expressibility of the model i.e. estimate the size of the space all possible functions which the model class can successfully mimic with the Fisher information as the metric \cite{berezniuk2020scale}. Using the above definition of the effective dimension, the authors show that the full QNN has the highest effective dimension compared to the easy quantum model (without entangling gates in the circuit encoding the features) and even the classical neural network for $(d=40, s_{i}=4, s_{o}=2) $ and size of data $n = 10^5-10^6$ (see Fig 3(a) in Ref \cite{abbas2021power}). They also demonstrated that the full QNN trains faster and achieves lesser loss function values within smaller number of iterations compared to the other two (see Fig. 3(b) in Ref \cite{abbas2021power}). The conclusion remains invariant to training even on the real hardware.

\item Recently, a new report has been published \cite{PhysRevLett.128.070501} which extends the famous no-free lunch theorem \cite{Adam2019,Wolf_lectures} to learning process in a quantum neural network (QNN) wherein the training data-set may be intrinsically entangled with a third accessible register. The no-free lunch theorem (NFL) for classical learning task is deduced for an unknown map , say $f : \chi \mapsto Y$ where the size of the set $\chi$ is $d_x$ and that of set $Y$ is $d_y$. One generates a training set $S$ consisting of $t$ points from this function defined as $S=\{(x_i,y_i)| x_i \in \chi, y_i =f(x_i) \in Y \}_{i=1}^t$. In general in the supervised learning setup, this set $S$ is used to construct a merit-function $\sum_i^t L(h_S(x_i), y_i, x_i)$ where $h_S(x_i)$ is the hypothesis function that is expected to mimic the unknown function $f$. The merit-function is minimized to obtain the parameters defining the hypothesis function $h_S(x)$ which can then be used to make predictions for unseen $x \in (\chi \cap S^c)$. 
To quantify, how well the approximate function $h_S(x)$ resembles the actual one $f$ one can define a risk function as follows:
\begin{eqnarray}
R_f(h_S) = \sum_{x \in \chi} P(x) P(h_S(x) \neq f(x)) \label{risk_defn}
\end{eqnarray}
where $P(x)$ is the prior probability distribution of sampling the input $x \in \chi$ and $P(h_S(x) \neq f(x))$ is the probability that the output of the hypothesis function differs from the actual output for the specific input. The statement of NFL which the authors used is the following 
\begin{eqnarray}
\langle \langle R_f(h_S) \rangle_S \rangle_f \ge (1-\frac{1}{d_y})(1-\frac{t}{d_x}) \label{Risk_cl}
\end{eqnarray}
where the averaging of Eq. \ref{risk_defn} has been done over many training sets $S$ and many different functional maps $f$. The result in Eq.\ref{Risk_cl}  indicates that the average risk can be minimized if the number of training samples $t=d_x$ and hence is entirely determined by the training set $S$ independent of the specific details of the optimization scheme. In the quantum setting, the authors deduce a version of NFL wherein the data-set contains entries of quantum states that are entangled with an accessible auxillary quantum system. The setup of the deduction involves a unitary map $U : H_x \mapsto H_y$ both of which are $d$-dimensional. The user herein has access to another auxillary quantum system (say $R \in H_R$). The training set $S_Q$ contains $t$ pairs of states which are entangled with $R$ as follows:
\begin{eqnarray}
S_Q &=& \{ (|\psi_{in}^i\rangle, |\psi_{out}^i\rangle)| \nonumber \\ &&|\psi_{in}^i\rangle \in H_X \otimes H_R , \nonumber \\
&&|\psi_{out}^i\rangle = (U\otimes I_R)|\psi_{in}^i\rangle \in H_Y \otimes H_R \}_{i=1}^t
\end{eqnarray}
All input states $|\psi_{in}^i\rangle \in H_X \otimes H_R$ are entangled with the same Schmidt rank $r=\{1,2...d\}$
The learning task is to design an unitary $V$ such that $|\phi^i_{hyp}\rangle = (V\otimes I_R)|\psi_{in}^i\rangle$ and $|\langle \phi^i_{hyp}|\psi_{out}^i\rangle| \approx 1 $. The risk function in this case is defined as 
\begin{eqnarray}
R_U(V) = \int d\mu \frac{1}{4} ||\rho_y - \rho_y^\prime||_1^2 
\end{eqnarray}
where $\rho_y = |y\rangle \langle y| = U|x\rangle \langle x| U^\dagger$ and $\rho_y^\prime = |y^\prime\rangle \langle y^\prime| = V|x\rangle \langle x|V^\dagger$ and $|x\rangle \in H_X$ (not necessarily within $S_Q$) and $|y\rangle, |y^\prime\rangle \in H_Y$ and the measure $d\mu$ is over the Haar measure of states. The averaging of the above risk function over all training sets $S_Q$ and unitary maps $U$ as before yields
\begin{eqnarray}
\langle \langle R_U(V) \rangle_{S_Q}\rangle_U \geq 1 - \frac{r^2t^2+d+1}{d^2 +d}
\end{eqnarray}

The bound actually holds for each $S_Q$ and hence averaging over $S_Q$ is unnecessary. It is derived under the assumption that over the training samples the outcomes of $V$ and $U$ match perfectly. Implication of the above bound is that for $r=1$, the bound vanishes and the average risk can be minimized only if $t=d=2^n$ where $2^n$ is the dimension of $H_X$. This indicates that if the input states are product states with the quantum system characterized by $H_R$ then exponentially many training samples might be required in $n$. This result was previously obtained in \cite{poland2020free}. However for $r\neq 1$ that is not the case. Specifically, in the limiting case of $r=d$, a single training example would suffice to saturate the lower bound on the average. For any $r$ in between one can easily see that the number of training samples $t$ can be set to be far fewer than $d$. The authors show numerical experiments on Rigetti's quantum processor \cite{} for $2 \times 2$ unitaries and demonstrate that the average risk can be minimized well below the classically accessible limit by controlling r which alters entanglement with the third quantum system $R$. Similar results were obtained even in the simulator. The concrete proof of the theorem restores hope that the size of the training set can be small yet a quantum advantage can be retained as long as entanglement is used as a useful resource which is not possible classically. This result joins the group of other results wherein entanglement have served similar roles like in superdense coding \cite{PhysRevLett.69.2881,PhysRevLett.92.187901} or quantum teleportation \cite{PhysRevLett.70.1895,PhysRevLett.123.070505}

\item This example is different in spirit than the previous ones as it demonstrates how the quantum convolutional neural network (QCNN) \cite{cong2019quantum} which is discussed in detail in Section \ref{CNN_section} can lead to efficient parameter reduction compared to other methods of quantum phase classification. The power of a quantum classifier over and beyond that of a classical classifier has already been demonstrated for certain tasks in some of the points above. The task which the authors tested the QCNN was to segregate quantum states belonging to a particular phase given a training data-set $S=\{|\psi\rangle_i, y_i\}_{i=1}^M$ where $y_i = \{0,1\} \forall i$ are the corresponding labels. The specific example was the symmetry-protected topological phase of a 1D spin-chain. The algorithm requires only $O(log(n))$ parameters to classify such an $n$-qubit quantum state which the authors claim is a two-fold exponential reduction in parameter space compared to other quantum classifiers. This essentially means that the QCNN is more expressive compared to other classifiers as it solves a relatively low-dimensional optimization problem without sacrificing the accuracy of the task at hand. Similarly, they also show that the sample complexity of QCNN which is defined as the number of copies of the input state that needs to be accessed by the algorithm for correctly identifying the phase, is lesser than from other techniques like direct evaluation of expectation values of certain operators which can also act as a marker of the phase. We shall return to this example in Section \ref{State_class_sec}.
Pesah $et al$ \cite{PhysRevX.11.041011} have demonstrated that in the conventional QCNN architecture there is absence of barren plateaus as the gradient estimates vanishes polynomially (and not exponentially) in the size of the system for random initialization . Recently MacCormack $et al$ have extended the concept to introduce a variant of QCNN which is called the branching quantum convolutional network (bQCNN) \cite{maccormack2022branching}. In QCNN as discussed in Section \ref{CNN_section}, the primary idea is reduction of the number of qubits while preserving important features. The convolutional layers perform multi-qubit unitaries for generating entanglement among the qubits whereas in pooling layers certain number of qubits are discarded through measurements and controlled rotations on nearby qubits are performed conditioned on the measurement results of the discarded qubits. The new reduced set of qubits is then fed into the convolutional unitaries again and the process is repeated. For bQCNN, the authors make a deviation at the pooling layer. Instead of using certain measurement outcomes only of the discarded qubits, the authors use all possible outcomes or binary bit combinations to design several different channels/branches for subsequent convolutional operation each of which is realized for a given bit string. This enhances the parameter requirement drastically as noted in the report. However the authors also demonstrate that the expressibility of the ansatz from bQCNN is higher than that of QCNN at similar circuit depth. 

\end{enumerate}

\subsection{Power of Quantum Computers for Tensor-Network based Machine Learning tasks}

In this section, we highlight an advantage that has been recently reported for a classification task performed using a tensor network ansatz on a quantum computer. Such classification can be extended to the physico-chemical domain like in ligand selectivity for structure-based drug designing (see Section \ref{St_drug_des}) with the quantum benefit reaped. An example of another tensor network based classification of phases of a spin-model can be found in Section \ref{State_class_sec}. Besides, Tensor networks on a classical processor have also been aggressively used for representing many-body states for a variety of applications like for spin-liquids \cite{lee2020tensor}, excitonic states in materials \cite{kuhn2020tensor} and molecules \cite{gunst2018t3ns}. Quantum advantages as has been noted for the example below can be extended to such applications. We shall discuss such prospects with concrete examples in Section \ref{QC_MBS}.

Recently, a report has illustrated the use of quantum architectures of Tree and Matrix Product state Tensor Networks \cite{Huggins_2019} to demonstrate the working of a discriminative machine-learning model on MNIST dataset \cite{lecun-mnisthandwrittendigit-2010}. 
The primary objective of the study was to perform classification and recognition of hand-written digits using a variational optimization procedure that can be efficiently carried out on a quantum hardware in the NISQ era. Classical data-entries (say $x \in \mathbb{R}^N$) from the data-set are mapped to a $N$-qubit quantum state using the data-encoding protocol described below in Eq. \ref{data_encod_cl}
\begin{eqnarray}
x \longrightarrow |\phi(x)\rangle = \begin{bmatrix} cos(\frac{\pi}{2} x_1)\\ sin(\frac{\pi}{2} x_1) \end{bmatrix} \otimes \begin{bmatrix} cos(\frac{\pi}{2} x_2)\\ sin(\frac{\pi}{2} x_2) \end{bmatrix} \otimes \cdots \otimes \begin{bmatrix} cos(\frac{\pi}{2} x_N)\\ sin(\frac{\pi}{2} x_N) \end{bmatrix} \label{data_encod_cl}
\end{eqnarray}

State $|\phi\rangle$ is a product state which can easily be prepared by applying single qubit rotation gates on $|0\rangle ^{\otimes N}$ state. After state preparation, each set of qubits (say $2V$ qubits where $V$ is the number of virtual states) is acted upon by a parameterized unitary gate. The scheme is inspired from coarse-graining techniques. After each parameterized unitary operation, $V$ qubits are discarded/reset and the other $V$ qubits proceed to the next step where they are merged with $V$ qubits coming from another set of $2V$ qubits. This process is continued until the last 2V qubits remain which are acted upon by a unitary gate to produce output qubits. One or more output qubits are measured to determine the probability distribution of the output labels. While the model is agnostic to the grouping of the qubit sets, it is usually wise to group qubits that represent local regions in the input data. This ansatz is motivated from a  Tree Tensor Network (see Fig. \ref{fig:Tree_TN_ansatz}). The optimization problem which they solve involves classification of hand-written digits using a loss-function that penalizes the difference in the probability of attaining the true-label from the measurement protocol as opposed to the most-probable incorrect label. The algorithm used is adapted from Simultaneous Perturbation Stochastic Approximation (SPSA) \cite{10.1145/324138.324170} with momentum being included inside the gradient estimates \cite{ruder2016overview}. The accuracy of the method as reported is extremely high i.e. with the lowest percentage error being 87\% and an average test accuracy of over 95\%. The authors noted that for usual quantum algorithms, encoding such a $N$-qubit state and applying tunable unitaries would require $N$-physical qubits. However, the tree-structure in the ansatz with sequential measurements allows the authors to use $\approx Vlog(N)$ physical qubits indicating that the scheme is qubit efficient. 
It must also be emphasized that merely using a tree-tensor network approach (TTN) on a classical processor would require a space complexity of $O(N2^{3V})$ \cite{doi:10.1137/090748330} where $V$ is as defined before the number of virtual states. This result indicates that the TTN ansatz on the quantum computer is more expressive than a corresponding classical implementation as similar accuracies are afforded even with a reduced bond-dimension (bond-dimension is usually denoted by $D$ where $D=2^V$ and as seen here scales logarithmically for the quantum TTN version whereas would have a cubic scaling for classical TTN). The authors also perform a systematic analysis of the effect of noise on the algorithm and concluded that the degradation of the performance was only 0.004 with a strongly enhanced ($\times 10$) noise parameters thereby indicating the resilience of the algorithm.
\begin{figure}
    \centering
    \includegraphics[width=8cm]{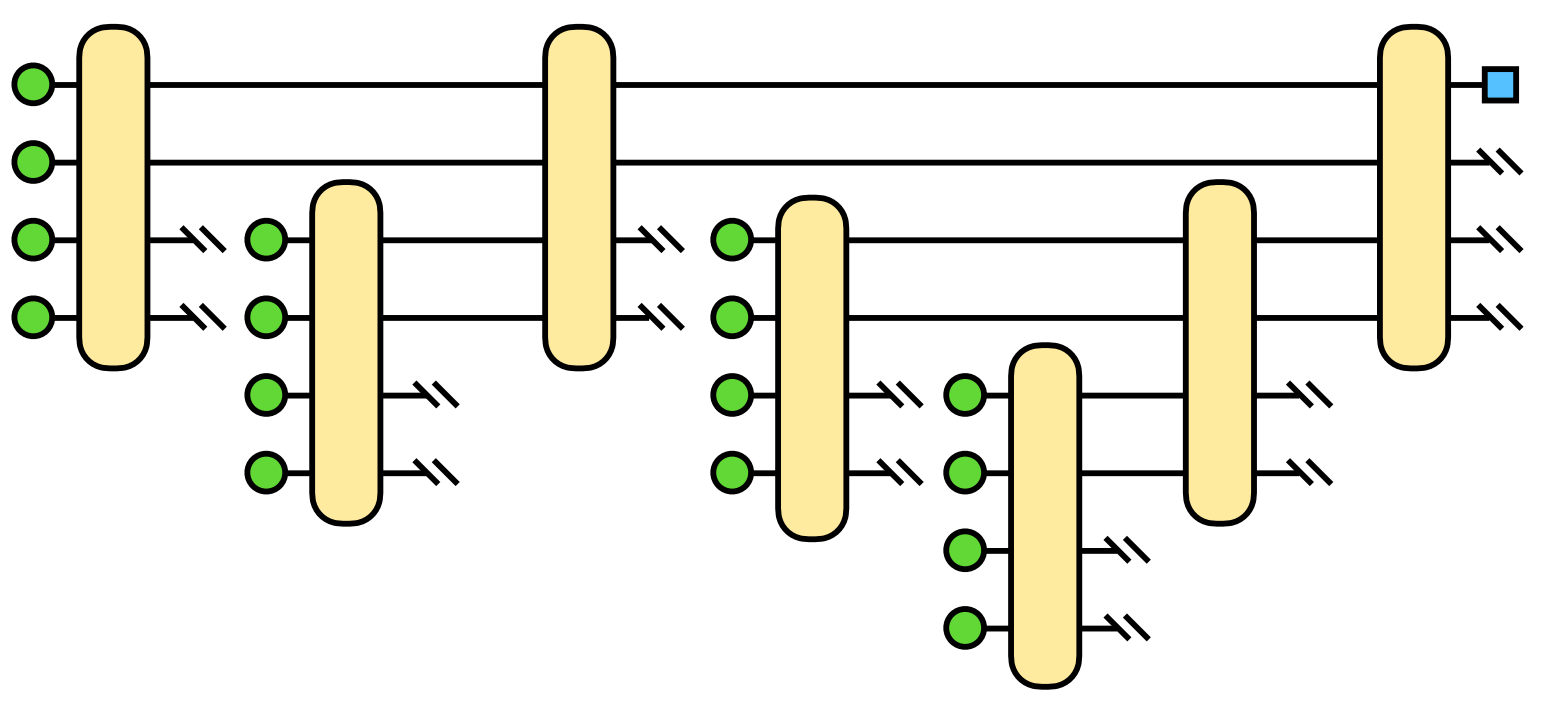}
    \caption{The architecture for evaluating Discriminative tree tensor network model using Qubit-efficient scheme with two virtual qubits (V) and 16 input states (N) as used in Ref.\cite{Huggins_2019}. The architecture requires $O(V\log(N))=8$ qubits for its operation. The qubits indicated with hash marks are measured and reset to accept input states in the next step. $\copyright$ IOP Publishing. Reproduced with permission from William Huggins, Piyush Patil, Bradley Mitchell, K Birgitta Whaley and E Miles Stoudenmire, Towards quantum machine learning with tensor networks, Quantum Science and Technology, Volume 4, Number 2, 9 January 2019, https://doi.org/10.1088/2058-9565/aaea94  All rights reserved.} \label{fig:Tree_TN_ansatz} 
\end{figure}
}

\section{Applications} \label{Applications}
\subsection{State preparation protocols and Quantum State Tomography} \label{state_prep}

With the rapid advancement of quantum technology in the past decade, there is an increased demand for efficient methods that can verify the generated quantum states according to specific needs. This is primarily important for the validation and benchmarking of the quantum devices. To address this, quantum state tomography (QST) aims to obtain the statistical inference of the unknown quantum state of the system based on the information of the expectation values of a complete set of observables \cite{nielson, kais, photonic,qubits,cramer}. However, the traditional approach to QST has been pushed to its limits for the large quantum platforms available today \cite{10-qubit}. This is because the number of measurements required for full reconstruction of the quantum state scales exponentially with the system size which poses a critical challenge for performing QST even for moderately sized systems. Also, in order to obtain valuable insight about the physical observables of the system and to estimate them accurately, the measurement outcomes need to be stored and processed for which exponential amounts of classical memory and computing power are required which makes the technique infeasible for practical applications. \\
Apart from the problem of exponential scaling for complex systems, another drawback for accurate estimation of a quantum state is the inherent noise in the present day noisy intermediate-scale quantum (NISQ) devices \cite{preskill2018quantum}. Because of this the measurements available are of limited fidelity and in certain cases, some of the measurements are even not accessible. Several methods have been proposed as a means of an alternative approach to the traditional QST such as matrix product state tomography \cite{cramer}, neural network tomography \cite{Torlai2018, Carrasquilla2019, Palmieri2020, xin2019local}, quantum overlapping tomography \cite{overlap},  shadow tomography \cite{aaronson, Huang2020}. Because of the noisy nature of the quantum systems since not all measurements are available at high fidelity, there have also been approaches that try to carry out QST based on incomplete measurements \cite{jaynes, wich, katz1967principles} such as maximum likelihood estimation (MLE) \cite{hradil, teo2011quantum, teo2012incomplete,blume2010hedged, smolin2012efficient,baumgratz2013scalable}, Bayesian mean estimation (BME) \cite{blume2010optimal, huszar2012adaptive,lukens2020practical,lukens2020bayesian}, and maximal entropy formalism based QST \cite{gupta2021maximal, gupta2021convergence}. \\
Quantum detection and estimation theory has been a prominent field of research in quantum information theory since the 1970s \cite{helstrom1976quantum, gudder1985holevo, peres1993quantum} and the rapid progress in quantum communication and computation in the past two decades motivated the use of big data for the classification of quantum systems through quantum learning and quantum matching machines \cite{sasaki2002quantum}. The notion of self-improvement of the performances of quantum machines via quantum learning was introduced by Ron Chrisley in 1995 in \cite{chrisley1995quantum} through the example of barrier/slit/plate feed-forward back-propagation network. In a feed-forward network, parameterized non-linear functions map an input state to an output state space. The interactions with the environment help the networks to modify those parameters such that each network can better approximate the resulting function. The proposed quantum implementation of such a network involved setting up a barrier with several slits in front of a particle beam. Some of the slits are designated as input slits and rest are the weight slits. Behind the barrier is a photo-sensitive plate on which the interference pattern is observed that serves as an output for each input slit configuration. Once an interference pattern of high resolution is obtained on the plate, the error, which is a function of the desired and actual output vectors, is calculated. The gradient descent method is applied by taking the partial derivative of the error function with respect to the control variables, the weights, and the slits are adjusted accordingly for the next reading so as to minimize the error function. After sufficient training, optimum values for the weight configuration are obtained that ensured minimal error on the training set. This feed-forward network established the correspondence between the neural networks and the quantum system and is one amongst the many diverse approaches for the practical implementation of quantum neural networks \cite{behrman1996quantum, behrman1999spatial}. \\
One of the most valuable applications of quantum tomography is the validation and testing of near-term quantum devices called the NISQ devices. Leveraging the capabilities of quantum hardware of the gate-based NISQ devices, Alejandro \textit{et al} in \cite{benedetti2019generative} proposed a hybrid quantum-classical framework called data-driven quantum circuit learning (DDQCL) algorithm for benchmarking and training shallow quantum circuits for generative modeling. Much like the various models encompassed within the Born machines \cite{cheng2018information, stoudenmire2016supervised, han2018unsupervised, gao2018quantum}, the authors captured the correlations in the data set using the 2$^N$ amplitudes of wave function obtained from an \textit{N}-qubit quantum circuit. However, the distinctive feature in their work is the use of quantum circuit as a model for the data set that naturally works as a Born machine, thereby avoiding the dependence on tensor networks for the same. They demonstrated their approach of training quantum circuits for preparation of the GHZ state and coherent thermal states thereby illustrating the power of Born machines for approximating Boltzmann machines. \\
Finding an effective quantum circuit that can optimally carry out a desired transformation between the input and output quantum states also constitutes an important aspect of QST. In \cite{arrazola2019machine} the authors proposed a machine learning based optimization algorithm for quantum state preparation and gate synthesis on photonic quantum computers. They used the continuous-variable quantum neural network architecture \cite{killoran2019continuous} as an ansatz whose optimization was carried out on the Strawberry Fields software platform for photonic quantum computation \cite{killoran2019strawberry}. Using the proposed method, the authors were able to achieve high fidelities of over 99$\%$ using short-depth circuits given only the target state as an input.  \\  
Another quantum state preparation method was presented in \cite{zhang2019does} based on reinforcement learning which is a machine learning training architecture framework that finds an optimal solution to a problem based on the principle of rewarding the desired actions and penalizing the negative actions. The authors made a comparative study of the performances of three reinforcement learning algorithms: tabular Q-learning (TQL), deep Q-learning (DQL), and policy gradient (PG), and two traditionally used non-machine-learning methods: stochastic gradient descent (SGD) and Krotov algorithms, demonstrating their efficiency with reference to quantum state preparations under certain control constraints. Their results illustrated the effectiveness of reinforcement learning algorithms in solving complex optimization problems as the algorithms, especially DQL and PG, performed better amongst the five algorithms considered for state preparation under different types of constraints. \\
{\color{black}Parameterized quantum circuits (PQC) are yet another machine learning model that utilizes the resources of both quantum and classical computation for applications in a variety of data-driven tasks. \cite{benedetti2019parameterized} presented a comprehensive review of various machine learning models involving PQC and also their applications in diverse fields including quantum tomography. Another class of algorithms within the domain of quantum machine learning is hybrid quantum-classical Variational Quantum Algorithms (VQAs) \cite{mcclean2016theory} that has been gaining popularity in the recent years with numerous applications \cite{huang2019near}, \cite{larose2019variational}, \cite{cerezo2020variational}, \cite{wang2021variational}, \cite{wang2021hybrid}, \cite{li2021vsql}, \cite{chen2021variational}. VQAs try to reduce the quantum resource allocation by using shallow quantum circuits for carrying out computations on a quantum device. One such algorithm was proposed in \cite{wang2021gibbs} by Wang \textit{et al} to prepare quantum Gibbs state on near-term quantum hardware using parameterized quantum circuits. On such devices it is in general quite difficult to prepare Gibbs state at arbitrary low temperature just like finding the ground states of Hamiltonians \cite{aharonov2013guest}. Preparation of quantum Gibbs state of a given Hamiltonian has its applications in a variety of fields like many-body physics, quantum simulations \cite{childs2018toward}, quantum optimization \cite{somma2008quantum}, etc. In the method proposed by Wang \textit{et al} minimization of free energy serves as the loss function. However, within the calculation of free energy estimation of entropy is the most challenging part \cite{gheorghiu2020estimating}. To tackle this problem they used truncation of Taylor series of the von Neumann entropy at order \textit{K} and thus, the truncated free energy was set as the loss function whose minimum would correspond to the optimal parameters of the quantum circuit giving the Gibbs state. The estimation of the Taylor series expansion terms of entropy can be practically carried out using the well-known swap test \cite{buhrman2001quantum}, \cite{gottesman2001quantum} and therefore, their method can be physically realized on a near-term quantum hardware. To validate the approach they numerically showed the preparation of high-fidelity Gibbs state for Ising chain and \textit{XY} spin-$\frac{1}{2}$ chain models and were able to achieve fidelity of at least 95$\%$ for a range of temperature. \\
Another kind of quantum state preparation commonly termed as Quantum State Tomography relies on accessing experimentally measured observables. We shall discuss, hereby, how various techniques within the domain of machine learning/deep learning have been applied to perform QST. }

\begin{figure*}[ht!]
  \centering 
\includegraphics[width=5.2in]{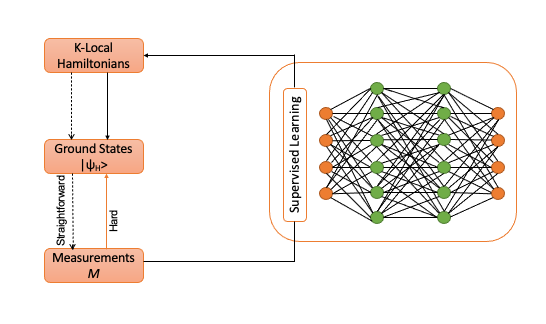}
\caption{Local measurement based quantum state tomography via neural networks: The procedure starts with generating the training and test dataset, represented by dashed arrows, through random k-local Hamiltonians H to obtain the local measurement results \textit{M} from the ground states $\psi_H$. The neural network is then trained using the training dataset which then produces the local measurements \textit{M}, represented by black arrows. The measurements are used to find the Hamiltonian H followed by obtaining the ground states. The normal QST process follows the red arrow which is computationally hard. Figure is a schematic of the protocol illustrated in Ref \cite{xin2019local}.}
\label{fig_2}
\end{figure*}

As machine learning and neural networks became increasingly popular with its applications in many diverse fields, the concoction of quantum mechanics and machine learning algorithms also started surfacing \cite{neven2008training, pudenz2013quantum, lloyd2013quantum, rebentrost2014quantum}. The development of quantum annealing processors \cite{johnson2011quantum} deemed a natural fit for testing the machine learning algorithms on a quantum hardware \cite{adachi2015application,benedetti2017quantum} to check for any quantum advantage. In the initial stages quantum mechanics was only used to facilitate the training for solving classical problems which, in fact for certain problems, did result in obtaining polynomial speed-ups relative to classical training methods \cite{wiebe2015quantum}. Although the training of the Boltzmann machines using quantum processors did result in accurate training at a lower cost, however, the success of machine learning algorithms based on classical Boltzmann distribution inspired the proposition of a quantum probabilistic model for machine learning called quantum Boltzmann machines (QBM) based on Boltzmann distribution for quantum Hamiltonian \cite{amin2018quantum}. In QBM, not only the training is performed utilizing the quantum nature of the processors but also the model in itself is inherently quantum. The data modeling and training of Boltzmann machine is carried out using the equilibrium thermal states of the transverse Ising type Hamiltonian. \\
However, the training procedure used in \cite{amin2018quantum} suffered from two limitations: a) brute force techniques are required to find out the transverse field terms as it cannot be learned through classical data making it very hard to find the full Hamiltonian, b) quantum Monte Carlo methods can be used to efficiently simulate the transverse Ising models and therefore, using the training method with transverse Ising models in thermal equilibrium did not show a clear quantum advantage. In \cite{kieferova2017tomography}, the authors proposed the quantum analog of the generative training model that included quantum data sets in the training set so that their QBM is capable of learning the quantum terms along with the classical ones. The training of the Boltzmann machine to incorporate quantum data was carried out through two methods: POVM-based Golden-Thompson training and state-based relative entropy training (quantum equivalence of KL divergence). The relative entropy training method allowed the QBM to clone the quantum states within certain level of approximation and given considerable number of copies of the density operator for training of QBM, it could reproduce approximate copies of the input state. Thus, although different from the traditional quantum state tomography wherein an explicit representation of the state operator is available at the output, the generative training of the QBM resulted in a quantum process that can learn Hamiltonian models for complex quantum states which in itself is a form of partial tomography. \\
Machine learning techniques offer a big advantage of representing high dimensional data in compressed form which can be really favourable for QST which in itself is a highly data-driven technique. In 2018, Torlai and colleagues \cite{Torlai2018} utilized this property of the neural networks for obtaining the complete quantum mechanical description of highly entangled many-body quantum systems based on the availability of a limited set of experimental measurements. Their method involves training of a restricted Boltzmann machine (RBM) using simulated measurement data for the true state. Their RBM architecture comprises of a layer of visible neurons for encoding the measurement degrees of freedom and a hidden layer of binary neurons for encoding the tunable parameters. The training of the RBM is performed such that the the generated probability distribution resembles closely to the given data distribution and parameters are tuned depending on the desired degree of accuracy that is required for reconstruction. They demonstrated their approach by first benchmarking the neural network tomography of the N-qubit multipartite entangled \textit{W} state. They also demonstrated QST for more complex systems consisting of quadratic spin Hamiltonians, namely the transverse-field Ising model and the XXZ spin-1/2 model. \\
According to the authors, their RBM model works really well for entangled many-body systems and quantum-optic states. However, when there is an unstructured data as in the case of states generated from random unitary operation, the performance of the RBM goes down. The paper also did not highlight some of the pertinent questions such as what should be the size of the training set, or the optimal number of hidden neurons, and the dominance of RBM over the other contemporary machine-learning approaches. Also, in the demonstration of their approach, both in the case of the tomographic reconstruction of the \textit{W} state or ground states of the quadratic-spin Hamiltonians, the wavefunctions considered are real which decreases the measurement degrees of freedom required for state reconstruction and thereby, dramatically reduces the tomographic overhead. \\
Another neural network based quantum state estimation method empowered by machine learning techniques was presented in \cite{xu2018neural} in 2018 for full quantum state tomography. Instead of reducing the number of mean measurements required for full state reconstruction, the work focuses on speeding up the data processing in full QST without the assumption of any prior knowledge of the quantum state. Their training model is based on standard supervised learning techniques and the state estimation is carried out using a regression process wherein a parameterized function is applied for the mapping of the measurement data onto the estimated states. For a single state reconstruction, the computational complexity of their model is $\mathcal{O}(d^3)$, where \textit{d} is the dimension of the Hilbert space, and was the fastest amongst the full QST algorithms such as MLE, LRE, BME, etc. at the time. \\
A further application of neural network in the field of state tomography was presented in \cite{xin2019local} where local measurements on reduced density matrices (RDMs)\cite{baldwin2016strictly, linden2002almost, linden2002parts, chen2012ground, chen2013uniqueness} were used to characterize the quantum state. Using the recent approach of measuring RDMs and thereby reconstructing the full state is a convenient alternative to the traditional QST approaches as the whole system can be characterized in polynomial number of parameters as opposed to the exponential parameters required for full reconstruction in the traditional QST techniques. However, QST via local measurements on RDMs is a computationally hard problem \cite{qi2013quantum}. In this work, the authors addressed this problem using machine learning techniques by building a fully connected feedforward neural network, as shown in Figure \ref{fig_2}, to demonstrate the full reconstruction of the states for up to 7-qubit in simulation and also reconstructed 4-qubit nuclear magnetic resonance (NMR) states in experiments. Their approach also had comparable fidelities with the MLE method but with the additional advantage in terms of speed up and better noise tolerance.   \\
Another machine learning assisted quantum state estimation technique based on convolutional neural networks (CNN) (Basic theoretical framework discussed in Section \ref{CNN_section}) was presented in \cite{lohani2020machine} in 2020 for reconstructing quantum states from a given set of coincidence measurements for both pure and mixed input states. Their approach involves feeding the noisy or incomplete set of simulated measurements to the CNN which then makes prediction of the $\tau$-matrix based on the decomposition method discussed in \cite{james2005measurement}. The predicted matrix, which is the output, is then inverted to give the final density matrix. In order to compare the fidelity of the reconstructed states, the authors also implemented the Stokes reconstruction method \cite{james2005measurement} and found a significant improvement in fidelity for not just noisy data sets but also when the projective measurements are incomplete and thereby, demonstrating the advantage of CNN over the typical reconstruction techniques.

As an alternative means of QST several methods have been proposed, such as MLE, BME, and least-squares (LS) inversion \cite{opatrny1997least}, for efficient reconstruction of the quantum state. Since measurements on many copies of the state are required for efficient reconstruction of the quantum state so in order to gain maximum information from the measurements on the next copy, the measurements are adjusted based on the already available information from the measurements made thus far. This method of reconstructing the quantum state through adaptive measurements is called adaptive quantum state tomography. One such approach was introduced in \cite{fischer2000quantum} where self-learning algorithm was used in combination with different optimization strategies such as random selection, maximization of average information gain, fidelity maximization for quantum state estimation. A generalization of self-learning algorithm was presented in \cite{huszar2012adaptive} in the form of adaptive Bayesian quantum tomography (ABQT) with the aim to optimally design quantum tomographic experiments based on full Bayesian inference and Shannon information. Through their adaptive tomography strategy the authors were able to achieve significant reduction in the number of measurements required for full state reconstruction in case of two qubits pure states in comparison to Monte Carlo simulation of the qubit systems. The experimental realization of the ABQT method was also carried out in \cite{struchalin2016experimental} for two qubit quantum system which did show a significant improvement in the accuracy of state estimation in comparison to the nonadaptive tomographic schemes. Recently, neural adaptive quantum state tomography (NAQT) was introduced in \cite{quek2021adaptive} that utilizes the neural network framework to replace the standard method of Bayes' update in the ABQT scheme and thereby obtained orders of magnitude faster processing in estimating the quantum state while retaining the accuracy of the standard model. Basically, in the adaptive Bayesian-type tomography, the quantum space is discretized into samples and with each sample there is an associated weight that gets updated with each new measurement according to the Bayes' rule in order to update the prior distribution to the posterior distribution of the quantum state space. However, with the increase in the number of measurements the likelihood function becomes sharply peaked with a tiny subset of the sample states having the majority weights. This can lead to a numerical singularity which is then avoided by resampling of the weights onto the space states which is computationally a very expensive process. In \cite{quek2021adaptive} the authors have used machine learning algorithm to map the Bayesian update rule on a neural network to replace the traditional approach and thereby eliminate the problem of resampling of the weights saving the computational cost significantly without compromising on the accuracy of the reconstruction process. In comparison to the ABQT technique, the NAQT approach was able to speed up the reconstruction process by a factor of a million for approximately $10^7$ measurements and is independent of the number of qubits involved and type of measurements used for estimating the quantum state.  
\begin{figure*}[ht!]
  \centering 
\includegraphics[width=6in]{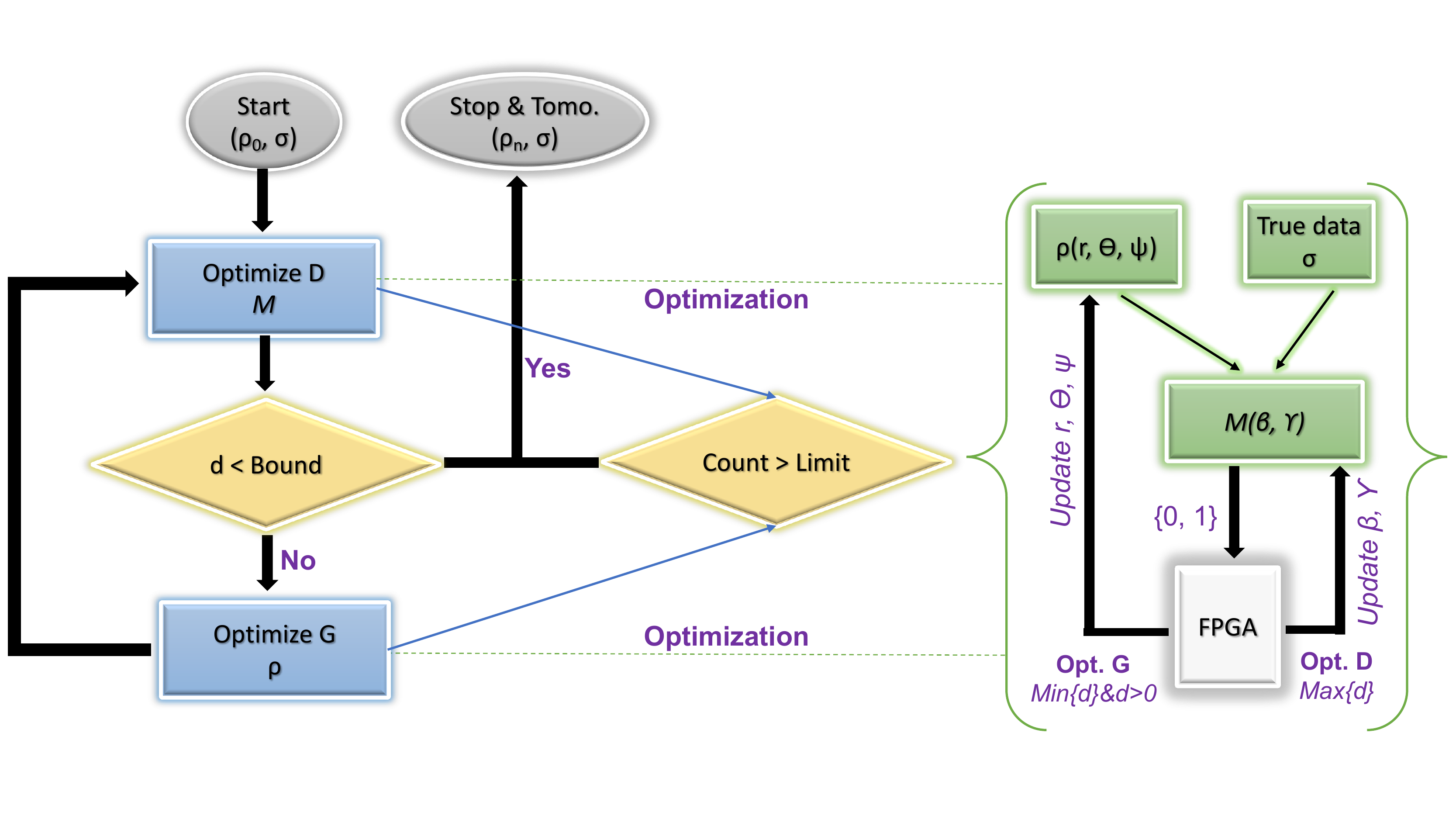}
\caption{Flowchart demonstrating the experimental protocol of the QGAN algorithm in Ref \cite{hu2019quantum} and the optimization scheme of the generator G and the discriminator D: G generates a random starting state $\rho_0$($r_0, \theta_0, \psi_0$) following which both D and G compete against each other by optimizing their strategies alternatively. The process is terminated when either D fails to discriminate the generated state $\rho$ from the true state $\sigma$ or the number of steps c$_{step}$ reaches the limit. The optimization scheme, based on gradient descent method, of D involves updating the parameters $\beta$ and $\gamma$ of the measurement operator \textit{M} whereas r, $\theta$, and $\psi$ are updated to optimize G. The gradients estimation is carried out on a classical computer whereas field programmable gate arrays (FPGAs) are used for the measurement and control of the quantum system.}
\label{fig_1}
\end{figure*}
In 2013, another useful method was proposed in \cite{qi2013quantum} wherein linear regression estimation (LRE) model was used to identify optimal measurement sets to reconstruct the quantum state efficiently. The model was relatively straightforward where the authors converted the state estimation problem into a parameter estimation problem of a linear regression model and the LS method was employed to determine the unknown parameters. For a \textit{d}-dimensional quantum state, the computational complexity of the method for state reconstruction is $\mathcal{O}(d^4)$ and thereby saves up the cost of computation in comparison with the MLE or BME method. A natural extension to this work, in terms of both theory and experiment, was presented in \cite{qi2017adaptive} in order to improve the tomography accuracy by better tomographic measurements via adaptive tomography protocol that does not necessarily require non-local measurements from experiments. The method is called recursively adaptive qauntum state tomography (RAQST) primarily because the parameters are updated using the recursive LRE proposed in \cite{qi2013quantum} and using the already collected data the measurement strategy is adaptively optimized for obtaining the state estimation. The authors also performed two-qubit state tomography experiments and show the superiority of RAQST method over nonadaptive methods for qauntum states with high level of purity which is an important criterion for most forms of information processing methods. 

Another significant machine learning approach is the Generative adversarial network (GAN) (Basic theoretical framework is discussed in Section \ref{GAN_section}) based QST  \cite{yang2020tomographic,liu2020tomogan} that basically involves learning the map between the data and the quantum state unlike the RBM-based QST  methods where the map yields a probability distribution. In the GAN method, two competing entities: generator G and discriminator D, engage with the objective to output a data distribution from some prior noisy distribution. Both G and D are parameterized non-linear functions consisting of multi-layered neural networks \cite{goodfellow2014generative}. With each step of optimization the generator becomes better at yielding outputs closer to the target data and the discriminator becomes better at detecting fake output. Inspired by the classical model, the quantum generative adversarial network (QGAN) was introduced in \cite{lloyd2018quantum,dallaire2018quantum} where quantum processors are used for running the neural nets of generator and discriminator as well as the data can be both quantum and classical. Thus, making the entire system quantum mechanical, at the end of the optimization process the generator can reproduce true ensemble of quantum states without the discriminator being able to distinguish between the true and generated ensemble.

The first proof-of-principle experimental demonstration of the QGAN algorithm was presented in \cite{hu2019quantum} in a superconducting quantum circuit on datasets that are inherently quantum for both the input and the output. The QGAN algorithm employed was able to reproduce the statistics of the quantum data generated from a quantum channel simulator with high level of fidelity (98.8$\%$ on average). Their experimental approach involves a superconducting quantum circuit for the generator G that outputs an ensemble of quantum states with a probability distribution to mimic the quantum true data whereas the discriminator D is used to carry out projective measurements on the true and the generated data in order to distinguish the two based on the measurement outcomes. The optimization process based on the gradient descent method consists of the adversarial learning by the discriminator and the generator in alternative steps that is terminated when the a Nash equilibrium point is reached i.e. G produces the statistics such that D can no longer differentiate between the fake and the true data. The experimental protocol of the implementation of the QGAN algorithm is shown in Fig. \ref{fig_1}.

In \cite{ahmed2020quantum} the authors introduced conditional generative adversarial network (CGAN) based QST as in the standard GAN approach there is no control over the output as the generator input is random which can be addressed using CGAN. Because of the improved control over the output, the CGAN led to many diverse applications in a variety of fields \cite{isola2017image, karras2018style, karras2020analyzing,yang2018unsupervised, subramanian2018towards}. With the CGAN based QST the authors were able to achieve higher fidelity of reconstruction, faster convergence and also reduced the number the measurements required for reconstructing a quantum state as compared to the standard model of state reconstruction using MLE. Also, with sufficient training on simulated data their model can even reconstruct the quantum states in a single shot. In their CGAN approach, the measurement operators ($\{O_i\}$) and the measurement statistics are the conditioning input to the generator which then outputs a density matrix $\rho_G$ that is used to generate the measurement statistics by calculating tr($O_i\rho_G$). These measurement statistics and the experimental measurement statistics serve as the input for the discriminator which then outputs a set of numbers to distinguish between the generated statistics and the true data.The standard gradient-based optimization techniques are then used to train the network which is completed when the discriminator is unable to differentiate between the generated statistics from the generator and the true data. With better control over the output the CGAN approach can further find its applications in efficiently eliminating noise from experimental data by training it on noisy simulated data as well as it can have potential advantages in adaptive tomography as well by improving on the choice of adaptive measurements for better reconstruction of quantum states. 

{\color{black} The huge success of attention mechanism-based neural network generative model \cite{cho2014learning}, \cite{cheng2016long}, \cite{parikh2016decomposable} to learn long-range correlations in natural language processing (NLP) \cite{vaswani2017attention} prompted for its applications in QST owing to the entanglement among qubits that can also be learnt through the self-attention mechanism used in the former case. The self-attention mechanism computes a representation of a sequence by relating different positions of a single sequence and has been very successful in a variety of sub-fields under NLP \cite{lin2017structured}, \cite{paulus2017deep}. In \cite{vaswani2017attention} the authors proposed the first purely self-attention based transduction model, the Transformer, for deriving global dependencies between input and output without the use of recurrent neural network (RNN) or convolutions. Just like other transduction models, the Transformer also uses an architecture of fully connected layers for both the encoder and decoder \cite{bahdanau2014neural}, \cite{cho2014learning}, \cite{sutskever2014sequence} using stacked self-attention that results in significantly faster training than architectures based on recurrent or convolutional layers. To the best of our knowledge the quantum-enhanced version of the Transformer has not been studied yet. However, the application of Transformer on quantum data has shown tremendous potential, as discussed below in the context of QST, for future research in this field.    \\
The long-range correlations exhibited by entangled quantum systems can be modeled analogous to the sentences in natural language using informationally complete positive operator-valued measurements (IC-POVM). Therefore, with this motivation, recently, Cha \textit{et al} proposed the 'attention-based quantum tomography' (AQT) \cite{cha2021attention} to reconstruct the mixed state density matrix of a noisy quantum system using the Transformer \cite{vaswani2017attention} architecture for the neural network. In this work they first benchmark their approach against previous neural network based QST using RNN \cite{carrasquilla2019reconstructing} that demonstrated a high fidelity classical description of a noisy many-body quantum state. To compare the performance of AQT with other state-of-art tomographic techniques they considered Greenberger–Horne–Zeilinger (GHZ) state as their target state for upto 90 qubits and a built-in simulated error resulting in a mixed state. They showed that the complexity of learning the GHZ state can be improved by an order of magnitude when compared with the RNN method. They also benchmark AQT against MLE for a 3-qubit system and found a superior quantum fidelity of reconstruction for the AQT with the additional advantage of being scalable to larger systems. Furthermore, they were also able to reconstruct the density matrix of a 6-qubit GHZ state using AQT, with a quantum fidelity of 0.977, which is currently beyond the tomographic capabilities of IBM Qiskit.
}

QST becomes a formidable task as the size of the quantum system increases. In order to address the problem, in this section of the review, we presented research based on several machine learning driven QST techniques such as RBM based QST for highly entangled many-body quantum systems, characterizing quantum systems using local measurements on RDMs, using CNN for state reconstruction, adaptive QST through self-learning and optimization based algorithms, VQAs for QST, generative models like GANs and attention based QST. Although the use of classical machine learning algorithms and deep neural networks have proven to be very effective in finding patterns in data but for implementation on a quantum computer, loading classical data onto quantum devices can present a serious bottleneck for the implementation of these algorithms \cite{aaronson2015read}. Since large number of independent parameters are required for reconstructing the quantum state that scales exponentially with the system size and therefore, using quantum machine learning algorithms directly on the quantum states of the system can help in the learning and optimization of these parameters much more efficiently, owing to their ability of handling larger Hilbert space. As mentioned in the above sections, generative machine learning models such as quantum Boltzmann machine (QBM) and quantum generative adversarial network (QGAN) provide valuable insights into exploiting the full capabilities of the present day quantum computers in order to reproduce the desired quantum states. The improvement of quantum hardware with time also calls for their validation and benchmarking and QML can play a major role in the design of cost-effective quantum tomography techniques and protocols that can be easily implemented on the NISQ devices.

\subsection{State classification protocols} \label{State_class_sec}
\begin{figure*}[ht!]
    \centering
    \includegraphics[width=3 in]{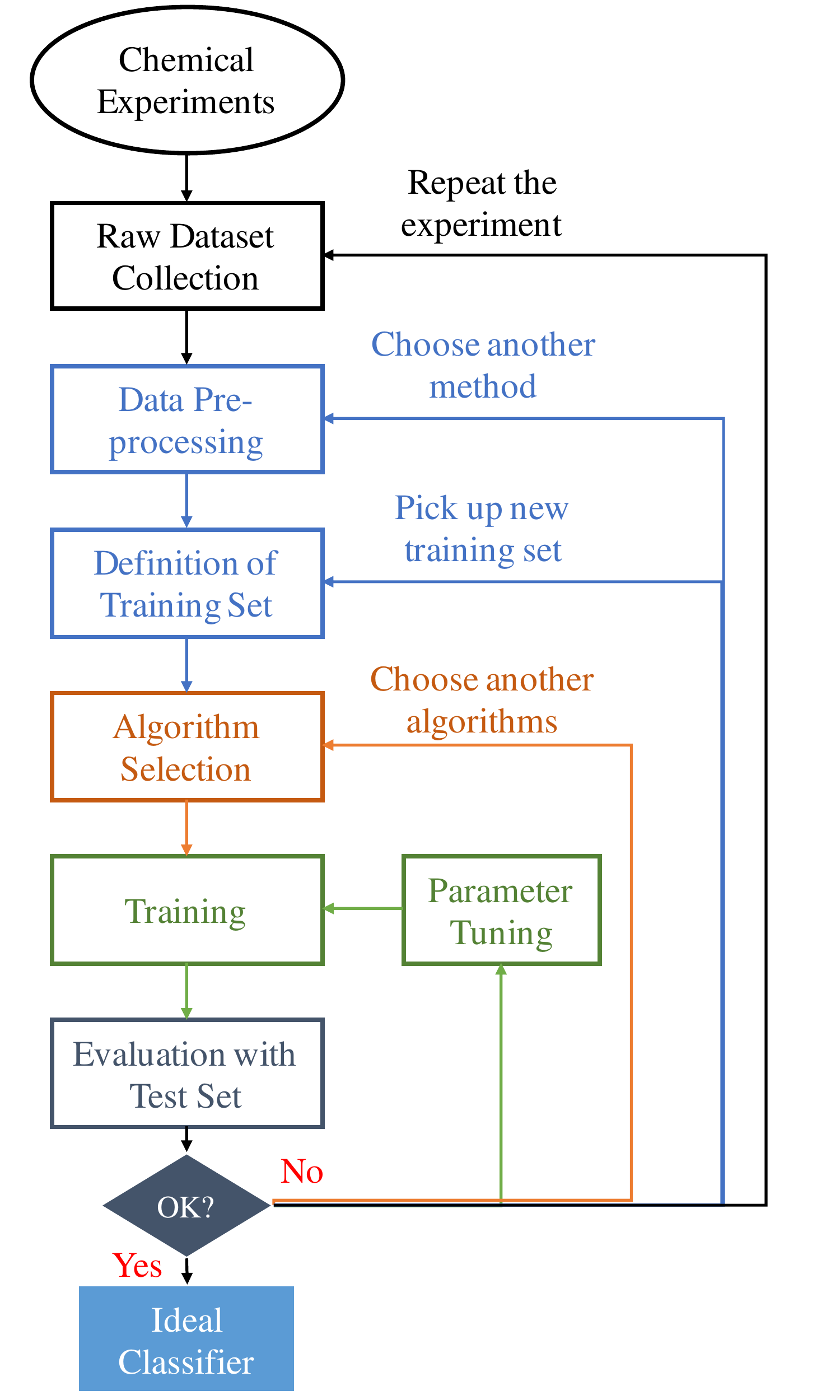}
    \caption{
    {\bf Working flowchart of supervised classification model to a chemical problem.}
    The first step is to collect raw data from chemical experiments, and clean the dataset with various pre-processing methods.
    The second step is to select appropriate algorithms and initialize the classifier model.
    Then, split the dataset using cross-validation and feed the classifier model with training data.
    Next, predict the label for test dataset, and evaluate error rate of the classifier model.
    }
    \label{fig_MLprocess}
\end{figure*}
Classification is always one of the most significant applications for classical machine learning(ML) or quantum machine learning(QML). Due to the fact that people are prone to make mistakes when establishing connections among various features, ML and QML are often able to improve the efficiency dealing with the classification problems.
Each instance in the dataset used by machine learning algorithms should be represented with the same set of features, which could be continuous, categorical or binary\cite{kotsiantis2007supervised}.
The learning process is then denoted as supervised machine learning if all instances are given with known labels. Generally, the classification problems can be divided as binary classification and multi-label classification. Binary classification is a classification with two possible outcomes.
For example, classify if an atom or molecule is excited or at ground state. Multi-label classification is a classification task with more than two possible outcomes. For example, classify the electrical resistivity and conductivity of materials as conductor, insulator, semiconductor or superconductor. We shall focus extensively on the supervised QML assisted classification problems in chemical systems, specifically, where the output of instances admits only discrete un-ordered values and then discuss unsupervised learning strategies too.

The process\cite{kotsiantis2007supervised,sen2020supervised} of applying supervised ML or QML to a typical classification problem with physico-chemical is demonstrated in Fig.(\ref{fig_MLprocess}).
The first step is to collect the dataset from chemical experiments. Due to the very existence of errors in measurement and impurities in reaction, in most cases the raw dataset contains noise and missing feature values, and therefore significant pre-processing is required\cite{zhang2003data}.
The second step is to select appropriate algorithms and initialize the classifier model. Then, split the dataset using cross-validation and feed the classifier model with training data. Next, predict the label for test dataset, and evaluate error rate of the classifier model. Additionally, parameter tuning should be repeated until an acceptable evaluation with the test set is obtained.

Classifiers based on decision trees (see Section \ref{Dec_trees}) play important roles in various chemical problems, such as the toxicity prediction\cite{karim2019efficient}, mutagenesis analysis\cite{chevaleyre2001solving}, and reaction prediction\cite{skoraczynski2017predicting}.
Even though, quantum decision trees are hardly applied independently dealing with intricate chemical classification problems, since simplicity, one of the key advantages of decision tree, could be eclipsed during the complicated mapping and learning process in QML.

\begin{figure*}[ht!]
    \centering
    \includegraphics[width=0.65\textwidth]{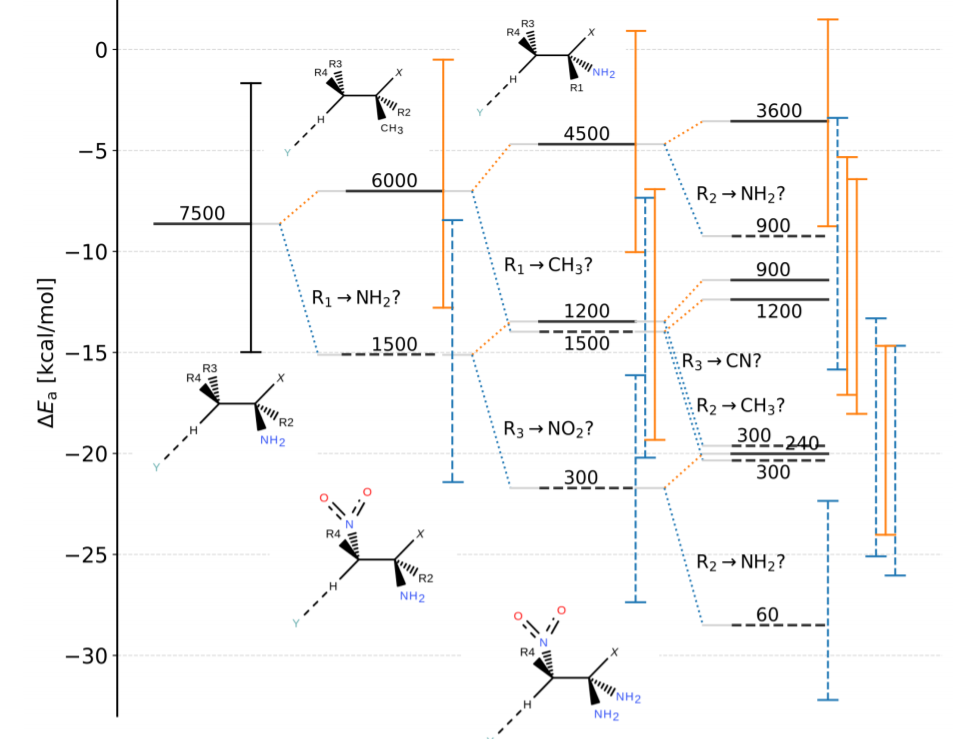}
    \caption{
    {\bf Decision tree using extracted rules and design guidelines.}
    The decision tree is generated using the reactant-to-barrier (R2B) method estimated activation barriers to predict changes in barrier heights by starting at all reactions (first energy level on the left) and subsequently apply changes by substituting functional groups, leaving groups and nucleophiles with E2 (see Ref\cite{heinen2020quantum}).
    Blue dotted lines refer to an accepted change meaning only compounds containing this substituents at the position are considered. Orange dotted lines refer to substitution declined, meaning all compounds except the decision are kept.
    Vertical lines on the right of energy levels denote the minimum first (lower limit), and the third (upper limit) quartile of a box plot over the energy range. 
    Numbers above energy levels correspond to the number of compounds left after the decision. 
    Lewis structures resemble the final decision.
    Reprinted from H.,Stefan  and Von Rudorff, Guido Falk  and Von Lilienfeld, O. Anatole, J. Chem. Phys., 155, 6, 064105, 2021 with the permission of AIP Publishing.
    }
    \label{fig_appDecisionTree}
\end{figure*}

Decision trees are sometimes applied along with other algorithms to analyze and demonstrate the key features of intricate classification process.
Recently, Heinen and coworkers studied two competing reactive processes mainly with a reactant-to-barrier (R2B) machine learning model\cite{heinen2020quantum},
where decision tree generated by the R2B method systematically extracts the information hidden in the data and the model. Fig.(\ref{fig_appDecisionTree}) is a scheme of the tree generated from the R2B method.
Blue dotted lines refer to an accepted change meaning only compounds containing this substituents at the position are considered. Orange dotted lines refer to substitution declined, meaning all compounds except the decision are kept. Vertical lines on the right of energy levels denote the minimum first (lower limit), and the third (upper limit) quartile of a box plot over the energy range. 
Numbers above energy levels correspond to the number of compounds left after the decision. 
Lewis structures resemble the final decision.

In the recent years there arise more than a few reports where Bayesian networks (BN) (see Section \ref{Bayesian_networks}) based methods are applied solving various chemical classification problems. The BN approach shows fruitful capability in the predictions of chemical shifts in NMR crystallography\cite{engel2019bayesian}, simulation of the  entropy driven phase transitions\cite{jinnouchi2019phase}, and particularly, the simulation of quantum
molecular dynamics simulation\cite{krems2019bayesian}.

\begin{figure*}[ht!]
    \centering
    \includegraphics[width=0.65\textwidth]{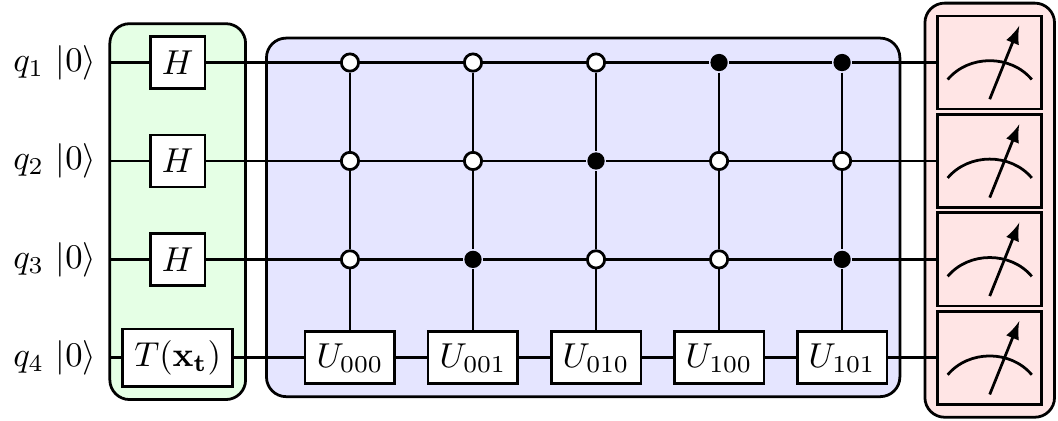}
    \caption{
    {\bf Scheme of the structure of the quantum circuit implementing classification process.}
    \\
    For training instances that are clustered into N sublabels, $\lceil{log_2N}\rceil$ qubits are required representing the sublabels in the classifier circuit.
    Meanwhile, $\lceil{log_2d}\rceil$ qubits are required to represent the test instance, where $d$ is denoted as the dimension of instance.
    For simplicity, here we assume that there are only 5 sublabels totally, and all instances are 2-d vectors.
    Thus, $q_{1,2,3}$ represent the possible sublabels and $q_4$ represents the test instance, meanwhile $U_{n}$ is operation corresponding to the centroid or mean value of the training instances under the same sublabel.
    Figure is reproduced from Ref\cite{li2021quantum}. (Under Creative Commons Attribution 4.0 International License)
    }
    \label{fig_circuit_vo2}
\end{figure*}

Quantum instance-based learning algorithms like k-NN (see Section \ref{k-NN_section}) are also applied in chemical classification problems. Recently, the authors have studied the phase transition of $VO_2$ based on a quantum instance-based learning algorithms\cite{li2021quantum}.
The training instances are firstly assigned into several sublabels via the quantum clustering algorithm, based on which a quantum circuit is constructed for classification. Fig.(\ref{fig_circuit_vo2}) is a scheme of the quantum circuit implementing classification process of the quantum clustering algorithm.
For training instances that are clustered into N sublabels, $\lceil{log_2N}\rceil$ qubits are required representing the sublabels in the classifier circuit.
Meanwhile, $\lceil{log_2d}\rceil$ qubits are required to represent the test instance, where $d$ is denoted as the dimension of instance.
For simplicity, here we assume that there are only 5 sublabels totally, and all instances are 2-d vectors.
Thus, $q_{1,2,3}$ represent the possible sublabels and $q_4$ represents the test instance, meanwhile $U_{n}$ is operation corresponding to the centroid or mean value of the training instances under the same sublabel.
Here Hadamard gates are applied on $q_{1,2,3}$, preparing a uniform distribution of the possible sublabels.
To improve the accuracy, the weighting scheme can be included by assigning $q_{1,2,3}$ as some certain quantum states corresponding to the weights.

\begin{figure*}[ht!]
    \begin{subfigure}[t]{0.32\textwidth}
        \centering
        \includegraphics[width=\textwidth]{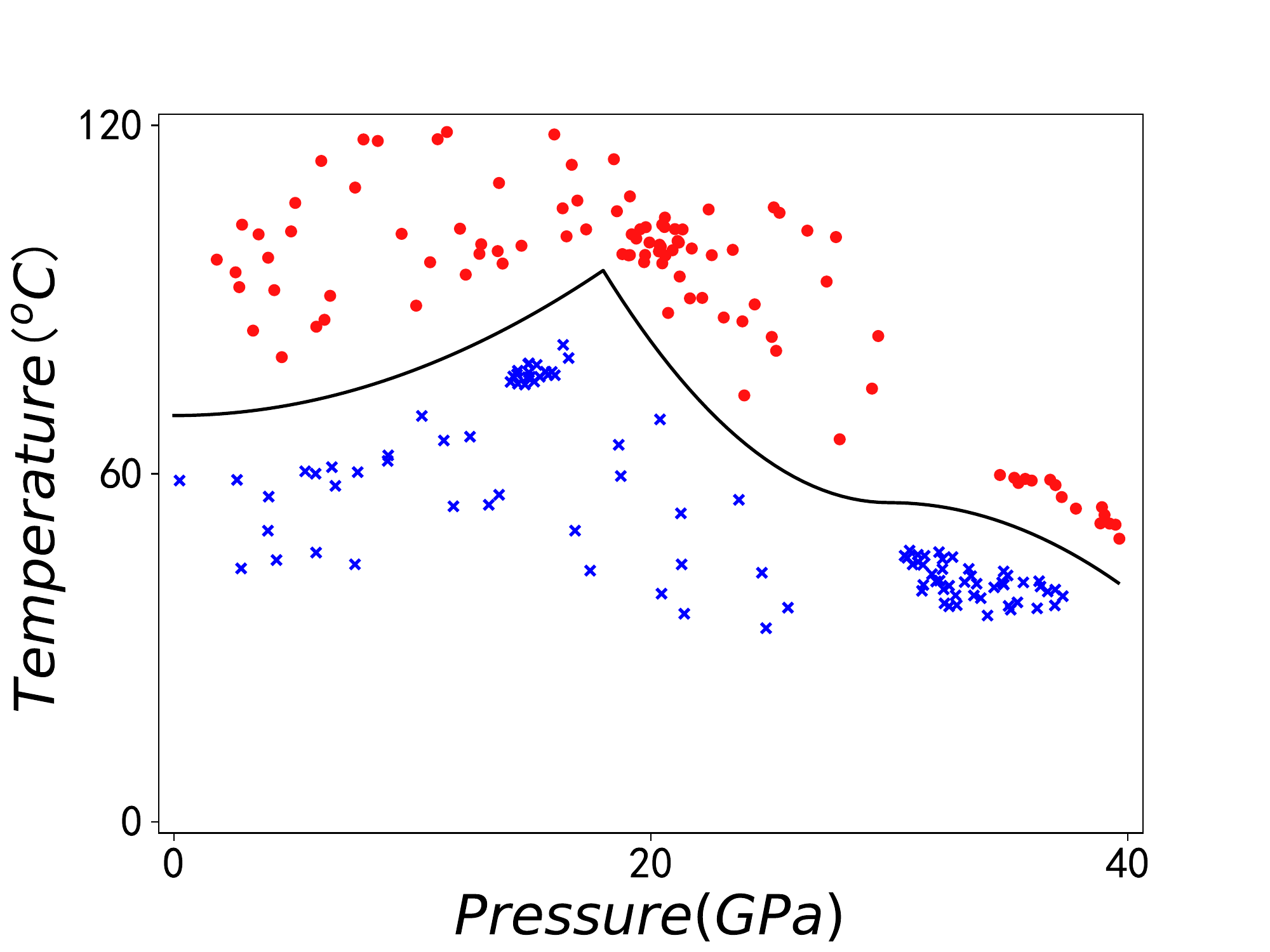}
        \caption{}
        \label{subfig_data_vo2}
    \end{subfigure}
    \centering
    \begin{subfigure}[t]{0.32\textwidth}
        \centering
        \includegraphics[width=\textwidth]{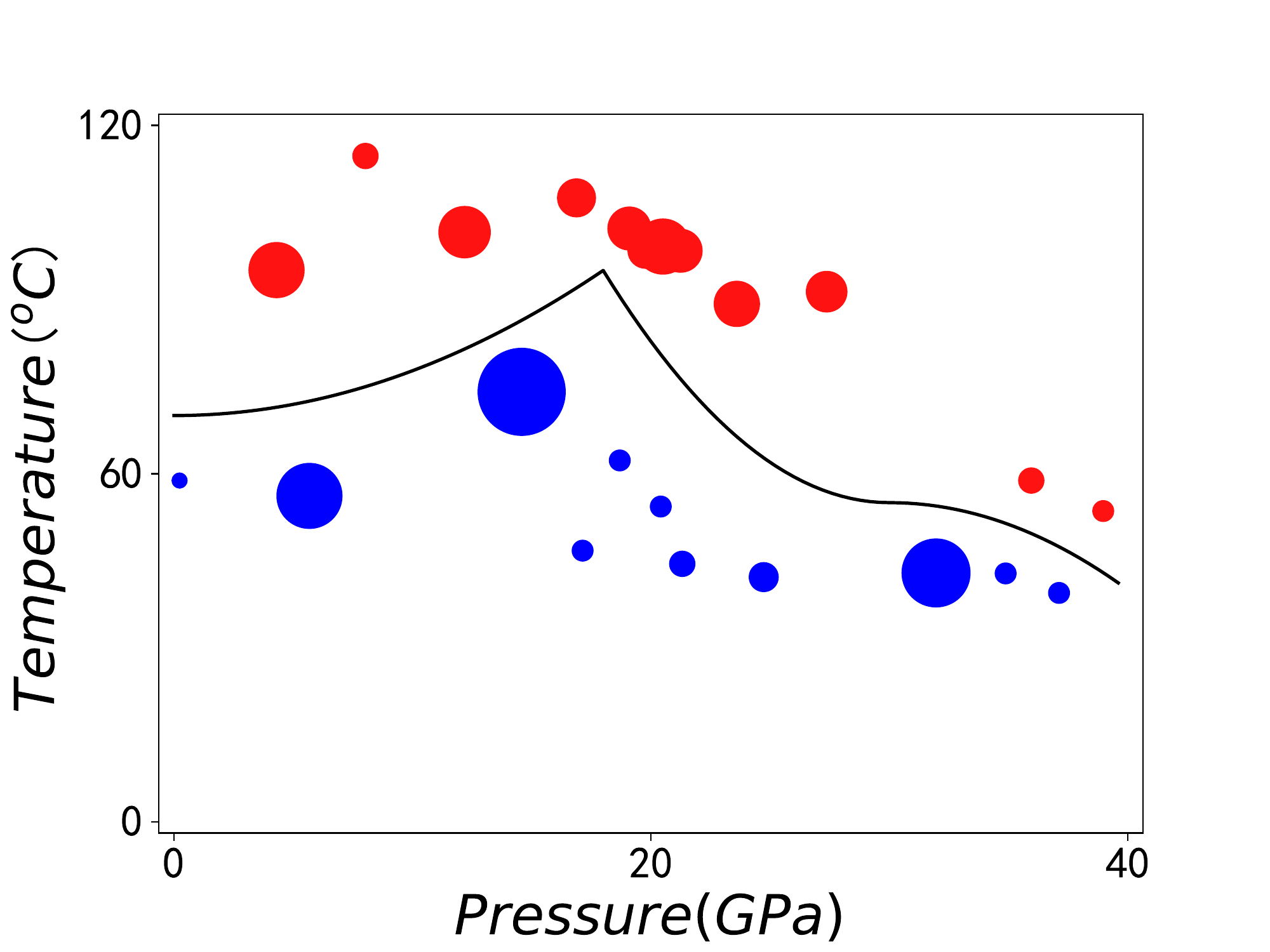}
        \caption{}
        \label{subfig_reduce_vo2}
    \end{subfigure}
    \centering
    \begin{subfigure}[t]{0.32\textwidth}
        \centering
        \includegraphics[width=\textwidth]{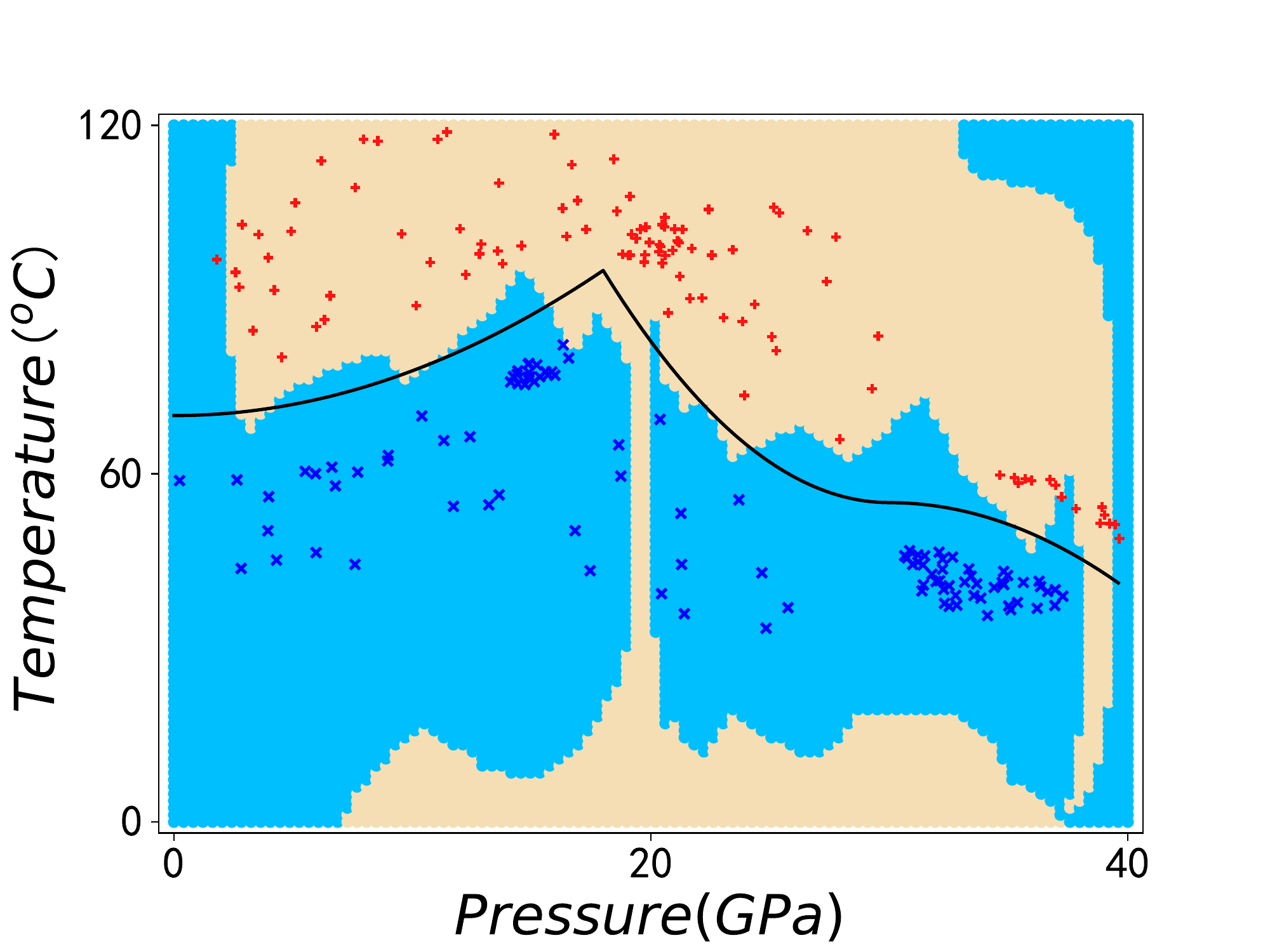}
        \caption{}
        \label{subfig_classify_vo2}
    \end{subfigure}
    \caption{
    {\bf Classification of of metallic and insulating states of $VO_2$ based on the quantum clustering algorithm.}
    \\
    Fig.(\ref{subfig_data_vo2}) demonstrates the original data used for classification.
    All training instances are 2-d vectors (pressure and temperature), while the label is denoted by color. 
    Red dots represent metallic state, and blue ones represent insulating state. Phase transition line indicated by the black solid curve.
    The sublabels left after quantum clustering algorithm is shown in Fig.(\ref{subfig_reduce_vo2}), where each sphere represents a sublabel, with center corresponding to the centroid or meanvalue, and radius corresponding to the number of instances.
    Prediction of test instances are shown in Fig.(\ref{subfig_classify_vo2}). 
    Test instances in the blue part will be recognized with label ’insulating’, and label of test instances in yellow part will be predicted as ’metallic’.
    Figure is reproduced from Ref\cite{li2021quantum} (Under Creative Commons Attribution 4.0 International License).}
    \label{fig_classification}
\end{figure*}

The process of classification of metallic and insulating states of $VO_2$ are shown in Fig.(\ref{fig_classification}).
Fig.(\ref{subfig_data_vo2}) demonstrates the original data used for classification.
All training instances are 2-d vectors (pressure and temperature), while the label is denoted by color. 
Red dots represent metallic state, and blue ones represent insulating state. Phase transition line indicated by the black solid curve.
The sublabels left after quantum clustering algorithm is shown in Fig.(\ref{subfig_reduce_vo2}), where each sphere represents a sublabel, with center corresponding to the centroid or mean-value, and radius corresponding to the number of instances.
Prediction of test instances are shown in Fig.(\ref{subfig_classify_vo2}). 
Test instances in the blue part will be recognized with label ’insulating’, and label of test instances in yellow part will be predicted as ’metallic’.

The quantum SVM (see Section \ref{Supp_vec_machine}) has also been applied into various classification problems, such as handwritten character recognition\cite{li2015experimental}, solar irradiation prediction\cite{senekane2016prediction}, and even the study of particle decays in high energy physics\cite{heredge2021quantum}. 
Additionally, in experiments provable quantum advantage has been demonstrated by the recent quantum classifiers based on a variational quantum classifier and a quantum kernel estimator—build on noisy intermediate-scale (NISQ) devices\cite{havlivcek2019supervised}.
Though there are only a few attempts to deal with specific chemical problems with quantum SVM methods, quantum SVM demonstrate great capacity in classification problems, while the optimization process of quantum SVM leads to exponential speedup comparing with the classical version. 
Therefore, there exists enormous potential for the quantum SVM methods to assist chemical classification and regression problems.

It is always one of the most crucial challenge in the study of many-body problems that the dimensionality of the Hilbert space grows exponentially with system size, which leads tremendous difficulty to solve the Shr\"{o}dinger equations. Among the modern numerical techniques designed to study the complicated systems, neural networks (NN) attracts enormous attention due to the remarkable abilities to extract features and classify or characterize complex sets of data. 
Modern ANN architectures, especially the feed-forward neural networks (see Section \ref{DNN_section}) and convolutional neural networks (see Section \ref{CNN_section}) have been playing significant roles in the classification problems of various fields.
It is reported that the neural network technologies can be used to discriminate phase transitions in correlated many-body systems\cite{carrasquilla2017machine}, to probe the localization in many-body systems\cite{schindler2017probing}, and even to classify the entangled states from separated ones\cite{gao2018experimental}.

On the other hand, dramatic success of classical neural networks as well provokes interest developing the quantum version. More than 20 years ago, pioneers attempted to build up quantum neural networks(QNN), particularly, the quantum version of feed-forward neural networks\cite{purushothaman1997quantum} (see Section \ref{DNN_section}).
There were also reports where the QNN were applied into real classification problems, such as the vehicle classification\cite{zhou2003automatic}.
The rapid development of hardware further provided more possibilities designing the QNN models.

Special-purpose quantum information processors such as quantum annealers and programmable photonic circuits are suitable fundamental implementation of deep quantum
learning networks\cite{biamonte2017quantum, dumoulin2014challenges}.
Researchers also developed recurrent quantum neural networks (RQNN) (see Section \ref{RNN_section}) that can characterize a nonstationary stochastic signals\cite{gandhi2013quantum, gao2018ima}.
Additionally, there are some other QNN models aiming to supervised learning, such as the classifiers based on QNN and measurements of entanglement\cite{zidan2019quantum}.
Even though, most of these models are actually hybrid models, as the activation functions are calculated classically, and the dataset are generally classical data.

In 2018 Farhi and Neven proposed a specific framework for building QNN that can be used to do supervised learning both on classical and quantum data\cite{farhi2018classification}.
For binary classification problems, the input instance can be represented by quantum state $|z,1\rangle$, where $z$ is a $n$-bit binary string carrying information of the inputs, and an auxiliary qubit is set as $|1\rangle$ initially.
Totally there are $n+1$ qubits, $n$ qubits representing the input instance, and 1 qubit representing the label.
After the unitary operation $U({\bf \theta})$, the auxiliary qubit is measured by a Pauli operator, denoted as $Y_{n+1}$, and the measurement result $1$ or $-1$ indicates the prediction of label. 
With multiple copies, the average of the observed outcomes can be written as $\langle z,1|U^\dagger({\bf \theta})Y_{n+1}U({\bf \theta})|z,1\rangle$.
Further, we can estimate the loss function
\begin{equation}
    loss({\bf \theta},z) = 1- l(z)\langle z,1|U^\dagger({\bf \theta})Y_{n+1}U({\bf \theta})|z,1\rangle
\end{equation}
where $l(z)$ is the label of instance $z$, which might be $1$ or $-1$.
For a training set ${z_j,l(z_j), j=1,\cdots,N}$, the training process is to find the optimal parameters ${\bf \theta}$ minimizing the loss function $\sum_{j=1}^Nloss({\bf \theta},z_j)$.
However, in the numerical simulations, they did not find any cases where the QNN could show speedup over classical competitors for supervised learning\cite{farhi2018classification}.

Meanwhile in 2018, researchers from Xanadu investigated the relationship between feature maps, kernel methods and quantum computing\cite{schuld2019quantum}.
There contains two main steps in the classification steps. They attempted to encode the inputs in a quantum state as a nonlinear feature map that maps data to quantum Hilbert space. Inner products of quantum states in the quantum Hilbert space can be used to evaluate a kernel function.
Then a variational quantum circuit is trained as an explicit classifier in feature space to learn a decision boundary. 
In the model, a vector $(x_1, x_2)^T$ from the input space $X$ is mapped into the feature space $F$ which is the infinite-dimensional space of the quantum system. 
The model circuit then implements a linear model in feature space and reduces the “infinite hidden layer” to two outputs. Though linear transformations are natural for quantum theory, nonlinearities are difficult to design in the quantum circuits.
Therefore the feature map approach offers an elegant solution. Classical machine learning took many years from the original inception until the construction of a general framework for supervised learning.
Therefore, towards the general implementation much efforts might be required as we are still at the exploratory stage in the design of quantum neural networks.


{
\color{black}
In 2019, Adhikary and coworkers proposed a quantum classifier using a quantum feature space\cite{adhikary2020supervised}, with both quantum variational algorithm and hybrid quantum-classical algorithm for training.
The input are encoded into a multi-level system, therefore the required number of training parameters is significantly less than the classical ones.
Simulation based on four benchmark datasets (CANCER, SONAR, IRIS and IRIS2) shows that the quantum classifier could lead to a better performance with respect to some classical machine learning classifiers.
In 2020, Wiebe's group proposed circuit-centric quantum classifier\cite{schuld2020circuit}, which is a class of variational circuits designed for supervised machine learning.
The quantum classifier contains relatively few trainable parameters, and constructed by only a small number of one- and two-qubit quantum gates,
as entanglement among the qubits plays a crucial role capturing patterns in the data.
The optimal parameters are obtained via a hybrid gradient descent method.
The circuit-centric quantum classifier shows significant model size reduction comparing the classical predictive models\cite{schuld2020circuit}.

The impressive success of these hybrid methods provide an alternative to study the chemistry classification problems.
There are plenty of attempts to study the phase diagrams classification with classical machine learning methods\cite{liu2020phase, deffrennes2022machine}. 
It would be of great interest to study these problems with quantum or hybrid classifiers.

}

{\color{black} 

Recently \cite{PhysRevA.102.012415} a report has proposed a variational quantum algorithm to classify phases of matter using a tensor network ansatz. The algorithm has been exemplified on XXZ model and transverse field Ising model. The classification circuit is composed of two parts: The first part prepares the approximate ground state of the system using Variational Quantum Eigensolver and feeds the state to the second part which is a quantum classifier which is used to label the phase of the state. Since the quantum state is fed directly into the classification circuit from the variational quantum eigensolver, it bypasses the data reading overhead which slows down many applications of quantum-enhanced machine learning. For both the parts, the quantum state $|\psi(\vec{\theta})\rangle$ is represented using a shallow tensor network which makes the algorithms realizable on NISQ devices. The first part of the algorithm represents the Hamiltonian matrix $H$ of the system in terms of Pauli strings and variationally minimize the average energy $\langle \psi(\vec{\theta})| H|\psi(\vec{\theta})\rangle$ to prepare the ground state. A checkerboard tensor network scheme with tunable number of layers $L$ is used for the representation. For an $n$-qubit state, the ansatz requires $O(n L)$ independent parameters, where $L$ is the number of layers in the circuit.  In this scheme, maximally entangled state would require $L=\lfloor n/2 \rfloor$ layers with periodic boundary conditions and for a critical one-dimensional systems $L=\lfloor log_2(n) \rfloor$ is enough. This can be contrasted with a UCCSD ansatz which requires typically $O(n^4)$ parameters \cite{RevModPhys.79.291} thereby necessitating a higher dimensional optimization. As mentioned before, the second part of the circuit i.e. the classifier receives the state from VQE part and applies a unitary $U_{class}(\phi)$. The unitary is again approximated using the Checkerboard tensor network ansatz. Then the output of the circuit is measured in Z-basis to determine the phase of the state using majority voting. For the transverse field Ising model the report demonstrated a $99\%$ accuracy with 4-layered classifier and for the XXZ model it was $94\%$ accuracy with a 6-layered classifier.

\begin{figure*}
    \centering
    \includegraphics[width=14cm]{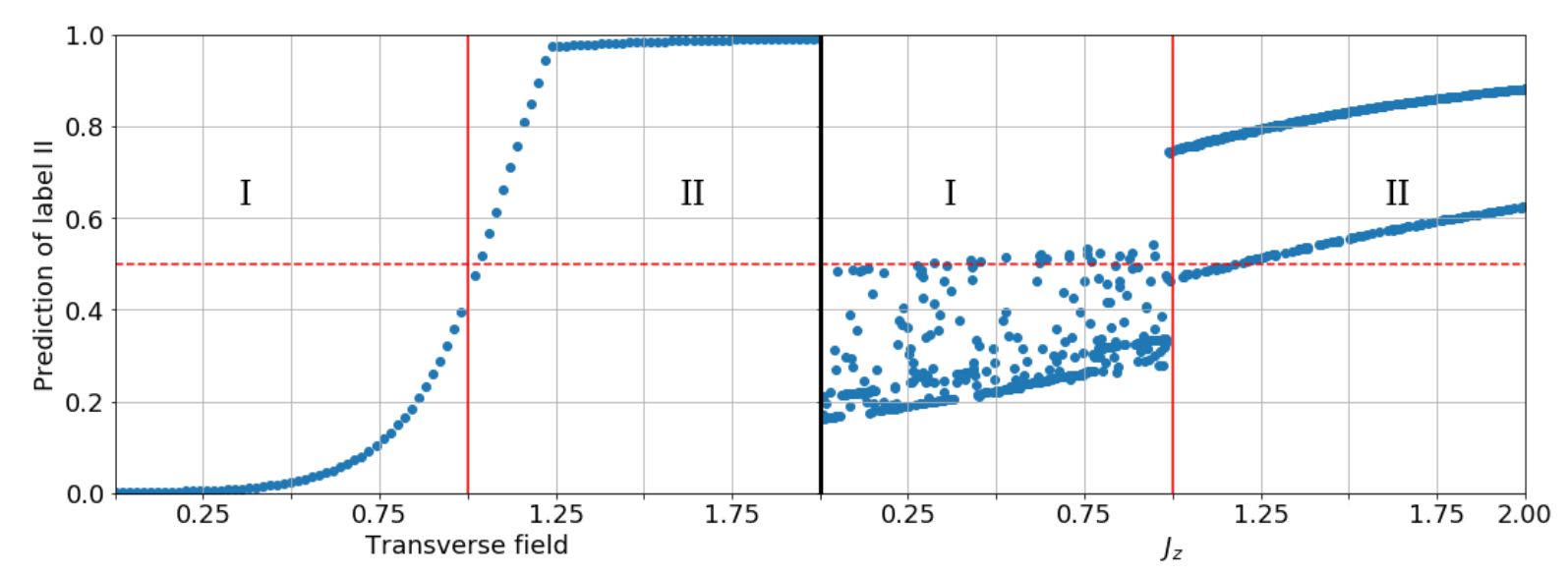}
    \caption{A figure showing the prediction of the phases using quantum classifier based on tensor network ansatz. The plot on the left (right) shows the prediction of phase II as a function of magnetic field ($J_z$) for Transverse-field Ising (XXZ) model. The Roman numbers I and II denote the phases the models. Reprinted figure with permission from A. V. Uvarov, A. S. Kardashin, and J. D. Biamonte, Machine learning phase transitions with a quantum processor, Phys. Rev. A 102, 012415, 2020. Copyright (2022) by the American Physical Society.}
\end{figure*}
}

Picking up an appropriate algorithm is always crucial dealing with the classification problems
The classifier’s evaluation is often based on prediction accuracy. Here we present three techniques estimating a classifier’s accuracy\cite{kotsiantis2007supervised}. 
One popular technique is to split the training instances into two groups, where two-thirds are regard as training data and the other third is regard as test data. Another technique is known as cross-validation.
The training set is manually divided into exclusive and equal-sized subsets in initial. Then for each subset the classifier is trained on all the other subsets. Thus, estimation of the error rate of the classifier is obtained by calculating the average of the error rate of each subset. The last one is denoted as Leave-one-out validation, which is a special case of cross validation. In Leave-one-out validation there is only a single instance in each subset. If the prediction accuracy can not reach the demand, another supervised learning algorithm should be selected, as shown in Fig.(\ref{fig_MLprocess}). 

Additionally, we will present some remarks about the techniques as follows. Even though the optimal solution always depends on the task at hand, these remarks can prevent the practitioners from selecting a wholly inappropriate algorithm. Logic-based systems often perform better when dealing with discrete features. Decision trees are in general resistant to noise because of their pruning strategies.
In contrast, most decision tree algorithms cannot perform well when diagonal partitioning is required.
Interference allows a class of quantum decision trees to be penetrated exponentially faster by quantum evolution than by a classical random walk.
Even though, these examples could also be solved in polynomial time by different classical algorithms\cite{farhi1998quantum}. BN methods are able to achieve its maximum prediction accuracy with a relatively a small dataset.
Besides, BN methods train very quickly since they require only a single pass on the data either to collect the frequencies or to compute the normal
probability density functions.
The graph structure of BN can efficiently construct a quantum state representing the intended classical distribution, and a square-root speedup time can be obtained per sample by implementing a quantum version of rejection sampling\cite{low2014quantum}.
Lazy learning methods require zero training time due the training instance is initially stored.
On the other hand, k-NN is quite sensitive to the irrelevant features, and is generally intolerant of noise. Quantum nearest neighbor algorithm and quantum nearest centroid algorithm both show significant speedup comparing the classical version.
In certain cases, there are exponential or even super-exponential reductions over the classical analog\cite{wiebe2014quantum}. SVM methods generally perform better when dealing with classification problems with multi-dimensions and continuous features. Moreover, SVMs are still able to perform well when there exists a nonlinear relationship between the input and output features.
Even though, a large sample size is required to achieve its maximum prediction accuracy. Optimization of quantum SVM is implemented by the quantum algorithm solving linear equations, leading to exponential speedup comparing to the classical version\cite{rebentrost2014quantum}.

Let us now focus on unsupervised approaches to classification briefly. The learning process is denoted as unsupervised when the given training instances are not assigned with the desired labels. Due to the absence of supervision, the unsupervised learning can hardly be applied to distinguish various types of chemicals or detect some certain structures. Instead, unsupervised learning process can find out the boundaries that divides the instances,
so that it could be beneficial in the recognition of phase transitions. For instance, unsupervised machine learning methods can be applied to identify the phase transition to non-trivial many-body phases such as superfluids\cite{broecker2017quantum}, or to detect the topological quantum phase transitions\cite{che2020topological}.

One important approach in the unsupervised QML is clustering methods. The clustering methods can be assigned as instance-based learning algorithms.
Consider the k-means problem of assigning given vectors to k clusters minimizing the average distance to the centroid of the cluster. The standard unsupervised learning method is Lloyd’s algorithm, which contains the following steps\cite{lloyd1982least, mackay2003information}:
\noindent(1)Pick up the initial centroid randomly; 
\noindent(2)Assign each vector to the cluster with the closest mean; 
\noindent(3)Recalculate the centroids of the clusters; 
\noindent(4)Repeat steps (1-2) until a stationary assignment is attained.
Based on the classical version, in 2013 Lloyd and coworkers proposed the quantum unsupervised machine learning method\cite{lloyd2013quantum}, rephrasing the the k-means problem as a quadratic programming problem which is amenable to solution by the quantum adiabatic algorithm. In 2018, Iordanis and coworkers proposed q-means, a new quantum algorithm for clustering problem, which provides substantial savings compared to the classical k-means algorithm.
In 2017, researchers implemented a hybrid quantum algorithm for clustering on a 19-qubit quantum computer\cite{otterbach2017unsupervised}, which shows robustness to realistic noise.

Inspired by the success of neural network-based machine learning, Iris and coworkers proposed the quantum convolutional neural networks (QCNN) in 2019\cite{cong2019quantum} (see Section \ref{CNN_section}). A paradigm is presented where the QCNN is applied to 1-D quantum phase recognition. There are less applications of unsupervised QML compared to supervised learning in chemical classification problems reported so far. Even though, the recent advancements suggest the future that unsupervised QML could take a place in the study of complex many-body systems, especially in the recognition of phases.

\subsection{Many-body structure and property prediction for molecules, materials, and lattice models and spin-systems} \label{MBS_sec}

\subsubsection{Machine learning techniques on a classical processor}
Obtaining an electronic structure description for material systems has been a problem with continued research in Chemistry and material science. Since this task is a many-body problem, solving it with high accuracy is crucial as numerous material properties and chemical reactions entails quantum many-body effects. For a significant duration, performing electronic structure calculations were done using the Density Functional Theory (DFT), which is based on the effective single-particle Kohn-Sham equations \cite{kohn1965self}. In DFT, the ground state energy of a many-electron system is written as a functional of the electron density, thereby reducing the many-body problem for an N particle wavefunction to just one. This has yielded accurate results with efficient computations compared to its predecessors, but the functional form of the exact solution is unknown and efficient approximations are made for practical purposes. Attempts have been therefore made to obtain such density-functionals using ML algorithms. One of the earliest study is due to Snyder $et al$ \cite{snyder2012finding} which constructed the kinetic energy functional for spinless fermions in a 1D box subjected to an external potential made from the linear combination of several gaussians defined on a dense spatial grid. Many such randomly selected external potentials were chosen as the training set with the actual labelled density and kinetic energy obtained by solving the Schroedinger equation as the training data. Thereafter kernel-ridge regression was used to construct the kinetic energy functional from the aforesaid known density with excellent accuracy. The oscillations in functional derivative of the so constructed kinetic energy functional were dampened by using the principal components. From the knowledge of this functional derivative a protocol to procure a self-consistent density field that minimizes the energy was presented. Since then many report exists which have attempted to construct density functionals especially the exchange-correlation functional \cite{schmidt2019machine, bogojeski2020quantum, nagai2020completing, li2016pure, borlido2020exchange, fritz2016optimization, liu2017improving, ryczko2019deep}. 

{\color{black} Kernel-ridge regression (see Section \ref{KKR_sec}) has been extensively used in chemistry for a variety of other purposes too like predicting the energy of the highest occupied molecular orbital from three different molecular datasets \cite{doi:10.1063/1.5086105} using two different technique to  encode structural information about the molecule or for the prediction of atomization energies \cite{PhysRevLett.108.058301,hansen2013assessment} through a Coulomb matrix representation of the molecule wherein the energy is expressed as a sum of weighted gaussian functions. Recently many new schemes to represent structural features have also been designed \cite{doi:10.1063/1.5020710,doi:10.1063/1.5126701} wherein information about the environment of each constituent atom is encoded within its $M$-body interaction elements each of which is a weighted sum of several gaussian functions. The distance metric between each such interaction representation between element $I$ and $J$ is considered to be the usual Euclidean norm. Atomization energies, energies for even non-bonded interaction like in water clusters predicted using such representations are extremely accurate\cite{doi:10.1063/1.5126701}.
More such examples can be found in topical reviews like in Ref \cite{DRAL2020291}.
}
Caetano et al. \cite{caetano2011using} used Artificial Neural Networks (ANNs) (Theoretical framework discussed in section. \ref{ANN_section}) trained using the Genetic Algorithm (GA) to solve the Kohn-Sham equations for He and Li atoms. They used a network comprising of one neuron in the input layer, one hidden layer with eight neurons, and two neurons in the output layer. For the GA based optimization, the number of individuals N in the population was kept to 50. By generating the initial orbitals randomly and building the electron density, an effective Hamiltonian is constructed. The ANN is trained using GA to find the orbitals that minimize the energy functional and then the total energy is calculated, which is repeated until a convergence point. The results from the ANN were shown to be in good agreement with the other numerical procedures.

Performing higher order calculations like CCSD provides accurate results but have very high computational cost. While, methods such as semi-empirical theory PM7, Hartree-Fock (HF), or (DFT) provide less accurate results but scale efficiently. The work by Ramakrishnan et al. \cite{ramakrishnan2015big} corrects the lower-order methods to provide accurate calculations by training their $\Delta$ model to to learn enthalpies, free energies, entropies, and electron correlation energies from a dataset consisting of organic molecules. The property of interest was corrected by expressing:
\begin{equation}
    P_t(R_t) \approx \Delta_b^t R_b = P'_b R_b + \sum_{i=1}^N \alpha_i k(R_b, R_i)
\end{equation}
where, $\alpha_i$ are regression coefficients, obtained through kernel ridge regression (Theoretical framework discussed in section. \ref{GPR_section}), $k(R_b, R_i) = \exp{\frac{|R_b - R_i|}{\sigma}}$, with $\sigma$ being a hyperparameter that is tuned, $|R_b - R_i|$ is the Manhattan norm \cite{rupp2012fast} measuring the similarity between the features of target molecule $R_b$ and molecule in the data $R_i$. 

Burak Himmetoglu \cite{himmetoglu2016tree} constructed a dataset incorporated from PubChem comprising of the electronic structures of 16,242 molecules composed of CHNOPS atoms. Having constructed the Coulomb matrices as in \cite{rupp2012fast} defined by:
\begin{equation}
    C_{ij} = \begin{cases}
             \text{$0.5 Z_{i}^{2.4};     i=j$}\\
             \text{$\frac{Z_i Z_j}{|R_i - R_j|};  i \neq j$}\\
             \end{cases}
\end{equation}
where, the atomic numbers are denoted by $Z$, and $R$ is their corresponding positions. Design matrices using the eigenvalues of the Coulomb matrices are constructed and two types of ML approaches are used to predict the molecular ground state energies. Firstly, boosted regression trees (Theoretical framework discussed in section. \ref{Dec_trees}) and then ANNs are used, and their performances are compared. 

Geometry optimization is a crucial task, which directly impacts the electronic structure calculations. The total energy calculations can prove to be quite expensive depending on the choice of electronic structure method. Hence, the number of evaluations of the potential energy surface has to be reduced considerably.  A novel gradient-based geometry optimizer was developed by Denzel and {\color{black}K\"{a}stner \cite{denzel2018gaussian}, that exploits Gaussian Process Regression or GPR (Theoretical framework discussed in section. \ref{GPR_section})} to find the minimum structures. By comparing Mat\'ern kernel with the squared exponential kernel, the authors show that the performance is better when Mat\'ern kernel is used. The performance of their optimizer was tested on a set of 25 test cases along with a higher dimensional molybdenum system, molybdenum amidato bisalkyl alkylidyne complex, and was shown that GPR based approach can handle the optimization quite well. 

\begin{figure*}[ht!]
    \centering
    \includegraphics[width=0.95\textwidth]{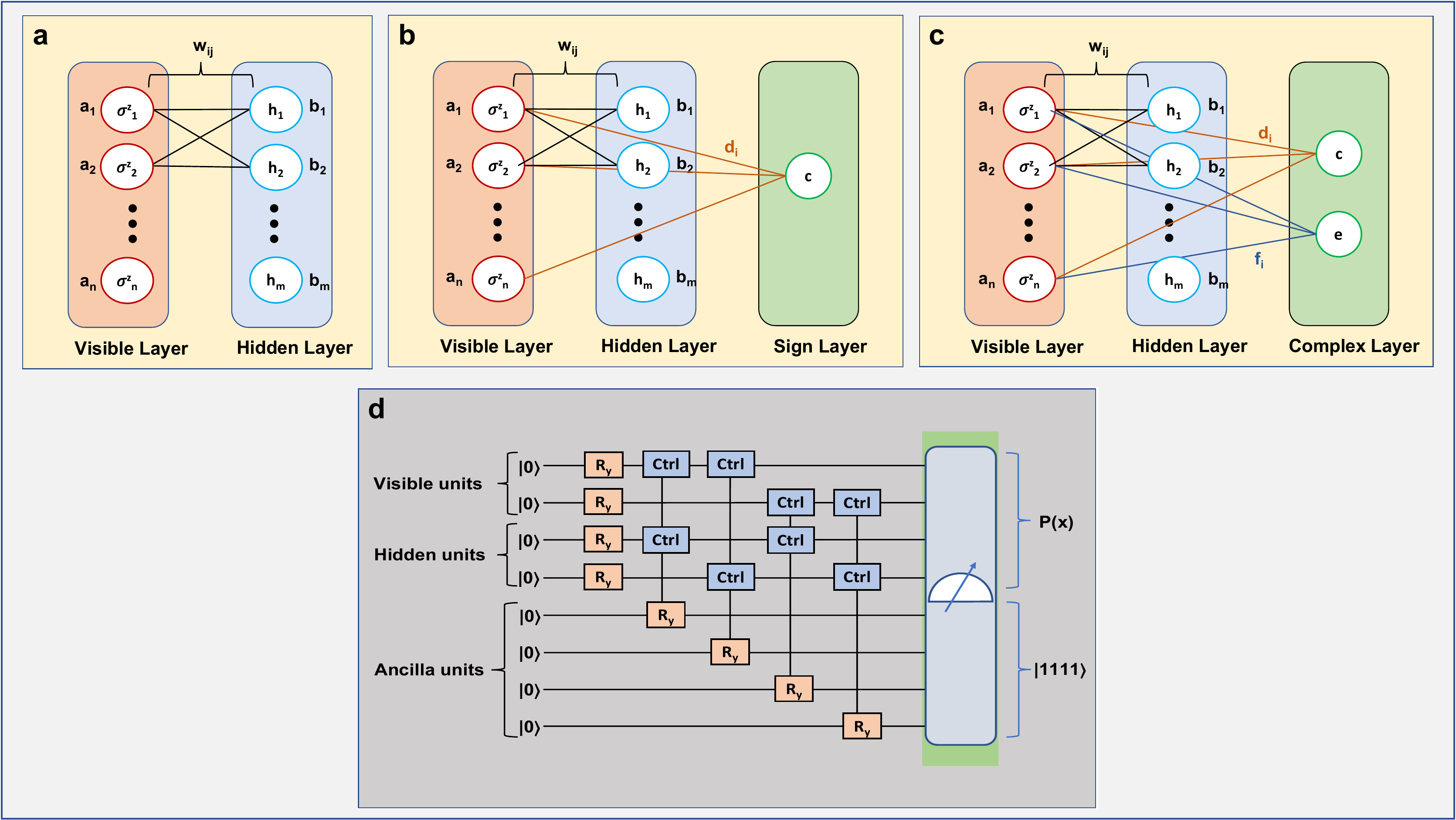}
    \caption{The RBM architecture evolution. \textbf{a} The RBM architecture consisting of a visible layer and a hidden layer. The neurons in the visible layer represented encode a variable $\sigma^{z}_i \in \{1,-1\}$ have biases denoted by $a_i$ and the neurons in the hidden layer encode a variable $h_j \in \{1,-1\}$ have biases denoted by $b_j$. The weights associated with each connection between the visible node and the hidden node is denoted by $w_{ij}$ (Schematic of RBM as illustrated in Ref. \cite{carleo2017solving}). \textbf{b} The three layered RBM architecture with the first two layers and their corresponding parameters same as in \textbf{a}. The third layer has one neuron which encodes a variable $s(x)$ (see Eq. \ref {phase_encod} for the case with $(e, f_i)\:\: = (0,0)\:\: \forall i$) and is known as the sign layer. The bias unit for the sign layer is represented by $c_i$ and the weights associated with each connection between the visible node and the sign node is denoted by $d_{i}$  (Schematic of RBM as illustrated in Ref. \cite{Xia_2018}). The sign layer do not share any connection with the hidden layer. \textbf{c} The three layered RBM architecture with the third layer consisting of two units. $c$, and $e$ denote the bias parameters of the unit representing the real part of the complex layer, and the unit representing the complex part of the complex layer, respectively. $f_i$ indicate the weights corresponding to the connections between $\sigma^z_i$ and the unit representing the complex part of the complex layer  (Schematic of RBM as illustrated in Ref. \cite{kanno2019manybody, sureshbabu2021implementation, sajjan2021quantum}). \textbf{d} The quantum circuit to sample Gibbs distribution on quantum hardware. In general for $n$ neurons in the visible layer and $m$ neurons in the hidden layer, this quantum circuit would require $m\times n$ ancilla qubits as well. The circuit thus shown is for a special case of $(n=m=2)$. The circuit is responsible for simulating the amplitude field $\sqrt{P(\bf x)}$ by sampling from the distribution $P(\bf x)$ as defined in Eq. \ref{RBM_amp_encoding} (Schematic of quantum circuit as illustrated in Ref \cite{sureshbabu2021implementation, sajjan2021quantum}).}. 
    \label{fig:rbm_evolution}
\end{figure*}

In the work by Carleo and Troyer \cite{carleo2017solving}, it was shown that by representing the many-body wavefunction in terms of ANNs, the quantum many-body problem can be solved. They used this idea to solve for the ground states and describe the time evolution of the quantum spin models, viz. transverse field Ising, and anti-ferromagnetic Heisenberg models. Expanding the quantum many-body state $\psi$ in the basis $\ket{x}$ :
\begin{equation}
    \ket{\psi} = \sum{{\psi(x)} \ket{x}}
\end{equation}
where, ${\psi(x)}$ is the wavefunction represented in terms of a Restricted Boltzmann Machine (RBM), which is the ANN architecture that was used in this study (Fig. \ref{fig:rbm_evolution} \textbf{a}). The description of Neural-Network Quantum States (NQS) results in the wavefunction ${\psi(x)}$ to be written as ${\psi(x; \theta)}$, where $\theta$ denotes the tunable parameters of the neural network. The quantum states can now be expressed as:

\begin{equation}
    {\psi_{M}(x;\theta)} \propto \sum_{h_i} e^{\frac{1}{2}\sum_j a_j \sigma^{z}_j + \sum_i b_i h_i + \sum_{ij} w_{ij} h_i \sigma^{z}_j}
\end{equation} \label{psi_RBM}
where, the values of $\sigma^{z}_j$ \& $h_i$ $\in$ \{+1,-1\}, $a_j$ \& $b_i$ are the bias parameters corresponding to the visible and hidden layer respectively, and $w_{ij}$ is the weight associated with the connections between $\sigma^{z}_j$ \& $h_i$ (see Fig. \ref{fig:rbm_evolution} (a)). For a given Hamiltonian, the average energy written as a statistical expectation value over $|\psi_{M}(x;\theta)|^2$ distribution is computed. For a specific set of parameters, samples from the $|\psi_{M}(x;\theta)|^2$ distribution are taken via Markov Chain Monte Carlo (MCMC) and the gradients of the expectation value are calculated. With the gradients known, the parameters are optimized in order to model the ground states of the Hamiltonian. The accuracy of the results were extremely high in comparison to the exact values. Recently Choo $et al$ \cite{choo2020fermionic} have also extended the method to model ground states of molecular hamiltonians. The protocol maps the molecular hamiltonian onto the basis of Pauli strings using any of the fermion-qubit mapping like Jordan-Wigner or Bravi-Kitaev encoding that is routinely employed in quantum computing \cite{Cao2019QuantumCI}. Thereafter the wave-function as described in Eq. \ref{psi_RBM} is used to variationally minimized the ground state energy by drawing samples using Variational Monte Carlo (VMC) \cite{choo2020fermionic}. Several insightful features like the fact that support of the probability distribution in the space of spin configurations is peaked over certain dominant configurations only near the Hartree-Fock state, out-performance of RBM ansatz over more traditional quantum chemistry methods like CCSD(T) and Jastrow ansatz, the efficacy of the various fermion-qubit mapping techniques in classical quantum chemistry simnulations etc were elucidated.

This NQS approach was then extended by Saito \cite{saito2017solving} to compute the ground states of the Bose-Hubbard model. Here, the quantum state is expanded by the Fock states. one-dimensional optical lattice with 11 sites and 9 particles, and a two-dimensional lattice with 9$\times$9 sites and 25 particles were studied. The optimization scheme is similar to that in \cite{carleo2017solving} and the ground states were calculated for the 1D and 2D cases were shown to be in good agreement with those obtained from exact diagonalization and Gutzwiller approximation respectively.

In the work by Coe \cite{coe2018machine}, NNs were used to select important configurations in the Configuration Interaction (CI) method. By using just a single hidden layer, with binary occupations of the spin orbitals in the desired configurations, the transformed co-efficient of the configuration that fitted to the co-efficients in the training set, are predicted. The important configurations of stretched $\rm{CO}$ and $\rm{Co}$ at its equilibrium geometry were shown to have accurate predictions. This work was extended in their follow-up paper \cite{coe2019machine}, in which the potential energy curves for $\rm{N_2}$, $\rm{H_2O}$, and, $\rm{CO}$ were computed with near-full CI accuracy.

Custódio et al. \cite{custodio2019artificial} developed a feedforward neural network (Theoretical framework discussed in section. \ref{ANN_section}) to obtain a functional form for calculations pertaining to inhomogeneous systems within DFT and Local Spin Density Approximation (LSDA) framework. The network consists of an input layer with 3 neurons, a hidden layer with 20 neurons, and an output layer consisting of one neuron. The network was trained on 20,891 numerically exact Lieb-Wu results for 1000 epochs. The authors test their network on non-magnetic and magnetic systems and through their results claim that the neural network functional provides is capable of capturing the qualitative behavior of energy and density profiles for all the inhomogeneous systems. In another closely related work \cite{moreno2020deep}, the authors attempt to construct the many-body wavefunction directly from 1D discretized electron density using a feed-forward neural network. The network was trained by a supervised learning scheme with the infidelity of the procured wave-function and the target wave-function within the training data. The model showed excellent performance for the Fermni-Hubbard hamiltonian in both the metallic and the Mott-insulating phases. To bypass the need to construct the exponentially scaling many-body wavefunction, the authors also construct the density-density two-point auto-correlation function with remarkable accuracy. From such auto-correlation function, the power-law scaling parameters of different phases can be obtained \cite{moreno2020deep}

A deep neural network titled SchNet introduced by \cite{schutt2018schnet} and SchNOrd introduced by \cite{schutt2019unifying}, to predict the wavefunction in a local basis of atomic orbitals. By treating a molecule as a collection of atoms and having a descriptor for each atom, the output property being predicted is a sum of all these atomic descriptions. The inputs to the network are the atom types and the position of these atoms in the Cartesian coordinates. The atom types are embedded in random initializations and are convoluted with continuous filters in order to encode the distances of these atoms. The interaction layers encode the interaction between different atoms, which essentially dictates the features of these atoms. These features are now input to a factorized tensor layer, which connects the features into pairwise combinations representing every pair of atomic orbitals. Multi-layer perceptrons (Theoretical framework discussed in section. \ref{ANN_section}) are then used to describe the pair-wise interactions within these pair of orbitals. This model was shown to predict with good accuracy the energies, Hamiltonians, and overlap matrices corresponding to water, ethanol, malondialdehyde, and uracil. 

Hermann et al. \cite{hermann2020deep} extended the work by Schutt et al. \cite{schutt2018schnet} by using SchNet in the representation of electrons in molecular environments. The representation of the wavefunction through their neural network titled PauliNet in association with the training done using Variational Monte Carlo approaches very high accuracy for energies of Hydrogen molecule $\rm{H_2}$, Lithium Hydride $\rm{LiH}$, Beryllium $\rm{Be}$, Boron $\rm{B}$, and a linear chain of hydrogen atoms $\rm{H_{10}}$. The authors also investigate the scaling of PauliNet with the number of determinants and with system size on $\rm{Li_2}$, $\rm{Be_2}$, $\rm{B_2}$, and $\rm{C_2}$ and state the high accuracy in comparison to Diffusion Monte Carlo (DMC) can be achieved quite fast. Having studied the energies for systems that have benchmark results, the authors move on to high accuracy prediction of the minimum and transition-state energies of cyclobutadiene. 

Faber et al. \cite{faber2018alchemical} introduced a representations of atoms as a sum of multidimensional Gaussians in a given chemical compound for a Kernel-Ridge Regression (KRR) (Theoretical framework discussed in section. \ref{GPR_section}) based model to predict the electronic properties having learnt them from several datasets. By deriving analytical expressions for the distances between chemical compounds, the authors use these distances to make the KRR based model to learn the electronic ground state properties. In a follow-up paper from the same group \cite{christensen2020fchl}, Christensen et al. provide a discretized representation as opposed to comparing atomic environments by solving the aforementioned analytical expression. In this work, the authors use KRR to learn the energy of chemical compounds and three other regressors viz. operator quantum machine learning (OQML), GPR, and gradient-domain machine learning (GDML), are reviewed for the learning of forces and energies of chemical compounds. 

In order to have a ML model that can generalize well in large datasets and have efficient tranferability, Huang and Lilienfeld \cite{huang2020quantum} introduced the atom-in-molecule (amon) approach where fragments of such amons with increasing size act as training data to predict molecular quantum properties. Considering only two and three-body interatomic potential based representations of amons, a subgraph matching procedure is adopted, which iterates over all the non-hydrogen atoms in the molecule, and identifies the relevant amons. The relevent amons are identified by converting the molecular geometry to a molecular graph with vertices specified by the nuclear charge of atoms and bond orders, then verifying if their hybridization state is preserved or not, and if such subgraphs are isomorphic to other identified subgraphs with some additional checks. Such amons are selected and sorted based on size. 

Obtaining the right material that can be explored experimentally in a large database can prove to be a daunting task but provides very rewarding results. Multilayer Perceptrons (MLP) were used by Pyzer-Knapp et al. \cite{pyzer2015learning} in the High Throughput Virtual Screening (HTVS) of the Harvard Clean Energy Project, which was developed for the discovery of organic photovoltaic materials. The network consisting of a linear input and output layers, three hidden layers with 128, 64, and 32 nodes, was trained on 200,000 molecules and an additional 50,000 molecules were considered to make up the validation set.

Having performed a FT based high throughput screening, Choudhary et al. \cite{choudhary2019accelerated} trained a Gradient Boosting Decision Tree (GBDT) (Theoretical framework discussed in section. \ref{Dec_trees}) based supervised learning model on 1557 descriptors obtained from classical force-field inspired descriptors {CFID}. The model was used to classify the materials based on spectroscopic limited maximum efficiency (SLME) to be greater than 10\% or not. The authors The authors use this classification to prescreen over a million materials available through large crystallographic and DFT databases, with the goal of accelerating the prediction of novel high-efficiency solar cell materials.  

By using a series of random forest models (Theoretical framework discussed in section. \ref{Dec_trees}) trained with different threshold temperatures, Stanev et al. \cite{stanev2018machine} show that the materials in the SuperCon database can be classified into two classes, above or below a critical temperature ($T_c$) of 10 K. The features required to be learnt by the model are obtained by using Materials Agnostic Platform for Informatics and Exploration (Magpie) based predictors from SuperCon along with a set of 26 predictors from AFLOW online Repositories. The authors also develop a regression model to predict the $T_c$ of cuprate, iron-based and low-$T_c$ materials. 

Barett et al. \cite{barrett2021autoregressive} showed the expressibility power of representing the wavefunction ansatz using an Autoregressive Neural Network (ARN). ARNs are a class of generative, sequential models that are feedforward, where observations from the previous time-steps are used to predict the value at the current time step. By developing an architecture with several sub-networks, each made up of a multi-level perceptron, to model the amplitudes and a separate sub-network to model the phase of the wavefunction, the authors go on to introduce a unique sampling procedure that scales with the number of unique configurations sampled and describe the efficiency of ARNs by computing the ground states of various molecules achieving standard full configurational interaction results.

{\color{black}In order to discover non-centrosymmetric oxides, Prasanna V. Balachandran et al. \cite{balachandran2017learning} developed a methodolofy combining group theroretic approach, ML, and DFT and applied it towards layered Ruddlesden-Popper oxides. Using group theory to establish a dataset consisting of 3253 total chemical compositions and performing PCA  (Theoretical framework discussed in section. \ref{PCA_sec}) and classification learning, the authors identify 242 relevant compositions that displayed potential for NCS ground state structures. Then, taking advantage of DFT, 19 compositions were predicted that were suggested for experimental  synthesis.
Autoencoders (Theoretical framework discussed in section. \ref{Auto_section}) are known for their ability to reduce the dimenisonality of the problem at hand and can be used to design molecules with a specific desirable property. This was used by Gómez-Bombarelli et al. \cite{gomez2018automatic} by coupling an autoencoder with a neural network to generate molecular structures along with predicting the properties of molecules, which were represented by points in the latent space. 
Neural networks have also branched into the domain of generating complex crystalline nanomaterials thanks to the work by Baekjun Kim et al. \cite {kim2020inverse}. In this work, the authors base their approach with the use of Wasserstein GAN (WGAN) (Theoretical framework discussed in section. \ref{GAN_section}) with gradient penalty on the loss function on the critic (which is a renamed discriminator specific to the WGANs). By considering a training set consisting of 31713 known zeolites, the network takes enery and material dimensions (with the materials grid subdivided into silicon and oxygen atom grids) as input, to produce 121 crystalline porous materials. Considering methane potential energy to be the energy dimension, and a user defined range from 18 kJ/mol to 22 kJ/mol, the authors were able to successfully demonstrate inverse design of zeolites. Since the combinatorial space of multi-principal element alloys (MPEAs) is extensive, it becomes necessary to have a reliable method that accurately and rapidly predicts the intermetallics and their formation enthalpies. In accordance with this necessity, an ML model using GPR  (Theoretical framework discussed in section. \ref{GPR_section}) with a sum kernel, which consists of the square exponential kernel and a white noise kernel, was developed \cite{zhang2020machine}. In this work, the ML model is trained on 1538 stable binary intermetallics from the Materials Project database and uses elemental properties as descriptors in order to learn the distribution that maps these descriptors to the formation enthalpies. By doing so, the authors show that stable intermetallics can be predicted using this method. They also perform transfer learning to predict ternary intermetallics in the aforementioned database, thereby informing about the generalizability of the model. With growing interest in superhard materials for various industrial applications, Efim Mazhni et el. \cite{mazhnik2020application} developed a neural network on graphs model, with a linear embedding layer, followed by three convolutional layers, a pooling layer, and two linear transformations with softmax activation layers, to calculate hardness and fracture toughness. By training their network on the database of crystal structures by Materials Project, consisting of 8033 structures, the authors predict the bulk modulus and the shear modulus, from which the Young’s modulus and the Poisson’s ratio is obtained.}

{\color{black}
Tensor networks have been discussed in Section \ref{Tensor_Network}. Tensor Network have been applied to solve numerous problems in physics, chemistry and material science. (To review tensor network algorithms for simulating strongly correlated systems refer \cite{bruognolo2017tensor}). Recently Kshetrimayum $et al$ \cite{KSHETRIMAYUM2020168292} published a report developing tensor network models for strongly correlated systems in two spatial dimensions and extending it to more realistic example. The specific material the authors chose is that 
of a quantum magnet $Ca_{10}Cr_7O_{28}$ which is known to show properties of a quantum spin liquid in inelastic neutron scattering experiments \cite{balz2016physical}. The material possesses a distorted breathing bilayer Kagome lattice crystal structure comprising of $Cr^{5+}$ ions with spin-1/2 moments. Despite having even number of spin-1/2 $Cr^{5+}$ ions, the interactions lack the tendency to form a singlet state. The description of the lattice is shown in Fig. \ref{spin_liq_Kagome}. The two Kagome layers are different from each other and each of them consists of two non-equivalent alternating triangles. 
\begin{figure}[h]
    \centering
    \includegraphics[width=8cm]{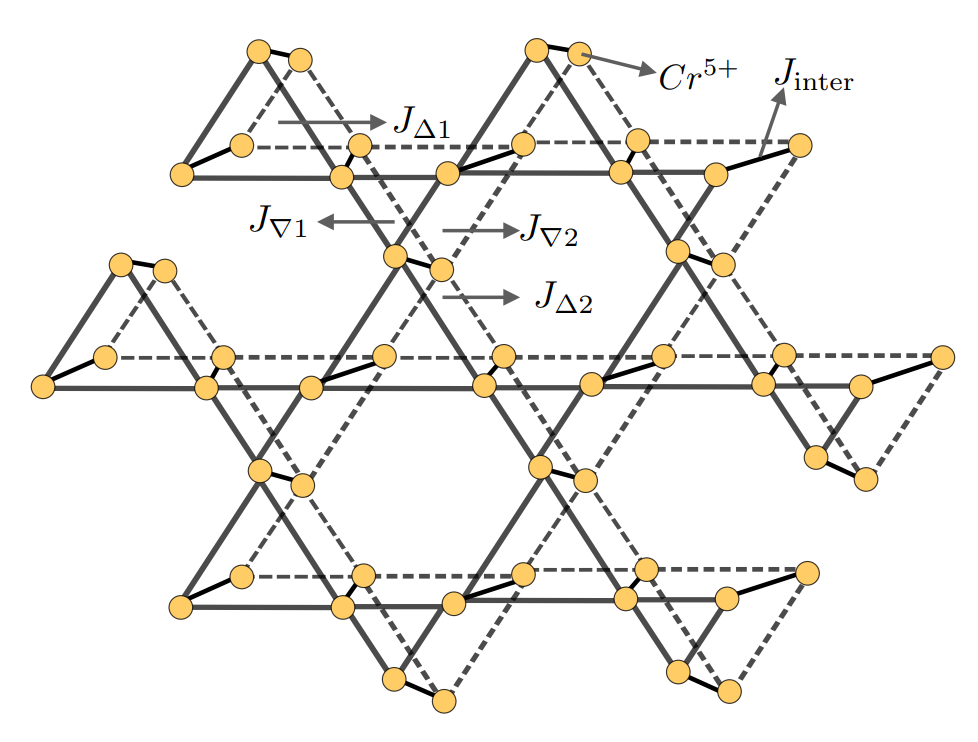}
    \caption{The crystal structure showing only the $Cr^{+5}$ ions in $Ca_{10}Cr_7O_{28}$. Five different interaction matrix elements as illustrated in Eq. \ref{Ham_spin_liq} is depicted. Reprinted from Annals of Physics, Volume 421, Augustine Kshetrimayum, Christian Balz, Bella Lake, Jens Eisert, Tensor network investigation of the double layer Kagome compound $Ca_{10}Cr_7O_{28}$, 168292, Copyright (2022), with permission from Elsevier.}  \label{spin_liq_Kagome}
\end{figure}
The Hamiltonian of this compound consists of five distinct Heisenberg type interactions and is as follows:

\begin{eqnarray}
H &=& J_{inter} \sum_{\langle i,j \rangle} \Vec{S}_i \cdot \vec{S}_j + J_{\Delta 1} \sum_{\langle i,j \rangle} \Vec{S}_i \cdot \vec{S}_j +  J_{\nabla 1} \sum_{\langle i,j \rangle} \Vec{S}_i \cdot \vec{S}_j \nonumber \\
&+& J_{\Delta 2} \sum_{\langle i,j \rangle} \Vec{S}_i \cdot \vec{S}_j + J_{\nabla 2} \sum_{\langle i,j \rangle} \Vec{S}_i \cdot \vec{S}_j
\label{Ham_spin_liq}
\end{eqnarray}

where, $\langle i, j \rangle$ corresponds to the nearest-neighbor interaction between $Cr^{+5}$ ions only in the lattice. The first term with an interaction strength of $J_{inter}$ is ferromagnetic and defines the coupling between two layers in the lattice (see Fig. \ref{spin_liq_Kagome}). The second term $J_\Delta 1$ is responsible for coupling spins within the `up'-triangles in first layer and is ferromagnetic too. The third term $J_{\nabla 1}$ is similar to the second term but for the `down' triangles and is anti-ferromagnetic. Terms with interaction matrix elements labelled by $J_{\Delta 2}$ and $J_{\nabla 2}$
are similar to the first two terms but for the second layer. The combination of ferromagnetic and anti-ferromagnetic interactions leads to spin-frustration. The magnetization curve of the material along the $z$-direction is obtained by adding the following term
\begin{eqnarray}
H_{field} = H + \sum_i g_s \mu_B B_z S_i^z 
\end{eqnarray}
where $g_s \approx 2$ and $\mu_B$ is the Bohr magneton and is equal to $5.7883818012 \times 10^{-5}$ eV per Tesla and $B_z$ is the strength of the external field along z-direction. The ground state of the above model was investigated using Projected Entangled Simplex State (PESS) algorithm in the thermodynamic limit. \cite{Biamonte2019-cp} as a function of bond dimensions. Trends in the ground state energy, heat capacity and magnetization and magnetic susceptibility indicated a gap-less spectrum of a spin-liquid.}

{\color{black}

The DMRG algorithm (see Section \ref{DMRG_sec}) has been extremely useful in finding the ground state of one-dimensional systems. Despite the success of DMRG, the method is not suitable for simulating high-dimensional system. It requires projecting the multi-dimensional state into one dimension which causes the computation time to grow many-fold. Murg et al. in their report \cite{murg2010simulating} demonstrate a general approach to simulate ground states of systems in higher dimension using the Tree Tensor Network ansatz. (See Section \ref{TTN}) By exploiting the advantages of correlation scaling of TTN (correlations in TTN only deviates polynomially from the mean-field value compared to MPS which deviates exponentially)\cite{murg2010simulating} they efficiently simulate the two-dimensional Heisenberg model and interacting spinless fermions on a square lattice. They also demonstrate its working on the simulation of the ground state of Beryllium atom. Another work by Barthel et al. \cite{barthel2009contraction} proposed an ansatz, Fermionic Operator circuits (FOC), to simulate fermionic systems in higher dimension by mapping the fermionic operators onto known Tensor Network architectures, namely, MPS, PEPS, and MERA. (See Sections \ref{MPS_sec}, \ref{PEPS}, and \ref{MERA_sec} respectively). FOC is composed of products of fermionic operators and are known to be parity symmetric. The biggest challenge in formulating a FOC is to manage the sign factors while reordering the fermionic operators. Authors propose an efficient scheme to contract FOC which computationally scale similar to the contraction of standard TN architectures. The efficiency of the scheme emerges from the fact that while contracting Jordan-Wigner string within a causal cone (specifically referring to the MERA based FOC), the strings outside it are not accounted. The scheme provides a method to formulate fermionic problems in the tensor network notation so that they can be benefited from the existing Tensor network algorithms. In another work\cite{Wille2017-dc}, authors propose a PEPS based algorithm to classify quantum phases of matter in fermionic system with a specific emphasis on topological phases. Authors introduce a Grassman number tensor network ansatz to study the exemplary Toric code model and Fermionic Twisted Quantum Double model which support topological phases. While working with fermionic problems in quantum chemistry, the choice of most suitable orbital is very crucial. Author in the report \cite{krumnow2016fermionic} propose a tool to optimize the orbitals using Tensor Network.

Thermal state or Gibbs state provide efficient description of systems in equilibrium. (See Ref. \cite{gogolin2016equilibration} for a detailed review) Simulating these states at finite temperature for higher dimensions is computationally challenging. Authors in the report \cite{kshetrimayum2019tensor} propose an algorithm based on Projected Entangled Pair States (see Section \ref{PEPS}) to compute the thermal state of two-dimensional Ising model and Bose-Hubbard model on infinite square lattice. They use a method akin to annealing, i.e. cool down the state from a high temperature to attain the desired finite-temperature state. A thermal state can be described as, $\rho = e^{-\beta H}$,
where H is the Hamiltonian of the system and $\beta$ is the inverse temperature. In the infinite temperature limit, the thermal state is described by just an Identity matrix. The evolution of state can performed by the operator $e^{-\Delta \beta H}$ which can be approximated by Suzuki-Trotter expansion. The final state's description reads as,
\begin{eqnarray}
\rho=(e^{-\Delta \beta H})^{m/2} \mathbb{I} (e^{-\Delta \beta H})^{m/2}
\end{eqnarray}
where $m \Delta \beta = \beta$ (m is the number of temperature steps). The state evolution operator is represented in the Projected Entangled Pair Operator (PEPO) notation. 

Moving a step further, there are states which do not thermalize due to many-body localization (MBL) and studying them is both interesting and difficult. In the report \cite{kshetrimayum2020time}, authors propose an algorithm based on infinite PEPS to simulate time evolution of disordered spin-1/2 XXZ Hamiltonian on a square lattice. They initialize the system in the Neel state $|\psi_0\rangle$ and update it with the time evolution operator as,
$$|\psi(t)\rangle = e^{-iHt} |\psi_0\rangle$$
They estimated the expectation value of the local particle number to determine the phases of the system for varying disorder dimension and disorder strength.

}

\subsubsection{Quantum-computing enhanced machine learning techniques}\label{QC_MBS}

The above works involved the usage of ML in its classical sense. Implementing neural networks where a quantum computer is involved to either in part (hybrid classical-quantum approach) or in entirety is suspected to achieve speed-ups that could be very beneficial to many fields of study in particular chemistry. A parameterized quantum circuit-based quantum-classical hybrid neural network was proposed by Xia et al. \cite{xia2020hybrid}. This study uses the fundamental idea that a neural network is composed of a linear part and a non-linear activation part (Theoretical framework discussed in section. \ref{DNN_section}). The linear part is now made of a quantum layer and the non-linear part is composed of measurements, which is also the classical part of the hybrid quantum-classical method. The quantum layer is constructed using parameterized quantum circuit (PQC). Having encoded the input data as quantum states, the PQC is optimized to approximate the output state, and the outputs are extracted as the expectation values by measurements. The encoding of input along with the usage of PQC and computing the outputs can be repeated several times to construct a hybrid multi-layer neural network and is trained in an unsupervised fashion for the ground state energies on a few bond lengths. The trained network is now used to predict the energies for other bond lengths. The authors show high accuracy for obtaining the ground state energies of $\rm{H_2}$, $\rm{LiH}$, and $\rm{BeH_2}$.

Xia and Kais \cite{Xia_2018} also developed a hybrid quantum machine learning algorithm, which involves a three-layered RBM approach as shown in Fig. \ref{fig:rbm_evolution} (b). The first two layers encode the amplitude field similar to Ref \cite{carleo2017solving}, while the third layer consisting of one unit is to deal with the lack of sign $(\pm 1)$ in the electronic structure problems. Now the ansatz for state-vector $\ket{\psi}$ is given by:
\begin{equation}
\ket{\Psi} = \sum_{\bf x}\sqrt{P(\bf x)}s(\bf x)\ket{x} \,,
\end{equation}
where
\begin{align}
P({\bf x}) &= \frac{\sum_{\{h\}}e^{\sum_{i}a_i\sigma^z_i + \sum_{j}b_j h_j + \sum_{ij}w_{ij}\sigma^z_i h_j}}
{\sum_{\bf x'}\sum_{\{h\}}e^{\sum_{i}a_i\sigma^{z'}_i + \sum_{j}b_j h_j + \sum_{ij}w_{ij}\sigma^{z'}_i h_j}} \label{RBM_amp_encoding}\\[11pt]
s(\bf x) &= \tanh\left[(c + \sum_{i}d_i\sigma_i) + i(e + \sum_{i}f_i\sigma_i)\right] \label{phase_encod}
\end{align}

\begin{figure}[h]
    \centering
    \includegraphics[width=0.41\textwidth]{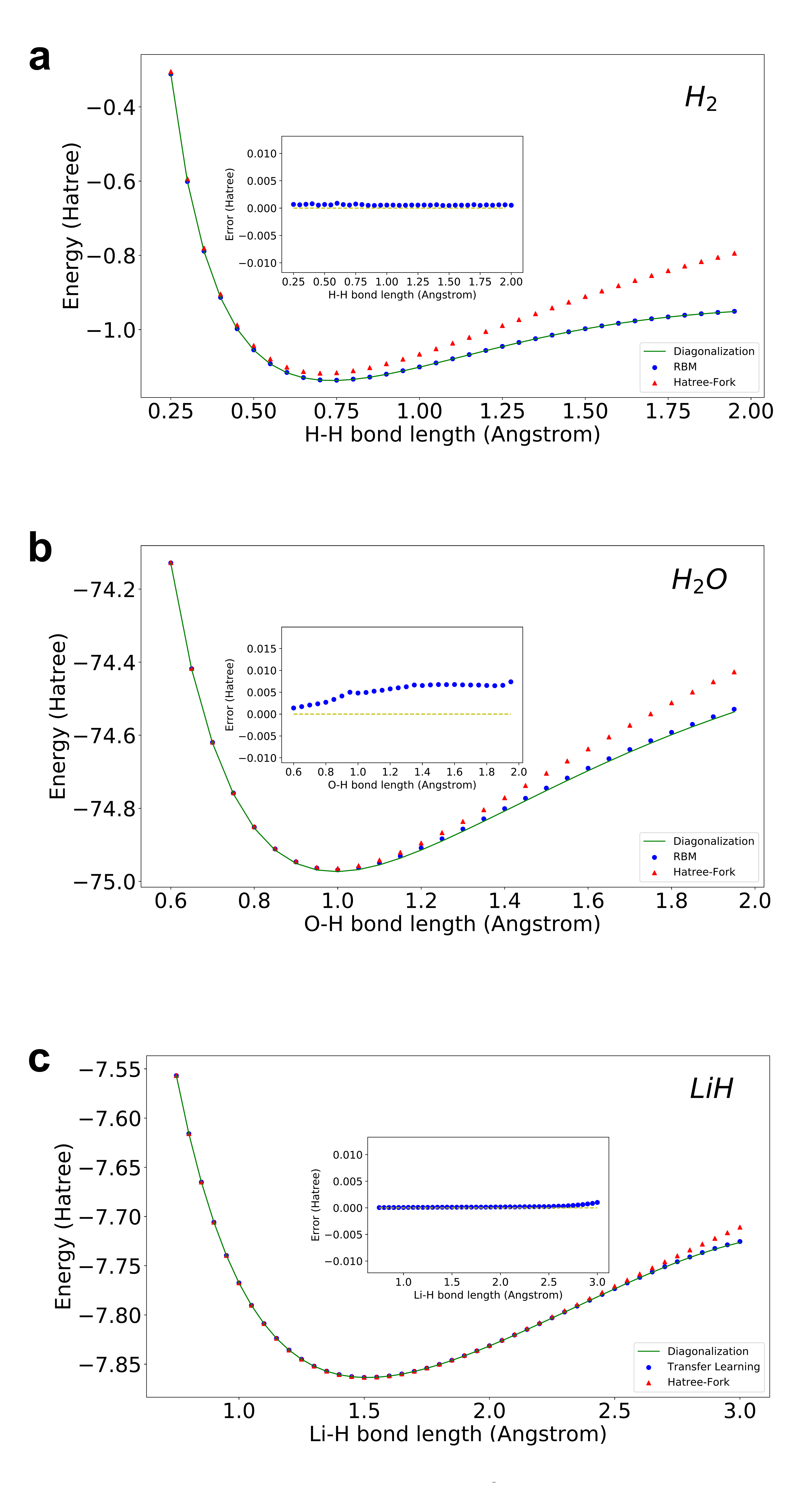}
    \caption{\textbf{a-c} Ground state energies of $\rm{H_2}$, $\rm{LiH}$, and $\rm{H_2O}$ respectively. \textbf{d} Ground state energy of $\rm{LiH}$ having used a warm starting procedure. The inner panels indicate the error with respect to exact diagonalization values. The panel is reproduced from Ref \cite{Xia_2018} with permission under Creative Commons CC BY license.} 
    \label{fig:molecular_sim}
\end{figure}

In order to simulate the distribution P({\bf x}), a quantum algorithm is proposed to sample from the Gibb's distribution. The quantum circuit mainly consists of two types of operations: 

\begin{itemize}
    \item Single-qubit \textbf{$R_y$} gates, that corresponds to a rotational operation whose angle is determined by the bias parameters $a_i$ (visible) and $b_j$(hidden) and is responsible for simulating the non-interacting of the distribution in Eq.\ref{RBM_amp_encoding}
    \item A three-qubit gates \textbf{$C1-C2-R_y$}(efficiently representable by two-qubit and single-qubit operations), that is a controlled-controlled-rotation whose angle is determined by the connection parameter $w_{ij}$ and is responsible for simulating the interaction between the visible and hidden layer of the distribution in Eq.\ref{RBM_amp_encoding}. The target qubit of each such controlled operations is an ancillary qubit which was re-initialized and re-used post-measurement.
\end{itemize}
A Boltzmann distribution for all configurations of the visible and hidden layers can be generated through the quantum circuit similar to as shown in Fig. \ref{fig:rbm_evolution} (d). This algorithm is based on sequential applications of controlled-rotation operations, which tries to calculate the interacting part of the distribution P({\bf x}) with an ancillary qubit. The ancillary qubit was thereafter measured and conditioned on the measurement results sampling from P({\bf x}) is deemed successful. With P({\bf x}) computed through the quantum circuit and $s(\bf x)$ computed classically, $\ket{\psi}$ can now be minimized by using gradient descent. Having described the method, the authors show the results corresponding to the ground states of $\rm{H_2}$, $\rm{LiH}$, and $\rm{H_2O}$ molecules (Fig. \ref{fig:molecular_sim} ). 

An extension to the work by Xia et al. \cite{Xia_2018} was proposed in the study by Kanno \cite{kanno2019manybody}, wherein an additional unit in the third layer of the RBM was introduced in order to tackle periodic systems and take into account the complex value of the wavefunction. So, the sign layer contains two units, one to encode the real part and the other to encode the complex part of the wavefunction as shown in Fig. \ref{fig:rbm_evolution} (c).
\begin{figure*}[ht!]
    \centering
   \includegraphics[width=0.77\textwidth]{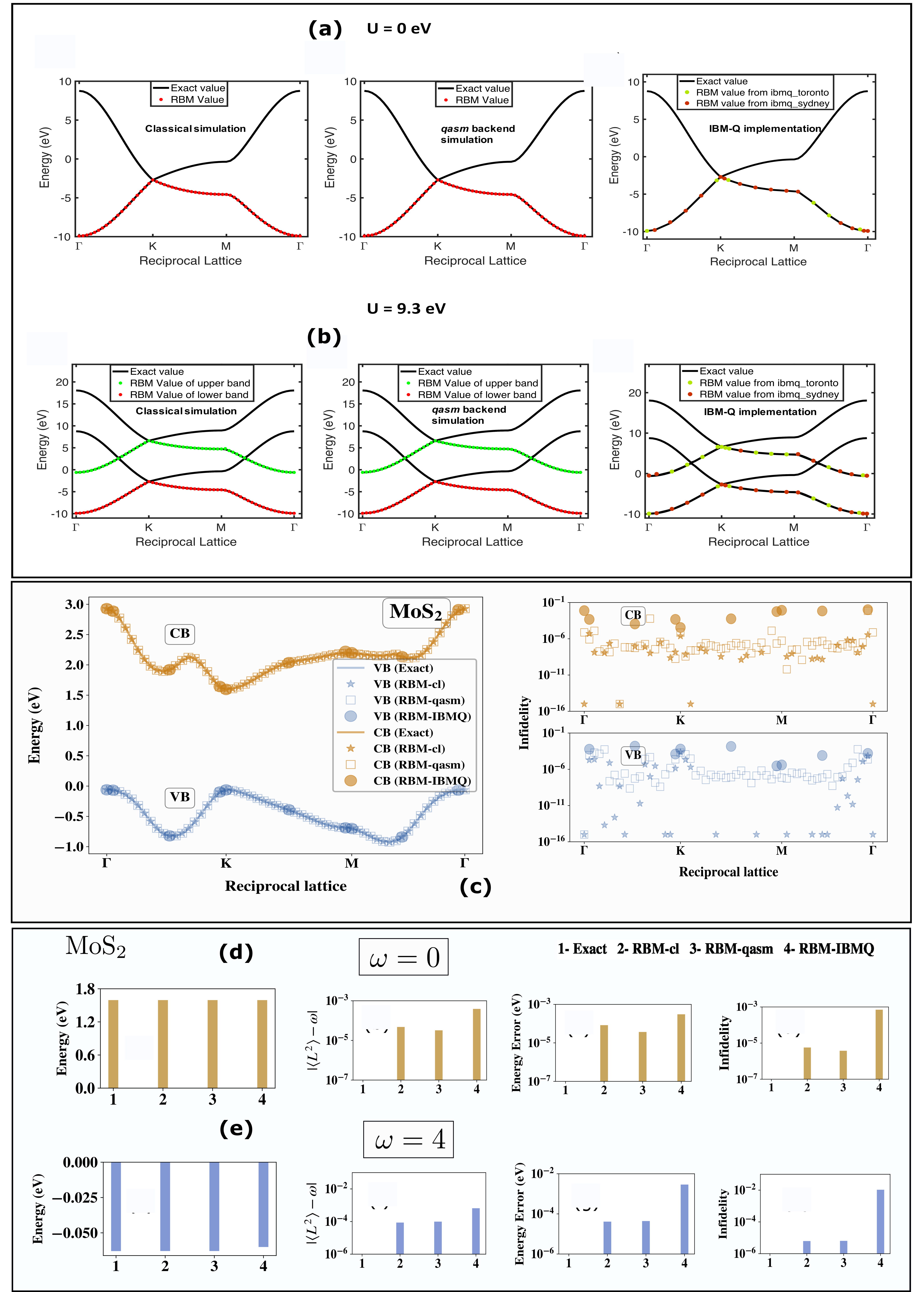}
   \caption{Band structures of 2D materials as computed from RBM implementation on classical computer, quantum simulator ($qasm$) and actual IBM-Q devices. \cite{sureshbabu2021implementation} (a) Valence band of graphene with the Hubbard $U$ interaction = 0 eV. (b) Valence band of graphene with the Hubbard $U$ interaction = 9.3 eV (Reprinted (adapted) with permission from S. H. Sureshbabu, M. Sajjan, S. Oh and S. Kais, Journal of Chemical Information and Modeling, 61, 6, 2667, 2021. Copyright 2021 American Chemical Society.) (c) Valence (VB) and Conduction band (CB) of $\rm{MoS_2}$ obtained along with the infidelity of the target state learnt by the network in each flavor of RBM implementation. CB is obtained through the constrained minimization procedure \cite{sajjan2021quantum} (d) The energy, constraint violation error, energy error and state infidelity comparison corresponding to symmetry filtering for $\rm{MoS_2}$ with the operator $(L^2)$ using the constrained minimization procedure \cite{sajjan2021quantum}. The eigenspace of this operator is labelled by $\omega=0$ (e) The energy, constraint violation error, energy error and state infidelity comparison corresponding to symmetry filtering for $\rm{MoS_2}$ with the operator $(L^2)$ and $\omega=4$ a.u. (Reprinted (adapted) with permission from M. Sajjan, S. H. Sureshbabu, and S. Kais, Journal of the American Chemical Society, 10.1021/jacs.1c06246, 2021. Copyright 2021 American Chemical Society.) } 
    \label{fig:2dmaterials}
\end{figure*}
The authors construct the global electronic structure using DFT, then an effective model with just the target band. By using the maximally localized Wannier function for basis construction for the effective model, a Hubbard Hamiltonian for the target bands is built. The choice of the material is graphene with the target band being the valence band contributed by the $2p_z$ orbitals of the two carbon atoms within an unit cell. Using the algorithm by Xia and Kais \cite{Xia_2018}, this study shows the pristine valence band of graphene in presence and absence of band splitting for a finite repulsion $U$ parameter within the Hubbard Hamiltonian \cite{kanno2019manybody}. 

In the above references \cite{Xia_2018, kanno2019manybody} efficacy of the quantum circuit was tested by simulating it on a classical computer. To benchmark the performance on an actual quantum device, repeated use of a single ancilla qubit would not be operationally convenient. A slightly modified variant of the quantum circuit with $(m\times n)$ qubits in the ancilla register has been used thereafter to act as targets of the controlled operations \cite{sureshbabu2021implementation, sajjan2021quantum} as shown in Fig. \ref{fig:rbm_evolution} (d). One must note that $m$ denotes the number of neurons in the hidden layer and $n$ denotes the number of neurons in the visible layer. Sureshbabu $et al$ in Ref \cite{sureshbabu2021implementation} uses this circuit to benchmark implementation on two 27 qubit IBM-Q devices for the valence bands (within the tight-binding framework) of hexagonal Boron Nitride (h-BN) and monolayer graphene respectively. A Hubbard Hamiltonian similar to Ref \cite{kanno2019manybody} was used to explore band-splitting as shown in Fig. \ref{fig:2dmaterials} (a-b). Excellent results were obtained even on an actual NISQ device through the use of measurement-error mitigation and repeated warm starting with well converged results for nearby $k$ points in the energy trajectory.

The classical-quantum hybrid algorithms described above focus their attention on only computing the ground states of molecules and materials. In the work by Sajjan et al. \cite{sajjan2021quantum}, the authors use the idea of constrained optimization to obtain any arbitrary energy eigenstates of molecules and materials through a user-defined choice. The authors define a quadratic minimization problem with a penalty procedure to achieve the target state \cite{sajjan2021quantum}. The procedure is capable of producing a minimum energy state 
in the orthogonal complement sub-space of a given user-defined state. The latter state can be obtained from a prior run of the same algorithm.
The authors also elucidate the protocol to systematically filter states using a symmetry operator (say $S$) of the Hamiltonian by sampling the symmetry eigenspace labelled by the eigenvalue (say $\omega$). Apart from this in the same reference Sajjan $et al$ \cite{sajjan2021quantum} also deduces a generic lower bound for the successful sampling of the quantum circuit and thoroughly discusses special limiting cases. The lower bound can be surpassed with a tunable parameter which the user can set to ensure the ancilla register collapses into the favorable state enough number of times on repeated measurements of the quantum circuit as shown in Fig. \ref{fig:rbm_evolution} (d). Only such measurements are important in constructing the distribution in Eq. \ref{RBM_amp_encoding}. The role of measurement error mitigation and warm-initialization on the algorithm, measurement statistics of the algorithm, transferability of the algorithm to related tasks, effect of hidden node density to name a few were thoroughly explored. Specific examples used were important category of 2D-materials like transition metal di-chalcogenides (TMDCs) whose entire set of frontier bands (both valence and conduction), band-splitting due to spin-orbit coupling (SOC) etc were accurately obtained even when implemented on 27-qubit IBMQ processors. Representative data for monolayer Molybdenum di-Sulfide ($\rm{MoS_2}$) for valence and conduction bands are shown in Fig. \ref{fig:2dmaterials} (c) and for symmetry filtering using squared-orbital angular momentum ($L^2$) operator in Fig. \ref{fig:2dmaterials} (d-e). Molecular examples to study the effect of multi-reference correlation was explored both in the ground and excited state. In each case the performance of the algorithm was benchmarked with metric like energy errors, infidelity of the target state trained on the neural network, constraint violation etc.

{\color{black} 
Tensor Network as described in Section \ref{Tensor_Network} has been used as an ansatz to classically simulate numerous quantum system with limited entanglement. One can map a quantum many-body states represented on tensor network to quantum circuits so that it can harness the quantum advantage. The goal is to prepare variational states using quantum circuits which are more expressive than tensor network or any other classical ansatz and also are difficult to simulate on a classical computers. In a recent report, the authors \cite{Haghshenas2022-db} use this idea to represent quantum states with variational parameters of quantum circuit defined on a few qubits instead of standard parameterized tensor used in DMRG (see Section \ref{DMRG_sec}). They show that sparsely parameterized quantum circuit tensor networks are capable of representing physical states more efficiently than the dense tensor networks. Author refer to the standard Tensor Networks with all variable elements as Dense with a prefix 'D' while the quantum circuit Tensor Networks are referred with their names prefixed with 'q'. In theory, the number of gates required to exactly recover a q-qubit unitary grows exponentially with q. The bond
dimensions (D) between local tensors of tensors networks are encoded into quantum circuit using an unitary operator defined on q qubits where $q=log_2(D)$. Contrary to this, the authors claim that only a  polynomial number of variational parameters of two qubit unitaries and isometries are sufficient to express the quantum state. Authors work with three types of local circuits to represent the unitary:
\begin{itemize}
    \item Brick-wall circuit: It has a layered structure with successive layers fully connected via two-body unitary gates. In [Fig \ref{fig: TN_structures_appln} (a)] a brick-wall circuit is shown with depth $\tau=6$. The effective correlation length is proportional to the depth, hence the correlations in brick-wall circuit are known to spread slowly with increasing depth.
    \item Ladder circuit: It also has a layered structure with denser connections. The first and the last qubits are entangled to each other in the first layer itself. This structure allows efficient propagation of correlations. [Fig \ref{fig: TN_structures_appln} (b)] shows a ladder circuit with depth $\tau=4$. 
    \item MERA circuit: Its architecture is inspired from the MERA Tensor Networks (Described in Section \ref{MERA}). It has isometric tensors and unitary disentanglers arranged in alternate layers. [Fig \ref{fig: TN_structures_appln} (c)] shows a MERA circuit with depth $\tau=5$. 
\end{itemize}

\begin{figure*}
    \centering
    \includegraphics[width=17cm]{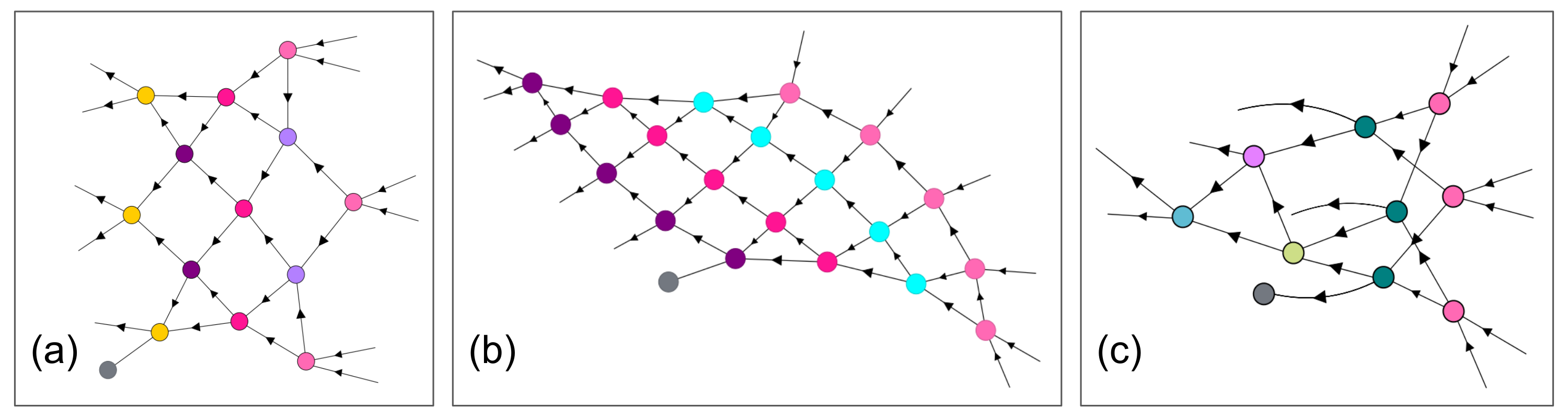}
    \caption{A schematic diagram showing architectures of different local circuits. (a) A brick-wall circuit, (b) Ladder circuit, and (c) MERA circuit. Each rank-4 tensor is represented by a circle denoting two-qubit unitary gates. The different colours represent different layers of the circuit. The arrows in the circuits show the direction of canonicalization. Depths of the circuits are six, four, and five respectively. Reproduced from Ref \cite{Haghshenas2022-db} under Creative Commons Attribution 4.0 International license.}
    \label{fig: TN_structures_appln}
\end{figure*}

Two different paradigmatic Hamiltonians - Heisenberg  and Fermi-Hubbard model has been used to test the efficacy of the technique using both local DMRG like optimization and global gradient based optimization.
The minimization of the energy ($E=\langle \psi| \hat{H} |\psi\rangle $) is performed using conjugate gradient \cite{shewchuk1994introduction} and LBFGS-B \cite{schraudolph2007stochastic}. 

Empirically the relative error $\delta E$ in the ground state energy was found to be inversely proportional to the polynomial of number of variational parameters $n$.
\begin{eqnarray}
\delta E(n) \sim an^{-b}
\end{eqnarray}
The results obtained by implementing different local quantum circuit tensor networks are shown in Fig. \ref{Fig: QCTN_Res}. Fitting the Eq. \eqref{119} on the data generates set of $(a,b)$ parameters which are summarized in Table \ref{tab:ab_vals}. 

\begin{center}
\begin{table}
\begin{tabular}{ |c|c|c| } 
 \hline
 Ansatz & Heisenberg (a, b) & Hubbard (a, b) \\ \hline
 qMPS-b & (20, 4) & (9, 1.9) \\
 qMPS-l & (14, 3.1) & (10, 1.9) \\
 qMERA-b & (15, 3.1) & (6.0, 1.4) \\
 dMPS (DMRG) & (15, 2.9) & (8.0, 1.5) \\
 dMERA & (3.5, 1.2) & (0.8, 0.6) \\
 \hline
\end{tabular}
\caption{\label{tab:ab_vals} The (a,b) values obtained numerically from the various ansatz employed in Ref \cite{Haghshenas2022-db}}
\end{table}
\end{center}

The parameter b gives the asymptotic behaviour of accuracy of the circuit depending on the number of parameters. A higher b indicates that for same number of parameters, the model is approximating the ground state better. Hence, it is quite evident that ansatz based on quantum circuit tensor networks are more expressive compared to the standard classical variants studied in the report as the former yields comparable accuracy to the latter with lower parameter cost. This clearly explicates the need for simulating tensor networks on a quantum-hardware to realize its optimum potential for understanding many-body quantum systems.  

\begin{figure*}
    \centering
    \includegraphics[width=17cm]{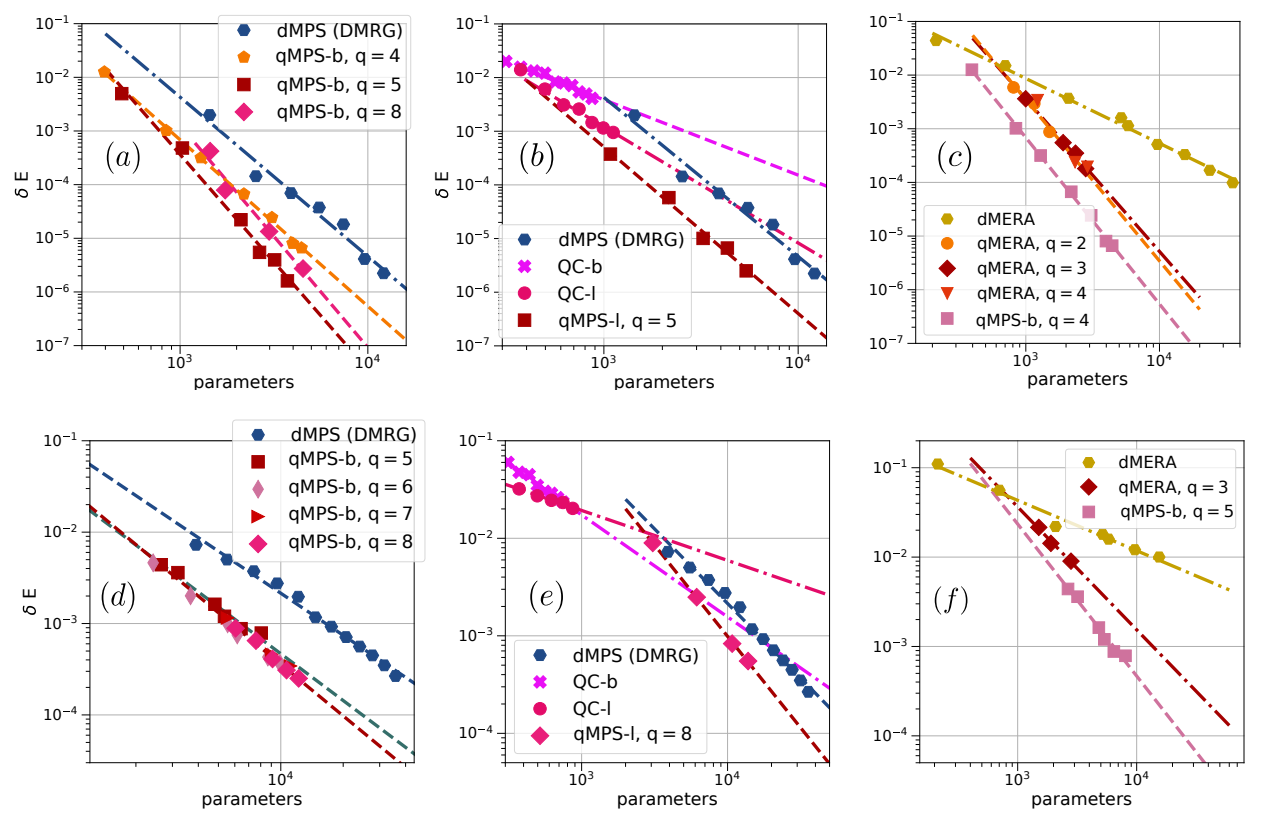}
    \caption{A comparison of expressibility of Quantum circuit tensor networks (qMPS, qMERA), standard tensor networks (dMPS and dMERA), and global quantum circuits (QC). The expressibility (or variational power) is measured by seeing the relation between relative energy error $\delta E$ versus the number of variational parameters in the ansatz. Figures (a,b,c) are for Heisenberg model while (d,e,f) are for Fermi-Hubbard model of lattice size 32. The post-fixes 'b' and 'l' denote the brick-wall and ladder circuit. (a, b, d, e) Show the comparison between brick-wall and ladder qMPS acting on different number of qubits (q), QC and dMPS. While (c and f) depict comparison between qMERA-b with varying q, qMPS-b, and dMERA. Reproduced from Ref \cite{Haghshenas2022-db} under Creative Commons Attribution 4.0 International license.}
    \label{Fig: QCTN_Results}
\end{figure*}

Being limited by the number of noiseless qubits available on any quantum hardware in the NISQ era, a recent report \cite{eddins2022doubling} have illustrated how to decompose a $2N$-qubit circuit into multiple $N$-qubit circuits. These $N$-qubit circuits can be run on NISQ devices while their results can be post-processed on a classical computer. In this formulation, the state $|\psi\rangle$ on $2N$ qubit system can be partitioned into smaller states defined on N qubits using Schmidt decomposition (Similar to MPS defined in Section \ref{MPS_sec}).
\begin{eqnarray}
|\psi\rangle = (U\otimes V) \sum_{n=1}^{2^N} \lambda_n |b_n \rangle \otimes |b_n\rangle
\end{eqnarray}
$|b_n\rangle$ are the N-qubits states in the computational basis, U and V are unitary operators acting on the two subsystems that transforms the computational basis to the desired state. The $\lambda_n$'s are the Schmidt coefficients which determine the degree of correlation present within the system. 
Using the above state, the expectation of a 2N-qubit operator defined as $O=O_1 \otimes O_2$ can be written as

\begin{eqnarray}
\langle O \rangle &=& \sum_{n=1}^{2^N} \Bigg(\lambda_n^2 \langle b_n| \Tilde{O}_1 |b_n\rangle \langle b_n| \Tilde{O}_2 |b_n\rangle  + \sum_{m=1}^{n-1} \lambda_n \lambda_m \nonumber \\
&\times& \sum_{p \in \mathbb{Z}_4} (-1)^p  \langle \phi_{b_n b_m}^p| \Tilde{O}_1 |\phi_{b_n b_m}^p\rangle \langle \phi_{b_n b_m}^p| \Tilde{O}_2 |\phi_{b_n b_m}^p\rangle \Bigg)
\end{eqnarray}
where $\Tilde{O}_1 = U^\dagger O_1 U$ and $\Tilde{O}_2 = V^\dagger O_2 V$, and $|\phi_{b_n b_m}^p\rangle = (|b_n\rangle + i^p |b_m\rangle)/\sqrt{2}$ with $p \in \mathbb{Z}_4$. 

Authors use this approach to simulate ground state of water molecule using VQE simulation. They use the frozen core approximation and to enforce spin symmetry set $U=V$. This yields 
ten spin orbitals of the water molecule in $STO-6G$ basis set which using the aforesaid scheme can be encoded into five qubits on the quantum processor. The formalism yields excellent values of energies for geometries which are distorted through stretching of the $O-H$ bond in $H_2O$ molecule by using just 3 Schmidt coefficients even though the results degrade from the exact value for deformations of the $H-O-H$ bond angle.

Another algorithm has been proposed based on tensor networks to solve for any eigenvector of a unitary matrix Q, given a black-box access to the matrix \cite{}. When working with a given Hamiltonian for a physical system, the black-box unitary can be prepared by applying the unitaries of the problem Hamiltonian and the mixing Hamiltonian alternatively for different times. In that case Q will be characterized by time-parameters. A parameterized unitary $U(\vec{\theta})$ is used to define the state ansatz.
The loss function for estimating the ground state of Q is simply maximizing the probability of projecting each qubit to $|0\rangle$. If $k$ denotes a k-ebit matrix product state (MPS) then the value  k value is iteratively increased until the desired accuracy is achieved. k ranges from 1 to $\lceil n/2 \rceil$ because the maximum entanglement produced by a n-qubit quantum circuit of depth m is $min\{ \lceil n/2 \rceil, m\}$ ebits.
The implementation complexity of the algorithm (measured in terms of number of CNOT gates used) scales as $O(l\cdot n \cdot r^2$ where n is the number of blocks and r is the rank of tensor network and $l$ is the number of steps required to terminate the optimization routine. This algorithm has a significant advantage over other Variational techniques in that it terminates when the state reaches the eigenstate of system.
}

\subsection{Estimation and parameterization of force fields in molecular dynamics} \label{FF_sec}

\subsubsection{Machine learning techniques on a classical processor}
The use of molecular dynamics (MD) simulations to unravel the motion of molecules dates back to 1977 by McCammon et. al.\cite{mccammon1977dynamics}. Their study simulated dynamics of folded protein for $\approx$ 10 ps which unraveled the fluid-like nature of the interior region of protein for the first time. Since then the field of molecular dynamics has seen various incarnations from increasing the simulations system size to million-atoms \cite{schulz2009scaling} to simulating it for a longer time scale using parallel computing \cite{shaw2008anton}. While performing MD simulations Newton's laws of motions are numerically integrated at each time step. Which requires a set of initial conditions (position and velocity of each particle) and a comprehensive understanding of atomic forces acting on the system. The best way to obtain these forces is via first-principles i.e. solving the Schr\"{o}dinger equation for a particular configuration of the nuclei. Unfortunately getting an exact analytical solution for the Schr\"{o}dinger equation (SE) is a herculean task for most of the chemical species. Thus some levels of approximations are considered while solving the exact SE. In this approach, the size of the problem increases exponentially as a function of degrees of freedom (DOFs) and thus increasing the computation cost. For example advance $ab$ $initio$ electronic structure calculations, such as coupled cluster singles-doubles (CCSD), it's perturbative triples variant CCSD(T) scales as $\mathcal{O}({n}^{7})$ where $n$ is the number of basis functions used. 

 Thus $ab$ $initio$ calculations are known for their precision but they are computationally expensive restricting their application to smaller systems in the gas phase or solid-state periodic materials.
 To model larger systems we have to use a higher level of approximations and use empirical force field (FF) methods.  Their computational efficiency allows the simulation of systems containing millions of atoms \cite{schulz2009scaling,mackerell2007empirical} and explore much longer simulation time scales $(100 ms)$ \cite{zimmerman2021sars}. A FF is an analytical expression that denotes interatomic potential energy as a function of the coordinates of the system (for a particular configuration) and set of parameters. Depending on the intricacy of the underlying system different FFs are employed and today's scientific literature provides a myriad choices. But a standard expression for a FF resembles like
 
 \begin{flalign}
           &U = \sum_{bonds}^{} \frac{1}{2} k_{bo} (r-r_{eq})^{2} + \sum_{angles}^{} \frac{1}{2} k_{an} (\theta-\theta_{eq})^{2} + \sum_{dihedral}^{} V_{dih} + \sum_{improper}^{} V_{imp}&&\\\nonumber 
          &\hspace{0.6cm}+ \sum_{LJ}^{} 4 \epsilon_{ij} \left( \frac{\sigma^{12}_{ij}}{r^{12}_{ij}}-\frac{\sigma^{6}_{ij}}{r^{6}_{ij}} \right) + \sum_{ij}^{} \frac{q_{i}q_{j}}{r_{ij}}
 &&    
 \end{flalign}
 
The first four terms are intramolecular contributions (the first term corresponds to bond stretching followed by bending, dihedral rotation, and improper torsion). The last two terms correspond to Van der Waals (12-6 Lennard-Jones potential) and coulombic interactions. The parameters in FF ($ k_{bo}, r_{eq}, k_{an}$ and $\theta_{eq}$) are usually obtained via $ab$ $initio$ or semi-empirical calculations or by fitting to experimental data such as X-ray and electron diffraction, NMR, infrared, Raman spectroscopy. A general review of the force fields for molecular dynamics can be found in Refs. \cite{gonzalez2011force,mackerell2004empirical}. The final aim of FF is to express all the quantum mechanical information in classical terms, splitting up the total electronic energy of the system into well-divided atom-atom contributions (coulombic, dispersion, etc). Although it's an arduous task to split up the total electronic energy even after using precise quantum mechanical calculations. Thus while determining inter-molecular forces we need to consider crucial physical approximations which limit the accuracy. So depending on the system one chooses a certain level of approximation and uses the input data to optimize parameters which makes this approach empirical. Basic steps to form a new Force field involve accessing the system to select a functional form for the system's energy. After that, we need to choose the data set which determines the parameters in the function defined earlier. Earlier people used to use experimental data from x-ray or neutron diffraction (for equilibrium bond lengths) and different spectroscopic techniques (for force constants). But in most of the cases, the experimental data used to be insufficient or inaccurate, thus nowadays $ab$ $initio$ data is preferred. Next, we optimize these parameters, in general, there exists colinearity between them i.e. these parameters are coupled (changing one would change another) so the optimization is done iteratively. The last step involves validation where we calculate different properties of the system which are not involved in the parametrization. 

Thus the underlying assumptions behind a FF eventually limit the accuracy of the physical insights gained from them. Since conventional FFs do not explicitly model multi-body interactions, polarizations and bond breakings during a chemical reaction make their predictions highly inaccurate. Although there are specially developed FFs (AMOEBA, ReaxFF, RMDFF, ARMD) \cite{shi2013polarizable,warshel2007polarizable,unke2017minimal,nagy2014multisurface} that include these effects at a certain computational cost, in most of these cases there exists ambiguity regarding the necessity of inclusion of these effects. Mixed Quantum Mechanics/Molecular Mechanics (QM/MM)\cite{senn:2009} becomes a handy tool while dealing with huge systems (bio-molecules).   As the name suggests it employees Quantum Mechanical treatment for a subset of the problem ("reactive region") and the rest of the problem ("environment") is being treated classically. This way QM/MM approach includes certain quantum correlations in bigger systems improving its accuracy (compared to FF). Although it may seem QM/MM approach is the most optimal way to simulate huge problems but one needs to consider huge "reactive region" to get converged results which eventually increases the computational cost. 

Machine Learning Force Fields (ML FFs) combines the accuracy of $ab$ $initio$ methods with the efficiency of classical FFs  and resolves the accuracy/efficiency dilemma. ML approaches evade solving equations and rely on recognizing the underlying pattern in the data and learning the functional correspondence between the input and the output data with unparalleled accuracy and efficiency\cite{hansen2013assessment,chmiela2019sgdml}. Unlike conventional FFs ML FFs does not require predetermined ideas about the bonding pattern assumptions. This distinctive feature makes ML approaches admirable in the chemical space and there are huge number of options ML models are available or MLFFs ranging from PhysNEt\cite{unke2019physnet}, sGDML\cite{chmiela2019sgdml}, GAP\cite{bartok2010gaussian}, SchNet\cite{schutt2018schnetpack}, HDNN\cite{behler2007generalized} and ANI\cite{smith2017ani}. However, it's not trivial to extend the classical ML formalism to generate FFs. The reason being exacting standards are imposed on the ML FFs which offers an alternative to the already established benchmark FFs. Additionally, classical ML techniques (Natural Language Processing (NLP), Computer Vision, etc) assumes huge reference data sets while optimizing thousands of training parameters, whereas it is very difficult to get such extensive data sets in case of natural sciences. Since each data set is generated either from an expensive $ab$ $initio$ calculation or from rigorous experimental measurement. Thus data efficiency becomes a key factor while developing ML FFs. Which is resolved by encoding the physical knowledge or laws directly into the architecture of the ML models\cite{chmiela2019sgdml,schutt2018schnetpack}. \textcolor{black}{Ref. \cite{zhang2018deep,westermayr2020machine,schutt2017quantum} discuss construction of potential energy surfaces using different supervised machine learning techniques for complex chemical systems and electronically excited molecular states}. Recent review by Unke et. al. \cite{unke2021machine} presents an in depth overview of ML FFs along with the step-by-step guideline for constructing and testing them. The rest of the section focuses on the application of ML to optimize the FF parameters in ReaxFF\cite{van2001reaxff} which eventually leads to more accurate physical insights. We will show specific example \cite{nakata2019development} where ML is used for parameter optimization. 

Refinement of parameters is essential while employing ReaxFF MD for chemical reactions. To get insights about static properties (energy, force or charge distribution) fitting of FF parameters is done by using specific Quantum Mechanical (QM) training data. Genetic algorithm\cite{angibaud2011parameter,dittner2015efficient} and its multi-objective variant\cite{jaramillo2014general} have been developed to ease out the parameter optimization using QM data. However application of ReaxFF for dynamic non-equilibrium chemical reactions (Chemical Vapour Deposition) is not straightforward, as it is unfeasible to gain QM training data set for fitting. In addition, the dynamical properties we would like to predict decides the computational cost and resources needed for parameter fitting. In such situations, ML-based approaches comes to our rescue. Hiroya and Shandan \cite{nakata2019development} recognized the flexibility that ML-based models offer and used it to efficiently optimize the parameters in ReaxFF. Their approach uses $k$ nearest neighbor algorithm to get several local minima and then optimization is done using ML. The most distinctive feature in this approach is that it can combine efficiency from ML-based optimization with accuracy from other optimization techniques (Monte Carlo/genetic) making itself a powerful tool. 

The first step in parameter optimization is creating a reference parameter data set ($P_{be1}$,$P_{be2}$,$\cdots$,$D_{ij}$,$R_{vdw}$,$\cdots$,$P_{val2}$,$\cdots$) for the given ReaxFF. The ReaxFF parameters encode the information about the system's potential energy's different terms (bonding, lone pairs, van der Waals etc). This reference data set is then used for random generation of $N$ ($\approx100$) different data sets. While generating random samples from the reference data two different strategies are implemented in the first strategy a random change is made in the ReaxFF parameter set which is followed by the second strategy where we exchange parameters between different ReaxFF data sets. During the random sampling process, it is important to find the sweet spot where the sampling process should not result in a huge deviation making the resultant random sample unphysical at the same time the resultant random sample should not be too close to the reference data (the model will be stuck at the local minima).

\begin{figure*}[ht!]
    \centering
   \includegraphics[width=0.99\textwidth]{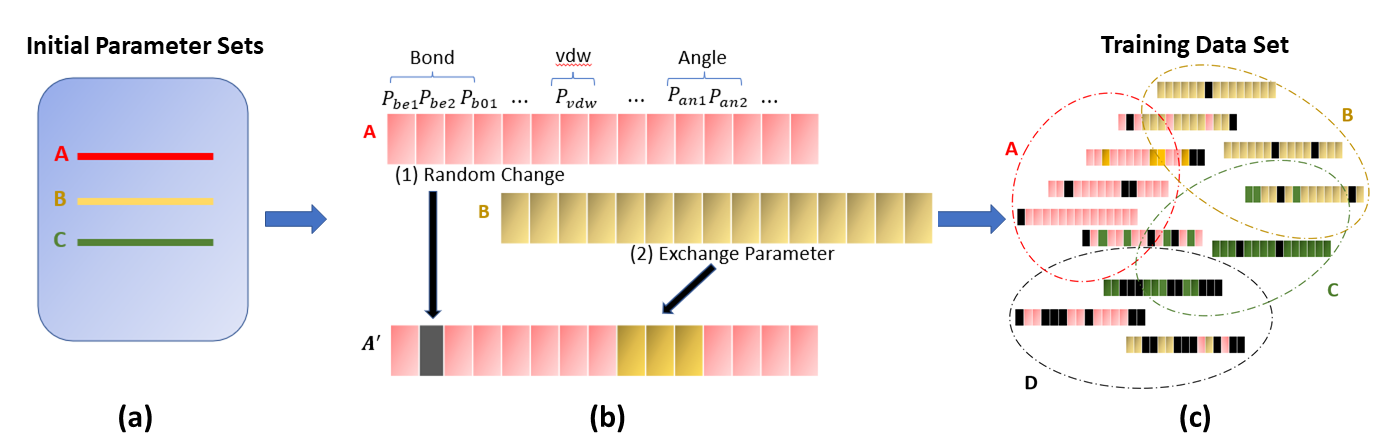}
   \caption{Schematic representation of Random sampling (a) Initial parameter references (A, B, and C inside denotes the initial parameter set to generate the reference ReaxFF parameter set) (b) Definition of parameter set and explanation about random modification scheme (c) Entire image of the sampling space using the random modification scheme. The open dashed circles A, B, and C on the right side denote the sampling space in the vicinity of the initial ReaxFF parameter sets A, B, and C. The open dashed circle D denotes a new sampling space, which is different from A, B, and C. Figure adapted with permission from Ref \cite{nakata2019development}.} 
    \label{random_samp_FF}
\end{figure*}

The next step involves score assessment where a score $S(p_{i})^{ReaxFF}$ is calculated for each training reference data set $p_{i}$ is calculated. 

\begin{equation}
    S(p_{i})^{ReaxFF} = \sum_{j}^{N_{QMtype}}\frac{w_{j} S_{j}(p_{i})}{N_{QMtype}}
    \label{score1}
\end{equation}
 
\begin{equation}
    S_{j}(p_{i})^{ReaxFF} = \sqrt{\sum_{k}^{N_{j}^{QMtype}}\frac{\left(Q_{k,j}^{ReaxFF}(p_{i})-Q_{k,j}^{QM}\right)^{2}}{N_{j}^{QMtype}}}
    \label{score2}
\end{equation}
 
Here $N_{QMtype}$ is the number of geometry sets and $N_{j}^{QMtype}$ corresponds to number of different structures in the $j^{th}$ geometry set. These structures constitute a specific physical property (Potential energy change as a function of volume change in $\alpha-Al_{2}O_{3}$ crystal). 

After evaluating the score of every training data ML is used for data analysis. There are three major steps (1) Use Random forest regression to extract important features (2) update initial parameters by k-nearest neighbor (k-NN) algorithm (3) use grid search to get the optimal parameters. In the first step, random forest regression is employed. Where the objective function for minimization can be written in the form of difference between the actual score calculated via ReaxFF and the estimated score via ML: 

\begin{equation}
    v_{rmse} = \sqrt{\sum_{i}^{N}\frac{S_(p_{i})^{ReaxFF}-S_(p_{i})^{ML}}{N}}
    \label{rmse}
\end{equation}

\begin{equation}
    v_{rmse}^{j} = \sqrt{\sum_{i}^{N}\frac{S_{j}(p_{i})^{ReaxFF}-S_{j}(p_{i})^{ML}}{N}}
    \label{jrmse}
\end{equation}

Here $N$ is the number of random samples created and $j$ denotes each structure group. The ML model is build by minimizing Eq. \ref{rmse} and Eq. \ref{jrmse}. At the end of ML training important features are extracted which will be used further during grid search parameter optimization. The second step consists of applying the k-NN algorithm to divide the training data into k different groups on the basis of closest distance. For each group scores (Eq. \ref{score2}) are calculated and the parameter set corresponding to the minimum score is selected eventually modifying the initial parameter set. The modified initial parameter set after the k-NN algorithm is not fully optimized the reason being in a way it is being chosen from the initial data set. Thus grid search parameter optimization is used in the third step which is combined with the extracted parameters from the trained ML model.    
\begin{figure*}[ht!]
    \centering
   \includegraphics[width=0.77\textwidth]{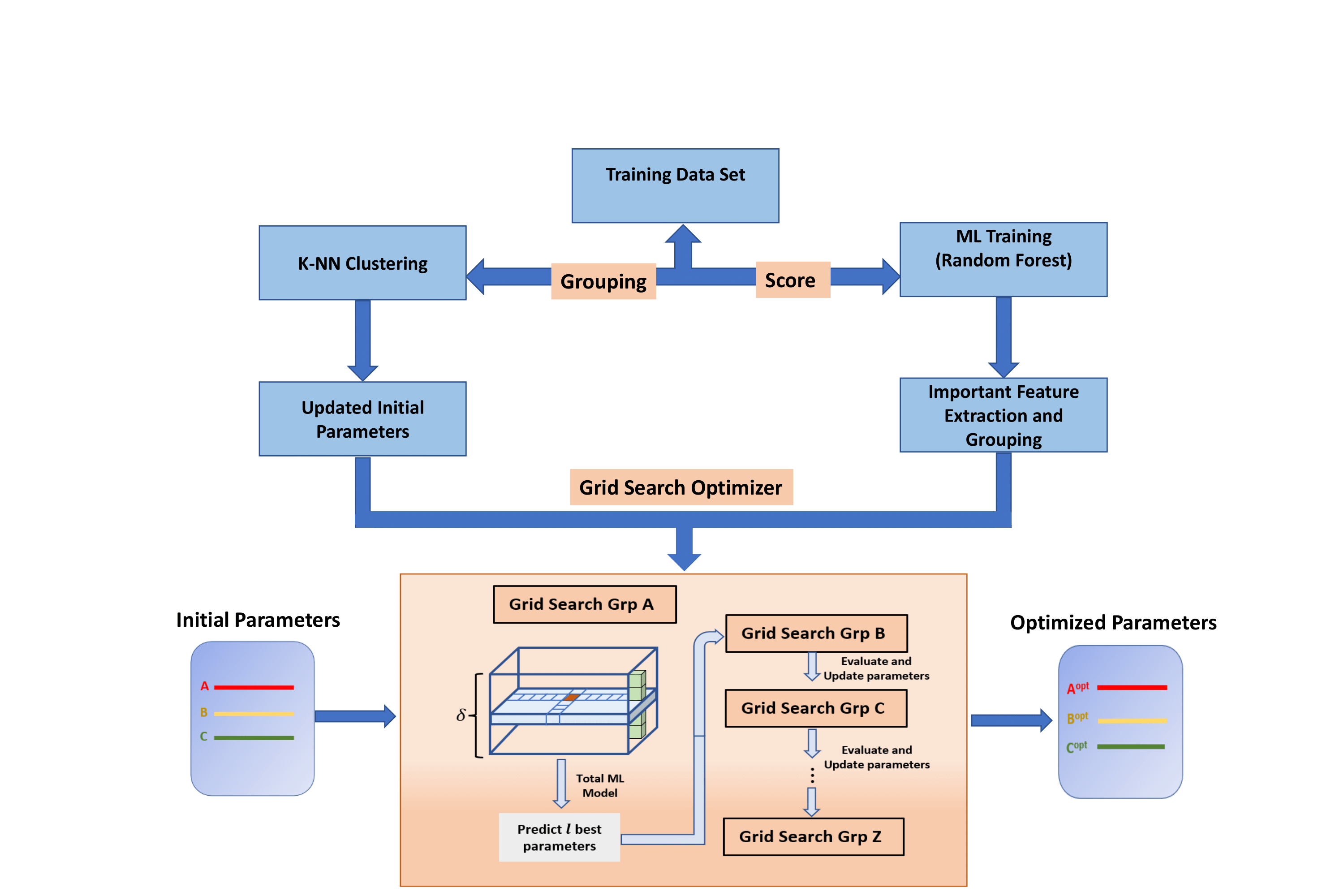}
   \caption{Illustration showing entire Machine learning scheme. Figure adapted from Ref \cite{nakata2019development}.} 
    \label{para_opt_FF}
\end{figure*}

Grid search parameter optimization (GSPO) is defined by three parameters $n$, $l$ and $m$. Every geometry set (containing $N_{QMtype}$ various physical properties) has a corresponding ML model with different variations of important features. Depending on the importance of particular feature groups are formed containing $n$ elements. In Fig. \ref{para_opt_FF}(b) we take $n=4$, the parameters corresponding to model A are sorted according to their importance (sensitivity). Since $n=4$ first four important parameters are clubbed together forming group A the next set of 4 will be named group B so on so forth. This process is repeated with all the ML models. Since each ML model has varying feature importance levels the groups will contain different parameters. GSPO is then carried out on individual groups by splitting it into m grid points. As each group contains n parameters number the total number of grid points becomes $m^{n}$. Getting scores Eq.\ref{score1} for every grid point for comparison is a laborious task. ML alleviates this issue by selecting $l$ test parameter sets from the haystack of $m^{n}$ different parameter sets. At every group, the scores of $l$ test parameters are compared with the Initial sets of parameters obtained from k-NN. If the score of a particular parameter set (in $l$) is less than the score of the Initial sets of parameters the parameters are updated. However, it's still a time-consuming task to repeat the process for all the groups ($A, B,\cdots, Z$). Since the score does not change much when the group is not sensitive enough. A new cutoff term $n_{layer}$ is introduced which determines when to stop computing score for a particular group and move on to the next. For instance, if $n_{layer}=2$ GSPO is carried out on group $A$ (most important) and group $B$ (nest most important). After this GSPO moves on to ML model B.

To summarize while optimizing the parameters of ReaxFF broadly there are three major steps involved (1) Use random forest regression to construct ML models and extract important features (2) Locating probable local minima using k-Nearest Neighbour algorithm (3) Use grid search optimization to optimize the parameters using information from developed ML models. It is important to note that different optimized parameters sets will predict the potential energy decently (in agreement with QM), although their estimation for different physical properties will diverge. The reason is an unavoidable uncertainty is carried out corresponding to a particular physical property during the parameter estimation. Hiroya and Shandan \cite{nakata2019development} used this approach on a pilot test in which the optimized parameters of ReaxFF were further used to simulate Chemical Vapour Deposition (CVD) of an $\alpha-Al_{2}O_{3}$ crystal. The optimized parameters decently predicted crystal structure of $\alpha-Al_{2}O_{3}$ even at high temperatures (2000K). Stochastic behavior or the random forest algorithm used in different ML approaches results in hundreds of error evaluations during training for complex training jobs. Where each error evaluation involves minimizing the energy for many molecules in the training data. Recently Mehmet et. al.\cite{kaymak2021jax} proposed a novel method that employs an automatic differentiation technique from the JAX library developed by Google to calculate the gradients of the loss function. It's impressive that the efficiency of the gradient-based local optimizer is independent of the initial approximation made for ReaxFF. Given reasonable computing resources, ML-assisted parameter optimization techniques are powerful tools that can be used to simulate wide spectra of reactive MD simulations. 

\textcolor{black}{ In Non-Adiabatic (NA) Molecular Dynamics (MD) simulations Born-Oppenheimer (BO) approximation breaks down and the electronic and nuclear degrees of freedom can't be treated independently. These simulations perform a pivotal role while understanding the dynamics of excited states. Recently, supervised machine learning techniques have been employed which interpolates NA Hamiltonian along the classical path approximated NA-MD trajectories\cite{akimov2013pyxaid,akimov2014advanced,nijjar2019ehrenfest} eventually speeding up NA MD simulations\cite{dral2018nonadiabatic,wang2021interpolating,hu2018inclusion,zhang2021doping}. NA MD simulations act as a robust resource especially while predicting macroscopic observables such as quantum yield without knowing the mechanism for the larger systems involving strong couplings in which it's difficult to choose a reaction coordinate\cite{li2021ab,zhang2021dynamics,olson2021band}. But they come with a cost of expensive $ab$ $initio$ calculations of geometry-dependent forces and energies or the different states. In such situations, ML techniques come to the rescue by predicting band gaps and NA coupling using a small fragment of $ab$ $initio$ training data\cite{wang2021interpolating,hu2018inclusion,zhang2021doping,westermayr2020combining,posenitskiy2021application}. To predict the physical properties of materials, \textcolor{black}{unsupervised ML techniques\cite{glielmo2021unsupervised,virshup2012nonlinear,li2017analysis,peng2021analysis}} has been employed on the trajectories of NA MD simulations while explaining the dominant structural factors \cite{zhou2020structural,tavadze2018machine,mangan2021dependence}. Many times Mutual Information (MI) is used to identify unanticipated correlations between many important features. Results from MI are easier to interpret and it is supported by information-theoretic bound which makes it not sensitive to the size of the data set\cite{mangan2021dependence,kraskov2004estimating}. These properties make it popular for its application in the chemical regime. For instance, a model of metal halide perovskites (MHPs) based on unsupervised MI unveiled the importance of geometric features as compared to the atomic velocities while predicting Non-Adiabatic coupling (NAC)\cite{zhou2020structural}.}

\textcolor{black}{In a recent study by How et al.\cite{how2021significance}, supervised and unsupervised ML techniques have been used for feature selection, prediction of Non-Adiabatic couplings, and excitation energies of NA MD simulations of $CsPbI_{3}$ metal halide perovskites (MHPs). MHPs have high optical absorption, low cost of manufacturing and long carrier diffusion \cite{xiang2021review,green2014emergence,ahn2015highly} which makes them an ideal candidate for their use in optoelectronics and solar energy harvesting materials. In order to improve the design of MHPs it is important  to develop a computationally efficient and a systematic NA MD which utilizes the theory as well as simulations to predict the physical properties of MHPs. How et al.\cite{how2021significance} fills this knowledge gap by employing MI on the NA MD trajectory data set of $CsPbI_{3}$ perovskite and extracting the most important features that determine the NA Hamiltonian. The ML model is then validated by checking the performance of the extracted important features to predict the band gap and NAC. Their model showed surprising and counterintuitive results suggesting that the NA Hamiltonian can be predicted by using a single most important feature of the chemical environment information of any of the three elements in $CsPbI_{3}$. This eventually leads to a drastic reduction in the dimensionality of the original 360-feature ML model developed from ML force fields to merely 12 featured ML models which can produce high-quality NA-MD simulation results which are further confirmed by the present theoretical knowledge about the electronic properties of $CsPbI_{3}$. This dimensionality reduction technique helps alleviating the high
computational cost of ab initio NA-MD and to extends NA-MD
simulations to larger, more complex systems and longer time
scales.}

\subsubsection{Quantum-computing enhanced machine learning techniques}
\textcolor{black}{Previously variational quantum algorithms (VQAs) have been applied for simulation of small systems \cite{bian2019quantum} with strongly bounded intramolecular forces \cite{kandala2017hardware}. Most of these approaches relies on complete electronic basis set of the hamiltionian. Smaller coherence times needed in VQA's makes them ideal fit for the current generation Noisy intermediate scale quantum (NISQ) processors devices. However it is difficult to employ them for simulation of weak intermolecular interactions. As it requires consideration of core electrons (for dispersive forces) leading to a bigger orbital sets eventually requiring large number for qubits. Recently Anderson et. al. \cite{anderson2021coarse} resolved this issue by developing a VQA compatible coarse grained model that scales linearly with the system size. Inspired by the maximally coarse grained method \cite{cipcigan2019electronic} their model represents the polarisable part of the molecular charge distribution as a quantum harmonic oscillator in which the parameters are finely tuned empirically to emulate reference polarisability values corresponding to a real molecule. The model contains zero point multipolar fluctuations by definition and thus dispersion interactions exists inherently. Additionally it does not define force laws between atoms and molecules (interaction potentials) which are predicted by using coarse grained electronic structures. Thus the model combines properties from empirical approach and first principle ab-initio approach. Another advantage that this approach has is the number of unique quantum circuits required to measure the given Hamiltonian after considering all possible dipole interactions scales linearly ($\mathcal{O}({n})$) with the number of quantum oscillators compared to $\mathcal{O}({n}^{3})$ scaling when we consider an orbital based VQE hamiltonian with $n$ orbitals \cite{gokhale2020n}.}

\textcolor{black}{Anderson et.al. \cite{anderson2021coarse} showed the solubility of the proposed model by presenting a proof of principle example by calculating London dispersion energies for interacting $I_{2}$ dimers on IBM quantum processor. While relaxing the harmonic approximation, classical methods heavily relies on path integral and Monte Carlo techniques for efficient sampling of two point correlators for Gaussian states created by harmonic Hamiltonians \cite{westbroek2018user}. Since sampling process for non-Gaussian ground states of anharmonic potentials is very expensive anharmonic electron-nuclear potentials remain unexplored using classical computational techniques. VQA base approach shows a quantum advantage as it suffers negligible experimental overhead while performing relaxation of the harmonic approximation. Thus the VQA based approach provides a road-map that includes anharmonicity and higher order terms for simulation of realistic systems which are inaccessible for current classical methods. In conclusion quantum machine learning provides an efficient and accurate way to model realistic systems and may show a quantum advantage in the future.
}

\textcolor{black}{ A detailed theoretical description of the nonadiabatic (NA) process is limited by intricate coupling between nuclear and electronic degrees of freedom (DOF) eventually leading towards adverse scaling of classical computational resources as the system size grows. Although in principle Quantum Computers can simulate real-time quantum dynamics within polynomial time and memory resource complexity. But quantum algorithms have not been extensively investigated for their application towards the simulation of NA processes. A recent study by Ollitrault et. al. \cite{PhysRevLett.125.260511} proposed a quantum algorithm for simulation of rapid NA chemical process which scales linearly as the system size. Specifically, they propagated the nuclear wave packet across $\kappa$ diabatic surfaces having nonlinear couplings (Marcus Model). The algorithm requires three quantum registers. First quantization formalism is used for DOF, so the space and momentum are discretized  and encoded in the $Position$ quantum register. The population transfer between $\kappa$ diabatic potentials is encoded in $Ancilla$ registers and the nonlinear coupling in $Coupling$ register. Scaling logarithmic with the precision the proposed algorithm is efficient in terms of the number of qubits requirements. This majestic memory compression while storing the propagated wave function denotes an exponential quantum advantage as compared to its classical counterparts.}

\begin{figure}[ht]
    \centering
    \includegraphics[width=1.\linewidth]{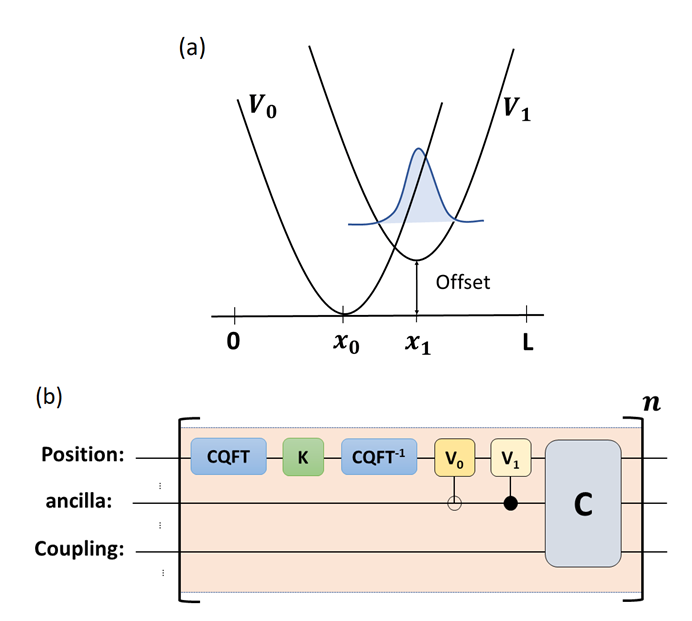}
    \caption{(a) Graphical description of the Marcus model (b) Quantum circuit representation for the time evolution of the wave packet. The blocks represents the evolution operators shown in the form of quantum gates. $ CQFT, V_{i}, K$ and $C$  corresponds to Centred Quantum Fourier Transform (used to switch from position to momentum space), $i^{th}$ potential, Kinetic and Coupling terms respectively. Figure adapted from Ref.  \cite{PhysRevLett.125.260511} }
    \label{marcus}
\end{figure}

\textcolor{black}{The algorithm was applied to simulate NA dynamics of a wave-packet on a simple two one-dimensional harmonic potentials which are shifted in energy by a variable offset. In total 18 qubits were required to simulate this simple model.
The characteristic feature of decline in the population transfer of 
the Marcus model in the inverted region was clearly observed in the simulation, showing an excellent agreement with the exact propagation. Although the simulation was not entirely implemented on the quantum processor the reason is the huge circuit depth of the quantum circuit corresponding to the dynamics parts which requires higher coherence time not attainable by the current generation of quantum processors. But the first part of the algorithm which prepares an initial Gaussian wave packet\cite{PhysRevLett.125.260511} was implemented on an IBM Q device. The extension of this algorithm to higher dimensions is straightforward and a $d$ dimensional polynomial potential energy surface (PES) can be encoded with $\mathcal{O}({d\log_{2}(N)})$ ($N$ is the number of discrete grid points). Current classical algorithms \cite{capano2014quantum,zhugayevych2015theoretical} are limited to simulations of molecular systems which are characterized by up to ten vibrational modes. Hence a quantum processor offering approximately 165 qubits with sufficient coherence time would alleviate this hurdle thus providing an immense quantum advantage while understanding fast chemical processes involving exciton formation, inter-system crossings, and charge separation. }

\textcolor{black}{Although Quantum Machine Learning (QML) shows quantum advantage in electronic structure calculation, its application towards force field generation remains unexplored. A recent study by Kiss et. al.\cite{kiss_quantum_2022} learns a neural network potential energy surface and generates a molecular force field via systematic application of parametrized Quantum Neural Networks (QNN) techniques on the data set generated from classical $ab$ $initio$ techniques. The proposed QNN model was applied on single molecules and the results show the competitive performance of the QNN model with respect to its classical counterparts. Additionally, the results suggest that a properly designed QNN model exhibits a larger effective dimension resulting in fast and stable training capabilities. This potentially hints towards a possible quantum advantage of QML's application in force field generation}

\subsection{Drug-Discovery Pipeline and pharmaceutical applications} \label{Drug_discovery}

\begin{figure*}[ht!]
    \centering
    \includegraphics[width=0.95\textwidth]{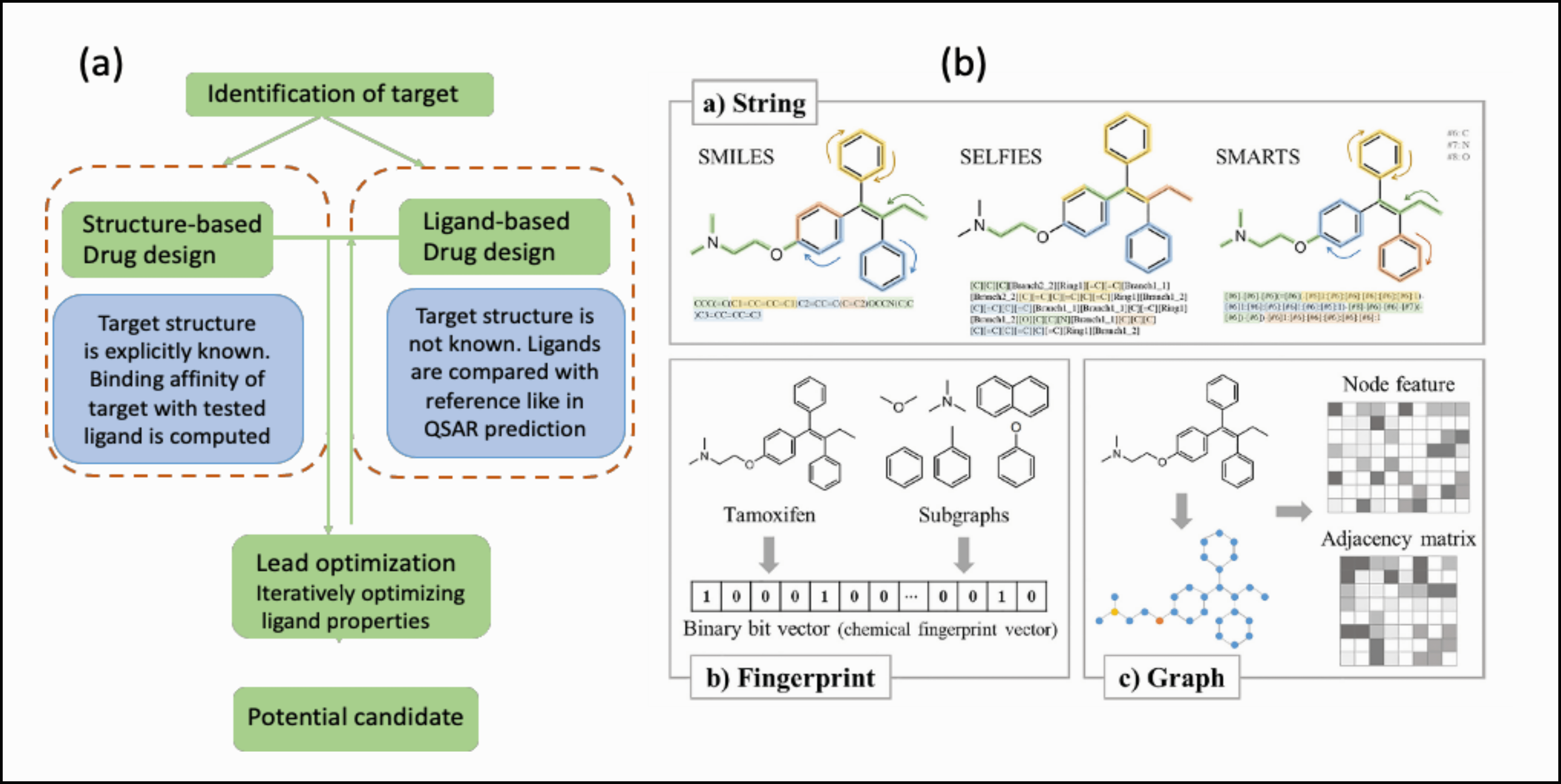}
    \caption{
    (a) The schematic overview of the key steps in computer-aided drug discovery (CADD).
    (b) Encoding the prospective drug/molecule into various representative formats \cite{Pic_fig2_drug} for machine learning algorithms to act on. Reprinted from -A review on compound-protein interaction prediction methods: Data, format, representation and model, 19, Sangsoo Lim and Yijingxiu Lu and Chang Yun Cho and Inyoung Sung and Jungwoo Kim and Youngkuk Kim and Sungjoon Park and Sun Kim, 1541-1556, Copyright (2021), with permission
    from Elsevier 
    }
    \label{fig_cadd}
\end{figure*}

To deploy machine learning or deep learning algorithms for computer-aided drug discovery (CADD) \cite{sliwoski2014computational,leelananda2016comp} it is important to access large molecular databases wherein such structural information about the target (receptor) and/or drug (ligand) candidates may be found. Such databases are screened to generate prospective hits based on either binding affinity with the target above a threshold for structure based protocols or by looking at chemical similarities with previously known bio-active candidates for ligand-based protocols. For the second category, chemical databases like PubChem \cite{kim2016pubchem,kim2019pubchem}, ChEMBL \cite{gaulton2017chembl}, Drug-Bank \cite{wishart2018drugbank}, DUD-E \cite{mysinger2012directory} etc may be good choices. DUD-E contains many decoys (ligands/drug candidates which shows similarity in properties but are not topologically similar in structure) which may be useful for testing and validation of the trained models. For the first category wherein physical information about the target protein is necessary, databases like UnitProt \cite{uniprot2017uniprot}, PDB\cite{berman2000protein,berman2007worldwide,burley2017protein}, PDBbind \cite{wang2005pdbbind} may be a useful resource. Combination databases  BindingDB \cite{gilson2016bindingdb} which contains experimental data for several ligands and targets together have also been employed extensively too.
Most of these databases do encode the structural/physico-chemical attributes of the candidate molecule and/or the target into computer-inputable format which are either numerically describable or string-based. These are called features or very simply molecular descriptors and often depending on the choice of the user other molecular descriptors can be generated using the accessed structural information with libraries like RDKit  etc. For the ligand/drug generic features like number of atoms, molecular weight, number of isomers etc are often called as 0D features as they do not describe the specific nature of the connectivity of atoms within the molecule and remain oblivious to conformational changes. 1D features like SMILES \cite{weininger1988smiles, weininger1989smiles, o2012towards}, SELFIES \cite{krenn2020self, krenn2019selfies}, SMARTS \cite{dalke2008parsers,sykora2008chemical} which encodes the connectivity pattern within the molecule using strings are quite commonly used. On the other hand, numeric features are based on fingerprints which usually represent the molecule as a binary vector with entry 1 (0) corresponding presence (absence) of certain prototypical substructures/functional groups. These are further divided into many categories like circular fingerprints like ECFP \cite{rogers2010extended} which are extremely popular as they are quickly generated, Morgan fingerprints \cite{cereto2015molecular}, Molecular ACCess System (MACCS)\cite{duan2010analysis}, Tree based fingerprints \cite{hert2004comparison}, Atom pairs \cite{perez2009apif,awale2014atom} to name a few. Graph based 2D descriptors\cite{de2018molgan, jiang2021could} are also commonly used with the atoms in the molecule represented as vertices and the bonds between them as connectivity pattern. Adjacency matrix derived from such a graph can act as a molecular descriptor. Fig. \ref{fig_cadd}(b) shows the an example for representing a given molecule in the commonly used 1D and 2D formats. However, these 1D and 2D descriptors even though widely used are often less sensitive to stereochemistry within the molecule which may be important for evaluating binding proclivity with the target. 3D descriptors are useful for this purpose as detailed structural information like dihedral angles are important for this encoding\cite{carracedo2021review, lin2020review}. Higher dimensional encoding with information like specific conformational state, interaction with the solvent or the background residues of the target may also be used to enhance the predictive capacity of the model\cite{carracedo2021review, lin2020review}.  


\subsubsection{Structure-based drug designing protocols using classical machine learning techniques} \label{St_drug_des}  

We first see the performance of machine learning techniques on the structure-based drug designing protocols (see Fig.\ref{fig_cadd} (a)). The primary goal for such studies is to determine whether the prospective candidate molecule can bind effectively (and thereafter evaluate the binding pose and compute the binding affinity) to the target receptor given that the structural information about the target protein is available from prior experimental (like X-ray, NMR etc) or theoretical studies \cite{sliwoski2014computational, lin2020review}. 
Even though experimental assurances are the most trusted means of evaluating such an interaction , yet simulation of the process gives a window into understanding the interaction between a prospective candidates with the desired target relegating the need for direct labor and money intensive experimental verification at a later stage before clinical trial thereby leading to efficient screening. Recently, convolutional neural-network (CNN) (Basic theoretical formalism discussed in Section \ref{CNN_section} and recurrent neural network based models (RNN) (Basic theoretical formalism discussed in Section \ref{RNN_section}) are being used to investigate the process. In a celebrated work by Ragoza $et al$ \cite{ragoza2017protein}, 3D conformational images of the protein and molecule (ligand/drug) is taken and trained with a CNN to identify which poses are suitable for binding and which are not. The SAR-NRC-HiQ dataset was used \cite{Dunbar2011CSARBE} containing 466 ligand-bound receptors (proteins). Two training sets are generated from it by re-docking using Smina \cite{Koes2013LessonsLI} and labelled using the Auto Dock Vina scoring function \cite{Trott2010AutoDockVI}. These re-docked 3D structures are discretized into grid near the binding-site with a length of 24 \AA $\:$ on either axis and 0.5 \AA $\:$ resolution. The ligand and protein atoms within each such grid points were differentiated. This served as input to the CNN. CNN model used had an architecture of five $3\times 3 \times 3$ hidden layers with ReLU activation and additional max pooling layers. The model was trained using Caffe Deep Learning framework \cite{jia2014caffe} minimizing multi-dimensional logistic loss function using gradient descent as the training algorithm. CNN outperformed AutoDock Vina scoring in pose-prediction ability i.e. grouping which poses affords a good binding affinity. The superior performance of the model was upheld for virtual screening too. Compounded datasets by combining training examples from both the tasks were used and the CNN model based training was found to be as effective as their separate counterparts. Even though the CNN was not trained on mutated protein datasets for binding affinity, yet it was able to register the amino acid residues within the protein critical for binding which afforded an easily interpretable visualization of the features that the network is learning. The work envisioned developing protocols to perform tasks like pose-ranking, binding affinity prediction , virtual screening using a highly multi-task network trained on much larger dataset which can ameliorate its performance even more. 

In a similar work Yang-Bin Wang $et al$ \cite{Wang_2020} developed a computational model with memory based on LSTM to construct a framework for drug-target interaction.
Information about the drug-target pairs were obtained using Kegg \cite{Kanehisa2008KEGGFL}, DrugBank \cite{Wishart2006DrugBankAC} and Super Target databases \cite{Gnther2008SuperTargetAM}. Four classes of targets were considered i.e. enzymes, ion-channels, GPCRs  and nuclear receptors. From the dataset curated from the above bases, the drug-target interaction pairs with known affinities are set as positive examples while the remaining are treated as negative examples. The structural features of the protein was described using Position Specific Scoring Matrix (PSSM) \cite{Wang_2020}. For a protein consisting of $N$ amino acids , PSSM is a $N \times 20$ matrix with the (i,j) th element of the PSSM denotes the probability of mutating the ith amino acid in the sequence with the native amino acid from the list of 20. The PSSM was constructed using PSI BLAST \cite{Wang2017PCVMZMUT}. Legendre moments using the elements of PSSM was then constructed to remove redundancy in features. At the end 961 features were obtained for each protein sequence. The structural features of the drug was encoded molecular fingerprints. PubChem database was used for this purpose which defines 881 sub-structural features. As a result each drug/ligand was represented by an 881 dimensional Boolean vector denoting the presence or absence of these tagged molecular substructures. The total 1842 dimensional vector (881 + 961) for the molecule and the receptor target was reduced to 400 size feature vector using sparse principal component analysis (SPCA). This combined feature vector was fed into the classifier. Multiple LSTM layers were stacked to get a deep LSTM setup i.e. 4 hidden layers with 36 neurons were used. Overfitting was compensated by using dropout of randomly chosen neurons during the training process. Different performance metrics were used like ACC, true positive rate or even the standard AUC as defined before \cite{hanley1982meaning}.
Both hyperbolic tangent and logistic sigmoid was used as the activation function depending on the case (see Section \ref{DNN_section}). The output layer being a classifier uses a softmax. The method attained great accuracy across all performance metrics for all the 4 classes of drug-target chosen in comparison to traditional machine learning techniques or even multi-layer perceptrons. The multi-layer perceptron they trained had the same number of hidden units as the LSTM network being used for a fair comparison. The method performed reasonably well even with small training sample size like 180 as was available for the nuclear-receptor family.
In another recent work by Zheng $et al$ \cite{Zheng2019PredictingDP}, a deep learning algorithm is developed with both CNN and LSTM. The target/receptor proteins is processed into a fixed length feature vector using a dynamic attentive CNN. 16-32 filters and 30 residual blocks were used in the construction of the dynamic attentive CNN \cite{he2016identity}. The drug candidate was represented as a 2D matrix similar to \cite{Olivecrona2017MolecularDD}. This is processed using a self-attentional LSTM network \cite{Zheng2019PredictingDP} known as BiLSTM which extracts features. The number of hidden layers in the BiLSTM network was 64. Three kind of databases were used for numerical experiments. The first is DUD-E \cite{Mysinger2012DirectoryOU}. This dataset has 102 target receptors across 8 protein families. Each receptor has 224 active drug molecules and 10,000 inactive ones. The final dataset curated from this database had 22,645 active drug-protein interaction pair examples and 1,407,145 negative ones. The second database used was Human \cite{Liu2015ImprovingCI} containing 1,998 unique proteins and 6,675 interactions. The third database used is Binding DB \cite{Gilson2016BindingDBI2} which contains experimental results of binding affinities. The dataset curated from this had 39, 747 positive examples and 31,218 negative binding interaction pairs. The two feature vector of the target and the protein are combined and fed into a classifier which generated a probabilitty vector from it using sigmoid activation. This probability vector was then minimized against the data label of positive and negative interaction pairs using cross-entropy with regularization. The metric used to evaluate the final performance of the model is 
area under the ROC (Receiver-Operating Characteristic)\cite{zou2007receiver} curve. This metric is often abbreviated as AUC in literature.
Receiver-operating characteristic curve (ROC) enrichment metric (RE) is also used for DUD-E dataset. For the Human dataset, the method achieved 98.7 \% for AUC outperforming all traditional methods like Random Forests, Support-Vector Machines etc. For DUD-E dataset, the method outperformed Vina \cite{Trott2010AutoDockVI}, AtomNet \cite{Wallach2015AtomNetAD} to name a few. On BindingDB with seen and unseen drug-target pairs too , the model outperformed all traditional competitive algorithms. Visual demonstration of which amino acids residues of the target and what structural features/moieties in the drug are important for interaction was also provided.

\subsubsection{Ligand-based drug designing protocols using classical machine learning techniques}
Next we move onto to ligand-based drug designing protocols (see Fig.\ref{fig_cadd} (a)). The objective of these methods is to analyze what kind of prospective drug/ligand candidates obtained by screening compound libraries share similar structural features with previously known drug candidates (used as reference) against the given target. In this paradigm one operates under the premise that ligands with such features will have similar bio-activity too against the said target. These methods are therefore useful when direct structural information of the target is not available \cite{sliwoski2014computational, lin2020review} but bio-activity of some reference compounds against the target is known from previous domain knowledge. One most commonly employed technique in this category is constructing quantitative relationship between structure and activity (QSAR). QSAR is the analytical quantification to the degree to which structural related molecules will share isomorphic bio-activity and hence allows prediction of the behavior of the newly screened compounds against the specified target. This facilitates the identification what structural features are responsible for the activity and hence provides insight into rational synthesis of drugs in future.

The first report for the use of deep-learning models in QSAR prediction after the publically accessible Merck challenge was due to Dahl $et al$ \cite{dahl2014multi}. QSAR studies mainly focus on understanding what chemical composition and structural features of the prospective molecule of choice (ligand) might have desired pharmacological activities against a chosen target (receptor). The study by Dahl $et al$ focused mainly on the effectiveness of multi-tasking while designing architectures for neural-networks (Basic theoretical formalism discussed in Section \ref{ANN_section}). Multi-tasking refers to the ability of the network design wherein different outputs of interest are simultaneously retrievable. For instance, in the aforesaid study a single-task network would be training using molecular descriptors obtained from compounds within a single assay (data-set) as input features and using the activity of the said molecule as a performance label for comparing the output. However, this model of training requires huge data assemblies from single assays alone which may not always be available. To circumvent the issue, the authors of the aforesaid study combined data from multiple assays using an architecture in the final output layer wherein individual neurons are dedicated to each assay. The same molecule may have appeared in different assays with different activity labels. For any given such molecule, the input feature vector is the molecular descriptor. The output at each of the neurons in the final layer are the activity/inactivity classification values learnt by the network for each assay. This output is then compared against the recorded activity label obtained from the corresponding assay for back-propagation.  The study used data from 19 such assays from PubChem database (see Table I in \cite{dahl2014multi}) with each assay containing $10^4-10^5$ compounds. The molecular descriptors used were generated using Dragon software\cite{Mauri2006DRAGONSA} as a feature vector of length 3764 for each compound. Although the model is a binary classification study, the performance metric used is AUC as defined before. The network was trained using Stochastic-Gradient Descent algorithm with momentum and the problem of over-fitting due to many tunable parameters was eschewed using drop-out \cite{srivastava2014dropout}. It was seen that the Deep-learning network used outperformed traditional machine learning models in 14 out of the 19 assays. Among these 14, multi-task networks outperformed single-task models in 12 of the assays. These favorable results were retained by grouping similar molecules across different assays into a customized composite data-set. The depth of the neural network/addition of more hidden layers did not always produce improvement in the performance of the multi-task network which the authors attribute to smallness of the data within each assay. All the deep-learning models used handled reasonably well correlated features in the input feature vector. In a related study Ramsundar $et al$.\cite{Ramsundar2015MassivelyMN} resolved certain questions about the efficacy of multi-task neural networks in virtual screening of candidate molecules against several targets using extremely large data-sets. 259 assays were used and divided into 4 groups - PCBA, DUD-E\cite{Mysinger2012DirectoryOU}, MUV, Tox21. Together all these databases had 1.6 M compounds. The validation scheme used is AUC as before and the feature vector of the studied compounds were ECPF4 fingerprints\cite{rogers2010extended}. All such collection of fingerprints for the molecule were hashed into a single bit vector. The trained network as before is a multi-task classifier with each neuron at the output layer having a softmax activation\cite{nwankpa2018activation} corresponding to each assay. The study found that such multi-task NN performed better than several ML models and the performance metric can be improved with increasing number of tasks and/or increasing the volume of data per task. Certain data-sets in the study showed better performance than others which were attributed to shared set of compounds among such databases. However the biological class of the target receptor did not affect the performance metric too much. The study concluded by saying that extensive data-sharing among proprietary databases needs to happen to benchmark the performance of such models with bigger data-sets.
Another study which thoroughly benchmarked the performance of deep neural networks against a commonly used machine learning model like Random Forest (RF) was due to Ma $et al$ \cite{ma2015deep}. The report used 15 Kaggle data-sets for training and another 15 for validation too. Each such data-set had around $10^4-10^5$ candidates as molecular designs for drugs each labelled with response activity against designated target(s). The type of descriptors used for each of these molecules included combination of atom pairs and global donor-acceptor pairs \cite{Carhart1985AtomPA} as input feature vectors. The report used the squared Pearson correlation coefficient (R$^2$) \cite{alexander2015beware} between observed and predicted activities in the testing set as the preferred metric of performance. The study showed that there was a mean improvement in R$^2$ of 0.043 when using deep-neural network as opposed to RF against arbitrarily selected parameters. 4 of the data-set showed dramatic improvement in favor of deep-neural network whereas one favored RF. When refined parameter set was used instead of arbitrarily chosen ones, the trend is retained with an expectedly higher mean improvement of 0.051. Increasing the number of hidden layers and also number of neurons in each such hidden layer displayed a R$^2$ in favor of deep neural network. Changing the activation function from sigmoid to ReLU \cite{nwankpa2018activation} also favored the deep network model for 8 data-sets. R$^2$ was found to also favor networks when it is not pre-trained. 

\subsubsection{Machine learning and drug-induced toxicity}

Another area wherein machine learning algorithms are important is identifying if a particular drug when administered in a biological medium would be toxic or not. Such adverse effects due to an administered drug may lead to serious health complications culminating in the eventual withdrawal of the drug during development/testing or even post-marketing thereby leading to wastage of resources.  Many such studies has been initiated like in \cite{vo2019overview}. The earlier investigations primarily used machine learning methods. This can be exemplified from the work of Rodgers $et al$ \cite{rodgers2010modeling} which created a model using k-nearest neighbor (kNN) (Basic theoretical formalism discussed in Section \ref{k-NN_section}) for identifying whether a drug candidate from Human Liver Adverse Effects Database (HLAED) belongs to two categories- hepatotoxic in humans or not based on labelled markers from five liver enzymes. With a dataset of over 400 compounds, the algorithm was successful in classifying with a sensitivity of $\ge$ 70 \% and a specificity $\ge$ 90 \%. The model was extended to test unseen cases in World Drug Index (WDI) database and Prestwick Chemical Library (PCL) Database with good success ratio. To do so, a compound similarity metric based on Euclidean norm between molecular candidates were used and an applicability domain threshold was constructed using the metric. Predictions for candidates with similarity scores outside the applicability domain threshold were considered unreliable. Chemical features like aromatic hydroxyl units in drugs like Methyldopa \cite{plaa1998toxicology} or pyrimidyl units in drugs like Trimethoprim \cite{guyton1993role} were identified to play a key role in the assignment of high hepatotoxic activity of the respective drugs as they are prone to oxidation and can form hapten adducts with cellular proteins. In yet another study \cite{sun2012structure} a classification task among molecular candidates were designed using Support-Vector Machines (Basic theoretical formalism discussed in Section \ref{Supp_vec_machine}) with the Gaussian Radial Basis Function (RBF) as the kernel to group molecules into active and inactive categories with respect to susceptibility to induce phospholipidosis (PLD). Phospholipidosis refers to intracellular accummulation of phospholipids induced by drug candidates when they bind to polar phospholipids in the lysosome \cite{Nonoyama2008DruginducedP}. The model was trained by curating dataset from three databases National Institutes of Health Chemical Genomics Center (NCGC) Pharmaceutical Collections (NPC)\cite{Huang2011TheNP}, the Library of Pharmacologically Active Compounds (LOPAC) and Tocris Biosciences collection and the target cell used was HepG2 \cite{bhandari2008phospholipidosis}. The model developed was found to accomplish the selection task with high sensitivity and specificity as seen from the AUC metric. The training was found to be sensitive to the nature of molecular descriptors used and also to the size of the dataset. Certain simple chemical attributes like size of hydrophillic moieties are often used as indicators to characterize if a drug can induce PLD. Such features even though showed correlation with the identified active compounds in some cases but did not agree on some others. On the contrary features like presence of positively charged nitrogen center correlated well across the entire dataset. Identification of such features may be important to chemist for avoiding or replacing such structural moieties during the early developmental stage of the drug. 

Deep learning models have also been deployed for the said purpose. In a recent one using artificial neural network, toxicity due to epoxide formation is thoroughly investigated\cite{hughes2015modeling}. The study identified among a given set of drugs/ligands and targets which drug is susceptible to be epoxidized with natural oxidants in the biological medium by oxidants like cytochrome P450. Formation of such epoxidized metabolities can be harmful for the body as has been explicitly noted for the case of an anti-epileptic drug like carbamazepine \cite{Pearce2005PATHWAYSOC} which after epoxidation binds to nucleophillic sites within a protein forming a hapten adduct thereby triggering immune response\cite{Yip2014CovalentAO}. The product of such reactions need not always be an epoxide as the study suggests from previous reports \cite{Alton1975BiotransformationOA} for drugs like N-desmethyl triflubazam wherein a DNA adduct is formed post a transient epoxidation. The neural-network model used in the study not only decides if a given drug is epoxidizable but also focuses on identifying if the site of epoxidation (SOE) is a double bond or an aromatic ring. It further delineates such sites from site of hydroxylation (SOH) which shares some key structural features with SOEs and can also potentially lead to harmful oxidation products. The Accelrys Metabolite Database (AMD) was used from which a dataset of 389 molecules were curated having toxicity labels. These molecules had 411 aromatic SOEs, 168 double bond SOEs and 20 even single bond SOEs. Non-epoxidizable molecules were also included in the set thereafter to afford a proper distinction. To describe each bond within the specific molecule 214 molecular descriptors/features were identified- 89 each for the left and right atom sharing the bond, 13 specific bond descriptor and 23 overall molecular descriptors. The neural network used had 1 input and 2 output layers . The top output layer was for the molecular epoxidation score whereas the last one for the specific bond epoxidation score. The training data involved using labelled binary vector within the molecule with the designated SOE marked as 1. Cross-entropy minimization was used as the cost function. The final metric of performance as before was AUC. The model outperformed logistic regression and other competitive algorithms in all departments like identification of SOEs, differentiation of SOEs and SOHs etc. The model could correctly identify SOE in carbamazepine which is absent in substitutes like oxcarbazepine with similar functionality, in the furan ring of furosemide \cite{Stepan2011StructuralAM} and in severely hepatotoxic sudoxicam vs its less problematic cousin like meloxicam \cite{Stepan2011StructuralAM}. More such examples can be found in specialized topical reviews like \cite{vo2019overview}. 

\subsubsection{Power of quantum computers and quantum-computing enhanced machine learning techniques} \label{drug_QC}
As discussed in other domains, quantum computing enhanced machine learning techniques are also beginning to gain attention in the overall drug production pipeline. An early review \cite{cao2018potential} identified the efficacy of quantum computers to the drug discovery process by noting the key areas wherein quantum computers can impact. The study reported that for structure based drug designing protocols, quantum computers may help in understanding the structure of target protein sequence better. It claimed that using both gate model of quantum computing and quantum annealers, simple problems like the Miyazawa-Jernigan model were investigated for understanding the dynamics of protein folding for smaller peptides \cite{PerdomoOrtiz2012FindingLC}. Unfortunately such model problems may not accurately assess the real situation in all cases especially for complicated situations like the presence of several protein conformations with minimal free energy difference thereby rendering them accessible via thermal fluctuations or how the native 3D conformation of the protein is sustained due to its interaction with the surrounding media. In most cases for biologically significant proteins, crystallized 3D structure is not available in the database due to sheer size of the protein and/or lack of solubility etc \cite{sliwoski2014computational}. As a result structure-based designing protocols are often thwarted. Better computational models for predicting the protein structure is therefore necessary and can influence the drug-discovery pipeline immensely. Banchi $et al$ \cite{banchi2020molecular} used Gaussian Boson Sampling to identify the highest affinity binding poses of a given ligand with the active centre of the target. The primary workhorse of the protocol is to map the active centre and the ligand onto a pharmacophoric feature space of few descriptors. Each such descriptor corresponded to a vertex in a graph and the edges defined the Euclidean distance between the corresponding structural motif in the lowest energy 3D geometry. The resultant encoded graphs were then used to construct a binding-configuration graph wherein structural features of the ligand and the target that are compatible to bind are represented by a weighted vertices. The maximum-weighted clique (a closed sub-graph) within the binding-configuration will be the preferred binding pose. Such configurations are identified with high-probability using a Gaussian Boson sampler with the input state of photons being in squeezed states and identifying the detector wherein the photon appears at the output. Such detectors corresponds to vertices on the binding-configuration graph. If the combination so-attained at the output is not a clique then the authors have defined greedy shrinking of vertices or expansion of vertices by probing the local environment to modify the search space for next iteration. 
Recently, an efficient hybrid-variational algorithm amenable to NISQ architecture has also been proposed which using an $N$-sequence amino acid can construct a sample of the lower energy 3D-conformations\cite{robert2021resource}. Previous reports tackling the same problem were either inefficient or offered problem specific solutions \cite{babej2018coarsegrained}. The work represented the given sequence by stacking monomeric units on a tetrahedral lattice. 4 qubits were assigned for each of the 4 different directions the lattice could grow from a given monomer. New ancillary qubits were also used to define the interactions between l nearest neighboring (l-$NN$) monomeric units. A graph hamiltonian was constructed with these interactions and the self-energy terms of each residue. Overlapping positions of the amino acids were avoided by including penalty terms corresponding to such assignments.
The ground state of this hamiltonian is the most stable conformation. To solve the problem, a variational circuit was constructed parameterized by tunable rotation angles of both the single-qubit and entangling gates involved. The various bit-strings obtained from the measurement of the circuit encoded the 3D structural arrangement of the amino acid residues in the tetrahedral lattice and the energy distribution would highlight the relative stability of these various arrangements/conformations. The circuit was optimized variationally to selectively enhance the chances of  sampling the lower energy conformations corresponding to tail of the aforesaid energy distribution. The number of qubits in the method scales as $O(N^2)$ and the number of terms in the hamiltonian is $O(N^4)$.
The model was tested on 10 amino acid peptide Angiotensin using 22 qubits and also on a designed 7 amino acid neuropeptide using 9 qubits on IBMQ processors. In the former the probability of collectively sampling all the lower energy conformations was reported to be 89.5 \% which augmented further with increase in the number of measurements. 

However recently a work from Google's DeepMind (UK) \cite{jumper2021highly}, have made enormous strides in predicting the 3D structure of a peptide from just a given sequence of amino acids solving this 50-year old grand challenge. Even though the algorithm uses a neural-network architecture trainable on a classical computer (and hence is not a quantum-computing enhanced algorithm), yet the performance of the method is so good that it deserves a special discussion. This novel algorithm has won the CASP14 challenge which involves a blind assessment of the efficacy of structure determination from amino acid sequence for proteins/peptides whose structure has been recently solved through explicit experimentation yet has not been publicly released in common databases. The neural network design is broken down into three components. The first component in the neural network architecture takes as input a 1D sequence of amino acid residues and searches multiple databases to generate multiple sequence alignments (MSAs) which are essentially sequence of previously identified amino acids closely resembling the target one and understanding the evolutionary history of the mutation of these MSAs. 
\begin{figure*}[ht!]
    \centering
    \includegraphics[width=1.0\textwidth]{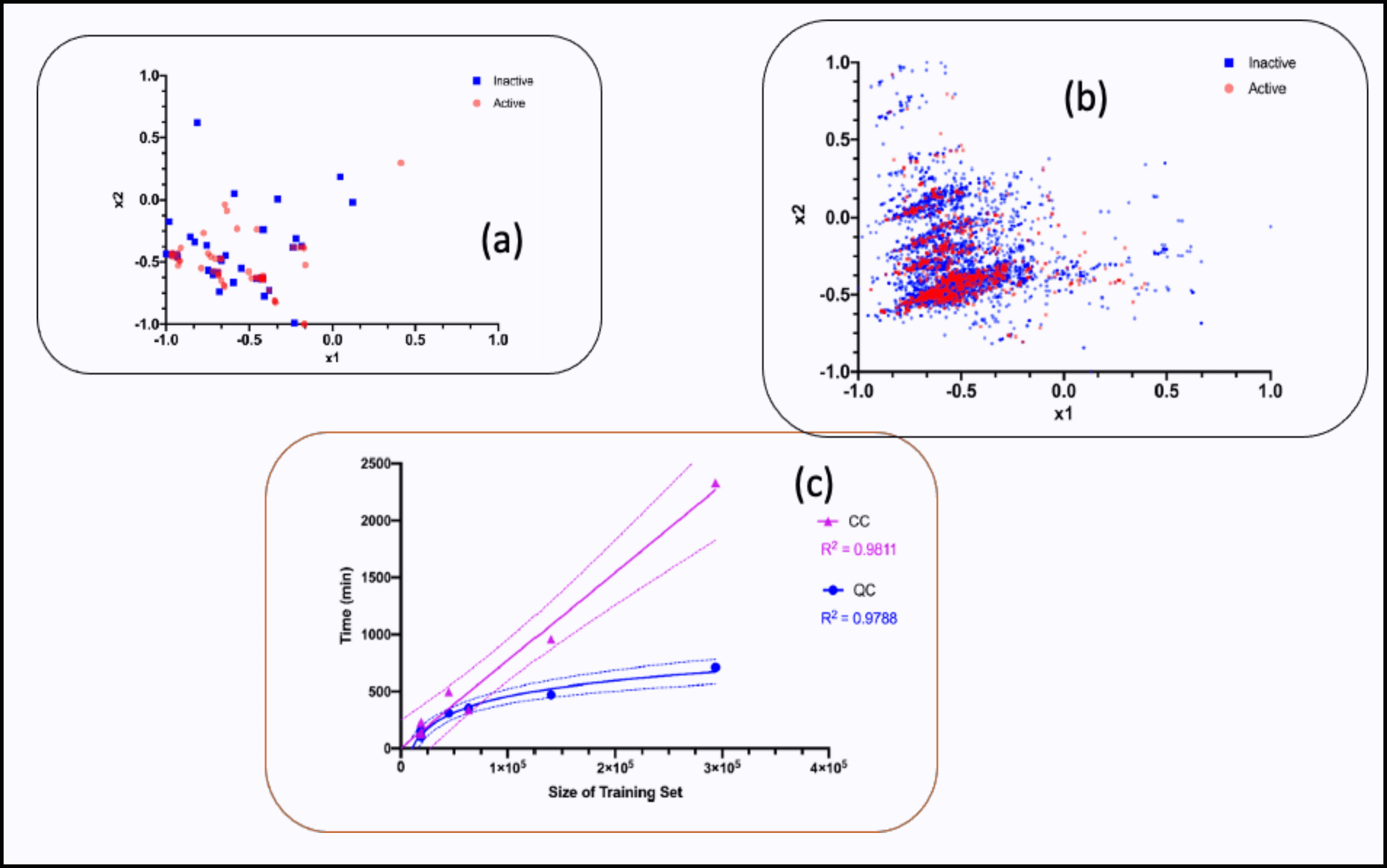}
    \caption{
    (a) Classification of ligands into active and inactive ones based on performance against SARS-CoV 2 in the space of the two most dominant features after application of two methods of feature extraction as in Ref \cite{batra2021quantum}
    (b) Classification of ligands into active and inactive ones based on performance based on performance against \textit {M.Tubercolosis} in the space of the two most dominant features \cite{batra2021quantum}
    (c) The run time required for screening training datasets of varying sizes on a quantum (QC) and a classical computer (CC). The training time shows sublinear scaling on QC displaying an advantage. \cite{batra2021quantum}. Reprinted (adapted) with permission from Batra, Kushal and Zorn, Kimberley M and Foil, Daniel H and Minerali, Eni and Gawriljuk, Victor O and Lane, Thomas R and Ekins, Sean, Journal of Chemical Information and Modeling, 61, 6, 2021. Copyright 2021 American Chemical Society.
    }
    \label{fig_QC_drug}
\end{figure*}
This is important to derive information about structural proximity between amino acid pairs which shows correlated mutation. This component then generates a template 3D structure also called pair representation as an initial hypothesis that is to be modified at later stages. The next functional unit is called the Evoformer and is the heart of the design. This takes as input both the pair representation and the MSA and subsequently refines the representation of each self-iteratively using the architecture of transformer \cite{vaswani2017attention}. The third block takes the refined MSA and the pair representation and generates a 3D structure which is essentially cartesian co-ordinates of the individual atoms i.e. the native 3D conformation of the protein/peptide. This process is repeated several times by feeding back the 3D structure into the Evoformer block until convergence. The final prediction seems to have surpassed all previously known methods with an accuracy of 0.95 \AA root-mean square error from the target structure for 95 \% of the residues. This will definitely be a landmark study and for years to come one has to investigate the efficacy of the method for antibodies , synthetic peptide sequences for which evolutionary data to generate the initial MSA will be scarce. This method will positively impact understanding protein-protein interactions and also bring in new insight in diseases like Alzheimer's and Parkinson's. For structure based drug-discovery since at least one method exist now which can determine the 3D conformation efficiently thereby massively speeding up the product development pipeline, quantum computers can now help in understanding the drug-protein affinity and molecular docking mechanisms. The review by Cao $et al$ \cite{cao2018potential} already identifies this possibility by noting that with algorithms like quantum phase estimation on fault-tolerant devices and variational eigensolvers for near term devices, we are capable of computing the potential energy surfaces of larger molecular systems efficiently and hence force-field calculations as is required for understanding molecular docking will also be greatly impacted. 

For QSAR studies too, benchmarking the performance of quantum computer against classical processors has been documented recently in the work of Batra $et al$ \cite{batra2021quantum}. The authors have used several molecular databases to identify prospective ligands for diseases like \textit {M.Tubercolosis}, Krabbe disease, SARS-CoV-2 in Vero cells , plague and hERG. With the curated data from the compound datasets feature vectors were constructed and used for binary classification of the compound in active or inactive using kernel-based SVM techniques (see Section \ref{Supp_vec_machine} for basic theoretical formalism). The authors used several techniques to reduce the size of the feature vector such that the classification can be performed on a NISQ device ($ibmq\_rochester$ was used) using Qiskit. It was seen that for most datasets comparable accuracy on a quantum computer was attained too using the feature reduction techniques employed by the authors. A hybrid quantum-classical approach was also used for high-throughput virtual screening data screening with good accuracy and slightly faster run time for processing the data on a quantum computer than on a classical computer. Representative data from the study is displayed in Fig. \ref{fig_QC_drug}.

Beyond the precincts of academic research, even the interest of industrial players on quantum-enabled technologies seems to be escalating rapidly. The report by Zinner $et al$ \cite{zinner2021quantum} have identified that 17 out of 21 established pharmaceutical companies and 38 start-ups are directly working on enhancing and ameliorating the technical challenges in the drug discovery pipeline using quantum computers. 75\% of such companies so far have been identified to be geographically in Europe and North America. The cumulative funding received by all the start-ups as per the report \cite{zinner2021quantum} is \texteuro 311 million with the top five funded start-ups being Cambridge Quantum Computing, Zapata, 1QBit, Quantum Biosystems and SeeQC. Most of the activity is directed towards virtual screening for ligand-based drug designing protocol and subsequent lead optimization.  

The report from Langione $et al$ \cite{Langione_page} and Evers $et al$ from McKinsey \cite{Evers_page} also systematically delineates what pharmaceutical industries should do to prepare themselves attain a favorable position in order to leverage the quantum revolution. Both the reports agree that bio-pharmaceutical companies should start now to reap the benefits of early movers advantage. In fact \cite{Langione_page} mentions that it might be possible that tech-giants equipped with quantum computers with higher number of qubits and better noise-tolerance might enter the race of \textit{in\:-silico} drug discovery in the future relegating the task of post-design synthesis, clinical trials and commercialization to pharmaceutical companies. That can lead to a situation wherein companies may race to patent the best molecule that are responsive to a particular disease. However pharmaceutical companies are at an advantage here due to years of experience with computational drug-designing protocols. So strictly they do not have to change the inherent model or the business goal they already follow. They will likely be using a more capable device like a quantum computer to attain that goal. In order to avoid such undue competition, pharmaceutical companies should start now by assessing and answering few key questions about the probable impact quantum computers are likely to have on the workflow and the specific product design the respective company is targeting. This can happen by understanding what are the key areas where development can be sought in the product design model the company is following  and more importantly if those areas fall into the category of problems that can be solved efficiently on a quantum computer. The key directions to think would be would a quantum-computer enabled business provide an unprecedented value to their supply-chain or an undue advantage to their competitors. If the analysis is positive, one needs to then think of the time scale of such developmental changes and whether the resources the company has access to would be enough to sustain entry into the domain. Some of above resource overheads can be solved through appropriate memberships through consortium like QuPharm that categorically specializes in the designated area. QuPharm, as an organization was developed by many member pharmaceutical companies for specifically understanding and optimizing the benefits the quantum revolution can have on the pharmaceutical industry. The consortium is known to have a collaboration with Quantum Economic Development Consortium (QED-C)\cite{zinner2021quantum}. Hardware needs can be sorted through collaborations with end-to-end technological providers like Google, IBM, Honeywell, Rigetti, Xanadu etc each of which have quantum computers of varying architecture and even on different platforms and have even promised to develop larger scale ones in near future. For example Amgen, a pharmaceutical company has declared collaboration with both Amazon Braket and IBMQ \cite{zinner2021quantum}. Boehringer, another pharmaceutical has established collaboration with Google QuantumAI to develop algorithms for molecular dynamics simulation \cite{zinner2021quantum}. New software solutions are necessary to interface with the quantum hardware for which commercially available varieties like Qiskit (IBM), OpenFermion (Google), tKet (Cambridge Quantum Computing) can be leveraged. Companies can also initiate partnerships with firms like ProteinQure, GTN, Rahko, Qulab etc which are developing softwares to specifically cater to advancing quantum-computing algorithms for drug discovery . Classical computing algorithms have also benefited from the ideas that were synthesized to initiate a rapid discovery of quantum algorithms over the last few years. Companies like Qubit Pharmaceuticals, Turbine etc are developing such approaches and combining them with machine learning. Such collaborations with global and local companies with specialized expertise can result in engineering custom solutions to tackle specific problems during the drug discovery pipeline. Pharmaceutical companies can thus immensely benefit from such collaborative ventures. 

Partnerships can be built with other pharmaceutical companies or start-ups too to share expertise and develop mutually beneficial quantum computing based drug-development protocols. The report by Zinner $et al$ \cite{zinner2021quantum} has identified 17 pharmaceutical companies with publically disclosed collaborations with at least 12 start-ups. All the pharmaceutical companies are members of QuPharm, NEASQC etc. Among the big pharma corporations Merck (Germany) has invested \texteuro 4 million on start-ups like SeeQC and has disclosed active collaboration with Rahko, HQC \cite{Merck_HQS, Merck_Rahko}. Merck (USA) has made financial investment in Zapata \cite{Zapata_collab}. Non-equity partnerships like that by Amgen with QSimulate to develop advanced molecular simulation tools or like that by AstraZeneca with ProteinQure to design amino-acid sequences have also been seen \cite{Protein_zeneca}. Some new collaborations are also announced which depicts not only the prospective promise associated with the technology but also the seriousness of the industry players. For example, Cambridge Quantum Computing have announced a collaboration with CrownBio and JSR LifeSciences about using quantum machine learning algorithms to identify multi-cancer genes biomarkers \cite{CQC_JSR} which can positively impact bioinformatics research. Apart from this, it would also be beneficial for pharmaceutical companies to collaborate with scientists in academia. Adopting a strategy that allows effective co-operation among all involved parties internal and/or external beyond the traditional organizational structure will accelerate growth and foster efficient and fast sharing of resources and knowledge which can otherwise be harder to access due to institutional barriers. The reports \cite{Langione_page, Evers_page} identify that recruitment of skilled technicians and professionals with training in developing algorithms on a quantum computer is necessary for pharmaceutical companies to enhance quantum-enabled research. The report by Zinner $et al$ \cite{zinner2021quantum} conducted a  thorough search across 21 pharmaceutical companies with a combined revenue of \texteuro 800 billion in 2019 and reported that only about 50 employees were found designated to quantum-enabled technology. This is partly because quantum computing is an emerging technology and hence such a skillset may not be readily available among the usually hired talent pool of the pharmaceutical industries\cite{Langione_page}. Companies like IBM which holds outreach programs and workshops in quantum computing interfaces like Qiskit can be a useful resource. The other alternative might be looking into developing specialized programs internally to train scientists and engineers once hired.

\section{Insight into learning mechanisms}
\label{Learnability}

Despite the enormous success of machine learning coming from network based models with large number of tunable parameters, little  progress has been made towards understanding the generalization capabilities displayed by them \cite{zhang2017understanding}. The choice of hyperparameters in these models have been  based on trial and error with no analytical guidance, despite it showing enormous potential in analyzing data sets.  Physics on the other has provided us with white box models of the universe around us that provides us with tools to predict and examine observed data. Intuition from statistical mechanics have helped provide understanding with respect to the learning limits of some network models. Seminal contributions in this regards include methods from spin glass theory, that have been used to extensively study associative memory of Hopfield networks \cite{Hopfield2554} and Valiants theory of learnable models that introduced statistical learning into the then existing logic based AI \cite{10.1145/1968.1972}. Another major contribution comes from Gardners usage of replica trick to calculate volume in the parameter space for feed forward neural networks in the case of both supervised and unsupervised models \cite{1987_gardner,1988_gardner}. The problem of learning was also shown to exhibit phase transitions in reference to generalization and training efficiency \cite{PhysRevA.41.7097}. A typical example of one such parameter is the ratio of input size to the number of model parameters. A diminishing value usually results in overfitting while a large allows for successful generalization. The reader is encouraged to refer \cite{144PhysRevLett.65.2312,145PhysRevA.45.4146,146PhysRevA.45.7590} for some of the early works that made use of statistical physics to understand multi layered network learning

\begin{figure}[ht!]
    \centering
    \includegraphics[width=0.5\textwidth]{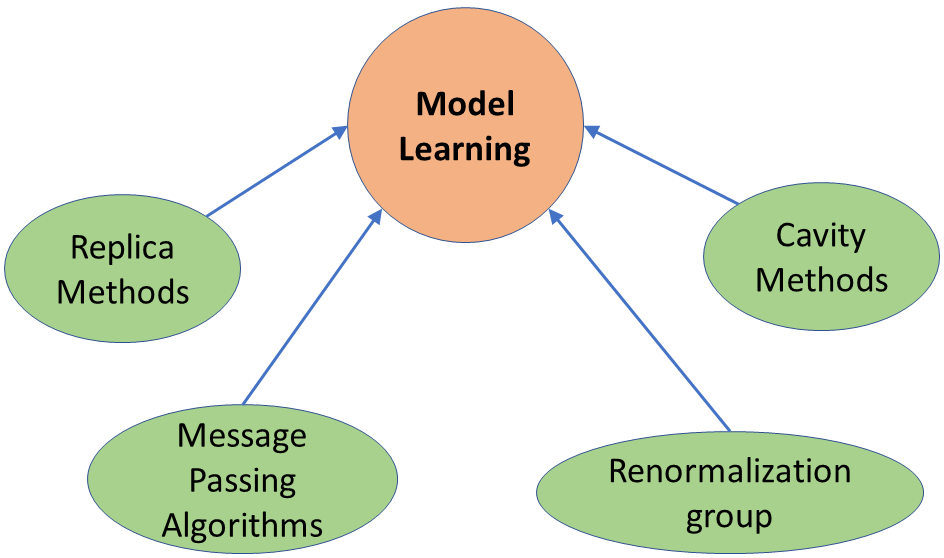}
    \caption{A schematic representation of the techniques borrowed from statistical physics that have been used to study model learning of networks in machine learning}
    
    \label{model_learn}
\end{figure}


We will see that some self averaging statistical properties in large random systems with microscopic heterogeneity give raise to macroscopic order that do not depend on the microscopic details. Learning these governing dynamics can play an important role in tuning the performance of machine learning techniques. One of the tools that provides an analytical handle in analyzing these details is the replica method \cite{morone2014replica}. Replica methods have been used to explore the teacher-student model to provide information theoretic best estimates of the latent variables that teacher uses in generating the data matrix handed over to the student \cite{Zdeborova2016}. This problem can be specialized to the case of providing a low rank matrix decomposition matrix  of underlying input data. Interest into statistical methods have stayed dormant since 1990, due to the limited tractability of algorithms used in learning. It was sparked again with the contribution of Decelle \cite{Decelle_2011} to use spin glass theory to study stochastic block model that played a major role in understanding community detection in sparse networks. They observed second order phase transitions in the models that separated regions of efficient clustering when solved using belief propagation algorithms \cite{Yedidia2003UnderstandingBP}. For a comprehensive list of physics inspired research in the machine learning community, refer \cite{Carleo_2019}.

Generative models are suitable for feature extraction apart from doing domain sampling. Within every layer of one such network, one could imagine some form of feature extraction being made to provide for a compact representation, which might later be used by generative models to learn the distribution of classifiers to do prediction. We point towards this feature extraction as the central idea for relating machine learning to renormalization group. We investigate to see if concepts like criticality and fixed points have something to reveal about the nature in which learning happens in the framework of deep learning and machine learning in general. Figure \ref{model_learn} provides a schematic sketch of the methods that have been primarily explored in studying network models. In here we shall restrict our attention to cavity methods, Renormalization group and Replica methods. Refer \cite{feng2021unifying} for a thorough exploration of Message Passing Algorithms.

A somewhat non rigorous argument for the remarkable working of these machine learning algorithm comes from noting the following two observations. Consider a vector of input size $n$ taking $v$ values and thus can span a space of $v^n$ possible configurations. Despite this being  exponentially a large space for image datasets, we have managed to build relatively small networks that learns to identify the features of the image accurately. This is to be attributed to the realization that the class of relevant images comes from a relatively small subspace that is efficiently learnt by the neural network with relatively fewer nodes that scale as $nv$ instead \cite{Lin_2017}. This is much similar to how low energy states of interest of hamiltonian mapping to small subspace of the Hilbert space. This simplification comes from the Hamiltonian having a low polynomial order, locality and symmetry built into it . Secondly, despite the vastness of all possible inputs that can be generated, most input data can be thought of coming from a Markovian process that identifies at each stage a select set of parameters. Thus the work of a deep learning machine would be to reverse some form of heirarchial markovian generative process using some minimal statistic function (A function $f$ is minimal statistic if for some random variables $y$ and $x$ we have $P(y|x) = P(y|T(x))$) that retains the features of the probability distribution by preserving mutual Information.


The methods in the following subsection describe its working in the context of Ising model, so we start by describing one briefly. Ising is a model for representing classical spins or magnets arranged in a 2d lattice whose interaction with one another is quantified by the strength of the coupling. Each spin takes a binary configuration $(+1,-1)$ of choosing to align up or down. At low temperatures spins prefer aligning in the same direction forming a magnet. At high temperatures the thermal fluctuations kill any order within the system causing them to arrange chaotically with no net observable magnetic field. A general Ising Hamiltonian is given by,

\begin{equation}
    H(\sigma) = -\sum_{<i j>} J_{ij}\sigma_i\sigma_j - \sum_j h_j\sigma_j
\end{equation}

where $<i j>$ indicates the sum over nearest neighbour pairs. The probability of any given configuration is determined by the Boltzmann distribution with inverse temperature of the system scaling the governing Hamiltonian. Expectation values of observables correspond to averages computed using the distribution. Given any observable $O$ the expectation value at a given inverse temperature $\beta$ is given by,
\begin{equation}
\braket{o}_\beta = \sum_{\{\sigma\}} \frac{e^{-H(\sigma)}}{Z} O(\sigma)
\end{equation}

\subsection{Replica Method}

\begin{figure}[ht!]
    \centering
    \includegraphics[width=0.5\textwidth]{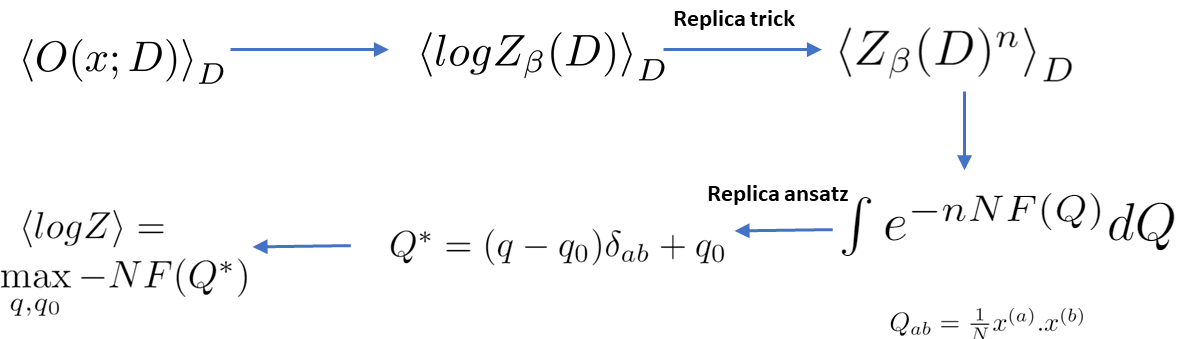}
    \caption{A schematic representation showing the use of replica replica trick and replica ansatz to compute the expectation value of self averaging observables }
    
    \label{replica}
\end{figure}

Replica method is a way of computing self averaging expectation values of observables $O(x)$ such that $x$ is a minimizer of $H(x,D)$ where $D$ is the distribution some input space. A typical example of $H$ would be the cost function of a learning problem and $D$ would be the dataset used for training in the problem. Figure \ref{replica} provides a schematic representation of the use of replica method. Here we shall be explore it in the context of Ising Hamiltonians. Consider the Ising model of N spins, given by the following Hamiltonian,

\begin{equation}
    H(s,J) = -\frac{1}{2} \sum_{ij} J_{ij}s_i s_j
\end{equation}

where the connectivity matrix entries $J_{ij}$ has been sampled independently from a Gaussian distribution with zero mean and variance $1/N$. The spins take values from ${+1,-1}$. In a bath of inverse temperature $\beta$ this results in an equilibrium distribution that is governed by the Gibbs distribution given by,
\begin{equation}
    P_J(s) = \frac{1}{Z[J]} e^{-\beta H(S,J)}
\end{equation}

where Z is the partition function. We would like to analyze the structure of low energy patterns that is independent of the realization of the couplings $J_{ij}$ in the large $N$ limit. Properties of disordered system can be learnt by studying self averaging properties (have zero relative variance when averaged over multiple realizations) . Steady states emerge in the large $N$ limit as a result of diverging free energy barriers causing time average activity patterns that no longer look like Gibbs averaging, due to broken ergodicity. 

To study the patterns encoded within the Gibbs distribution, we start with computing the free energy average over all the realizations of $J$. This involves computing expectations of a logarithm which can be simplified using the following Replica trick.
\begin{equation}
    \beta\braket{F[J]}_J = \braket{ln Z[J]}_J = \braket{\lim_{n\longrightarrow 0}\frac{Z^n -1}{n}}_J = \lim_{n\longrightarrow 0}\frac{\partial}{\partial n} \braket{Z^n}_J
\end{equation}
Evaluating $\braket{Z^n}$ is much simpler as this can be expressed as an average over replicated neuronal activity, i.e,
\begin{equation}
    \braket{Z^n}_J = \Big\langle \sum_{S^a} e^{\beta\sum_{a=1}^{n} \sum_{ij} J_{ij} s^{a}_{i}s^{a}_{j}} \Big\rangle_J
\end{equation}
where $s^a$ denotes the set of replicated spins over which the averaging is done. The Gaussian integrals can be easily evaluated resulting in,
\begin{equation}
    \braket{Z^n}_J = \sum_{S^a} e^{\frac{1}{4}{N\beta^2}\sum_{ab}Q^{2}_{ab}}
\end{equation}
where the overlap matrix $Q_{ab} = \frac{1}{N} \sum_{j=1}^{N} s^{a}_{j}s^{b}_j $. Minimizing the free energy amounts to retaining patterns with maximal overlap. The non zero averaging can be interpreted as certain patterns being promoted within replicas and enforced across different realizations of $J$. The minimization of free energy gives a self consistent equation for $Q$,
\begin{equation}
Q_{ab} = \braket{s^as^b}
\end{equation}
where $\braket{.}$ refers to averaging over the distribution $P(s^1,s^2,... s^n) =  \frac{1}{Z} e^{-\beta \Tilde{H}}$, where $ \tilde{H}= \sum_{ab}s^a Q_{ab}s^b $. The hamiltonian $\Tilde{H}$ is symmetric with respect to permutation of indices and thus $Q_{ab} = q$ for all $a \neq b$. Minimizing the free energy with respect to the variable $q$ gives a self consistent equation,

\begin{equation} \label{outeq}
    q = \braket{tanh^2(\beta\sqrt{q})z}_z
\end{equation}

where $z$ is a random variable with Gaussian distribution of mean zero and variance $1/N$. For $\beta <1$ i.e, high temperature, $q=0$ is the only solution, representing a paramagnetic phase while for $\beta >1 $, i.e, low temperatures we have a continuously raising $q$ value from zero, suggesting a phase transition. However the replica symmetric saddle point solution for $Q_{ab}$ derived for the Ising Hamiltonian is unstable and is thus inconsistent with physical predictions \cite{1978_stability}.

\subsection{Cavity method}
Cavity method \cite{1986_spinglass} provides for an alternative analysis to the results derived from Replica method. Consider the Ising Hamiltonian over N neurons reordered in the following fashion,
\begin{equation}
    H(s,J) = -s_1h_1 + H_{-1}
\end{equation}
where $h_1 = \sum_{i=2}^{N}J_{1i}s_{i}$ is the local field at site $1$ and $H_{-1} = -\frac{1}{2} \sum_{ij=2}^N J_{ij} s_is_j$ is the remaining Hamiltonian that defines the interaction over other spins. The distribution of $h_1$ in the system of the remaining $N-1$ neurons is given by
\begin{equation}
    P_{-1}(h_1) = \frac{1}{Z_{-1}} \sum_{s_2,..s_N} \delta(h_1 - \sum_{i=2}^{N}J_{1i}s_{i} ) e^{-\beta H_{-1}}
\end{equation}
The joint distribution of the $h_1$, $s_1$ is thus given by,
\begin{equation}
    P_N(s_1,h_1) = \frac{1}{Z} e^{-\beta s_1h_1} P_{-1}(h_1) 
\end{equation}
Since the cavity field in this method decouples with the remaining neurons, we can approximate the distribution for $h_1$ with a Gaussian of mean $\sum_{i=2}^{N} J_{1i}\braket{s_i}_{-1}$ and variance $1 - \frac{1}{N} \sum_{i=1}^{N} \braket{s_i}^{2}_{N}$. The variance has inbuilt into it an approximation of vanishing correlations $\braket{s_is_j}_{-1}$ that is equivalent to the single energy well approximation made in the Replica solution. Under this approximation we can write,
\begin{equation} \label{selfcons}
    P_N(s_1,h_1) \propto \exp(-\beta[s_1h_1 - \frac{1}{2-q}(h_1-\braket{h_1}_{-1}])^2
\end{equation}

We expect $\braket{h_i}_{-i} = \sum_{k \neq i} J_{ik}\braket{s_k}_{-i}$ to be self averaging  and have a Gaussian distribution (as $J_{ik}$ is uncorrelated with $\braket{s_k}_{-i}$) with a $0$ mean and variance $q$ over random realizations of $J_{ik}$ in the large $N$ limit . Replacing the averaging over the neurons with an average over the gaussian distribution we get,
\begin{equation} \label{subs_1}
    \braket{s_1 | \sqrt{q}z, 1-q}_N = tanh\beta \sqrt{q}z
\end{equation}
 
Since all the neurons are equivalent, neuron 1 replaced with any other neuron in equation \ref{selfcons}. The mean activity of neuron $i$ is thus given by,
\begin{equation}
 \braket{s_i| \sqrt{q}z, 1-q}_N = \sum_{s_i} s_i P_N(s_i,h_i)
\end{equation}

We can average over the above expression to write a self consistency condition on $q$. We thus get,
\begin{equation}
    q= \frac{1}{N} \sum_{i=1}^{N} \braket{s_i| \sqrt{q}z, 1-q}_{N}^{2}
\end{equation}
Substituting a generalized version of equation \ref{subs_1} for each neuron $i$ in the above equation we derive \ref{outeq}, obtained from the replica method. 

\subsection{Renormalization group and RBM}

Renormalization group (RG) \cite{PhysRevB.4.3174} is based on the idea that physics describing long range interactions can be obtained by coarse graining degrees of freedom at the short scale. Under this scheme small scale fluctuations get averaged out iteratively and certain relevant features becomes increasingly more prominent. This helps in building effective low energy physics, starting from microscopic description of the system. Despite its exactness and wide usage within the fields of quantum field theory and condensed matter, any form of exact RG computations in large systems is limited by computational power. RG was introduced within the context of Quantum Electrodynamics(QED) \cite{PhysRev.95.1300} and played a crucial role in addressing the problem of infinities. A proper physical understanding was given by Kadanoff within condensed matter systems while proposing the idea of block spin renormalization group \cite{PhysicsPhysiqueFizika.2.263}. This formed the ground for the later seminal work of Kenneth Wilson in producing the scaling laws of correlations near the critical point \cite{wilson}. 

RG can be analytically studied for 1d Ising model, as decimation does not produce additional interaction terms,leaving the hierarchy of effective Hamiltonians tractable. Consider the following Ising spin Hamiltonian with N spins whose interaction is given by,
\begin{equation}
    H_0(\sigma) = J_0 \sum_{i \in [N]} \sigma_{i}\sigma_{i+1} 
\end{equation}
where i runs over all the spins of the current description. Here $J_0$ is the strength of the uniform coupling and no external magnetic field. We study how the couplings transforms by doing a decimation over the odd spins (summing over the degrees of freedom labelled odd). This results in the following hamiltonian that only depends on the remaining $N/2$ spins with no new interaction terms generated,
\begin{equation}
H_1 = J_1 \sum_i \sigma_{i}\sigma_{i+2} \\
\end{equation}
where $J_1 =\frac{1}{2} ln(cosh(2J_0))$. This can be repeated recursively $k$ times giving raise to $H_k$ that depends on the remaining $N/2^k$ spins. 

For the 2d Ising model doing renormalization using spin decimation is not feasible as this produces higher order interactions that are not tractable. Approximations of higher order interactions have been introduced to allow for analytical extensions \cite{RevModPhys.47.773}. At the critical temperature the system exhibits conformal symmetry and this fixes the 2 point and higher point correlations along with the scaling dimensions. To verify that an observable $A'$ defined over the renormalized degrees of freedom remains invariant under renormalization, we will compute the expectation value over the initial probability distribution. Let $p$ refer to the probability distribution generated by the initial hamiltonian $H$ over spins $\sigma$ and $p'$ be the probability distribution generated by the renormalized hamiltonian $H'$ over spins $\sigma'$. Thus,

\begin{equation}
\begin{aligned}
    \braket{A'}_{p}  
    & = \frac{1}{Z}\sum_{\{\sigma\}} e^{-H(\sigma)}A'(\sigma'(\sigma)) \\
    & = \frac{1}{Z} \sum_{\{\sigma'\}} A'(\sigma') \sum_{\{\sigma_{\bot}} e^-H(\sigma_{\bot}) \\
    & = \frac{1}{Z} \sum_{\{\sigma'\}} A'(\sigma') e^{-H'(\sigma')} \\
    & = \braket{A'}_{p'}
\end{aligned}
\end{equation}

Notice that this only for true for observables that are defined on the coarse grained degrees and does not work for those defined on the observables that are defined on the microscopic degrees as these correlations are washed out during renormalization. In the remainder of this section we shall talk about methods of generating RG flows using RBM on a uniform 2d Ising Hamiltonian. Any indication of RBM generating flows that approach criticality like RG should be indicated through correlators that follow the behavior of conformal fields.

RG flows can be well described in the space of parameters that weighs different operators that make up the Hamiltonian. As one coarse grains the Hamiltonian from UV (microscopic description) to an IR (macroscopic description) prescription, we observe that certain parameter weights flow to zero (monotonically decrease). These are termed as irrelevant operators as they play no role in the flow . Operators which regulate the flow with monotonically increasing weights are relevant operators. Within the space of all possible Hamiltonians lies a critical surface where the theory respects conformal symmetry with length scales that run to infinity. When the RG flow meets such a surface it results in a fixed point referred to as critical point. Critical points are usually associated with phase transitions. For example, the critical temperature of uniform 2d Ising with no external magnetic field is given by $T_c = 2.269$ and marks the demarcation between low temperature ferromagnetic and high temperature paramagnetic phases. We shall describe three different methods of generating RBM flows using: (a) Learned weights (b) Variational RG (c) Real Space Mututal Information.  

\subsubsection{From learned weights}
In this method flows are generated through a markov chain of alternatively sampling the hidden and visible layer starting from a given distribution of initial configurations $q_0(v)$ that corresponds to a Hamiltonian with parameters ${\lambda_0}$. The RBM then generates a flow as follows,

\begin{equation}
\begin{aligned}
    & q_0(v) \longrightarrow \Tilde{q}_0(h) = \sum_{v} p(h|v)q_0(v) \\
    & \Tilde{q}_0(h) \longrightarrow q_1(v) = \sum_{h} p(v|h)q_0(h)
\end{aligned}
\end{equation}

This produces a new distribution $q_1(v)$ that corresponds to a flow within the probability distribution and can be seen as the distribution generated by some Hamiltonian of the same statistical model with parameters ${\lambda_1}$. This would correspond to a flow within the parameter space ($\lambda_0 \longrightarrow \lambda_1$). We would like to verify if such an transformation on the parameter space actually corresponds to an RG flow. We do that by implicitly computing correlation functions of certain operators and comparing against known results from RG.

The number of nodes in the successive layers is kept the same as the same RBM is used to produce the flow, unlike RG with reducing degrees of freedom as one flows. Its observed that the RBM flow generated in this method approaches the critical point, despite the RG flow moving away from unstable points. Despite these differences the authors still manage to provide accurate predictions of the cirtical exponents in [15],[16]. At the critical temperature the Ising model enjoy conformal symmetry, giving rise to operators whose correlations scale by well known power laws. The authors of the paper have used $s_{ij} = \sigma_{ij} - \Bar{\sigma}$ that has a weight of $\Delta_s =1/8$ and $\epsilon = s_{ij}(s_{i+1,j} + s_{i-1,j} + s_{i,j+1}+s{i,j-1}) - \Bar{\epsilon}$ that has a weight of $\Delta_{\epsilon} = 1$. The former operator $s_{ij}$ acts as a estimator of reproducing long range correlations, as it dies off faster, while $\epsilon_{ij}$ acts as a estimator for being able to reproduce short correlations when testing on the RBM.

Monte Carlo is used to generate 20000 samples of 10x10 square lattice Ising configurations according Boltzmann distribution at each temperature over 0 to 6 with increments of 0.1. A neural network is trained over these samples with supervised learning to predict the temperature. The RBM is then used to generate flows for temperatures close to the critical temperature. Samples collected from flow lengths of greater than 26 allows for predicting the proportionality constant $A/T_c$ and scaling dimension $\delta_m$ with a very high accuracy by fitting against,

\begin{equation}
    m \propto \frac{|T-T_c|^{\Delta_m}}{T_c}
\end{equation}

The above computation of $\delta_m$ involves using data across flows 2 different temperatures. We could rather compute the scaling dimension $\Delta_s$ and $\Delta_{\epsilon}$ from a single flow of different temperatures. This then allows us to interpolate and predict the dimension for the critical temperature. $\Delta_s$ is reproduced with a very high precision, indicating the RBM flow preserves long range correlation, while high errors in predicting $\delta_{\epsilon}$ shows that short range correlations are usually lost. 

\subsubsection{Variational RG}

Here the hidden layer of an RBM is used to construct the output of a single step of a variational RG \cite{mehta2014exact}. This is unlike the previous method where the number of spins were kept fixed with every iteration. To generate a flow several RBM's are stacked with each one using the output from the previous RBM hidden layers. The correlation pattern between the visible and hidden nodes are studied to check for any RG like connection. The quantity $<v_ih_j>$ as defined below is computed on the block spin renormalization procedure.
\begin{equation}
    \braket{v_ih_a} = \frac{1}{N} \sum_k v_i^{(k)} h_a^{(k)}
\end{equation}

where $v_i$ is a node within the visible layer, $h_j$ a node in the hidden layer, $a$ indexes the samples against which the correlation has been computed and $N$ refers to the total number of samples. This is then used to compute the following correlation function,
\begin{equation}
    \braket{x_ix_j} = \frac{1}{N_h} \sum_{a=1}{N_h} \braket{v_ih_a}{v_jh_a}
\end{equation}

The above correlation is plotted with against $|i-j|$ for renormalization over lattices of different sizes at the critical temperature for a RBM trained on data from a 2d Ising Hamiltonian with nearest neighbour interaction. A fall in correlation with the separation is noticed for large lattices and no pattern is obtained for small Ising lattice keeping the decimation ratio fixed. The RBM thus has managed to preserve the correlations with nearest neighbours showing some reminiscent behaviour of RG under some circumstance. 

\subsubsection{Real Space Mutual Information}

An alternative representation \cite{Koch_Janusz_2018} of block spin renormalization can be defined to capture relevant features by using information theoretic expressions. Lets consider a spin system, where a subset of spins $V$ (visible) are to be effectively represented using spins $H$ (hidden) such that the replacement retains the maximal mutual information with remaining spins (environment) $E$ of the system.  Refer \cite{apenko2011information} for a detailed study about mutual information in the context of RG flow. Thus we would like to maximize 
\begin{equation}
    I(p(e) | p(h)) = \sum_{e,h} p(e,h) log(\frac{p(e,h)}{p(e)p(h)})
\end{equation}

where $p(e)$ is the probability distribution over the environment and p(h) is the probability distribution over the hidden layer. The choice of maximization is motivated from the fact that the coarse grained effective hamiltonian be compact and short ranged (See supplementary in \cite{Koch_Janusz_2018}). We construct p(h) by marginalizing the joint probability distribution over an RBM that provides for $p(v,h)$. The samples for learning $p(v,h)$, $p(e)$ can come from a markov chain with 2 RBMs employed to learn these distributions. The updates to the RBM that learns the distribution $p(e,h)$ comes from minimizing $-I(p(e) | p(h))$. Note that this process needs to be repeated on every iteration of the renormalization procedure.

This procedure reproduces the Kadanoff renormalization when tested on a 2d lattice with 4 visible spins. Samples are generated from a square lattice of size 128x128. The effective temperature can be computed against a neural network trained with samples at different temperature or as described in the earlier section or by plotting the temperature against the mutual information. The procedure reveals a clear separation in the phase transition while predicting the critical temperature with a very high accuracy.

{\color{black}
\subsection{Learnability of Quantum Neural networks} \label{learnability_QNN}
The examples discussed in previous sections demonstrate the power of neural networks with regards to generalization, for problems related to classification and generative modelling. We also seen how some of these classically inspired models can be understood from the standpoint of classical physics. Here we would like to address the question of learnability of quantum models with regards to expressibility, trainability and generalization for Quantum Neural Networks (QNN). The working of QNN involves 2 components:

\begin{itemize}
    \item Feature Encoding layer: Feature extraction is performed on raw classical data to extract relevant features for the model to train on. This might for example include denoising, data compression, privacy filtration. Unlike classical ML, for QNN we need to have an efficient method of encoding the feature output as part of quantum states. One might go for qubit encoding (phase encoding), basis encoding (initialize starting state in the computation basis state) or amplitude encoding (using QRAM).
    \item A Function: In neural networks a generic non linear function this is implemented using layers of fully connected nodes, that seem to resemble multiple layers of RBM stacked over. In the case of QNN, a general unitary constructed from paramaterized quantum circuits is used. The ansatz that defines the unitary could be inspired by the problem at hand or could be a very general multilayered hardware efficient one.
\end{itemize}

Following this an optimizer is enrolled to compute gradients on the cost function of interest. The gradients of these operators may not directly correspond to another unitary operator, in which one needs to re-express them as a sum of terms with term corresponding to an efficient computation of expectation value of some observable for the corresponding unitary that's being output. An example of this would be to use parameters shift rule to re-express gradients as differences.

\subsubsection{Expressibility of QNN}
Just like in DNN where the depth adds to the networks capacity to fit the training data, we would like to quantify the circuits ability to generate states from the Hilbert space. When very little is known about the system that one is dealing with, one might choose to work with generic random ansatz that is agnostic to the system built from hardware efficient elementary gates as layers with repeating elements. The expressibility of such an ansatz \cite{expressibility}  can be expressed in terms of the deviation from the the Haar \cite{Haar} integral,

\begin{equation}
    A^{t} = \Vert \int_{Haar}(V\ket{0}\bra{0}V^{\dag})^{\otimes t} dV - \int_{\theta} (U(\theta)\ket{0}\bra{0}U^{\dag}(\theta))^{\otimes t} d\theta \Vert
\end{equation}

where $\Vert.\Vert$ refers to the Hilbert Schmidt norm and $t$ the moment up to which one would like to approximate. The above definition forms the basis for verifying if a given circuit is a t-design \cite{tdesign} approximation and quantifies the extent to which the ansatz can sample the hilbert space uniformly. Hubregtsen et al \cite{hubregtsen2021evaluation} showed that this correlates to the classification accuracy of the circuits  on MNIST dataset. We would like to next point out that despite expressibility being a good thing to achieve better approximations, the trainability of such ansatz is prone barren plateaus.

\subsubsection{Trainability of QNN}
Let $L(\theta,z)$ represent the loss function we would like to optimize to build the learning model, where $\theta$ represent the parameters to be optimized and $z=\cup_{j=1}^{n}\{(\vec{x}_j,y_j)\}$. Here $\vec{x_j}$ represents the input vector and $y_j$ represents the label assigned to it. Thus the optimization procedure solves the following empirical minimization problem,
\begin{equation}
    \theta^{*} = \underset{\theta}{\mathrm{argmin}} \; L(\theta,z) = \frac{1}{n} \sum_{j=1}^{n} \; l(y_j,\Tilde{y}_j) + \lambda \Vert\theta\Vert^2 
\end{equation}
here $\Tilde{y}_j$ represents the label predicted by the classifier and $\lambda \Vert\theta\Vert^2 $ is a regularization term added to prevent over-fitting. Some of the major sources for errors include noisy quantum gates, decoherence of qubits (ex:depolorizing noise), errors in measurement and errors coming from finite measurement statistics. Having defined a loss function, one can then define the following metrics,

\begin{equation}
\begin{aligned}
        R_1(\theta^T) := \braket{{\Vert \nabla L(\theta) \Vert}} \\
    \quad R_2(\theta^T) := \braket{L(\theta^T)} - L(\theta^{*})
\end{aligned}
\end{equation}

where $\theta^T$ denotes the parameters in the training iteration $T$ and the averaging is done over randomness in the noisy quantum gates and multiple measurements. Here $R_1$ quantifies the rate of convergence to a stationary point  and $R_2$ quantifies the rate of convergence and excess error in the loss function. Yuxuan et al \cite{PRXQuantum.2.040337} showed that $R_1$ and $R_2$ ( for $ \lambda\in [0,(1/3\pi)]\cup[1/\pi,\infty$ satisfy the following bounds,

\begin{equation}
\begin{aligned}
    \quad \quad \quad R_1 \leq \tilde{O} \left[ \mathrm{poly} \; \left( \frac{d}{T(1-p)^L}, \frac{d}{BK(1-p)^L}, \frac{d}{(1-p)^L} \right)\right] \\
    R_2 \leq \tilde{O} \left[ \mathrm{poly} \; \left( \frac{d}{BK^2(1-p)^L} + \frac{d}{(1-p)^L} \right)\right] \\
\end{aligned}
\end{equation}

where $D$ is the number of parameters, $T$ the number of iterations to be executed, $K$ number of measurements made, $B$ batch size used for computing gradients,  $p$ is the gate noise and $L$ is the circuit depth. One key result in establishing these bounds was to show that the empirically estimated gradients via measurements is biased. A multiplicative bias that depends on $(1-p)^L$ and an additive bias that comes from a distribution that depends on the labels, K and $(1-p)^L$. Functionally this marks another distinction between DNN and QNN. The noise models explicitly added to DNN as are bias free and help with the convergence, where as the intrinsic noise that come from gate and measurement errors, results in a bias that degrades learning. The bounds on $R_1$ and $R_2$ 
indicate that increasing $K$,$B$ and reducing $p$, $d$ and $L$ can result in better trainability of the quantum circuit model. We notice that the exponential powering of the noise by the circuit depth $L$, indicates that training deep circuits will be infeasible in the NISQ era.

\subsubsection{Generalizability of QNN}\label{Generalization}
Generalizability is an important aspect of an ML model that caters to the ability of that model to generalize a given task. One way to speak about the generalizability of a model is by looking if it capable of transfering the knowledge learnt from a task to perform a similar task with just a little additional training as opposed to training the model from scratch for the second task. In the work by Andrea Mari et al. \cite{mari2020transfer}, it was shown that a QML model is indeed capable of performing transfer learning. The generic transfer learning approach can be summarized as considering a network trained on a dataset for a particular task, using only the first few layers of this network as a feature extraction network and appending a new network to it that can be trained on a new dataset for a related new task. One can consider the first network to either be classical or quantum and subsequently the second appendable network to also be either classical or quantum, resulting in four possible combinations. The classical-classical network is a common framework, while in this work, the authors provide relevant examples for the other three cases corresponding to classical-quantum, quantum-classical, and quantum-quantum networks, thereby, providing evidence that QML models can be generalized for tasks using transfer learning. 
Generalizability is also the ability for the model to perform well when new data is shown having trained on a given set of data. There have been studies that show the performance of QML models on the testing set for their respective models \cite{abbas2021power, cheng2015learnability}. However, a general framework to study the generalization abilities of QML models was introduced in \cite{banchi2021generalization}. In this work, the authors establish a quantitative metric on the generalization of QML models for classification tasks with an error bound based on the R\'{e}nyi mutual information between the quantum state space and the classical input space, thereafter showing that overfitting does not occur if the number of training pairs considered is greater than base 2 exponentiation of the mutual information.

\subsubsection{Barren Plateaus in training Variational Circuits} \label{barren}

Barren plateaus are characterized by vanishing variance of sample gradients, causing the optimizer to perform random walks in regions on diminishing gradients, with a very low probability of leaving them. McClean \cite{McClean} in 2018 first showed that on an average the variance of the gradient is exponentially vanishing in the number of qubits for any t-design circuit leading to barren plateaus anytime the gradient vanishes. Fig. \ref{mcclean_barren} shows the rapidly falling variance of sample gradients with increasing number of qubits.

\begin{figure}[ht!]
    \centering
    \includegraphics[width=0.4\textwidth]{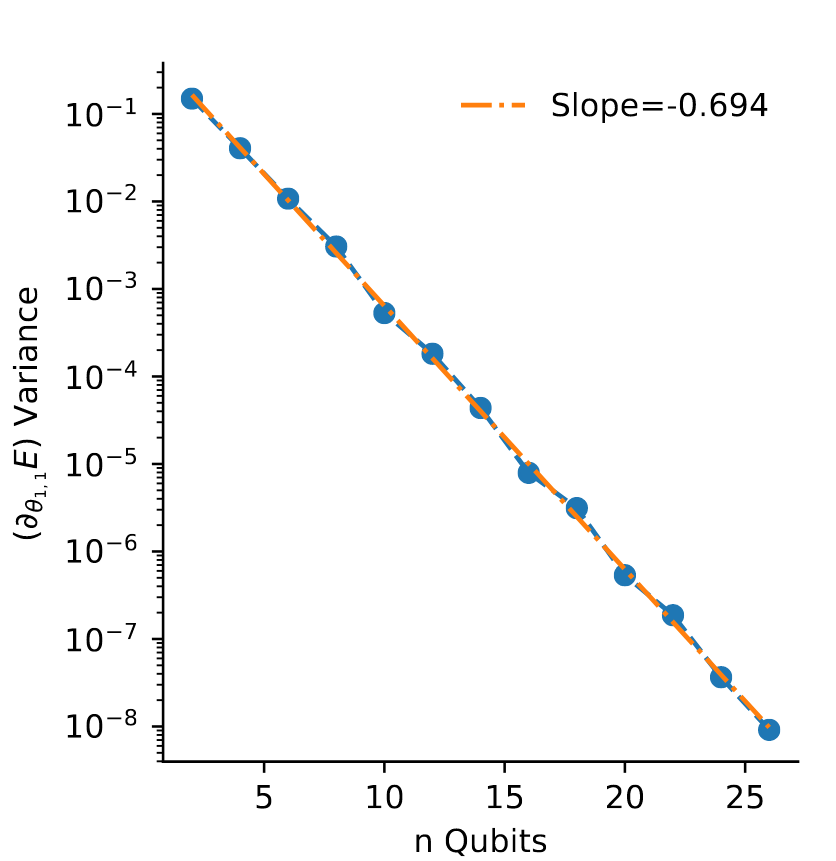}
    \caption{The plot shows sample variance of the gradient of a two-local Pauli term plotted as a function of the number of qubits on a semi-log plot. Reprinted from Ref \cite{McClean} with permission under Creative Commons Attribution 4.0 International License}
    
    \label{mcclean_barren}
\end{figure}

As polynomially deep (in terms of the number of qubits) ansatz built out of 2 qubit unitaries  as form 2-design circuits \cite{2design}, one is likely to encounter barren plateaus while training circuits with sufficiently large number of qubits. Barren plateaus can also arise from the use of global cost functions \cite{globalcost}. Examples of this include computing ground state energies of highly non local hamiltonians and preparing a given density matrix with high fidelity. Lorocca et. al. \cite{barrendiag} shows that for controllable systems (roughly refers to ansatz with highly expressibility) the gradients decay by the dimension of the symmetry subspace to which the initial state prepared by the circuit belongs to (In the lack of any symmetry subspaces, it will be the dimension of the Hilbert space). However one can tailor the ansatz to remain in the uncontrollable regime and sacrifice on expressibility to achieve approximately good results. Another unavoidable source of barren plateaus is the presence of gate noise in NISQ devices. Samson et. al. \cite{noiseinduced} show rigorously that for a noisy circuit the gradients vanish exponentially in the number of qubits with increasing depth.
}

\section{Conclusions} \label{conclusion}

In this review, we have explored some of the popular algorithms in machine learning that is used frequently for many physico-chemical applications. We discussed in detail not only the vanilla protocol implementable on a classical computer but also the quantum computing enhanced variants  wherever applicable. Equipped with this underlying theoretical framework, we thereafter ventured to investigate five distinct domains of applications which includes tomographic state-reconstruction, state-classification methodologies, electronic structure and property prediction, paramaterizing force-fields for molecular dynamics and even drug discovery pipeline. Such an organizational paradigm places equal emphasis on the methodologies and the applications unlike in most other reviews, and is expected to be be beneficial to new entrants in the field especially when supplemented with domain-specific examples as is the case in this review. Last but not the least, we offered an insight into the learning mechanisms, using tools from statistical physics and computer science, that have been used by researchers in recent years to understand the operational mechanism behind the training and feature-extracting ability of the deep-learning algorithms. In particular in the context of Ising Hamiltonains, the Replica and Cavity method provides for calculating observable expectation values that correspond to the least free energy (cost function). We followed this with a discussion on renormalization group searching for connections within deep learning and presented some methods that have been used in exploring the same. The kind of insight we believe reduces the obscurity of these models and the common reluctance associated with the fact that the learning dynamics of these protocols are unreliant on hard-coded physical principles or domain intuition. 

The applications explicated in this review covers a wide spectrum. As discussed even though the quantum-computing enhanced protocols are beginning to be duly recognized, we anticipate that we are still at a nascent stage scratching just the surface. For example, for modeling electronic structure of molecules and materials there already exists a huge variety of methods ranging from learning the functional form of density functionals, approximating wavefunction, to learning the atomic environment descriptors to predict the atom types and properties which have shown great accuracy. Applying these ML methods with the help of quantum computers can further augment our capabilities especially when solving the electronic Schrodinger equation of large and strongly correlated systems are concerned. One can use the variational quantum generator (VQG) based hybrid quantum-classical architecture developed by Romero and Guzik \cite{romero2021variational} in order to generate continuous classical probability distributions to perform chemical analysis, quantum state preparation, etc. The critical point and the critical exponents for a quantum phase transition can be determined using a QML approach through the finite-size scaling (FSS) analysis \cite{bilal2021}. Therefore, the development of QML approaches has a huge role to play in the coming years in the domain of electronic structure, material, and property prediction methods. Similar statements can also be extended to computation of force-fields wherein classical ML techniques even though successful have only efficiently modelled small systems. For the drug discovery pipeline, as has been indicated in Section \ref{Drug_discovery}, key players in both industry and academia are recognizing the potential of quantum computers through investments and collaborative ventures. In the context of interpretability, analytically understanding how generalization of some of the models presented works while increasing the number of parameters with respect to size of input space and the number of learning samples is still open. We would like to have an handle over the limit points of a given deep learning model and perturb to understand the neighborhood space much akin to having a conformal field theory that describes the critical surface and is used to explore the space of quantum field theories. Studies have been conducted in the context of 2d Ising model along the lines of analyzing information flow for RG and deep neural networks \cite{erdmenger2021quantifying}. Convergence towards critical points in these models is in stark contrast with it being an unstable point in the RG flow. It is important that further studies be conducted in models that have self organized criticality to probe if there exists a definitive relation and if the fixed points have anything to tell us about how choices are to be made in the models that we study, with respect to hyperparameters, cost function, optimizer choice and learning ability.

{\color{black}
Since machine learning tasks are data-intensive efficient protocols for loading entries from a high-dimensional classical vector onto a quantum state without smarter preprocessing for feature extraction continues to be a significant challenge. The early domain of QML algorithms included HHL \cite{harrow2009quantum} for PCA and clustering, required the assumption about the oracular presence of qRAM to enable efficient encoding of data. While the development of qRAMs is still an ongoing field of research, recent results claims that the exponential speedup in a subset of such algorithms is only due to the assumption that the encoding of data is efficient \cite{tang2021quantum}. Quantum inspired classical algorithms \cite{tang2019quantum} that manipulate $l^2$ norm sampling distributions provide an exponential speedups in the case of recommendation systems imply the lack of provability concerning the quantum speedups of certain early QML algorithms. Another primary concern for the development of any quantum algorithms even beyond ML applications is the inherent presence of noise manifested from shorter coherence times of qubits and greater gate-infidelities especially of multi-qubit operations. The fundamental research related to the development of better qubits, improving gate fidelities in unitary operations, and improving the qubit connectivity is very much an active field of investigation among hardware engineers and physicists. New reports have been demonstrated with protected qubits resilient against certain kind of hardware noises \cite{Dou_ot_2012}. Fault-tolerant quantum computation wherein logical qubits are protected using more physical qubits like in stabilizer codes \cite{PhysRevA.86.032324} or qubit configurations based on topological properties of the underlying interactions \cite{yao2012experimental, RevModPhys.87.307} have been proposed and is actively under development. First-ever such operation has been recently demonstrated on a trapped-ion platform \cite{egan2020fault}. The process of such error correction can itself suffer from noise which can be mitigated by the quantum fault-tolerant threshold theorem \cite{aharonov2008fault} provided noise levels are low. Partial suppression of bit and phase-flip errors have also been demonstrated \cite{chen2021exponential}. On the algorithmic side, algorithms that utilize the specific problem structure smartly have also been proposed\cite{parra2020digital}. One also needs to thoroughly understand the noise resilience of some of the existing methods and investigate how much of hardware noise can be tolerated before the results are corrupted beyond a certain threshold and the proclaimed quantum advantages are lost. Proper certification schemes and figures of merit for benchmarking such algorithms are beginning to gain  attention \cite{eisert2020quantum}. With the increased activity on developing quantum ML algorithms underway, creating a provision for generalizability of these models is an important consideration and this aspect has been already discussed in the Section \ref{Generalization}.
Some of the key open questions in this area would be a proper theoretical demonstration of asymptotic universality (as was discussed in Section \ref{Case_QML}) for the function class which quantum models can learn in presence of trainable unitaries of finite circuit depth \cite{schuld2021effect} thereby relaxing the assumptions used thereof. Another interesting question would be proposing real-life applications tailored to take advantages of the universality in such function classes such that quantum benefits over classical learning can be seen. Resource dependance of such algorithms from the perspective foundational aspects of quantum mechanics is also an interesting avenue for research. With regards to trainability of ML models one of the major menace to tackle is the presence of barren plateaus (see Section \ref{barren}) in exploring high dimensional feature spaces to find optimal parameters that minimize the cost function. Much of the questions concerning how the possibility of such exponentially vanishing gradients needs to be handled and mitigated are essentially open to further investigation.}

One must also note that there are other applications which have not been discussed in this review at all. Perhaps the most important one from the point of view of chemistry is modelling chemical reactions and computer aided rational design of molecules and synthetic strategies. In this technique one considers retro-synthetic pathways arising  from a given product until a set of precursors which are commercially available or synthetically known in literature is obtained. Such pathways are scored on efficacy based on number of reactions involved, intermediates, reaction conditions etc. Two different kinds of strategies are known in this regard. The first involves retrosynthetic disconnection based on domain knowledge or commonly used chemistry-inspired rules followed by subsequent ranking of the precursor steps. This can suffer for unknown or rare reactions where such intuition may not be available. The second category uses simple translation of the molecular descriptors of the reactants into products as is used in machine-induced linguistic translations. Promising results have been obtained for either category using ML/DL algorithms \cite{schreck2019learning, baylon2019enhancing, dai2020retrosynthesis, lin2019automatic, liu2017retrosynthetic}. For further information, the reader may consult already existing topical reviews like \cite{JOHANSSON201965, strieth2020machine}. To the best of our knowledge, the role of quantum computing in this area have not been explored.  Another area which is very important is understanding non-unitary dynamical evolution of quantum systems and the role of coupling to the environment and emergence of decoherence \cite{breuer2002theory, Hu2020AQA}. Such open system dynamics have also begun to receive attention from the point of view of machine learning wherein the density matrix of the state is encoded as within an efficiently constructible ansatz. 
{\color{black} In a recent report \cite{ullah2021speeding} Kernel-Ridge Regression (see Section \ref{KKR_sec}) has been used to faithfully recover long-time dynamical averages of the spin-boson model when linearly coupled to a harmonic bath characterized by the Drude-Lorentz spectral density. Hierarchical equation of motion approach (HEOM) was used to train the model using short-time trajectories but the results when extrapolated beyond the training time intervals using Gaussian kernels leads to unprecedented accuracy.  LSTM networks (see Section \ref{RNN_section}) have been used to model dynamical evolution of density operators for a coupled two-level system vibronically coupled to a harmonic bath \cite{lin2021simulation}. The population difference between the two levels and the real and imaginary part of the coherence was used as time series data for training at shorter times from the numerically exact multi-layer multi-configurational Time Dependant Hartree method (ML-MCTDH). Remarkable accuracy was seen being preserved even in the long-time limit. A similar result was also obtained with CNN \cite{herrera2021convolutional} (see Section \ref{CNN_section}) where input training data was the density matrix elements at various time steps and the prediction of the network through successive series of convolutions and max-pooling yielded accurate values of averages of the system operators (like the Pauli-z or $\sigma_z(t)$). For further elaboration on other such methods, the interested reader is referred to \cite{PhysRevResearch.3.023084, luchnikov2020machine, hase2017machine, PhysRevResearch.3.023095}. 
}

Yet another promising area which is left untouched here is the use of physics-inspired machine learning algorithms which even though is beginning to gain attention in problems of physical or technological interests \cite{KHAN2020135628, villmann2020quantum, bellinger2020reinforcement, trenti2020quantum, tiwari2019towards} but has been sparsely adopted in chemistry \cite{musil2021physics}. Reader may consult a recent review for further discussion \cite{karniadakis2021physics}. We thus see that the road ahead is ripe with possibilities that can be explored in future especially for the quantum-computing based ML variants. Hopefully with better error mitigating strategies \cite{bitflip_suppress} and large scale devices with over 1000 qubits being promised in recent future by tech-giants \cite{IBM_roadmap}, this burgeoning field will pick up momentum with enhanced capabilities to conduct many pioneering investigation.



\section*{Conflicts of interest}

There are no conflicts of interest to declare.

\section*{Acknowledgements}
The authors acknowledge funding by the U.S. Department of Energy (Office of Basic Energy Sciences) under Award No. DE-SC0019215,  the National Science Foundation under Award No. 1955907 and the  U.S. Department of Energy, Office of Science, National Quantum Information Science Research Centers, Quantum Science Center.



\balance



\begin{mcitethebibliography}{768}
\providecommand*{\natexlab}[1]{#1}
\providecommand*{\mciteSetBstSublistMode}[1]{}
\providecommand*{\mciteSetBstMaxWidthForm}[2]{}
\providecommand*{\mciteBstWouldAddEndPuncttrue}
  {\def\EndOfBibitem{\unskip.}}
\providecommand*{\mciteBstWouldAddEndPunctfalse}
  {\let\EndOfBibitem\relax}
\providecommand*{\mciteSetBstMidEndSepPunct}[3]{}
\providecommand*{\mciteSetBstSublistLabelBeginEnd}[3]{}
\providecommand*{\EndOfBibitem}{}
\mciteSetBstSublistMode{f}
\mciteSetBstMaxWidthForm{subitem}
{(\emph{\alph{mcitesubitemcount}})}
\mciteSetBstSublistLabelBeginEnd{\mcitemaxwidthsubitemform\space}
{\relax}{\relax}

\bibitem[Li(2017)]{li2017deep}
H.~Li, \emph{National Science Review}, 2017\relax
\mciteBstWouldAddEndPuncttrue
\mciteSetBstMidEndSepPunct{\mcitedefaultmidpunct}
{\mcitedefaultendpunct}{\mcitedefaultseppunct}\relax
\EndOfBibitem
\bibitem[Torfi \emph{et~al.}(2021)Torfi, Shirvani, Keneshloo, Tavaf, and
  Fox]{torfi2021natural}
A.~Torfi, R.~A. Shirvani, Y.~Keneshloo, N.~Tavaf and E.~A. Fox, \emph{Natural
  Language Processing Advancements By Deep Learning: A Survey}, 2021\relax
\mciteBstWouldAddEndPuncttrue
\mciteSetBstMidEndSepPunct{\mcitedefaultmidpunct}
{\mcitedefaultendpunct}{\mcitedefaultseppunct}\relax
\EndOfBibitem
\bibitem[Nagarhalli \emph{et~al.}(2021)Nagarhalli, Vaze, and
  Rana]{nagarhalli2021impact}
T.~P. Nagarhalli, V.~Vaze and N.~Rana, 2021 Third International Conference on
  Intelligent Communication Technologies and Virtual Mobile Networks (ICICV),
  2021, pp. 1529--1534\relax
\mciteBstWouldAddEndPuncttrue
\mciteSetBstMidEndSepPunct{\mcitedefaultmidpunct}
{\mcitedefaultendpunct}{\mcitedefaultseppunct}\relax
\EndOfBibitem
\bibitem[Wu \emph{et~al.}(2016)Wu, Schuster, Chen, Le, Norouzi, Macherey,
  Krikun, Cao, Gao, Macherey,\emph{et~al.}]{wu2016google}
Y.~Wu, M.~Schuster, Z.~Chen, Q.~V. Le, M.~Norouzi, W.~Macherey, M.~Krikun,
  Y.~Cao, Q.~Gao, K.~Macherey \emph{et~al.}, \emph{arXiv preprint
  arXiv:1609.08144}, 2016\relax
\mciteBstWouldAddEndPuncttrue
\mciteSetBstMidEndSepPunct{\mcitedefaultmidpunct}
{\mcitedefaultendpunct}{\mcitedefaultseppunct}\relax
\EndOfBibitem
\bibitem[Lopez(2008)]{lopez2008statistical}
A.~Lopez, \emph{ACM Computing Surveys (CSUR)}, 2008, \textbf{40}, 1--49\relax
\mciteBstWouldAddEndPuncttrue
\mciteSetBstMidEndSepPunct{\mcitedefaultmidpunct}
{\mcitedefaultendpunct}{\mcitedefaultseppunct}\relax
\EndOfBibitem
\bibitem[Tuncali \emph{et~al.}(2018)Tuncali, Fainekos, Ito, and
  Kapinski]{tuncali2018simulation}
C.~E. Tuncali, G.~Fainekos, H.~Ito and J.~Kapinski, 2018 IEEE Intelligent
  Vehicles Symposium (IV), 2018, pp. 1555--1562\relax
\mciteBstWouldAddEndPuncttrue
\mciteSetBstMidEndSepPunct{\mcitedefaultmidpunct}
{\mcitedefaultendpunct}{\mcitedefaultseppunct}\relax
\EndOfBibitem
\bibitem[Janai \emph{et~al.}(2020)Janai, G{\"u}ney, Behl,
  Geiger,\emph{et~al.}]{janai2020computer}
J.~Janai, F.~G{\"u}ney, A.~Behl, A.~Geiger \emph{et~al.}, \emph{Foundations and
  Trends{\textregistered} in Computer Graphics and Vision}, 2020, \textbf{12},
  1--308\relax
\mciteBstWouldAddEndPuncttrue
\mciteSetBstMidEndSepPunct{\mcitedefaultmidpunct}
{\mcitedefaultendpunct}{\mcitedefaultseppunct}\relax
\EndOfBibitem
\bibitem[Daily \emph{et~al.}(2017)Daily, Medasani, Behringer, and
  Trivedi]{daily2017self}
M.~Daily, S.~Medasani, R.~Behringer and M.~Trivedi, \emph{Computer}, 2017,
  \textbf{50}, 18--23\relax
\mciteBstWouldAddEndPuncttrue
\mciteSetBstMidEndSepPunct{\mcitedefaultmidpunct}
{\mcitedefaultendpunct}{\mcitedefaultseppunct}\relax
\EndOfBibitem
\bibitem[Siau and Wang(2018)]{siau2018building}
K.~Siau and W.~Wang, \emph{Cutter business technology journal}, 2018,
  \textbf{31}, 47--53\relax
\mciteBstWouldAddEndPuncttrue
\mciteSetBstMidEndSepPunct{\mcitedefaultmidpunct}
{\mcitedefaultendpunct}{\mcitedefaultseppunct}\relax
\EndOfBibitem
\bibitem[Pierson and Gashler(2017)]{pierson2017deep}
H.~A. Pierson and M.~S. Gashler, \emph{Advanced Robotics}, 2017, \textbf{31},
  821--835\relax
\mciteBstWouldAddEndPuncttrue
\mciteSetBstMidEndSepPunct{\mcitedefaultmidpunct}
{\mcitedefaultendpunct}{\mcitedefaultseppunct}\relax
\EndOfBibitem
\bibitem[He \emph{et~al.}(2015)He, Zhang, Ren, and Sun]{he2015delving}
K.~He, X.~Zhang, S.~Ren and J.~Sun, Proceedings of the IEEE international
  conference on computer vision, 2015, pp. 1026--1034\relax
\mciteBstWouldAddEndPuncttrue
\mciteSetBstMidEndSepPunct{\mcitedefaultmidpunct}
{\mcitedefaultendpunct}{\mcitedefaultseppunct}\relax
\EndOfBibitem
\bibitem[He \emph{et~al.}(2016)He, Zhang, Ren, and Sun]{he2016deep}
K.~He, X.~Zhang, S.~Ren and J.~Sun, Proceedings of the IEEE conference on
  computer vision and pattern recognition, 2016, pp. 770--778\relax
\mciteBstWouldAddEndPuncttrue
\mciteSetBstMidEndSepPunct{\mcitedefaultmidpunct}
{\mcitedefaultendpunct}{\mcitedefaultseppunct}\relax
\EndOfBibitem
\bibitem[Wu and Chen(2015)]{wu2015image}
M.~Wu and L.~Chen, 2015 Chinese Automation Congress (CAC), 2015, pp.
  542--546\relax
\mciteBstWouldAddEndPuncttrue
\mciteSetBstMidEndSepPunct{\mcitedefaultmidpunct}
{\mcitedefaultendpunct}{\mcitedefaultseppunct}\relax
\EndOfBibitem
\bibitem[Covington \emph{et~al.}(2016)Covington, Adams, and
  Sargin]{covington2016deep}
P.~Covington, J.~Adams and E.~Sargin, Proceedings of the 10th ACM conference on
  recommender systems, 2016, pp. 191--198\relax
\mciteBstWouldAddEndPuncttrue
\mciteSetBstMidEndSepPunct{\mcitedefaultmidpunct}
{\mcitedefaultendpunct}{\mcitedefaultseppunct}\relax
\EndOfBibitem
\bibitem[Pazzani and Billsus(1997)]{pazzani1997learning}
M.~Pazzani and D.~Billsus, \emph{Machine learning}, 1997, \textbf{27},
  313--331\relax
\mciteBstWouldAddEndPuncttrue
\mciteSetBstMidEndSepPunct{\mcitedefaultmidpunct}
{\mcitedefaultendpunct}{\mcitedefaultseppunct}\relax
\EndOfBibitem
\bibitem[Dada \emph{et~al.}(2019)Dada, Bassi, Chiroma, Adetunmbi,
  Ajibuwa,\emph{et~al.}]{dada2019machine}
E.~G. Dada, J.~S. Bassi, H.~Chiroma, A.~O. Adetunmbi, O.~E. Ajibuwa
  \emph{et~al.}, \emph{Heliyon}, 2019, \textbf{5}, e01802\relax
\mciteBstWouldAddEndPuncttrue
\mciteSetBstMidEndSepPunct{\mcitedefaultmidpunct}
{\mcitedefaultendpunct}{\mcitedefaultseppunct}\relax
\EndOfBibitem
\bibitem[Guzella and Caminhas(2009)]{guzella2009review}
T.~S. Guzella and W.~M. Caminhas, \emph{Expert Systems with Applications},
  2009, \textbf{36}, 10206--10222\relax
\mciteBstWouldAddEndPuncttrue
\mciteSetBstMidEndSepPunct{\mcitedefaultmidpunct}
{\mcitedefaultendpunct}{\mcitedefaultseppunct}\relax
\EndOfBibitem
\bibitem[Lengauer \emph{et~al.}(2007)Lengauer, Sander, Sierra, Thielen, and
  Kaiser]{lengauer2007bioinformatics}
T.~Lengauer, O.~Sander, S.~Sierra, A.~Thielen and R.~Kaiser, \emph{Nature
  biotechnology}, 2007, \textbf{25}, 1407--1410\relax
\mciteBstWouldAddEndPuncttrue
\mciteSetBstMidEndSepPunct{\mcitedefaultmidpunct}
{\mcitedefaultendpunct}{\mcitedefaultseppunct}\relax
\EndOfBibitem
\bibitem[Larranaga \emph{et~al.}(2006)Larranaga, Calvo, Santana, Bielza,
  Galdiano, Inza, Lozano, Armananzas, Santaf{\'e},
  P{\'e}rez,\emph{et~al.}]{larranaga2006machine}
P.~Larranaga, B.~Calvo, R.~Santana, C.~Bielza, J.~Galdiano, I.~Inza, J.~A.
  Lozano, R.~Armananzas, G.~Santaf{\'e}, A.~P{\'e}rez \emph{et~al.},
  \emph{Briefings in bioinformatics}, 2006, \textbf{7}, 86--112\relax
\mciteBstWouldAddEndPuncttrue
\mciteSetBstMidEndSepPunct{\mcitedefaultmidpunct}
{\mcitedefaultendpunct}{\mcitedefaultseppunct}\relax
\EndOfBibitem
\bibitem[Erickson \emph{et~al.}(2017)Erickson, Korfiatis, Akkus, and
  Kline]{erickson2017machine}
B.~J. Erickson, P.~Korfiatis, Z.~Akkus and T.~L. Kline, \emph{Radiographics},
  2017, \textbf{37}, 505--515\relax
\mciteBstWouldAddEndPuncttrue
\mciteSetBstMidEndSepPunct{\mcitedefaultmidpunct}
{\mcitedefaultendpunct}{\mcitedefaultseppunct}\relax
\EndOfBibitem
\bibitem[Sakhavi \emph{et~al.}(2018)Sakhavi, Guan, and
  Yan]{sakhavi2018learning}
S.~Sakhavi, C.~Guan and S.~Yan, \emph{IEEE transactions on neural networks and
  learning systems}, 2018, \textbf{29}, 5619--5629\relax
\mciteBstWouldAddEndPuncttrue
\mciteSetBstMidEndSepPunct{\mcitedefaultmidpunct}
{\mcitedefaultendpunct}{\mcitedefaultseppunct}\relax
\EndOfBibitem
\bibitem[Grimmer \emph{et~al.}(2021)Grimmer, Roberts, and
  Stewart]{grimmer2021machine}
J.~Grimmer, M.~E. Roberts and B.~M. Stewart, \emph{Annual Review of Political
  Science}, 2021, \textbf{24}, 395--419\relax
\mciteBstWouldAddEndPuncttrue
\mciteSetBstMidEndSepPunct{\mcitedefaultmidpunct}
{\mcitedefaultendpunct}{\mcitedefaultseppunct}\relax
\EndOfBibitem
\bibitem[Heaton \emph{et~al.}(2017)Heaton, Polson, and Witte]{heaton2017deep}
J.~B. Heaton, N.~G. Polson and J.~H. Witte, \emph{Applied Stochastic Models in
  Business and Industry}, 2017, \textbf{33}, 3--12\relax
\mciteBstWouldAddEndPuncttrue
\mciteSetBstMidEndSepPunct{\mcitedefaultmidpunct}
{\mcitedefaultendpunct}{\mcitedefaultseppunct}\relax
\EndOfBibitem
\bibitem[Bansak \emph{et~al.}(2018)Bansak, Ferwerda, Hainmueller, Dillon,
  Hangartner, Lawrence, and Weinstein]{bansak2018improving}
K.~Bansak, J.~Ferwerda, J.~Hainmueller, A.~Dillon, D.~Hangartner, D.~Lawrence
  and J.~Weinstein, \emph{Science}, 2018, \textbf{359}, 325--329\relax
\mciteBstWouldAddEndPuncttrue
\mciteSetBstMidEndSepPunct{\mcitedefaultmidpunct}
{\mcitedefaultendpunct}{\mcitedefaultseppunct}\relax
\EndOfBibitem
\bibitem[Van Der~Maaten \emph{et~al.}(2009)Van Der~Maaten, Postma, Van~den
  Herik,\emph{et~al.}]{van2009dimensionality}
L.~Van Der~Maaten, E.~Postma, J.~Van~den Herik \emph{et~al.}, \emph{J Mach
  Learn Res}, 2009, \textbf{10}, 13\relax
\mciteBstWouldAddEndPuncttrue
\mciteSetBstMidEndSepPunct{\mcitedefaultmidpunct}
{\mcitedefaultendpunct}{\mcitedefaultseppunct}\relax
\EndOfBibitem
\bibitem[Carleo \emph{et~al.}(2019)Carleo, Cirac, Cranmer, Daudet, Schuld,
  Tishby, Vogt-Maranto, and Zdeborov{\'a}]{carleo2019machine}
G.~Carleo, I.~Cirac, K.~Cranmer, L.~Daudet, M.~Schuld, N.~Tishby,
  L.~Vogt-Maranto and L.~Zdeborov{\'a}, \emph{Reviews of Modern Physics}, 2019,
  \textbf{91}, 045002\relax
\mciteBstWouldAddEndPuncttrue
\mciteSetBstMidEndSepPunct{\mcitedefaultmidpunct}
{\mcitedefaultendpunct}{\mcitedefaultseppunct}\relax
\EndOfBibitem
\bibitem[Ball and Brunner(2010)]{ball2010data}
N.~M. Ball and R.~J. Brunner, \emph{International Journal of Modern Physics D},
  2010, \textbf{19}, 1049--1106\relax
\mciteBstWouldAddEndPuncttrue
\mciteSetBstMidEndSepPunct{\mcitedefaultmidpunct}
{\mcitedefaultendpunct}{\mcitedefaultseppunct}\relax
\EndOfBibitem
\bibitem[Ivezi{\'c} \emph{et~al.}(2014)Ivezi{\'c}, Connolly, VanderPlas, and
  Gray]{ivezic2014statistics}
{\v{Z}}.~Ivezi{\'c}, A.~J. Connolly, J.~T. VanderPlas and A.~Gray,
  \emph{Statistics, data mining, and machine learning in astronomy}, Princeton
  University Press, 2014\relax
\mciteBstWouldAddEndPuncttrue
\mciteSetBstMidEndSepPunct{\mcitedefaultmidpunct}
{\mcitedefaultendpunct}{\mcitedefaultseppunct}\relax
\EndOfBibitem
\bibitem[Radovic \emph{et~al.}(2018)Radovic, Williams, Rousseau, Kagan,
  Bonacorsi, Himmel, Aurisano, Terao, and Wongjirad]{radovic2018machine}
A.~Radovic, M.~Williams, D.~Rousseau, M.~Kagan, D.~Bonacorsi, A.~Himmel,
  A.~Aurisano, K.~Terao and T.~Wongjirad, \emph{Nature}, 2018, \textbf{560},
  41--48\relax
\mciteBstWouldAddEndPuncttrue
\mciteSetBstMidEndSepPunct{\mcitedefaultmidpunct}
{\mcitedefaultendpunct}{\mcitedefaultseppunct}\relax
\EndOfBibitem
\bibitem[Wigley \emph{et~al.}(2016)Wigley, Everitt, van~den Hengel, Bastian,
  Sooriyabandara, McDonald, Hardman, Quinlivan, Manju,
  Kuhn,\emph{et~al.}]{wigley2016fast}
P.~B. Wigley, P.~J. Everitt, A.~van~den Hengel, J.~W. Bastian, M.~A.
  Sooriyabandara, G.~D. McDonald, K.~S. Hardman, C.~D. Quinlivan, P.~Manju,
  C.~C. Kuhn \emph{et~al.}, \emph{Scientific reports}, 2016, \textbf{6},
  1--6\relax
\mciteBstWouldAddEndPuncttrue
\mciteSetBstMidEndSepPunct{\mcitedefaultmidpunct}
{\mcitedefaultendpunct}{\mcitedefaultseppunct}\relax
\EndOfBibitem
\bibitem[Zhou \emph{et~al.}(2019)Zhou, Huang, Yan, and
  B{\"u}nzli]{zhou2019emerging}
J.~Zhou, B.~Huang, Z.~Yan and J.-C.~G. B{\"u}nzli, \emph{Light: Science \&
  Applications}, 2019, \textbf{8}, 1--7\relax
\mciteBstWouldAddEndPuncttrue
\mciteSetBstMidEndSepPunct{\mcitedefaultmidpunct}
{\mcitedefaultendpunct}{\mcitedefaultseppunct}\relax
\EndOfBibitem
\bibitem[Chattopadhyay \emph{et~al.}(2020)Chattopadhyay, Nabizadeh, and
  Hassanzadeh]{chattopadhyay2020analog}
A.~Chattopadhyay, E.~Nabizadeh and P.~Hassanzadeh, \emph{Journal of Advances in
  Modeling Earth Systems}, 2020, \textbf{12}, e2019MS001958\relax
\mciteBstWouldAddEndPuncttrue
\mciteSetBstMidEndSepPunct{\mcitedefaultmidpunct}
{\mcitedefaultendpunct}{\mcitedefaultseppunct}\relax
\EndOfBibitem
\bibitem[Ren \emph{et~al.}(2021)Ren, Li, Ren, Song, Xu, Deng, and
  Wang]{ren2021deep}
X.~Ren, X.~Li, K.~Ren, J.~Song, Z.~Xu, K.~Deng and X.~Wang, \emph{Big Data
  Research}, 2021, \textbf{23}, 100178\relax
\mciteBstWouldAddEndPuncttrue
\mciteSetBstMidEndSepPunct{\mcitedefaultmidpunct}
{\mcitedefaultendpunct}{\mcitedefaultseppunct}\relax
\EndOfBibitem
\bibitem[Monson \emph{et~al.}(2018)Monson, Armitage, and
  Hlusko]{monson2018using}
T.~A. Monson, D.~W. Armitage and L.~J. Hlusko, \emph{PaleoBios}, 2018,
  \textbf{35}, 1--20\relax
\mciteBstWouldAddEndPuncttrue
\mciteSetBstMidEndSepPunct{\mcitedefaultmidpunct}
{\mcitedefaultendpunct}{\mcitedefaultseppunct}\relax
\EndOfBibitem
\bibitem[Spradley \emph{et~al.}(2019)Spradley, Glazer, and
  Kay]{spradley2019mammalian}
J.~P. Spradley, B.~J. Glazer and R.~F. Kay, \emph{Palaeogeography,
  Palaeoclimatology, Palaeoecology}, 2019, \textbf{518}, 155--171\relax
\mciteBstWouldAddEndPuncttrue
\mciteSetBstMidEndSepPunct{\mcitedefaultmidpunct}
{\mcitedefaultendpunct}{\mcitedefaultseppunct}\relax
\EndOfBibitem
\bibitem[Romero \emph{et~al.}(2020)Romero, Kong, Fowlkes, Jaramillo, Urban,
  Oboh-Ikuenobe, D’Apolito, and Punyasena]{romero2020improving}
I.~C. Romero, S.~Kong, C.~C. Fowlkes, C.~Jaramillo, M.~A. Urban,
  F.~Oboh-Ikuenobe, C.~D’Apolito and S.~W. Punyasena, \emph{Proceedings of
  the National Academy of Sciences}, 2020, \textbf{117}, 28496--28505\relax
\mciteBstWouldAddEndPuncttrue
\mciteSetBstMidEndSepPunct{\mcitedefaultmidpunct}
{\mcitedefaultendpunct}{\mcitedefaultseppunct}\relax
\EndOfBibitem
\bibitem[Schleder \emph{et~al.}(2019)Schleder, Padilha, Acosta, Costa, and
  Fazzio]{schleder2019dft}
G.~R. Schleder, A.~C. Padilha, C.~M. Acosta, M.~Costa and A.~Fazzio,
  \emph{Journal of Physics: Materials}, 2019, \textbf{2}, 032001\relax
\mciteBstWouldAddEndPuncttrue
\mciteSetBstMidEndSepPunct{\mcitedefaultmidpunct}
{\mcitedefaultendpunct}{\mcitedefaultseppunct}\relax
\EndOfBibitem
\bibitem[Behler(2016)]{behler2016perspective}
J.~Behler, \emph{The Journal of chemical physics}, 2016, \textbf{145},
  170901\relax
\mciteBstWouldAddEndPuncttrue
\mciteSetBstMidEndSepPunct{\mcitedefaultmidpunct}
{\mcitedefaultendpunct}{\mcitedefaultseppunct}\relax
\EndOfBibitem
\bibitem[von Lilienfeld and Burke(2020)]{von2020retrospective}
O.~A. von Lilienfeld and K.~Burke, \emph{Nature communications}, 2020,
  \textbf{11}, 1--4\relax
\mciteBstWouldAddEndPuncttrue
\mciteSetBstMidEndSepPunct{\mcitedefaultmidpunct}
{\mcitedefaultendpunct}{\mcitedefaultseppunct}\relax
\EndOfBibitem
\bibitem[Liu \emph{et~al.}(2020)Liu, Niu, Wang, Gan, Zhu, Sun, and
  Shen]{liu2020machine}
Y.~Liu, C.~Niu, Z.~Wang, Y.~Gan, Y.~Zhu, S.~Sun and T.~Shen, \emph{Journal of
  Materials Science \& Technology}, 2020, \textbf{57}, 113--122\relax
\mciteBstWouldAddEndPuncttrue
\mciteSetBstMidEndSepPunct{\mcitedefaultmidpunct}
{\mcitedefaultendpunct}{\mcitedefaultseppunct}\relax
\EndOfBibitem
\bibitem[Strieth-Kalthoff \emph{et~al.}(2020)Strieth-Kalthoff, Sandfort,
  Segler, and Glorius]{strieth2020machine}
F.~Strieth-Kalthoff, F.~Sandfort, M.~H. Segler and F.~Glorius, \emph{Chemical
  Society Reviews}, 2020, \textbf{49}, 6154--6168\relax
\mciteBstWouldAddEndPuncttrue
\mciteSetBstMidEndSepPunct{\mcitedefaultmidpunct}
{\mcitedefaultendpunct}{\mcitedefaultseppunct}\relax
\EndOfBibitem
\bibitem[Coley \emph{et~al.}(2018)Coley, Green, and Jensen]{coley2018machine}
C.~W. Coley, W.~H. Green and K.~F. Jensen, \emph{Accounts of chemical
  research}, 2018, \textbf{51}, 1281--1289\relax
\mciteBstWouldAddEndPuncttrue
\mciteSetBstMidEndSepPunct{\mcitedefaultmidpunct}
{\mcitedefaultendpunct}{\mcitedefaultseppunct}\relax
\EndOfBibitem
\bibitem[Fu \emph{et~al.}(2020)Fu, Li, Wang, Li, Liu, Wu, Zhao, Ding, Wan,
  Zhong,\emph{et~al.}]{fu2020optimizing}
Z.~Fu, X.~Li, Z.~Wang, Z.~Li, X.~Liu, X.~Wu, J.~Zhao, X.~Ding, X.~Wan, F.~Zhong
  \emph{et~al.}, \emph{Organic Chemistry Frontiers}, 2020, \textbf{7},
  2269--2277\relax
\mciteBstWouldAddEndPuncttrue
\mciteSetBstMidEndSepPunct{\mcitedefaultmidpunct}
{\mcitedefaultendpunct}{\mcitedefaultseppunct}\relax
\EndOfBibitem
\bibitem[Fooshee \emph{et~al.}(2018)Fooshee, Mood, Gutman, Tavakoli, Urban,
  Liu, Huynh, Van~Vranken, and Baldi]{C7ME00107J}
D.~Fooshee, A.~Mood, E.~Gutman, M.~Tavakoli, G.~Urban, F.~Liu, N.~Huynh,
  D.~Van~Vranken and P.~Baldi, \emph{Mol. Syst. Des. Eng.}, 2018, \textbf{3},
  442--452\relax
\mciteBstWouldAddEndPuncttrue
\mciteSetBstMidEndSepPunct{\mcitedefaultmidpunct}
{\mcitedefaultendpunct}{\mcitedefaultseppunct}\relax
\EndOfBibitem
\bibitem[Hu \emph{et~al.}(2018)Hu, Xie, Li, Li, and Lan]{hu2018inclusion}
D.~Hu, Y.~Xie, X.~Li, L.~Li and Z.~Lan, \emph{The journal of physical chemistry
  letters}, 2018, \textbf{9}, 2725--2732\relax
\mciteBstWouldAddEndPuncttrue
\mciteSetBstMidEndSepPunct{\mcitedefaultmidpunct}
{\mcitedefaultendpunct}{\mcitedefaultseppunct}\relax
\EndOfBibitem
\bibitem[Amabilino \emph{et~al.}(2019)Amabilino, Bratholm, Bennie, Vaucher,
  Reiher, and Glowacki]{amabilino2019training}
S.~Amabilino, L.~A. Bratholm, S.~J. Bennie, A.~C. Vaucher, M.~Reiher and D.~R.
  Glowacki, \emph{The Journal of Physical Chemistry A}, 2019, \textbf{123},
  4486--4499\relax
\mciteBstWouldAddEndPuncttrue
\mciteSetBstMidEndSepPunct{\mcitedefaultmidpunct}
{\mcitedefaultendpunct}{\mcitedefaultseppunct}\relax
\EndOfBibitem
\bibitem[Kumar \emph{et~al.}(2021)Kumar, Loharch, Kumar, Ringe, and
  Parkesh]{kumar2021exploiting}
A.~Kumar, S.~Loharch, S.~Kumar, R.~P. Ringe and R.~Parkesh, \emph{Computational
  and structural biotechnology journal}, 2021, \textbf{19}, 424--438\relax
\mciteBstWouldAddEndPuncttrue
\mciteSetBstMidEndSepPunct{\mcitedefaultmidpunct}
{\mcitedefaultendpunct}{\mcitedefaultseppunct}\relax
\EndOfBibitem
\bibitem[Keith \emph{et~al.}(2021)Keith, Vassilev-Galindo, Cheng, Chmiela,
  Gastegger, M{\"u}ller, and Tkatchenko]{keith2021combining}
J.~A. Keith, V.~Vassilev-Galindo, B.~Cheng, S.~Chmiela, M.~Gastegger, K.-R.
  M{\"u}ller and A.~Tkatchenko, \emph{arXiv preprint arXiv:2102.06321},
  2021\relax
\mciteBstWouldAddEndPuncttrue
\mciteSetBstMidEndSepPunct{\mcitedefaultmidpunct}
{\mcitedefaultendpunct}{\mcitedefaultseppunct}\relax
\EndOfBibitem
\bibitem[Nielsen and Chuang(2011)]{nielson}
M.~A. Nielsen and I.~L. Chuang, \emph{Quantum Computation and Quantum
  Information: 10th Anniversary Edition}, Cambridge University Press, USA, 10th
  edn, 2011\relax
\mciteBstWouldAddEndPuncttrue
\mciteSetBstMidEndSepPunct{\mcitedefaultmidpunct}
{\mcitedefaultendpunct}{\mcitedefaultseppunct}\relax
\EndOfBibitem
\bibitem[Bennett \emph{et~al.}(1993)Bennett, Brassard, Cr{\'e}peau, Jozsa,
  Peres, and Wootters]{bennett1993teleporting}
C.~H. Bennett, G.~Brassard, C.~Cr{\'e}peau, R.~Jozsa, A.~Peres and W.~K.
  Wootters, \emph{Physical review letters}, 1993, \textbf{70}, 1895\relax
\mciteBstWouldAddEndPuncttrue
\mciteSetBstMidEndSepPunct{\mcitedefaultmidpunct}
{\mcitedefaultendpunct}{\mcitedefaultseppunct}\relax
\EndOfBibitem
\bibitem[Harrow \emph{et~al.}(2004)Harrow, Hayden, and
  Leung]{harrow2004superdense}
A.~Harrow, P.~Hayden and D.~Leung, \emph{Physical review letters}, 2004,
  \textbf{92}, 187901\relax
\mciteBstWouldAddEndPuncttrue
\mciteSetBstMidEndSepPunct{\mcitedefaultmidpunct}
{\mcitedefaultendpunct}{\mcitedefaultseppunct}\relax
\EndOfBibitem
\bibitem[Abrams and Lloyd(1999)]{PhysRevLett.83.5162}
D.~S. Abrams and S.~Lloyd, \emph{Phys. Rev. Lett.}, 1999, \textbf{83},
  5162--5165\relax
\mciteBstWouldAddEndPuncttrue
\mciteSetBstMidEndSepPunct{\mcitedefaultmidpunct}
{\mcitedefaultendpunct}{\mcitedefaultseppunct}\relax
\EndOfBibitem
\bibitem[Shor(1999)]{shor1999polynomial}
P.~W. Shor, \emph{SIAM review}, 1999, \textbf{41}, 303--332\relax
\mciteBstWouldAddEndPuncttrue
\mciteSetBstMidEndSepPunct{\mcitedefaultmidpunct}
{\mcitedefaultendpunct}{\mcitedefaultseppunct}\relax
\EndOfBibitem
\bibitem[Grover(1996)]{grover1996fast}
L.~K. Grover, Proceedings of the twenty-eighth annual ACM symposium on Theory
  of computing, 1996, pp. 212--219\relax
\mciteBstWouldAddEndPuncttrue
\mciteSetBstMidEndSepPunct{\mcitedefaultmidpunct}
{\mcitedefaultendpunct}{\mcitedefaultseppunct}\relax
\EndOfBibitem
\bibitem[Biamonte \emph{et~al.}(2017)Biamonte, Wittek, Pancotti, Rebentrost,
  Wiebe, and Lloyd]{biamonte2017quantum}
J.~Biamonte, P.~Wittek, N.~Pancotti, P.~Rebentrost, N.~Wiebe and S.~Lloyd,
  \emph{Nature}, 2017, \textbf{549}, 195--202\relax
\mciteBstWouldAddEndPuncttrue
\mciteSetBstMidEndSepPunct{\mcitedefaultmidpunct}
{\mcitedefaultendpunct}{\mcitedefaultseppunct}\relax
\EndOfBibitem
\bibitem[Wiebe \emph{et~al.}(2014)Wiebe, Kapoor, and Svore]{wiebe2014quantum}
N.~Wiebe, A.~Kapoor and K.~M. Svore, \emph{arXiv preprint arXiv:1412.3489},
  2014\relax
\mciteBstWouldAddEndPuncttrue
\mciteSetBstMidEndSepPunct{\mcitedefaultmidpunct}
{\mcitedefaultendpunct}{\mcitedefaultseppunct}\relax
\EndOfBibitem
\bibitem[Alpaydin(2010)]{10.5555/1734076}
E.~Alpaydin, \emph{Introduction to Machine Learning}, The MIT Press, 2nd edn,
  2010\relax
\mciteBstWouldAddEndPuncttrue
\mciteSetBstMidEndSepPunct{\mcitedefaultmidpunct}
{\mcitedefaultendpunct}{\mcitedefaultseppunct}\relax
\EndOfBibitem
\bibitem[Goodfellow \emph{et~al.}(2016)Goodfellow, Bengio, and
  Courville]{Goodfellow-et-al-2016}
I.~Goodfellow, Y.~Bengio and A.~Courville, \emph{Deep Learning}, MIT Press,
  2016\relax
\mciteBstWouldAddEndPuncttrue
\mciteSetBstMidEndSepPunct{\mcitedefaultmidpunct}
{\mcitedefaultendpunct}{\mcitedefaultseppunct}\relax
\EndOfBibitem
\bibitem[Kubat(2017)]{ML_kubat_book}
M.~Kubat, \emph{An Introduction to Machine Learning}, 2017\relax
\mciteBstWouldAddEndPuncttrue
\mciteSetBstMidEndSepPunct{\mcitedefaultmidpunct}
{\mcitedefaultendpunct}{\mcitedefaultseppunct}\relax
\EndOfBibitem
\bibitem[Von~Lilienfeld(2018)]{von2018quantum}
O.~A. Von~Lilienfeld, \emph{Angewandte Chemie International Edition}, 2018,
  \textbf{57}, 4164--4169\relax
\mciteBstWouldAddEndPuncttrue
\mciteSetBstMidEndSepPunct{\mcitedefaultmidpunct}
{\mcitedefaultendpunct}{\mcitedefaultseppunct}\relax
\EndOfBibitem
\bibitem[Huang \emph{et~al.}(2018)Huang, Symonds, and Lilienfeld]{Huang2018}
B.~Huang, N.~O. Symonds and O.~A.~v. Lilienfeld, \emph{Handbook of Materials
  Modeling}, 2018,  1–27\relax
\mciteBstWouldAddEndPuncttrue
\mciteSetBstMidEndSepPunct{\mcitedefaultmidpunct}
{\mcitedefaultendpunct}{\mcitedefaultseppunct}\relax
\EndOfBibitem
\bibitem[H{\"a}ffner \emph{et~al.}(2008)H{\"a}ffner, Roos, and
  Blatt]{haffner2008quantum}
H.~H{\"a}ffner, C.~F. Roos and R.~Blatt, \emph{Physics reports}, 2008,
  \textbf{469}, 155--203\relax
\mciteBstWouldAddEndPuncttrue
\mciteSetBstMidEndSepPunct{\mcitedefaultmidpunct}
{\mcitedefaultendpunct}{\mcitedefaultseppunct}\relax
\EndOfBibitem
\bibitem[Bruzewicz \emph{et~al.}(2019)Bruzewicz, Chiaverini, McConnell, and
  Sage]{bruzewicz2019trapped}
C.~D. Bruzewicz, J.~Chiaverini, R.~McConnell and J.~M. Sage, \emph{Applied
  Physics Reviews}, 2019, \textbf{6}, 021314\relax
\mciteBstWouldAddEndPuncttrue
\mciteSetBstMidEndSepPunct{\mcitedefaultmidpunct}
{\mcitedefaultendpunct}{\mcitedefaultseppunct}\relax
\EndOfBibitem
\bibitem[Kjaergaard \emph{et~al.}(2020)Kjaergaard, Schwartz, Braum{\"u}ller,
  Krantz, Wang, Gustavsson, and Oliver]{kjaergaard2020superconducting}
M.~Kjaergaard, M.~E. Schwartz, J.~Braum{\"u}ller, P.~Krantz, J.~I.-J. Wang,
  S.~Gustavsson and W.~D. Oliver, \emph{Annual Review of Condensed Matter
  Physics}, 2020, \textbf{11}, 369--395\relax
\mciteBstWouldAddEndPuncttrue
\mciteSetBstMidEndSepPunct{\mcitedefaultmidpunct}
{\mcitedefaultendpunct}{\mcitedefaultseppunct}\relax
\EndOfBibitem
\bibitem[Saffman \emph{et~al.}(2010)Saffman, Walker, and
  M{\o}lmer]{saffman2010quantum}
M.~Saffman, T.~G. Walker and K.~M{\o}lmer, \emph{Reviews of modern physics},
  2010, \textbf{82}, 2313\relax
\mciteBstWouldAddEndPuncttrue
\mciteSetBstMidEndSepPunct{\mcitedefaultmidpunct}
{\mcitedefaultendpunct}{\mcitedefaultseppunct}\relax
\EndOfBibitem
\bibitem[Saffman(2016)]{saffman2016quantum}
M.~Saffman, \emph{Journal of Physics B: Atomic, Molecular and Optical Physics},
  2016, \textbf{49}, 202001\relax
\mciteBstWouldAddEndPuncttrue
\mciteSetBstMidEndSepPunct{\mcitedefaultmidpunct}
{\mcitedefaultendpunct}{\mcitedefaultseppunct}\relax
\EndOfBibitem
\bibitem[Wei \emph{et~al.}(2016)Wei, Cao, Kais, Friedrich, and
  Herschbach]{Wei2016QuantumCU}
Q.~Wei, Y.~Cao, S.~Kais, B.~Friedrich and D.~R. Herschbach, \emph{Chemphyschem
  : a European journal of chemical physics and physical chemistry}, 2016,
  \textbf{17 22}, 3714--3722\relax
\mciteBstWouldAddEndPuncttrue
\mciteSetBstMidEndSepPunct{\mcitedefaultmidpunct}
{\mcitedefaultendpunct}{\mcitedefaultseppunct}\relax
\EndOfBibitem
\bibitem[Karra \emph{et~al.}(2016)Karra, Sharma, Friedrich, Kais, and
  Herschbach]{Karra2016ProspectsFQ}
M.~Karra, K.~Sharma, B.~Friedrich, S.~Kais and D.~R. Herschbach, \emph{The
  Journal of chemical physics}, 2016, \textbf{144 9}, 094301\relax
\mciteBstWouldAddEndPuncttrue
\mciteSetBstMidEndSepPunct{\mcitedefaultmidpunct}
{\mcitedefaultendpunct}{\mcitedefaultseppunct}\relax
\EndOfBibitem
\bibitem[Zhu \emph{et~al.}(2013)Zhu, Kais, Wei, Herschbach, and
  Friedrich]{Impl_pend_gates}
J.~Zhu, S.~Kais, Q.~Wei, D.~Herschbach and B.~Friedrich, \emph{The Journal of
  Chemical Physics}, 2013, \textbf{138}, 024104\relax
\mciteBstWouldAddEndPuncttrue
\mciteSetBstMidEndSepPunct{\mcitedefaultmidpunct}
{\mcitedefaultendpunct}{\mcitedefaultseppunct}\relax
\EndOfBibitem
\bibitem[Wei \emph{et~al.}(2011)Wei, Kais, Friedrich, and
  Herschbach]{entangle_pen_states}
Q.~Wei, S.~Kais, B.~Friedrich and D.~Herschbach, \emph{The Journal of Chemical
  Physics}, 2011, \textbf{134}, 124107\relax
\mciteBstWouldAddEndPuncttrue
\mciteSetBstMidEndSepPunct{\mcitedefaultmidpunct}
{\mcitedefaultendpunct}{\mcitedefaultseppunct}\relax
\EndOfBibitem
\bibitem[Wei \emph{et~al.}(2011)Wei, Kais, Friedrich, and
  Herschbach]{polar_sym_top}
Q.~Wei, S.~Kais, B.~Friedrich and D.~Herschbach, \emph{The Journal of Chemical
  Physics}, 2011, \textbf{135}, 154102\relax
\mciteBstWouldAddEndPuncttrue
\mciteSetBstMidEndSepPunct{\mcitedefaultmidpunct}
{\mcitedefaultendpunct}{\mcitedefaultseppunct}\relax
\EndOfBibitem
\bibitem[Watrous(2018)]{watrous2018theory}
J.~Watrous, \emph{The theory of quantum information}, Cambridge university
  press, 2018\relax
\mciteBstWouldAddEndPuncttrue
\mciteSetBstMidEndSepPunct{\mcitedefaultmidpunct}
{\mcitedefaultendpunct}{\mcitedefaultseppunct}\relax
\EndOfBibitem
\bibitem[Preskill()]{Preskill_notes}
J.~Preskill, \emph{Lecture Notes for Ph219/CS219},
  \url{http://theory.caltech.edu/~preskill/ph229/}, Accessed: 2021-10-22\relax
\mciteBstWouldAddEndPuncttrue
\mciteSetBstMidEndSepPunct{\mcitedefaultmidpunct}
{\mcitedefaultendpunct}{\mcitedefaultseppunct}\relax
\EndOfBibitem
\bibitem[Lin \emph{et~al.}(2021)Lin, Dilip, Green, Smith, and
  Pollmann]{PRXQuantum.2.010342}
S.-H. Lin, R.~Dilip, A.~G. Green, A.~Smith and F.~Pollmann, \emph{PRX Quantum},
  2021, \textbf{2}, 010342\relax
\mciteBstWouldAddEndPuncttrue
\mciteSetBstMidEndSepPunct{\mcitedefaultmidpunct}
{\mcitedefaultendpunct}{\mcitedefaultseppunct}\relax
\EndOfBibitem
\bibitem[Yeter-Aydeniz \emph{et~al.}(2022)Yeter-Aydeniz, Moschandreou, and
  Siopsis]{PhysRevA.105.012412}
K.~Yeter-Aydeniz, E.~Moschandreou and G.~Siopsis, \emph{Phys. Rev. A}, 2022,
  \textbf{105}, 012412\relax
\mciteBstWouldAddEndPuncttrue
\mciteSetBstMidEndSepPunct{\mcitedefaultmidpunct}
{\mcitedefaultendpunct}{\mcitedefaultseppunct}\relax
\EndOfBibitem
\bibitem[Yi(2021)]{PhysRevA.104.052603}
C.~Yi, \emph{Phys. Rev. A}, 2021, \textbf{104}, 052603\relax
\mciteBstWouldAddEndPuncttrue
\mciteSetBstMidEndSepPunct{\mcitedefaultmidpunct}
{\mcitedefaultendpunct}{\mcitedefaultseppunct}\relax
\EndOfBibitem
\bibitem[Endo \emph{et~al.}(2020)Endo, Sun, Li, Benjamin, and
  Yuan]{PhysRevLett.125.010501}
S.~Endo, J.~Sun, Y.~Li, S.~C. Benjamin and X.~Yuan, \emph{Phys. Rev. Lett.},
  2020, \textbf{125}, 010501\relax
\mciteBstWouldAddEndPuncttrue
\mciteSetBstMidEndSepPunct{\mcitedefaultmidpunct}
{\mcitedefaultendpunct}{\mcitedefaultseppunct}\relax
\EndOfBibitem
\bibitem[Deutsch(1985)]{deutsch1985quantum}
D.~Deutsch, \emph{Proceedings of the Royal Society of London. A. Mathematical
  and Physical Sciences}, 1985, \textbf{400}, 97--117\relax
\mciteBstWouldAddEndPuncttrue
\mciteSetBstMidEndSepPunct{\mcitedefaultmidpunct}
{\mcitedefaultendpunct}{\mcitedefaultseppunct}\relax
\EndOfBibitem
\bibitem[Benioff(1980)]{benioff1980computer}
P.~Benioff, \emph{Journal of statistical physics}, 1980, \textbf{22},
  563--591\relax
\mciteBstWouldAddEndPuncttrue
\mciteSetBstMidEndSepPunct{\mcitedefaultmidpunct}
{\mcitedefaultendpunct}{\mcitedefaultseppunct}\relax
\EndOfBibitem
\bibitem[Hu and Sarma(2002)]{hu2002gate}
X.~Hu and S.~D. Sarma, \emph{Physical Review A}, 2002, \textbf{66},
  012312\relax
\mciteBstWouldAddEndPuncttrue
\mciteSetBstMidEndSepPunct{\mcitedefaultmidpunct}
{\mcitedefaultendpunct}{\mcitedefaultseppunct}\relax
\EndOfBibitem
\bibitem[Terhal(2015)]{RevModPhys.87.307}
B.~M. Terhal, \emph{Rev. Mod. Phys.}, 2015, \textbf{87}, 307--346\relax
\mciteBstWouldAddEndPuncttrue
\mciteSetBstMidEndSepPunct{\mcitedefaultmidpunct}
{\mcitedefaultendpunct}{\mcitedefaultseppunct}\relax
\EndOfBibitem
\bibitem[Kitaev(2003)]{KITAEV20032}
A.~Kitaev, \emph{Annals of Physics}, 2003, \textbf{303}, 2--30\relax
\mciteBstWouldAddEndPuncttrue
\mciteSetBstMidEndSepPunct{\mcitedefaultmidpunct}
{\mcitedefaultendpunct}{\mcitedefaultseppunct}\relax
\EndOfBibitem
\bibitem[Knill and Laflamme(1997)]{knill1997theory}
E.~Knill and R.~Laflamme, \emph{Physical Review A}, 1997, \textbf{55},
  900\relax
\mciteBstWouldAddEndPuncttrue
\mciteSetBstMidEndSepPunct{\mcitedefaultmidpunct}
{\mcitedefaultendpunct}{\mcitedefaultseppunct}\relax
\EndOfBibitem
\bibitem[Preskill(2018)]{preskill2018quantum}
J.~Preskill, \emph{Quantum}, 2018, \textbf{2}, 79\relax
\mciteBstWouldAddEndPuncttrue
\mciteSetBstMidEndSepPunct{\mcitedefaultmidpunct}
{\mcitedefaultendpunct}{\mcitedefaultseppunct}\relax
\EndOfBibitem
\bibitem[Bharti \emph{et~al.}(2022)Bharti, Cervera-Lierta, Kyaw, Haug,
  Alperin-Lea, Anand, Degroote, Heimonen, Kottmann,
  Menke,\emph{et~al.}]{bharti2022noisy}
K.~Bharti, A.~Cervera-Lierta, T.~H. Kyaw, T.~Haug, S.~Alperin-Lea, A.~Anand,
  M.~Degroote, H.~Heimonen, J.~S. Kottmann, T.~Menke \emph{et~al.},
  \emph{Reviews of Modern Physics}, 2022, \textbf{94}, 015004\relax
\mciteBstWouldAddEndPuncttrue
\mciteSetBstMidEndSepPunct{\mcitedefaultmidpunct}
{\mcitedefaultendpunct}{\mcitedefaultseppunct}\relax
\EndOfBibitem
\bibitem[Quantum \emph{et~al.}(2020)Quantum, Collaborators*†, Arute, Arya,
  Babbush, Bacon, Bardin, Barends, Boixo, Broughton,
  Buckley,\emph{et~al.}]{google2020hartree}
G.~A. Quantum, Collaborators*†, F.~Arute, K.~Arya, R.~Babbush, D.~Bacon,
  J.~C. Bardin, R.~Barends, S.~Boixo, M.~Broughton, B.~B. Buckley
  \emph{et~al.}, \emph{Science}, 2020, \textbf{369}, 1084--1089\relax
\mciteBstWouldAddEndPuncttrue
\mciteSetBstMidEndSepPunct{\mcitedefaultmidpunct}
{\mcitedefaultendpunct}{\mcitedefaultseppunct}\relax
\EndOfBibitem
\bibitem[Cao \emph{et~al.}(2019)Cao, Romero, Olson, Degroote, Johnson,
  Kieferov{\'a}, Kivlichan, Menke, Peropadre,
  Sawaya,\emph{et~al.}]{cao2019quantum}
Y.~Cao, J.~Romero, J.~P. Olson, M.~Degroote, P.~D. Johnson, M.~Kieferov{\'a},
  I.~D. Kivlichan, T.~Menke, B.~Peropadre, N.~P. Sawaya \emph{et~al.},
  \emph{Chemical reviews}, 2019, \textbf{119}, 10856--10915\relax
\mciteBstWouldAddEndPuncttrue
\mciteSetBstMidEndSepPunct{\mcitedefaultmidpunct}
{\mcitedefaultendpunct}{\mcitedefaultseppunct}\relax
\EndOfBibitem
\bibitem[Head-Marsden \emph{et~al.}(2020)Head-Marsden, Flick, Ciccarino, and
  Narang]{head2020quantum}
K.~Head-Marsden, J.~Flick, C.~J. Ciccarino and P.~Narang, \emph{Chemical
  Reviews}, 2020, \textbf{121}, 3061--3120\relax
\mciteBstWouldAddEndPuncttrue
\mciteSetBstMidEndSepPunct{\mcitedefaultmidpunct}
{\mcitedefaultendpunct}{\mcitedefaultseppunct}\relax
\EndOfBibitem
\bibitem[Yuan \emph{et~al.}(2019)Yuan, Endo, Zhao, Li, and
  Benjamin]{yuan2019theory}
X.~Yuan, S.~Endo, Q.~Zhao, Y.~Li and S.~C. Benjamin, \emph{Quantum}, 2019,
  \textbf{3}, 191\relax
\mciteBstWouldAddEndPuncttrue
\mciteSetBstMidEndSepPunct{\mcitedefaultmidpunct}
{\mcitedefaultendpunct}{\mcitedefaultseppunct}\relax
\EndOfBibitem
\bibitem[Cerezo \emph{et~al.}(2021)Cerezo, Arrasmith, Babbush, Benjamin, Endo,
  Fujii, McClean, Mitarai, Yuan, Cincio,\emph{et~al.}]{cerezo2021variational}
M.~Cerezo, A.~Arrasmith, R.~Babbush, S.~C. Benjamin, S.~Endo, K.~Fujii, J.~R.
  McClean, K.~Mitarai, X.~Yuan, L.~Cincio \emph{et~al.}, \emph{Nature Reviews
  Physics}, 2021, \textbf{3}, 625--644\relax
\mciteBstWouldAddEndPuncttrue
\mciteSetBstMidEndSepPunct{\mcitedefaultmidpunct}
{\mcitedefaultendpunct}{\mcitedefaultseppunct}\relax
\EndOfBibitem
\bibitem[Bauer \emph{et~al.}(2020)Bauer, Bravyi, Motta, and
  Chan]{bauer2020quantum}
B.~Bauer, S.~Bravyi, M.~Motta and G.~K.-L. Chan, \emph{Chemical Reviews}, 2020,
  \textbf{120}, 12685--12717\relax
\mciteBstWouldAddEndPuncttrue
\mciteSetBstMidEndSepPunct{\mcitedefaultmidpunct}
{\mcitedefaultendpunct}{\mcitedefaultseppunct}\relax
\EndOfBibitem
\bibitem[Dumitrescu \emph{et~al.}(2018)Dumitrescu, McCaskey, Hagen, Jansen,
  Morris, Papenbrock, Pooser, Dean, and Lougovski]{dumitrescu2018cloud}
E.~F. Dumitrescu, A.~J. McCaskey, G.~Hagen, G.~R. Jansen, T.~D. Morris,
  T.~Papenbrock, R.~C. Pooser, D.~J. Dean and P.~Lougovski, \emph{Physical
  review letters}, 2018, \textbf{120}, 210501\relax
\mciteBstWouldAddEndPuncttrue
\mciteSetBstMidEndSepPunct{\mcitedefaultmidpunct}
{\mcitedefaultendpunct}{\mcitedefaultseppunct}\relax
\EndOfBibitem
\bibitem[Wu \emph{et~al.}(2021)Wu, Chan, Guan, Sun, Wang, Zhou, Livny,
  Carminati, Di~Meglio, Li,\emph{et~al.}]{wu2021application}
S.~L. Wu, J.~Chan, W.~Guan, S.~Sun, A.~Wang, C.~Zhou, M.~Livny, F.~Carminati,
  A.~Di~Meglio, A.~C. Li \emph{et~al.}, \emph{Journal of Physics G: Nuclear and
  Particle Physics}, 2021, \textbf{48}, 125003\relax
\mciteBstWouldAddEndPuncttrue
\mciteSetBstMidEndSepPunct{\mcitedefaultmidpunct}
{\mcitedefaultendpunct}{\mcitedefaultseppunct}\relax
\EndOfBibitem
\bibitem[Guan \emph{et~al.}(2021)Guan, Perdue, Pesah, Schuld, Terashi,
  Vallecorsa, and Vlimant]{guan2021quantum}
W.~Guan, G.~Perdue, A.~Pesah, M.~Schuld, K.~Terashi, S.~Vallecorsa and J.-R.
  Vlimant, \emph{Machine Learning: Science and Technology}, 2021, \textbf{2},
  011003\relax
\mciteBstWouldAddEndPuncttrue
\mciteSetBstMidEndSepPunct{\mcitedefaultmidpunct}
{\mcitedefaultendpunct}{\mcitedefaultseppunct}\relax
\EndOfBibitem
\bibitem[Cheng \emph{et~al.}(2020)Cheng, Deumens, Freericks, Li, and
  Sanders]{cheng2020application}
H.-P. Cheng, E.~Deumens, J.~K. Freericks, C.~Li and B.~A. Sanders,
  \emph{Frontiers in Chemistry}, 2020,  1066\relax
\mciteBstWouldAddEndPuncttrue
\mciteSetBstMidEndSepPunct{\mcitedefaultmidpunct}
{\mcitedefaultendpunct}{\mcitedefaultseppunct}\relax
\EndOfBibitem
\bibitem[Orus \emph{et~al.}(2019)Orus, Mugel, and Lizaso]{orus2019quantum}
R.~Orus, S.~Mugel and E.~Lizaso, \emph{Reviews in Physics}, 2019, \textbf{4},
  100028\relax
\mciteBstWouldAddEndPuncttrue
\mciteSetBstMidEndSepPunct{\mcitedefaultmidpunct}
{\mcitedefaultendpunct}{\mcitedefaultseppunct}\relax
\EndOfBibitem
\bibitem[Lloyd and Braunstein(2003)]{Lloyd2003}
S.~Lloyd and S.~L. Braunstein, in \emph{Quantum Computation Over Continuous
  Variables}, ed. S.~L. Braunstein and A.~K. Pati, Springer Netherlands,
  Dordrecht, 2003, pp. 9--17\relax
\mciteBstWouldAddEndPuncttrue
\mciteSetBstMidEndSepPunct{\mcitedefaultmidpunct}
{\mcitedefaultendpunct}{\mcitedefaultseppunct}\relax
\EndOfBibitem
\bibitem[Aspuru-Guzik \emph{et~al.}(2005)Aspuru-Guzik, Dutoi, Love, and
  Head-Gordon]{aspuru2005simulated}
A.~Aspuru-Guzik, A.~D. Dutoi, P.~J. Love and M.~Head-Gordon, \emph{Science},
  2005, \textbf{309}, 1704--1707\relax
\mciteBstWouldAddEndPuncttrue
\mciteSetBstMidEndSepPunct{\mcitedefaultmidpunct}
{\mcitedefaultendpunct}{\mcitedefaultseppunct}\relax
\EndOfBibitem
\bibitem[Albash and Lidar(2018)]{albash2018adiabatic}
T.~Albash and D.~A. Lidar, \emph{Reviews of Modern Physics}, 2018, \textbf{90},
  015002\relax
\mciteBstWouldAddEndPuncttrue
\mciteSetBstMidEndSepPunct{\mcitedefaultmidpunct}
{\mcitedefaultendpunct}{\mcitedefaultseppunct}\relax
\EndOfBibitem
\bibitem[Dwa()]{Dwave_sys}
\emph{D-Wave System Documentation},
  \url{https://docs.dwavesys.com/docs/latest/doc_getting_started.html},
  Accessed: 2021-10-22\relax
\mciteBstWouldAddEndPuncttrue
\mciteSetBstMidEndSepPunct{\mcitedefaultmidpunct}
{\mcitedefaultendpunct}{\mcitedefaultseppunct}\relax
\EndOfBibitem
\bibitem[Hauke \emph{et~al.}(2020)Hauke, Katzgraber, Lechner, Nishimori, and
  Oliver]{2020_pers_ann}
P.~Hauke, H.~G. Katzgraber, W.~Lechner, H.~Nishimori and W.~D. Oliver,
  \emph{Reports on Progress in Physics}, 2020, \textbf{83}, 054401\relax
\mciteBstWouldAddEndPuncttrue
\mciteSetBstMidEndSepPunct{\mcitedefaultmidpunct}
{\mcitedefaultendpunct}{\mcitedefaultseppunct}\relax
\EndOfBibitem
\bibitem[Djidjev \emph{et~al.}(2018)Djidjev, Chapuis, Hahn, and
  Rizk]{djidjev2018efficient}
H.~N. Djidjev, G.~Chapuis, G.~Hahn and G.~Rizk, \emph{Efficient Combinatorial
  Optimization Using Quantum Annealing}, 2018\relax
\mciteBstWouldAddEndPuncttrue
\mciteSetBstMidEndSepPunct{\mcitedefaultmidpunct}
{\mcitedefaultendpunct}{\mcitedefaultseppunct}\relax
\EndOfBibitem
\bibitem[Li \emph{et~al.}(2018)Li, Di~Felice, Rohs, and Lidar]{li2018quantum}
R.~Y. Li, R.~Di~Felice, R.~Rohs and D.~A. Lidar, \emph{NPJ quantum
  information}, 2018, \textbf{4}, 1--10\relax
\mciteBstWouldAddEndPuncttrue
\mciteSetBstMidEndSepPunct{\mcitedefaultmidpunct}
{\mcitedefaultendpunct}{\mcitedefaultseppunct}\relax
\EndOfBibitem
\bibitem[Neukart \emph{et~al.}(2017)Neukart, Compostella, Seidel, Von~Dollen,
  Yarkoni, and Parney]{neukart2017traffic}
F.~Neukart, G.~Compostella, C.~Seidel, D.~Von~Dollen, S.~Yarkoni and B.~Parney,
  \emph{Frontiers in ICT}, 2017, \textbf{4}, 29\relax
\mciteBstWouldAddEndPuncttrue
\mciteSetBstMidEndSepPunct{\mcitedefaultmidpunct}
{\mcitedefaultendpunct}{\mcitedefaultseppunct}\relax
\EndOfBibitem
\bibitem[Nath \emph{et~al.}(2021)Nath, Thapliyal, and Humble]{nath2021review}
R.~K. Nath, H.~Thapliyal and T.~S. Humble, \emph{arXiv preprint
  arXiv:2106.02964}, 2021\relax
\mciteBstWouldAddEndPuncttrue
\mciteSetBstMidEndSepPunct{\mcitedefaultmidpunct}
{\mcitedefaultendpunct}{\mcitedefaultseppunct}\relax
\EndOfBibitem
\bibitem[Ruder(2016)]{gradient_methods}
S.~Ruder, \emph{CoRR}, 2016, \textbf{abs/1609.04747}, year\relax
\mciteBstWouldAddEndPuncttrue
\mciteSetBstMidEndSepPunct{\mcitedefaultmidpunct}
{\mcitedefaultendpunct}{\mcitedefaultseppunct}\relax
\EndOfBibitem
\bibitem[Breuel(2015)]{DBLP:journals/corr/Breuel15a}
T.~M. Breuel, \emph{CoRR}, 2015, \textbf{abs/1508.02788}, year\relax
\mciteBstWouldAddEndPuncttrue
\mciteSetBstMidEndSepPunct{\mcitedefaultmidpunct}
{\mcitedefaultendpunct}{\mcitedefaultseppunct}\relax
\EndOfBibitem
\bibitem[Wang(2008)]{entropy-types}
Q.~A. Wang, \emph{Journal of Physics A: Mathematical and Theoretical}, 2008,
  \textbf{41}, 065004\relax
\mciteBstWouldAddEndPuncttrue
\mciteSetBstMidEndSepPunct{\mcitedefaultmidpunct}
{\mcitedefaultendpunct}{\mcitedefaultseppunct}\relax
\EndOfBibitem
\bibitem[Ouali \emph{et~al.}(2020)Ouali, Hudelot, and
  Tami]{learning_semisupervised}
Y.~Ouali, C.~Hudelot and M.~Tami, \emph{CoRR}, 2020, \textbf{abs/2006.05278},
  year\relax
\mciteBstWouldAddEndPuncttrue
\mciteSetBstMidEndSepPunct{\mcitedefaultmidpunct}
{\mcitedefaultendpunct}{\mcitedefaultseppunct}\relax
\EndOfBibitem
\bibitem[Garg and Kalai(2017)]{supervising_unsupervised}
V.~K. Garg and A.~Kalai, \emph{CoRR}, 2017, \textbf{abs/1709.05262}, year\relax
\mciteBstWouldAddEndPuncttrue
\mciteSetBstMidEndSepPunct{\mcitedefaultmidpunct}
{\mcitedefaultendpunct}{\mcitedefaultseppunct}\relax
\EndOfBibitem
\bibitem[Powell(2019)]{DBLP:journals/corr/abs-1912-03513}
W.~B. Powell, \emph{CoRR}, 2019, \textbf{abs/1912.03513}, year\relax
\mciteBstWouldAddEndPuncttrue
\mciteSetBstMidEndSepPunct{\mcitedefaultmidpunct}
{\mcitedefaultendpunct}{\mcitedefaultseppunct}\relax
\EndOfBibitem
\bibitem[François-Lavet \emph{et~al.}(2018)François-Lavet, Henderson, Islam,
  Bellemare, and Pineau]{2018_reinforcement}
V.~François-Lavet, P.~Henderson, R.~Islam, M.~G. Bellemare and J.~Pineau,
  \emph{Foundations and Trends® in Machine Learning}, 2018, \textbf{11},
  219–354\relax
\mciteBstWouldAddEndPuncttrue
\mciteSetBstMidEndSepPunct{\mcitedefaultmidpunct}
{\mcitedefaultendpunct}{\mcitedefaultseppunct}\relax
\EndOfBibitem
\bibitem[Scholkopf and Smola(2001)]{10.5555/559923}
B.~Scholkopf and A.~J. Smola, \emph{Learning with Kernels: Support Vector
  Machines, Regularization, Optimization, and Beyond}, MIT Press, Cambridge,
  MA, USA, 2001\relax
\mciteBstWouldAddEndPuncttrue
\mciteSetBstMidEndSepPunct{\mcitedefaultmidpunct}
{\mcitedefaultendpunct}{\mcitedefaultseppunct}\relax
\EndOfBibitem
\bibitem[Genton(2001)]{genton2001classes}
M.~G. Genton, \emph{Journal of machine learning research}, 2001, \textbf{2},
  299--312\relax
\mciteBstWouldAddEndPuncttrue
\mciteSetBstMidEndSepPunct{\mcitedefaultmidpunct}
{\mcitedefaultendpunct}{\mcitedefaultseppunct}\relax
\EndOfBibitem
\bibitem[Azim and Ahmed(2018)]{azim2018kernel}
T.~Azim and S.~Ahmed, \emph{Composing Fisher Kernels from Deep Neural Models},
  Springer, 2018, pp. 1--7\relax
\mciteBstWouldAddEndPuncttrue
\mciteSetBstMidEndSepPunct{\mcitedefaultmidpunct}
{\mcitedefaultendpunct}{\mcitedefaultseppunct}\relax
\EndOfBibitem
\bibitem[Schuld and Petruccione(2018)]{schuld2018supervised}
M.~Schuld and F.~Petruccione, \emph{Supervised learning with quantum
  computers}, Springer, 2018, vol.~17\relax
\mciteBstWouldAddEndPuncttrue
\mciteSetBstMidEndSepPunct{\mcitedefaultmidpunct}
{\mcitedefaultendpunct}{\mcitedefaultseppunct}\relax
\EndOfBibitem
\bibitem[Schuld and Killoran(2019)]{PhysRevLett.122.040504}
M.~Schuld and N.~Killoran, \emph{Phys. Rev. Lett.}, 2019, \textbf{122},
  040504\relax
\mciteBstWouldAddEndPuncttrue
\mciteSetBstMidEndSepPunct{\mcitedefaultmidpunct}
{\mcitedefaultendpunct}{\mcitedefaultseppunct}\relax
\EndOfBibitem
\bibitem[Ghojogh \emph{et~al.}(2021)Ghojogh, Ghodsi, Karray, and
  Crowley]{ghojogh2021reproducing}
B.~Ghojogh, A.~Ghodsi, F.~Karray and M.~Crowley, \emph{arXiv preprint
  arXiv:2106.08443}, 2021\relax
\mciteBstWouldAddEndPuncttrue
\mciteSetBstMidEndSepPunct{\mcitedefaultmidpunct}
{\mcitedefaultendpunct}{\mcitedefaultseppunct}\relax
\EndOfBibitem
\bibitem[Brunton and Kutz(2019)]{brunton2019data}
S.~L. Brunton and J.~N. Kutz, \emph{Data-driven science and engineering:
  Machine learning, dynamical systems, and control}, Cambridge University
  Press, 2019\relax
\mciteBstWouldAddEndPuncttrue
\mciteSetBstMidEndSepPunct{\mcitedefaultmidpunct}
{\mcitedefaultendpunct}{\mcitedefaultseppunct}\relax
\EndOfBibitem
\bibitem[Marquardt and Snee(1975)]{marquardt1975ridge}
D.~W. Marquardt and R.~D. Snee, \emph{The American Statistician}, 1975,
  \textbf{29}, 3--20\relax
\mciteBstWouldAddEndPuncttrue
\mciteSetBstMidEndSepPunct{\mcitedefaultmidpunct}
{\mcitedefaultendpunct}{\mcitedefaultseppunct}\relax
\EndOfBibitem
\bibitem[McDonald(2009)]{mcdonald2009ridge}
G.~C. McDonald, \emph{Wiley Interdisciplinary Reviews: Computational
  Statistics}, 2009, \textbf{1}, 93--100\relax
\mciteBstWouldAddEndPuncttrue
\mciteSetBstMidEndSepPunct{\mcitedefaultmidpunct}
{\mcitedefaultendpunct}{\mcitedefaultseppunct}\relax
\EndOfBibitem
\bibitem[Hoerl \emph{et~al.}(1975)Hoerl, Kannard, and Baldwin]{hoerl1975ridge}
A.~E. Hoerl, R.~W. Kannard and K.~F. Baldwin, \emph{Communications in
  Statistics-Theory and Methods}, 1975, \textbf{4}, 105--123\relax
\mciteBstWouldAddEndPuncttrue
\mciteSetBstMidEndSepPunct{\mcitedefaultmidpunct}
{\mcitedefaultendpunct}{\mcitedefaultseppunct}\relax
\EndOfBibitem
\bibitem[Ostertagov{\'a}(2012)]{ostertagova2012modelling}
E.~Ostertagov{\'a}, \emph{Procedia Engineering}, 2012, \textbf{48},
  500--506\relax
\mciteBstWouldAddEndPuncttrue
\mciteSetBstMidEndSepPunct{\mcitedefaultmidpunct}
{\mcitedefaultendpunct}{\mcitedefaultseppunct}\relax
\EndOfBibitem
\bibitem[Rawlings \emph{et~al.}(1998)Rawlings, Pantula, and
  Dickey]{poly_reg_book}
\emph{Polynomial Regression}, ed. J.~O. Rawlings, S.~G. Pantula and D.~A.
  Dickey, Springer New York, New York, NY, 1998, pp. 235--268\relax
\mciteBstWouldAddEndPuncttrue
\mciteSetBstMidEndSepPunct{\mcitedefaultmidpunct}
{\mcitedefaultendpunct}{\mcitedefaultseppunct}\relax
\EndOfBibitem
\bibitem[Vovk(2013)]{vovk2013kernel}
V.~Vovk, \emph{Empirical inference}, Springer, 2013, pp. 105--116\relax
\mciteBstWouldAddEndPuncttrue
\mciteSetBstMidEndSepPunct{\mcitedefaultmidpunct}
{\mcitedefaultendpunct}{\mcitedefaultseppunct}\relax
\EndOfBibitem
\bibitem[Vu \emph{et~al.}(2015)Vu, Snyder, Li, Rupp, Chen, Khelif, M{\"u}ller,
  and Burke]{vu2015understanding}
K.~Vu, J.~C. Snyder, L.~Li, M.~Rupp, B.~F. Chen, T.~Khelif, K.-R. M{\"u}ller
  and K.~Burke, \emph{International Journal of Quantum Chemistry}, 2015,
  \textbf{115}, 1115--1128\relax
\mciteBstWouldAddEndPuncttrue
\mciteSetBstMidEndSepPunct{\mcitedefaultmidpunct}
{\mcitedefaultendpunct}{\mcitedefaultseppunct}\relax
\EndOfBibitem
\bibitem[Saunders \emph{et~al.}(1998)Saunders, Gammerman, and
  Vovk]{saunders1998ridge}
C.~Saunders, A.~Gammerman and V.~Vovk, 1998\relax
\mciteBstWouldAddEndPuncttrue
\mciteSetBstMidEndSepPunct{\mcitedefaultmidpunct}
{\mcitedefaultendpunct}{\mcitedefaultseppunct}\relax
\EndOfBibitem
\bibitem[Ullah and Dral(2021)]{ullah2021speeding}
A.~Ullah and P.~O. Dral, \emph{New Journal of Physics}, 2021, \textbf{23},
  113019\relax
\mciteBstWouldAddEndPuncttrue
\mciteSetBstMidEndSepPunct{\mcitedefaultmidpunct}
{\mcitedefaultendpunct}{\mcitedefaultseppunct}\relax
\EndOfBibitem
\bibitem[Westermayr \emph{et~al.}(2020)Westermayr, Faber, Christensen, von
  Lilienfeld, and Marquetand]{Westermayr_2020}
J.~Westermayr, F.~A. Faber, A.~S. Christensen, O.~A. von Lilienfeld and
  P.~Marquetand, \emph{Machine Learning: Science and Technology}, 2020,
  \textbf{1}, 025009\relax
\mciteBstWouldAddEndPuncttrue
\mciteSetBstMidEndSepPunct{\mcitedefaultmidpunct}
{\mcitedefaultendpunct}{\mcitedefaultseppunct}\relax
\EndOfBibitem
\bibitem[Wiebe \emph{et~al.}(2012)Wiebe, Braun, and
  Lloyd]{PhysRevLett.109.050505}
N.~Wiebe, D.~Braun and S.~Lloyd, \emph{Phys. Rev. Lett.}, 2012, \textbf{109},
  050505\relax
\mciteBstWouldAddEndPuncttrue
\mciteSetBstMidEndSepPunct{\mcitedefaultmidpunct}
{\mcitedefaultendpunct}{\mcitedefaultseppunct}\relax
\EndOfBibitem
\bibitem[Harrow \emph{et~al.}(2009)Harrow, Hassidim, and
  Lloyd]{PhysRevLett.103.150502}
A.~W. Harrow, A.~Hassidim and S.~Lloyd, \emph{Phys. Rev. Lett.}, 2009,
  \textbf{103}, 150502\relax
\mciteBstWouldAddEndPuncttrue
\mciteSetBstMidEndSepPunct{\mcitedefaultmidpunct}
{\mcitedefaultendpunct}{\mcitedefaultseppunct}\relax
\EndOfBibitem
\bibitem[Lee \emph{et~al.}(2019)Lee, Joo, and Lee]{lee2019hybrid}
Y.~Lee, J.~Joo and S.~Lee, \emph{Scientific reports}, 2019, \textbf{9},
  1--12\relax
\mciteBstWouldAddEndPuncttrue
\mciteSetBstMidEndSepPunct{\mcitedefaultmidpunct}
{\mcitedefaultendpunct}{\mcitedefaultseppunct}\relax
\EndOfBibitem
\bibitem[Liu and Zhang(2017)]{LIU201738}
Y.~Liu and S.~Zhang, \emph{Theoretical Computer Science}, 2017, \textbf{657},
  38--47\relax
\mciteBstWouldAddEndPuncttrue
\mciteSetBstMidEndSepPunct{\mcitedefaultmidpunct}
{\mcitedefaultendpunct}{\mcitedefaultseppunct}\relax
\EndOfBibitem
\bibitem[Wang(2017)]{PhysRevA.96.012335}
G.~Wang, \emph{Phys. Rev. A}, 2017, \textbf{96}, 012335\relax
\mciteBstWouldAddEndPuncttrue
\mciteSetBstMidEndSepPunct{\mcitedefaultmidpunct}
{\mcitedefaultendpunct}{\mcitedefaultseppunct}\relax
\EndOfBibitem
\bibitem[Pan \emph{et~al.}(2014)Pan, Cao, Yao, Li, Ju, Chen, Peng, Kais, and
  Du]{PhysRevA.89.022313}
J.~Pan, Y.~Cao, X.~Yao, Z.~Li, C.~Ju, H.~Chen, X.~Peng, S.~Kais and J.~Du,
  \emph{Phys. Rev. A}, 2014, \textbf{89}, 022313\relax
\mciteBstWouldAddEndPuncttrue
\mciteSetBstMidEndSepPunct{\mcitedefaultmidpunct}
{\mcitedefaultendpunct}{\mcitedefaultseppunct}\relax
\EndOfBibitem
\bibitem[Zheng \emph{et~al.}(2017)Zheng, Song, Chen, Xia, Liu, Guo, Zhang, Xu,
  Deng, Huang, Wu, Yan, Zheng, Lu, Pan, Wang, Lu, and
  Zhu]{PhysRevLett.118.210504}
Y.~Zheng, C.~Song, M.-C. Chen, B.~Xia, W.~Liu, Q.~Guo, L.~Zhang, D.~Xu,
  H.~Deng, K.~Huang, Y.~Wu, Z.~Yan, D.~Zheng, L.~Lu, J.-W. Pan, H.~Wang, C.-Y.
  Lu and X.~Zhu, \emph{Phys. Rev. Lett.}, 2017, \textbf{118}, 210504\relax
\mciteBstWouldAddEndPuncttrue
\mciteSetBstMidEndSepPunct{\mcitedefaultmidpunct}
{\mcitedefaultendpunct}{\mcitedefaultseppunct}\relax
\EndOfBibitem
\bibitem[Slussarenko and Pryde(2019)]{doi:10.1063/1.5115814}
S.~Slussarenko and G.~J. Pryde, \emph{Applied Physics Reviews}, 2019,
  \textbf{6}, 041303\relax
\mciteBstWouldAddEndPuncttrue
\mciteSetBstMidEndSepPunct{\mcitedefaultmidpunct}
{\mcitedefaultendpunct}{\mcitedefaultseppunct}\relax
\EndOfBibitem
\bibitem[Schuld \emph{et~al.}(2016)Schuld, Sinayskiy, and
  Petruccione]{PhysRevA.94.022342}
M.~Schuld, I.~Sinayskiy and F.~Petruccione, \emph{Phys. Rev. A}, 2016,
  \textbf{94}, 022342\relax
\mciteBstWouldAddEndPuncttrue
\mciteSetBstMidEndSepPunct{\mcitedefaultmidpunct}
{\mcitedefaultendpunct}{\mcitedefaultseppunct}\relax
\EndOfBibitem
\bibitem[Suba{\c{s}}{\i} \emph{et~al.}(2019)Suba{\c{s}}{\i}, Somma, and
  Orsucci]{subacsi2019quantum}
Y.~Suba{\c{s}}{\i}, R.~D. Somma and D.~Orsucci, \emph{Physical review letters},
  2019, \textbf{122}, 060504\relax
\mciteBstWouldAddEndPuncttrue
\mciteSetBstMidEndSepPunct{\mcitedefaultmidpunct}
{\mcitedefaultendpunct}{\mcitedefaultseppunct}\relax
\EndOfBibitem
\bibitem[Bravo-Prieto \emph{et~al.}(2019)Bravo-Prieto, LaRose, Cerezo, Subasi,
  Cincio, and Coles]{bravo2019variational}
C.~Bravo-Prieto, R.~LaRose, M.~Cerezo, Y.~Subasi, L.~Cincio and P.~J. Coles,
  \emph{arXiv preprint arXiv:1909.05820}, 2019\relax
\mciteBstWouldAddEndPuncttrue
\mciteSetBstMidEndSepPunct{\mcitedefaultmidpunct}
{\mcitedefaultendpunct}{\mcitedefaultseppunct}\relax
\EndOfBibitem
\bibitem[Yu \emph{et~al.}(2019)Yu, Gao, and Wen]{yu2019improved}
C.-H. Yu, F.~Gao and Q.-Y. Wen, \emph{IEEE Transactions on Knowledge and Data
  Engineering}, 2019, \textbf{33}, 858--866\relax
\mciteBstWouldAddEndPuncttrue
\mciteSetBstMidEndSepPunct{\mcitedefaultmidpunct}
{\mcitedefaultendpunct}{\mcitedefaultseppunct}\relax
\EndOfBibitem
\bibitem[Berrar(2019)]{BERRAR2019542}
D.~Berrar, \emph{Encyclopedia of Bioinformatics and Computational Biology},
  Academic Press, Oxford, 2019, pp. 542--545\relax
\mciteBstWouldAddEndPuncttrue
\mciteSetBstMidEndSepPunct{\mcitedefaultmidpunct}
{\mcitedefaultendpunct}{\mcitedefaultseppunct}\relax
\EndOfBibitem
\bibitem[Jolliffe and Cadima(2016)]{jolliffe2016principal}
I.~T. Jolliffe and J.~Cadima, \emph{Philosophical Transactions of the Royal
  Society A: Mathematical, Physical and Engineering Sciences}, 2016,
  \textbf{374}, 20150202\relax
\mciteBstWouldAddEndPuncttrue
\mciteSetBstMidEndSepPunct{\mcitedefaultmidpunct}
{\mcitedefaultendpunct}{\mcitedefaultseppunct}\relax
\EndOfBibitem
\bibitem[Jolliffe(2002)]{jolliffe2002principal}
I.~T. Jolliffe, \emph{Principal component analysis for special types of data},
  Springer, 2002\relax
\mciteBstWouldAddEndPuncttrue
\mciteSetBstMidEndSepPunct{\mcitedefaultmidpunct}
{\mcitedefaultendpunct}{\mcitedefaultseppunct}\relax
\EndOfBibitem
\bibitem[Sch{\"o}lkopf \emph{et~al.}(1997)Sch{\"o}lkopf, Smola, and
  M{\"u}ller]{scholkopf1997kernel}
B.~Sch{\"o}lkopf, A.~Smola and K.-R. M{\"u}ller, International conference on
  artificial neural networks, 1997, pp. 583--588\relax
\mciteBstWouldAddEndPuncttrue
\mciteSetBstMidEndSepPunct{\mcitedefaultmidpunct}
{\mcitedefaultendpunct}{\mcitedefaultseppunct}\relax
\EndOfBibitem
\bibitem[Minh \emph{et~al.}(2006)Minh, Niyogi, and Yao]{minh2006mercer}
H.~Q. Minh, P.~Niyogi and Y.~Yao, International Conference on Computational
  Learning Theory, 2006, pp. 154--168\relax
\mciteBstWouldAddEndPuncttrue
\mciteSetBstMidEndSepPunct{\mcitedefaultmidpunct}
{\mcitedefaultendpunct}{\mcitedefaultseppunct}\relax
\EndOfBibitem
\bibitem[Campbell(2002)]{campbell2002kernel}
C.~Campbell, \emph{Neurocomputing}, 2002, \textbf{48}, 63--84\relax
\mciteBstWouldAddEndPuncttrue
\mciteSetBstMidEndSepPunct{\mcitedefaultmidpunct}
{\mcitedefaultendpunct}{\mcitedefaultseppunct}\relax
\EndOfBibitem
\bibitem[Wang(2012)]{wang2012kernel}
Q.~Wang, \emph{arXiv preprint arXiv:1207.3538}, 2012\relax
\mciteBstWouldAddEndPuncttrue
\mciteSetBstMidEndSepPunct{\mcitedefaultmidpunct}
{\mcitedefaultendpunct}{\mcitedefaultseppunct}\relax
\EndOfBibitem
\bibitem[Lloyd \emph{et~al.}(2014)Lloyd, Mohseni, and
  Rebentrost]{lloyd2014quantum}
S.~Lloyd, M.~Mohseni and P.~Rebentrost, \emph{Nature Physics}, 2014,
  \textbf{10}, 631--633\relax
\mciteBstWouldAddEndPuncttrue
\mciteSetBstMidEndSepPunct{\mcitedefaultmidpunct}
{\mcitedefaultendpunct}{\mcitedefaultseppunct}\relax
\EndOfBibitem
\bibitem[Li \emph{et~al.}(2021)Li, Chai, Guo, Ji, Wang, Shi, Wang, Lloyd, and
  Du]{li2021resonant}
Z.~Li, Z.~Chai, Y.~Guo, W.~Ji, M.~Wang, F.~Shi, Y.~Wang, S.~Lloyd and J.~Du,
  \emph{Science Advances}, 2021, \textbf{7}, eabg2589\relax
\mciteBstWouldAddEndPuncttrue
\mciteSetBstMidEndSepPunct{\mcitedefaultmidpunct}
{\mcitedefaultendpunct}{\mcitedefaultseppunct}\relax
\EndOfBibitem
\bibitem[Li \emph{et~al.}(2020)Li, Zhou, Xu, Hu, and Fan]{li2020quantum}
Y.~Li, R.-G. Zhou, R.~Xu, W.~Hu and P.~Fan, \emph{Quantum Science and
  Technology}, 2020, \textbf{6}, 014001\relax
\mciteBstWouldAddEndPuncttrue
\mciteSetBstMidEndSepPunct{\mcitedefaultmidpunct}
{\mcitedefaultendpunct}{\mcitedefaultseppunct}\relax
\EndOfBibitem
\bibitem[Cover and Hart(1967)]{cover1967nearest}
T.~Cover and P.~Hart, \emph{IEEE transactions on information theory}, 1967,
  \textbf{13}, 21--27\relax
\mciteBstWouldAddEndPuncttrue
\mciteSetBstMidEndSepPunct{\mcitedefaultmidpunct}
{\mcitedefaultendpunct}{\mcitedefaultseppunct}\relax
\EndOfBibitem
\bibitem[Ruan \emph{et~al.}(2017)Ruan, Xue, Liu, Tan, and Li]{ruan2017quantum}
Y.~Ruan, X.~Xue, H.~Liu, J.~Tan and X.~Li, \emph{International Journal of
  Theoretical Physics}, 2017, \textbf{56}, 3496--3507\relax
\mciteBstWouldAddEndPuncttrue
\mciteSetBstMidEndSepPunct{\mcitedefaultmidpunct}
{\mcitedefaultendpunct}{\mcitedefaultseppunct}\relax
\EndOfBibitem
\bibitem[Gou \emph{et~al.}(2012)Gou, Yi, Du, and Xiong]{gou2012local}
J.~Gou, Z.~Yi, L.~Du and T.~Xiong, \emph{The Computer Journal}, 2012,
  \textbf{55}, 1058--1071\relax
\mciteBstWouldAddEndPuncttrue
\mciteSetBstMidEndSepPunct{\mcitedefaultmidpunct}
{\mcitedefaultendpunct}{\mcitedefaultseppunct}\relax
\EndOfBibitem
\bibitem[Geler \emph{et~al.}(2016)Geler, Kurbalija, Radovanovi{\'c}, and
  Ivanovi{\'c}]{geler2016comparison}
Z.~Geler, V.~Kurbalija, M.~Radovanovi{\'c} and M.~Ivanovi{\'c}, \emph{Knowledge
  and Information Systems}, 2016, \textbf{48}, 331--378\relax
\mciteBstWouldAddEndPuncttrue
\mciteSetBstMidEndSepPunct{\mcitedefaultmidpunct}
{\mcitedefaultendpunct}{\mcitedefaultseppunct}\relax
\EndOfBibitem
\bibitem[Lloyd \emph{et~al.}(2013)Lloyd, Mohseni, and
  Rebentrost]{lloyd2013quantum}
S.~Lloyd, M.~Mohseni and P.~Rebentrost, \emph{arXiv preprint arXiv:1307.0411},
  2013\relax
\mciteBstWouldAddEndPuncttrue
\mciteSetBstMidEndSepPunct{\mcitedefaultmidpunct}
{\mcitedefaultendpunct}{\mcitedefaultseppunct}\relax
\EndOfBibitem
\bibitem[Aaronson(2010)]{aaronson2010bqp}
S.~Aaronson, Proceedings of the forty-second ACM symposium on Theory of
  computing, 2010, pp. 141--150\relax
\mciteBstWouldAddEndPuncttrue
\mciteSetBstMidEndSepPunct{\mcitedefaultmidpunct}
{\mcitedefaultendpunct}{\mcitedefaultseppunct}\relax
\EndOfBibitem
\bibitem[Wi{\'s}niewska and Sawerwain(2018)]{wisniewska2018recognizing}
J.~Wi{\'s}niewska and M.~Sawerwain, \emph{Vietnam Journal of Computer Science},
  2018, \textbf{5}, 197--204\relax
\mciteBstWouldAddEndPuncttrue
\mciteSetBstMidEndSepPunct{\mcitedefaultmidpunct}
{\mcitedefaultendpunct}{\mcitedefaultseppunct}\relax
\EndOfBibitem
\bibitem[Wang \emph{et~al.}(2019)Wang, Wang, Li, Adu-Gyamfi, Tian, and
  Zhu]{wang2019improved}
Y.~Wang, R.~Wang, D.~Li, D.~Adu-Gyamfi, K.~Tian and Y.~Zhu, \emph{International
  Journal of Theoretical Physics}, 2019, \textbf{58}, 2331--2340\relax
\mciteBstWouldAddEndPuncttrue
\mciteSetBstMidEndSepPunct{\mcitedefaultmidpunct}
{\mcitedefaultendpunct}{\mcitedefaultseppunct}\relax
\EndOfBibitem
\bibitem[Murthy(1998)]{murthy1998automatic}
S.~K. Murthy, \emph{Data mining and knowledge discovery}, 1998, \textbf{2},
  345--389\relax
\mciteBstWouldAddEndPuncttrue
\mciteSetBstMidEndSepPunct{\mcitedefaultmidpunct}
{\mcitedefaultendpunct}{\mcitedefaultseppunct}\relax
\EndOfBibitem
\bibitem[Kotsiantis \emph{et~al.}(2007)Kotsiantis, Zaharakis,
  Pintelas,\emph{et~al.}]{kotsiantis2007supervised}
S.~B. Kotsiantis, I.~Zaharakis, P.~Pintelas \emph{et~al.}, \emph{Emerging
  artificial intelligence applications in computer engineering}, 2007,
  \textbf{160}, 3--24\relax
\mciteBstWouldAddEndPuncttrue
\mciteSetBstMidEndSepPunct{\mcitedefaultmidpunct}
{\mcitedefaultendpunct}{\mcitedefaultseppunct}\relax
\EndOfBibitem
\bibitem[Hunt \emph{et~al.}(1966)Hunt, Marin, and Stone]{hunt1966experiments}
E.~B. Hunt, J.~Marin and P.~J. Stone, 1966\relax
\mciteBstWouldAddEndPuncttrue
\mciteSetBstMidEndSepPunct{\mcitedefaultmidpunct}
{\mcitedefaultendpunct}{\mcitedefaultseppunct}\relax
\EndOfBibitem
\bibitem[Breiman \emph{et~al.}(2017)Breiman, Friedman, Olshen, and
  Stone]{breiman2017classification}
L.~Breiman, J.~H. Friedman, R.~A. Olshen and C.~J. Stone, \emph{Classification
  and regression trees}, Routledge, 2017\relax
\mciteBstWouldAddEndPuncttrue
\mciteSetBstMidEndSepPunct{\mcitedefaultmidpunct}
{\mcitedefaultendpunct}{\mcitedefaultseppunct}\relax
\EndOfBibitem
\bibitem[Farhi and Gutmann(1998)]{farhi1998quantum}
E.~Farhi and S.~Gutmann, \emph{Physical Review A}, 1998, \textbf{58}, 915\relax
\mciteBstWouldAddEndPuncttrue
\mciteSetBstMidEndSepPunct{\mcitedefaultmidpunct}
{\mcitedefaultendpunct}{\mcitedefaultseppunct}\relax
\EndOfBibitem
\bibitem[Lu and Braunstein(2014)]{lu2014quantum}
S.~Lu and S.~L. Braunstein, \emph{Quantum information processing}, 2014,
  \textbf{13}, 757--770\relax
\mciteBstWouldAddEndPuncttrue
\mciteSetBstMidEndSepPunct{\mcitedefaultmidpunct}
{\mcitedefaultendpunct}{\mcitedefaultseppunct}\relax
\EndOfBibitem
\bibitem[Khadiev \emph{et~al.}(2019)Khadiev, Mannapov, and
  Safina]{khadiev2019quantum}
K.~Khadiev, I.~Mannapov and L.~Safina, \emph{arXiv preprint arXiv:1907.06840},
  2019\relax
\mciteBstWouldAddEndPuncttrue
\mciteSetBstMidEndSepPunct{\mcitedefaultmidpunct}
{\mcitedefaultendpunct}{\mcitedefaultseppunct}\relax
\EndOfBibitem
\bibitem[Wu \emph{et~al.}(2008)Wu, Kumar, Quinlan, Ghosh, Yang, Motoda,
  McLachlan, Ng, Liu, Philip,\emph{et~al.}]{wu2008top}
X.~Wu, V.~Kumar, J.~R. Quinlan, J.~Ghosh, Q.~Yang, H.~Motoda, G.~J. McLachlan,
  A.~Ng, B.~Liu, S.~Y. Philip \emph{et~al.}, \emph{Knowledge and information
  systems}, 2008, \textbf{14}, 1--37\relax
\mciteBstWouldAddEndPuncttrue
\mciteSetBstMidEndSepPunct{\mcitedefaultmidpunct}
{\mcitedefaultendpunct}{\mcitedefaultseppunct}\relax
\EndOfBibitem
\bibitem[Jensen and Nielsen(2007)]{jensen2007bayesian}
F.~V. Jensen and T.~D. Nielsen, \emph{Bayesian networks and decision graphs},
  Springer, 2007, vol.~2\relax
\mciteBstWouldAddEndPuncttrue
\mciteSetBstMidEndSepPunct{\mcitedefaultmidpunct}
{\mcitedefaultendpunct}{\mcitedefaultseppunct}\relax
\EndOfBibitem
\bibitem[Heckerman(2008)]{heckerman2008tutorial}
D.~Heckerman, \emph{Innovations in Bayesian networks}, 2008,  33--82\relax
\mciteBstWouldAddEndPuncttrue
\mciteSetBstMidEndSepPunct{\mcitedefaultmidpunct}
{\mcitedefaultendpunct}{\mcitedefaultseppunct}\relax
\EndOfBibitem
\bibitem[Tucci(1995)]{tucci1995quantum}
R.~R. Tucci, \emph{International Journal of Modern Physics B}, 1995,
  \textbf{9}, 295--337\relax
\mciteBstWouldAddEndPuncttrue
\mciteSetBstMidEndSepPunct{\mcitedefaultmidpunct}
{\mcitedefaultendpunct}{\mcitedefaultseppunct}\relax
\EndOfBibitem
\bibitem[Leifer and Poulin(2008)]{leifer2008quantum}
M.~S. Leifer and D.~Poulin, \emph{Annals of Physics}, 2008, \textbf{323},
  1899--1946\relax
\mciteBstWouldAddEndPuncttrue
\mciteSetBstMidEndSepPunct{\mcitedefaultmidpunct}
{\mcitedefaultendpunct}{\mcitedefaultseppunct}\relax
\EndOfBibitem
\bibitem[Moreira and Wichert(2016)]{moreira2016quantum}
C.~Moreira and A.~Wichert, \emph{Frontiers in psychology}, 2016, \textbf{7},
  11\relax
\mciteBstWouldAddEndPuncttrue
\mciteSetBstMidEndSepPunct{\mcitedefaultmidpunct}
{\mcitedefaultendpunct}{\mcitedefaultseppunct}\relax
\EndOfBibitem
\bibitem[Low \emph{et~al.}(2014)Low, Yoder, and Chuang]{low2014quantum}
G.~H. Low, T.~J. Yoder and I.~L. Chuang, \emph{Physical Review A}, 2014,
  \textbf{89}, 062315\relax
\mciteBstWouldAddEndPuncttrue
\mciteSetBstMidEndSepPunct{\mcitedefaultmidpunct}
{\mcitedefaultendpunct}{\mcitedefaultseppunct}\relax
\EndOfBibitem
\bibitem[Borujeni \emph{et~al.}(2021)Borujeni, Nannapaneni, Nguyen, Behrman,
  and Steck]{borujeni2021quantum}
S.~E. Borujeni, S.~Nannapaneni, N.~H. Nguyen, E.~C. Behrman and J.~E. Steck,
  \emph{Expert Systems with Applications}, 2021, \textbf{176}, 114768\relax
\mciteBstWouldAddEndPuncttrue
\mciteSetBstMidEndSepPunct{\mcitedefaultmidpunct}
{\mcitedefaultendpunct}{\mcitedefaultseppunct}\relax
\EndOfBibitem
\bibitem[Vapnik(2013)]{vapnik2013nature}
V.~Vapnik, \emph{The nature of statistical learning theory}, Springer science
  \& business media, 2013\relax
\mciteBstWouldAddEndPuncttrue
\mciteSetBstMidEndSepPunct{\mcitedefaultmidpunct}
{\mcitedefaultendpunct}{\mcitedefaultseppunct}\relax
\EndOfBibitem
\bibitem[Li \emph{et~al.}(2015)Li, Liu, Xu, and Du]{li2015experimental}
Z.~Li, X.~Liu, N.~Xu and J.~Du, \emph{Physical review letters}, 2015,
  \textbf{114}, 140504\relax
\mciteBstWouldAddEndPuncttrue
\mciteSetBstMidEndSepPunct{\mcitedefaultmidpunct}
{\mcitedefaultendpunct}{\mcitedefaultseppunct}\relax
\EndOfBibitem
\bibitem[Rebentrost \emph{et~al.}(2014)Rebentrost, Mohseni, and
  Lloyd]{rebentrost2014quantum}
P.~Rebentrost, M.~Mohseni and S.~Lloyd, \emph{Physical review letters}, 2014,
  \textbf{113}, 130503\relax
\mciteBstWouldAddEndPuncttrue
\mciteSetBstMidEndSepPunct{\mcitedefaultmidpunct}
{\mcitedefaultendpunct}{\mcitedefaultseppunct}\relax
\EndOfBibitem
\bibitem[Harrow \emph{et~al.}(2009)Harrow, Hassidim, and
  Lloyd]{harrow2009quantum}
A.~W. Harrow, A.~Hassidim and S.~Lloyd, \emph{Physical review letters}, 2009,
  \textbf{103}, 150502\relax
\mciteBstWouldAddEndPuncttrue
\mciteSetBstMidEndSepPunct{\mcitedefaultmidpunct}
{\mcitedefaultendpunct}{\mcitedefaultseppunct}\relax
\EndOfBibitem
\bibitem[Schuld and Killoran(2019)]{schuld2019quantum}
M.~Schuld and N.~Killoran, \emph{Physical review letters}, 2019, \textbf{122},
  040504\relax
\mciteBstWouldAddEndPuncttrue
\mciteSetBstMidEndSepPunct{\mcitedefaultmidpunct}
{\mcitedefaultendpunct}{\mcitedefaultseppunct}\relax
\EndOfBibitem
\bibitem[Liu \emph{et~al.}(2021)Liu, Arunachalam, and Temme]{liu2021rigorous}
Y.~Liu, S.~Arunachalam and K.~Temme, \emph{Nature Physics}, 2021, \textbf{17},
  1013--1017\relax
\mciteBstWouldAddEndPuncttrue
\mciteSetBstMidEndSepPunct{\mcitedefaultmidpunct}
{\mcitedefaultendpunct}{\mcitedefaultseppunct}\relax
\EndOfBibitem
\bibitem[Otten \emph{et~al.}(2020)Otten, Goumiri, Priest, Chapline, and
  Schneider]{otten2020quantum}
M.~Otten, I.~R. Goumiri, B.~W. Priest, G.~F. Chapline and M.~D. Schneider,
  \emph{arXiv preprint arXiv:2004.11280}, 2020\relax
\mciteBstWouldAddEndPuncttrue
\mciteSetBstMidEndSepPunct{\mcitedefaultmidpunct}
{\mcitedefaultendpunct}{\mcitedefaultseppunct}\relax
\EndOfBibitem
\bibitem[Williams and Rasmussen(2006)]{williams2006gaussian}
C.~K. Williams and C.~E. Rasmussen, \emph{Gaussian processes for machine
  learning}, MIT press Cambridge, MA, 2006, vol.~2\relax
\mciteBstWouldAddEndPuncttrue
\mciteSetBstMidEndSepPunct{\mcitedefaultmidpunct}
{\mcitedefaultendpunct}{\mcitedefaultseppunct}\relax
\EndOfBibitem
\bibitem[Deringer \emph{et~al.}(2021)Deringer, Bart{\'o}k, Bernstein, Wilkins,
  Ceriotti, and Cs{\'a}nyi]{deringer2021gaussian}
V.~L. Deringer, A.~P. Bart{\'o}k, N.~Bernstein, D.~M. Wilkins, M.~Ceriotti and
  G.~Cs{\'a}nyi, \emph{Chemical Reviews}, 2021, \textbf{121},
  10073--10141\relax
\mciteBstWouldAddEndPuncttrue
\mciteSetBstMidEndSepPunct{\mcitedefaultmidpunct}
{\mcitedefaultendpunct}{\mcitedefaultseppunct}\relax
\EndOfBibitem
\bibitem[Bishop(2006)]{bishop2006pattern}
C.~M. Bishop, \emph{Pattern Recognition and Machine Learning (Information
  Science and Statistics)}, Springer-Verlag, Berlin, Heidelberg, 2006\relax
\mciteBstWouldAddEndPuncttrue
\mciteSetBstMidEndSepPunct{\mcitedefaultmidpunct}
{\mcitedefaultendpunct}{\mcitedefaultseppunct}\relax
\EndOfBibitem
\bibitem[Freund and Schapire(1999)]{Freuend_percept}
Y.~Freund and R.~Schapire, \emph{Machine Learning}, 1999, \textbf{37}, \relax
\mciteBstWouldAddEndPuncttrue
\mciteSetBstMidEndSepPunct{\mcitedefaultmidpunct}
{\mcitedefaultendpunct}{\mcitedefaultseppunct}\relax
\EndOfBibitem
\bibitem[Al-Mahasneh \emph{et~al.}(2017)Al-Mahasneh, Anavatti, and
  Garratt]{al2017development}
A.~J. Al-Mahasneh, S.~G. Anavatti and M.~A. Garratt, 2017 International
  Conference on Advanced Mechatronics, Intelligent Manufacture, and Industrial
  Automation (ICAMIMIA), 2017, pp. 1--6\relax
\mciteBstWouldAddEndPuncttrue
\mciteSetBstMidEndSepPunct{\mcitedefaultmidpunct}
{\mcitedefaultendpunct}{\mcitedefaultseppunct}\relax
\EndOfBibitem
\bibitem[Rosenblatt(1958)]{rosenblatt1958perceptron}
F.~Rosenblatt, \emph{Psychological review}, 1958, \textbf{65}, 386\relax
\mciteBstWouldAddEndPuncttrue
\mciteSetBstMidEndSepPunct{\mcitedefaultmidpunct}
{\mcitedefaultendpunct}{\mcitedefaultseppunct}\relax
\EndOfBibitem
\bibitem[Haykin and Network(2004)]{haykin2004comprehensive}
S.~Haykin and N.~Network, \emph{Neural networks}, 2004, \textbf{2}, 41\relax
\mciteBstWouldAddEndPuncttrue
\mciteSetBstMidEndSepPunct{\mcitedefaultmidpunct}
{\mcitedefaultendpunct}{\mcitedefaultseppunct}\relax
\EndOfBibitem
\bibitem[Lau and Lim(2018)]{lau2018review}
M.~M. Lau and K.~H. Lim, 2018 IEEE-EMBS Conference on Biomedical Engineering
  and Sciences (IECBES), 2018, pp. 686--690\relax
\mciteBstWouldAddEndPuncttrue
\mciteSetBstMidEndSepPunct{\mcitedefaultmidpunct}
{\mcitedefaultendpunct}{\mcitedefaultseppunct}\relax
\EndOfBibitem
\bibitem[Sharma and Sharma(2017)]{sharma2017activation}
S.~Sharma and S.~Sharma, \emph{Towards Data Science}, 2017, \textbf{6},
  310--316\relax
\mciteBstWouldAddEndPuncttrue
\mciteSetBstMidEndSepPunct{\mcitedefaultmidpunct}
{\mcitedefaultendpunct}{\mcitedefaultseppunct}\relax
\EndOfBibitem
\bibitem[Maas \emph{et~al.}(2013)Maas, Hannun,
  Ng,\emph{et~al.}]{maas2013rectifier}
A.~L. Maas, A.~Y. Hannun, A.~Y. Ng \emph{et~al.}, Proc. icml, 2013, p.~3\relax
\mciteBstWouldAddEndPuncttrue
\mciteSetBstMidEndSepPunct{\mcitedefaultmidpunct}
{\mcitedefaultendpunct}{\mcitedefaultseppunct}\relax
\EndOfBibitem
\bibitem[Jankowski \emph{et~al.}(1996)Jankowski, Lozowski, and
  Zurada]{jankowski1996complex}
S.~Jankowski, A.~Lozowski and J.~M. Zurada, \emph{IEEE Transactions on Neural
  Networks}, 1996, \textbf{7}, 1491--1496\relax
\mciteBstWouldAddEndPuncttrue
\mciteSetBstMidEndSepPunct{\mcitedefaultmidpunct}
{\mcitedefaultendpunct}{\mcitedefaultseppunct}\relax
\EndOfBibitem
\bibitem[Tanaka and Aihara(2009)]{tanaka2009complex}
G.~Tanaka and K.~Aihara, \emph{IEEE Transactions on Neural Networks}, 2009,
  \textbf{20}, 1463--1473\relax
\mciteBstWouldAddEndPuncttrue
\mciteSetBstMidEndSepPunct{\mcitedefaultmidpunct}
{\mcitedefaultendpunct}{\mcitedefaultseppunct}\relax
\EndOfBibitem
\bibitem[Karlik and Olgac(2011)]{karlik2011performance}
B.~Karlik and A.~V. Olgac, \emph{International Journal of Artificial
  Intelligence and Expert Systems}, 2011, \textbf{1}, 111--122\relax
\mciteBstWouldAddEndPuncttrue
\mciteSetBstMidEndSepPunct{\mcitedefaultmidpunct}
{\mcitedefaultendpunct}{\mcitedefaultseppunct}\relax
\EndOfBibitem
\bibitem[Agostinelli \emph{et~al.}(2014)Agostinelli, Hoffman, Sadowski, and
  Baldi]{agostinelli2014learning}
F.~Agostinelli, M.~Hoffman, P.~Sadowski and P.~Baldi, \emph{arXiv preprint
  arXiv:1412.6830}, 2014\relax
\mciteBstWouldAddEndPuncttrue
\mciteSetBstMidEndSepPunct{\mcitedefaultmidpunct}
{\mcitedefaultendpunct}{\mcitedefaultseppunct}\relax
\EndOfBibitem
\bibitem[De~Boer \emph{et~al.}(2005)De~Boer, Kroese, Mannor, and
  Rubinstein]{de2005tutorial}
P.-T. De~Boer, D.~P. Kroese, S.~Mannor and R.~Y. Rubinstein, \emph{Annals of
  operations research}, 2005, \textbf{134}, 19--67\relax
\mciteBstWouldAddEndPuncttrue
\mciteSetBstMidEndSepPunct{\mcitedefaultmidpunct}
{\mcitedefaultendpunct}{\mcitedefaultseppunct}\relax
\EndOfBibitem
\bibitem[Rumelhart \emph{et~al.}(1986)Rumelhart, Hinton, and
  Williams]{rumelhart1986learning}
D.~E. Rumelhart, G.~E. Hinton and R.~J. Williams, \emph{nature}, 1986,
  \textbf{323}, 533--536\relax
\mciteBstWouldAddEndPuncttrue
\mciteSetBstMidEndSepPunct{\mcitedefaultmidpunct}
{\mcitedefaultendpunct}{\mcitedefaultseppunct}\relax
\EndOfBibitem
\bibitem[LeCun \emph{et~al.}(2015)LeCun, Bengio, and Hinton]{lecun2015deep}
Y.~LeCun, Y.~Bengio and G.~Hinton, \emph{nature}, 2015, \textbf{521},
  436--444\relax
\mciteBstWouldAddEndPuncttrue
\mciteSetBstMidEndSepPunct{\mcitedefaultmidpunct}
{\mcitedefaultendpunct}{\mcitedefaultseppunct}\relax
\EndOfBibitem
\bibitem[Ruder(2016)]{ruder2016overview}
S.~Ruder, \emph{arXiv preprint arXiv:1609.04747}, 2016\relax
\mciteBstWouldAddEndPuncttrue
\mciteSetBstMidEndSepPunct{\mcitedefaultmidpunct}
{\mcitedefaultendpunct}{\mcitedefaultseppunct}\relax
\EndOfBibitem
\bibitem[Duchi \emph{et~al.}(2011)Duchi, Hazan, and Singer]{duchi2011adaptive}
J.~Duchi, E.~Hazan and Y.~Singer, \emph{Journal of machine learning research},
  2011, \textbf{12}, 2121--2159\relax
\mciteBstWouldAddEndPuncttrue
\mciteSetBstMidEndSepPunct{\mcitedefaultmidpunct}
{\mcitedefaultendpunct}{\mcitedefaultseppunct}\relax
\EndOfBibitem
\bibitem[Kingma and Ba(2014)]{kingma2014adam}
D.~P. Kingma and J.~Ba, \emph{arXiv preprint arXiv:1412.6980}, 2014\relax
\mciteBstWouldAddEndPuncttrue
\mciteSetBstMidEndSepPunct{\mcitedefaultmidpunct}
{\mcitedefaultendpunct}{\mcitedefaultseppunct}\relax
\EndOfBibitem
\bibitem[Dozat(2016)]{dozat2016incorporating}
T.~Dozat, 2016\relax
\mciteBstWouldAddEndPuncttrue
\mciteSetBstMidEndSepPunct{\mcitedefaultmidpunct}
{\mcitedefaultendpunct}{\mcitedefaultseppunct}\relax
\EndOfBibitem
\bibitem[Wan \emph{et~al.}(2013)Wan, Zeiler, Zhang, Le~Cun, and
  Fergus]{wan2013regularization}
L.~Wan, M.~Zeiler, S.~Zhang, Y.~Le~Cun and R.~Fergus, International conference
  on machine learning, 2013, pp. 1058--1066\relax
\mciteBstWouldAddEndPuncttrue
\mciteSetBstMidEndSepPunct{\mcitedefaultmidpunct}
{\mcitedefaultendpunct}{\mcitedefaultseppunct}\relax
\EndOfBibitem
\bibitem[Girosi \emph{et~al.}(1995)Girosi, Jones, and
  Poggio]{girosi1995regularization}
F.~Girosi, M.~Jones and T.~Poggio, \emph{Neural computation}, 1995, \textbf{7},
  219--269\relax
\mciteBstWouldAddEndPuncttrue
\mciteSetBstMidEndSepPunct{\mcitedefaultmidpunct}
{\mcitedefaultendpunct}{\mcitedefaultseppunct}\relax
\EndOfBibitem
\bibitem[Srivastava \emph{et~al.}(2014)Srivastava, Hinton, Krizhevsky,
  Sutskever, and Salakhutdinov]{srivastava2014dropout}
N.~Srivastava, G.~Hinton, A.~Krizhevsky, I.~Sutskever and R.~Salakhutdinov,
  \emph{The journal of machine learning research}, 2014, \textbf{15},
  1929--1958\relax
\mciteBstWouldAddEndPuncttrue
\mciteSetBstMidEndSepPunct{\mcitedefaultmidpunct}
{\mcitedefaultendpunct}{\mcitedefaultseppunct}\relax
\EndOfBibitem
\bibitem[Xia and Kais(2020)]{xia2020hybrid}
R.~Xia and S.~Kais, \emph{Entropy}, 2020, \textbf{22}, 828\relax
\mciteBstWouldAddEndPuncttrue
\mciteSetBstMidEndSepPunct{\mcitedefaultmidpunct}
{\mcitedefaultendpunct}{\mcitedefaultseppunct}\relax
\EndOfBibitem
\bibitem[McClean \emph{et~al.}(2018)McClean, Boixo, Smelyanskiy, Babbush, and
  Neven]{mcclean2018barren}
J.~R. McClean, S.~Boixo, V.~N. Smelyanskiy, R.~Babbush and H.~Neven,
  \emph{Nature communications}, 2018, \textbf{9}, 1--6\relax
\mciteBstWouldAddEndPuncttrue
\mciteSetBstMidEndSepPunct{\mcitedefaultmidpunct}
{\mcitedefaultendpunct}{\mcitedefaultseppunct}\relax
\EndOfBibitem
\bibitem[Mari \emph{et~al.}(2020)Mari, Bromley, Izaac, Schuld, and
  Killoran]{mari2020transfer}
A.~Mari, T.~R. Bromley, J.~Izaac, M.~Schuld and N.~Killoran, \emph{Quantum},
  2020, \textbf{4}, 340\relax
\mciteBstWouldAddEndPuncttrue
\mciteSetBstMidEndSepPunct{\mcitedefaultmidpunct}
{\mcitedefaultendpunct}{\mcitedefaultseppunct}\relax
\EndOfBibitem
\bibitem[Li and Kais(2021)]{10.1088/1367-2630/ac2cb4}
J.~Li and S.~Kais, \emph{New Journal of Physics}, 2021\relax
\mciteBstWouldAddEndPuncttrue
\mciteSetBstMidEndSepPunct{\mcitedefaultmidpunct}
{\mcitedefaultendpunct}{\mcitedefaultseppunct}\relax
\EndOfBibitem
\bibitem[Wiebe and Kliuchnikov(2013)]{wiebe2013floating}
N.~Wiebe and V.~Kliuchnikov, \emph{New Journal of Physics}, 2013, \textbf{15},
  093041\relax
\mciteBstWouldAddEndPuncttrue
\mciteSetBstMidEndSepPunct{\mcitedefaultmidpunct}
{\mcitedefaultendpunct}{\mcitedefaultseppunct}\relax
\EndOfBibitem
\bibitem[Paetznick and Svore(2013)]{paetznick2013repeat}
A.~Paetznick and K.~M. Svore, \emph{arXiv preprint arXiv:1311.1074}, 2013\relax
\mciteBstWouldAddEndPuncttrue
\mciteSetBstMidEndSepPunct{\mcitedefaultmidpunct}
{\mcitedefaultendpunct}{\mcitedefaultseppunct}\relax
\EndOfBibitem
\bibitem[Cao \emph{et~al.}(2017)Cao, Guerreschi, and
  Aspuru-Guzik]{cao2017quantum}
Y.~Cao, G.~G. Guerreschi and A.~Aspuru-Guzik, \emph{arXiv preprint
  arXiv:1711.11240}, 2017\relax
\mciteBstWouldAddEndPuncttrue
\mciteSetBstMidEndSepPunct{\mcitedefaultmidpunct}
{\mcitedefaultendpunct}{\mcitedefaultseppunct}\relax
\EndOfBibitem
\bibitem[Daskin(2018)]{daskin2018simple}
A.~Daskin, 2018 IEEE International Conference on Systems, Man, and Cybernetics
  (SMC), 2018, pp. 2887--2891\relax
\mciteBstWouldAddEndPuncttrue
\mciteSetBstMidEndSepPunct{\mcitedefaultmidpunct}
{\mcitedefaultendpunct}{\mcitedefaultseppunct}\relax
\EndOfBibitem
\bibitem[Schuld \emph{et~al.}(2015)Schuld, Sinayskiy, and
  Petruccione]{schuld2015simulating}
M.~Schuld, I.~Sinayskiy and F.~Petruccione, \emph{Physics Letters A}, 2015,
  \textbf{379}, 660--663\relax
\mciteBstWouldAddEndPuncttrue
\mciteSetBstMidEndSepPunct{\mcitedefaultmidpunct}
{\mcitedefaultendpunct}{\mcitedefaultseppunct}\relax
\EndOfBibitem
\bibitem[Cong \emph{et~al.}(2019)Cong, Choi, and Lukin]{cong2019quantum}
I.~Cong, S.~Choi and M.~D. Lukin, \emph{Nature Physics}, 2019, \textbf{15},
  1273--1278\relax
\mciteBstWouldAddEndPuncttrue
\mciteSetBstMidEndSepPunct{\mcitedefaultmidpunct}
{\mcitedefaultendpunct}{\mcitedefaultseppunct}\relax
\EndOfBibitem
\bibitem[Rawat and Wang(2017)]{rawat2017deep}
W.~Rawat and Z.~Wang, \emph{Neural computation}, 2017, \textbf{29},
  2352--2449\relax
\mciteBstWouldAddEndPuncttrue
\mciteSetBstMidEndSepPunct{\mcitedefaultmidpunct}
{\mcitedefaultendpunct}{\mcitedefaultseppunct}\relax
\EndOfBibitem
\bibitem[Voulodimos \emph{et~al.}(2018)Voulodimos, Doulamis, Doulamis, and
  Protopapadakis]{voulodimos2018deep}
A.~Voulodimos, N.~Doulamis, A.~Doulamis and E.~Protopapadakis,
  \emph{Computational intelligence and neuroscience}, 2018, \textbf{2018},
  13\relax
\mciteBstWouldAddEndPuncttrue
\mciteSetBstMidEndSepPunct{\mcitedefaultmidpunct}
{\mcitedefaultendpunct}{\mcitedefaultseppunct}\relax
\EndOfBibitem
\bibitem[Albawi \emph{et~al.}(2017)Albawi, Mohammed, and
  Al-Zawi]{albawi2017understanding}
S.~Albawi, T.~A. Mohammed and S.~Al-Zawi, 2017 International Conference on
  Engineering and Technology (ICET), 2017, pp. 1--6\relax
\mciteBstWouldAddEndPuncttrue
\mciteSetBstMidEndSepPunct{\mcitedefaultmidpunct}
{\mcitedefaultendpunct}{\mcitedefaultseppunct}\relax
\EndOfBibitem
\bibitem[LeCun \emph{et~al.}(1999)LeCun, Haffner, Bottou, and
  Bengio]{lecun1999object}
Y.~LeCun, P.~Haffner, L.~Bottou and Y.~Bengio, \emph{Shape, contour and
  grouping in computer vision}, Springer, 1999, pp. 319--345\relax
\mciteBstWouldAddEndPuncttrue
\mciteSetBstMidEndSepPunct{\mcitedefaultmidpunct}
{\mcitedefaultendpunct}{\mcitedefaultseppunct}\relax
\EndOfBibitem
\bibitem[Dhillon and Verma(2020)]{dhillon2020convolutional}
A.~Dhillon and G.~K. Verma, \emph{Progress in Artificial Intelligence}, 2020,
  \textbf{9}, 85--112\relax
\mciteBstWouldAddEndPuncttrue
\mciteSetBstMidEndSepPunct{\mcitedefaultmidpunct}
{\mcitedefaultendpunct}{\mcitedefaultseppunct}\relax
\EndOfBibitem
\bibitem[Aloysius and Geetha(2017)]{aloysius2017review}
N.~Aloysius and M.~Geetha, 2017 International Conference on Communication and
  Signal Processing (ICCSP), 2017, pp. 0588--0592\relax
\mciteBstWouldAddEndPuncttrue
\mciteSetBstMidEndSepPunct{\mcitedefaultmidpunct}
{\mcitedefaultendpunct}{\mcitedefaultseppunct}\relax
\EndOfBibitem
\bibitem[Kong and Lucey(2017)]{kong2017take}
C.~Kong and S.~Lucey, \emph{arXiv preprint arXiv:1712.02502}, 2017\relax
\mciteBstWouldAddEndPuncttrue
\mciteSetBstMidEndSepPunct{\mcitedefaultmidpunct}
{\mcitedefaultendpunct}{\mcitedefaultseppunct}\relax
\EndOfBibitem
\bibitem[Alzubaidi \emph{et~al.}(2021)Alzubaidi, Zhang, Humaidi, Al-Dujaili,
  Duan, Al-Shamma, Santamar{\'\i}a, Fadhel, Al-Amidie, and
  Farhan]{alzubaidi2021review}
L.~Alzubaidi, J.~Zhang, A.~J. Humaidi, A.~Al-Dujaili, Y.~Duan, O.~Al-Shamma,
  J.~Santamar{\'\i}a, M.~A. Fadhel, M.~Al-Amidie and L.~Farhan, \emph{Journal
  of big Data}, 2021, \textbf{8}, 1--74\relax
\mciteBstWouldAddEndPuncttrue
\mciteSetBstMidEndSepPunct{\mcitedefaultmidpunct}
{\mcitedefaultendpunct}{\mcitedefaultseppunct}\relax
\EndOfBibitem
\bibitem[Yu \emph{et~al.}(2014)Yu, Wang, Chen, and Wei]{yu2014mixed}
D.~Yu, H.~Wang, P.~Chen and Z.~Wei, International conference on rough sets and
  knowledge technology, 2014, pp. 364--375\relax
\mciteBstWouldAddEndPuncttrue
\mciteSetBstMidEndSepPunct{\mcitedefaultmidpunct}
{\mcitedefaultendpunct}{\mcitedefaultseppunct}\relax
\EndOfBibitem
\bibitem[Li \emph{et~al.}(2019)Li, Wang, Fan, Cao, Zhang, and Guo]{li2019teeth}
Z.~Li, S.-H. Wang, R.-R. Fan, G.~Cao, Y.-D. Zhang and T.~Guo,
  \emph{International Journal of Imaging Systems and Technology}, 2019,
  \textbf{29}, 577--583\relax
\mciteBstWouldAddEndPuncttrue
\mciteSetBstMidEndSepPunct{\mcitedefaultmidpunct}
{\mcitedefaultendpunct}{\mcitedefaultseppunct}\relax
\EndOfBibitem
\bibitem[Yin \emph{et~al.}(2017)Yin, Kann, Yu, and
  Sch{\"u}tze]{yin2017comparative}
W.~Yin, K.~Kann, M.~Yu and H.~Sch{\"u}tze, \emph{arXiv preprint
  arXiv:1702.01923}, 2017\relax
\mciteBstWouldAddEndPuncttrue
\mciteSetBstMidEndSepPunct{\mcitedefaultmidpunct}
{\mcitedefaultendpunct}{\mcitedefaultseppunct}\relax
\EndOfBibitem
\bibitem[Selvin \emph{et~al.}(2017)Selvin, Vinayakumar, Gopalakrishnan, Menon,
  and Soman]{selvin2017stock}
S.~Selvin, R.~Vinayakumar, E.~Gopalakrishnan, V.~K. Menon and K.~Soman, 2017
  international conference on advances in computing, communications and
  informatics (icacci), 2017, pp. 1643--1647\relax
\mciteBstWouldAddEndPuncttrue
\mciteSetBstMidEndSepPunct{\mcitedefaultmidpunct}
{\mcitedefaultendpunct}{\mcitedefaultseppunct}\relax
\EndOfBibitem
\bibitem[Cho \emph{et~al.}(2014)Cho, Van~Merri{\"e}nboer, Gulcehre, Bahdanau,
  Bougares, Schwenk, and Bengio]{cho2014learning}
K.~Cho, B.~Van~Merri{\"e}nboer, C.~Gulcehre, D.~Bahdanau, F.~Bougares,
  H.~Schwenk and Y.~Bengio, \emph{arXiv preprint arXiv:1406.1078}, 2014\relax
\mciteBstWouldAddEndPuncttrue
\mciteSetBstMidEndSepPunct{\mcitedefaultmidpunct}
{\mcitedefaultendpunct}{\mcitedefaultseppunct}\relax
\EndOfBibitem
\bibitem[Qin \emph{et~al.}(2017)Qin, Song, Chen, Cheng, Jiang, and
  Cottrell]{qin2017dual}
Y.~Qin, D.~Song, H.~Chen, W.~Cheng, G.~Jiang and G.~Cottrell, \emph{arXiv
  preprint arXiv:1704.02971}, 2017\relax
\mciteBstWouldAddEndPuncttrue
\mciteSetBstMidEndSepPunct{\mcitedefaultmidpunct}
{\mcitedefaultendpunct}{\mcitedefaultseppunct}\relax
\EndOfBibitem
\bibitem[Yu \emph{et~al.}(2019)Yu, Si, Hu, and Zhang]{yu2019review}
Y.~Yu, X.~Si, C.~Hu and J.~Zhang, \emph{Neural computation}, 2019, \textbf{31},
  1235--1270\relax
\mciteBstWouldAddEndPuncttrue
\mciteSetBstMidEndSepPunct{\mcitedefaultmidpunct}
{\mcitedefaultendpunct}{\mcitedefaultseppunct}\relax
\EndOfBibitem
\bibitem[Takaki \emph{et~al.}(2021)Takaki, Mitarai, Negoro, Fujii, and
  Kitagawa]{takaki2021learning}
Y.~Takaki, K.~Mitarai, M.~Negoro, K.~Fujii and M.~Kitagawa, \emph{Physical
  Review A}, 2021, \textbf{103}, 052414\relax
\mciteBstWouldAddEndPuncttrue
\mciteSetBstMidEndSepPunct{\mcitedefaultmidpunct}
{\mcitedefaultendpunct}{\mcitedefaultseppunct}\relax
\EndOfBibitem
\bibitem[Sherstinsky(2020)]{sherstinsky2020fundamentals}
A.~Sherstinsky, \emph{Physica D: Nonlinear Phenomena}, 2020, \textbf{404},
  132306\relax
\mciteBstWouldAddEndPuncttrue
\mciteSetBstMidEndSepPunct{\mcitedefaultmidpunct}
{\mcitedefaultendpunct}{\mcitedefaultseppunct}\relax
\EndOfBibitem
\bibitem[Yang \emph{et~al.}(2020)Yang, Yu, and Zhou]{yang2020lstm}
S.~Yang, X.~Yu and Y.~Zhou, 2020 International workshop on electronic
  communication and artificial intelligence (IWECAI), 2020, pp. 98--101\relax
\mciteBstWouldAddEndPuncttrue
\mciteSetBstMidEndSepPunct{\mcitedefaultmidpunct}
{\mcitedefaultendpunct}{\mcitedefaultseppunct}\relax
\EndOfBibitem
\bibitem[Dey and Salem(2017)]{dey2017gate}
R.~Dey and F.~M. Salem, 2017 IEEE 60th international midwest symposium on
  circuits and systems (MWSCAS), 2017, pp. 1597--1600\relax
\mciteBstWouldAddEndPuncttrue
\mciteSetBstMidEndSepPunct{\mcitedefaultmidpunct}
{\mcitedefaultendpunct}{\mcitedefaultseppunct}\relax
\EndOfBibitem
\bibitem[Chung \emph{et~al.}(2014)Chung, Gulcehre, Cho, and
  Bengio]{chung2014empirical}
J.~Chung, C.~Gulcehre, K.~Cho and Y.~Bengio, \emph{arXiv preprint
  arXiv:1412.3555}, 2014\relax
\mciteBstWouldAddEndPuncttrue
\mciteSetBstMidEndSepPunct{\mcitedefaultmidpunct}
{\mcitedefaultendpunct}{\mcitedefaultseppunct}\relax
\EndOfBibitem
\bibitem[Hochreiter and Schmidhuber(1997)]{hochreiter1997long}
S.~Hochreiter and J.~Schmidhuber, \emph{Neural computation}, 1997, \textbf{9},
  1735--1780\relax
\mciteBstWouldAddEndPuncttrue
\mciteSetBstMidEndSepPunct{\mcitedefaultmidpunct}
{\mcitedefaultendpunct}{\mcitedefaultseppunct}\relax
\EndOfBibitem
\bibitem[Chen \emph{et~al.}(2020)Chen, Yoo, and Fang]{chen2020quantum}
S.~Y.-C. Chen, S.~Yoo and Y.-L.~L. Fang, \emph{arXiv preprint
  arXiv:2009.01783}, 2020\relax
\mciteBstWouldAddEndPuncttrue
\mciteSetBstMidEndSepPunct{\mcitedefaultmidpunct}
{\mcitedefaultendpunct}{\mcitedefaultseppunct}\relax
\EndOfBibitem
\bibitem[Makhzani and Frey(2014)]{makhzani2014ksparse}
A.~Makhzani and B.~Frey, \emph{k-Sparse Autoencoders}, 2014\relax
\mciteBstWouldAddEndPuncttrue
\mciteSetBstMidEndSepPunct{\mcitedefaultmidpunct}
{\mcitedefaultendpunct}{\mcitedefaultseppunct}\relax
\EndOfBibitem
\bibitem[Vincent \emph{et~al.}(2010)Vincent, Larochelle, Lajoie, Bengio, and
  Manzagol]{10.5555/1756006.1953039}
P.~Vincent, H.~Larochelle, I.~Lajoie, Y.~Bengio and P.-A. Manzagol, \emph{J.
  Mach. Learn. Res.}, 2010, \textbf{11}, 3371–3408\relax
\mciteBstWouldAddEndPuncttrue
\mciteSetBstMidEndSepPunct{\mcitedefaultmidpunct}
{\mcitedefaultendpunct}{\mcitedefaultseppunct}\relax
\EndOfBibitem
\bibitem[Salakhutdinov and Hinton(2009)]{SALAKHUTDINOV2009969}
R.~Salakhutdinov and G.~Hinton, \emph{International Journal of Approximate
  Reasoning}, 2009, \textbf{50}, 969--978\relax
\mciteBstWouldAddEndPuncttrue
\mciteSetBstMidEndSepPunct{\mcitedefaultmidpunct}
{\mcitedefaultendpunct}{\mcitedefaultseppunct}\relax
\EndOfBibitem
\bibitem[Ribeiro \emph{et~al.}(2018)Ribeiro, Lazzaretti, and
  Lopes]{RIBEIRO201813}
M.~Ribeiro, A.~E. Lazzaretti and H.~S. Lopes, \emph{Pattern Recognition
  Letters}, 2018, \textbf{105}, 13--22\relax
\mciteBstWouldAddEndPuncttrue
\mciteSetBstMidEndSepPunct{\mcitedefaultmidpunct}
{\mcitedefaultendpunct}{\mcitedefaultseppunct}\relax
\EndOfBibitem
\bibitem[Romero \emph{et~al.}(2017)Romero, Olson, and
  Aspuru-Guzik]{2017_romero}
J.~Romero, J.~P. Olson and A.~Aspuru-Guzik, \emph{Quantum Science and
  Technology}, 2017, \textbf{2}, 045001\relax
\mciteBstWouldAddEndPuncttrue
\mciteSetBstMidEndSepPunct{\mcitedefaultmidpunct}
{\mcitedefaultendpunct}{\mcitedefaultseppunct}\relax
\EndOfBibitem
\bibitem[Shende \emph{et~al.}(2006)Shende, Bullock, and Markov]{841961}
V.~Shende, S.~Bullock and I.~Markov, 2006\relax
\mciteBstWouldAddEndPuncttrue
\mciteSetBstMidEndSepPunct{\mcitedefaultmidpunct}
{\mcitedefaultendpunct}{\mcitedefaultseppunct}\relax
\EndOfBibitem
\bibitem[Lamata \emph{et~al.}(2018)Lamata, Alvarez-Rodriguez, Martín-Guerrero,
  Sanz, and Solano]{2018_qaegen}
L.~Lamata, U.~Alvarez-Rodriguez, J.~D. Martín-Guerrero, M.~Sanz and E.~Solano,
  \emph{Quantum Science and Technology}, 2018, \textbf{4}, 014007\relax
\mciteBstWouldAddEndPuncttrue
\mciteSetBstMidEndSepPunct{\mcitedefaultmidpunct}
{\mcitedefaultendpunct}{\mcitedefaultseppunct}\relax
\EndOfBibitem
\bibitem[Wan \emph{et~al.}(2017)Wan, Dahlsten, Kristjánsson, Gardner, and
  Kim]{2017_ffnn}
K.~H. Wan, O.~Dahlsten, H.~Kristjánsson, R.~Gardner and M.~S. Kim, \emph{npj
  Quantum Information}, 2017, \textbf{3}, 36\relax
\mciteBstWouldAddEndPuncttrue
\mciteSetBstMidEndSepPunct{\mcitedefaultmidpunct}
{\mcitedefaultendpunct}{\mcitedefaultseppunct}\relax
\EndOfBibitem
\bibitem[Dilokthanakul \emph{et~al.}(2017)Dilokthanakul, Mediano, Garnelo, Lee,
  Salimbeni, Arulkumaran, and Shanahan]{dilokthanakul2017deep}
N.~Dilokthanakul, P.~A.~M. Mediano, M.~Garnelo, M.~C.~H. Lee, H.~Salimbeni,
  K.~Arulkumaran and M.~Shanahan, \emph{Deep Unsupervised Clustering with
  Gaussian Mixture Variational Autoencoders}, 2017\relax
\mciteBstWouldAddEndPuncttrue
\mciteSetBstMidEndSepPunct{\mcitedefaultmidpunct}
{\mcitedefaultendpunct}{\mcitedefaultseppunct}\relax
\EndOfBibitem
\bibitem[Xu \emph{et~al.}(2017)Xu, Sun, Deng, and Tan]{Xu_Sun_Deng_Tan_2017}
W.~Xu, H.~Sun, C.~Deng and Y.~Tan, \emph{Proceedings of the AAAI Conference on
  Artificial Intelligence}, 2017, \textbf{31}, 3358\relax
\mciteBstWouldAddEndPuncttrue
\mciteSetBstMidEndSepPunct{\mcitedefaultmidpunct}
{\mcitedefaultendpunct}{\mcitedefaultseppunct}\relax
\EndOfBibitem
\bibitem[Dallaire-Demers and Killoran(2018)]{2018_qgan}
P.-L. Dallaire-Demers and N.~Killoran, \emph{Physical Review A}, 2018,
  \textbf{98}, 012324\relax
\mciteBstWouldAddEndPuncttrue
\mciteSetBstMidEndSepPunct{\mcitedefaultmidpunct}
{\mcitedefaultendpunct}{\mcitedefaultseppunct}\relax
\EndOfBibitem
\bibitem[Khoshaman \emph{et~al.}(2018)Khoshaman, Vinci, Denis, Andriyash,
  Sadeghi, and Amin]{2018_qvae}
A.~Khoshaman, W.~Vinci, B.~Denis, E.~Andriyash, H.~Sadeghi and M.~H. Amin,
  \emph{Quantum Science and Technology}, 2018, \textbf{4}, 014001\relax
\mciteBstWouldAddEndPuncttrue
\mciteSetBstMidEndSepPunct{\mcitedefaultmidpunct}
{\mcitedefaultendpunct}{\mcitedefaultseppunct}\relax
\EndOfBibitem
\bibitem[Amin \emph{et~al.}(2018)Amin, Andriyash, Rolfe, Kulchytskyy, and
  Melko]{2018_qbm}
M.~H. Amin, E.~Andriyash, J.~Rolfe, B.~Kulchytskyy and R.~Melko, \emph{Physical
  Review X}, 2018, \textbf{8}, 021050\relax
\mciteBstWouldAddEndPuncttrue
\mciteSetBstMidEndSepPunct{\mcitedefaultmidpunct}
{\mcitedefaultendpunct}{\mcitedefaultseppunct}\relax
\EndOfBibitem
\bibitem[Goodfellow \emph{et~al.}(2014)Goodfellow, Pouget-Abadie, Mirza, Xu,
  Warde-Farley, Ozair, Courville, and Bengio]{goodfellow2014generative}
I.~J. Goodfellow, J.~Pouget-Abadie, M.~Mirza, B.~Xu, D.~Warde-Farley, S.~Ozair,
  A.~Courville and Y.~Bengio, \emph{Generative Adversarial Networks},
  2014\relax
\mciteBstWouldAddEndPuncttrue
\mciteSetBstMidEndSepPunct{\mcitedefaultmidpunct}
{\mcitedefaultendpunct}{\mcitedefaultseppunct}\relax
\EndOfBibitem
\bibitem[Yu \emph{et~al.}(2018)Yu, Lin, Yang, Shen, Lu, and
  Huang]{yu2018generative}
J.~Yu, Z.~Lin, J.~Yang, X.~Shen, X.~Lu and T.~S. Huang, \emph{Generative Image
  Inpainting with Contextual Attention}, 2018\relax
\mciteBstWouldAddEndPuncttrue
\mciteSetBstMidEndSepPunct{\mcitedefaultmidpunct}
{\mcitedefaultendpunct}{\mcitedefaultseppunct}\relax
\EndOfBibitem
\bibitem[Schawinski \emph{et~al.}(2017)Schawinski, Zhang, Zhang, Fowler, and
  Santhanam]{2017_gen}
K.~Schawinski, C.~Zhang, H.~Zhang, L.~Fowler and G.~K. Santhanam, \emph{Monthly
  Notices of the Royal Astronomical Society: Letters}, 2017,  slx008\relax
\mciteBstWouldAddEndPuncttrue
\mciteSetBstMidEndSepPunct{\mcitedefaultmidpunct}
{\mcitedefaultendpunct}{\mcitedefaultseppunct}\relax
\EndOfBibitem
\bibitem[Wang \emph{et~al.}(2018)Wang, Yu, Wu, Gu, Liu, Dong, Loy, Qiao, and
  Tang]{wang2018esrgan}
X.~Wang, K.~Yu, S.~Wu, J.~Gu, Y.~Liu, C.~Dong, C.~C. Loy, Y.~Qiao and X.~Tang,
  \emph{ESRGAN: Enhanced Super-Resolution Generative Adversarial Networks},
  2018\relax
\mciteBstWouldAddEndPuncttrue
\mciteSetBstMidEndSepPunct{\mcitedefaultmidpunct}
{\mcitedefaultendpunct}{\mcitedefaultseppunct}\relax
\EndOfBibitem
\bibitem[Li \emph{et~al.}(2021)Li, François-Lavet, Doan, and
  Pineau]{li2021domain}
B.~Li, V.~François-Lavet, T.~Doan and J.~Pineau, \emph{Domain Adversarial
  Reinforcement Learning}, 2021\relax
\mciteBstWouldAddEndPuncttrue
\mciteSetBstMidEndSepPunct{\mcitedefaultmidpunct}
{\mcitedefaultendpunct}{\mcitedefaultseppunct}\relax
\EndOfBibitem
\bibitem[Lloyd and Weedbrook(2018)]{2018_seth}
S.~Lloyd and C.~Weedbrook, \emph{Physical Review Letters}, 2018, \textbf{121},
  040502\relax
\mciteBstWouldAddEndPuncttrue
\mciteSetBstMidEndSepPunct{\mcitedefaultmidpunct}
{\mcitedefaultendpunct}{\mcitedefaultseppunct}\relax
\EndOfBibitem
\bibitem[Schuld \emph{et~al.}(2019)Schuld, Bergholm, Gogolin, Izaac, and
  Killoran]{2019_maria}
M.~Schuld, V.~Bergholm, C.~Gogolin, J.~Izaac and N.~Killoran, \emph{Physical
  Review A}, 2019, \textbf{99}, 032331\relax
\mciteBstWouldAddEndPuncttrue
\mciteSetBstMidEndSepPunct{\mcitedefaultmidpunct}
{\mcitedefaultendpunct}{\mcitedefaultseppunct}\relax
\EndOfBibitem
\bibitem[Biamonte and Bergholm(2017)]{Biamonte2017-ip}
J.~Biamonte and V.~Bergholm, \emph{arXiv preprint arXiv:1708.00006}, 2017\relax
\mciteBstWouldAddEndPuncttrue
\mciteSetBstMidEndSepPunct{\mcitedefaultmidpunct}
{\mcitedefaultendpunct}{\mcitedefaultseppunct}\relax
\EndOfBibitem
\bibitem[Bridgeman and Chubb(2017)]{Bridgeman2017-kq}
J.~C. Bridgeman and C.~T. Chubb, \emph{Hand-waving and interpretive dance: an
  introductory course on tensor networks}, 2017\relax
\mciteBstWouldAddEndPuncttrue
\mciteSetBstMidEndSepPunct{\mcitedefaultmidpunct}
{\mcitedefaultendpunct}{\mcitedefaultseppunct}\relax
\EndOfBibitem
\bibitem[Vidal(2004)]{vidal2004efficient}
G.~Vidal, \emph{Physical review letters}, 2004, \textbf{93}, 040502\relax
\mciteBstWouldAddEndPuncttrue
\mciteSetBstMidEndSepPunct{\mcitedefaultmidpunct}
{\mcitedefaultendpunct}{\mcitedefaultseppunct}\relax
\EndOfBibitem
\bibitem[Vidal(2003)]{Vidal2003-on}
G.~Vidal, \emph{Physical review letters}, 2003, \textbf{91}, 147902\relax
\mciteBstWouldAddEndPuncttrue
\mciteSetBstMidEndSepPunct{\mcitedefaultmidpunct}
{\mcitedefaultendpunct}{\mcitedefaultseppunct}\relax
\EndOfBibitem
\bibitem[Eisert \emph{et~al.}(2010)Eisert, Cramer, and Plenio]{Eisert2010-xu}
J.~Eisert, M.~Cramer and M.~B. Plenio, \emph{Reviews of modern physics}, 2010,
  \textbf{82}, 277\relax
\mciteBstWouldAddEndPuncttrue
\mciteSetBstMidEndSepPunct{\mcitedefaultmidpunct}
{\mcitedefaultendpunct}{\mcitedefaultseppunct}\relax
\EndOfBibitem
\bibitem[Hastings(2007)]{Hastings2007-pl}
M.~B. Hastings, \emph{Journal of statistical mechanics: theory and experiment},
  2007, \textbf{2007}, P08024\relax
\mciteBstWouldAddEndPuncttrue
\mciteSetBstMidEndSepPunct{\mcitedefaultmidpunct}
{\mcitedefaultendpunct}{\mcitedefaultseppunct}\relax
\EndOfBibitem
\bibitem[Or{\'u}s(2014)]{Orus2014-bg}
R.~Or{\'u}s, \emph{Annals of physics}, 2014, \textbf{349}, 117--158\relax
\mciteBstWouldAddEndPuncttrue
\mciteSetBstMidEndSepPunct{\mcitedefaultmidpunct}
{\mcitedefaultendpunct}{\mcitedefaultseppunct}\relax
\EndOfBibitem
\bibitem[Vidal(2009)]{Vidal2007-by}
G.~Vidal, \emph{arXiv preprint arXiv:0912.1651}, 2009\relax
\mciteBstWouldAddEndPuncttrue
\mciteSetBstMidEndSepPunct{\mcitedefaultmidpunct}
{\mcitedefaultendpunct}{\mcitedefaultseppunct}\relax
\EndOfBibitem
\bibitem[Schollw{\"o}ck(2011)]{Schollwock2011-al}
U.~Schollw{\"o}ck, \emph{Annals of physics}, 2011, \textbf{326}, 96--192\relax
\mciteBstWouldAddEndPuncttrue
\mciteSetBstMidEndSepPunct{\mcitedefaultmidpunct}
{\mcitedefaultendpunct}{\mcitedefaultseppunct}\relax
\EndOfBibitem
\bibitem[Ose()]{Oseledets2009-xd}
\relax
\mciteBstWouldAddEndPunctfalse
\mciteSetBstMidEndSepPunct{\mcitedefaultmidpunct}
{}{\mcitedefaultseppunct}\relax
\EndOfBibitem
\bibitem[Oseledets(2011)]{Oseledets2011-te}
I.~V. Oseledets, \emph{SIAM Journal on Scientific Computing}, 2011,
  \textbf{33}, 2295--2317\relax
\mciteBstWouldAddEndPuncttrue
\mciteSetBstMidEndSepPunct{\mcitedefaultmidpunct}
{\mcitedefaultendpunct}{\mcitedefaultseppunct}\relax
\EndOfBibitem
\bibitem[Shi \emph{et~al.}(2006)Shi, Duan, and Vidal]{Shi2005-gm}
Y.-Y. Shi, L.-M. Duan and G.~Vidal, \emph{Physical review a}, 2006,
  \textbf{74}, 022320\relax
\mciteBstWouldAddEndPuncttrue
\mciteSetBstMidEndSepPunct{\mcitedefaultmidpunct}
{\mcitedefaultendpunct}{\mcitedefaultseppunct}\relax
\EndOfBibitem
\bibitem[Nagaj \emph{et~al.}(2008)Nagaj, Farhi, Goldstone, Shor, and
  Sylvester]{Nagaj2008-kr}
D.~Nagaj, E.~Farhi, J.~Goldstone, P.~Shor and I.~Sylvester, \emph{Physical
  Review B}, 2008, \textbf{77}, 214431\relax
\mciteBstWouldAddEndPuncttrue
\mciteSetBstMidEndSepPunct{\mcitedefaultmidpunct}
{\mcitedefaultendpunct}{\mcitedefaultseppunct}\relax
\EndOfBibitem
\bibitem[Friedman(1997)]{Friedman1997-zp}
B.~Friedman, \emph{Journal of Physics: Condensed Matter}, 1997, \textbf{9},
  9021\relax
\mciteBstWouldAddEndPuncttrue
\mciteSetBstMidEndSepPunct{\mcitedefaultmidpunct}
{\mcitedefaultendpunct}{\mcitedefaultseppunct}\relax
\EndOfBibitem
\bibitem[Nakatani and Chan(2013)]{doi:10.1063/1.4798639}
N.~Nakatani and G.~K.-L. Chan, \emph{The Journal of Chemical Physics}, 2013,
  \textbf{138}, 134113\relax
\mciteBstWouldAddEndPuncttrue
\mciteSetBstMidEndSepPunct{\mcitedefaultmidpunct}
{\mcitedefaultendpunct}{\mcitedefaultseppunct}\relax
\EndOfBibitem
\bibitem[Castellana and
  Bacciu(2020)]{https://doi.org/10.48550/arxiv.2011.00860}
D.~Castellana and D.~Bacciu, \emph{Learning from Non-Binary Constituency Trees
  via Tensor Decomposition}, 2020, \url{https://arxiv.org/abs/2011.00860}\relax
\mciteBstWouldAddEndPuncttrue
\mciteSetBstMidEndSepPunct{\mcitedefaultmidpunct}
{\mcitedefaultendpunct}{\mcitedefaultseppunct}\relax
\EndOfBibitem
\bibitem[Fisher(1998)]{Fisher1998-xg}
M.~E. Fisher, \emph{Reviews of Modern Physics}, 1998, \textbf{70}, 653\relax
\mciteBstWouldAddEndPuncttrue
\mciteSetBstMidEndSepPunct{\mcitedefaultmidpunct}
{\mcitedefaultendpunct}{\mcitedefaultseppunct}\relax
\EndOfBibitem
\bibitem[Tagliacozzo \emph{et~al.}(2009)Tagliacozzo, Evenbly, and
  Vidal]{Tagliacozzo2009-xp}
L.~Tagliacozzo, G.~Evenbly and G.~Vidal, \emph{Physical Review B}, 2009,
  \textbf{80}, 235127\relax
\mciteBstWouldAddEndPuncttrue
\mciteSetBstMidEndSepPunct{\mcitedefaultmidpunct}
{\mcitedefaultendpunct}{\mcitedefaultseppunct}\relax
\EndOfBibitem
\bibitem[Larsson(2019)]{larsson2019computing}
H.~R. Larsson, \emph{The Journal of chemical physics}, 2019, \textbf{151},
  204102\relax
\mciteBstWouldAddEndPuncttrue
\mciteSetBstMidEndSepPunct{\mcitedefaultmidpunct}
{\mcitedefaultendpunct}{\mcitedefaultseppunct}\relax
\EndOfBibitem
\bibitem[Kurashige(2018)]{kurashige2018matrix}
Y.~Kurashige, \emph{The Journal of Chemical Physics}, 2018, \textbf{149},
  194114\relax
\mciteBstWouldAddEndPuncttrue
\mciteSetBstMidEndSepPunct{\mcitedefaultmidpunct}
{\mcitedefaultendpunct}{\mcitedefaultseppunct}\relax
\EndOfBibitem
\bibitem[Kliesch \emph{et~al.}(2014)Kliesch, Gross, and Eisert]{Kliesch2014-rd}
M.~Kliesch, D.~Gross and J.~Eisert, \emph{Physical review letters}, 2014,
  \textbf{113}, 160503\relax
\mciteBstWouldAddEndPuncttrue
\mciteSetBstMidEndSepPunct{\mcitedefaultmidpunct}
{\mcitedefaultendpunct}{\mcitedefaultseppunct}\relax
\EndOfBibitem
\bibitem[Raussendorf and Briegel(2001)]{Raussendorf2001-ib}
R.~Raussendorf and H.~J. Briegel, \emph{Physical Review Letters}, 2001,
  \textbf{86}, 5188\relax
\mciteBstWouldAddEndPuncttrue
\mciteSetBstMidEndSepPunct{\mcitedefaultmidpunct}
{\mcitedefaultendpunct}{\mcitedefaultseppunct}\relax
\EndOfBibitem
\bibitem[Verstraete \emph{et~al.}(2006)Verstraete, Wolf, Perez-Garcia, and
  Cirac]{Verstraete2006-rp}
F.~Verstraete, M.~M. Wolf, D.~Perez-Garcia and J.~I. Cirac, \emph{Physical
  review letters}, 2006, \textbf{96}, 220601\relax
\mciteBstWouldAddEndPuncttrue
\mciteSetBstMidEndSepPunct{\mcitedefaultmidpunct}
{\mcitedefaultendpunct}{\mcitedefaultseppunct}\relax
\EndOfBibitem
\bibitem[Kitaev(2003)]{Kitaev2003-pu}
A.~Y. Kitaev, \emph{Annals of Physics}, 2003, \textbf{303}, 2--30\relax
\mciteBstWouldAddEndPuncttrue
\mciteSetBstMidEndSepPunct{\mcitedefaultmidpunct}
{\mcitedefaultendpunct}{\mcitedefaultseppunct}\relax
\EndOfBibitem
\bibitem[Schwarz \emph{et~al.}(2017)Schwarz, Buerschaper, and
  Eisert]{Schwarz2017-wy}
M.~Schwarz, O.~Buerschaper and J.~Eisert, \emph{Phys. Rev. A}, 2017,
  \textbf{95}, 060102\relax
\mciteBstWouldAddEndPuncttrue
\mciteSetBstMidEndSepPunct{\mcitedefaultmidpunct}
{\mcitedefaultendpunct}{\mcitedefaultseppunct}\relax
\EndOfBibitem
\bibitem[Jordan \emph{et~al.}(2008)Jordan, Or{\'u}s, Vidal, Verstraete, and
  Cirac]{Jordan2008-qw}
J.~Jordan, R.~Or{\'u}s, G.~Vidal, F.~Verstraete and J.~I. Cirac,
  \emph{Classical Simulation of {Infinite-Size} Quantum Lattice Systems in Two
  Spatial Dimensions}, 2008\relax
\mciteBstWouldAddEndPuncttrue
\mciteSetBstMidEndSepPunct{\mcitedefaultmidpunct}
{\mcitedefaultendpunct}{\mcitedefaultseppunct}\relax
\EndOfBibitem
\bibitem[Gu \emph{et~al.}(2008)Gu, Levin, and Wen]{Gu2008-rz}
Z.-C. Gu, M.~Levin and X.-G. Wen, \emph{Phys. Rev. B Condens. Matter}, 2008,
  \textbf{78}, 205116\relax
\mciteBstWouldAddEndPuncttrue
\mciteSetBstMidEndSepPunct{\mcitedefaultmidpunct}
{\mcitedefaultendpunct}{\mcitedefaultseppunct}\relax
\EndOfBibitem
\bibitem[Schwarz \emph{et~al.}(2012)Schwarz, Temme, and
  Verstraete]{Schwarz2012-oz}
M.~Schwarz, K.~Temme and F.~Verstraete, \emph{Physical review letters}, 2012,
  \textbf{108}, 110502\relax
\mciteBstWouldAddEndPuncttrue
\mciteSetBstMidEndSepPunct{\mcitedefaultmidpunct}
{\mcitedefaultendpunct}{\mcitedefaultseppunct}\relax
\EndOfBibitem
\bibitem[Schwarz \emph{et~al.}(2013)Schwarz, Temme, Verstraete, Perez-Garcia,
  and Cubitt]{Schwarz2013-rl}
M.~Schwarz, K.~Temme, F.~Verstraete, D.~Perez-Garcia and T.~S. Cubitt,
  \emph{Physical Review A}, 2013, \textbf{88}, 032321\relax
\mciteBstWouldAddEndPuncttrue
\mciteSetBstMidEndSepPunct{\mcitedefaultmidpunct}
{\mcitedefaultendpunct}{\mcitedefaultseppunct}\relax
\EndOfBibitem
\bibitem[Corboz \emph{et~al.}(2018)Corboz, Czarnik, Kapteijns, and
  Tagliacozzo]{Corboz2018-ut}
P.~Corboz, P.~Czarnik, G.~Kapteijns and L.~Tagliacozzo, \emph{Finite
  Correlation Length Scaling with Infinite Projected {Entangled-Pair} States},
  2018\relax
\mciteBstWouldAddEndPuncttrue
\mciteSetBstMidEndSepPunct{\mcitedefaultmidpunct}
{\mcitedefaultendpunct}{\mcitedefaultseppunct}\relax
\EndOfBibitem
\bibitem[Milsted and Vidal(2018)]{Milsted_undated-fs}
A.~Milsted and G.~Vidal, \emph{arXiv preprint arXiv:1812.00529}, 2018\relax
\mciteBstWouldAddEndPuncttrue
\mciteSetBstMidEndSepPunct{\mcitedefaultmidpunct}
{\mcitedefaultendpunct}{\mcitedefaultseppunct}\relax
\EndOfBibitem
\bibitem[Vidal(2008)]{Vidal2008-vt}
G.~Vidal, \emph{Physical review letters}, 2008, \textbf{101}, 110501\relax
\mciteBstWouldAddEndPuncttrue
\mciteSetBstMidEndSepPunct{\mcitedefaultmidpunct}
{\mcitedefaultendpunct}{\mcitedefaultseppunct}\relax
\EndOfBibitem
\bibitem[Evenbly and Vidal(2009)]{Evenbly2009-am}
G.~Evenbly and G.~Vidal, \emph{Entanglement Renormalization in Two Spatial
  Dimensions}, 2009\relax
\mciteBstWouldAddEndPuncttrue
\mciteSetBstMidEndSepPunct{\mcitedefaultmidpunct}
{\mcitedefaultendpunct}{\mcitedefaultseppunct}\relax
\EndOfBibitem
\bibitem[White(1993)]{White1993-cj}
S.~R. White, \emph{Physical review b}, 1993, \textbf{48}, 10345\relax
\mciteBstWouldAddEndPuncttrue
\mciteSetBstMidEndSepPunct{\mcitedefaultmidpunct}
{\mcitedefaultendpunct}{\mcitedefaultseppunct}\relax
\EndOfBibitem
\bibitem[Kempe \emph{et~al.}(2006)Kempe, Kitaev, and Regev]{Kempe2006-li}
J.~Kempe, A.~Kitaev and O.~Regev, \emph{Siam journal on computing}, 2006,
  \textbf{35}, 1070--1097\relax
\mciteBstWouldAddEndPuncttrue
\mciteSetBstMidEndSepPunct{\mcitedefaultmidpunct}
{\mcitedefaultendpunct}{\mcitedefaultseppunct}\relax
\EndOfBibitem
\bibitem[Kawaguchi \emph{et~al.}(2004)Kawaguchi, Shimizu, Tokura, and
  Imoto]{Kawaguchi2004-kj}
A.~Kawaguchi, K.~Shimizu, Y.~Tokura and N.~Imoto, \emph{arXiv preprint
  quant-ph/0411205}, 2004\relax
\mciteBstWouldAddEndPuncttrue
\mciteSetBstMidEndSepPunct{\mcitedefaultmidpunct}
{\mcitedefaultendpunct}{\mcitedefaultseppunct}\relax
\EndOfBibitem
\bibitem[Dang \emph{et~al.}(2019)Dang, Hill, and Hollenberg]{Dang2019-yp}
A.~Dang, C.~D. Hill and L.~C. Hollenberg, \emph{Quantum}, 2019, \textbf{3},
  116\relax
\mciteBstWouldAddEndPuncttrue
\mciteSetBstMidEndSepPunct{\mcitedefaultmidpunct}
{\mcitedefaultendpunct}{\mcitedefaultseppunct}\relax
\EndOfBibitem
\bibitem[Dumitrescu(2017)]{Dumitrescu2017-yk}
E.~Dumitrescu, \emph{Phys. Rev. A}, 2017, \textbf{96}, 062322\relax
\mciteBstWouldAddEndPuncttrue
\mciteSetBstMidEndSepPunct{\mcitedefaultmidpunct}
{\mcitedefaultendpunct}{\mcitedefaultseppunct}\relax
\EndOfBibitem
\bibitem[Huang \emph{et~al.}(2020)Huang, Zhang, Newman, Cai, Gao, Tian, Wu, Xu,
  Yu, Yuan,\emph{et~al.}]{Huang2020-nz}
C.~Huang, F.~Zhang, M.~Newman, J.~Cai, X.~Gao, Z.~Tian, J.~Wu, H.~Xu, H.~Yu,
  B.~Yuan \emph{et~al.}, \emph{arXiv preprint arXiv:2005.06787}, 2020\relax
\mciteBstWouldAddEndPuncttrue
\mciteSetBstMidEndSepPunct{\mcitedefaultmidpunct}
{\mcitedefaultendpunct}{\mcitedefaultseppunct}\relax
\EndOfBibitem
\bibitem[Arute \emph{et~al.}(2019)Arute, Arya, Babbush, Bacon, Bardin, Barends,
  Biswas, Boixo, Brandao, Buell,\emph{et~al.}]{Arute2019-mu}
F.~Arute, K.~Arya, R.~Babbush, D.~Bacon, J.~C. Bardin, R.~Barends, R.~Biswas,
  S.~Boixo, F.~G. Brandao, D.~A. Buell \emph{et~al.}, \emph{Nature}, 2019,
  \textbf{574}, 505--510\relax
\mciteBstWouldAddEndPuncttrue
\mciteSetBstMidEndSepPunct{\mcitedefaultmidpunct}
{\mcitedefaultendpunct}{\mcitedefaultseppunct}\relax
\EndOfBibitem
\bibitem[Huang \emph{et~al.}(2021)Huang, Zhang, Newman, Ni, Ding, Cai, Gao,
  Wang, Wu, Zhang,\emph{et~al.}]{Huang2021-uw}
C.~Huang, F.~Zhang, M.~Newman, X.~Ni, D.~Ding, J.~Cai, X.~Gao, T.~Wang, F.~Wu,
  G.~Zhang \emph{et~al.}, \emph{Nature Computational Science}, 2021,
  \textbf{1}, 578--587\relax
\mciteBstWouldAddEndPuncttrue
\mciteSetBstMidEndSepPunct{\mcitedefaultmidpunct}
{\mcitedefaultendpunct}{\mcitedefaultseppunct}\relax
\EndOfBibitem
\bibitem[Markov and Shi(2008)]{Markov2008-kk}
I.~L. Markov and Y.~Shi, \emph{SIAM Journal on Computing}, 2008, \textbf{38},
  963--981\relax
\mciteBstWouldAddEndPuncttrue
\mciteSetBstMidEndSepPunct{\mcitedefaultmidpunct}
{\mcitedefaultendpunct}{\mcitedefaultseppunct}\relax
\EndOfBibitem
\bibitem[Liu \emph{et~al.}(2019)Liu, Zhang, Wan, and Wang]{Liu2019-mi}
J.-G. Liu, Y.-H. Zhang, Y.~Wan and L.~Wang, \emph{Phys. Rev. Research}, 2019,
  \textbf{1}, 023025\relax
\mciteBstWouldAddEndPuncttrue
\mciteSetBstMidEndSepPunct{\mcitedefaultmidpunct}
{\mcitedefaultendpunct}{\mcitedefaultseppunct}\relax
\EndOfBibitem
\bibitem[Ran(2020)]{Ran2020-hr}
S.-J. Ran, \emph{Encoding of matrix product states into quantum circuits of
  one- and two-qubit gates}, 2020\relax
\mciteBstWouldAddEndPuncttrue
\mciteSetBstMidEndSepPunct{\mcitedefaultmidpunct}
{\mcitedefaultendpunct}{\mcitedefaultseppunct}\relax
\EndOfBibitem
\bibitem[Huggins \emph{et~al.}(2019)Huggins, Patil, Mitchell, Whaley, and
  Stoudenmire]{Huggins_2019}
W.~Huggins, P.~Patil, B.~Mitchell, K.~B. Whaley and E.~M. Stoudenmire,
  \emph{Quantum Science and technology}, 2019, \textbf{4}, 024001\relax
\mciteBstWouldAddEndPuncttrue
\mciteSetBstMidEndSepPunct{\mcitedefaultmidpunct}
{\mcitedefaultendpunct}{\mcitedefaultseppunct}\relax
\EndOfBibitem
\bibitem[Kardashin \emph{et~al.}(2021)Kardashin, Uvarov, and
  Biamonte]{Kardashin2018-ds}
A.~Kardashin, A.~Uvarov and J.~Biamonte, \emph{Frontiers in Physics}, 2021,
  644\relax
\mciteBstWouldAddEndPuncttrue
\mciteSetBstMidEndSepPunct{\mcitedefaultmidpunct}
{\mcitedefaultendpunct}{\mcitedefaultseppunct}\relax
\EndOfBibitem
\bibitem[Kim and Swingle(2017)]{Kim2017-cc}
I.~H. Kim and B.~Swingle, \emph{arXiv preprint arXiv:1711.07500}, 2017\relax
\mciteBstWouldAddEndPuncttrue
\mciteSetBstMidEndSepPunct{\mcitedefaultmidpunct}
{\mcitedefaultendpunct}{\mcitedefaultseppunct}\relax
\EndOfBibitem
\bibitem[Yuan \emph{et~al.}(2021)Yuan, Sun, Liu, Zhao, and Zhou]{Yuan2021-kv}
X.~Yuan, J.~Sun, J.~Liu, Q.~Zhao and Y.~Zhou, \emph{Physical Review Letters},
  2021, \textbf{127}, 040501\relax
\mciteBstWouldAddEndPuncttrue
\mciteSetBstMidEndSepPunct{\mcitedefaultmidpunct}
{\mcitedefaultendpunct}{\mcitedefaultseppunct}\relax
\EndOfBibitem
\bibitem[Mitarai \emph{et~al.}(2018)Mitarai, Negoro, Kitagawa, and
  Fujii]{mitarai2018quantum}
K.~Mitarai, M.~Negoro, M.~Kitagawa and K.~Fujii, \emph{Physical Review A},
  2018, \textbf{98}, 032309\relax
\mciteBstWouldAddEndPuncttrue
\mciteSetBstMidEndSepPunct{\mcitedefaultmidpunct}
{\mcitedefaultendpunct}{\mcitedefaultseppunct}\relax
\EndOfBibitem
\bibitem[Arrighi(2019)]{arrighi2019overview}
P.~Arrighi, \emph{Natural Computing}, 2019, \textbf{18}, 885--899\relax
\mciteBstWouldAddEndPuncttrue
\mciteSetBstMidEndSepPunct{\mcitedefaultmidpunct}
{\mcitedefaultendpunct}{\mcitedefaultseppunct}\relax
\EndOfBibitem
\bibitem[P{\'e}rez-Salinas \emph{et~al.}(2020)P{\'e}rez-Salinas,
  Cervera-Lierta, Gil-Fuster, and Latorre]{perez2020data}
A.~P{\'e}rez-Salinas, A.~Cervera-Lierta, E.~Gil-Fuster and J.~I. Latorre,
  \emph{Quantum}, 2020, \textbf{4}, 226\relax
\mciteBstWouldAddEndPuncttrue
\mciteSetBstMidEndSepPunct{\mcitedefaultmidpunct}
{\mcitedefaultendpunct}{\mcitedefaultseppunct}\relax
\EndOfBibitem
\bibitem[Schuld \emph{et~al.}(2021)Schuld, Sweke, and Meyer]{schuld2021effect}
M.~Schuld, R.~Sweke and J.~J. Meyer, \emph{Physical Review A}, 2021,
  \textbf{103}, 032430\relax
\mciteBstWouldAddEndPuncttrue
\mciteSetBstMidEndSepPunct{\mcitedefaultmidpunct}
{\mcitedefaultendpunct}{\mcitedefaultseppunct}\relax
\EndOfBibitem
\bibitem[Chen \emph{et~al.}(2021)Chen, Wossnig, Severini, Neven, and
  Mohseni]{chen2021universal}
H.~Chen, L.~Wossnig, S.~Severini, H.~Neven and M.~Mohseni, \emph{Quantum
  Machine Intelligence}, 2021, \textbf{3}, 1--11\relax
\mciteBstWouldAddEndPuncttrue
\mciteSetBstMidEndSepPunct{\mcitedefaultmidpunct}
{\mcitedefaultendpunct}{\mcitedefaultseppunct}\relax
\EndOfBibitem
\bibitem[Sim \emph{et~al.}(2019)Sim, Johnson, and
  Aspuru-Guzik]{https://doi.org/10.1002/qute.201900070}
S.~Sim, P.~D. Johnson and A.~Aspuru-Guzik, \emph{Advanced Quantum
  Technologies}, 2019, \textbf{2}, 1900070\relax
\mciteBstWouldAddEndPuncttrue
\mciteSetBstMidEndSepPunct{\mcitedefaultmidpunct}
{\mcitedefaultendpunct}{\mcitedefaultseppunct}\relax
\EndOfBibitem
\bibitem[Snyder \emph{et~al.}(2012)Snyder, Rupp, Hansen, M{\"u}ller, and
  Burke]{snyder2012finding}
J.~C. Snyder, M.~Rupp, K.~Hansen, K.-R. M{\"u}ller and K.~Burke, \emph{Physical
  review letters}, 2012, \textbf{108}, 253002\relax
\mciteBstWouldAddEndPuncttrue
\mciteSetBstMidEndSepPunct{\mcitedefaultmidpunct}
{\mcitedefaultendpunct}{\mcitedefaultseppunct}\relax
\EndOfBibitem
\bibitem[Rupp \emph{et~al.}(2012)Rupp, Tkatchenko, M\"uller, and von
  Lilienfeld]{PhysRevLett.108.058301}
M.~Rupp, A.~Tkatchenko, K.-R. M\"uller and O.~A. von Lilienfeld, \emph{Phys.
  Rev. Lett.}, 2012, \textbf{108}, 058301\relax
\mciteBstWouldAddEndPuncttrue
\mciteSetBstMidEndSepPunct{\mcitedefaultmidpunct}
{\mcitedefaultendpunct}{\mcitedefaultseppunct}\relax
\EndOfBibitem
\bibitem[Christensen \emph{et~al.}(2020)Christensen, Bratholm, Faber, and
  Anatole~von Lilienfeld]{doi:10.1063/1.5126701}
A.~S. Christensen, L.~A. Bratholm, F.~A. Faber and O.~Anatole~von Lilienfeld,
  \emph{The Journal of Chemical Physics}, 2020, \textbf{152}, 044107\relax
\mciteBstWouldAddEndPuncttrue
\mciteSetBstMidEndSepPunct{\mcitedefaultmidpunct}
{\mcitedefaultendpunct}{\mcitedefaultseppunct}\relax
\EndOfBibitem
\bibitem[Huang \emph{et~al.}(2017)Huang, Chen, Lin, Ke, and Tsai]{huang2017svm}
M.-W. Huang, C.-W. Chen, W.-C. Lin, S.-W. Ke and C.-F. Tsai, \emph{PloS one},
  2017, \textbf{12}, e0161501\relax
\mciteBstWouldAddEndPuncttrue
\mciteSetBstMidEndSepPunct{\mcitedefaultmidpunct}
{\mcitedefaultendpunct}{\mcitedefaultseppunct}\relax
\EndOfBibitem
\bibitem[Sun \emph{et~al.}(2012)Sun, Shahane, Xia, Austin, and
  Huang]{sun2012structure}
H.~Sun, S.~Shahane, M.~Xia, C.~P. Austin and R.~Huang, \emph{Journal of
  chemical information and modeling}, 2012, \textbf{52}, 1798--1805\relax
\mciteBstWouldAddEndPuncttrue
\mciteSetBstMidEndSepPunct{\mcitedefaultmidpunct}
{\mcitedefaultendpunct}{\mcitedefaultseppunct}\relax
\EndOfBibitem
\bibitem[Batra \emph{et~al.}(2021)Batra, Zorn, Foil, Minerali, Gawriljuk, Lane,
  and Ekins]{batra2021quantum}
K.~Batra, K.~M. Zorn, D.~H. Foil, E.~Minerali, V.~O. Gawriljuk, T.~R. Lane and
  S.~Ekins, \emph{Journal of Chemical Information and Modeling}, 2021\relax
\mciteBstWouldAddEndPuncttrue
\mciteSetBstMidEndSepPunct{\mcitedefaultmidpunct}
{\mcitedefaultendpunct}{\mcitedefaultseppunct}\relax
\EndOfBibitem
\bibitem[Li \emph{et~al.}(2021)Li, Zhang, and Wang]{li2021quantum}
L.~H. Li, D.-B. Zhang and Z.~Wang, \emph{arXiv preprint arXiv:2108.11114},
  2021\relax
\mciteBstWouldAddEndPuncttrue
\mciteSetBstMidEndSepPunct{\mcitedefaultmidpunct}
{\mcitedefaultendpunct}{\mcitedefaultseppunct}\relax
\EndOfBibitem
\bibitem[Bartlett \emph{et~al.}(2003)Bartlett, Sanders, Braunstein, and
  Nemoto]{Bartlett2003}
S.~D. Bartlett, B.~C. Sanders, S.~L. Braunstein and K.~Nemoto, in
  \emph{Efficient Classical Simulation of Continuous Variable Quantum
  Information Processes}, ed. S.~L. Braunstein and A.~K. Pati, Springer
  Netherlands, Dordrecht, 2003, pp. 47--55\relax
\mciteBstWouldAddEndPuncttrue
\mciteSetBstMidEndSepPunct{\mcitedefaultmidpunct}
{\mcitedefaultendpunct}{\mcitedefaultseppunct}\relax
\EndOfBibitem
\bibitem[Braunstein and van Loock(2005)]{RevModPhys.77.513}
S.~L. Braunstein and P.~van Loock, \emph{Rev. Mod. Phys.}, 2005, \textbf{77},
  513--577\relax
\mciteBstWouldAddEndPuncttrue
\mciteSetBstMidEndSepPunct{\mcitedefaultmidpunct}
{\mcitedefaultendpunct}{\mcitedefaultseppunct}\relax
\EndOfBibitem
\bibitem[Bartlett and Sanders(2002)]{PhysRevA.65.042304}
S.~D. Bartlett and B.~C. Sanders, \emph{Phys. Rev. A}, 2002, \textbf{65},
  042304\relax
\mciteBstWouldAddEndPuncttrue
\mciteSetBstMidEndSepPunct{\mcitedefaultmidpunct}
{\mcitedefaultendpunct}{\mcitedefaultseppunct}\relax
\EndOfBibitem
\bibitem[Douce \emph{et~al.}(2017)Douce, Markham, Kashefi, Diamanti, Coudreau,
  Milman, van Loock, and Ferrini]{PhysRevLett.118.070503}
T.~Douce, D.~Markham, E.~Kashefi, E.~Diamanti, T.~Coudreau, P.~Milman, P.~van
  Loock and G.~Ferrini, \emph{Phys. Rev. Lett.}, 2017, \textbf{118},
  070503\relax
\mciteBstWouldAddEndPuncttrue
\mciteSetBstMidEndSepPunct{\mcitedefaultmidpunct}
{\mcitedefaultendpunct}{\mcitedefaultseppunct}\relax
\EndOfBibitem
\bibitem[Pedregosa \emph{et~al.}(2011)Pedregosa, Varoquaux, Gramfort, Michel,
  Thirion, Grisel, Blondel, Prettenhofer, Weiss,
  Dubourg,\emph{et~al.}]{pedregosa2011scikit}
F.~Pedregosa, G.~Varoquaux, A.~Gramfort, V.~Michel, B.~Thirion, O.~Grisel,
  M.~Blondel, P.~Prettenhofer, R.~Weiss, V.~Dubourg \emph{et~al.}, \emph{the
  Journal of machine Learning research}, 2011, \textbf{12}, 2825--2830\relax
\mciteBstWouldAddEndPuncttrue
\mciteSetBstMidEndSepPunct{\mcitedefaultmidpunct}
{\mcitedefaultendpunct}{\mcitedefaultseppunct}\relax
\EndOfBibitem
\bibitem[Havl{\'\i}{\v{c}}ek \emph{et~al.}(2019)Havl{\'\i}{\v{c}}ek,
  C{\'o}rcoles, Temme, Harrow, Kandala, Chow, and
  Gambetta]{havlivcek2019supervised}
V.~Havl{\'\i}{\v{c}}ek, A.~D. C{\'o}rcoles, K.~Temme, A.~W. Harrow, A.~Kandala,
  J.~M. Chow and J.~M. Gambetta, \emph{Nature}, 2019, \textbf{567},
  209--212\relax
\mciteBstWouldAddEndPuncttrue
\mciteSetBstMidEndSepPunct{\mcitedefaultmidpunct}
{\mcitedefaultendpunct}{\mcitedefaultseppunct}\relax
\EndOfBibitem
\bibitem[Kusumoto \emph{et~al.}(2021)Kusumoto, Mitarai, Fujii, Kitagawa, and
  Negoro]{kusumoto2021experimental}
T.~Kusumoto, K.~Mitarai, K.~Fujii, M.~Kitagawa and M.~Negoro, \emph{npj Quantum
  Information}, 2021, \textbf{7}, 1--7\relax
\mciteBstWouldAddEndPuncttrue
\mciteSetBstMidEndSepPunct{\mcitedefaultmidpunct}
{\mcitedefaultendpunct}{\mcitedefaultseppunct}\relax
\EndOfBibitem
\bibitem[Bartkiewicz \emph{et~al.}(2020)Bartkiewicz, Gneiting, {\v{C}}ernoch,
  Jir{\'a}kov{\'a}, Lemr, and Nori]{bartkiewicz2020experimental}
K.~Bartkiewicz, C.~Gneiting, A.~{\v{C}}ernoch, K.~Jir{\'a}kov{\'a}, K.~Lemr and
  F.~Nori, \emph{Scientific reports}, 2020, \textbf{10}, 1--9\relax
\mciteBstWouldAddEndPuncttrue
\mciteSetBstMidEndSepPunct{\mcitedefaultmidpunct}
{\mcitedefaultendpunct}{\mcitedefaultseppunct}\relax
\EndOfBibitem
\bibitem[Guo and Weng(2022)]{guo2022where}
M.~Guo and Y.~Weng, \emph{Where can quantum kernel methods make a big
  difference?}, 2022, \url{https://openreview.net/forum?id=NoE4RfaOOa}\relax
\mciteBstWouldAddEndPuncttrue
\mciteSetBstMidEndSepPunct{\mcitedefaultmidpunct}
{\mcitedefaultendpunct}{\mcitedefaultseppunct}\relax
\EndOfBibitem
\bibitem[Matsumoto and Nishimura(1998)]{10.1145/272991.272995}
M.~Matsumoto and T.~Nishimura, \emph{ACM Trans. Model. Comput. Simul.}, 1998,
  \textbf{8}, 3–30\relax
\mciteBstWouldAddEndPuncttrue
\mciteSetBstMidEndSepPunct{\mcitedefaultmidpunct}
{\mcitedefaultendpunct}{\mcitedefaultseppunct}\relax
\EndOfBibitem
\bibitem[Wang \emph{et~al.}(2021)Wang, Du, Luo, and Tao]{Wang2021towards}
X.~Wang, Y.~Du, Y.~Luo and D.~Tao, \emph{{Quantum}}, 2021, \textbf{5},
  531\relax
\mciteBstWouldAddEndPuncttrue
\mciteSetBstMidEndSepPunct{\mcitedefaultmidpunct}
{\mcitedefaultendpunct}{\mcitedefaultseppunct}\relax
\EndOfBibitem
\bibitem[Vedaie \emph{et~al.}(2020)Vedaie, Noori, Oberoi, Sanders, and
  Zahedinejad]{vedaie2020quantum}
S.~S. Vedaie, M.~Noori, J.~S. Oberoi, B.~C. Sanders and E.~Zahedinejad,
  \emph{arXiv preprint arXiv:2011.09694}, 2020\relax
\mciteBstWouldAddEndPuncttrue
\mciteSetBstMidEndSepPunct{\mcitedefaultmidpunct}
{\mcitedefaultendpunct}{\mcitedefaultseppunct}\relax
\EndOfBibitem
\bibitem[Blank \emph{et~al.}(2020)Blank, Park, Rhee, and
  Petruccione]{blank2020quantum}
C.~Blank, D.~K. Park, J.-K.~K. Rhee and F.~Petruccione, \emph{npj Quantum
  Information}, 2020, \textbf{6}, 1--7\relax
\mciteBstWouldAddEndPuncttrue
\mciteSetBstMidEndSepPunct{\mcitedefaultmidpunct}
{\mcitedefaultendpunct}{\mcitedefaultseppunct}\relax
\EndOfBibitem
\bibitem[Lloyd \emph{et~al.}(2020)Lloyd, Schuld, Ijaz, Izaac, and
  Killoran]{lloyd2020quantum}
S.~Lloyd, M.~Schuld, A.~Ijaz, J.~Izaac and N.~Killoran, \emph{arXiv preprint
  arXiv:2001.03622}, 2020\relax
\mciteBstWouldAddEndPuncttrue
\mciteSetBstMidEndSepPunct{\mcitedefaultmidpunct}
{\mcitedefaultendpunct}{\mcitedefaultseppunct}\relax
\EndOfBibitem
\bibitem[Peters \emph{et~al.}(2021)Peters, Caldeira, Ho, Leichenauer, Mohseni,
  Neven, Spentzouris, Strain, and Perdue]{peters2021machine}
E.~Peters, J.~Caldeira, A.~Ho, S.~Leichenauer, M.~Mohseni, H.~Neven,
  P.~Spentzouris, D.~Strain and G.~N. Perdue, \emph{npj Quantum Information},
  2021, \textbf{7}, 1--5\relax
\mciteBstWouldAddEndPuncttrue
\mciteSetBstMidEndSepPunct{\mcitedefaultmidpunct}
{\mcitedefaultendpunct}{\mcitedefaultseppunct}\relax
\EndOfBibitem
\bibitem[Yang \emph{et~al.}(2020)Yang, Kahnt, Br{\"u}ckner, Schropp, Fam,
  Becher, Grunwaldt, Sheppard, and Schroer]{yang2020tomographic}
X.~Yang, M.~Kahnt, D.~Br{\"u}ckner, A.~Schropp, Y.~Fam, J.~Becher, J.-D.
  Grunwaldt, T.~L. Sheppard and C.~G. Schroer, \emph{Journal of synchrotron
  radiation}, 2020, \textbf{27}, 486--493\relax
\mciteBstWouldAddEndPuncttrue
\mciteSetBstMidEndSepPunct{\mcitedefaultmidpunct}
{\mcitedefaultendpunct}{\mcitedefaultseppunct}\relax
\EndOfBibitem
\bibitem[Liu \emph{et~al.}(2020)Liu, Bicer, Kettimuthu, Gursoy, De~Carlo, and
  Foster]{liu2020tomogan}
Z.~Liu, T.~Bicer, R.~Kettimuthu, D.~Gursoy, F.~De~Carlo and I.~Foster,
  \emph{JOSA A}, 2020, \textbf{37}, 422--434\relax
\mciteBstWouldAddEndPuncttrue
\mciteSetBstMidEndSepPunct{\mcitedefaultmidpunct}
{\mcitedefaultendpunct}{\mcitedefaultseppunct}\relax
\EndOfBibitem
\bibitem[Carrasquilla and Melko(2017)]{carrasquilla2017machine}
J.~Carrasquilla and R.~G. Melko, \emph{Nature Physics}, 2017, \textbf{13},
  431--434\relax
\mciteBstWouldAddEndPuncttrue
\mciteSetBstMidEndSepPunct{\mcitedefaultmidpunct}
{\mcitedefaultendpunct}{\mcitedefaultseppunct}\relax
\EndOfBibitem
\bibitem[Gao \emph{et~al.}(2018)Gao, Qiao, Jiao, Ma, Hu, Ren, Yang, Tang, Yung,
  and Jin]{gao2018experimental}
J.~Gao, L.-F. Qiao, Z.-Q. Jiao, Y.-C. Ma, C.-Q. Hu, R.-J. Ren, A.-L. Yang,
  H.~Tang, M.-H. Yung and X.-M. Jin, \emph{Physical review letters}, 2018,
  \textbf{120}, 240501\relax
\mciteBstWouldAddEndPuncttrue
\mciteSetBstMidEndSepPunct{\mcitedefaultmidpunct}
{\mcitedefaultendpunct}{\mcitedefaultseppunct}\relax
\EndOfBibitem
\bibitem[Lohani \emph{et~al.}(2020)Lohani, Kirby, Brodsky, Danaci, and
  Glasser]{lohani2020machine}
S.~Lohani, B.~T. Kirby, M.~Brodsky, O.~Danaci and R.~T. Glasser, \emph{Machine
  Learning: Science and Technology}, 2020, \textbf{1}, 035007\relax
\mciteBstWouldAddEndPuncttrue
\mciteSetBstMidEndSepPunct{\mcitedefaultmidpunct}
{\mcitedefaultendpunct}{\mcitedefaultseppunct}\relax
\EndOfBibitem
\bibitem[Ragoza \emph{et~al.}(2017)Ragoza, Hochuli, Idrobo, Sunseri, and
  Koes]{ragoza2017protein}
M.~Ragoza, J.~Hochuli, E.~Idrobo, J.~Sunseri and D.~R. Koes, \emph{Journal of
  chemical information and modeling}, 2017, \textbf{57}, 942--957\relax
\mciteBstWouldAddEndPuncttrue
\mciteSetBstMidEndSepPunct{\mcitedefaultmidpunct}
{\mcitedefaultendpunct}{\mcitedefaultseppunct}\relax
\EndOfBibitem
\bibitem[Yao \emph{et~al.}(2018)Yao, Herr, Toth, Mckintyre, and
  Parkhill]{yao2018tensormol}
K.~Yao, J.~E. Herr, D.~W. Toth, R.~Mckintyre and J.~Parkhill, \emph{Chemical
  science}, 2018, \textbf{9}, 2261--2269\relax
\mciteBstWouldAddEndPuncttrue
\mciteSetBstMidEndSepPunct{\mcitedefaultmidpunct}
{\mcitedefaultendpunct}{\mcitedefaultseppunct}\relax
\EndOfBibitem
\bibitem[Hughes \emph{et~al.}(2015)Hughes, Miller, and
  Swamidass]{hughes2015modeling}
T.~B. Hughes, G.~P. Miller and S.~J. Swamidass, \emph{ACS central science},
  2015, \textbf{1}, 168--180\relax
\mciteBstWouldAddEndPuncttrue
\mciteSetBstMidEndSepPunct{\mcitedefaultmidpunct}
{\mcitedefaultendpunct}{\mcitedefaultseppunct}\relax
\EndOfBibitem
\bibitem[Caetano \emph{et~al.}(2011)Caetano, Reis~Jr, Amorim, Lemes, and
  Pino~Jr]{caetano2011using}
C.~Caetano, J.~Reis~Jr, J.~Amorim, M.~R. Lemes and A.~D. Pino~Jr,
  \emph{International Journal of Quantum Chemistry}, 2011, \textbf{111},
  2732--2740\relax
\mciteBstWouldAddEndPuncttrue
\mciteSetBstMidEndSepPunct{\mcitedefaultmidpunct}
{\mcitedefaultendpunct}{\mcitedefaultseppunct}\relax
\EndOfBibitem
\bibitem[Sch{\"u}tt \emph{et~al.}(2019)Sch{\"u}tt, Gastegger, Tkatchenko,
  M{\"u}ller, and Maurer]{schutt2019unifying}
K.~T. Sch{\"u}tt, M.~Gastegger, A.~Tkatchenko, K.-R. M{\"u}ller and R.~J.
  Maurer, \emph{Nature communications}, 2019, \textbf{10}, 1--10\relax
\mciteBstWouldAddEndPuncttrue
\mciteSetBstMidEndSepPunct{\mcitedefaultmidpunct}
{\mcitedefaultendpunct}{\mcitedefaultseppunct}\relax
\EndOfBibitem
\bibitem[Galvelis and Sugita(2017)]{galvelis2017neural}
R.~Galvelis and Y.~Sugita, \emph{Journal of chemical theory and computation},
  2017, \textbf{13}, 2489--2500\relax
\mciteBstWouldAddEndPuncttrue
\mciteSetBstMidEndSepPunct{\mcitedefaultmidpunct}
{\mcitedefaultendpunct}{\mcitedefaultseppunct}\relax
\EndOfBibitem
\bibitem[Zeng \emph{et~al.}(2020)Zeng, Cao, Xu, Zhu, and
  Zhang]{zeng2020complex}
J.~Zeng, L.~Cao, M.~Xu, T.~Zhu and J.~Z. Zhang, \emph{Nature communications},
  2020, \textbf{11}, 1--9\relax
\mciteBstWouldAddEndPuncttrue
\mciteSetBstMidEndSepPunct{\mcitedefaultmidpunct}
{\mcitedefaultendpunct}{\mcitedefaultseppunct}\relax
\EndOfBibitem
\bibitem[Carleo and Troyer(2017)]{carleo2017solving}
G.~Carleo and M.~Troyer, \emph{Science}, 2017, \textbf{355}, 602--606\relax
\mciteBstWouldAddEndPuncttrue
\mciteSetBstMidEndSepPunct{\mcitedefaultmidpunct}
{\mcitedefaultendpunct}{\mcitedefaultseppunct}\relax
\EndOfBibitem
\bibitem[Sajjan \emph{et~al.}(2021)Sajjan, Sureshbabu, and
  Kais]{sajjan2021quantum}
M.~Sajjan, S.~H. Sureshbabu and S.~Kais, \emph{J. Am. Chem. Soc}, 2021\relax
\mciteBstWouldAddEndPuncttrue
\mciteSetBstMidEndSepPunct{\mcitedefaultmidpunct}
{\mcitedefaultendpunct}{\mcitedefaultseppunct}\relax
\EndOfBibitem
\bibitem[Abbas \emph{et~al.}(2021)Abbas, Sutter, Zoufal, Lucchi, Figalli, and
  Woerner]{abbas2021power}
A.~Abbas, D.~Sutter, C.~Zoufal, A.~Lucchi, A.~Figalli and S.~Woerner,
  \emph{Nature Computational Science}, 2021, \textbf{1}, 403--409\relax
\mciteBstWouldAddEndPuncttrue
\mciteSetBstMidEndSepPunct{\mcitedefaultmidpunct}
{\mcitedefaultendpunct}{\mcitedefaultseppunct}\relax
\EndOfBibitem
\bibitem[Kunstner \emph{et~al.}(2019)Kunstner, Hennig, and
  Balles]{kunstner2019limitations}
F.~Kunstner, P.~Hennig and L.~Balles, \emph{Advances in neural information
  processing systems}, 2019, \textbf{32}, \relax
\mciteBstWouldAddEndPuncttrue
\mciteSetBstMidEndSepPunct{\mcitedefaultmidpunct}
{\mcitedefaultendpunct}{\mcitedefaultseppunct}\relax
\EndOfBibitem
\bibitem[Karakida \emph{et~al.}(2019)Karakida, Akaho, and
  Amari]{karakida2019universal}
R.~Karakida, S.~Akaho and S.-i. Amari, The 22nd International Conference on
  Artificial Intelligence and Statistics, 2019, pp. 1032--1041\relax
\mciteBstWouldAddEndPuncttrue
\mciteSetBstMidEndSepPunct{\mcitedefaultmidpunct}
{\mcitedefaultendpunct}{\mcitedefaultseppunct}\relax
\EndOfBibitem
\bibitem[Berezniuk \emph{et~al.}(2020)Berezniuk, Figalli, Ghigliazza, and
  Musaelian]{berezniuk2020scale}
O.~Berezniuk, A.~Figalli, R.~Ghigliazza and K.~Musaelian, \emph{arXiv preprint
  arXiv:2001.10872}, 2020\relax
\mciteBstWouldAddEndPuncttrue
\mciteSetBstMidEndSepPunct{\mcitedefaultmidpunct}
{\mcitedefaultendpunct}{\mcitedefaultseppunct}\relax
\EndOfBibitem
\bibitem[Sharma \emph{et~al.}(2022)Sharma, Cerezo, Holmes, Cincio, Sornborger,
  and Coles]{PhysRevLett.128.070501}
K.~Sharma, M.~Cerezo, Z.~Holmes, L.~Cincio, A.~Sornborger and P.~J. Coles,
  \emph{Phys. Rev. Lett.}, 2022, \textbf{128}, 070501\relax
\mciteBstWouldAddEndPuncttrue
\mciteSetBstMidEndSepPunct{\mcitedefaultmidpunct}
{\mcitedefaultendpunct}{\mcitedefaultseppunct}\relax
\EndOfBibitem
\bibitem[Adam \emph{et~al.}(2019)Adam, Alexandropoulos, Pardalos, and
  Vrahatis]{Adam2019}
S.~P. Adam, S.-A.~N. Alexandropoulos, P.~M. Pardalos and M.~N. Vrahatis, in
  \emph{No Free Lunch Theorem: A Review}, ed. I.~C. Demetriou and P.~M.
  Pardalos, Springer International Publishing, Cham, 2019, pp. 57--82\relax
\mciteBstWouldAddEndPuncttrue
\mciteSetBstMidEndSepPunct{\mcitedefaultmidpunct}
{\mcitedefaultendpunct}{\mcitedefaultseppunct}\relax
\EndOfBibitem
\bibitem[M.M.Wolf(2018)]{Wolf_lectures}
M.M.Wolf, \emph{Mathematical Foundations of Supervised Learning}, 2018,
  \url{https://www-m5.ma.tum.de/Allgemeines/MA4801_2016S}\relax
\mciteBstWouldAddEndPuncttrue
\mciteSetBstMidEndSepPunct{\mcitedefaultmidpunct}
{\mcitedefaultendpunct}{\mcitedefaultseppunct}\relax
\EndOfBibitem
\bibitem[Poland \emph{et~al.}(2020)Poland, Beer, and Osborne]{poland2020free}
K.~Poland, K.~Beer and T.~J. Osborne, \emph{No Free Lunch for Quantum Machine
  Learning}, 2020\relax
\mciteBstWouldAddEndPuncttrue
\mciteSetBstMidEndSepPunct{\mcitedefaultmidpunct}
{\mcitedefaultendpunct}{\mcitedefaultseppunct}\relax
\EndOfBibitem
\bibitem[Bennett and Wiesner(1992)]{PhysRevLett.69.2881}
C.~H. Bennett and S.~J. Wiesner, \emph{Phys. Rev. Lett.}, 1992, \textbf{69},
  2881--2884\relax
\mciteBstWouldAddEndPuncttrue
\mciteSetBstMidEndSepPunct{\mcitedefaultmidpunct}
{\mcitedefaultendpunct}{\mcitedefaultseppunct}\relax
\EndOfBibitem
\bibitem[Harrow \emph{et~al.}(2004)Harrow, Hayden, and
  Leung]{PhysRevLett.92.187901}
A.~Harrow, P.~Hayden and D.~Leung, \emph{Phys. Rev. Lett.}, 2004, \textbf{92},
  187901\relax
\mciteBstWouldAddEndPuncttrue
\mciteSetBstMidEndSepPunct{\mcitedefaultmidpunct}
{\mcitedefaultendpunct}{\mcitedefaultseppunct}\relax
\EndOfBibitem
\bibitem[Bennett \emph{et~al.}(1993)Bennett, Brassard, Cr\'epeau, Jozsa, Peres,
  and Wootters]{PhysRevLett.70.1895}
C.~H. Bennett, G.~Brassard, C.~Cr\'epeau, R.~Jozsa, A.~Peres and W.~K.
  Wootters, \emph{Phys. Rev. Lett.}, 1993, \textbf{70}, 1895--1899\relax
\mciteBstWouldAddEndPuncttrue
\mciteSetBstMidEndSepPunct{\mcitedefaultmidpunct}
{\mcitedefaultendpunct}{\mcitedefaultseppunct}\relax
\EndOfBibitem
\bibitem[Luo \emph{et~al.}(2019)Luo, Zhong, Erhard, Wang, Peng, Krenn, Jiang,
  Li, Liu, Lu, Zeilinger, and Pan]{PhysRevLett.123.070505}
Y.-H. Luo, H.-S. Zhong, M.~Erhard, X.-L. Wang, L.-C. Peng, M.~Krenn, X.~Jiang,
  L.~Li, N.-L. Liu, C.-Y. Lu, A.~Zeilinger and J.-W. Pan, \emph{Phys. Rev.
  Lett.}, 2019, \textbf{123}, 070505\relax
\mciteBstWouldAddEndPuncttrue
\mciteSetBstMidEndSepPunct{\mcitedefaultmidpunct}
{\mcitedefaultendpunct}{\mcitedefaultseppunct}\relax
\EndOfBibitem
\bibitem[Pesah \emph{et~al.}(2021)Pesah, Cerezo, Wang, Volkoff, Sornborger, and
  Coles]{PhysRevX.11.041011}
A.~Pesah, M.~Cerezo, S.~Wang, T.~Volkoff, A.~T. Sornborger and P.~J. Coles,
  \emph{Phys. Rev. X}, 2021, \textbf{11}, 041011\relax
\mciteBstWouldAddEndPuncttrue
\mciteSetBstMidEndSepPunct{\mcitedefaultmidpunct}
{\mcitedefaultendpunct}{\mcitedefaultseppunct}\relax
\EndOfBibitem
\bibitem[MacCormack \emph{et~al.}(2022)MacCormack, Delaney, Galda, Aggarwal,
  and Narang]{maccormack2022branching}
I.~MacCormack, C.~Delaney, A.~Galda, N.~Aggarwal and P.~Narang, \emph{Physical
  Review Research}, 2022, \textbf{4}, 013117\relax
\mciteBstWouldAddEndPuncttrue
\mciteSetBstMidEndSepPunct{\mcitedefaultmidpunct}
{\mcitedefaultendpunct}{\mcitedefaultseppunct}\relax
\EndOfBibitem
\bibitem[Lee \emph{et~al.}(2020)Lee, Kawashima, and Kim]{lee2020tensor}
H.-Y. Lee, N.~Kawashima and Y.~B. Kim, \emph{Physical Review Research}, 2020,
  \textbf{2}, 033318\relax
\mciteBstWouldAddEndPuncttrue
\mciteSetBstMidEndSepPunct{\mcitedefaultmidpunct}
{\mcitedefaultendpunct}{\mcitedefaultseppunct}\relax
\EndOfBibitem
\bibitem[Kuhn and Richter(2020)]{kuhn2020tensor}
S.~C. Kuhn and M.~Richter, \emph{Physical Review B}, 2020, \textbf{101},
  075302\relax
\mciteBstWouldAddEndPuncttrue
\mciteSetBstMidEndSepPunct{\mcitedefaultmidpunct}
{\mcitedefaultendpunct}{\mcitedefaultseppunct}\relax
\EndOfBibitem
\bibitem[Gunst \emph{et~al.}(2018)Gunst, Verstraete, Wouters, Legeza, and
  Van~Neck]{gunst2018t3ns}
K.~Gunst, F.~Verstraete, S.~Wouters, O.~Legeza and D.~Van~Neck, \emph{Journal
  of chemical theory and computation}, 2018, \textbf{14}, 2026--2033\relax
\mciteBstWouldAddEndPuncttrue
\mciteSetBstMidEndSepPunct{\mcitedefaultmidpunct}
{\mcitedefaultendpunct}{\mcitedefaultseppunct}\relax
\EndOfBibitem
\bibitem[LeCun and Cortes(2010)]{lecun-mnisthandwrittendigit-2010}
Y.~LeCun and C.~Cortes, 2010\relax
\mciteBstWouldAddEndPuncttrue
\mciteSetBstMidEndSepPunct{\mcitedefaultmidpunct}
{\mcitedefaultendpunct}{\mcitedefaultseppunct}\relax
\EndOfBibitem
\bibitem[Spall(1999)]{10.1145/324138.324170}
J.~C. Spall, Proceedings of the 31st conference on Winter simulation:
  Simulation---a bridge to the future-Volume 1, 1999, pp. 101--109\relax
\mciteBstWouldAddEndPuncttrue
\mciteSetBstMidEndSepPunct{\mcitedefaultmidpunct}
{\mcitedefaultendpunct}{\mcitedefaultseppunct}\relax
\EndOfBibitem
\bibitem[Oseledets and Tyrtyshnikov(2009)]{doi:10.1137/090748330}
I.~V. Oseledets and E.~E. Tyrtyshnikov, \emph{SIAM Journal on Scientific
  Computing}, 2009, \textbf{31}, 3744--3759\relax
\mciteBstWouldAddEndPuncttrue
\mciteSetBstMidEndSepPunct{\mcitedefaultmidpunct}
{\mcitedefaultendpunct}{\mcitedefaultseppunct}\relax
\EndOfBibitem
\bibitem[Kais(2014)]{kais}
S.~Kais, \emph{Quantum Information and Computation for Chemistry}, Wiley and
  Sons: Hoboken, NJ, 2014, vol. 154\relax
\mciteBstWouldAddEndPuncttrue
\mciteSetBstMidEndSepPunct{\mcitedefaultmidpunct}
{\mcitedefaultendpunct}{\mcitedefaultseppunct}\relax
\EndOfBibitem
\bibitem[Altepeter \emph{et~al.}(2005)Altepeter, Jeffrey, and Kwiat]{photonic}
J.~Altepeter, E.~Jeffrey and P.~Kwiat, \emph{Photonic State Tomography},
  Academic Press, 2005, vol.~52, pp. 105--159\relax
\mciteBstWouldAddEndPuncttrue
\mciteSetBstMidEndSepPunct{\mcitedefaultmidpunct}
{\mcitedefaultendpunct}{\mcitedefaultseppunct}\relax
\EndOfBibitem
\bibitem[James \emph{et~al.}(2001)James, Kwiat, Munro, and White]{qubits}
D.~F.~V. James, P.~G. Kwiat, W.~J. Munro and A.~G. White, \emph{Phys. Rev. A},
  2001, \textbf{64}, 052312\relax
\mciteBstWouldAddEndPuncttrue
\mciteSetBstMidEndSepPunct{\mcitedefaultmidpunct}
{\mcitedefaultendpunct}{\mcitedefaultseppunct}\relax
\EndOfBibitem
\bibitem[Banaszek \emph{et~al.}(2013)Banaszek, Cramer, and Gross]{cramer}
K.~Banaszek, M.~Cramer and D.~Gross, \emph{New J. Phys.}, 2013, \textbf{15},
  125020\relax
\mciteBstWouldAddEndPuncttrue
\mciteSetBstMidEndSepPunct{\mcitedefaultmidpunct}
{\mcitedefaultendpunct}{\mcitedefaultseppunct}\relax
\EndOfBibitem
\bibitem[Song \emph{et~al.}(2017)Song, Xu, Liu, Yang, Zheng, Deng, Xie, Huang,
  Guo, Zhang, Zhang, Xu, Zheng, Zhu, Wang, Chen, Lu, Han, and Pan]{10-qubit}
C.~Song, K.~Xu, W.~Liu, C.-p. Yang, S.-B. Zheng, H.~Deng, Q.~Xie, K.~Huang,
  Q.~Guo, L.~Zhang, P.~Zhang, D.~Xu, D.~Zheng, X.~Zhu, H.~Wang, Y.-A. Chen,
  C.-Y. Lu, S.~Han and J.-W. Pan, \emph{Phys. Rev. Lett.}, 2017, \textbf{119},
  180511\relax
\mciteBstWouldAddEndPuncttrue
\mciteSetBstMidEndSepPunct{\mcitedefaultmidpunct}
{\mcitedefaultendpunct}{\mcitedefaultseppunct}\relax
\EndOfBibitem
\bibitem[Torlai \emph{et~al.}(2018)Torlai, Mazzola, Carrasquilla, Troyer,
  Melko, and Carleo]{Torlai2018}
G.~Torlai, G.~Mazzola, J.~Carrasquilla, M.~Troyer, R.~Melko and G.~Carleo,
  \emph{Nature Physics}, 2018, \textbf{14}, 447--450\relax
\mciteBstWouldAddEndPuncttrue
\mciteSetBstMidEndSepPunct{\mcitedefaultmidpunct}
{\mcitedefaultendpunct}{\mcitedefaultseppunct}\relax
\EndOfBibitem
\bibitem[Carrasquilla \emph{et~al.}(2019)Carrasquilla, Torlai, Melko, and
  Aolita]{Carrasquilla2019}
J.~Carrasquilla, G.~Torlai, R.~G. Melko and L.~Aolita, \emph{Nature Machine
  Intelligence}, 2019, \textbf{1}, 155--161\relax
\mciteBstWouldAddEndPuncttrue
\mciteSetBstMidEndSepPunct{\mcitedefaultmidpunct}
{\mcitedefaultendpunct}{\mcitedefaultseppunct}\relax
\EndOfBibitem
\bibitem[Palmieri \emph{et~al.}(2020)Palmieri, Kovlakov, Bianchi, Yudin,
  Straupe, Biamonte, and Kulik]{Palmieri2020}
A.~M. Palmieri, E.~Kovlakov, F.~Bianchi, D.~Yudin, S.~Straupe, J.~D. Biamonte
  and S.~Kulik, \emph{npj Quantum Information}, 2020, \textbf{6}, 20\relax
\mciteBstWouldAddEndPuncttrue
\mciteSetBstMidEndSepPunct{\mcitedefaultmidpunct}
{\mcitedefaultendpunct}{\mcitedefaultseppunct}\relax
\EndOfBibitem
\bibitem[Xin \emph{et~al.}(2019)Xin, Lu, Cao, Anikeeva, Lu, Li, Long, and
  Zeng]{xin2019local}
T.~Xin, S.~Lu, N.~Cao, G.~Anikeeva, D.~Lu, J.~Li, G.~Long and B.~Zeng,
  \emph{npj Quantum Information}, 2019, \textbf{5}, 1--8\relax
\mciteBstWouldAddEndPuncttrue
\mciteSetBstMidEndSepPunct{\mcitedefaultmidpunct}
{\mcitedefaultendpunct}{\mcitedefaultseppunct}\relax
\EndOfBibitem
\bibitem[Cotler and Wilczek(2020)]{overlap}
J.~Cotler and F.~Wilczek, \emph{Phys. Rev. Lett.}, 2020, \textbf{124},
  100401\relax
\mciteBstWouldAddEndPuncttrue
\mciteSetBstMidEndSepPunct{\mcitedefaultmidpunct}
{\mcitedefaultendpunct}{\mcitedefaultseppunct}\relax
\EndOfBibitem
\bibitem[Aaronson(2018)]{aaronson}
S.~Aaronson, Proceedings of the 50th Annual ACM SIGACT Symposium on Theory of
  Computing, New York, NY, USA, 2018, p. 325–338\relax
\mciteBstWouldAddEndPuncttrue
\mciteSetBstMidEndSepPunct{\mcitedefaultmidpunct}
{\mcitedefaultendpunct}{\mcitedefaultseppunct}\relax
\EndOfBibitem
\bibitem[Huang \emph{et~al.}(2020)Huang, Kueng, and Preskill]{Huang2020}
H.-Y. Huang, R.~Kueng and J.~Preskill, \emph{Nature Physics}, 2020,
  \textbf{16}, 1050--1057\relax
\mciteBstWouldAddEndPuncttrue
\mciteSetBstMidEndSepPunct{\mcitedefaultmidpunct}
{\mcitedefaultendpunct}{\mcitedefaultseppunct}\relax
\EndOfBibitem
\bibitem[Jaynes(1957)]{jaynes}
E.~T. Jaynes, \emph{Phys. Rev.}, 1957, \textbf{108}, 171--190\relax
\mciteBstWouldAddEndPuncttrue
\mciteSetBstMidEndSepPunct{\mcitedefaultmidpunct}
{\mcitedefaultendpunct}{\mcitedefaultseppunct}\relax
\EndOfBibitem
\bibitem[Wichmann(1963)]{wich}
E.~H. Wichmann, \emph{J. Math. Phys.}, 1963, \textbf{4}, 884--896\relax
\mciteBstWouldAddEndPuncttrue
\mciteSetBstMidEndSepPunct{\mcitedefaultmidpunct}
{\mcitedefaultendpunct}{\mcitedefaultseppunct}\relax
\EndOfBibitem
\bibitem[Katz(1967)]{katz1967principles}
A.~Katz, \emph{Principles of Statistical Mechanics: The Information Theory
  Approach}, W. H. Freeman, 1967\relax
\mciteBstWouldAddEndPuncttrue
\mciteSetBstMidEndSepPunct{\mcitedefaultmidpunct}
{\mcitedefaultendpunct}{\mcitedefaultseppunct}\relax
\EndOfBibitem
\bibitem[Hradil(1997)]{hradil}
Z.~Hradil, \emph{Phys. Rev. A}, 1997, \textbf{55}, R1561--R1564\relax
\mciteBstWouldAddEndPuncttrue
\mciteSetBstMidEndSepPunct{\mcitedefaultmidpunct}
{\mcitedefaultendpunct}{\mcitedefaultseppunct}\relax
\EndOfBibitem
\bibitem[Teo \emph{et~al.}(2011)Teo, Zhu, Englert, {\v{R}}eh{\'a}{\v{c}}ek, and
  Hradil]{teo2011quantum}
Y.~S. Teo, H.~Zhu, B.-G. Englert, J.~{\v{R}}eh{\'a}{\v{c}}ek and Z.~Hradil,
  \emph{Physical review letters}, 2011, \textbf{107}, 020404\relax
\mciteBstWouldAddEndPuncttrue
\mciteSetBstMidEndSepPunct{\mcitedefaultmidpunct}
{\mcitedefaultendpunct}{\mcitedefaultseppunct}\relax
\EndOfBibitem
\bibitem[Teo \emph{et~al.}(2012)Teo, Stoklasa, Englert,
  {\v{R}}eh{\'a}{\v{c}}ek, and Hradil]{teo2012incomplete}
Y.~S. Teo, B.~Stoklasa, B.-G. Englert, J.~{\v{R}}eh{\'a}{\v{c}}ek and
  Z.~Hradil, \emph{Physical Review A}, 2012, \textbf{85}, 042317\relax
\mciteBstWouldAddEndPuncttrue
\mciteSetBstMidEndSepPunct{\mcitedefaultmidpunct}
{\mcitedefaultendpunct}{\mcitedefaultseppunct}\relax
\EndOfBibitem
\bibitem[Blume-Kohout(2010)]{blume2010hedged}
R.~Blume-Kohout, \emph{Physical review letters}, 2010, \textbf{105},
  200504\relax
\mciteBstWouldAddEndPuncttrue
\mciteSetBstMidEndSepPunct{\mcitedefaultmidpunct}
{\mcitedefaultendpunct}{\mcitedefaultseppunct}\relax
\EndOfBibitem
\bibitem[Smolin \emph{et~al.}(2012)Smolin, Gambetta, and
  Smith]{smolin2012efficient}
J.~A. Smolin, J.~M. Gambetta and G.~Smith, \emph{Physical review letters},
  2012, \textbf{108}, 070502\relax
\mciteBstWouldAddEndPuncttrue
\mciteSetBstMidEndSepPunct{\mcitedefaultmidpunct}
{\mcitedefaultendpunct}{\mcitedefaultseppunct}\relax
\EndOfBibitem
\bibitem[Baumgratz \emph{et~al.}(2013)Baumgratz, N{\"u}{\ss}eler, Cramer, and
  Plenio]{baumgratz2013scalable}
T.~Baumgratz, A.~N{\"u}{\ss}eler, M.~Cramer and M.~B. Plenio, \emph{New Journal
  of Physics}, 2013, \textbf{15}, 125004\relax
\mciteBstWouldAddEndPuncttrue
\mciteSetBstMidEndSepPunct{\mcitedefaultmidpunct}
{\mcitedefaultendpunct}{\mcitedefaultseppunct}\relax
\EndOfBibitem
\bibitem[Blume-Kohout(2010)]{blume2010optimal}
R.~Blume-Kohout, \emph{New Journal of Physics}, 2010, \textbf{12}, 043034\relax
\mciteBstWouldAddEndPuncttrue
\mciteSetBstMidEndSepPunct{\mcitedefaultmidpunct}
{\mcitedefaultendpunct}{\mcitedefaultseppunct}\relax
\EndOfBibitem
\bibitem[Husz{\'a}r and Houlsby(2012)]{huszar2012adaptive}
F.~Husz{\'a}r and N.~M. Houlsby, \emph{Physical Review A}, 2012, \textbf{85},
  052120\relax
\mciteBstWouldAddEndPuncttrue
\mciteSetBstMidEndSepPunct{\mcitedefaultmidpunct}
{\mcitedefaultendpunct}{\mcitedefaultseppunct}\relax
\EndOfBibitem
\bibitem[Lukens \emph{et~al.}(2020)Lukens, Law, Jasra, and
  Lougovski]{lukens2020practical}
J.~M. Lukens, K.~J. Law, A.~Jasra and P.~Lougovski, \emph{New Journal of
  Physics}, 2020, \textbf{22}, 063038\relax
\mciteBstWouldAddEndPuncttrue
\mciteSetBstMidEndSepPunct{\mcitedefaultmidpunct}
{\mcitedefaultendpunct}{\mcitedefaultseppunct}\relax
\EndOfBibitem
\bibitem[Lukens \emph{et~al.}(2020)Lukens, Law, and
  Bennink]{lukens2020bayesian}
J.~M. Lukens, K.~J. Law and R.~S. Bennink, \emph{arXiv preprint
  arXiv:2012.08997}, 2020\relax
\mciteBstWouldAddEndPuncttrue
\mciteSetBstMidEndSepPunct{\mcitedefaultmidpunct}
{\mcitedefaultendpunct}{\mcitedefaultseppunct}\relax
\EndOfBibitem
\bibitem[Gupta \emph{et~al.}(2021)Gupta, Xia, Levine, and
  Kais]{gupta2021maximal}
R.~Gupta, R.~Xia, R.~D. Levine and S.~Kais, \emph{PRX Quantum}, 2021,
  \textbf{2}, 010318\relax
\mciteBstWouldAddEndPuncttrue
\mciteSetBstMidEndSepPunct{\mcitedefaultmidpunct}
{\mcitedefaultendpunct}{\mcitedefaultseppunct}\relax
\EndOfBibitem
\bibitem[Gupta \emph{et~al.}(2021)Gupta, Levine, and
  Kais]{gupta2021convergence}
R.~Gupta, R.~D. Levine and S.~Kais, \emph{The Journal of Physical Chemistry A},
  2021,  7588--7594\relax
\mciteBstWouldAddEndPuncttrue
\mciteSetBstMidEndSepPunct{\mcitedefaultmidpunct}
{\mcitedefaultendpunct}{\mcitedefaultseppunct}\relax
\EndOfBibitem
\bibitem[Helstrom(1976)]{helstrom1976quantum}
C.~Helstrom, \emph{Mathematics in Science and Engineering. New York: Academic
  Press}, 1976, \textbf{123}, 1572--9613\relax
\mciteBstWouldAddEndPuncttrue
\mciteSetBstMidEndSepPunct{\mcitedefaultmidpunct}
{\mcitedefaultendpunct}{\mcitedefaultseppunct}\relax
\EndOfBibitem
\bibitem[Gudder \emph{et~al.}(1985)Gudder\emph{et~al.}]{gudder1985holevo}
S.~Gudder \emph{et~al.}, \emph{Bulletin (New Series) of the American
  Mathematical Society}, 1985, \textbf{13}, 80--85\relax
\mciteBstWouldAddEndPuncttrue
\mciteSetBstMidEndSepPunct{\mcitedefaultmidpunct}
{\mcitedefaultendpunct}{\mcitedefaultseppunct}\relax
\EndOfBibitem
\bibitem[Peres(1993)]{peres1993quantum}
A.~Peres, \emph{Publ., Boston}, 1993\relax
\mciteBstWouldAddEndPuncttrue
\mciteSetBstMidEndSepPunct{\mcitedefaultmidpunct}
{\mcitedefaultendpunct}{\mcitedefaultseppunct}\relax
\EndOfBibitem
\bibitem[Sasaki and Carlini(2002)]{sasaki2002quantum}
M.~Sasaki and A.~Carlini, \emph{Physical Review A}, 2002, \textbf{66},
  022303\relax
\mciteBstWouldAddEndPuncttrue
\mciteSetBstMidEndSepPunct{\mcitedefaultmidpunct}
{\mcitedefaultendpunct}{\mcitedefaultseppunct}\relax
\EndOfBibitem
\bibitem[Chrisley(1995)]{chrisley1995quantum}
R.~Chrisley, New directions in cognitive science: Proceedings of the
  international symposium, Saariselka, 1995\relax
\mciteBstWouldAddEndPuncttrue
\mciteSetBstMidEndSepPunct{\mcitedefaultmidpunct}
{\mcitedefaultendpunct}{\mcitedefaultseppunct}\relax
\EndOfBibitem
\bibitem[Behrman \emph{et~al.}(1996)Behrman, Niemel, Steck, and
  Skinner]{behrman1996quantum}
E.~C. Behrman, J.~Niemel, J.~E. Steck and S.~R. Skinner, Proceedings of the 4th
  Workshop on Physics of Computation, 1996, pp. 22--24\relax
\mciteBstWouldAddEndPuncttrue
\mciteSetBstMidEndSepPunct{\mcitedefaultmidpunct}
{\mcitedefaultendpunct}{\mcitedefaultseppunct}\relax
\EndOfBibitem
\bibitem[Behrman \emph{et~al.}(1999)Behrman, Steck, and
  Skinner]{behrman1999spatial}
E.~C. Behrman, J.~E. Steck and S.~R. Skinner, IJCNN'99. International Joint
  Conference on Neural Networks. Proceedings (Cat. No. 99CH36339), 1999, pp.
  874--877\relax
\mciteBstWouldAddEndPuncttrue
\mciteSetBstMidEndSepPunct{\mcitedefaultmidpunct}
{\mcitedefaultendpunct}{\mcitedefaultseppunct}\relax
\EndOfBibitem
\bibitem[Benedetti \emph{et~al.}(2019)Benedetti, Garcia-Pintos, Perdomo,
  Leyton-Ortega, Nam, and Perdomo-Ortiz]{benedetti2019generative}
M.~Benedetti, D.~Garcia-Pintos, O.~Perdomo, V.~Leyton-Ortega, Y.~Nam and
  A.~Perdomo-Ortiz, \emph{npj Quantum Information}, 2019, \textbf{5},
  1--9\relax
\mciteBstWouldAddEndPuncttrue
\mciteSetBstMidEndSepPunct{\mcitedefaultmidpunct}
{\mcitedefaultendpunct}{\mcitedefaultseppunct}\relax
\EndOfBibitem
\bibitem[Cheng \emph{et~al.}(2018)Cheng, Chen, and Wang]{cheng2018information}
S.~Cheng, J.~Chen and L.~Wang, \emph{Entropy}, 2018, \textbf{20}, 583\relax
\mciteBstWouldAddEndPuncttrue
\mciteSetBstMidEndSepPunct{\mcitedefaultmidpunct}
{\mcitedefaultendpunct}{\mcitedefaultseppunct}\relax
\EndOfBibitem
\bibitem[Stoudenmire and Schwab(2016)]{stoudenmire2016supervised}
E.~M. Stoudenmire and D.~J. Schwab, \emph{arXiv preprint arXiv:1605.05775},
  2016\relax
\mciteBstWouldAddEndPuncttrue
\mciteSetBstMidEndSepPunct{\mcitedefaultmidpunct}
{\mcitedefaultendpunct}{\mcitedefaultseppunct}\relax
\EndOfBibitem
\bibitem[Han \emph{et~al.}(2018)Han, Wang, Fan, Wang, and
  Zhang]{han2018unsupervised}
Z.-Y. Han, J.~Wang, H.~Fan, L.~Wang and P.~Zhang, \emph{Physical Review X},
  2018, \textbf{8}, 031012\relax
\mciteBstWouldAddEndPuncttrue
\mciteSetBstMidEndSepPunct{\mcitedefaultmidpunct}
{\mcitedefaultendpunct}{\mcitedefaultseppunct}\relax
\EndOfBibitem
\bibitem[Gao \emph{et~al.}(2018)Gao, Zhang, and Duan]{gao2018quantum}
X.~Gao, Z.-Y. Zhang and L.-M. Duan, \emph{Science advances}, 2018, \textbf{4},
  eaat9004\relax
\mciteBstWouldAddEndPuncttrue
\mciteSetBstMidEndSepPunct{\mcitedefaultmidpunct}
{\mcitedefaultendpunct}{\mcitedefaultseppunct}\relax
\EndOfBibitem
\bibitem[Arrazola \emph{et~al.}(2019)Arrazola, Bromley, Izaac, Myers,
  Br{\'a}dler, and Killoran]{arrazola2019machine}
J.~M. Arrazola, T.~R. Bromley, J.~Izaac, C.~R. Myers, K.~Br{\'a}dler and
  N.~Killoran, \emph{Quantum Science and Technology}, 2019, \textbf{4},
  024004\relax
\mciteBstWouldAddEndPuncttrue
\mciteSetBstMidEndSepPunct{\mcitedefaultmidpunct}
{\mcitedefaultendpunct}{\mcitedefaultseppunct}\relax
\EndOfBibitem
\bibitem[Killoran \emph{et~al.}(2019)Killoran, Bromley, Arrazola, Schuld,
  Quesada, and Lloyd]{killoran2019continuous}
N.~Killoran, T.~R. Bromley, J.~M. Arrazola, M.~Schuld, N.~Quesada and S.~Lloyd,
  \emph{Physical Review Research}, 2019, \textbf{1}, 033063\relax
\mciteBstWouldAddEndPuncttrue
\mciteSetBstMidEndSepPunct{\mcitedefaultmidpunct}
{\mcitedefaultendpunct}{\mcitedefaultseppunct}\relax
\EndOfBibitem
\bibitem[Killoran \emph{et~al.}(2019)Killoran, Izaac, Quesada, Bergholm, Amy,
  and Weedbrook]{killoran2019strawberry}
N.~Killoran, J.~Izaac, N.~Quesada, V.~Bergholm, M.~Amy and C.~Weedbrook,
  \emph{Quantum}, 2019, \textbf{3}, 129\relax
\mciteBstWouldAddEndPuncttrue
\mciteSetBstMidEndSepPunct{\mcitedefaultmidpunct}
{\mcitedefaultendpunct}{\mcitedefaultseppunct}\relax
\EndOfBibitem
\bibitem[Zhang \emph{et~al.}(2019)Zhang, Wei, Asad, Yang, and
  Wang]{zhang2019does}
X.-M. Zhang, Z.~Wei, R.~Asad, X.-C. Yang and X.~Wang, \emph{npj Quantum
  Information}, 2019, \textbf{5}, 1--7\relax
\mciteBstWouldAddEndPuncttrue
\mciteSetBstMidEndSepPunct{\mcitedefaultmidpunct}
{\mcitedefaultendpunct}{\mcitedefaultseppunct}\relax
\EndOfBibitem
\bibitem[Benedetti \emph{et~al.}(2019)Benedetti, Lloyd, Sack, and
  Fiorentini]{benedetti2019parameterized}
M.~Benedetti, E.~Lloyd, S.~Sack and M.~Fiorentini, \emph{Quantum Science and
  Technology}, 2019, \textbf{4}, 043001\relax
\mciteBstWouldAddEndPuncttrue
\mciteSetBstMidEndSepPunct{\mcitedefaultmidpunct}
{\mcitedefaultendpunct}{\mcitedefaultseppunct}\relax
\EndOfBibitem
\bibitem[McClean \emph{et~al.}(2016)McClean, Romero, Babbush, and
  Aspuru-Guzik]{mcclean2016theory}
J.~R. McClean, J.~Romero, R.~Babbush and A.~Aspuru-Guzik, \emph{New Journal of
  Physics}, 2016, \textbf{18}, 023023\relax
\mciteBstWouldAddEndPuncttrue
\mciteSetBstMidEndSepPunct{\mcitedefaultmidpunct}
{\mcitedefaultendpunct}{\mcitedefaultseppunct}\relax
\EndOfBibitem
\bibitem[Huang \emph{et~al.}(2019)Huang, Bharti, and Rebentrost]{huang2019near}
H.-Y. Huang, K.~Bharti and P.~Rebentrost, \emph{arXiv preprint
  arXiv:1909.07344}, 2019\relax
\mciteBstWouldAddEndPuncttrue
\mciteSetBstMidEndSepPunct{\mcitedefaultmidpunct}
{\mcitedefaultendpunct}{\mcitedefaultseppunct}\relax
\EndOfBibitem
\bibitem[LaRose \emph{et~al.}(2019)LaRose, Tikku, O’Neel-Judy, Cincio, and
  Coles]{larose2019variational}
R.~LaRose, A.~Tikku, {\'E}.~O’Neel-Judy, L.~Cincio and P.~J. Coles, \emph{npj
  Quantum Information}, 2019, \textbf{5}, 1--10\relax
\mciteBstWouldAddEndPuncttrue
\mciteSetBstMidEndSepPunct{\mcitedefaultmidpunct}
{\mcitedefaultendpunct}{\mcitedefaultseppunct}\relax
\EndOfBibitem
\bibitem[Cerezo \emph{et~al.}(2020)Cerezo, Sharma, Arrasmith, and
  Coles]{cerezo2020variational}
M.~Cerezo, K.~Sharma, A.~Arrasmith and P.~J. Coles, \emph{arXiv preprint
  arXiv:2004.01372}, 2020\relax
\mciteBstWouldAddEndPuncttrue
\mciteSetBstMidEndSepPunct{\mcitedefaultmidpunct}
{\mcitedefaultendpunct}{\mcitedefaultseppunct}\relax
\EndOfBibitem
\bibitem[Wang \emph{et~al.}(2021)Wang, Song, and Wang]{wang2021variational}
X.~Wang, Z.~Song and Y.~Wang, \emph{Quantum}, 2021, \textbf{5}, 483\relax
\mciteBstWouldAddEndPuncttrue
\mciteSetBstMidEndSepPunct{\mcitedefaultmidpunct}
{\mcitedefaultendpunct}{\mcitedefaultseppunct}\relax
\EndOfBibitem
\bibitem[Wang \emph{et~al.}(2021)Wang, Li, and Wang]{wang2021hybrid}
Y.~Wang, G.~Li and X.~Wang, \emph{arXiv preprint arXiv:2103.01061}, 2021\relax
\mciteBstWouldAddEndPuncttrue
\mciteSetBstMidEndSepPunct{\mcitedefaultmidpunct}
{\mcitedefaultendpunct}{\mcitedefaultseppunct}\relax
\EndOfBibitem
\bibitem[Li \emph{et~al.}(2021)Li, Song, and Wang]{li2021vsql}
G.~Li, Z.~Song and X.~Wang, Proceedings of the AAAI Conference on Artificial
  Intelligence, 2021, pp. 8357--8365\relax
\mciteBstWouldAddEndPuncttrue
\mciteSetBstMidEndSepPunct{\mcitedefaultmidpunct}
{\mcitedefaultendpunct}{\mcitedefaultseppunct}\relax
\EndOfBibitem
\bibitem[Chen \emph{et~al.}(2021)Chen, Song, Zhao, and
  Wang]{chen2021variational}
R.~Chen, Z.~Song, X.~Zhao and X.~Wang, \emph{Quantum Science and Technology},
  2021, \textbf{7}, 015019\relax
\mciteBstWouldAddEndPuncttrue
\mciteSetBstMidEndSepPunct{\mcitedefaultmidpunct}
{\mcitedefaultendpunct}{\mcitedefaultseppunct}\relax
\EndOfBibitem
\bibitem[Wang \emph{et~al.}(2021)Wang, Li, and Wang]{wang2021gibbs}
Y.~Wang, G.~Li and X.~Wang, \emph{Physical Review Applied}, 2021, \textbf{16},
  054035\relax
\mciteBstWouldAddEndPuncttrue
\mciteSetBstMidEndSepPunct{\mcitedefaultmidpunct}
{\mcitedefaultendpunct}{\mcitedefaultseppunct}\relax
\EndOfBibitem
\bibitem[Aharonov \emph{et~al.}(2013)Aharonov, Arad, and
  Vidick]{aharonov2013guest}
D.~Aharonov, I.~Arad and T.~Vidick, \emph{Acm sigact news}, 2013, \textbf{44},
  47--79\relax
\mciteBstWouldAddEndPuncttrue
\mciteSetBstMidEndSepPunct{\mcitedefaultmidpunct}
{\mcitedefaultendpunct}{\mcitedefaultseppunct}\relax
\EndOfBibitem
\bibitem[Childs \emph{et~al.}(2018)Childs, Maslov, Nam, Ross, and
  Su]{childs2018toward}
A.~M. Childs, D.~Maslov, Y.~Nam, N.~J. Ross and Y.~Su, \emph{Proceedings of the
  National Academy of Sciences}, 2018, \textbf{115}, 9456--9461\relax
\mciteBstWouldAddEndPuncttrue
\mciteSetBstMidEndSepPunct{\mcitedefaultmidpunct}
{\mcitedefaultendpunct}{\mcitedefaultseppunct}\relax
\EndOfBibitem
\bibitem[Somma \emph{et~al.}(2008)Somma, Boixo, Barnum, and
  Knill]{somma2008quantum}
R.~D. Somma, S.~Boixo, H.~Barnum and E.~Knill, \emph{Physical review letters},
  2008, \textbf{101}, 130504\relax
\mciteBstWouldAddEndPuncttrue
\mciteSetBstMidEndSepPunct{\mcitedefaultmidpunct}
{\mcitedefaultendpunct}{\mcitedefaultseppunct}\relax
\EndOfBibitem
\bibitem[Gheorghiu and Hoban(2020)]{gheorghiu2020estimating}
A.~Gheorghiu and M.~J. Hoban, \emph{arXiv preprint arXiv:2002.12814},
  2020\relax
\mciteBstWouldAddEndPuncttrue
\mciteSetBstMidEndSepPunct{\mcitedefaultmidpunct}
{\mcitedefaultendpunct}{\mcitedefaultseppunct}\relax
\EndOfBibitem
\bibitem[Buhrman \emph{et~al.}(2001)Buhrman, Cleve, Watrous, and
  De~Wolf]{buhrman2001quantum}
H.~Buhrman, R.~Cleve, J.~Watrous and R.~De~Wolf, \emph{Physical Review
  Letters}, 2001, \textbf{87}, 167902\relax
\mciteBstWouldAddEndPuncttrue
\mciteSetBstMidEndSepPunct{\mcitedefaultmidpunct}
{\mcitedefaultendpunct}{\mcitedefaultseppunct}\relax
\EndOfBibitem
\bibitem[Gottesman and Chuang(2001)]{gottesman2001quantum}
D.~Gottesman and I.~Chuang, \emph{arXiv preprint quant-ph/0105032}, 2001\relax
\mciteBstWouldAddEndPuncttrue
\mciteSetBstMidEndSepPunct{\mcitedefaultmidpunct}
{\mcitedefaultendpunct}{\mcitedefaultseppunct}\relax
\EndOfBibitem
\bibitem[Neven \emph{et~al.}(2008)Neven, Denchev, Rose, and
  Macready]{neven2008training}
H.~Neven, V.~S. Denchev, G.~Rose and W.~G. Macready, \emph{arXiv preprint
  arXiv:0811.0416}, 2008\relax
\mciteBstWouldAddEndPuncttrue
\mciteSetBstMidEndSepPunct{\mcitedefaultmidpunct}
{\mcitedefaultendpunct}{\mcitedefaultseppunct}\relax
\EndOfBibitem
\bibitem[Pudenz and Lidar(2013)]{pudenz2013quantum}
K.~L. Pudenz and D.~A. Lidar, \emph{Quantum information processing}, 2013,
  \textbf{12}, 2027--2070\relax
\mciteBstWouldAddEndPuncttrue
\mciteSetBstMidEndSepPunct{\mcitedefaultmidpunct}
{\mcitedefaultendpunct}{\mcitedefaultseppunct}\relax
\EndOfBibitem
\bibitem[Johnson \emph{et~al.}(2011)Johnson, Amin, Gildert, Lanting, Hamze,
  Dickson, Harris, Berkley, Johansson, Bunyk,\emph{et~al.}]{johnson2011quantum}
M.~W. Johnson, M.~H. Amin, S.~Gildert, T.~Lanting, F.~Hamze, N.~Dickson,
  R.~Harris, A.~J. Berkley, J.~Johansson, P.~Bunyk \emph{et~al.},
  \emph{Nature}, 2011, \textbf{473}, 194--198\relax
\mciteBstWouldAddEndPuncttrue
\mciteSetBstMidEndSepPunct{\mcitedefaultmidpunct}
{\mcitedefaultendpunct}{\mcitedefaultseppunct}\relax
\EndOfBibitem
\bibitem[Adachi and Henderson(2015)]{adachi2015application}
S.~H. Adachi and M.~P. Henderson, \emph{arXiv preprint arXiv:1510.06356},
  2015\relax
\mciteBstWouldAddEndPuncttrue
\mciteSetBstMidEndSepPunct{\mcitedefaultmidpunct}
{\mcitedefaultendpunct}{\mcitedefaultseppunct}\relax
\EndOfBibitem
\bibitem[Benedetti \emph{et~al.}(2017)Benedetti, Realpe-G{\'o}mez, Biswas, and
  Perdomo-Ortiz]{benedetti2017quantum}
M.~Benedetti, J.~Realpe-G{\'o}mez, R.~Biswas and A.~Perdomo-Ortiz,
  \emph{Physical Review X}, 2017, \textbf{7}, 041052\relax
\mciteBstWouldAddEndPuncttrue
\mciteSetBstMidEndSepPunct{\mcitedefaultmidpunct}
{\mcitedefaultendpunct}{\mcitedefaultseppunct}\relax
\EndOfBibitem
\bibitem[Wiebe \emph{et~al.}(2015)Wiebe, Kapoor, Granade, and
  Svore]{wiebe2015quantum}
N.~Wiebe, A.~Kapoor, C.~Granade and K.~M. Svore, \emph{arXiv preprint
  arXiv:1507.02642}, 2015\relax
\mciteBstWouldAddEndPuncttrue
\mciteSetBstMidEndSepPunct{\mcitedefaultmidpunct}
{\mcitedefaultendpunct}{\mcitedefaultseppunct}\relax
\EndOfBibitem
\bibitem[Amin \emph{et~al.}(2018)Amin, Andriyash, Rolfe, Kulchytskyy, and
  Melko]{amin2018quantum}
M.~H. Amin, E.~Andriyash, J.~Rolfe, B.~Kulchytskyy and R.~Melko, \emph{Physical
  Review X}, 2018, \textbf{8}, 021050\relax
\mciteBstWouldAddEndPuncttrue
\mciteSetBstMidEndSepPunct{\mcitedefaultmidpunct}
{\mcitedefaultendpunct}{\mcitedefaultseppunct}\relax
\EndOfBibitem
\bibitem[Kieferov{\'a} and Wiebe(2017)]{kieferova2017tomography}
M.~Kieferov{\'a} and N.~Wiebe, \emph{Physical Review A}, 2017, \textbf{96},
  062327\relax
\mciteBstWouldAddEndPuncttrue
\mciteSetBstMidEndSepPunct{\mcitedefaultmidpunct}
{\mcitedefaultendpunct}{\mcitedefaultseppunct}\relax
\EndOfBibitem
\bibitem[Xu and Xu(2018)]{xu2018neural}
Q.~Xu and S.~Xu, \emph{arXiv preprint arXiv:1811.06654}, 2018\relax
\mciteBstWouldAddEndPuncttrue
\mciteSetBstMidEndSepPunct{\mcitedefaultmidpunct}
{\mcitedefaultendpunct}{\mcitedefaultseppunct}\relax
\EndOfBibitem
\bibitem[Baldwin \emph{et~al.}(2016)Baldwin, Deutsch, and
  Kalev]{baldwin2016strictly}
C.~H. Baldwin, I.~H. Deutsch and A.~Kalev, \emph{Physical Review A}, 2016,
  \textbf{93}, 052105\relax
\mciteBstWouldAddEndPuncttrue
\mciteSetBstMidEndSepPunct{\mcitedefaultmidpunct}
{\mcitedefaultendpunct}{\mcitedefaultseppunct}\relax
\EndOfBibitem
\bibitem[Linden \emph{et~al.}(2002)Linden, Popescu, and
  Wootters]{linden2002almost}
N.~Linden, S.~Popescu and W.~Wootters, \emph{Physical review letters}, 2002,
  \textbf{89}, 207901\relax
\mciteBstWouldAddEndPuncttrue
\mciteSetBstMidEndSepPunct{\mcitedefaultmidpunct}
{\mcitedefaultendpunct}{\mcitedefaultseppunct}\relax
\EndOfBibitem
\bibitem[Linden and Wootters(2002)]{linden2002parts}
N.~Linden and W.~Wootters, \emph{Physical review letters}, 2002, \textbf{89},
  277906\relax
\mciteBstWouldAddEndPuncttrue
\mciteSetBstMidEndSepPunct{\mcitedefaultmidpunct}
{\mcitedefaultendpunct}{\mcitedefaultseppunct}\relax
\EndOfBibitem
\bibitem[Chen \emph{et~al.}(2012)Chen, Ji, Zeng, and Zhou]{chen2012ground}
J.~Chen, Z.~Ji, B.~Zeng and D.~Zhou, \emph{Physical Review A}, 2012,
  \textbf{86}, 022339\relax
\mciteBstWouldAddEndPuncttrue
\mciteSetBstMidEndSepPunct{\mcitedefaultmidpunct}
{\mcitedefaultendpunct}{\mcitedefaultseppunct}\relax
\EndOfBibitem
\bibitem[Chen \emph{et~al.}(2013)Chen, Dawkins, Ji, Johnston, Kribs, Shultz,
  and Zeng]{chen2013uniqueness}
J.~Chen, H.~Dawkins, Z.~Ji, N.~Johnston, D.~Kribs, F.~Shultz and B.~Zeng,
  \emph{Physical Review A}, 2013, \textbf{88}, 012109\relax
\mciteBstWouldAddEndPuncttrue
\mciteSetBstMidEndSepPunct{\mcitedefaultmidpunct}
{\mcitedefaultendpunct}{\mcitedefaultseppunct}\relax
\EndOfBibitem
\bibitem[Qi \emph{et~al.}(2013)Qi, Hou, Li, Dong, Xiang, and
  Guo]{qi2013quantum}
B.~Qi, Z.~Hou, L.~Li, D.~Dong, G.~Xiang and G.~Guo, \emph{Scientific reports},
  2013, \textbf{3}, 1--6\relax
\mciteBstWouldAddEndPuncttrue
\mciteSetBstMidEndSepPunct{\mcitedefaultmidpunct}
{\mcitedefaultendpunct}{\mcitedefaultseppunct}\relax
\EndOfBibitem
\bibitem[James \emph{et~al.}(2005)James, Kwiat, Munro, and
  White]{james2005measurement}
D.~F. James, P.~G. Kwiat, W.~J. Munro and A.~G. White, \emph{Asymptotic Theory
  of Quantum Statistical Inference: Selected Papers}, World Scientific, 2005,
  pp. 509--538\relax
\mciteBstWouldAddEndPuncttrue
\mciteSetBstMidEndSepPunct{\mcitedefaultmidpunct}
{\mcitedefaultendpunct}{\mcitedefaultseppunct}\relax
\EndOfBibitem
\bibitem[Opatrn{\`y} \emph{et~al.}(1997)Opatrn{\`y}, Welsch, and
  Vogel]{opatrny1997least}
T.~Opatrn{\`y}, D.-G. Welsch and W.~Vogel, \emph{Physical Review A}, 1997,
  \textbf{56}, 1788\relax
\mciteBstWouldAddEndPuncttrue
\mciteSetBstMidEndSepPunct{\mcitedefaultmidpunct}
{\mcitedefaultendpunct}{\mcitedefaultseppunct}\relax
\EndOfBibitem
\bibitem[Fischer \emph{et~al.}(2000)Fischer, Kienle, and
  Freyberger]{fischer2000quantum}
D.~G. Fischer, S.~H. Kienle and M.~Freyberger, \emph{Physical Review A}, 2000,
  \textbf{61}, 032306\relax
\mciteBstWouldAddEndPuncttrue
\mciteSetBstMidEndSepPunct{\mcitedefaultmidpunct}
{\mcitedefaultendpunct}{\mcitedefaultseppunct}\relax
\EndOfBibitem
\bibitem[Struchalin \emph{et~al.}(2016)Struchalin, Pogorelov, Straupe,
  Kravtsov, Radchenko, and Kulik]{struchalin2016experimental}
G.~I. Struchalin, I.~A. Pogorelov, S.~S. Straupe, K.~S. Kravtsov, I.~V.
  Radchenko and S.~P. Kulik, \emph{Physical Review A}, 2016, \textbf{93},
  012103\relax
\mciteBstWouldAddEndPuncttrue
\mciteSetBstMidEndSepPunct{\mcitedefaultmidpunct}
{\mcitedefaultendpunct}{\mcitedefaultseppunct}\relax
\EndOfBibitem
\bibitem[Quek \emph{et~al.}(2021)Quek, Fort, and Ng]{quek2021adaptive}
Y.~Quek, S.~Fort and H.~K. Ng, \emph{npj Quantum Information}, 2021,
  \textbf{7}, 1--7\relax
\mciteBstWouldAddEndPuncttrue
\mciteSetBstMidEndSepPunct{\mcitedefaultmidpunct}
{\mcitedefaultendpunct}{\mcitedefaultseppunct}\relax
\EndOfBibitem
\bibitem[Hu \emph{et~al.}(2019)Hu, Wu, Cai, Ma, Mu, Xu, Wang, Song, Deng,
  Zou,\emph{et~al.}]{hu2019quantum}
L.~Hu, S.-H. Wu, W.~Cai, Y.~Ma, X.~Mu, Y.~Xu, H.~Wang, Y.~Song, D.-L. Deng,
  C.-L. Zou \emph{et~al.}, \emph{Science advances}, 2019, \textbf{5},
  eaav2761\relax
\mciteBstWouldAddEndPuncttrue
\mciteSetBstMidEndSepPunct{\mcitedefaultmidpunct}
{\mcitedefaultendpunct}{\mcitedefaultseppunct}\relax
\EndOfBibitem
\bibitem[Qi \emph{et~al.}(2017)Qi, Hou, Wang, Dong, Zhong, Li, Xiang, Wiseman,
  Li, and Guo]{qi2017adaptive}
B.~Qi, Z.~Hou, Y.~Wang, D.~Dong, H.-S. Zhong, L.~Li, G.-Y. Xiang, H.~M.
  Wiseman, C.-F. Li and G.-C. Guo, \emph{npj Quantum Information}, 2017,
  \textbf{3}, 1--7\relax
\mciteBstWouldAddEndPuncttrue
\mciteSetBstMidEndSepPunct{\mcitedefaultmidpunct}
{\mcitedefaultendpunct}{\mcitedefaultseppunct}\relax
\EndOfBibitem
\bibitem[Lloyd and Weedbrook(2018)]{lloyd2018quantum}
S.~Lloyd and C.~Weedbrook, \emph{Physical review letters}, 2018, \textbf{121},
  040502\relax
\mciteBstWouldAddEndPuncttrue
\mciteSetBstMidEndSepPunct{\mcitedefaultmidpunct}
{\mcitedefaultendpunct}{\mcitedefaultseppunct}\relax
\EndOfBibitem
\bibitem[Dallaire-Demers and Killoran(2018)]{dallaire2018quantum}
P.-L. Dallaire-Demers and N.~Killoran, \emph{Physical Review A}, 2018,
  \textbf{98}, 012324\relax
\mciteBstWouldAddEndPuncttrue
\mciteSetBstMidEndSepPunct{\mcitedefaultmidpunct}
{\mcitedefaultendpunct}{\mcitedefaultseppunct}\relax
\EndOfBibitem
\bibitem[Ahmed \emph{et~al.}(2020)Ahmed, Mu{\~n}oz, Nori, and
  Kockum]{ahmed2020quantum}
S.~Ahmed, C.~S. Mu{\~n}oz, F.~Nori and A.~F. Kockum, \emph{arXiv preprint
  arXiv:2008.03240}, 2020\relax
\mciteBstWouldAddEndPuncttrue
\mciteSetBstMidEndSepPunct{\mcitedefaultmidpunct}
{\mcitedefaultendpunct}{\mcitedefaultseppunct}\relax
\EndOfBibitem
\bibitem[Isola \emph{et~al.}(2017)Isola, Zhu, Zhou, and Efros]{isola2017image}
P.~Isola, J.-Y. Zhu, T.~Zhou and A.~A. Efros, Proceedings of the IEEE
  conference on computer vision and pattern recognition, 2017, pp.
  1125--1134\relax
\mciteBstWouldAddEndPuncttrue
\mciteSetBstMidEndSepPunct{\mcitedefaultmidpunct}
{\mcitedefaultendpunct}{\mcitedefaultseppunct}\relax
\EndOfBibitem
\bibitem[Karras \emph{et~al.}(2018)Karras, Laine, and Aila]{karras2018style}
T.~Karras, S.~Laine and T.~Aila, \emph{arXiv preprint arXiv:1812.04948},
  2018\relax
\mciteBstWouldAddEndPuncttrue
\mciteSetBstMidEndSepPunct{\mcitedefaultmidpunct}
{\mcitedefaultendpunct}{\mcitedefaultseppunct}\relax
\EndOfBibitem
\bibitem[Karras \emph{et~al.}(2020)Karras, Laine, Aittala, Hellsten, Lehtinen,
  and Aila]{karras2020analyzing}
T.~Karras, S.~Laine, M.~Aittala, J.~Hellsten, J.~Lehtinen and T.~Aila,
  Proceedings of the IEEE/CVF Conference on Computer Vision and Pattern
  Recognition, 2020, pp. 8110--8119\relax
\mciteBstWouldAddEndPuncttrue
\mciteSetBstMidEndSepPunct{\mcitedefaultmidpunct}
{\mcitedefaultendpunct}{\mcitedefaultseppunct}\relax
\EndOfBibitem
\bibitem[Yang \emph{et~al.}(2018)Yang, Hu, Dyer, Xing, and
  Berg-Kirkpatrick]{yang2018unsupervised}
Z.~Yang, Z.~Hu, C.~Dyer, E.~P. Xing and T.~Berg-Kirkpatrick, Proceedings of the
  32nd International Conference on Neural Information Processing Systems, 2018,
  pp. 7298--7309\relax
\mciteBstWouldAddEndPuncttrue
\mciteSetBstMidEndSepPunct{\mcitedefaultmidpunct}
{\mcitedefaultendpunct}{\mcitedefaultseppunct}\relax
\EndOfBibitem
\bibitem[Subramanian \emph{et~al.}(2018)Subramanian, Rajeswar, Sordoni,
  Trischler, Courville, and Pal]{subramanian2018towards}
S.~Subramanian, S.~Rajeswar, A.~Sordoni, A.~Trischler, A.~Courville and C.~Pal,
  Proceedings of the 32nd International Conference on Neural Information
  Processing Systems, 2018, pp. 7562--7574\relax
\mciteBstWouldAddEndPuncttrue
\mciteSetBstMidEndSepPunct{\mcitedefaultmidpunct}
{\mcitedefaultendpunct}{\mcitedefaultseppunct}\relax
\EndOfBibitem
\bibitem[Cheng \emph{et~al.}(2016)Cheng, Dong, and Lapata]{cheng2016long}
J.~Cheng, L.~Dong and M.~Lapata, \emph{arXiv preprint arXiv:1601.06733},
  2016\relax
\mciteBstWouldAddEndPuncttrue
\mciteSetBstMidEndSepPunct{\mcitedefaultmidpunct}
{\mcitedefaultendpunct}{\mcitedefaultseppunct}\relax
\EndOfBibitem
\bibitem[Parikh \emph{et~al.}(2016)Parikh, T{\"a}ckstr{\"o}m, Das, and
  Uszkoreit]{parikh2016decomposable}
A.~P. Parikh, O.~T{\"a}ckstr{\"o}m, D.~Das and J.~Uszkoreit, \emph{arXiv
  preprint arXiv:1606.01933}, 2016\relax
\mciteBstWouldAddEndPuncttrue
\mciteSetBstMidEndSepPunct{\mcitedefaultmidpunct}
{\mcitedefaultendpunct}{\mcitedefaultseppunct}\relax
\EndOfBibitem
\bibitem[Vaswani \emph{et~al.}(2017)Vaswani, Shazeer, Parmar, Uszkoreit, Jones,
  Gomez, Kaiser, and Polosukhin]{vaswani2017attention}
A.~Vaswani, N.~Shazeer, N.~Parmar, J.~Uszkoreit, L.~Jones, A.~N. Gomez,
  {\L}.~Kaiser and I.~Polosukhin, \emph{Advances in neural information
  processing systems}, 2017, \textbf{30}, year\relax
\mciteBstWouldAddEndPuncttrue
\mciteSetBstMidEndSepPunct{\mcitedefaultmidpunct}
{\mcitedefaultendpunct}{\mcitedefaultseppunct}\relax
\EndOfBibitem
\bibitem[Lin \emph{et~al.}(2017)Lin, Feng, Santos, Yu, Xiang, Zhou, and
  Bengio]{lin2017structured}
Z.~Lin, M.~Feng, C.~N.~d. Santos, M.~Yu, B.~Xiang, B.~Zhou and Y.~Bengio,
  \emph{arXiv preprint arXiv:1703.03130}, 2017\relax
\mciteBstWouldAddEndPuncttrue
\mciteSetBstMidEndSepPunct{\mcitedefaultmidpunct}
{\mcitedefaultendpunct}{\mcitedefaultseppunct}\relax
\EndOfBibitem
\bibitem[Paulus \emph{et~al.}(2017)Paulus, Xiong, and Socher]{paulus2017deep}
R.~Paulus, C.~Xiong and R.~Socher, \emph{arXiv preprint arXiv:1705.04304},
  2017\relax
\mciteBstWouldAddEndPuncttrue
\mciteSetBstMidEndSepPunct{\mcitedefaultmidpunct}
{\mcitedefaultendpunct}{\mcitedefaultseppunct}\relax
\EndOfBibitem
\bibitem[Bahdanau \emph{et~al.}(2014)Bahdanau, Cho, and
  Bengio]{bahdanau2014neural}
D.~Bahdanau, K.~Cho and Y.~Bengio, \emph{arXiv preprint arXiv:1409.0473},
  2014\relax
\mciteBstWouldAddEndPuncttrue
\mciteSetBstMidEndSepPunct{\mcitedefaultmidpunct}
{\mcitedefaultendpunct}{\mcitedefaultseppunct}\relax
\EndOfBibitem
\bibitem[Sutskever \emph{et~al.}(2014)Sutskever, Vinyals, and
  Le]{sutskever2014sequence}
I.~Sutskever, O.~Vinyals and Q.~V. Le, \emph{Advances in neural information
  processing systems}, 2014, \textbf{27}, year\relax
\mciteBstWouldAddEndPuncttrue
\mciteSetBstMidEndSepPunct{\mcitedefaultmidpunct}
{\mcitedefaultendpunct}{\mcitedefaultseppunct}\relax
\EndOfBibitem
\bibitem[Cha \emph{et~al.}(2021)Cha, Ginsparg, Wu, Carrasquilla, McMahon, and
  Kim]{cha2021attention}
P.~Cha, P.~Ginsparg, F.~Wu, J.~Carrasquilla, P.~L. McMahon and E.-A. Kim,
  \emph{Machine Learning: Science and Technology}, 2021, \textbf{3},
  01LT01\relax
\mciteBstWouldAddEndPuncttrue
\mciteSetBstMidEndSepPunct{\mcitedefaultmidpunct}
{\mcitedefaultendpunct}{\mcitedefaultseppunct}\relax
\EndOfBibitem
\bibitem[Carrasquilla \emph{et~al.}(2019)Carrasquilla, Torlai, Melko, and
  Aolita]{carrasquilla2019reconstructing}
J.~Carrasquilla, G.~Torlai, R.~G. Melko and L.~Aolita, \emph{Nature Machine
  Intelligence}, 2019, \textbf{1}, 155--161\relax
\mciteBstWouldAddEndPuncttrue
\mciteSetBstMidEndSepPunct{\mcitedefaultmidpunct}
{\mcitedefaultendpunct}{\mcitedefaultseppunct}\relax
\EndOfBibitem
\bibitem[Aaronson(2015)]{aaronson2015read}
S.~Aaronson, \emph{Nature Physics}, 2015, \textbf{11}, 291--293\relax
\mciteBstWouldAddEndPuncttrue
\mciteSetBstMidEndSepPunct{\mcitedefaultmidpunct}
{\mcitedefaultendpunct}{\mcitedefaultseppunct}\relax
\EndOfBibitem
\bibitem[Sen \emph{et~al.}(2020)Sen, Hajra, and Ghosh]{sen2020supervised}
P.~C. Sen, M.~Hajra and M.~Ghosh, \emph{Emerging technology in modelling and
  graphics}, Springer, 2020, pp. 99--111\relax
\mciteBstWouldAddEndPuncttrue
\mciteSetBstMidEndSepPunct{\mcitedefaultmidpunct}
{\mcitedefaultendpunct}{\mcitedefaultseppunct}\relax
\EndOfBibitem
\bibitem[Zhang \emph{et~al.}(2003)Zhang, Zhang, and Yang]{zhang2003data}
S.~Zhang, C.~Zhang and Q.~Yang, \emph{Applied artificial intelligence}, 2003,
  \textbf{17}, 375--381\relax
\mciteBstWouldAddEndPuncttrue
\mciteSetBstMidEndSepPunct{\mcitedefaultmidpunct}
{\mcitedefaultendpunct}{\mcitedefaultseppunct}\relax
\EndOfBibitem
\bibitem[Karim \emph{et~al.}(2019)Karim, Mishra, Newton, and
  Sattar]{karim2019efficient}
A.~Karim, A.~Mishra, M.~H. Newton and A.~Sattar, \emph{Acs Omega}, 2019,
  \textbf{4}, 1874--1888\relax
\mciteBstWouldAddEndPuncttrue
\mciteSetBstMidEndSepPunct{\mcitedefaultmidpunct}
{\mcitedefaultendpunct}{\mcitedefaultseppunct}\relax
\EndOfBibitem
\bibitem[Chevaleyre and Zucker(2001)]{chevaleyre2001solving}
Y.~Chevaleyre and J.-D. Zucker, Conference of the Canadian Society for
  Computational Studies of Intelligence, 2001, pp. 204--214\relax
\mciteBstWouldAddEndPuncttrue
\mciteSetBstMidEndSepPunct{\mcitedefaultmidpunct}
{\mcitedefaultendpunct}{\mcitedefaultseppunct}\relax
\EndOfBibitem
\bibitem[Skoraczy{\'n}ski \emph{et~al.}(2017)Skoraczy{\'n}ski, Dittwald,
  Miasojedow, Szymku{\'c}, Gajewska, Grzybowski, and
  Gambin]{skoraczynski2017predicting}
G.~Skoraczy{\'n}ski, P.~Dittwald, B.~Miasojedow, S.~Szymku{\'c}, E.~Gajewska,
  B.~A. Grzybowski and A.~Gambin, \emph{Scientific reports}, 2017, \textbf{7},
  1--9\relax
\mciteBstWouldAddEndPuncttrue
\mciteSetBstMidEndSepPunct{\mcitedefaultmidpunct}
{\mcitedefaultendpunct}{\mcitedefaultseppunct}\relax
\EndOfBibitem
\bibitem[Heinen \emph{et~al.}(2021)Heinen, von Rudorff, and von
  Lilienfeld]{heinen2020quantum}
S.~Heinen, G.~F. von Rudorff and O.~A. von Lilienfeld, \emph{The Journal of
  Chemical Physics}, 2021, \textbf{155}, 064105\relax
\mciteBstWouldAddEndPuncttrue
\mciteSetBstMidEndSepPunct{\mcitedefaultmidpunct}
{\mcitedefaultendpunct}{\mcitedefaultseppunct}\relax
\EndOfBibitem
\bibitem[Engel \emph{et~al.}(2019)Engel, Anelli, Hofstetter, Paruzzo, Emsley,
  and Ceriotti]{engel2019bayesian}
E.~A. Engel, A.~Anelli, A.~Hofstetter, F.~Paruzzo, L.~Emsley and M.~Ceriotti,
  \emph{Physical Chemistry Chemical Physics}, 2019, \textbf{21},
  23385--23400\relax
\mciteBstWouldAddEndPuncttrue
\mciteSetBstMidEndSepPunct{\mcitedefaultmidpunct}
{\mcitedefaultendpunct}{\mcitedefaultseppunct}\relax
\EndOfBibitem
\bibitem[Jinnouchi \emph{et~al.}(2019)Jinnouchi, Lahnsteiner, Karsai, Kresse,
  and Bokdam]{jinnouchi2019phase}
R.~Jinnouchi, J.~Lahnsteiner, F.~Karsai, G.~Kresse and M.~Bokdam,
  \emph{Physical review letters}, 2019, \textbf{122}, 225701\relax
\mciteBstWouldAddEndPuncttrue
\mciteSetBstMidEndSepPunct{\mcitedefaultmidpunct}
{\mcitedefaultendpunct}{\mcitedefaultseppunct}\relax
\EndOfBibitem
\bibitem[Krems(2019)]{krems2019bayesian}
R.~Krems, \emph{Physical Chemistry Chemical Physics}, 2019, \textbf{21},
  13392--13410\relax
\mciteBstWouldAddEndPuncttrue
\mciteSetBstMidEndSepPunct{\mcitedefaultmidpunct}
{\mcitedefaultendpunct}{\mcitedefaultseppunct}\relax
\EndOfBibitem
\bibitem[Senekane \emph{et~al.}(2016)Senekane,
  Taele,\emph{et~al.}]{senekane2016prediction}
M.~Senekane, B.~M. Taele \emph{et~al.}, \emph{Smart Grid and Renewable Energy},
  2016, \textbf{7}, 293\relax
\mciteBstWouldAddEndPuncttrue
\mciteSetBstMidEndSepPunct{\mcitedefaultmidpunct}
{\mcitedefaultendpunct}{\mcitedefaultseppunct}\relax
\EndOfBibitem
\bibitem[Heredge \emph{et~al.}(2021)Heredge, Hill, Hollenberg, and
  Sevior]{heredge2021quantum}
J.~Heredge, C.~Hill, L.~Hollenberg and M.~Sevior, \emph{arXiv preprint
  arXiv:2103.12257}, 2021\relax
\mciteBstWouldAddEndPuncttrue
\mciteSetBstMidEndSepPunct{\mcitedefaultmidpunct}
{\mcitedefaultendpunct}{\mcitedefaultseppunct}\relax
\EndOfBibitem
\bibitem[Schindler \emph{et~al.}(2017)Schindler, Regnault, and
  Neupert]{schindler2017probing}
F.~Schindler, N.~Regnault and T.~Neupert, \emph{Physical Review B}, 2017,
  \textbf{95}, 245134\relax
\mciteBstWouldAddEndPuncttrue
\mciteSetBstMidEndSepPunct{\mcitedefaultmidpunct}
{\mcitedefaultendpunct}{\mcitedefaultseppunct}\relax
\EndOfBibitem
\bibitem[Purushothaman and Karayiannis(1997)]{purushothaman1997quantum}
G.~Purushothaman and N.~B. Karayiannis, \emph{IEEE Transactions on neural
  networks}, 1997, \textbf{8}, 679--693\relax
\mciteBstWouldAddEndPuncttrue
\mciteSetBstMidEndSepPunct{\mcitedefaultmidpunct}
{\mcitedefaultendpunct}{\mcitedefaultseppunct}\relax
\EndOfBibitem
\bibitem[Zhou(2003)]{zhou2003automatic}
J.~Zhou, Third IEEE Symposium on Bioinformatics and Bioengineering, 2003.
  Proceedings., 2003, pp. 169--173\relax
\mciteBstWouldAddEndPuncttrue
\mciteSetBstMidEndSepPunct{\mcitedefaultmidpunct}
{\mcitedefaultendpunct}{\mcitedefaultseppunct}\relax
\EndOfBibitem
\bibitem[Dumoulin \emph{et~al.}(2014)Dumoulin, Goodfellow, Courville, and
  Bengio]{dumoulin2014challenges}
V.~Dumoulin, I.~Goodfellow, A.~Courville and Y.~Bengio, Proceedings of the AAAI
  Conference on Artificial Intelligence, 2014\relax
\mciteBstWouldAddEndPuncttrue
\mciteSetBstMidEndSepPunct{\mcitedefaultmidpunct}
{\mcitedefaultendpunct}{\mcitedefaultseppunct}\relax
\EndOfBibitem
\bibitem[Gandhi \emph{et~al.}(2013)Gandhi, Prasad, Coyle, Behera, and
  McGinnity]{gandhi2013quantum}
V.~Gandhi, G.~Prasad, D.~Coyle, L.~Behera and T.~M. McGinnity, \emph{IEEE
  transactions on neural networks and learning systems}, 2013, \textbf{25},
  278--288\relax
\mciteBstWouldAddEndPuncttrue
\mciteSetBstMidEndSepPunct{\mcitedefaultmidpunct}
{\mcitedefaultendpunct}{\mcitedefaultseppunct}\relax
\EndOfBibitem
\bibitem[Gao \emph{et~al.}(2018)Gao, Ma, Luo, and Liu]{gao2018ima}
Z.~Gao, C.~Ma, Y.~Luo and Z.~Liu, \emph{Engineering Applications of Artificial
  Intelligence}, 2018, \textbf{76}, 119--129\relax
\mciteBstWouldAddEndPuncttrue
\mciteSetBstMidEndSepPunct{\mcitedefaultmidpunct}
{\mcitedefaultendpunct}{\mcitedefaultseppunct}\relax
\EndOfBibitem
\bibitem[Zidan \emph{et~al.}(2019)Zidan, Abdel-Aty, El-shafei, Feraig, Al-Sbou,
  Eleuch, and Abdel-Aty]{zidan2019quantum}
M.~Zidan, A.-H. Abdel-Aty, M.~El-shafei, M.~Feraig, Y.~Al-Sbou, H.~Eleuch and
  M.~Abdel-Aty, \emph{Applied Sciences}, 2019, \textbf{9}, 1277\relax
\mciteBstWouldAddEndPuncttrue
\mciteSetBstMidEndSepPunct{\mcitedefaultmidpunct}
{\mcitedefaultendpunct}{\mcitedefaultseppunct}\relax
\EndOfBibitem
\bibitem[Farhi and Neven(2018)]{farhi2018classification}
E.~Farhi and H.~Neven, \emph{arXiv preprint arXiv:1802.06002}, 2018\relax
\mciteBstWouldAddEndPuncttrue
\mciteSetBstMidEndSepPunct{\mcitedefaultmidpunct}
{\mcitedefaultendpunct}{\mcitedefaultseppunct}\relax
\EndOfBibitem
\bibitem[Adhikary \emph{et~al.}(2020)Adhikary, Dangwal, and
  Bhowmik]{adhikary2020supervised}
S.~Adhikary, S.~Dangwal and D.~Bhowmik, \emph{Quantum Information Processing},
  2020, \textbf{19}, 1--12\relax
\mciteBstWouldAddEndPuncttrue
\mciteSetBstMidEndSepPunct{\mcitedefaultmidpunct}
{\mcitedefaultendpunct}{\mcitedefaultseppunct}\relax
\EndOfBibitem
\bibitem[Schuld \emph{et~al.}(2020)Schuld, Bocharov, Svore, and
  Wiebe]{schuld2020circuit}
M.~Schuld, A.~Bocharov, K.~M. Svore and N.~Wiebe, \emph{Physical Review A},
  2020, \textbf{101}, 032308\relax
\mciteBstWouldAddEndPuncttrue
\mciteSetBstMidEndSepPunct{\mcitedefaultmidpunct}
{\mcitedefaultendpunct}{\mcitedefaultseppunct}\relax
\EndOfBibitem
\bibitem[Liu \emph{et~al.}(2020)Liu, Bai, and Gao]{liu2020phase}
D.~Liu, G.~Bai and C.~Gao, \emph{Journal of Applied Physics}, 2020,
  \textbf{127}, 154101\relax
\mciteBstWouldAddEndPuncttrue
\mciteSetBstMidEndSepPunct{\mcitedefaultmidpunct}
{\mcitedefaultendpunct}{\mcitedefaultseppunct}\relax
\EndOfBibitem
\bibitem[Deffrennes \emph{et~al.}(2022)Deffrennes, Terayama, Abe, and
  Tamura]{deffrennes2022machine}
G.~Deffrennes, K.~Terayama, T.~Abe and R.~Tamura, \emph{arXiv preprint
  arXiv:2201.01932}, 2022\relax
\mciteBstWouldAddEndPuncttrue
\mciteSetBstMidEndSepPunct{\mcitedefaultmidpunct}
{\mcitedefaultendpunct}{\mcitedefaultseppunct}\relax
\EndOfBibitem
\bibitem[Uvarov \emph{et~al.}(2020)Uvarov, Kardashin, and
  Biamonte]{PhysRevA.102.012415}
A.~V. Uvarov, A.~S. Kardashin and J.~D. Biamonte, \emph{Phys. Rev. A}, 2020,
  \textbf{102}, 012415\relax
\mciteBstWouldAddEndPuncttrue
\mciteSetBstMidEndSepPunct{\mcitedefaultmidpunct}
{\mcitedefaultendpunct}{\mcitedefaultseppunct}\relax
\EndOfBibitem
\bibitem[Bartlett and Musia\l{}(2007)]{RevModPhys.79.291}
R.~J. Bartlett and M.~Musia\l{}, \emph{Rev. Mod. Phys.}, 2007, \textbf{79},
  291--352\relax
\mciteBstWouldAddEndPuncttrue
\mciteSetBstMidEndSepPunct{\mcitedefaultmidpunct}
{\mcitedefaultendpunct}{\mcitedefaultseppunct}\relax
\EndOfBibitem
\bibitem[Broecker \emph{et~al.}(2017)Broecker, Assaad, and
  Trebst]{broecker2017quantum}
P.~Broecker, F.~F. Assaad and S.~Trebst, \emph{arXiv preprint
  arXiv:1707.00663}, 2017\relax
\mciteBstWouldAddEndPuncttrue
\mciteSetBstMidEndSepPunct{\mcitedefaultmidpunct}
{\mcitedefaultendpunct}{\mcitedefaultseppunct}\relax
\EndOfBibitem
\bibitem[Che \emph{et~al.}(2020)Che, Gneiting, Liu, and
  Nori]{che2020topological}
Y.~Che, C.~Gneiting, T.~Liu and F.~Nori, \emph{Physical Review B}, 2020,
  \textbf{102}, 134213\relax
\mciteBstWouldAddEndPuncttrue
\mciteSetBstMidEndSepPunct{\mcitedefaultmidpunct}
{\mcitedefaultendpunct}{\mcitedefaultseppunct}\relax
\EndOfBibitem
\bibitem[Lloyd(1982)]{lloyd1982least}
S.~Lloyd, \emph{IEEE transactions on information theory}, 1982, \textbf{28},
  129--137\relax
\mciteBstWouldAddEndPuncttrue
\mciteSetBstMidEndSepPunct{\mcitedefaultmidpunct}
{\mcitedefaultendpunct}{\mcitedefaultseppunct}\relax
\EndOfBibitem
\bibitem[MacKay and Mac~Kay(2003)]{mackay2003information}
D.~J. MacKay and D.~J. Mac~Kay, \emph{Information theory, inference and
  learning algorithms}, Cambridge university press, 2003\relax
\mciteBstWouldAddEndPuncttrue
\mciteSetBstMidEndSepPunct{\mcitedefaultmidpunct}
{\mcitedefaultendpunct}{\mcitedefaultseppunct}\relax
\EndOfBibitem
\bibitem[Otterbach \emph{et~al.}(2017)Otterbach, Manenti, Alidoust, Bestwick,
  Block, Bloom, Caldwell, Didier, Fried,
  Hong,\emph{et~al.}]{otterbach2017unsupervised}
J.~Otterbach, R.~Manenti, N.~Alidoust, A.~Bestwick, M.~Block, B.~Bloom,
  S.~Caldwell, N.~Didier, E.~S. Fried, S.~Hong \emph{et~al.}, \emph{arXiv
  preprint arXiv:1712.05771}, 2017\relax
\mciteBstWouldAddEndPuncttrue
\mciteSetBstMidEndSepPunct{\mcitedefaultmidpunct}
{\mcitedefaultendpunct}{\mcitedefaultseppunct}\relax
\EndOfBibitem
\bibitem[Kohn and Sham(1965)]{kohn1965self}
W.~Kohn and L.~J. Sham, \emph{Physical review}, 1965, \textbf{140}, A1133\relax
\mciteBstWouldAddEndPuncttrue
\mciteSetBstMidEndSepPunct{\mcitedefaultmidpunct}
{\mcitedefaultendpunct}{\mcitedefaultseppunct}\relax
\EndOfBibitem
\bibitem[Schmidt \emph{et~al.}(2019)Schmidt, Benavides-Riveros, and
  Marques]{schmidt2019machine}
J.~Schmidt, C.~L. Benavides-Riveros and M.~A. Marques, \emph{The journal of
  physical chemistry letters}, 2019, \textbf{10}, 6425--6431\relax
\mciteBstWouldAddEndPuncttrue
\mciteSetBstMidEndSepPunct{\mcitedefaultmidpunct}
{\mcitedefaultendpunct}{\mcitedefaultseppunct}\relax
\EndOfBibitem
\bibitem[Bogojeski \emph{et~al.}(2020)Bogojeski, Vogt-Maranto, Tuckerman,
  M{\"u}ller, and Burke]{bogojeski2020quantum}
M.~Bogojeski, L.~Vogt-Maranto, M.~E. Tuckerman, K.-R. M{\"u}ller and K.~Burke,
  \emph{Nature communications}, 2020, \textbf{11}, 1--11\relax
\mciteBstWouldAddEndPuncttrue
\mciteSetBstMidEndSepPunct{\mcitedefaultmidpunct}
{\mcitedefaultendpunct}{\mcitedefaultseppunct}\relax
\EndOfBibitem
\bibitem[Nagai \emph{et~al.}(2020)Nagai, Akashi, and
  Sugino]{nagai2020completing}
R.~Nagai, R.~Akashi and O.~Sugino, \emph{npj Computational Materials}, 2020,
  \textbf{6}, 1--8\relax
\mciteBstWouldAddEndPuncttrue
\mciteSetBstMidEndSepPunct{\mcitedefaultmidpunct}
{\mcitedefaultendpunct}{\mcitedefaultseppunct}\relax
\EndOfBibitem
\bibitem[Li \emph{et~al.}(2016)Li, Baker, White,
  Burke,\emph{et~al.}]{li2016pure}
L.~Li, T.~E. Baker, S.~R. White, K.~Burke \emph{et~al.}, \emph{Physical Review
  B}, 2016, \textbf{94}, 245129\relax
\mciteBstWouldAddEndPuncttrue
\mciteSetBstMidEndSepPunct{\mcitedefaultmidpunct}
{\mcitedefaultendpunct}{\mcitedefaultseppunct}\relax
\EndOfBibitem
\bibitem[Borlido \emph{et~al.}(2020)Borlido, Schmidt, Huran, Tran, Marques, and
  Botti]{borlido2020exchange}
P.~Borlido, J.~Schmidt, A.~W. Huran, F.~Tran, M.~A. Marques and S.~Botti,
  \emph{npj Computational Materials}, 2020, \textbf{6}, 1--17\relax
\mciteBstWouldAddEndPuncttrue
\mciteSetBstMidEndSepPunct{\mcitedefaultmidpunct}
{\mcitedefaultendpunct}{\mcitedefaultseppunct}\relax
\EndOfBibitem
\bibitem[Fritz \emph{et~al.}(2016)Fritz, Fern{\'a}ndez-Serra, and
  Soler]{fritz2016optimization}
M.~Fritz, M.~Fern{\'a}ndez-Serra and J.~M. Soler, \emph{The Journal of chemical
  physics}, 2016, \textbf{144}, 224101\relax
\mciteBstWouldAddEndPuncttrue
\mciteSetBstMidEndSepPunct{\mcitedefaultmidpunct}
{\mcitedefaultendpunct}{\mcitedefaultseppunct}\relax
\EndOfBibitem
\bibitem[Liu \emph{et~al.}(2017)Liu, Wang, Du, Hu, Zheng, and
  Chen]{liu2017improving}
Q.~Liu, J.~Wang, P.~Du, L.~Hu, X.~Zheng and G.~Chen, \emph{The Journal of
  Physical Chemistry A}, 2017, \textbf{121}, 7273--7281\relax
\mciteBstWouldAddEndPuncttrue
\mciteSetBstMidEndSepPunct{\mcitedefaultmidpunct}
{\mcitedefaultendpunct}{\mcitedefaultseppunct}\relax
\EndOfBibitem
\bibitem[Ryczko \emph{et~al.}(2019)Ryczko, Strubbe, and
  Tamblyn]{ryczko2019deep}
K.~Ryczko, D.~A. Strubbe and I.~Tamblyn, \emph{Physical Review A}, 2019,
  \textbf{100}, 022512\relax
\mciteBstWouldAddEndPuncttrue
\mciteSetBstMidEndSepPunct{\mcitedefaultmidpunct}
{\mcitedefaultendpunct}{\mcitedefaultseppunct}\relax
\EndOfBibitem
\bibitem[Stuke \emph{et~al.}(2019)Stuke, Todorović, Rupp, Kunkel, Ghosh,
  Himanen, and Rinke]{doi:10.1063/1.5086105}
A.~Stuke, M.~Todorović, M.~Rupp, C.~Kunkel, K.~Ghosh, L.~Himanen and P.~Rinke,
  \emph{The Journal of Chemical Physics}, 2019, \textbf{150}, 204121\relax
\mciteBstWouldAddEndPuncttrue
\mciteSetBstMidEndSepPunct{\mcitedefaultmidpunct}
{\mcitedefaultendpunct}{\mcitedefaultseppunct}\relax
\EndOfBibitem
\bibitem[Hansen \emph{et~al.}(2013)Hansen, Montavon, Biegler, Fazli, Rupp,
  Scheffler, Von~Lilienfeld, Tkatchenko, and Muller]{hansen2013assessment}
K.~Hansen, G.~Montavon, F.~Biegler, S.~Fazli, M.~Rupp, M.~Scheffler, O.~A.
  Von~Lilienfeld, A.~Tkatchenko and K.-R. Muller, \emph{Journal of Chemical
  Theory and Computation}, 2013, \textbf{9}, 3404--3419\relax
\mciteBstWouldAddEndPuncttrue
\mciteSetBstMidEndSepPunct{\mcitedefaultmidpunct}
{\mcitedefaultendpunct}{\mcitedefaultseppunct}\relax
\EndOfBibitem
\bibitem[Faber \emph{et~al.}(2018)Faber, Christensen, Huang, and von
  Lilienfeld]{doi:10.1063/1.5020710}
F.~A. Faber, A.~S. Christensen, B.~Huang and O.~A. von Lilienfeld, \emph{The
  Journal of Chemical Physics}, 2018, \textbf{148}, 241717\relax
\mciteBstWouldAddEndPuncttrue
\mciteSetBstMidEndSepPunct{\mcitedefaultmidpunct}
{\mcitedefaultendpunct}{\mcitedefaultseppunct}\relax
\EndOfBibitem
\bibitem[Dral(2020)]{DRAL2020291}
P.~O. Dral, \emph{Chemical Physics and Quantum Chemistry}, Academic Press,
  2020, vol.~81, pp. 291--324\relax
\mciteBstWouldAddEndPuncttrue
\mciteSetBstMidEndSepPunct{\mcitedefaultmidpunct}
{\mcitedefaultendpunct}{\mcitedefaultseppunct}\relax
\EndOfBibitem
\bibitem[Ramakrishnan \emph{et~al.}(2015)Ramakrishnan, Dral, Rupp, and von
  Lilienfeld]{ramakrishnan2015big}
R.~Ramakrishnan, P.~O. Dral, M.~Rupp and O.~A. von Lilienfeld, \emph{Journal of
  chemical theory and computation}, 2015, \textbf{11}, 2087--2096\relax
\mciteBstWouldAddEndPuncttrue
\mciteSetBstMidEndSepPunct{\mcitedefaultmidpunct}
{\mcitedefaultendpunct}{\mcitedefaultseppunct}\relax
\EndOfBibitem
\bibitem[Rupp \emph{et~al.}(2012)Rupp, Tkatchenko, M{\"u}ller, and
  Von~Lilienfeld]{rupp2012fast}
M.~Rupp, A.~Tkatchenko, K.-R. M{\"u}ller and O.~A. Von~Lilienfeld,
  \emph{Physical review letters}, 2012, \textbf{108}, 058301\relax
\mciteBstWouldAddEndPuncttrue
\mciteSetBstMidEndSepPunct{\mcitedefaultmidpunct}
{\mcitedefaultendpunct}{\mcitedefaultseppunct}\relax
\EndOfBibitem
\bibitem[Himmetoglu(2016)]{himmetoglu2016tree}
B.~Himmetoglu, \emph{The Journal of chemical physics}, 2016, \textbf{145},
  134101\relax
\mciteBstWouldAddEndPuncttrue
\mciteSetBstMidEndSepPunct{\mcitedefaultmidpunct}
{\mcitedefaultendpunct}{\mcitedefaultseppunct}\relax
\EndOfBibitem
\bibitem[Denzel and K{\"a}stner(2018)]{denzel2018gaussian}
A.~Denzel and J.~K{\"a}stner, \emph{The Journal of Chemical Physics}, 2018,
  \textbf{148}, 094114\relax
\mciteBstWouldAddEndPuncttrue
\mciteSetBstMidEndSepPunct{\mcitedefaultmidpunct}
{\mcitedefaultendpunct}{\mcitedefaultseppunct}\relax
\EndOfBibitem
\bibitem[Choo \emph{et~al.}(2020)Choo, Mezzacapo, and
  Carleo]{choo2020fermionic}
K.~Choo, A.~Mezzacapo and G.~Carleo, \emph{Nature communications}, 2020,
  \textbf{11}, 1--7\relax
\mciteBstWouldAddEndPuncttrue
\mciteSetBstMidEndSepPunct{\mcitedefaultmidpunct}
{\mcitedefaultendpunct}{\mcitedefaultseppunct}\relax
\EndOfBibitem
\bibitem[Cao \emph{et~al.}(2019)Cao, Romero, Olson, Degroote, Johnson,
  Kieferov{\'a}, Kivlichan, Menke, Peropadre, Sawaya, Sim, Veis, and
  Aspuru-Guzik]{Cao2019QuantumCI}
Y.~Cao, J.~Romero, J.~Olson, M.~Degroote, P.~D. Johnson, M.~Kieferov{\'a},
  I.~D. Kivlichan, T.~Menke, B.~Peropadre, N.~P.~D. Sawaya, S.~Sim, L.~Veis and
  A.~Aspuru-Guzik, \emph{Chemical reviews}, 2019\relax
\mciteBstWouldAddEndPuncttrue
\mciteSetBstMidEndSepPunct{\mcitedefaultmidpunct}
{\mcitedefaultendpunct}{\mcitedefaultseppunct}\relax
\EndOfBibitem
\bibitem[Saito(2017)]{saito2017solving}
H.~Saito, \emph{Journal of the Physical Society of Japan}, 2017, \textbf{86},
  093001\relax
\mciteBstWouldAddEndPuncttrue
\mciteSetBstMidEndSepPunct{\mcitedefaultmidpunct}
{\mcitedefaultendpunct}{\mcitedefaultseppunct}\relax
\EndOfBibitem
\bibitem[Xia and Kais(2018)]{Xia_2018}
R.~Xia and S.~Kais, \emph{Nature communications}, 2018, \textbf{9}, 1--6\relax
\mciteBstWouldAddEndPuncttrue
\mciteSetBstMidEndSepPunct{\mcitedefaultmidpunct}
{\mcitedefaultendpunct}{\mcitedefaultseppunct}\relax
\EndOfBibitem
\bibitem[Kanno and Tada(2021)]{kanno2019manybody}
S.~Kanno and T.~Tada, \emph{Quantum Science and Technology}, 2021, \textbf{6},
  025015\relax
\mciteBstWouldAddEndPuncttrue
\mciteSetBstMidEndSepPunct{\mcitedefaultmidpunct}
{\mcitedefaultendpunct}{\mcitedefaultseppunct}\relax
\EndOfBibitem
\bibitem[Sureshbabu \emph{et~al.}(2021)Sureshbabu, Sajjan, Oh, and
  Kais]{sureshbabu2021implementation}
S.~H. Sureshbabu, M.~Sajjan, S.~Oh and S.~Kais, \emph{Journal of Chemical
  Information and Modeling}, 2021\relax
\mciteBstWouldAddEndPuncttrue
\mciteSetBstMidEndSepPunct{\mcitedefaultmidpunct}
{\mcitedefaultendpunct}{\mcitedefaultseppunct}\relax
\EndOfBibitem
\bibitem[Coe(2018)]{coe2018machine}
J.~P. Coe, \emph{Journal of chemical theory and computation}, 2018,
  \textbf{14}, 5739--5749\relax
\mciteBstWouldAddEndPuncttrue
\mciteSetBstMidEndSepPunct{\mcitedefaultmidpunct}
{\mcitedefaultendpunct}{\mcitedefaultseppunct}\relax
\EndOfBibitem
\bibitem[Coe(2019)]{coe2019machine}
J.~P. Coe, \emph{Journal of chemical theory and computation}, 2019,
  \textbf{15}, 6179--6189\relax
\mciteBstWouldAddEndPuncttrue
\mciteSetBstMidEndSepPunct{\mcitedefaultmidpunct}
{\mcitedefaultendpunct}{\mcitedefaultseppunct}\relax
\EndOfBibitem
\bibitem[Cust{\'o}dio \emph{et~al.}(2019)Cust{\'o}dio, Filletti, and
  Fran{\c{c}}a]{custodio2019artificial}
C.~A. Cust{\'o}dio, {\'E}.~R. Filletti and V.~V. Fran{\c{c}}a, \emph{Scientific
  reports}, 2019, \textbf{9}, 1--7\relax
\mciteBstWouldAddEndPuncttrue
\mciteSetBstMidEndSepPunct{\mcitedefaultmidpunct}
{\mcitedefaultendpunct}{\mcitedefaultseppunct}\relax
\EndOfBibitem
\bibitem[Moreno \emph{et~al.}(2020)Moreno, Carleo, and Georges]{moreno2020deep}
J.~R. Moreno, G.~Carleo and A.~Georges, \emph{Physical Review Letters}, 2020,
  \textbf{125}, 076402\relax
\mciteBstWouldAddEndPuncttrue
\mciteSetBstMidEndSepPunct{\mcitedefaultmidpunct}
{\mcitedefaultendpunct}{\mcitedefaultseppunct}\relax
\EndOfBibitem
\bibitem[Sch{\"u}tt \emph{et~al.}(2018)Sch{\"u}tt, Sauceda, Kindermans,
  Tkatchenko, and M{\"u}ller]{schutt2018schnet}
K.~T. Sch{\"u}tt, H.~E. Sauceda, P.-J. Kindermans, A.~Tkatchenko and K.-R.
  M{\"u}ller, \emph{The Journal of Chemical Physics}, 2018, \textbf{148},
  241722\relax
\mciteBstWouldAddEndPuncttrue
\mciteSetBstMidEndSepPunct{\mcitedefaultmidpunct}
{\mcitedefaultendpunct}{\mcitedefaultseppunct}\relax
\EndOfBibitem
\bibitem[Hermann \emph{et~al.}(2020)Hermann, Sch{\"a}tzle, and
  No{\'e}]{hermann2020deep}
J.~Hermann, Z.~Sch{\"a}tzle and F.~No{\'e}, \emph{Nature Chemistry}, 2020,
  \textbf{12}, 891--897\relax
\mciteBstWouldAddEndPuncttrue
\mciteSetBstMidEndSepPunct{\mcitedefaultmidpunct}
{\mcitedefaultendpunct}{\mcitedefaultseppunct}\relax
\EndOfBibitem
\bibitem[Faber \emph{et~al.}(2018)Faber, Christensen, Huang, and
  Von~Lilienfeld]{faber2018alchemical}
F.~A. Faber, A.~S. Christensen, B.~Huang and O.~A. Von~Lilienfeld, \emph{The
  Journal of chemical physics}, 2018, \textbf{148}, 241717\relax
\mciteBstWouldAddEndPuncttrue
\mciteSetBstMidEndSepPunct{\mcitedefaultmidpunct}
{\mcitedefaultendpunct}{\mcitedefaultseppunct}\relax
\EndOfBibitem
\bibitem[Christensen \emph{et~al.}(2020)Christensen, Bratholm, Faber, and
  Anatole~von Lilienfeld]{christensen2020fchl}
A.~S. Christensen, L.~A. Bratholm, F.~A. Faber and O.~Anatole~von Lilienfeld,
  \emph{The Journal of chemical physics}, 2020, \textbf{152}, 044107\relax
\mciteBstWouldAddEndPuncttrue
\mciteSetBstMidEndSepPunct{\mcitedefaultmidpunct}
{\mcitedefaultendpunct}{\mcitedefaultseppunct}\relax
\EndOfBibitem
\bibitem[Huang and von Lilienfeld(2020)]{huang2020quantum}
B.~Huang and O.~A. von Lilienfeld, \emph{Nature Chemistry}, 2020, \textbf{12},
  945--951\relax
\mciteBstWouldAddEndPuncttrue
\mciteSetBstMidEndSepPunct{\mcitedefaultmidpunct}
{\mcitedefaultendpunct}{\mcitedefaultseppunct}\relax
\EndOfBibitem
\bibitem[Pyzer-Knapp \emph{et~al.}(2015)Pyzer-Knapp, Li, and
  Aspuru-Guzik]{pyzer2015learning}
E.~O. Pyzer-Knapp, K.~Li and A.~Aspuru-Guzik, \emph{Advanced Functional
  Materials}, 2015, \textbf{25}, 6495--6502\relax
\mciteBstWouldAddEndPuncttrue
\mciteSetBstMidEndSepPunct{\mcitedefaultmidpunct}
{\mcitedefaultendpunct}{\mcitedefaultseppunct}\relax
\EndOfBibitem
\bibitem[Choudhary \emph{et~al.}(2019)Choudhary, Bercx, Jiang, Pachter, Lamoen,
  and Tavazza]{choudhary2019accelerated}
K.~Choudhary, M.~Bercx, J.~Jiang, R.~Pachter, D.~Lamoen and F.~Tavazza,
  \emph{Chemistry of Materials}, 2019, \textbf{31}, 5900--5908\relax
\mciteBstWouldAddEndPuncttrue
\mciteSetBstMidEndSepPunct{\mcitedefaultmidpunct}
{\mcitedefaultendpunct}{\mcitedefaultseppunct}\relax
\EndOfBibitem
\bibitem[Stanev \emph{et~al.}(2018)Stanev, Oses, Kusne, Rodriguez, Paglione,
  Curtarolo, and Takeuchi]{stanev2018machine}
V.~Stanev, C.~Oses, A.~G. Kusne, E.~Rodriguez, J.~Paglione, S.~Curtarolo and
  I.~Takeuchi, \emph{npj Computational Materials}, 2018, \textbf{4},
  1--14\relax
\mciteBstWouldAddEndPuncttrue
\mciteSetBstMidEndSepPunct{\mcitedefaultmidpunct}
{\mcitedefaultendpunct}{\mcitedefaultseppunct}\relax
\EndOfBibitem
\bibitem[Barrett \emph{et~al.}(2021)Barrett, Malyshev, and
  Lvovsky]{barrett2021autoregressive}
T.~D. Barrett, A.~Malyshev and A.~Lvovsky, \emph{arXiv preprint
  arXiv:2109.12606}, 2021\relax
\mciteBstWouldAddEndPuncttrue
\mciteSetBstMidEndSepPunct{\mcitedefaultmidpunct}
{\mcitedefaultendpunct}{\mcitedefaultseppunct}\relax
\EndOfBibitem
\bibitem[Balachandran \emph{et~al.}(2017)Balachandran, Young, Lookman, and
  Rondinelli]{balachandran2017learning}
P.~V. Balachandran, J.~Young, T.~Lookman and J.~M. Rondinelli, \emph{Nature
  communications}, 2017, \textbf{8}, 1--13\relax
\mciteBstWouldAddEndPuncttrue
\mciteSetBstMidEndSepPunct{\mcitedefaultmidpunct}
{\mcitedefaultendpunct}{\mcitedefaultseppunct}\relax
\EndOfBibitem
\bibitem[G{\'o}mez-Bombarelli \emph{et~al.}(2018)G{\'o}mez-Bombarelli, Wei,
  Duvenaud, Hern{\'a}ndez-Lobato, S{\'a}nchez-Lengeling, Sheberla,
  Aguilera-Iparraguirre, Hirzel, Adams, and Aspuru-Guzik]{gomez2018automatic}
R.~G{\'o}mez-Bombarelli, J.~N. Wei, D.~Duvenaud, J.~M. Hern{\'a}ndez-Lobato,
  B.~S{\'a}nchez-Lengeling, D.~Sheberla, J.~Aguilera-Iparraguirre, T.~D.
  Hirzel, R.~P. Adams and A.~Aspuru-Guzik, \emph{ACS central science}, 2018,
  \textbf{4}, 268--276\relax
\mciteBstWouldAddEndPuncttrue
\mciteSetBstMidEndSepPunct{\mcitedefaultmidpunct}
{\mcitedefaultendpunct}{\mcitedefaultseppunct}\relax
\EndOfBibitem
\bibitem[Kim \emph{et~al.}(2020)Kim, Lee, and Kim]{kim2020inverse}
B.~Kim, S.~Lee and J.~Kim, \emph{Science advances}, 2020, \textbf{6},
  eaax9324\relax
\mciteBstWouldAddEndPuncttrue
\mciteSetBstMidEndSepPunct{\mcitedefaultmidpunct}
{\mcitedefaultendpunct}{\mcitedefaultseppunct}\relax
\EndOfBibitem
\bibitem[Zhang \emph{et~al.}(2020)Zhang, Li, Flores, and
  Mishra]{zhang2020machine}
Z.~Zhang, M.~Li, K.~Flores and R.~Mishra, \emph{Journal of Applied Physics},
  2020, \textbf{128}, 105103\relax
\mciteBstWouldAddEndPuncttrue
\mciteSetBstMidEndSepPunct{\mcitedefaultmidpunct}
{\mcitedefaultendpunct}{\mcitedefaultseppunct}\relax
\EndOfBibitem
\bibitem[Mazhnik and Oganov(2020)]{mazhnik2020application}
E.~Mazhnik and A.~R. Oganov, \emph{Journal of Applied Physics}, 2020,
  \textbf{128}, 075102\relax
\mciteBstWouldAddEndPuncttrue
\mciteSetBstMidEndSepPunct{\mcitedefaultmidpunct}
{\mcitedefaultendpunct}{\mcitedefaultseppunct}\relax
\EndOfBibitem
\bibitem[Bruognolo(2017)]{bruognolo2017tensor}
B.~Bruognolo, \emph{PhD thesis}, lmu, 2017\relax
\mciteBstWouldAddEndPuncttrue
\mciteSetBstMidEndSepPunct{\mcitedefaultmidpunct}
{\mcitedefaultendpunct}{\mcitedefaultseppunct}\relax
\EndOfBibitem
\bibitem[Kshetrimayum \emph{et~al.}(2020)Kshetrimayum, Balz, Lake, and
  Eisert]{KSHETRIMAYUM2020168292}
A.~Kshetrimayum, C.~Balz, B.~Lake and J.~Eisert, \emph{Annals of Physics},
  2020, \textbf{421}, 168292\relax
\mciteBstWouldAddEndPuncttrue
\mciteSetBstMidEndSepPunct{\mcitedefaultmidpunct}
{\mcitedefaultendpunct}{\mcitedefaultseppunct}\relax
\EndOfBibitem
\bibitem[Balz \emph{et~al.}(2016)Balz, Lake, Reuther, Luetkens, Sch{\"o}nemann,
  Herrmannsd{\"o}rfer, Singh, Nazmul~Islam, Wheeler,
  Rodriguez-Rivera,\emph{et~al.}]{balz2016physical}
C.~Balz, B.~Lake, J.~Reuther, H.~Luetkens, R.~Sch{\"o}nemann,
  T.~Herrmannsd{\"o}rfer, Y.~Singh, A.~Nazmul~Islam, E.~M. Wheeler, J.~A.
  Rodriguez-Rivera \emph{et~al.}, \emph{Nature Physics}, 2016, \textbf{12},
  942--949\relax
\mciteBstWouldAddEndPuncttrue
\mciteSetBstMidEndSepPunct{\mcitedefaultmidpunct}
{\mcitedefaultendpunct}{\mcitedefaultseppunct}\relax
\EndOfBibitem
\bibitem[Biamonte(2019)]{Biamonte2019-cp}
J.~Biamonte, \emph{arXiv preprint arXiv:1912.10049}, 2019\relax
\mciteBstWouldAddEndPuncttrue
\mciteSetBstMidEndSepPunct{\mcitedefaultmidpunct}
{\mcitedefaultendpunct}{\mcitedefaultseppunct}\relax
\EndOfBibitem
\bibitem[Murg \emph{et~al.}(2010)Murg, Verstraete, Legeza, and
  Noack]{murg2010simulating}
V.~Murg, F.~Verstraete, {\"O}.~Legeza and R.~M. Noack, \emph{Physical Review
  B}, 2010, \textbf{82}, 205105\relax
\mciteBstWouldAddEndPuncttrue
\mciteSetBstMidEndSepPunct{\mcitedefaultmidpunct}
{\mcitedefaultendpunct}{\mcitedefaultseppunct}\relax
\EndOfBibitem
\bibitem[Barthel \emph{et~al.}(2009)Barthel, Pineda, and
  Eisert]{barthel2009contraction}
T.~Barthel, C.~Pineda and J.~Eisert, \emph{Physical Review A}, 2009,
  \textbf{80}, 042333\relax
\mciteBstWouldAddEndPuncttrue
\mciteSetBstMidEndSepPunct{\mcitedefaultmidpunct}
{\mcitedefaultendpunct}{\mcitedefaultseppunct}\relax
\EndOfBibitem
\bibitem[Wille \emph{et~al.}(2017)Wille, Buerschaper, and Eisert]{Wille2017-dc}
C.~Wille, O.~Buerschaper and J.~Eisert, \emph{Physical Review B}, 2017,
  \textbf{95}, 245127\relax
\mciteBstWouldAddEndPuncttrue
\mciteSetBstMidEndSepPunct{\mcitedefaultmidpunct}
{\mcitedefaultendpunct}{\mcitedefaultseppunct}\relax
\EndOfBibitem
\bibitem[Krumnow \emph{et~al.}(2016)Krumnow, Veis, Legeza, and
  Eisert]{krumnow2016fermionic}
C.~Krumnow, L.~Veis, {\"O}.~Legeza and J.~Eisert, \emph{Physical review
  letters}, 2016, \textbf{117}, 210402\relax
\mciteBstWouldAddEndPuncttrue
\mciteSetBstMidEndSepPunct{\mcitedefaultmidpunct}
{\mcitedefaultendpunct}{\mcitedefaultseppunct}\relax
\EndOfBibitem
\bibitem[Gogolin and Eisert(2016)]{gogolin2016equilibration}
C.~Gogolin and J.~Eisert, \emph{Reports on Progress in Physics}, 2016,
  \textbf{79}, 056001\relax
\mciteBstWouldAddEndPuncttrue
\mciteSetBstMidEndSepPunct{\mcitedefaultmidpunct}
{\mcitedefaultendpunct}{\mcitedefaultseppunct}\relax
\EndOfBibitem
\bibitem[Kshetrimayum \emph{et~al.}(2019)Kshetrimayum, Rizzi, Eisert, and
  Or{\'u}s]{kshetrimayum2019tensor}
A.~Kshetrimayum, M.~Rizzi, J.~Eisert and R.~Or{\'u}s, \emph{Physical Review
  Letters}, 2019, \textbf{122}, 070502\relax
\mciteBstWouldAddEndPuncttrue
\mciteSetBstMidEndSepPunct{\mcitedefaultmidpunct}
{\mcitedefaultendpunct}{\mcitedefaultseppunct}\relax
\EndOfBibitem
\bibitem[Kshetrimayum \emph{et~al.}(2020)Kshetrimayum, Goihl, and
  Eisert]{kshetrimayum2020time}
A.~Kshetrimayum, M.~Goihl and J.~Eisert, \emph{Physical Review B}, 2020,
  \textbf{102}, 235132\relax
\mciteBstWouldAddEndPuncttrue
\mciteSetBstMidEndSepPunct{\mcitedefaultmidpunct}
{\mcitedefaultendpunct}{\mcitedefaultseppunct}\relax
\EndOfBibitem
\bibitem[Haghshenas \emph{et~al.}(2022)Haghshenas, Gray, Potter, and
  Chan]{Haghshenas2022-db}
R.~Haghshenas, J.~Gray, A.~C. Potter and G.~K.-L. Chan, \emph{Variational Power
  of Quantum Circuit Tensor Networks}, 2022\relax
\mciteBstWouldAddEndPuncttrue
\mciteSetBstMidEndSepPunct{\mcitedefaultmidpunct}
{\mcitedefaultendpunct}{\mcitedefaultseppunct}\relax
\EndOfBibitem
\bibitem[Shewchuk
  \emph{et~al.}(1994)Shewchuk\emph{et~al.}]{shewchuk1994introduction}
J.~R. Shewchuk \emph{et~al.}, \emph{An introduction to the conjugate gradient
  method without the agonizing pain}, 1994\relax
\mciteBstWouldAddEndPuncttrue
\mciteSetBstMidEndSepPunct{\mcitedefaultmidpunct}
{\mcitedefaultendpunct}{\mcitedefaultseppunct}\relax
\EndOfBibitem
\bibitem[Schraudolph \emph{et~al.}(2007)Schraudolph, Yu, and
  G{\"u}nter]{schraudolph2007stochastic}
N.~N. Schraudolph, J.~Yu and S.~G{\"u}nter, Artificial intelligence and
  statistics, 2007, pp. 436--443\relax
\mciteBstWouldAddEndPuncttrue
\mciteSetBstMidEndSepPunct{\mcitedefaultmidpunct}
{\mcitedefaultendpunct}{\mcitedefaultseppunct}\relax
\EndOfBibitem
\bibitem[Eddins \emph{et~al.}(2022)Eddins, Motta, Gujarati, Bravyi, Mezzacapo,
  Hadfield, and Sheldon]{eddins2022doubling}
A.~Eddins, M.~Motta, T.~P. Gujarati, S.~Bravyi, A.~Mezzacapo, C.~Hadfield and
  S.~Sheldon, \emph{PRX Quantum}, 2022, \textbf{3}, 010309\relax
\mciteBstWouldAddEndPuncttrue
\mciteSetBstMidEndSepPunct{\mcitedefaultmidpunct}
{\mcitedefaultendpunct}{\mcitedefaultseppunct}\relax
\EndOfBibitem
\bibitem[McCammon \emph{et~al.}(1977)McCammon, Gelin, and
  Karplus]{mccammon1977dynamics}
J.~A. McCammon, B.~R. Gelin and M.~Karplus, \emph{Nature}, 1977, \textbf{267},
  585--590\relax
\mciteBstWouldAddEndPuncttrue
\mciteSetBstMidEndSepPunct{\mcitedefaultmidpunct}
{\mcitedefaultendpunct}{\mcitedefaultseppunct}\relax
\EndOfBibitem
\bibitem[Schulz \emph{et~al.}(2009)Schulz, Lindner, Petridis, and
  Smith]{schulz2009scaling}
R.~Schulz, B.~Lindner, L.~Petridis and J.~C. Smith, \emph{J. Chem. Theory
  Comput.}, 2009, \textbf{5}, 2798--2808\relax
\mciteBstWouldAddEndPuncttrue
\mciteSetBstMidEndSepPunct{\mcitedefaultmidpunct}
{\mcitedefaultendpunct}{\mcitedefaultseppunct}\relax
\EndOfBibitem
\bibitem[Shaw \emph{et~al.}(2008)Shaw, Deneroff, Dror, Kuskin, Larson, Salmon,
  Young, Batson, Bowers, Chao,\emph{et~al.}]{shaw2008anton}
D.~E. Shaw, M.~M. Deneroff, R.~O. Dror, J.~S. Kuskin, R.~H. Larson, J.~K.
  Salmon, C.~Young, B.~Batson, K.~J. Bowers, J.~C. Chao \emph{et~al.},
  \emph{Commun. ACM}, 2008, \textbf{51}, 91--97\relax
\mciteBstWouldAddEndPuncttrue
\mciteSetBstMidEndSepPunct{\mcitedefaultmidpunct}
{\mcitedefaultendpunct}{\mcitedefaultseppunct}\relax
\EndOfBibitem
\bibitem[MacKerell(2007)]{mackerell2007empirical}
A.~D. MacKerell, \emph{Computational Methods for Protein Structure Prediction
  and Modeling}, Springer, 2007, pp. 45--69\relax
\mciteBstWouldAddEndPuncttrue
\mciteSetBstMidEndSepPunct{\mcitedefaultmidpunct}
{\mcitedefaultendpunct}{\mcitedefaultseppunct}\relax
\EndOfBibitem
\bibitem[Zimmerman and Bowman(2021)]{zimmerman2021sars}
M.~I. Zimmerman and G.~Bowman, \emph{Biophysical Journal}, 2021, \textbf{120},
  299a\relax
\mciteBstWouldAddEndPuncttrue
\mciteSetBstMidEndSepPunct{\mcitedefaultmidpunct}
{\mcitedefaultendpunct}{\mcitedefaultseppunct}\relax
\EndOfBibitem
\bibitem[Gonz~{\ 'a} lez(2011)]{gonzalez2011force}
M.~Gonz~{\ 'a} lez, \emph{{\ 'E} th {\' e} matic school of the Soci {\ 'e} t
  {\' e} Fran {\ c {c}} ease of Neutronics}, 2011, \textbf{12}, 169--200\relax
\mciteBstWouldAddEndPuncttrue
\mciteSetBstMidEndSepPunct{\mcitedefaultmidpunct}
{\mcitedefaultendpunct}{\mcitedefaultseppunct}\relax
\EndOfBibitem
\bibitem[MacKerell~Jr(2004)]{mackerell2004empirical}
A.~D. MacKerell~Jr, \emph{Journal of computational chemistry}, 2004,
  \textbf{25}, 1584--1604\relax
\mciteBstWouldAddEndPuncttrue
\mciteSetBstMidEndSepPunct{\mcitedefaultmidpunct}
{\mcitedefaultendpunct}{\mcitedefaultseppunct}\relax
\EndOfBibitem
\bibitem[Shi \emph{et~al.}(2013)Shi, Xia, Zhang, Best, Wu, Ponder, and
  Ren]{shi2013polarizable}
Y.~Shi, Z.~Xia, J.~Zhang, R.~Best, C.~Wu, J.~W. Ponder and P.~Ren,
  \emph{Journal of chemical theory and computation}, 2013, \textbf{9},
  4046--4063\relax
\mciteBstWouldAddEndPuncttrue
\mciteSetBstMidEndSepPunct{\mcitedefaultmidpunct}
{\mcitedefaultendpunct}{\mcitedefaultseppunct}\relax
\EndOfBibitem
\bibitem[Warshel \emph{et~al.}(2007)Warshel, Kato, and
  Pisliakov]{warshel2007polarizable}
A.~Warshel, M.~Kato and A.~V. Pisliakov, \emph{Journal of Chemical Theory and
  Computation}, 2007, \textbf{3}, 2034--2045\relax
\mciteBstWouldAddEndPuncttrue
\mciteSetBstMidEndSepPunct{\mcitedefaultmidpunct}
{\mcitedefaultendpunct}{\mcitedefaultseppunct}\relax
\EndOfBibitem
\bibitem[Unke \emph{et~al.}(2017)Unke, Devereux, and Meuwly]{unke2017minimal}
O.~T. Unke, M.~Devereux and M.~Meuwly, \emph{The Journal of chemical physics},
  2017, \textbf{147}, 161712\relax
\mciteBstWouldAddEndPuncttrue
\mciteSetBstMidEndSepPunct{\mcitedefaultmidpunct}
{\mcitedefaultendpunct}{\mcitedefaultseppunct}\relax
\EndOfBibitem
\bibitem[Nagy \emph{et~al.}(2014)Nagy, Yosa~Reyes, and
  Meuwly]{nagy2014multisurface}
T.~Nagy, J.~Yosa~Reyes and M.~Meuwly, \emph{Journal of chemical theory and
  computation}, 2014, \textbf{10}, 1366--1375\relax
\mciteBstWouldAddEndPuncttrue
\mciteSetBstMidEndSepPunct{\mcitedefaultmidpunct}
{\mcitedefaultendpunct}{\mcitedefaultseppunct}\relax
\EndOfBibitem
\bibitem[Senn and Thiel({2009})]{senn:2009}
H.~M. Senn and W.~Thiel, \emph{anie}, {2009}, \textbf{{48}}, {1198--1229}\relax
\mciteBstWouldAddEndPuncttrue
\mciteSetBstMidEndSepPunct{\mcitedefaultmidpunct}
{\mcitedefaultendpunct}{\mcitedefaultseppunct}\relax
\EndOfBibitem
\bibitem[Chmiela \emph{et~al.}(2019)Chmiela, Sauceda, Poltavsky, Muller, and
  Tkatchenko]{chmiela2019sgdml}
S.~Chmiela, H.~E. Sauceda, I.~Poltavsky, K.-R. Muller and A.~Tkatchenko,
  \emph{Computer Physics Communications}, 2019, \textbf{240}, 38--45\relax
\mciteBstWouldAddEndPuncttrue
\mciteSetBstMidEndSepPunct{\mcitedefaultmidpunct}
{\mcitedefaultendpunct}{\mcitedefaultseppunct}\relax
\EndOfBibitem
\bibitem[Unke and Meuwly(2019)]{unke2019physnet}
O.~T. Unke and M.~Meuwly, \emph{Journal of chemical theory and computation},
  2019, \textbf{15}, 3678--3693\relax
\mciteBstWouldAddEndPuncttrue
\mciteSetBstMidEndSepPunct{\mcitedefaultmidpunct}
{\mcitedefaultendpunct}{\mcitedefaultseppunct}\relax
\EndOfBibitem
\bibitem[Bartok \emph{et~al.}(2010)Bartok, Payne, Kondor, and
  Csanyi]{bartok2010gaussian}
A.~P. Bartok, M.~C. Payne, R.~Kondor and G.~Csanyi, \emph{Physical review
  letters}, 2010, \textbf{104}, 136403\relax
\mciteBstWouldAddEndPuncttrue
\mciteSetBstMidEndSepPunct{\mcitedefaultmidpunct}
{\mcitedefaultendpunct}{\mcitedefaultseppunct}\relax
\EndOfBibitem
\bibitem[Schutt \emph{et~al.}(2018)Schutt, Kessel, Gastegger, Nicoli,
  Tkatchenko, and Muller]{schutt2018schnetpack}
K.~Schutt, P.~Kessel, M.~Gastegger, K.~Nicoli, A.~Tkatchenko and K.-R. Muller,
  \emph{Journal of chemical theory and computation}, 2018, \textbf{15},
  448--455\relax
\mciteBstWouldAddEndPuncttrue
\mciteSetBstMidEndSepPunct{\mcitedefaultmidpunct}
{\mcitedefaultendpunct}{\mcitedefaultseppunct}\relax
\EndOfBibitem
\bibitem[Behler and Parrinello(2007)]{behler2007generalized}
J.~Behler and M.~Parrinello, \emph{Physical review letters}, 2007, \textbf{98},
  146401\relax
\mciteBstWouldAddEndPuncttrue
\mciteSetBstMidEndSepPunct{\mcitedefaultmidpunct}
{\mcitedefaultendpunct}{\mcitedefaultseppunct}\relax
\EndOfBibitem
\bibitem[Smith \emph{et~al.}(2017)Smith, Isayev, and Roitberg]{smith2017ani}
J.~S. Smith, O.~Isayev and A.~E. Roitberg, \emph{Chemical science}, 2017,
  \textbf{8}, 3192--3203\relax
\mciteBstWouldAddEndPuncttrue
\mciteSetBstMidEndSepPunct{\mcitedefaultmidpunct}
{\mcitedefaultendpunct}{\mcitedefaultseppunct}\relax
\EndOfBibitem
\bibitem[Zhang \emph{et~al.}(2018)Zhang, Han, Wang, Car, and
  Weinan]{zhang2018deep}
L.~Zhang, J.~Han, H.~Wang, R.~Car and E.~Weinan, \emph{Physical review
  letters}, 2018, \textbf{120}, 143001\relax
\mciteBstWouldAddEndPuncttrue
\mciteSetBstMidEndSepPunct{\mcitedefaultmidpunct}
{\mcitedefaultendpunct}{\mcitedefaultseppunct}\relax
\EndOfBibitem
\bibitem[Westermayr and Marquetand(2020)]{westermayr2020machine}
J.~Westermayr and P.~Marquetand, \emph{Chemical Reviews}, 2020, \textbf{121},
  9873--9926\relax
\mciteBstWouldAddEndPuncttrue
\mciteSetBstMidEndSepPunct{\mcitedefaultmidpunct}
{\mcitedefaultendpunct}{\mcitedefaultseppunct}\relax
\EndOfBibitem
\bibitem[Schutt \emph{et~al.}(2017)Schutt, Arbabzadah, Chmiela, Muller, and
  Tkatchenko]{schutt2017quantum}
K.~T. Schutt, F.~Arbabzadah, S.~Chmiela, K.~R. Muller and A.~Tkatchenko,
  \emph{Nature communications}, 2017, \textbf{8}, 1--8\relax
\mciteBstWouldAddEndPuncttrue
\mciteSetBstMidEndSepPunct{\mcitedefaultmidpunct}
{\mcitedefaultendpunct}{\mcitedefaultseppunct}\relax
\EndOfBibitem
\bibitem[Unke \emph{et~al.}(2021)Unke, Chmiela, Sauceda, Gastegger, Poltavsky,
  Schutt, Tkatchenko, and Muller]{unke2021machine}
O.~T. Unke, S.~Chmiela, H.~E. Sauceda, M.~Gastegger, I.~Poltavsky, K.~T.
  Schutt, A.~Tkatchenko and K.-R. Muller, \emph{Chemical Reviews}, 2021\relax
\mciteBstWouldAddEndPuncttrue
\mciteSetBstMidEndSepPunct{\mcitedefaultmidpunct}
{\mcitedefaultendpunct}{\mcitedefaultseppunct}\relax
\EndOfBibitem
\bibitem[Van~Duin \emph{et~al.}(2001)Van~Duin, Dasgupta, Lorant, and
  Goddard]{van2001reaxff}
A.~C. Van~Duin, S.~Dasgupta, F.~Lorant and W.~A. Goddard, \emph{The Journal of
  Physical Chemistry A}, 2001, \textbf{105}, 9396--9409\relax
\mciteBstWouldAddEndPuncttrue
\mciteSetBstMidEndSepPunct{\mcitedefaultmidpunct}
{\mcitedefaultendpunct}{\mcitedefaultseppunct}\relax
\EndOfBibitem
\bibitem[Nakata and Bai(2019)]{nakata2019development}
H.~Nakata and S.~Bai, \emph{Journal of computational chemistry}, 2019,
  \textbf{40}, 2000--2012\relax
\mciteBstWouldAddEndPuncttrue
\mciteSetBstMidEndSepPunct{\mcitedefaultmidpunct}
{\mcitedefaultendpunct}{\mcitedefaultseppunct}\relax
\EndOfBibitem
\bibitem[Angibaud \emph{et~al.}(2011)Angibaud, Briquet, Philipp, Wirtz, and
  Kieffer]{angibaud2011parameter}
L.~Angibaud, L.~Briquet, P.~Philipp, T.~Wirtz and J.~Kieffer, \emph{Nuclear
  Instruments and Methods in Physics Research Section B: Beam Interactions with
  Materials and Atoms}, 2011, \textbf{269}, 1559--1563\relax
\mciteBstWouldAddEndPuncttrue
\mciteSetBstMidEndSepPunct{\mcitedefaultmidpunct}
{\mcitedefaultendpunct}{\mcitedefaultseppunct}\relax
\EndOfBibitem
\bibitem[Dittner \emph{et~al.}(2015)Dittner, Muller, Aktulga, and
  Hartke]{dittner2015efficient}
M.~Dittner, J.~Muller, H.~M. Aktulga and B.~Hartke, \emph{Journal of
  computational chemistry}, 2015, \textbf{36}, 1550--1561\relax
\mciteBstWouldAddEndPuncttrue
\mciteSetBstMidEndSepPunct{\mcitedefaultmidpunct}
{\mcitedefaultendpunct}{\mcitedefaultseppunct}\relax
\EndOfBibitem
\bibitem[Jaramillo-Botero \emph{et~al.}(2014)Jaramillo-Botero, Naserifar, and
  Goddard~III]{jaramillo2014general}
A.~Jaramillo-Botero, S.~Naserifar and W.~A. Goddard~III, \emph{Journal of
  Chemical Theory and Computation}, 2014, \textbf{10}, 1426--1439\relax
\mciteBstWouldAddEndPuncttrue
\mciteSetBstMidEndSepPunct{\mcitedefaultmidpunct}
{\mcitedefaultendpunct}{\mcitedefaultseppunct}\relax
\EndOfBibitem
\bibitem[Kaymak \emph{et~al.}(2021)Kaymak, Rahnamoun, O'Hearn, van Duin,
  Merz~Jr, and Aktulga]{kaymak2021jax}
M.~C. Kaymak, A.~Rahnamoun, K.~A. O'Hearn, A.~C. van Duin, K.~M. Merz~Jr and
  H.~M. Aktulga, 2021\relax
\mciteBstWouldAddEndPuncttrue
\mciteSetBstMidEndSepPunct{\mcitedefaultmidpunct}
{\mcitedefaultendpunct}{\mcitedefaultseppunct}\relax
\EndOfBibitem
\bibitem[Akimov and Prezhdo(2013)]{akimov2013pyxaid}
A.~V. Akimov and O.~V. Prezhdo, \emph{Journal of chemical theory and
  computation}, 2013, \textbf{9}, 4959--4972\relax
\mciteBstWouldAddEndPuncttrue
\mciteSetBstMidEndSepPunct{\mcitedefaultmidpunct}
{\mcitedefaultendpunct}{\mcitedefaultseppunct}\relax
\EndOfBibitem
\bibitem[Akimov and Prezhdo(2014)]{akimov2014advanced}
A.~V. Akimov and O.~V. Prezhdo, \emph{Journal of chemical theory and
  computation}, 2014, \textbf{10}, 789--804\relax
\mciteBstWouldAddEndPuncttrue
\mciteSetBstMidEndSepPunct{\mcitedefaultmidpunct}
{\mcitedefaultendpunct}{\mcitedefaultseppunct}\relax
\EndOfBibitem
\bibitem[Nijjar \emph{et~al.}(2019)Nijjar, Jankowska, and
  Prezhdo]{nijjar2019ehrenfest}
P.~Nijjar, J.~Jankowska and O.~V. Prezhdo, \emph{The Journal of Chemical
  Physics}, 2019, \textbf{150}, 204124\relax
\mciteBstWouldAddEndPuncttrue
\mciteSetBstMidEndSepPunct{\mcitedefaultmidpunct}
{\mcitedefaultendpunct}{\mcitedefaultseppunct}\relax
\EndOfBibitem
\bibitem[Dral \emph{et~al.}(2018)Dral, Barbatti, and
  Thiel]{dral2018nonadiabatic}
P.~O. Dral, M.~Barbatti and W.~Thiel, \emph{The journal of physical chemistry
  letters}, 2018, \textbf{9}, 5660--5663\relax
\mciteBstWouldAddEndPuncttrue
\mciteSetBstMidEndSepPunct{\mcitedefaultmidpunct}
{\mcitedefaultendpunct}{\mcitedefaultseppunct}\relax
\EndOfBibitem
\bibitem[Wang \emph{et~al.}(2021)Wang, Chu, Tkatchenko, and
  Prezhdo]{wang2021interpolating}
B.~Wang, W.~Chu, A.~Tkatchenko and O.~V. Prezhdo, \emph{The Journal of Physical
  Chemistry Letters}, 2021, \textbf{12}, 6070--6077\relax
\mciteBstWouldAddEndPuncttrue
\mciteSetBstMidEndSepPunct{\mcitedefaultmidpunct}
{\mcitedefaultendpunct}{\mcitedefaultseppunct}\relax
\EndOfBibitem
\bibitem[Zhang \emph{et~al.}(2021)Zhang, Zhang, Wang, Xu, and
  Long]{zhang2021doping}
Z.~Zhang, Y.~Zhang, J.~Wang, J.~Xu and R.~Long, \emph{The Journal of Physical
  Chemistry Letters}, 2021, \textbf{12}, 835--842\relax
\mciteBstWouldAddEndPuncttrue
\mciteSetBstMidEndSepPunct{\mcitedefaultmidpunct}
{\mcitedefaultendpunct}{\mcitedefaultseppunct}\relax
\EndOfBibitem
\bibitem[Li \emph{et~al.}(2021)Li, She, Vasenko, and Prezhdo]{li2021ab}
W.~Li, Y.~She, A.~S. Vasenko and O.~V. Prezhdo, \emph{Nanoscale}, 2021,
  \textbf{13}, 10239--10265\relax
\mciteBstWouldAddEndPuncttrue
\mciteSetBstMidEndSepPunct{\mcitedefaultmidpunct}
{\mcitedefaultendpunct}{\mcitedefaultseppunct}\relax
\EndOfBibitem
\bibitem[Zhang \emph{et~al.}(2021)Zhang, Chu, Zhao, Zheng, Prezhdo, and
  Zhao]{zhang2021dynamics}
L.~Zhang, W.~Chu, C.~Zhao, Q.~Zheng, O.~V. Prezhdo and J.~Zhao, \emph{The
  Journal of Physical Chemistry Letters}, 2021, \textbf{12}, 2191--2198\relax
\mciteBstWouldAddEndPuncttrue
\mciteSetBstMidEndSepPunct{\mcitedefaultmidpunct}
{\mcitedefaultendpunct}{\mcitedefaultseppunct}\relax
\EndOfBibitem
\bibitem[Olson \emph{et~al.}(2021)Olson, Sales, Tomko, Lu, Prezhdo, McDonnell,
  and Hopkins]{olson2021band}
D.~H. Olson, M.~G. Sales, J.~A. Tomko, T.-F. Lu, O.~V. Prezhdo, S.~J. McDonnell
  and P.~E. Hopkins, \emph{Applied Physics Letters}, 2021, \textbf{118},
  163503\relax
\mciteBstWouldAddEndPuncttrue
\mciteSetBstMidEndSepPunct{\mcitedefaultmidpunct}
{\mcitedefaultendpunct}{\mcitedefaultseppunct}\relax
\EndOfBibitem
\bibitem[Westermayr \emph{et~al.}(2020)Westermayr, Gastegger, and
  Marquetand]{westermayr2020combining}
J.~Westermayr, M.~Gastegger and P.~Marquetand, \emph{The journal of physical
  chemistry letters}, 2020, \textbf{11}, 3828--3834\relax
\mciteBstWouldAddEndPuncttrue
\mciteSetBstMidEndSepPunct{\mcitedefaultmidpunct}
{\mcitedefaultendpunct}{\mcitedefaultseppunct}\relax
\EndOfBibitem
\bibitem[Posenitskiy \emph{et~al.}(2021)Posenitskiy, Spiegelman, and
  Lemoine]{posenitskiy2021application}
E.~Posenitskiy, F.~Spiegelman and D.~Lemoine, \emph{Machine Learning: Science
  and Technology}, 2021, \textbf{2}, 035039\relax
\mciteBstWouldAddEndPuncttrue
\mciteSetBstMidEndSepPunct{\mcitedefaultmidpunct}
{\mcitedefaultendpunct}{\mcitedefaultseppunct}\relax
\EndOfBibitem
\bibitem[Glielmo \emph{et~al.}(2021)Glielmo, Husic, Rodriguez, Clementi,
  No{\'e}, and Laio]{glielmo2021unsupervised}
A.~Glielmo, B.~E. Husic, A.~Rodriguez, C.~Clementi, F.~No{\'e} and A.~Laio,
  \emph{Chemical Reviews}, 2021, \textbf{121}, 9722--9758\relax
\mciteBstWouldAddEndPuncttrue
\mciteSetBstMidEndSepPunct{\mcitedefaultmidpunct}
{\mcitedefaultendpunct}{\mcitedefaultseppunct}\relax
\EndOfBibitem
\bibitem[Virshup \emph{et~al.}(2012)Virshup, Chen, and
  Mart{\'\i}nez]{virshup2012nonlinear}
A.~M. Virshup, J.~Chen and T.~J. Mart{\'\i}nez, \emph{The Journal of chemical
  physics}, 2012, \textbf{137}, 22A519\relax
\mciteBstWouldAddEndPuncttrue
\mciteSetBstMidEndSepPunct{\mcitedefaultmidpunct}
{\mcitedefaultendpunct}{\mcitedefaultseppunct}\relax
\EndOfBibitem
\bibitem[Li \emph{et~al.}(2017)Li, Xie, Hu, and Lan]{li2017analysis}
X.~Li, Y.~Xie, D.~Hu and Z.~Lan, \emph{Journal of chemical theory and
  computation}, 2017, \textbf{13}, 4611--4623\relax
\mciteBstWouldAddEndPuncttrue
\mciteSetBstMidEndSepPunct{\mcitedefaultmidpunct}
{\mcitedefaultendpunct}{\mcitedefaultseppunct}\relax
\EndOfBibitem
\bibitem[Peng \emph{et~al.}(2021)Peng, Xie, Hu, and Lan]{peng2021analysis}
J.~Peng, Y.~Xie, D.~Hu and Z.~Lan, \emph{The Journal of Chemical Physics},
  2021, \textbf{154}, 094122\relax
\mciteBstWouldAddEndPuncttrue
\mciteSetBstMidEndSepPunct{\mcitedefaultmidpunct}
{\mcitedefaultendpunct}{\mcitedefaultseppunct}\relax
\EndOfBibitem
\bibitem[Zhou \emph{et~al.}(2020)Zhou, Chu, and Prezhdo]{zhou2020structural}
G.~Zhou, W.~Chu and O.~V. Prezhdo, \emph{ACS Energy Letters}, 2020, \textbf{5},
  1930--1938\relax
\mciteBstWouldAddEndPuncttrue
\mciteSetBstMidEndSepPunct{\mcitedefaultmidpunct}
{\mcitedefaultendpunct}{\mcitedefaultseppunct}\relax
\EndOfBibitem
\bibitem[Tavadze \emph{et~al.}(2018)Tavadze, Avendano~Franco, Ren, Wen, Li, and
  Lewis]{tavadze2018machine}
P.~Tavadze, G.~Avendano~Franco, P.~Ren, X.~Wen, Y.~Li and J.~P. Lewis,
  \emph{Journal of the American Chemical Society}, 2018, \textbf{140},
  285--290\relax
\mciteBstWouldAddEndPuncttrue
\mciteSetBstMidEndSepPunct{\mcitedefaultmidpunct}
{\mcitedefaultendpunct}{\mcitedefaultseppunct}\relax
\EndOfBibitem
\bibitem[Mangan \emph{et~al.}(2021)Mangan, Zhou, Chu, and
  Prezhdo]{mangan2021dependence}
S.~M. Mangan, G.~Zhou, W.~Chu and O.~V. Prezhdo, \emph{The Journal of Physical
  Chemistry Letters}, 2021, \textbf{12}, 8672--8678\relax
\mciteBstWouldAddEndPuncttrue
\mciteSetBstMidEndSepPunct{\mcitedefaultmidpunct}
{\mcitedefaultendpunct}{\mcitedefaultseppunct}\relax
\EndOfBibitem
\bibitem[Kraskov \emph{et~al.}(2004)Kraskov, St{\"o}gbauer, and
  Grassberger]{kraskov2004estimating}
A.~Kraskov, H.~St{\"o}gbauer and P.~Grassberger, \emph{Physical review E},
  2004, \textbf{69}, 066138\relax
\mciteBstWouldAddEndPuncttrue
\mciteSetBstMidEndSepPunct{\mcitedefaultmidpunct}
{\mcitedefaultendpunct}{\mcitedefaultseppunct}\relax
\EndOfBibitem
\bibitem[How \emph{et~al.}(2021)How, Wang, Chu, Tkatchenko, and
  Prezhdo]{how2021significance}
W.~B. How, B.~Wang, W.~Chu, A.~Tkatchenko and O.~V. Prezhdo, \emph{The Journal
  of Physical Chemistry Letters}, 2021, \textbf{12}, 12026--12032\relax
\mciteBstWouldAddEndPuncttrue
\mciteSetBstMidEndSepPunct{\mcitedefaultmidpunct}
{\mcitedefaultendpunct}{\mcitedefaultseppunct}\relax
\EndOfBibitem
\bibitem[Xiang \emph{et~al.}(2021)Xiang, Liu, and Tress]{xiang2021review}
W.~Xiang, S.~F. Liu and W.~Tress, \emph{Energy \& Environmental Science}, 2021,
  \textbf{14}, 2090--2113\relax
\mciteBstWouldAddEndPuncttrue
\mciteSetBstMidEndSepPunct{\mcitedefaultmidpunct}
{\mcitedefaultendpunct}{\mcitedefaultseppunct}\relax
\EndOfBibitem
\bibitem[Green \emph{et~al.}(2014)Green, Ho-Baillie, and
  Snaith]{green2014emergence}
M.~A. Green, A.~Ho-Baillie and H.~J. Snaith, \emph{Nature photonics}, 2014,
  \textbf{8}, 506--514\relax
\mciteBstWouldAddEndPuncttrue
\mciteSetBstMidEndSepPunct{\mcitedefaultmidpunct}
{\mcitedefaultendpunct}{\mcitedefaultseppunct}\relax
\EndOfBibitem
\bibitem[Ahn \emph{et~al.}(2015)Ahn, Son, Jang, Kang, Choi, and
  Park]{ahn2015highly}
N.~Ahn, D.-Y. Son, I.-H. Jang, S.~M. Kang, M.~Choi and N.-G. Park,
  \emph{Journal of the American Chemical Society}, 2015, \textbf{137},
  8696--8699\relax
\mciteBstWouldAddEndPuncttrue
\mciteSetBstMidEndSepPunct{\mcitedefaultmidpunct}
{\mcitedefaultendpunct}{\mcitedefaultseppunct}\relax
\EndOfBibitem
\bibitem[Bian \emph{et~al.}(2019)Bian, Murphy, Xia, Daskin, and
  Kais]{bian2019quantum}
T.~Bian, D.~Murphy, R.~Xia, A.~Daskin and S.~Kais, \emph{Molecular Physics},
  2019, \textbf{117}, 2069--2082\relax
\mciteBstWouldAddEndPuncttrue
\mciteSetBstMidEndSepPunct{\mcitedefaultmidpunct}
{\mcitedefaultendpunct}{\mcitedefaultseppunct}\relax
\EndOfBibitem
\bibitem[Kandala \emph{et~al.}(2017)Kandala, Mezzacapo, Temme, Takita, Brink,
  Chow, and Gambetta]{kandala2017hardware}
A.~Kandala, A.~Mezzacapo, K.~Temme, M.~Takita, M.~Brink, J.~M. Chow and J.~M.
  Gambetta, \emph{Nature}, 2017, \textbf{549}, 242--246\relax
\mciteBstWouldAddEndPuncttrue
\mciteSetBstMidEndSepPunct{\mcitedefaultmidpunct}
{\mcitedefaultendpunct}{\mcitedefaultseppunct}\relax
\EndOfBibitem
\bibitem[Anderson \emph{et~al.}(2021)Anderson, Kiffner, Barkoutsos, Tavernelli,
  Crain, and Jaksch]{anderson2021coarse}
L.~W. Anderson, M.~Kiffner, P.~K. Barkoutsos, I.~Tavernelli, J.~Crain and
  D.~Jaksch, \emph{arXiv preprint arXiv:2110.00968}, 2021\relax
\mciteBstWouldAddEndPuncttrue
\mciteSetBstMidEndSepPunct{\mcitedefaultmidpunct}
{\mcitedefaultendpunct}{\mcitedefaultseppunct}\relax
\EndOfBibitem
\bibitem[Cipcigan \emph{et~al.}(2019)Cipcigan, Crain, Sokhan, and
  Martyna]{cipcigan2019electronic}
F.~Cipcigan, J.~Crain, V.~Sokhan and G.~Martyna, \emph{Reviews of Modern
  Physics}, 2019, \textbf{91}, 025003\relax
\mciteBstWouldAddEndPuncttrue
\mciteSetBstMidEndSepPunct{\mcitedefaultmidpunct}
{\mcitedefaultendpunct}{\mcitedefaultseppunct}\relax
\EndOfBibitem
\bibitem[Gokhale \emph{et~al.}(2020)Gokhale, Angiuli, Ding, Gui, Tomesh,
  Suchara, Martonosi, and Chong]{gokhale2020n}
P.~Gokhale, O.~Angiuli, Y.~Ding, K.~Gui, T.~Tomesh, M.~Suchara, M.~Martonosi
  and F.~T. Chong, \emph{IEEE Transactions on Quantum Engineering}, 2020,
  \textbf{1}, 1--24\relax
\mciteBstWouldAddEndPuncttrue
\mciteSetBstMidEndSepPunct{\mcitedefaultmidpunct}
{\mcitedefaultendpunct}{\mcitedefaultseppunct}\relax
\EndOfBibitem
\bibitem[Westbroek \emph{et~al.}(2018)Westbroek, King, Vvedensky, and
  D{\"u}rr]{westbroek2018user}
M.~J. Westbroek, P.~R. King, D.~D. Vvedensky and S.~D{\"u}rr, \emph{American
  Journal of Physics}, 2018, \textbf{86}, 293--304\relax
\mciteBstWouldAddEndPuncttrue
\mciteSetBstMidEndSepPunct{\mcitedefaultmidpunct}
{\mcitedefaultendpunct}{\mcitedefaultseppunct}\relax
\EndOfBibitem
\bibitem[Ollitrault \emph{et~al.}(2020)Ollitrault, Mazzola, and
  Tavernelli]{PhysRevLett.125.260511}
P.~J. Ollitrault, G.~Mazzola and I.~Tavernelli, \emph{Phys. Rev. Lett.}, 2020,
  \textbf{125}, 260511\relax
\mciteBstWouldAddEndPuncttrue
\mciteSetBstMidEndSepPunct{\mcitedefaultmidpunct}
{\mcitedefaultendpunct}{\mcitedefaultseppunct}\relax
\EndOfBibitem
\bibitem[Capano \emph{et~al.}(2014)Capano, Chergui, Rothlisberger, Tavernelli,
  and Penfold]{capano2014quantum}
G.~Capano, M.~Chergui, U.~Rothlisberger, I.~Tavernelli and T.~J. Penfold,
  \emph{The Journal of Physical Chemistry A}, 2014, \textbf{118},
  9861--9869\relax
\mciteBstWouldAddEndPuncttrue
\mciteSetBstMidEndSepPunct{\mcitedefaultmidpunct}
{\mcitedefaultendpunct}{\mcitedefaultseppunct}\relax
\EndOfBibitem
\bibitem[Zhugayevych and Tretiak(2015)]{zhugayevych2015theoretical}
A.~Zhugayevych and S.~Tretiak, \emph{Annual Review of Physical Chemistry},
  2015, \textbf{66}, 305--330\relax
\mciteBstWouldAddEndPuncttrue
\mciteSetBstMidEndSepPunct{\mcitedefaultmidpunct}
{\mcitedefaultendpunct}{\mcitedefaultseppunct}\relax
\EndOfBibitem
\bibitem[Kiss \emph{et~al.}(2022)Kiss, Tacchino, Vallecorsa, and
  Tavernelli]{kiss_quantum_2022}
O.~Kiss, F.~Tacchino, S.~Vallecorsa and I.~Tavernelli, \emph{arXiv:2203.04666
  [physics, physics:quant-ph]}, 2022\relax
\mciteBstWouldAddEndPuncttrue
\mciteSetBstMidEndSepPunct{\mcitedefaultmidpunct}
{\mcitedefaultendpunct}{\mcitedefaultseppunct}\relax
\EndOfBibitem
\bibitem[Lim \emph{et~al.}(2021)Lim, Lu, Cho, Sung, Kim, Kim, Park, and
  Kim]{Pic_fig2_drug}
S.~Lim, Y.~Lu, C.~Y. Cho, I.~Sung, J.~Kim, Y.~Kim, S.~Park and S.~Kim,
  \emph{Computational and Structural Biotechnology Journal}, 2021, \textbf{19},
  1541--1556\relax
\mciteBstWouldAddEndPuncttrue
\mciteSetBstMidEndSepPunct{\mcitedefaultmidpunct}
{\mcitedefaultendpunct}{\mcitedefaultseppunct}\relax
\EndOfBibitem
\bibitem[Sliwoski \emph{et~al.}(2014)Sliwoski, Kothiwale, Meiler, and
  Lowe]{sliwoski2014computational}
G.~Sliwoski, S.~Kothiwale, J.~Meiler and E.~W. Lowe, \emph{Pharmacological
  reviews}, 2014, \textbf{66}, 334--395\relax
\mciteBstWouldAddEndPuncttrue
\mciteSetBstMidEndSepPunct{\mcitedefaultmidpunct}
{\mcitedefaultendpunct}{\mcitedefaultseppunct}\relax
\EndOfBibitem
\bibitem[Leelananda and Lindert(2016)]{leelananda2016comp}
S.~P. Leelananda and S.~Lindert, \emph{Beilstein journal of organic chemistry},
  2016, \textbf{12}, 2694--2718\relax
\mciteBstWouldAddEndPuncttrue
\mciteSetBstMidEndSepPunct{\mcitedefaultmidpunct}
{\mcitedefaultendpunct}{\mcitedefaultseppunct}\relax
\EndOfBibitem
\bibitem[Kim \emph{et~al.}(2016)Kim, Thiessen, Bolton, Chen, Fu, Gindulyte,
  Han, He, He, Shoemaker,\emph{et~al.}]{kim2016pubchem}
S.~Kim, P.~A. Thiessen, E.~E. Bolton, J.~Chen, G.~Fu, A.~Gindulyte, L.~Han,
  J.~He, S.~He, B.~A. Shoemaker \emph{et~al.}, \emph{Nucleic acids research},
  2016, \textbf{44}, D1202--D1213\relax
\mciteBstWouldAddEndPuncttrue
\mciteSetBstMidEndSepPunct{\mcitedefaultmidpunct}
{\mcitedefaultendpunct}{\mcitedefaultseppunct}\relax
\EndOfBibitem
\bibitem[Kim \emph{et~al.}(2019)Kim, Chen, Cheng, Gindulyte, He, He, Li,
  Shoemaker, Thiessen, Yu,\emph{et~al.}]{kim2019pubchem}
S.~Kim, J.~Chen, T.~Cheng, A.~Gindulyte, J.~He, S.~He, Q.~Li, B.~A. Shoemaker,
  P.~A. Thiessen, B.~Yu \emph{et~al.}, \emph{Nucleic acids research}, 2019,
  \textbf{47}, D1102--D1109\relax
\mciteBstWouldAddEndPuncttrue
\mciteSetBstMidEndSepPunct{\mcitedefaultmidpunct}
{\mcitedefaultendpunct}{\mcitedefaultseppunct}\relax
\EndOfBibitem
\bibitem[Gaulton \emph{et~al.}(2017)Gaulton, Hersey, Nowotka, Bento, Chambers,
  Mendez, Mutowo, Atkinson, Bellis,
  Cibri{\'a}n-Uhalte,\emph{et~al.}]{gaulton2017chembl}
A.~Gaulton, A.~Hersey, M.~Nowotka, A.~P. Bento, J.~Chambers, D.~Mendez,
  P.~Mutowo, F.~Atkinson, L.~J. Bellis, E.~Cibri{\'a}n-Uhalte \emph{et~al.},
  \emph{Nucleic acids research}, 2017, \textbf{45}, D945--D954\relax
\mciteBstWouldAddEndPuncttrue
\mciteSetBstMidEndSepPunct{\mcitedefaultmidpunct}
{\mcitedefaultendpunct}{\mcitedefaultseppunct}\relax
\EndOfBibitem
\bibitem[Wishart \emph{et~al.}(2018)Wishart, Feunang, Guo, Lo, Marcu, Grant,
  Sajed, Johnson, Li, Sayeeda,\emph{et~al.}]{wishart2018drugbank}
D.~S. Wishart, Y.~D. Feunang, A.~C. Guo, E.~J. Lo, A.~Marcu, J.~R. Grant,
  T.~Sajed, D.~Johnson, C.~Li, Z.~Sayeeda \emph{et~al.}, \emph{Nucleic acids
  research}, 2018, \textbf{46}, D1074--D1082\relax
\mciteBstWouldAddEndPuncttrue
\mciteSetBstMidEndSepPunct{\mcitedefaultmidpunct}
{\mcitedefaultendpunct}{\mcitedefaultseppunct}\relax
\EndOfBibitem
\bibitem[Mysinger \emph{et~al.}(2012)Mysinger, Carchia, Irwin, and
  Shoichet]{mysinger2012directory}
M.~M. Mysinger, M.~Carchia, J.~J. Irwin and B.~K. Shoichet, \emph{Journal of
  medicinal chemistry}, 2012, \textbf{55}, 6582--6594\relax
\mciteBstWouldAddEndPuncttrue
\mciteSetBstMidEndSepPunct{\mcitedefaultmidpunct}
{\mcitedefaultendpunct}{\mcitedefaultseppunct}\relax
\EndOfBibitem
\bibitem[uni(2017)]{uniprot2017uniprot}
\emph{Nucleic acids research}, 2017, \textbf{45}, D158--D169\relax
\mciteBstWouldAddEndPuncttrue
\mciteSetBstMidEndSepPunct{\mcitedefaultmidpunct}
{\mcitedefaultendpunct}{\mcitedefaultseppunct}\relax
\EndOfBibitem
\bibitem[Berman \emph{et~al.}(2000)Berman, Westbrook, Feng, Gilliland, Bhat,
  Weissig, Shindyalov, and Bourne]{berman2000protein}
H.~M. Berman, J.~Westbrook, Z.~Feng, G.~Gilliland, T.~N. Bhat, H.~Weissig,
  I.~N. Shindyalov and P.~E. Bourne, \emph{Nucleic acids research}, 2000,
  \textbf{28}, 235--242\relax
\mciteBstWouldAddEndPuncttrue
\mciteSetBstMidEndSepPunct{\mcitedefaultmidpunct}
{\mcitedefaultendpunct}{\mcitedefaultseppunct}\relax
\EndOfBibitem
\bibitem[Berman \emph{et~al.}(2007)Berman, Henrick, Nakamura, and
  Markley]{berman2007worldwide}
H.~Berman, K.~Henrick, H.~Nakamura and J.~L. Markley, \emph{Nucleic acids
  research}, 2007, \textbf{35}, D301--D303\relax
\mciteBstWouldAddEndPuncttrue
\mciteSetBstMidEndSepPunct{\mcitedefaultmidpunct}
{\mcitedefaultendpunct}{\mcitedefaultseppunct}\relax
\EndOfBibitem
\bibitem[Burley \emph{et~al.}(2017)Burley, Berman, Kleywegt, Markley, Nakamura,
  and Velankar]{burley2017protein}
S.~K. Burley, H.~M. Berman, G.~J. Kleywegt, J.~L. Markley, H.~Nakamura and
  S.~Velankar, \emph{Protein Crystallography}, 2017,  627--641\relax
\mciteBstWouldAddEndPuncttrue
\mciteSetBstMidEndSepPunct{\mcitedefaultmidpunct}
{\mcitedefaultendpunct}{\mcitedefaultseppunct}\relax
\EndOfBibitem
\bibitem[Wang \emph{et~al.}(2005)Wang, Fang, Lu, Yang, and
  Wang]{wang2005pdbbind}
R.~Wang, X.~Fang, Y.~Lu, C.-Y. Yang and S.~Wang, \emph{Journal of medicinal
  chemistry}, 2005, \textbf{48}, 4111--4119\relax
\mciteBstWouldAddEndPuncttrue
\mciteSetBstMidEndSepPunct{\mcitedefaultmidpunct}
{\mcitedefaultendpunct}{\mcitedefaultseppunct}\relax
\EndOfBibitem
\bibitem[Gilson \emph{et~al.}(2016)Gilson, Liu, Baitaluk, Nicola, Hwang, and
  Chong]{gilson2016bindingdb}
M.~K. Gilson, T.~Liu, M.~Baitaluk, G.~Nicola, L.~Hwang and J.~Chong,
  \emph{Nucleic acids research}, 2016, \textbf{44}, D1045--D1053\relax
\mciteBstWouldAddEndPuncttrue
\mciteSetBstMidEndSepPunct{\mcitedefaultmidpunct}
{\mcitedefaultendpunct}{\mcitedefaultseppunct}\relax
\EndOfBibitem
\bibitem[Weininger(1988)]{weininger1988smiles}
D.~Weininger, \emph{Journal of chemical information and computer sciences},
  1988, \textbf{28}, 31--36\relax
\mciteBstWouldAddEndPuncttrue
\mciteSetBstMidEndSepPunct{\mcitedefaultmidpunct}
{\mcitedefaultendpunct}{\mcitedefaultseppunct}\relax
\EndOfBibitem
\bibitem[Weininger \emph{et~al.}(1989)Weininger, Weininger, and
  Weininger]{weininger1989smiles}
D.~Weininger, A.~Weininger and J.~L. Weininger, \emph{Journal of chemical
  information and computer sciences}, 1989, \textbf{29}, 97--101\relax
\mciteBstWouldAddEndPuncttrue
\mciteSetBstMidEndSepPunct{\mcitedefaultmidpunct}
{\mcitedefaultendpunct}{\mcitedefaultseppunct}\relax
\EndOfBibitem
\bibitem[O’Boyle(2012)]{o2012towards}
N.~M. O’Boyle, \emph{Journal of cheminformatics}, 2012, \textbf{4},
  1--14\relax
\mciteBstWouldAddEndPuncttrue
\mciteSetBstMidEndSepPunct{\mcitedefaultmidpunct}
{\mcitedefaultendpunct}{\mcitedefaultseppunct}\relax
\EndOfBibitem
\bibitem[Krenn \emph{et~al.}(2020)Krenn, H{\"a}se, Nigam, Friederich, and
  Aspuru-Guzik]{krenn2020self}
M.~Krenn, F.~H{\"a}se, A.~Nigam, P.~Friederich and A.~Aspuru-Guzik,
  \emph{Machine Learning: Science and Technology}, 2020, \textbf{1},
  045024\relax
\mciteBstWouldAddEndPuncttrue
\mciteSetBstMidEndSepPunct{\mcitedefaultmidpunct}
{\mcitedefaultendpunct}{\mcitedefaultseppunct}\relax
\EndOfBibitem
\bibitem[Krenn \emph{et~al.}(2019)Krenn, H{\"a}se, Nigam, Friederich, and
  Aspuru-Guzik]{krenn2019selfies}
M.~Krenn, F.~H{\"a}se, A.~Nigam, P.~Friederich and A.~Aspuru-Guzik, \emph{arXiv
  preprint arXiv:1905.13741}, 2019\relax
\mciteBstWouldAddEndPuncttrue
\mciteSetBstMidEndSepPunct{\mcitedefaultmidpunct}
{\mcitedefaultendpunct}{\mcitedefaultseppunct}\relax
\EndOfBibitem
\bibitem[Dalke(2008)]{dalke2008parsers}
A.~Dalke, \emph{Chemistry Central Journal}, 2008, \textbf{2}, 1--1\relax
\mciteBstWouldAddEndPuncttrue
\mciteSetBstMidEndSepPunct{\mcitedefaultmidpunct}
{\mcitedefaultendpunct}{\mcitedefaultseppunct}\relax
\EndOfBibitem
\bibitem[Sykora and Leahy(2008)]{sykora2008chemical}
V.~J. Sykora and D.~E. Leahy, \emph{Journal of chemical information and
  modeling}, 2008, \textbf{48}, 1931--1942\relax
\mciteBstWouldAddEndPuncttrue
\mciteSetBstMidEndSepPunct{\mcitedefaultmidpunct}
{\mcitedefaultendpunct}{\mcitedefaultseppunct}\relax
\EndOfBibitem
\bibitem[Rogers and Hahn(2010)]{rogers2010extended}
D.~Rogers and M.~Hahn, \emph{Journal of chemical information and modeling},
  2010, \textbf{50}, 742--754\relax
\mciteBstWouldAddEndPuncttrue
\mciteSetBstMidEndSepPunct{\mcitedefaultmidpunct}
{\mcitedefaultendpunct}{\mcitedefaultseppunct}\relax
\EndOfBibitem
\bibitem[Cereto-Massagu{\'e} \emph{et~al.}(2015)Cereto-Massagu{\'e}, Ojeda,
  Valls, Mulero, Garcia-Vallv{\'e}, and Pujadas]{cereto2015molecular}
A.~Cereto-Massagu{\'e}, M.~J. Ojeda, C.~Valls, M.~Mulero, S.~Garcia-Vallv{\'e}
  and G.~Pujadas, \emph{Methods}, 2015, \textbf{71}, 58--63\relax
\mciteBstWouldAddEndPuncttrue
\mciteSetBstMidEndSepPunct{\mcitedefaultmidpunct}
{\mcitedefaultendpunct}{\mcitedefaultseppunct}\relax
\EndOfBibitem
\bibitem[Duan \emph{et~al.}(2010)Duan, Dixon, Lowrie, and
  Sherman]{duan2010analysis}
J.~Duan, S.~L. Dixon, J.~F. Lowrie and W.~Sherman, \emph{Journal of Molecular
  Graphics and Modelling}, 2010, \textbf{29}, 157--170\relax
\mciteBstWouldAddEndPuncttrue
\mciteSetBstMidEndSepPunct{\mcitedefaultmidpunct}
{\mcitedefaultendpunct}{\mcitedefaultseppunct}\relax
\EndOfBibitem
\bibitem[Hert \emph{et~al.}(2004)Hert, Willett, Wilton, Acklin, Azzaoui,
  Jacoby, and Schuffenhauer]{hert2004comparison}
J.~Hert, P.~Willett, D.~J. Wilton, P.~Acklin, K.~Azzaoui, E.~Jacoby and
  A.~Schuffenhauer, \emph{Journal of chemical information and computer
  sciences}, 2004, \textbf{44}, 1177--1185\relax
\mciteBstWouldAddEndPuncttrue
\mciteSetBstMidEndSepPunct{\mcitedefaultmidpunct}
{\mcitedefaultendpunct}{\mcitedefaultseppunct}\relax
\EndOfBibitem
\bibitem[P{\'e}rez-Nueno \emph{et~al.}(2009)P{\'e}rez-Nueno, Rabal, Borrell,
  and Teixid{\'o}]{perez2009apif}
V.~I. P{\'e}rez-Nueno, O.~Rabal, J.~I. Borrell and J.~Teixid{\'o},
  \emph{Journal of chemical information and modeling}, 2009, \textbf{49},
  1245--1260\relax
\mciteBstWouldAddEndPuncttrue
\mciteSetBstMidEndSepPunct{\mcitedefaultmidpunct}
{\mcitedefaultendpunct}{\mcitedefaultseppunct}\relax
\EndOfBibitem
\bibitem[Awale and Reymond(2014)]{awale2014atom}
M.~Awale and J.-L. Reymond, \emph{Journal of chemical information and
  modeling}, 2014, \textbf{54}, 1892--1907\relax
\mciteBstWouldAddEndPuncttrue
\mciteSetBstMidEndSepPunct{\mcitedefaultmidpunct}
{\mcitedefaultendpunct}{\mcitedefaultseppunct}\relax
\EndOfBibitem
\bibitem[De~Cao and Kipf(2018)]{de2018molgan}
N.~De~Cao and T.~Kipf, \emph{arXiv preprint arXiv:1805.11973}, 2018\relax
\mciteBstWouldAddEndPuncttrue
\mciteSetBstMidEndSepPunct{\mcitedefaultmidpunct}
{\mcitedefaultendpunct}{\mcitedefaultseppunct}\relax
\EndOfBibitem
\bibitem[Jiang \emph{et~al.}(2021)Jiang, Wu, Hsieh, Chen, Liao, Wang, Shen,
  Cao, Wu, and Hou]{jiang2021could}
D.~Jiang, Z.~Wu, C.-Y. Hsieh, G.~Chen, B.~Liao, Z.~Wang, C.~Shen, D.~Cao, J.~Wu
  and T.~Hou, \emph{Journal of cheminformatics}, 2021, \textbf{13}, 1--23\relax
\mciteBstWouldAddEndPuncttrue
\mciteSetBstMidEndSepPunct{\mcitedefaultmidpunct}
{\mcitedefaultendpunct}{\mcitedefaultseppunct}\relax
\EndOfBibitem
\bibitem[Carracedo-Reboredo \emph{et~al.}(2021)Carracedo-Reboredo,
  Li{\~n}ares-Blanco, Rodr{\'\i}guez-Fern{\'a}ndez, Cedr{\'o}n, Novoa,
  Carballal, Maojo, Pazos, and Fernandez-Lozano]{carracedo2021review}
P.~Carracedo-Reboredo, J.~Li{\~n}ares-Blanco, N.~Rodr{\'\i}guez-Fern{\'a}ndez,
  F.~Cedr{\'o}n, F.~J. Novoa, A.~Carballal, V.~Maojo, A.~Pazos and
  C.~Fernandez-Lozano, \emph{Computational and Structural Biotechnology
  Journal}, 2021, \textbf{19}, 4538\relax
\mciteBstWouldAddEndPuncttrue
\mciteSetBstMidEndSepPunct{\mcitedefaultmidpunct}
{\mcitedefaultendpunct}{\mcitedefaultseppunct}\relax
\EndOfBibitem
\bibitem[Lin \emph{et~al.}(2020)Lin, Li, and Lin]{lin2020review}
X.~Lin, X.~Li and X.~Lin, \emph{Molecules}, 2020, \textbf{25}, 1375\relax
\mciteBstWouldAddEndPuncttrue
\mciteSetBstMidEndSepPunct{\mcitedefaultmidpunct}
{\mcitedefaultendpunct}{\mcitedefaultseppunct}\relax
\EndOfBibitem
\bibitem[Dunbar \emph{et~al.}(2011)Dunbar, Smith, Yang, Ung, Lexa, Khazanov,
  Stuckey, Wang, and Carlson]{Dunbar2011CSARBE}
J.~B. Dunbar, R.~D. Smith, C.-Y. Yang, P.~M.-U. Ung, K.~W. Lexa, N.~A.
  Khazanov, J.~A. Stuckey, S.~Wang and H.~A. Carlson, \emph{Journal of Chemical
  Information and Modeling}, 2011, \textbf{51}, 2036 -- 2046\relax
\mciteBstWouldAddEndPuncttrue
\mciteSetBstMidEndSepPunct{\mcitedefaultmidpunct}
{\mcitedefaultendpunct}{\mcitedefaultseppunct}\relax
\EndOfBibitem
\bibitem[Koes \emph{et~al.}(2013)Koes, Baumgartner, and
  Camacho]{Koes2013LessonsLI}
D.~R. Koes, M.~P. Baumgartner and C.~J. Camacho, \emph{Journal of chemical
  information and modeling}, 2013, \textbf{53 8}, 1893--904\relax
\mciteBstWouldAddEndPuncttrue
\mciteSetBstMidEndSepPunct{\mcitedefaultmidpunct}
{\mcitedefaultendpunct}{\mcitedefaultseppunct}\relax
\EndOfBibitem
\bibitem[Trott and Olson(2010)]{Trott2010AutoDockVI}
O.~Trott and A.~J. Olson, \emph{Journal of Computational Chemistry}, 2010,
  \textbf{31}, 455--461\relax
\mciteBstWouldAddEndPuncttrue
\mciteSetBstMidEndSepPunct{\mcitedefaultmidpunct}
{\mcitedefaultendpunct}{\mcitedefaultseppunct}\relax
\EndOfBibitem
\bibitem[Jia \emph{et~al.}(2014)Jia, Shelhamer, Donahue, Karayev, Long,
  Girshick, Guadarrama, and Darrell]{jia2014caffe}
Y.~Jia, E.~Shelhamer, J.~Donahue, S.~Karayev, J.~Long, R.~Girshick,
  S.~Guadarrama and T.~Darrell, Proceedings of the 22nd ACM international
  conference on Multimedia, 2014, pp. 675--678\relax
\mciteBstWouldAddEndPuncttrue
\mciteSetBstMidEndSepPunct{\mcitedefaultmidpunct}
{\mcitedefaultendpunct}{\mcitedefaultseppunct}\relax
\EndOfBibitem
\bibitem[Wang \emph{et~al.}(2020)Wang, You, Yang, Yi, Chen, and
  Zheng]{Wang_2020}
Y.-B. Wang, Z.-H. You, S.~Yang, H.-C. Yi, Z.-H. Chen and K.~Zheng, \emph{{BMC}
  Medical Informatics and Decision Making}, 2020, \textbf{20}, 1--9\relax
\mciteBstWouldAddEndPuncttrue
\mciteSetBstMidEndSepPunct{\mcitedefaultmidpunct}
{\mcitedefaultendpunct}{\mcitedefaultseppunct}\relax
\EndOfBibitem
\bibitem[Kanehisa \emph{et~al.}(2008)Kanehisa, Araki, Goto, Hattori, Hirakawa,
  Itoh, Katayama, Kawashima, Okuda, Tokimatsu, and
  Yamanishi]{Kanehisa2008KEGGFL}
M.~Kanehisa, M.~Araki, S.~Goto, M.~Hattori, M.~Hirakawa, M.~Itoh, T.~Katayama,
  S.~Kawashima, S.~Okuda, T.~Tokimatsu and Y.~Yamanishi, \emph{Nucleic Acids
  Research}, 2008, \textbf{36}, D480 -- D484\relax
\mciteBstWouldAddEndPuncttrue
\mciteSetBstMidEndSepPunct{\mcitedefaultmidpunct}
{\mcitedefaultendpunct}{\mcitedefaultseppunct}\relax
\EndOfBibitem
\bibitem[Wishart \emph{et~al.}(2006)Wishart, Knox, Guo, Shrivastava, Hassanali,
  Stothard, Chang, and Woolsey]{Wishart2006DrugBankAC}
D.~S. Wishart, C.~Knox, A.~Guo, S.~Shrivastava, M.~Hassanali, P.~Stothard,
  Z.~Chang and J.~Woolsey, \emph{Nucleic Acids Research}, 2006, \textbf{34},
  D668 -- D672\relax
\mciteBstWouldAddEndPuncttrue
\mciteSetBstMidEndSepPunct{\mcitedefaultmidpunct}
{\mcitedefaultendpunct}{\mcitedefaultseppunct}\relax
\EndOfBibitem
\bibitem[G{\"u}nther \emph{et~al.}(2008)G{\"u}nther, Kuhn, Dunkel, Campillos,
  Senger, Petsalaki, Ahmed, Urdiales, Gewiess, Jensen, Schneider, Skoblo,
  Russell, Bourne, Bork, and Preissner]{Gnther2008SuperTargetAM}
S.~G{\"u}nther, M.~Kuhn, M.~Dunkel, M.~Campillos, C.~Senger, E.~Petsalaki,
  J.~Ahmed, E.~G. Urdiales, A.~Gewiess, L.~J. Jensen, R.~Schneider, R.~Skoblo,
  R.~B. Russell, P.~E. Bourne, P.~Bork and R.~Preissner, \emph{Nucleic Acids
  Research}, 2008, \textbf{36}, D919 -- D922\relax
\mciteBstWouldAddEndPuncttrue
\mciteSetBstMidEndSepPunct{\mcitedefaultmidpunct}
{\mcitedefaultendpunct}{\mcitedefaultseppunct}\relax
\EndOfBibitem
\bibitem[Wang \emph{et~al.}(2017)Wang, You, Li, Chen, Jiang, and
  Zhang]{Wang2017PCVMZMUT}
Y.~Wang, Z.-H. You, X.~Li, X.~Chen, T.~Jiang and J.~Zhang, \emph{International
  Journal of Molecular Sciences}, 2017, \textbf{18}, 1--13\relax
\mciteBstWouldAddEndPuncttrue
\mciteSetBstMidEndSepPunct{\mcitedefaultmidpunct}
{\mcitedefaultendpunct}{\mcitedefaultseppunct}\relax
\EndOfBibitem
\bibitem[Hanley and McNeil(1982)]{hanley1982meaning}
J.~A. Hanley and B.~J. McNeil, \emph{Radiology}, 1982, \textbf{143},
  29--36\relax
\mciteBstWouldAddEndPuncttrue
\mciteSetBstMidEndSepPunct{\mcitedefaultmidpunct}
{\mcitedefaultendpunct}{\mcitedefaultseppunct}\relax
\EndOfBibitem
\bibitem[Zheng \emph{et~al.}(2019)Zheng, Li, Chen, Xu, and
  Yang]{Zheng2019PredictingDP}
S.~Zheng, Y.~Li, S.~Chen, J.~Xu and Y.~Yang, \emph{bioRxiv}, 2019\relax
\mciteBstWouldAddEndPuncttrue
\mciteSetBstMidEndSepPunct{\mcitedefaultmidpunct}
{\mcitedefaultendpunct}{\mcitedefaultseppunct}\relax
\EndOfBibitem
\bibitem[He \emph{et~al.}(2016)He, Zhang, Ren, and Sun]{he2016identity}
K.~He, X.~Zhang, S.~Ren and J.~Sun, European conference on computer vision,
  2016, pp. 630--645\relax
\mciteBstWouldAddEndPuncttrue
\mciteSetBstMidEndSepPunct{\mcitedefaultmidpunct}
{\mcitedefaultendpunct}{\mcitedefaultseppunct}\relax
\EndOfBibitem
\bibitem[Olivecrona \emph{et~al.}(2017)Olivecrona, Blaschke, Engkvist, and
  Chen]{Olivecrona2017MolecularDD}
M.~Olivecrona, T.~Blaschke, O.~Engkvist and H.~Chen, \emph{Journal of
  Cheminformatics}, 2017, \textbf{9}, 1--14\relax
\mciteBstWouldAddEndPuncttrue
\mciteSetBstMidEndSepPunct{\mcitedefaultmidpunct}
{\mcitedefaultendpunct}{\mcitedefaultseppunct}\relax
\EndOfBibitem
\bibitem[Mysinger \emph{et~al.}(2012)Mysinger, Carchia, Irwin, and
  Shoichet]{Mysinger2012DirectoryOU}
M.~M. Mysinger, M.~Carchia, J.~J. Irwin and B.~K. Shoichet, \emph{Journal of
  Medicinal Chemistry}, 2012, \textbf{55}, 6582 -- 6594\relax
\mciteBstWouldAddEndPuncttrue
\mciteSetBstMidEndSepPunct{\mcitedefaultmidpunct}
{\mcitedefaultendpunct}{\mcitedefaultseppunct}\relax
\EndOfBibitem
\bibitem[Liu \emph{et~al.}(2015)Liu, Sun, Guan, Zheng, and
  Zhou]{Liu2015ImprovingCI}
H.~Liu, J.~Sun, J.~Guan, J.~Zheng and S.~Zhou, \emph{Bioinformatics}, 2015,
  \textbf{31}, i221 -- i229\relax
\mciteBstWouldAddEndPuncttrue
\mciteSetBstMidEndSepPunct{\mcitedefaultmidpunct}
{\mcitedefaultendpunct}{\mcitedefaultseppunct}\relax
\EndOfBibitem
\bibitem[Gilson \emph{et~al.}(2016)Gilson, Liu, Baitaluk, Nicola, Hwang, and
  Chong]{Gilson2016BindingDBI2}
M.~K. Gilson, T.~Liu, M.~Baitaluk, G.~Nicola, L.~Hwang and J.~Chong,
  \emph{Nucleic Acids Research}, 2016, \textbf{44}, D1045 -- D1053\relax
\mciteBstWouldAddEndPuncttrue
\mciteSetBstMidEndSepPunct{\mcitedefaultmidpunct}
{\mcitedefaultendpunct}{\mcitedefaultseppunct}\relax
\EndOfBibitem
\bibitem[Zou \emph{et~al.}(2007)Zou, O’Malley, and Mauri]{zou2007receiver}
K.~H. Zou, A.~J. O’Malley and L.~Mauri, \emph{Circulation}, 2007,
  \textbf{115}, 654--657\relax
\mciteBstWouldAddEndPuncttrue
\mciteSetBstMidEndSepPunct{\mcitedefaultmidpunct}
{\mcitedefaultendpunct}{\mcitedefaultseppunct}\relax
\EndOfBibitem
\bibitem[Wallach \emph{et~al.}(2015)Wallach, Dzamba, and
  Heifets]{Wallach2015AtomNetAD}
I.~Wallach, M.~Dzamba and A.~Heifets, \emph{CoRR}, 2015,
  \textbf{abs/1510.02855}, 1--9\relax
\mciteBstWouldAddEndPuncttrue
\mciteSetBstMidEndSepPunct{\mcitedefaultmidpunct}
{\mcitedefaultendpunct}{\mcitedefaultseppunct}\relax
\EndOfBibitem
\bibitem[Dahl \emph{et~al.}(2014)Dahl, Jaitly, and
  Salakhutdinov]{dahl2014multi}
G.~E. Dahl, N.~Jaitly and R.~Salakhutdinov, \emph{arXiv preprint
  arXiv:1406.1231}, 2014\relax
\mciteBstWouldAddEndPuncttrue
\mciteSetBstMidEndSepPunct{\mcitedefaultmidpunct}
{\mcitedefaultendpunct}{\mcitedefaultseppunct}\relax
\EndOfBibitem
\bibitem[Mauri \emph{et~al.}(2006)Mauri, Consonni, Pavan, Todeschini, and
  Chemometrics]{Mauri2006DRAGONSA}
A.~Mauri, V.~Consonni, M.~Pavan, R.~Todeschini and M.~Chemometrics,
  \emph{Commun. Math. Comput. Chem.}, 2006, \textbf{56}, 237--248\relax
\mciteBstWouldAddEndPuncttrue
\mciteSetBstMidEndSepPunct{\mcitedefaultmidpunct}
{\mcitedefaultendpunct}{\mcitedefaultseppunct}\relax
\EndOfBibitem
\bibitem[Ramsundar \emph{et~al.}(2015)Ramsundar, Kearnes, Riley, Webster,
  Konerding, and Pande]{Ramsundar2015MassivelyMN}
B.~Ramsundar, S.~M. Kearnes, P.~F. Riley, D.~R. Webster, D.~E. Konerding and
  V.~S. Pande, \emph{ArXiv}, 2015, \textbf{abs/1502.02072}, 1--9\relax
\mciteBstWouldAddEndPuncttrue
\mciteSetBstMidEndSepPunct{\mcitedefaultmidpunct}
{\mcitedefaultendpunct}{\mcitedefaultseppunct}\relax
\EndOfBibitem
\bibitem[Nwankpa \emph{et~al.}(2018)Nwankpa, Ijomah, Gachagan, and
  Marshall]{nwankpa2018activation}
C.~Nwankpa, W.~Ijomah, A.~Gachagan and S.~Marshall, \emph{arXiv preprint
  arXiv:1811.03378}, 2018\relax
\mciteBstWouldAddEndPuncttrue
\mciteSetBstMidEndSepPunct{\mcitedefaultmidpunct}
{\mcitedefaultendpunct}{\mcitedefaultseppunct}\relax
\EndOfBibitem
\bibitem[Ma \emph{et~al.}(2015)Ma, Sheridan, Liaw, Dahl, and
  Svetnik]{ma2015deep}
J.~Ma, R.~P. Sheridan, A.~Liaw, G.~E. Dahl and V.~Svetnik, \emph{Journal of
  chemical information and modeling}, 2015, \textbf{55}, 263--274\relax
\mciteBstWouldAddEndPuncttrue
\mciteSetBstMidEndSepPunct{\mcitedefaultmidpunct}
{\mcitedefaultendpunct}{\mcitedefaultseppunct}\relax
\EndOfBibitem
\bibitem[Carhart \emph{et~al.}(1985)Carhart, Smith, and
  Venkataraghavan]{Carhart1985AtomPA}
R.~E. Carhart, D.~H. Smith and R.~Venkataraghavan, \emph{J. Chem. Inf. Comput.
  Sci.}, 1985, \textbf{25}, 64--73\relax
\mciteBstWouldAddEndPuncttrue
\mciteSetBstMidEndSepPunct{\mcitedefaultmidpunct}
{\mcitedefaultendpunct}{\mcitedefaultseppunct}\relax
\EndOfBibitem
\bibitem[Alexander \emph{et~al.}(2015)Alexander, Tropsha, and
  Winkler]{alexander2015beware}
D.~L. Alexander, A.~Tropsha and D.~A. Winkler, \emph{Journal of chemical
  information and modeling}, 2015, \textbf{55}, 1316--1322\relax
\mciteBstWouldAddEndPuncttrue
\mciteSetBstMidEndSepPunct{\mcitedefaultmidpunct}
{\mcitedefaultendpunct}{\mcitedefaultseppunct}\relax
\EndOfBibitem
\bibitem[Vo \emph{et~al.}(2019)Vo, Van~Vleet, Gupta, Liguori, and
  Rao]{vo2019overview}
A.~H. Vo, T.~R. Van~Vleet, R.~R. Gupta, M.~J. Liguori and M.~S. Rao,
  \emph{Chemical research in toxicology}, 2019, \textbf{33}, 20--37\relax
\mciteBstWouldAddEndPuncttrue
\mciteSetBstMidEndSepPunct{\mcitedefaultmidpunct}
{\mcitedefaultendpunct}{\mcitedefaultseppunct}\relax
\EndOfBibitem
\bibitem[Rodgers \emph{et~al.}(2010)Rodgers, Zhu, Fourches, Rusyn, and
  Tropsha]{rodgers2010modeling}
A.~D. Rodgers, H.~Zhu, D.~Fourches, I.~Rusyn and A.~Tropsha, \emph{Chemical
  research in toxicology}, 2010, \textbf{23}, 724--732\relax
\mciteBstWouldAddEndPuncttrue
\mciteSetBstMidEndSepPunct{\mcitedefaultmidpunct}
{\mcitedefaultendpunct}{\mcitedefaultseppunct}\relax
\EndOfBibitem
\bibitem[Plaa and Hewitt(1998)]{plaa1998toxicology}
G.~L. Plaa and W.~R. Hewitt, \emph{Toxicology of the Liver}, CRC Press,
  1998\relax
\mciteBstWouldAddEndPuncttrue
\mciteSetBstMidEndSepPunct{\mcitedefaultmidpunct}
{\mcitedefaultendpunct}{\mcitedefaultseppunct}\relax
\EndOfBibitem
\bibitem[Guyton \emph{et~al.}(1993)Guyton, Thompson, and
  Kensler]{guyton1993role}
K.~Z. Guyton, J.~A. Thompson and T.~W. Kensler, \emph{Chemical research in
  toxicology}, 1993, \textbf{6}, 731--738\relax
\mciteBstWouldAddEndPuncttrue
\mciteSetBstMidEndSepPunct{\mcitedefaultmidpunct}
{\mcitedefaultendpunct}{\mcitedefaultseppunct}\relax
\EndOfBibitem
\bibitem[Nonoyama and Fukuda(2008)]{Nonoyama2008DruginducedP}
T.~Nonoyama and R.~Fukuda, \emph{Journal of Toxicologic Pathology}, 2008,
  \textbf{21}, 9--24\relax
\mciteBstWouldAddEndPuncttrue
\mciteSetBstMidEndSepPunct{\mcitedefaultmidpunct}
{\mcitedefaultendpunct}{\mcitedefaultseppunct}\relax
\EndOfBibitem
\bibitem[Huang \emph{et~al.}(2011)Huang, Southall, Wang, Yasgar, Shinn, Jadhav,
  Nguyen, and Austin]{Huang2011TheNP}
R.~Huang, N.~Southall, Y.~Wang, A.~Yasgar, P.~Shinn, A.~Jadhav, D.-T. Nguyen
  and C.~P. Austin, \emph{Science Translational Medicine}, 2011, \textbf{3},
  80ps16 -- 80ps16\relax
\mciteBstWouldAddEndPuncttrue
\mciteSetBstMidEndSepPunct{\mcitedefaultmidpunct}
{\mcitedefaultendpunct}{\mcitedefaultseppunct}\relax
\EndOfBibitem
\bibitem[Bhandari \emph{et~al.}(2008)Bhandari, Figueroa, Lawrence, and
  Gerhold]{bhandari2008phospholipidosis}
N.~Bhandari, D.~J. Figueroa, J.~W. Lawrence and D.~L. Gerhold, \emph{Assay and
  drug development technologies}, 2008, \textbf{6}, 407--419\relax
\mciteBstWouldAddEndPuncttrue
\mciteSetBstMidEndSepPunct{\mcitedefaultmidpunct}
{\mcitedefaultendpunct}{\mcitedefaultseppunct}\relax
\EndOfBibitem
\bibitem[Pearce \emph{et~al.}(2005)Pearce, Uetrecht, and
  Leeder]{Pearce2005PATHWAYSOC}
R.~E. Pearce, J.~Uetrecht and J.~S. Leeder, \emph{Drug Metabolism and
  Disposition}, 2005, \textbf{33}, 1819 -- 1826\relax
\mciteBstWouldAddEndPuncttrue
\mciteSetBstMidEndSepPunct{\mcitedefaultmidpunct}
{\mcitedefaultendpunct}{\mcitedefaultseppunct}\relax
\EndOfBibitem
\bibitem[Yip \emph{et~al.}(2014)Yip, Maggs, Meng, Marson, Park, and
  Pirmohamed]{Yip2014CovalentAO}
V.~L.~M. Yip, J.~L. Maggs, X.~Meng, A.~G. Marson, K.~B. Park and M.~Pirmohamed,
  \emph{The Lancet}, 2014, \textbf{383}, S114\relax
\mciteBstWouldAddEndPuncttrue
\mciteSetBstMidEndSepPunct{\mcitedefaultmidpunct}
{\mcitedefaultendpunct}{\mcitedefaultseppunct}\relax
\EndOfBibitem
\bibitem[Alton \emph{et~al.}(1975)Alton, Grimes, Shaw, Patrick, and
  Mcguire]{Alton1975BiotransformationOA}
K.~B. Alton, R.~M. Grimes, C.~J. Shaw, J.~E. Patrick and J.~L. Mcguire,
  \emph{Drug metabolism and disposition: the biological fate of chemicals},
  1975, \textbf{3 5}, 352--60\relax
\mciteBstWouldAddEndPuncttrue
\mciteSetBstMidEndSepPunct{\mcitedefaultmidpunct}
{\mcitedefaultendpunct}{\mcitedefaultseppunct}\relax
\EndOfBibitem
\bibitem[Stepan \emph{et~al.}(2011)Stepan, Walker, Bauman, Price, Baillie,
  Kalgutkar, and Aleo]{Stepan2011StructuralAM}
A.~F. Stepan, D.~P. Walker, J.~N. Bauman, D.~Price, T.~A. Baillie, A.~S.
  Kalgutkar and M.~D. Aleo, \emph{Chemical research in toxicology}, 2011,
  \textbf{24 9}, 1345--410\relax
\mciteBstWouldAddEndPuncttrue
\mciteSetBstMidEndSepPunct{\mcitedefaultmidpunct}
{\mcitedefaultendpunct}{\mcitedefaultseppunct}\relax
\EndOfBibitem
\bibitem[Cao \emph{et~al.}(2018)Cao, Romero, and
  Aspuru-Guzik]{cao2018potential}
Y.~Cao, J.~Romero and A.~Aspuru-Guzik, \emph{IBM Journal of Research and
  Development}, 2018, \textbf{62}, 6--1\relax
\mciteBstWouldAddEndPuncttrue
\mciteSetBstMidEndSepPunct{\mcitedefaultmidpunct}
{\mcitedefaultendpunct}{\mcitedefaultseppunct}\relax
\EndOfBibitem
\bibitem[Perdomo-Ortiz \emph{et~al.}(2012)Perdomo-Ortiz, Dickson, Drew-Brook,
  Rose, and Aspuru-Guzik]{PerdomoOrtiz2012FindingLC}
A.~Perdomo-Ortiz, N.~Dickson, M.~Drew-Brook, G.~Rose and A.~Aspuru-Guzik,
  \emph{Scientific reports}, 2012, \textbf{2}, 1--7\relax
\mciteBstWouldAddEndPuncttrue
\mciteSetBstMidEndSepPunct{\mcitedefaultmidpunct}
{\mcitedefaultendpunct}{\mcitedefaultseppunct}\relax
\EndOfBibitem
\bibitem[Banchi \emph{et~al.}(2020)Banchi, Fingerhuth, Babej, Ing, and
  Arrazola]{banchi2020molecular}
L.~Banchi, M.~Fingerhuth, T.~Babej, C.~Ing and J.~M. Arrazola, \emph{Science
  advances}, 2020, \textbf{6}, eaax1950\relax
\mciteBstWouldAddEndPuncttrue
\mciteSetBstMidEndSepPunct{\mcitedefaultmidpunct}
{\mcitedefaultendpunct}{\mcitedefaultseppunct}\relax
\EndOfBibitem
\bibitem[Robert \emph{et~al.}(2021)Robert, Barkoutsos, Woerner, and
  Tavernelli]{robert2021resource}
A.~Robert, P.~K. Barkoutsos, S.~Woerner and I.~Tavernelli, \emph{npj Quantum
  Information}, 2021, \textbf{7}, 1--5\relax
\mciteBstWouldAddEndPuncttrue
\mciteSetBstMidEndSepPunct{\mcitedefaultmidpunct}
{\mcitedefaultendpunct}{\mcitedefaultseppunct}\relax
\EndOfBibitem
\bibitem[Babej \emph{et~al.}(2018)Babej, Ing, and
  Fingerhuth]{babej2018coarsegrained}
T.~Babej, C.~Ing and M.~Fingerhuth, \emph{Coarse-grained lattice protein
  folding on a quantum annealer}, 2018\relax
\mciteBstWouldAddEndPuncttrue
\mciteSetBstMidEndSepPunct{\mcitedefaultmidpunct}
{\mcitedefaultendpunct}{\mcitedefaultseppunct}\relax
\EndOfBibitem
\bibitem[Jumper \emph{et~al.}(2021)Jumper, Evans, Pritzel, Green, Figurnov,
  Ronneberger, Tunyasuvunakool, Bates, {\v{Z}}{\'\i}dek,
  Potapenko,\emph{et~al.}]{jumper2021highly}
J.~Jumper, R.~Evans, A.~Pritzel, T.~Green, M.~Figurnov, O.~Ronneberger,
  K.~Tunyasuvunakool, R.~Bates, A.~{\v{Z}}{\'\i}dek, A.~Potapenko
  \emph{et~al.}, \emph{Nature}, 2021, \textbf{596}, 583--589\relax
\mciteBstWouldAddEndPuncttrue
\mciteSetBstMidEndSepPunct{\mcitedefaultmidpunct}
{\mcitedefaultendpunct}{\mcitedefaultseppunct}\relax
\EndOfBibitem
\bibitem[Zinner \emph{et~al.}(2021)Zinner, Dahlhausen, Boehme, Ehlers, Bieske,
  and Fehring]{zinner2021quantum}
M.~Zinner, F.~Dahlhausen, P.~Boehme, J.~Ehlers, L.~Bieske and L.~Fehring,
  \emph{Drug Discovery Today}, 2021\relax
\mciteBstWouldAddEndPuncttrue
\mciteSetBstMidEndSepPunct{\mcitedefaultmidpunct}
{\mcitedefaultendpunct}{\mcitedefaultseppunct}\relax
\EndOfBibitem
\bibitem[Langione \emph{et~al.}()Langione, Bobier, Meier, Hasenfuss, and
  Schulze]{Langione_page}
M.~Langione, F.~Bobier, C.~Meier, S.~Hasenfuss and U.~Schulze, \emph{Will
  Quantum Computing Transform Biopharma R\&D?}, Accessed: 2021-10-12\relax
\mciteBstWouldAddEndPuncttrue
\mciteSetBstMidEndSepPunct{\mcitedefaultmidpunct}
{\mcitedefaultendpunct}{\mcitedefaultseppunct}\relax
\EndOfBibitem
\bibitem[Evers \emph{et~al.}()Evers, Heid, and Ostojic]{Evers_page}
M.~Evers, A.~Heid and E.~Ostojic, \emph{Pharma’s digital Rx: Quantum
  computing in drug research and development}, Accessed: 2021-10-12\relax
\mciteBstWouldAddEndPuncttrue
\mciteSetBstMidEndSepPunct{\mcitedefaultmidpunct}
{\mcitedefaultendpunct}{\mcitedefaultseppunct}\relax
\EndOfBibitem
\bibitem[Mer()]{Merck_HQS}
\emph{Merck KGaA, Darmstadt, Germany, and HQS Quantum Simulations Cooperate in
  Quantum Computing}, Accessed: 2021-10-12\relax
\mciteBstWouldAddEndPuncttrue
\mciteSetBstMidEndSepPunct{\mcitedefaultmidpunct}
{\mcitedefaultendpunct}{\mcitedefaultseppunct}\relax
\EndOfBibitem
\bibitem[Mer()]{Merck_Rahko}
\emph{Rahko announces Merck collaboration}, Accessed: 2021-10-12\relax
\mciteBstWouldAddEndPuncttrue
\mciteSetBstMidEndSepPunct{\mcitedefaultmidpunct}
{\mcitedefaultendpunct}{\mcitedefaultseppunct}\relax
\EndOfBibitem
\bibitem[Metinko()]{Zapata_collab}
C.~Metinko, \emph{Zapata Computing Raises \$ 38M As Quantum Computing Nears},
  Accessed: 2021-10-12\relax
\mciteBstWouldAddEndPuncttrue
\mciteSetBstMidEndSepPunct{\mcitedefaultmidpunct}
{\mcitedefaultendpunct}{\mcitedefaultseppunct}\relax
\EndOfBibitem
\bibitem[Siow()]{Protein_zeneca}
L.~Siow, \emph{ProteinQure Collaborates with AstraZeneca to Design Novel
  Peptide Therapeutics}, Accessed: 2021-10-12\relax
\mciteBstWouldAddEndPuncttrue
\mciteSetBstMidEndSepPunct{\mcitedefaultmidpunct}
{\mcitedefaultendpunct}{\mcitedefaultseppunct}\relax
\EndOfBibitem
\bibitem[Beyer()]{CQC_JSR}
M.~Beyer, \emph{CrownBio and JSR Life Sciences Partner with Cambridge Quantum
  Computing to Leverage Quantum Machine Learning for Novel Cancer Treatment
  Biomarker Discovery}, Accessed: 2021-10-12\relax
\mciteBstWouldAddEndPuncttrue
\mciteSetBstMidEndSepPunct{\mcitedefaultmidpunct}
{\mcitedefaultendpunct}{\mcitedefaultseppunct}\relax
\EndOfBibitem
\bibitem[Zhang \emph{et~al.}(2017)Zhang, Bengio, Hardt, Recht, and
  Vinyals]{zhang2017understanding}
C.~Zhang, S.~Bengio, M.~Hardt, B.~Recht and O.~Vinyals, \emph{Understanding
  deep learning requires rethinking generalization}, 2017\relax
\mciteBstWouldAddEndPuncttrue
\mciteSetBstMidEndSepPunct{\mcitedefaultmidpunct}
{\mcitedefaultendpunct}{\mcitedefaultseppunct}\relax
\EndOfBibitem
\bibitem[Hopfield(1982)]{Hopfield2554}
J.~J. Hopfield, \emph{Proceedings of the National Academy of Sciences}, 1982,
  \textbf{79}, 2554--2558\relax
\mciteBstWouldAddEndPuncttrue
\mciteSetBstMidEndSepPunct{\mcitedefaultmidpunct}
{\mcitedefaultendpunct}{\mcitedefaultseppunct}\relax
\EndOfBibitem
\bibitem[Valiant(1984)]{10.1145/1968.1972}
L.~G. Valiant, \emph{Commun. ACM}, 1984, \textbf{27}, 1134–1142\relax
\mciteBstWouldAddEndPuncttrue
\mciteSetBstMidEndSepPunct{\mcitedefaultmidpunct}
{\mcitedefaultendpunct}{\mcitedefaultseppunct}\relax
\EndOfBibitem
\bibitem[Gardner(1987)]{1987_gardner}
E.~Gardner, 1987, \textbf{4}, 481--485\relax
\mciteBstWouldAddEndPuncttrue
\mciteSetBstMidEndSepPunct{\mcitedefaultmidpunct}
{\mcitedefaultendpunct}{\mcitedefaultseppunct}\relax
\EndOfBibitem
\bibitem[Gardner(1988)]{1988_gardner}
E.~Gardner, 1988, \textbf{21}, 257--270\relax
\mciteBstWouldAddEndPuncttrue
\mciteSetBstMidEndSepPunct{\mcitedefaultmidpunct}
{\mcitedefaultendpunct}{\mcitedefaultseppunct}\relax
\EndOfBibitem
\bibitem[Gy\"orgyi(1990)]{PhysRevA.41.7097}
G.~Gy\"orgyi, \emph{Phys. Rev. A}, 1990, \textbf{41}, 7097--7100\relax
\mciteBstWouldAddEndPuncttrue
\mciteSetBstMidEndSepPunct{\mcitedefaultmidpunct}
{\mcitedefaultendpunct}{\mcitedefaultseppunct}\relax
\EndOfBibitem
\bibitem[Barkai \emph{et~al.}(1990)Barkai, Hansel, and
  Kanter]{144PhysRevLett.65.2312}
E.~Barkai, D.~Hansel and I.~Kanter, \emph{Phys. Rev. Lett.}, 1990, \textbf{65},
  2312--2315\relax
\mciteBstWouldAddEndPuncttrue
\mciteSetBstMidEndSepPunct{\mcitedefaultmidpunct}
{\mcitedefaultendpunct}{\mcitedefaultseppunct}\relax
\EndOfBibitem
\bibitem[Barkai \emph{et~al.}(1992)Barkai, Hansel, and
  Sompolinsky]{145PhysRevA.45.4146}
E.~Barkai, D.~Hansel and H.~Sompolinsky, \emph{Phys. Rev. A}, 1992,
  \textbf{45}, 4146--4161\relax
\mciteBstWouldAddEndPuncttrue
\mciteSetBstMidEndSepPunct{\mcitedefaultmidpunct}
{\mcitedefaultendpunct}{\mcitedefaultseppunct}\relax
\EndOfBibitem
\bibitem[Engel \emph{et~al.}(1992)Engel, K\"ohler, Tschepke, Vollmayr, and
  Zippelius]{146PhysRevA.45.7590}
A.~Engel, H.~M. K\"ohler, F.~Tschepke, H.~Vollmayr and A.~Zippelius,
  \emph{Phys. Rev. A}, 1992, \textbf{45}, 7590--7609\relax
\mciteBstWouldAddEndPuncttrue
\mciteSetBstMidEndSepPunct{\mcitedefaultmidpunct}
{\mcitedefaultendpunct}{\mcitedefaultseppunct}\relax
\EndOfBibitem
\bibitem[Morone \emph{et~al.}(2014)Morone, Caltagirone, Harrison, and
  Parisi]{morone2014replica}
F.~Morone, F.~Caltagirone, E.~Harrison and G.~Parisi, \emph{Replica Theory and
  Spin Glasses}, 2014\relax
\mciteBstWouldAddEndPuncttrue
\mciteSetBstMidEndSepPunct{\mcitedefaultmidpunct}
{\mcitedefaultendpunct}{\mcitedefaultseppunct}\relax
\EndOfBibitem
\bibitem[Zdeborov{\'{a}} and Krzakala(2016)]{Zdeborova2016}
L.~Zdeborov{\'{a}} and F.~Krzakala, \emph{Advances in Physics}, 2016,
  \textbf{65}, 453--552\relax
\mciteBstWouldAddEndPuncttrue
\mciteSetBstMidEndSepPunct{\mcitedefaultmidpunct}
{\mcitedefaultendpunct}{\mcitedefaultseppunct}\relax
\EndOfBibitem
\bibitem[Decelle \emph{et~al.}(2011)Decelle, Krzakala, Moore, and
  Zdeborová]{Decelle_2011}
A.~Decelle, F.~Krzakala, C.~Moore and L.~Zdeborová, \emph{Physical Review E},
  2011, \textbf{84}, 066106\relax
\mciteBstWouldAddEndPuncttrue
\mciteSetBstMidEndSepPunct{\mcitedefaultmidpunct}
{\mcitedefaultendpunct}{\mcitedefaultseppunct}\relax
\EndOfBibitem
\bibitem[Yedidia \emph{et~al.}(2003)Yedidia, Freeman,
  Weiss,\emph{et~al.}]{Yedidia2003UnderstandingBP}
J.~S. Yedidia, W.~T. Freeman, Y.~Weiss \emph{et~al.}, \emph{Exploring
  artificial intelligence in the new millennium}, 2003, \textbf{8},
  0018--9448\relax
\mciteBstWouldAddEndPuncttrue
\mciteSetBstMidEndSepPunct{\mcitedefaultmidpunct}
{\mcitedefaultendpunct}{\mcitedefaultseppunct}\relax
\EndOfBibitem
\bibitem[Carleo \emph{et~al.}(2019)Carleo, Cirac, Cranmer, Daudet, Schuld,
  Tishby, Vogt-Maranto, and Zdeborová]{Carleo_2019}
G.~Carleo, I.~Cirac, K.~Cranmer, L.~Daudet, M.~Schuld, N.~Tishby,
  L.~Vogt-Maranto and L.~Zdeborová, \emph{Reviews of Modern Physics}, 2019,
  \textbf{91}, 045002\relax
\mciteBstWouldAddEndPuncttrue
\mciteSetBstMidEndSepPunct{\mcitedefaultmidpunct}
{\mcitedefaultendpunct}{\mcitedefaultseppunct}\relax
\EndOfBibitem
\bibitem[Feng \emph{et~al.}(2021)Feng, Venkataramanan, Rush, and
  Samworth]{feng2021unifying}
O.~Y. Feng, R.~Venkataramanan, C.~Rush and R.~J. Samworth, \emph{A unifying
  tutorial on Approximate Message Passing}, 2021\relax
\mciteBstWouldAddEndPuncttrue
\mciteSetBstMidEndSepPunct{\mcitedefaultmidpunct}
{\mcitedefaultendpunct}{\mcitedefaultseppunct}\relax
\EndOfBibitem
\bibitem[Lin \emph{et~al.}(2017)Lin, Tegmark, and Rolnick]{Lin_2017}
H.~W. Lin, M.~Tegmark and D.~Rolnick, \emph{Journal of Statistical Physics},
  2017, \textbf{168}, 1223–1247\relax
\mciteBstWouldAddEndPuncttrue
\mciteSetBstMidEndSepPunct{\mcitedefaultmidpunct}
{\mcitedefaultendpunct}{\mcitedefaultseppunct}\relax
\EndOfBibitem
\bibitem[de~Almeida and Thouless(1978)]{1978_stability}
J.~R.~L. de~Almeida and D.~J. Thouless, 1978, \textbf{11}, 983--990\relax
\mciteBstWouldAddEndPuncttrue
\mciteSetBstMidEndSepPunct{\mcitedefaultmidpunct}
{\mcitedefaultendpunct}{\mcitedefaultseppunct}\relax
\EndOfBibitem
\bibitem[M{\'{e}}zard \emph{et~al.}(1986)M{\'{e}}zard, Parisi, and
  Virasoro]{1986_spinglass}
M.~M{\'{e}}zard, G.~Parisi and M.~A. Virasoro, 1986, \textbf{1}, 77--82\relax
\mciteBstWouldAddEndPuncttrue
\mciteSetBstMidEndSepPunct{\mcitedefaultmidpunct}
{\mcitedefaultendpunct}{\mcitedefaultseppunct}\relax
\EndOfBibitem
\bibitem[Wilson(1971)]{PhysRevB.4.3174}
K.~G. Wilson, \emph{Phys. Rev. B}, 1971, \textbf{4}, 3174--3183\relax
\mciteBstWouldAddEndPuncttrue
\mciteSetBstMidEndSepPunct{\mcitedefaultmidpunct}
{\mcitedefaultendpunct}{\mcitedefaultseppunct}\relax
\EndOfBibitem
\bibitem[Gell-Mann and Low(1954)]{PhysRev.95.1300}
M.~Gell-Mann and F.~E. Low, \emph{Phys. Rev.}, 1954, \textbf{95},
  1300--1312\relax
\mciteBstWouldAddEndPuncttrue
\mciteSetBstMidEndSepPunct{\mcitedefaultmidpunct}
{\mcitedefaultendpunct}{\mcitedefaultseppunct}\relax
\EndOfBibitem
\bibitem[Kadanoff(1966)]{PhysicsPhysiqueFizika.2.263}
L.~P. Kadanoff, \emph{Physics Physique Fizika}, 1966, \textbf{2},
  263--272\relax
\mciteBstWouldAddEndPuncttrue
\mciteSetBstMidEndSepPunct{\mcitedefaultmidpunct}
{\mcitedefaultendpunct}{\mcitedefaultseppunct}\relax
\EndOfBibitem
\bibitem[Wilson(1971)]{wilson}
K.~G. Wilson, \emph{Phys. Rev. B}, 1971, \textbf{4}, 3174--3183\relax
\mciteBstWouldAddEndPuncttrue
\mciteSetBstMidEndSepPunct{\mcitedefaultmidpunct}
{\mcitedefaultendpunct}{\mcitedefaultseppunct}\relax
\EndOfBibitem
\bibitem[Wilson(1975)]{RevModPhys.47.773}
K.~G. Wilson, \emph{Rev. Mod. Phys.}, 1975, \textbf{47}, 773--840\relax
\mciteBstWouldAddEndPuncttrue
\mciteSetBstMidEndSepPunct{\mcitedefaultmidpunct}
{\mcitedefaultendpunct}{\mcitedefaultseppunct}\relax
\EndOfBibitem
\bibitem[Mehta and Schwab(2014)]{mehta2014exact}
P.~Mehta and D.~J. Schwab, \emph{An exact mapping between the Variational
  Renormalization Group and Deep Learning}, 2014\relax
\mciteBstWouldAddEndPuncttrue
\mciteSetBstMidEndSepPunct{\mcitedefaultmidpunct}
{\mcitedefaultendpunct}{\mcitedefaultseppunct}\relax
\EndOfBibitem
\bibitem[Koch-Janusz and Ringel(2018)]{Koch_Janusz_2018}
M.~Koch-Janusz and Z.~Ringel, \emph{Nature Physics}, 2018, \textbf{14},
  578–582\relax
\mciteBstWouldAddEndPuncttrue
\mciteSetBstMidEndSepPunct{\mcitedefaultmidpunct}
{\mcitedefaultendpunct}{\mcitedefaultseppunct}\relax
\EndOfBibitem
\bibitem[Apenko(2011)]{apenko2011information}
S.~M. Apenko, \emph{Information theory and renormalization group flows},
  2011\relax
\mciteBstWouldAddEndPuncttrue
\mciteSetBstMidEndSepPunct{\mcitedefaultmidpunct}
{\mcitedefaultendpunct}{\mcitedefaultseppunct}\relax
\EndOfBibitem
\bibitem[Sim \emph{et~al.}(2019)Sim, Johnson, and
  Aspuru‐Guzik]{expressibility}
S.~Sim, P.~D. Johnson and A.~Aspuru‐Guzik, \emph{Advanced Quantum
  Technologies}, 2019, \textbf{2}, 1900070\relax
\mciteBstWouldAddEndPuncttrue
\mciteSetBstMidEndSepPunct{\mcitedefaultmidpunct}
{\mcitedefaultendpunct}{\mcitedefaultseppunct}\relax
\EndOfBibitem
\bibitem[Collins and Śniady(2006)]{Haar}
B.~Collins and P.~Śniady, \emph{Communications in Mathematical Physics}, 2006,
  \textbf{264}, 773–795\relax
\mciteBstWouldAddEndPuncttrue
\mciteSetBstMidEndSepPunct{\mcitedefaultmidpunct}
{\mcitedefaultendpunct}{\mcitedefaultseppunct}\relax
\EndOfBibitem
\bibitem[Ambainis and Emerson(2007)]{tdesign}
A.~Ambainis and J.~Emerson, \emph{Quantum t-designs: t-wise independence in the
  quantum world}, 2007\relax
\mciteBstWouldAddEndPuncttrue
\mciteSetBstMidEndSepPunct{\mcitedefaultmidpunct}
{\mcitedefaultendpunct}{\mcitedefaultseppunct}\relax
\EndOfBibitem
\bibitem[Hubregtsen \emph{et~al.}(2021)Hubregtsen, Pichlmeier, Stecher, and
  Bertels]{hubregtsen2021evaluation}
T.~Hubregtsen, J.~Pichlmeier, P.~Stecher and K.~Bertels, \emph{Quantum Machine
  Intelligence}, 2021, \textbf{3}, 1--19\relax
\mciteBstWouldAddEndPuncttrue
\mciteSetBstMidEndSepPunct{\mcitedefaultmidpunct}
{\mcitedefaultendpunct}{\mcitedefaultseppunct}\relax
\EndOfBibitem
\bibitem[Du \emph{et~al.}(2021)Du, Hsieh, Liu, You, and
  Tao]{PRXQuantum.2.040337}
Y.~Du, M.-H. Hsieh, T.~Liu, S.~You and D.~Tao, \emph{PRX Quantum}, 2021,
  \textbf{2}, 040337\relax
\mciteBstWouldAddEndPuncttrue
\mciteSetBstMidEndSepPunct{\mcitedefaultmidpunct}
{\mcitedefaultendpunct}{\mcitedefaultseppunct}\relax
\EndOfBibitem
\bibitem[Cheng \emph{et~al.}(2015)Cheng, Hsieh, and Yeh]{cheng2015learnability}
H.-C. Cheng, M.-H. Hsieh and P.-C. Yeh, \emph{arXiv preprint arXiv:1501.00559},
  2015\relax
\mciteBstWouldAddEndPuncttrue
\mciteSetBstMidEndSepPunct{\mcitedefaultmidpunct}
{\mcitedefaultendpunct}{\mcitedefaultseppunct}\relax
\EndOfBibitem
\bibitem[Banchi \emph{et~al.}(2021)Banchi, Pereira, and
  Pirandola]{banchi2021generalization}
L.~Banchi, J.~Pereira and S.~Pirandola, \emph{PRX Quantum}, 2021, \textbf{2},
  040321\relax
\mciteBstWouldAddEndPuncttrue
\mciteSetBstMidEndSepPunct{\mcitedefaultmidpunct}
{\mcitedefaultendpunct}{\mcitedefaultseppunct}\relax
\EndOfBibitem
\bibitem[McClean \emph{et~al.}(2018)McClean, Boixo, Smelyanskiy, Babbush, and
  Neven]{McClean}
J.~R. McClean, S.~Boixo, V.~N. Smelyanskiy, R.~Babbush and H.~Neven,
  \emph{Nature Communications}, 2018, \textbf{9}, year\relax
\mciteBstWouldAddEndPuncttrue
\mciteSetBstMidEndSepPunct{\mcitedefaultmidpunct}
{\mcitedefaultendpunct}{\mcitedefaultseppunct}\relax
\EndOfBibitem
\bibitem[Harrow and Low(2009)]{2design}
A.~W. Harrow and R.~A. Low, \emph{Communications in Mathematical Physics},
  2009, \textbf{291}, 257–302\relax
\mciteBstWouldAddEndPuncttrue
\mciteSetBstMidEndSepPunct{\mcitedefaultmidpunct}
{\mcitedefaultendpunct}{\mcitedefaultseppunct}\relax
\EndOfBibitem
\bibitem[Cerezo \emph{et~al.}(2021)Cerezo, Sone, Volkoff, Cincio, and
  Coles]{globalcost}
M.~Cerezo, A.~Sone, T.~Volkoff, L.~Cincio and P.~J. Coles, \emph{Nature
  Communications}, 2021, \textbf{12}, year\relax
\mciteBstWouldAddEndPuncttrue
\mciteSetBstMidEndSepPunct{\mcitedefaultmidpunct}
{\mcitedefaultendpunct}{\mcitedefaultseppunct}\relax
\EndOfBibitem
\bibitem[Larocca \emph{et~al.}(2021)Larocca, Czarnik, Sharma, Muraleedharan,
  Coles, and Cerezo]{barrendiag}
M.~Larocca, P.~Czarnik, K.~Sharma, G.~Muraleedharan, P.~J. Coles and M.~Cerezo,
  \emph{arXiv preprint arXiv:2105.14377}, 2021\relax
\mciteBstWouldAddEndPuncttrue
\mciteSetBstMidEndSepPunct{\mcitedefaultmidpunct}
{\mcitedefaultendpunct}{\mcitedefaultseppunct}\relax
\EndOfBibitem
\bibitem[Wang \emph{et~al.}(2021)Wang, Fontana, Cerezo, Sharma, Sone, Cincio,
  and Coles]{noiseinduced}
S.~Wang, E.~Fontana, M.~Cerezo, K.~Sharma, A.~Sone, L.~Cincio and P.~J. Coles,
  \emph{Nature Communications}, 2021, \textbf{12}, year\relax
\mciteBstWouldAddEndPuncttrue
\mciteSetBstMidEndSepPunct{\mcitedefaultmidpunct}
{\mcitedefaultendpunct}{\mcitedefaultseppunct}\relax
\EndOfBibitem
\bibitem[Romero and Aspuru-Guzik(2021)]{romero2021variational}
J.~Romero and A.~Aspuru-Guzik, \emph{Advanced Quantum Technologies}, 2021,
  \textbf{4}, 2000003\relax
\mciteBstWouldAddEndPuncttrue
\mciteSetBstMidEndSepPunct{\mcitedefaultmidpunct}
{\mcitedefaultendpunct}{\mcitedefaultseppunct}\relax
\EndOfBibitem
\bibitem[Khalid \emph{et~al.}(2022)Khalid, Sureshbabu, Banerjee, and
  Kais]{bilal2021}
B.~Khalid, S.~H. Sureshbabu, A.~Banerjee and S.~Kais, \emph{arXiv preprint
  arXiv:2202.00112}, 2022\relax
\mciteBstWouldAddEndPuncttrue
\mciteSetBstMidEndSepPunct{\mcitedefaultmidpunct}
{\mcitedefaultendpunct}{\mcitedefaultseppunct}\relax
\EndOfBibitem
\bibitem[Erdmenger \emph{et~al.}(2021)Erdmenger, Grosvenor, and
  Jefferson]{erdmenger2021quantifying}
J.~Erdmenger, K.~T. Grosvenor and R.~Jefferson, \emph{Towards quantifying
  information flows: relative entropy in deep neural networks and the
  renormalization group}, 2021\relax
\mciteBstWouldAddEndPuncttrue
\mciteSetBstMidEndSepPunct{\mcitedefaultmidpunct}
{\mcitedefaultendpunct}{\mcitedefaultseppunct}\relax
\EndOfBibitem
\bibitem[Tang(2021)]{tang2021quantum}
E.~Tang, \emph{Physical Review Letters}, 2021, \textbf{127}, 060503\relax
\mciteBstWouldAddEndPuncttrue
\mciteSetBstMidEndSepPunct{\mcitedefaultmidpunct}
{\mcitedefaultendpunct}{\mcitedefaultseppunct}\relax
\EndOfBibitem
\bibitem[Tang(2019)]{tang2019quantum}
E.~Tang, Proceedings of the 51st Annual ACM SIGACT Symposium on Theory of
  Computing, 2019, pp. 217--228\relax
\mciteBstWouldAddEndPuncttrue
\mciteSetBstMidEndSepPunct{\mcitedefaultmidpunct}
{\mcitedefaultendpunct}{\mcitedefaultseppunct}\relax
\EndOfBibitem
\bibitem[Dou{\c{c}}ot and Ioffe(2012)]{Dou_ot_2012}
B.~Dou{\c{c}}ot and L.~B. Ioffe, \emph{Reports on Progress in Physics}, 2012,
  \textbf{75}, 072001\relax
\mciteBstWouldAddEndPuncttrue
\mciteSetBstMidEndSepPunct{\mcitedefaultmidpunct}
{\mcitedefaultendpunct}{\mcitedefaultseppunct}\relax
\EndOfBibitem
\bibitem[Fowler \emph{et~al.}(2012)Fowler, Mariantoni, Martinis, and
  Cleland]{PhysRevA.86.032324}
A.~G. Fowler, M.~Mariantoni, J.~M. Martinis and A.~N. Cleland, \emph{Phys. Rev.
  A}, 2012, \textbf{86}, 032324\relax
\mciteBstWouldAddEndPuncttrue
\mciteSetBstMidEndSepPunct{\mcitedefaultmidpunct}
{\mcitedefaultendpunct}{\mcitedefaultseppunct}\relax
\EndOfBibitem
\bibitem[Yao \emph{et~al.}(2012)Yao, Wang, Chen, Gao, Fowler, Raussendorf,
  Chen, Liu, Lu, Deng,\emph{et~al.}]{yao2012experimental}
X.-C. Yao, T.-X. Wang, H.-Z. Chen, W.-B. Gao, A.~G. Fowler, R.~Raussendorf,
  Z.-B. Chen, N.-L. Liu, C.-Y. Lu, Y.-J. Deng \emph{et~al.}, \emph{Nature},
  2012, \textbf{482}, 489--494\relax
\mciteBstWouldAddEndPuncttrue
\mciteSetBstMidEndSepPunct{\mcitedefaultmidpunct}
{\mcitedefaultendpunct}{\mcitedefaultseppunct}\relax
\EndOfBibitem
\bibitem[Egan \emph{et~al.}(2020)Egan, Debroy, Noel, Risinger, Zhu, Biswas,
  Newman, Li, Brown, Cetina,\emph{et~al.}]{egan2020fault}
L.~Egan, D.~M. Debroy, C.~Noel, A.~Risinger, D.~Zhu, D.~Biswas, M.~Newman,
  M.~Li, K.~R. Brown, M.~Cetina \emph{et~al.}, \emph{arXiv preprint
  arXiv:2009.11482}, 2020\relax
\mciteBstWouldAddEndPuncttrue
\mciteSetBstMidEndSepPunct{\mcitedefaultmidpunct}
{\mcitedefaultendpunct}{\mcitedefaultseppunct}\relax
\EndOfBibitem
\bibitem[Aharonov and Ben-Or(2008)]{aharonov2008fault}
D.~Aharonov and M.~Ben-Or, \emph{SIAM Journal on Computing}, 2008\relax
\mciteBstWouldAddEndPuncttrue
\mciteSetBstMidEndSepPunct{\mcitedefaultmidpunct}
{\mcitedefaultendpunct}{\mcitedefaultseppunct}\relax
\EndOfBibitem
\bibitem[Chen \emph{et~al.}(2021)Chen, Satzinger, Atalaya, Korotkov, Dunsworth,
  Sank, Quintana, McEwen, Barends, Klimov,\emph{et~al.}]{chen2021exponential}
Z.~Chen, K.~J. Satzinger, J.~Atalaya, A.~N. Korotkov, A.~Dunsworth, D.~Sank,
  C.~Quintana, M.~McEwen, R.~Barends, P.~V. Klimov \emph{et~al.},
  \emph{Nature}, 2021, \textbf{595}, 383--387\relax
\mciteBstWouldAddEndPuncttrue
\mciteSetBstMidEndSepPunct{\mcitedefaultmidpunct}
{\mcitedefaultendpunct}{\mcitedefaultseppunct}\relax
\EndOfBibitem
\bibitem[Parra-Rodriguez \emph{et~al.}(2020)Parra-Rodriguez, Lougovski, Lamata,
  Solano, and Sanz]{parra2020digital}
A.~Parra-Rodriguez, P.~Lougovski, L.~Lamata, E.~Solano and M.~Sanz,
  \emph{Physical Review A}, 2020, \textbf{101}, 022305\relax
\mciteBstWouldAddEndPuncttrue
\mciteSetBstMidEndSepPunct{\mcitedefaultmidpunct}
{\mcitedefaultendpunct}{\mcitedefaultseppunct}\relax
\EndOfBibitem
\bibitem[Eisert \emph{et~al.}(2020)Eisert, Hangleiter, Walk, Roth, Markham,
  Parekh, Chabaud, and Kashefi]{eisert2020quantum}
J.~Eisert, D.~Hangleiter, N.~Walk, I.~Roth, D.~Markham, R.~Parekh, U.~Chabaud
  and E.~Kashefi, \emph{Nature Reviews Physics}, 2020, \textbf{2},
  382--390\relax
\mciteBstWouldAddEndPuncttrue
\mciteSetBstMidEndSepPunct{\mcitedefaultmidpunct}
{\mcitedefaultendpunct}{\mcitedefaultseppunct}\relax
\EndOfBibitem
\bibitem[Schreck \emph{et~al.}(2019)Schreck, Coley, and
  Bishop]{schreck2019learning}
J.~S. Schreck, C.~W. Coley and K.~J. Bishop, \emph{ACS central science}, 2019,
  \textbf{5}, 970--981\relax
\mciteBstWouldAddEndPuncttrue
\mciteSetBstMidEndSepPunct{\mcitedefaultmidpunct}
{\mcitedefaultendpunct}{\mcitedefaultseppunct}\relax
\EndOfBibitem
\bibitem[Baylon \emph{et~al.}(2019)Baylon, Cilfone, Gulcher, and
  Chittenden]{baylon2019enhancing}
J.~L. Baylon, N.~A. Cilfone, J.~R. Gulcher and T.~W. Chittenden, \emph{Journal
  of chemical information and modeling}, 2019, \textbf{59}, 673--688\relax
\mciteBstWouldAddEndPuncttrue
\mciteSetBstMidEndSepPunct{\mcitedefaultmidpunct}
{\mcitedefaultendpunct}{\mcitedefaultseppunct}\relax
\EndOfBibitem
\bibitem[Dai \emph{et~al.}(2020)Dai, Li, Coley, Dai, and
  Song]{dai2020retrosynthesis}
H.~Dai, C.~Li, C.~W. Coley, B.~Dai and L.~Song, \emph{arXiv preprint
  arXiv:2001.01408}, 2020\relax
\mciteBstWouldAddEndPuncttrue
\mciteSetBstMidEndSepPunct{\mcitedefaultmidpunct}
{\mcitedefaultendpunct}{\mcitedefaultseppunct}\relax
\EndOfBibitem
\bibitem[Lin \emph{et~al.}(2019)Lin, Xu, Pei, and Lai]{lin2019automatic}
K.~Lin, Y.~Xu, J.~Pei and L.~Lai, \emph{Automatic Retrosynthetic Pathway
  Planning Using Template-free Models}, 2019\relax
\mciteBstWouldAddEndPuncttrue
\mciteSetBstMidEndSepPunct{\mcitedefaultmidpunct}
{\mcitedefaultendpunct}{\mcitedefaultseppunct}\relax
\EndOfBibitem
\bibitem[Liu \emph{et~al.}(2017)Liu, Ramsundar, Kawthekar, Shi, Gomes,
  Luu~Nguyen, Ho, Sloane, Wender, and Pande]{liu2017retrosynthetic}
B.~Liu, B.~Ramsundar, P.~Kawthekar, J.~Shi, J.~Gomes, Q.~Luu~Nguyen, S.~Ho,
  J.~Sloane, P.~Wender and V.~Pande, \emph{ACS central science}, 2017,
  \textbf{3}, 1103--1113\relax
\mciteBstWouldAddEndPuncttrue
\mciteSetBstMidEndSepPunct{\mcitedefaultmidpunct}
{\mcitedefaultendpunct}{\mcitedefaultseppunct}\relax
\EndOfBibitem
\bibitem[Johansson \emph{et~al.}(2019)Johansson, Thakkar, Kogej, Bjerrum,
  Genheden, Bastys, Kannas, Schliep, Chen, and Engkvist]{JOHANSSON201965}
S.~Johansson, A.~Thakkar, T.~Kogej, E.~Bjerrum, S.~Genheden, T.~Bastys,
  C.~Kannas, A.~Schliep, H.~Chen and O.~Engkvist, \emph{Drug Discovery Today:
  Technologies}, 2019, \textbf{32-33}, 65--72\relax
\mciteBstWouldAddEndPuncttrue
\mciteSetBstMidEndSepPunct{\mcitedefaultmidpunct}
{\mcitedefaultendpunct}{\mcitedefaultseppunct}\relax
\EndOfBibitem
\bibitem[Breuer \emph{et~al.}(2002)Breuer,
  Petruccione,\emph{et~al.}]{breuer2002theory}
H.-P. Breuer, F.~Petruccione \emph{et~al.}, \emph{The theory of open quantum
  systems}, Oxford University Press on Demand, 2002\relax
\mciteBstWouldAddEndPuncttrue
\mciteSetBstMidEndSepPunct{\mcitedefaultmidpunct}
{\mcitedefaultendpunct}{\mcitedefaultseppunct}\relax
\EndOfBibitem
\bibitem[Hu \emph{et~al.}(2020)Hu, Xia, and Kais]{Hu2020AQA}
Z.~Hu, R.~Xia and S.~Kais, \emph{Scientific Reports}, 2020, \textbf{10},
  year\relax
\mciteBstWouldAddEndPuncttrue
\mciteSetBstMidEndSepPunct{\mcitedefaultmidpunct}
{\mcitedefaultendpunct}{\mcitedefaultseppunct}\relax
\EndOfBibitem
\bibitem[Lin \emph{et~al.}(2021)Lin, Peng, Gu, and Lan]{lin2021simulation}
K.~Lin, J.~Peng, F.~L. Gu and Z.~Lan, \emph{The Journal of Physical Chemistry
  Letters}, 2021, \textbf{12}, 10225--10234\relax
\mciteBstWouldAddEndPuncttrue
\mciteSetBstMidEndSepPunct{\mcitedefaultmidpunct}
{\mcitedefaultendpunct}{\mcitedefaultseppunct}\relax
\EndOfBibitem
\bibitem[Herrera~Rodriguez and Kananenka(2021)]{herrera2021convolutional}
L.~E. Herrera~Rodriguez and A.~A. Kananenka, \emph{The Journal of Physical
  Chemistry Letters}, 2021, \textbf{12}, 2476--2483\relax
\mciteBstWouldAddEndPuncttrue
\mciteSetBstMidEndSepPunct{\mcitedefaultmidpunct}
{\mcitedefaultendpunct}{\mcitedefaultseppunct}\relax
\EndOfBibitem
\bibitem[Mazza \emph{et~al.}(2021)Mazza, Zietlow, Carollo, Andergassen,
  Martius, and Lesanovsky]{PhysRevResearch.3.023084}
P.~P. Mazza, D.~Zietlow, F.~Carollo, S.~Andergassen, G.~Martius and
  I.~Lesanovsky, \emph{Phys. Rev. Research}, 2021, \textbf{3}, 023084\relax
\mciteBstWouldAddEndPuncttrue
\mciteSetBstMidEndSepPunct{\mcitedefaultmidpunct}
{\mcitedefaultendpunct}{\mcitedefaultseppunct}\relax
\EndOfBibitem
\bibitem[Luchnikov \emph{et~al.}(2020)Luchnikov, Vintskevich, Grigoriev, and
  Filippov]{luchnikov2020machine}
I.~Luchnikov, S.~Vintskevich, D.~Grigoriev and S.~Filippov, \emph{Physical
  review letters}, 2020, \textbf{124}, 140502\relax
\mciteBstWouldAddEndPuncttrue
\mciteSetBstMidEndSepPunct{\mcitedefaultmidpunct}
{\mcitedefaultendpunct}{\mcitedefaultseppunct}\relax
\EndOfBibitem
\bibitem[H{\"a}se \emph{et~al.}(2017)H{\"a}se, Kreisbeck, and
  Aspuru-Guzik]{hase2017machine}
F.~H{\"a}se, C.~Kreisbeck and A.~Aspuru-Guzik, \emph{Chemical science}, 2017,
  \textbf{8}, 8419--8426\relax
\mciteBstWouldAddEndPuncttrue
\mciteSetBstMidEndSepPunct{\mcitedefaultmidpunct}
{\mcitedefaultendpunct}{\mcitedefaultseppunct}\relax
\EndOfBibitem
\bibitem[Lee \emph{et~al.}(2021)Lee, Patil, Zhang, and
  Hsieh]{PhysRevResearch.3.023095}
C.~K. Lee, P.~Patil, S.~Zhang and C.~Y. Hsieh, \emph{Phys. Rev. Research},
  2021, \textbf{3}, 023095\relax
\mciteBstWouldAddEndPuncttrue
\mciteSetBstMidEndSepPunct{\mcitedefaultmidpunct}
{\mcitedefaultendpunct}{\mcitedefaultseppunct}\relax
\EndOfBibitem
\bibitem[Khan \emph{et~al.}(2020)Khan, Huerta, and Das]{KHAN2020135628}
A.~Khan, E.~Huerta and A.~Das, \emph{Physics Letters B}, 2020, \textbf{808},
  135628\relax
\mciteBstWouldAddEndPuncttrue
\mciteSetBstMidEndSepPunct{\mcitedefaultmidpunct}
{\mcitedefaultendpunct}{\mcitedefaultseppunct}\relax
\EndOfBibitem
\bibitem[Villmann \emph{et~al.}(2020)Villmann, Engelsberger, Ravichandran,
  Villmann, and Kaden]{villmann2020quantum}
T.~Villmann, A.~Engelsberger, J.~Ravichandran, A.~Villmann and M.~Kaden,
  \emph{Neural Computing and Applications}, 2020,  1--10\relax
\mciteBstWouldAddEndPuncttrue
\mciteSetBstMidEndSepPunct{\mcitedefaultmidpunct}
{\mcitedefaultendpunct}{\mcitedefaultseppunct}\relax
\EndOfBibitem
\bibitem[Bellinger \emph{et~al.}(2020)Bellinger, Coles, Crowley, and
  Tamblyn]{bellinger2020reinforcement}
C.~Bellinger, R.~Coles, M.~Crowley and I.~Tamblyn, \emph{Reinforcement Learning
  in a Physics-Inspired Semi-Markov Environment}, 2020\relax
\mciteBstWouldAddEndPuncttrue
\mciteSetBstMidEndSepPunct{\mcitedefaultmidpunct}
{\mcitedefaultendpunct}{\mcitedefaultseppunct}\relax
\EndOfBibitem
\bibitem[Trenti \emph{et~al.}(2020)Trenti, Sestini, Gianelle, Zuliani, Felser,
  Lucchesi, and Montangero]{trenti2020quantum}
M.~Trenti, L.~Sestini, A.~Gianelle, D.~Zuliani, T.~Felser, D.~Lucchesi and
  S.~Montangero, \emph{arXiv preprint arXiv:2004.13747}, 2020\relax
\mciteBstWouldAddEndPuncttrue
\mciteSetBstMidEndSepPunct{\mcitedefaultmidpunct}
{\mcitedefaultendpunct}{\mcitedefaultseppunct}\relax
\EndOfBibitem
\bibitem[Tiwari and Melucci(2019)]{tiwari2019towards}
P.~Tiwari and M.~Melucci, \emph{IEEE Access}, 2019, \textbf{7},
  42354--42372\relax
\mciteBstWouldAddEndPuncttrue
\mciteSetBstMidEndSepPunct{\mcitedefaultmidpunct}
{\mcitedefaultendpunct}{\mcitedefaultseppunct}\relax
\EndOfBibitem
\bibitem[Musil \emph{et~al.}(2021)Musil, Grisafi, Bart{\'o}k, Ortner,
  Cs{\'a}nyi, and Ceriotti]{musil2021physics}
F.~Musil, A.~Grisafi, A.~P. Bart{\'o}k, C.~Ortner, G.~Cs{\'a}nyi and
  M.~Ceriotti, \emph{Chemical Reviews}, 2021, \textbf{121}, 9759--9815\relax
\mciteBstWouldAddEndPuncttrue
\mciteSetBstMidEndSepPunct{\mcitedefaultmidpunct}
{\mcitedefaultendpunct}{\mcitedefaultseppunct}\relax
\EndOfBibitem
\bibitem[Karniadakis \emph{et~al.}(2021)Karniadakis, Kevrekidis, Lu,
  Perdikaris, Wang, and Yang]{karniadakis2021physics}
G.~E. Karniadakis, I.~G. Kevrekidis, L.~Lu, P.~Perdikaris, S.~Wang and L.~Yang,
  \emph{Nature Reviews Physics}, 2021, \textbf{3}, 422--440\relax
\mciteBstWouldAddEndPuncttrue
\mciteSetBstMidEndSepPunct{\mcitedefaultmidpunct}
{\mcitedefaultendpunct}{\mcitedefaultseppunct}\relax
\EndOfBibitem
\bibitem[Chen \emph{et~al.}(2021)Chen, Satzinger, Atalaya, Korotkov, Dunsworth,
  Sank, Quintana, McEwen, Barends, Klimov, Hong, Jones, Petukhov, Kafri,
  Demura, Burkett, Gidney, Fowler, Paler, and Kelly]{bitflip_suppress}
Z.~Chen, K.~Satzinger, J.~Atalaya, A.~Korotkov, A.~Dunsworth, D.~Sank,
  C.~Quintana, M.~McEwen, R.~Barends, P.~Klimov, S.~Hong, C.~Jones,
  A.~Petukhov, D.~Kafri, S.~Demura, B.~Burkett, C.~Gidney, A.~Fowler, A.~Paler
  and J.~Kelly, \emph{Nature}, 2021, \textbf{595}, 383--387\relax
\mciteBstWouldAddEndPuncttrue
\mciteSetBstMidEndSepPunct{\mcitedefaultmidpunct}
{\mcitedefaultendpunct}{\mcitedefaultseppunct}\relax
\EndOfBibitem
\bibitem[IBM()]{IBM_roadmap}
\emph{IBM’s roadmap for scaling quantum technology},
  \url{https://research.ibm.com/blog/ibm-quantum-roadmap}, Accessed:
  2021-10-12\relax
\mciteBstWouldAddEndPuncttrue
\mciteSetBstMidEndSepPunct{\mcitedefaultmidpunct}
{\mcitedefaultendpunct}{\mcitedefaultseppunct}\relax
\EndOfBibitem
\end{mcitethebibliography}

\providecommand*{\mcitethebibliography}{\thebibliography}
\csname @ifundefined\endcsname{endmcitethebibliography}
{\let\endmcitethebibliography\endthebibliography}{}

\bibliographystyle{unsrt} 

\end{document}